%% file: thesis.tex
\documentclass[a4paper,12pt,twoside,openright,headings=standardclasses,numbers=noenddot]{scrbook}

\usepackage[%includehead, %
 top=3.5cm, %
 bottom=3.5cm, %
 left=2.5cm, %
 right=2.5cm]{geometry}

\usepackage{scrhack}  % otherwise listings or float give warning "float@addtolists implementation became (scrbook) deprecated" and more
\usepackage[automark,headsepline,plainheadsepline]{scrlayer-scrpage} % for fancy headers
\usepackage[utf8]{inputenc}
\usepackage[ngerman,english]{babel}
\usepackage{csquotes}
\usepackage{graphicx}
\usepackage{amsmath,amsfonts,amssymb}
\usepackage[colorlinks=false]{hyperref}
\usepackage{braket}
\usepackage{color}
\usepackage{enumerate}
\usepackage{qcircuit}
\usepackage[normalem]{ulem}
\usepackage{url}
\usepackage{dsfont} % bbm is a bitmap and prints badly!
\usepackage{breqn}
\usepackage{float}
\usepackage[toc,page]{appendix}
\usepackage{booktabs}
\usepackage[format=plain]{caption}
\usepackage[caption=false,font=footnotesize,labelformat=simple]{subfig}
\usepackage{qcircuit}
\usepackage{rotating} % for sidewayfigure
\usepackage{blkarray}
\usepackage{listings}
\usepackage{tabularx}
\usepackage{siunitx} % to make \micro seconds work
\usepackage{import}
\usepackage{enumitem}
\usepackage{multirow}

\pdfpageattr {/Group << /S /Transparency /I true /CS /DeviceRGB>>} % fix the transparency acroread-9 simulate overprinting issue caused by these figures with /Transparency (GST chapter 6)
\addtokomafont{disposition}{\rmfamily} % nice font for headings
\newcolumntype{Y}{>{\centering\arraybackslash}X}  % for centered extended column in tabularx
\usepackage[style=alphabetic,maxnames=100,maxalphanames=1,alldates=short,giveninits=true,backref=true,hyperref=true,
url=false,doi=false,isbn=false,eprint=false,date=year,backend=biber]{biblatex}
\DeclareNolabel{ % to replace the spaces in [De 87]
  \nolabel{\regexp{[\p{Z}\p{P}\p{S}\p{C}]+}}
}
\renewcommand*{\labelalphaothers}{} % remove the + sign for more than 1 author
\DeclareLabelalphaTemplate{ % for a 4-digit year
  \labelelement{
    \field[final]{shorthand}
    \field{label}
    \field[strwidth=3,strside=left,ifnames=1]{labelname}
    \field[strwidth=1,strside=left]{labelname}
  }
  \labelelement{
    \field{year}
  }
}
\def\bibinitdelim#1\bibinitperiod{} % remove all middle names (needs giveninits=true)
\DeclareBibliographyDriver{article}{%
  \usebibmacro{bibindex}%
  \usebibmacro{begentry}%
  \usebibmacro{author/translator+others}%
  \setunit{\addcomma\space}\newblock%
  \usebibmacro{title}
  \setunit{\addcomma\space}\newblock%
  \usebibmacro{journal}%
  \setunit*{\addspace}%
  \printfield[bold]{volume}%
  \setunit{\addcomma\linebreak[1]\addspace}% % NOTE: needed to do this "drastic solution" for the Pal2017 linebreak
  \printfield{pages}%
  \setunit*{\addspace}%
  \usebibmacro{articledate}%
  \setunit{\bibpagerefpunct}\newblock%
  \usebibmacro{pageref}%
  \newunit\newblock%
  \usebibmacro{finentry}%
}
\DeclareBibliographyDriver{unpublished}{%
  \usebibmacro{bibindex}%
  \usebibmacro{begentry}%
  \usebibmacro{author}%
  \setunit{\addcomma\space}\newblock%
  \usebibmacro{title}
  \setunit{\addcomma\space}\newblock%
  \iffieldundef{eprinttype}
    {\printtext{arXiv}\setunit{\addcolon\linebreak[3]}}%
    {\printtext{\thefield{eprinttype}}\setunit{\addcolon\linebreak[3]}}%
  \iffieldequalstr{eprint}{1811.04909}{\setunit{\addcolon\linebreak[4]}}{} % for Gilyen
  \printtext{\thefield{eprint}}%
  \setunit*{\addspace}%
  \usebibmacro{articledate}
  \setunit{\bibpagerefpunct}\newblock%
  \usebibmacro{pageref}%
  \newunit\newblock%
  \usebibmacro{finentry}%
}
\DeclareBibliographyDriver{book}{%
  \usebibmacro{bibindex}%
  \usebibmacro{begentry}%
  \usebibmacro{author/editor+others/translator+others}%
  \setunit{\addcomma\space}\newblock%
  \usebibmacro{maintitle+title}
  \setunit{\addcomma\space}\newblock%
  \printlist{publisher}%
  \setunit*{\addcomma\space}%
  \usebibmacro{date}%
  \setunit{\bibpagerefpunct}\newblock%
  \usebibmacro{pageref}%
  \newunit\newblock%
  \usebibmacro{finentry}%
}
\DeclareBibliographyDriver{thesis}{% %also for PhDThesis and MastersThesis
  \usebibmacro{bibindex}%
  \usebibmacro{begentry}%
  \usebibmacro{author}%
  \setunit{\addcomma\space}\newblock%
  \usebibmacro{title}%
  \setunit{\addcomma\space}\newblock%
  \printfield{type}%
  \setunit{\addcomma\space}\newblock%
  \usebibmacro{institution+location+date}%}%
  \iffieldundef{addendum}
    {}%
    {\setunit{\addcomma\space}%
    \printfield{addendum}}% % for master thesis url
  \setunit{\bibpagerefpunct}\newblock
  \usebibmacro{pageref}%
  \newunit\newblock%
  \usebibmacro{finentry}
}
\DeclareBibliographyDriver{online}{% also for electronic (ibm q)
  \usebibmacro{bibindex}%
  \usebibmacro{begentry}%
  \usebibmacro{author}%
  \setunit{\addcomma\space}\newblock%
  \usebibmacro{title}%
  \setunit{\addcomma\space}\newblock%
  \usebibmacro{date}%
  \setunit{\addcomma\space}\newblock%
  \usebibmacro{url}%
  \setunit{\bibpagerefpunct}\newblock
  \usebibmacro{pageref}%
  \newunit\newblock
  \usebibmacro{finentry}
}
\DeclareBibliographyDriver{inproceedings}{%
  \usebibmacro{bibindex}%
  \usebibmacro{begentry}%
  \usebibmacro{author/translator+others}%
  \setunit{\addcomma\space}\newblock%
  \usebibmacro{title}%
  \setunit{\addcomma\space}\newblock%
  \printfield{booktitle}%
  \iffieldundef{series}%
    {%
      \iffieldundef{volume}%
      {}%
      {\setunit{\addcomma\space}%
      \printtext{Vol.}
      \printfield[bold]{volume}}%
    }%
    {%
      \setunit{\addcomma\space}%
      \printfield{series}%
      \iffieldundef{volume}%
      {}%
      {%
        \setunit*{\addspace}%
        \printfield[bold]{volume}%
      }%
    }%
  \iffieldundef{pages}
    {}%
    {\setunit{\addcomma\space}%
    \printfield{pages}}%
  \ifnameundef{editor}
    {}%
    {\setunit{\addcomma\space}%
    \usebibmacro{byeditor+others}}%
  \setunit*{\addspace}%
  \usebibmacro{articledate}%
  \setunit{\bibpagerefpunct}\newblock%
  \usebibmacro{pageref}%
  \newunit\newblock%
  \usebibmacro{finentry}%
}
\DeclareBibliographyDriver{incollection}{%
  \usebibmacro{bibindex}%
  \usebibmacro{begentry}%
  \usebibmacro{author/translator+others}%
  \setunit{\addcomma\space}\newblock%
  \usebibmacro{title}%
  \setunit{\addcomma\space}\newblock%
  \printtext{in:}%
  \setunit*{\addspace}%
  \printfield{booktitle}%
  \iffieldundef{series}%
    {%
      \iffieldundef{volume}%
      {}%
      {\setunit{\addcomma\space}%
      \printfield[bold]{volume}}%
    }%
    {%
      \setunit{\addcomma\space}%
      \printfield{series}%
      \iffieldundef{volume}%
      {}%
      {%
        \setunit*{\addspace}%
        \printfield[bold]{volume}%
      }%
    }%
  \iffieldundef{pages}
    {}%
    {\setunit{\addcomma\space}%
    \printtext{pp.}
    \printfield{pages}}%
  \ifnameundef{editor}
    {}%
    {\setunit{\addcomma\space}%
    \usebibmacro{byeditor+others}}%
  \setunit*{\addspace}%
  \usebibmacro{articledate}%
  \setunit{\bibpagerefpunct}\newblock%
  \usebibmacro{pageref}%
  \newunit\newblock%
  \usebibmacro{finentry}%
}
\DeclareBibliographyDriver{inbook}{%
  \usebibmacro{bibindex}%
  \usebibmacro{begentry}%
  \usebibmacro{author/translator+others}%
  \setunit{\addcomma\space}\newblock%
  \usebibmacro{title}%
  \setunit{\addcomma\space}\newblock%
  \printtext{in:}%
  \setunit*{\addspace}%
  \printfield{booktitle}%
  \setunit{\addcomma\space}\newblock%
  \printlist{publisher}%
  \setunit*{\addcomma\space}%
  \usebibmacro{date}%
  \setunit{\bibpagerefpunct}\newblock%
  \usebibmacro{pageref}%
  \newunit\newblock%
  \usebibmacro{finentry}%
}
\DeclareBibliographyDriver{report}{% %also techreport
  \usebibmacro{bibindex}%
  \usebibmacro{begentry}%
  \usebibmacro{author/translator+others}%
  \setunit{\addcomma\space}\newblock%
  \usebibmacro{title}%
  \setunit{\addcomma\space}\newblock%
  \printfield{booktitle}%
  \iffieldundef{series}%
    {%
      \iffieldundef{volume}%
      {}%
      {\setunit{\addcomma\space}%
      \printfield[bold]{volume}}%
    }%
    {%
      \setunit{\addcomma\space}%
      \printfield{series}%
      \iffieldundef{volume}%
      {}%
      {%
        \setunit*{\addspace}%
        \printfield[bold]{volume}%
      }%
    }%
  \iffieldundef{addendum}
    {}%
    {\setunit{\addcomma\space}%
    \printfield{addendum}}% % for "report to the ..."
  \iffieldundef{pages}
    {}%
    {\setunit{\addcomma\space}%
    \printtext{pp.}
    \printfield{pages}}%
  \ifnameundef{editor}
    {}%
    {\setunit{\addcomma\space}%
    \usebibmacro{byeditor+others}}%
  \setunit*{\addspace}%
  \usebibmacro{articledate}%
  \setunit{\bibpagerefpunct}\newblock%
  \usebibmacro{pageref}%
  \newunit\newblock%
  \usebibmacro{finentry}%
}
\DeclareBibliographyDriver{misc}{% %for "personal correspondence" and such
  \usebibmacro{bibindex}%
  \usebibmacro{begentry}%
  \usebibmacro{author/editor+others/translator+others}%
  \setunit{\addcomma\space}\newblock%
  \printfield{howpublished}% put "in preparation" or "personal correspondence" here
  \setunit{\addcomma\space}\newblock%
  \usebibmacro{organization+location+date}%
  \setunit{\bibpagerefpunct}\newblock
  \usebibmacro{pageref}%
  \newunit\newblock
  \usebibmacro{finentry}%
}
\newbibmacro{string+doiurlisbn}[1]{%
  \iffieldundef{doi}{%
    \iffieldundef{url}{%
      \iffieldundef{isbn}{%
        \iffieldundef{issn}{%
          \iffieldundef{eprint}{%
            {#1}%
          }{\href{https://arxiv.org/abs/\thefield{eprint}}{#1}}%
        }{\href{https://books.google.com/books?vid=ISSN\thefield{issn}}{#1}}%
      }{\href{https://books.google.com/books?vid=ISBN\thefield{isbn}}{#1}}%
    }{\href{\thefield{url}}{#1}}%
  }{\href{https://dx.doi.org/\thefield{doi}}{#1}}%
}
\DeclareFieldFormat{linked}{\usebibmacro{string+doiurlisbn}{#1}}
\renewbibmacro*{date}{\printtext[linked]{\printdate}}
\newbibmacro*{articledate}{\printtext[parens]{\printtext[linked]{\printdate}}}
\renewcommand{\bibpagerefpunct}{\addspace}
\renewbibmacro*{pageref}{%
  \iflistundef{pageref}
    {}
    {\printtext[brackets]{%<--- here we had parens before
       \ifnumgreater{\value{pageref}}{1}
         {\bibstring{backrefpages}\ppspace}
         {\bibstring{backrefpage}\ppspace}%
       \printlist[pageref][-\value{listtotal}]{pageref}}}
}
\renewbibmacro{in:}{} % remove the 'In:'
\DefineBibliographyStrings{english}{ % remove the 'p.' (inproceedings it is added if necessary)
  page             = {},
}
\renewbibmacro*{author}{%
  \ifboolexpr{
    test \ifuseauthor
    and
    not test {\ifnameundef{author}}
  }
    {\textsc{\printnames{author}}%
     \iffieldundef{authortype}
       {}
       {\setunit{\printdelim{authortypedelim}}%
        \usebibmacro{authorstrg}}}
    {}}
\DeclareFieldFormat{pages}{\mkfirstpage[{\mkpageprefix[bookpagination]}]{#1}} % only start page
\DeclareFieldFormat[inproceedings]{series}{\mkbibemph{#1}}
\DeclareFieldFormat{url}{\url{#1}} % for electronic (ibm q)

\newcommand{\appref}[1]{Appendix~\ref{#1}}
\newcommand{\chapref}[1]{Chapter~\ref{#1}}
\newcommand{\secref}[1]{Section~\ref{#1}}
\newcommand{\secaref}[2]{Sections~\ref{#1} and \ref{#2}}
\newcommand{\ssecref}[1]{Section~\ref{#1}}
\newcommand{\figref}[1]{Fig.~\ref{#1}}
\newcommand{\figaref}[2]{Figs.~\ref{#1} and \ref{#2}}
\newcommand{\figsref}[2]{Figs.\ \ref{#1}--\ref{#2}}
\newcommand{\sfigref}[1]{Figure~\ref{#1}}
\newcommand{\tabref}[1]{Tab.~\ref{#1}}
\newcommand{\stabref}[1]{Table~\ref{#1}}
\newcommand{\equref}[1]{Eq.~\eqref{#1}}
\newcommand{\equaref}[2]{Eqs.~\eqref{#1} and \eqref{#2}}
\newcommand{\equsref}[2]{Eqs.\ \eqref{#1}--\eqref{#2}}
\newcommand{\sequref}[1]{Equation~\eqref{#1}}
\providecommand{\abs}[1]{\lvert#1\rvert}
\providecommand{\norm}[1]{\lVert#1\rVert}
\providecommand{\e}[1]{\ensuremath{\times 10^{#1}}}
\newcommand{\ketbra}[2]{\ket{#1}\!\!\bra{#2}}
\newcommand{\sket}[1]{\vert #1 \rangle\!\rangle}
\newcommand{\sbra}[1]{\langle\!\langle #1 \vert}
\newcommand{\sbraket}[2]{\langle\!\langle #1 \vert #2 \rangle\!\rangle}

\newcommand{\matindex}[1]{\mbox{\scriptsize\ensuremath{#1}}}
\newcommand{\overarrow}[2]{%
  \overset{\makebox[0pt]{\begin{tabular}{@{}c@{}}\ensuremath{#2}\\[0pt]\ensuremath{\uparrow}\end{tabular}}}{\ensuremath{#1}}
}

\newlength\stextwidth
\newcommand\makesamesize[3][c]{\settowidth{\stextwidth}{\ensuremath{#2}}\makebox[\stextwidth][#1]{\ensuremath{#3}}}

\floatstyle{ruled}
\newfloat{lstfloat}{htbp}{lop}[chapter]
\floatname{lstfloat}{Listing}
\lstdefinestyle{mycodestyle} {
  language=C++,
  tabsize=4,
  numbers=left,
  basicstyle=\scriptsize\ttfamily,
  identifierstyle=\color{black},
  stringstyle=\color{orange},
  numberstyle=\scriptsize\color[rgb]{0.6,0.6,0.6},
  commentstyle=\color[rgb]{0,0.6,0},
  keywordstyle=\bfseries\color{blue},
  morekeywords={uint64_t,pragma,omp},
  breaklines=true,
  showtabs=false,
  showstringspaces=false,
  aboveskip=0pt,
  belowskip=0pt,
}
\lstdefinestyle{myconfigstyle} {
  language=python,
  tabsize=4,
  numbers=left,
  basicstyle=\scriptsize\ttfamily,
  identifierstyle=\color{black},
  stringstyle=\color{orange},
  numberstyle=\scriptsize\color[rgb]{0.6,0.6,0.6},
  commentstyle=\color[rgb]{0,0.6,0},
  keywordstyle=\bfseries\color{blue},
  morekeywords={uint64_t,pragma,omp},
  breaklines=true,
  showtabs=false,
  showstringspaces=false,
  aboveskip=0pt,
  belowskip=0pt,
}
\lstdefinestyle{fullcircuitstyle} {
  language=C,
  literate={0}{{\textcolor{red}{0}}}{1}%
           {1}{{\textcolor{red}{1}}}{1}%
           {2}{{\textcolor{red}{2}}}{1}%
           {3}{{\textcolor{red}{3}}}{1}%
           {4}{{\textcolor{red}{4}}}{1}%
           {5}{{\textcolor{red}{5}}}{1}%
           {6}{{\textcolor{red}{6}}}{1}%
           {7}{{\textcolor{red}{7}}}{1}%
           {8}{{\textcolor{red}{8}}}{1}%
           {9}{{\textcolor{red}{9}}}{1}
           {X1}{{\textcolor{blue}{X1}}}{2}%
           {X2}{{\textcolor{blue}{X2}}}{2}%
           {Z1}{{\textcolor{blue}{Z1}}}{2}%
           {Z2}{{\textcolor{blue}{Z2}}}{2}%
           {CZ}{{\textcolor{blue}{CZ}}}{2}%
           {HHS}{{\textcolor{blue}{HHS}}}{3},
  basicstyle=\tiny\ttfamily,
  identifierstyle=\color{black},
  breaklines=false,
  showtabs=false,
  showstringspaces=false,
  aboveskip=0pt,
  belowskip=0pt,
}

\begin{document}

\frontmatter
\include{titleafter}

\include{abstract}

\tableofcontents
\mainmatter
\include{chap1}

\include{chap2}

\include{chap3}

\include{chap4}

\include{chap5}

\include{chap6}

\include{chap7}

\include{conclusion}

\appendix
\include{app}

\backmatter
\include{bib}

\include{publications}

\include{declaration}

\include{ack}

\end{document}

%% file: titleafter.tex
\raggedbottom
\pagenumbering{Roman} % roman is done automatically by \frontmatter but we want capitals
\thispagestyle{empty}
\begin{titlepage}
  \centering
	\vspace*{\baselineskip}
  \rule{\textwidth}{1.6pt}\vspace*{-\baselineskip}\vspace*{2pt}
  \rule{\textwidth}{0.4pt}\\[\baselineskip]
	{\huge\textsc{Supercomputer simulations of transmon quantum computers} \par}
	\vspace*{\baselineskip}
  \rule{\textwidth}{0.4pt}\vspace*{-\baselineskip}\vspace*{3.2pt}
  \rule{\textwidth}{1.6pt}\\[\baselineskip]
	\vspace{2cm}
	{Von der Fakultät für Mathematik, Informatik und Naturwissenschaften der\par
  RWTH Aachen University zur Erlangung des akademischen Grades eines\par
  Doktors der Naturwissenschaften genehmigte Dissertation\par}
	\vspace{1cm}
	vorgelegt von\par
  \vspace{2cm}
  {\textsc{Dennis Willsch, M.Sc.}}\par
	\vspace{1cm}
	aus\par
  \vspace{1cm}
  {\textsc{Köln}}\par
  \vspace{2cm}
  \begin{flushleft}
    \begin{tabular}{@{}ll}
      Berichter: & Prof.~Dr.~Kristel Michielsen\\
                 & Prof.~Dr.~David DiVincenzo\\
    \end{tabular}\par
    \vspace{1cm}
    Tag der mündlichen Prüfung: 7.~Juli 2020\par
    \vspace{1cm}
    Diese Dissertation ist auf den Internetseiten der Universitätsbibliothek verfügbar.
  \end{flushleft}
	\vfill
\end{titlepage}
%\pagenumbering{arabic}

%% file: abstract.tex
\chapter*{Abstract}

We develop a simulator for quantum computers composed of superconducting
transmon qubits. The simulation model supports an arbitrary number of transmons
and resonators. Quantum gates are implemented by time-dependent pulses.
Nontrivial effects such as crosstalk, leakage to non-computational states,
entanglement between transmons  and resonators, and control errors due to the
pulses are inherently included.

The time evolution of the quantum computer is obtained by solving the
time-dependent Schr\"odinger equation.  The simulation algorithm shows excellent
scalability on high-performance supercomputers. We present results for the
simulation of up to 16 transmons and resonators. Additionally, the model can
be used to simulate environments, and we demonstrate the transition from an
isolated system to an open quantum system governed by a Lindblad master
equation. We also describe a procedure to extract model parameters from
electromagnetic simulations or experiments.

We compare simulation results to experiments on several NISQ processors of the
IBM Q Experience. We find nearly perfect agreement between simulation and
experiment for quantum circuits designed to probe crosstalk in transmon systems.
By studying common gate metrics such as the fidelity or the diamond distance, we
find that they cannot reliably predict the performance of repeated gate
applications or practical quantum algorithms. As an alternative, we find that
the results from two-transmon gate set tomography have an exceptional predictive
power. Finally, we test a protocol from the theory of quantum error correction
and fault tolerance. We find that the protocol systematically improves the
performance of transmon quantum computers in the presence of characteristic
control and measurement errors.

\chapter*{Zusammenfassung}

\begin{otherlanguage}{ngerman}

Wir entwickeln einen Simulator für Quantencomputer, die aus supraleitenden
Transmon-Qubits bestehen. Das Simulationsmodell unterstützt eine beliebige
Anzahl von Transmons und Resonatoren. Quantengatter werden durch zeitabhängige
Pulse realisiert. Nicht-triviale Effekte wie Crosstalk, Verlust in nicht
rechnerische Zustände, Verschränkung zwischen Transmons und Resonatoren sowie
Steuerungsfehler verursacht durch die Pulse sind automatisch miteinbezogen.

Die Zeitentwicklung des Quantencomputers wird durch Lösung der zeitabhängigen
Schrödingergleichung bestimmt.  Der Simulationsalgorithmus zeigt
ausgezeichnete Skalierbarkeit auf Hochleistungs-Supercomputern. Wir präsentieren
Ergebnisse für die Simulation von bis zu 16 Transmons und Resonatoren.
Zusätzlich kann das Modell zur Simulation von Umgebungen verwendet werden.
Wir demonstrieren den Übergang von einem isolierten System zu einem offenen
Quantensystem, das von einer Lindblad-Mastergleichung bestimmt wird. Wir
beschreiben außerdem ein Verfahren zur Extraktion von Modellparametern aus
elektromagnetischen Simulationen oder Experimenten.

Wir vergleichen Simulationsergebnisse mit Experimenten auf mehreren
NISQ-Pro\-zes\-so\-ren der IBM Q Experience. Wir finden eine nahezu perfekte
Übereinstimmung zwischen Simulation und Experiment für Quantenschaltungen zur
Untersuchung von Cross\-talk in Transmon-Systemen. Durch Untersuchung gängiger
Gatter-Metriken wie der Fidelity oder der Diamant-Distanz finden wir, dass sie
die Leistung von wiederholten Gatteranwendungen oder praktischen
Quantenalgorithmen nicht zuverlässig vorhersagen können. Als Alternative finden
wir, dass die Ergebnisse einer Zwei-Transmon-Gattermengen\-to\-mo\-gra\-phie
eine außergewöhnlich gute Vorhersagekraft aufweisen. Zum Schluss testen wir ein
Protokoll aus der Theorie der Quantenfehlerkorrektur und Fehlertoleranz. Wir
stellen fest, dass das Protokoll systematisch die Leistung von
Transmon-Quantencomputern bei charakteristischen Steuerungs- und Messfehlern
verbessert.

\end{otherlanguage}

%% file: chap1.tex
\chapter{Introduction}
\label{cha:introduction}

For over a century, humans have designed and built digital computing machines.
The initial ideas can be traced back to the mid-1800s \cite{Babbage1837,
boole1847logic}, but the actual construction started less than a century ago. In
1936, Zuse designed a floating point general-purpose computer \cite{Zuse1936}
that led to the first programmable floating point machine in 1941, the Z3
\cite{ModernHistoryOfComputing}. Turing formalized the \emph{universal computing
machine} \cite{Turing1937ComputingMachine} that influenced the construction of
the Colossus in 1943 \cite{Randell1973OriginsDigitalComputer}, which was used to
perform Boolean operations for cryptanalysis. Other computers of that time were
the ABC \cite{Atanasoff1940} built in 1942 and the ENIAC built in 1945
\cite{Randell1973OriginsDigitalComputer}.

Most of these early computers were based on vacuum tubes which made them large
and unreliable. Universal digital computing only became \emph{scalable} after
the vacuum tubes were replaced by semiconductor devices such as transistors.
Nowadays, computers are ubiquitous in everyday life; every mobile phone contains
a general-purpose digital computer, and large-scale high-performance
supercomputers are used routinely to solve some of the most difficult
computational problems.

Similarly, for about a century, humans have developed and studied quantum
theory. This physical theory has extraordinary descriptive power, also in
numerous fields beyond physics \cite{Khrennikov2010UbiquitousQuantumStructure}.
The predictions of quantum theory are fundamentally stochastic, meaning that
quantum theory can only predict probabilities for observable events
\cite{ballentine1998quantum}. In this sense, quantum theory is inherently linked
to probability theory \cite{jaynes2003probability}. The mathematical framework
of quantum theory is based on linear algebra and can be reduced to a few axioms
\cite{vonneumann1955, ballentine1998quantum, NielsenChuang}.

\subsubsection{Quantum computing}

The essential idea of a \emph{quantum computer} is to combine these two
concepts, i.e., the universal computing machine and quantum theory. The goal is
to build a computing machine that implements the equations of quantum theory for
two-level systems. The elementary two-level systems of digital computers, the
bits, are replaced by quantum bits, commonly known as \emph{qubits}. As quantum
theory only predicts probabilities, a program for a quantum computer basically
determines a set of probabilities for the qubits.

An actual device, however, does not produce probabilities but individual bits.
Therefore, a program is typically repeated multiple times to make the connection
to the predicted probabilities. This is a common principle of all types of
quantum computing: From the user perspective, a quantum computer produces a
large ensemble of individual results.

Two basic approaches to quantum computing are currently pursued by science and
industry \cite{NationalAcademyOfSciences2019QuantumComputing}. The first is
called the \emph{gate-based quantum computer}. Inspired by the gate model of
digital computing, programs for a gate-based quantum computer are specified in
terms of elementary \emph{quantum gates} \cite{NielsenChuang}.  As every
algorithm can be decomposed into a sequence of such elementary gates, the
gate-based quantum computer is considered universal
\cite{barenco1995u2andCNOTareuniversal,
Deutsch95universality, divincenzo1995twoqubitgates}.

The other approach to quantum computing is \emph{quantum annealing}
\cite{Finnila1994QuantumAnnealing, KadowakiNishimori1998QuantumAnnealing,
Fahri2000AdiabaticQuantumComputation, Harris2010DWave, Johnson2011DWave}.
Quantum annealers have turned out to be very useful for quickly producing  an
ensemble of close-to-optimal solutions to a given optimization or  machine
learning problem \cite{Pudenz2012QML,
PerdomoOrtiz2019ReadinessQuantumOptimizationMethodsPUBOQUBO,
Orus2019QuantumComputingForFinance, Willsch2020QSVM}.

Over the past decades, research in gate-based quantum computing has evolved from
an abstract, mathematical model of a computing machine
\cite{Benioff1980QuantumTuringMachine, Deutsch85QuantumComputer} to a broad
range of experimental devices. All of these pursue the idea of implementing the
mathematical framework of quantum theory to gain a computational advantage over
the mature technology of digital computers \cite{ekert1996quantumalgorithms}.
The strongest advantage is envisioned as an exponential speedup for a special
set of mathematical problems, such as (1) the simulation of quantum mechanical
systems which is believed to significantly aid in research and development
\cite{Feynman1982Simulating,
Babbush2018QuantumSimulationOfMaterialsElectronicStructureTheory,
Kuehn2019QuantumChemistryBASF}, (2) the approximate solution of sparse linear
systems in logarithmic time \cite{Harro2009HHLalgorithm} (albeit with a few
caveats \cite{Aaronson2015fineprint}), or (3) the polynomial-time factorization
of integers \cite{shor94factoring, shor1997algorithm} which might form a
potential threat to the security of widely-used asymmetric cryptosystems such as
RSA \cite{RSA1978RSA}.

Currently, the two most advanced technologies for gate-based quantum computers
use superconducting circuits \cite{Raimond2001CavityCircuitQEDQuantumComputing,
Vion2002CPBqubitsQuantronium, blais2004circuitqed,
Wendin2017SuperconductingReview} and trapped ions
\cite{Cirac1995TrappedIonQuantumComputer, Monroe1995DemoTrappedIonQC}. IBM and
several other companies have made small superconducting quantum processors
available to the community to explore the technology
\cite{ibmquantumexperience2016, rigetti2017computing, china2018quantumcomputing,
dwave2019leap, LaRose2019OverviewGateLevelQuantumSoftware}. Additionally, a
superconducting quantum processor manufactured by Google has produced results
for a well-defined class of problems that are beyond the reach of digital
supercomputers \cite{Google2019QuantumSupremacy}, thereby achieving
\emph{quantum supremacy} \cite{Preskill2012quantumsupremacy,
Boixo2018quantumsupremacy}.

\subsubsection{Objectives}

Despite the recent progress, it is still an open question if a universal, fully
error-corrected quantum computer can be built. All current quantum processors
belong to the class of \emph{noisy intermediate-scale quantum} (NISQ) devices
\cite{Preskill2018NISQ, NationalAcademyOfSciences2019QuantumComputing}. And
although a fully error-corrected device is possible in theory
\cite{Shor1996FaultTolerantQC, Aharonov1997ThresholdTheoremOriginalSTOC,
aliferis2007FTQCwithLeakage}, there is a priori no guarantee that it can also
be built in practice. The essential questions are: How close do current NISQ
devices come to the ideal, mathematical qubit model of a quantum computer?
What are the main errors and limitations that would need to be overcome?

To address these questions, we carry out detailed supercomputer simulations of
current NISQ devices. Additionally, we perform experiments on such devices to
compare simulation results with experimental observations. We identify and
analyze the main limitations, i.e.,  \emph{leakage} and \emph{crosstalk}, and
study to what extent the induced errors can be corrected.

\subsubsection{Supercomputer simulations}

The aim of this thesis is to utilize the power of digital supercomputers to
study the emerging technology of quantum computers. Supercomputer simulations
are vital for the development and verification of quantum computers. For
instance, massively parallel simulators such as the J\"ulich universal quantum
computer simulator (JUQCS) \cite{DeRaedt2007MassivelyParallel,
DeRaedt2018MassivelyParallel, Willsch2020BenchmarkingWithJUQCS} have been essential
to verify Google's demonstration of quantum supremacy
\cite{Google2019QuantumSupremacy}.

In this work, we develop a simulator of superconducting quantum
processors.  We focus on  gate-based quantum computers with superconducting
transmon qubits \cite{koch2007transmon} because of the tremendous progress that
has been reported recently for the transmon architecture
\cite{ibmquantumexperience2016, rigetti2017computing, china2018quantumcomputing,
Google2019QuantumSupremacy}. We devise a scalable method to simulate transmon
systems with an arbitrary number of qubits and couplers, limited only by the
available amount of physical memory and computing time. The simulation model
includes the effects of higher transmon levels and generic time-dependent
pulses. All model and pulse parameters are freely configurable.

\subsubsection{Simulation method}

The simulator solves the time-dependent Schr\"odinger equation (TDSE) with
$\hbar=1$,
\begin{align}
  i \frac{\partial}{\partial t} \ket{\Psi(t)} = H(t) \ket{\Psi(t)},
  \label{eq:tdse1}
\end{align}
where $H(t)$ is a generic, time-dependent model Hamiltonian representing the
hardware of the quantum processor, including the transmon system and their
electromagnetic environment.

From the solution $\ket{\Psi(t)}$, we can compute any physically relevant
quantity of the system such as reduced density matrices with non-unitary dynamics
describing the actual qubits. In this sense, the TDSE approach can be related to
other common approaches based on master equations, perturbative studies, and
completely positive trace-preserving maps. We study each of these relations in
the course of this thesis.

The time dependence of $H(t)$ in \equref{eq:tdse1} represents external microwave
control pulses that are applied to the system. Each pulse is a time-dependent
voltage signal designed to implement a certain quantum gate. We use an
optimization procedure to find optimal pulse parameters for each simulated
transmon system.

The simulator is based on the Suzuki-Trotter product-formula algorithm
\cite{deraedt1987productformula, DeRaedt2000QCE, deraedt2004computational}. This
allows the TDSE given by \equref{eq:tdse1} to be solved on a sub-picosecond scale
for time evolutions over several hundred microseconds, without making additional
approximations. We formulate the TDSE in an appropriate basis that makes its
solution amenable to large-scale supercomputer simulations. Most of the
simulations presented in this thesis were performed on the supercomputers JURECA
\cite{JURECA} and JUWELS \cite{JUWELS}.

\subsubsection{Outline}

In this thesis, we present results from the simulation of transmon systems with
up to 16 transmons and couplers. As real NISQ devices of the same size and
architecture are publicly accessible on the IBM Q Experience
\cite{ibmquantumexperience2016}, we also perform some of the experiments on
these quantum processors. This offers a great opportunity to relate the
simulation results directly to experiments. We find that the main limitations
revealed by the simulation, i.e., leakage and crosstalk due to additional
transmon and resonator states, capture most of the errors observed in the
corresponding NISQ devices.
\\\\
This thesis is organized as follows. \chapref{cha:quantumcomputing} reviews the
mathematical model of a gate-based quantum computer, i.e., qubits, gates,
circuits, quantum operations, and leakage.
In \chapref{cha:simulation}, we
define the supercomputer simulation method. After specifying the full model
Hamiltonian, we describe in detail the numerical algorithm to solve the TDSE
given by \equref{eq:tdse1}. We then define the primary model systems. Finally,
we present a method to obtain model parameters for the simulation of
electromagnetic environments.

\chapref{cha:freeevolution} focuses on free,
undriven time evolutions. This includes accuracy and performance benchmarks.
Additionally, we relate our simulation approach to perturbative results and
simulations of a Lindblad master equation.

In \chapref{cha:optimization}, we define the elementary single-qubit and
two-qubit pulses used to implement quantum gates. We describe the optimization
procedure used to find optimal pulse parameters and present optimization results
for the larger transmon systems. Finally, we discuss the compilation process
used to translate quantum circuits to pulse information for the
simulator.

In \chapref{cha:gateerrors}, we characterize the optimized quantum
gates in detail. After introducing the most prominent gate metrics, we study
repeated gate applications on both the simulated systems and experimental
devices. The chapter concludes with an application of gate set tomography (GST)
and an assessment of its predictive power.

\chapref{cha:fullcircuitsimulations}
combines the results from the previous chapters and applies them to a selected
class of quantum circuits, executed using both the transmon simulator and
experimental processors. We first design and implement a class of quantum
circuits to study crosstalk in transmon systems. Secondly, we study quantum
circuits designed to characterize the singlet state. Finally, we implement a
full protocol from the theory of quantum fault tolerance to assess its potential
to improve quantum computation in transmon architectures.

\chapref{sec:conclusion} contains our conclusions and an outlook on many
interesting paths to continue the present work.
Implementation details and
separate mathematical proofs are given in
Appendices~\ref{app:visualization}--\ref{app:gatedecomposition}.
Some results
presented in this thesis have previously been published in
\cite{Willsch2017GateErrorAnalysis, Willsch2018TestingFaultTolerance}.

%% file: chap2.tex
\chapter{Ideal gate-based quantum computing}
\label{cha:quantumcomputing}

In this chapter, we review the computational architecture of an ideal gate-based
quantum computer and related concepts as formally defined in the literature
\cite{yanofsky2008quantumcomputing, NielsenChuang, watrous2018theoryofQI}. We
start by introducing the quantum bit as the fundamental unit of computation in
\secref{sec:qubits}. In \secref{sec:quantumgates}, we define quantum gates as
the basic operations that can be performed on a qubit. A combination of these
operations is called a quantum circuit, which is introduced in
\secref{sec:quantumcircuits}. Qubits, quantum gates, and quantum circuits are
the basic building blocks that are required to define algorithms for a
gate-based quantum computer. Finally, in \secref{sec:quantumoperations}, we
introduce the concept of quantum operations, which are used in a more general
description of gate-based quantum computers in terms of mixed states.

Although it is still unclear when a large universal gate-based quantum computer
can be built or if the envisioned exponential speedup can be delivered,  the
mathematical model of an ideal quantum computer is an interesting model to
study. Advances in quantum algorithms with a  theoretical speedup can inspire
valuable discoveries in other areas. For instance, a quantum algorithm for
recommendation systems (i.e., systems that are supposed to provide product
suggestions to users based on past purchases) is known to provide an exponential
speedup over previous classical algorithms
\cite{kerenidis2016quantumrecommendationsystems}. Recently, this discovery has
led to the development of novel, similarly efficient algorithms for digital
computers \cite{TangAaronson2018RecommendationSystems}. Further examples of such
\emph{quantum-inspired} algorithms are given in
\cite{Tang2018QuantumInspiredAlg2, Gilyen2018QuantumInspiredAlg3,
Chia2018QuantumInspiredAlg4}.

\section{Quantum bits}
\label{sec:qubits}

A \emph{quantum bit} or \emph{qubit} is the generalization of a digital bit to
the mathematical framework of quantum theory. The goal of this section is to
understand the precise meaning of \emph{generalization} in this context. This
helps to understand why quantum computers have the theoretical potential to be
more powerful than digital computers. We approach the concept of generalization
by first defining the digital bit, and then highlighting the mathematical
difference to the qubit.

\subsection{Single qubits}
\label{sec:singlequbits}

A (digital) bit $j$ is the fundamental unit of computation in every digital
computer. The name ``bit'' stands for \emph{binary digit}. It means that, at
each point in a computation, $j$ can only be either 0 or 1, i.e., $j\in\{0,1\}$.
If $n$ bits are combined, the state of the computation is described by  a bit
string $J=j_{0}\cdots j_{n-1}$. Again, each bit can only be either 0 or 1. The
algebraic structure to describe all possible $n$-bit states $J$  is given by the
$n$-fold Cartesian product
\begin{align}
    J=j_{0}\cdots j_{n-1} \in \{0,1\}^n = \{0\cdots00,0\cdots01,\ldots,1\cdots11\},
    \label{eq:multiplebits}
\end{align}
which has a finite number of elements $\abs{\{0,1\}^n}=2^n$. A
computation can be formally expressed as a function
$f:\{0,1\}^n\to\{0,1\}^m$ for $m,n\in\mathbb N$. This formalism is known as
Boolean algebra and was first introduced in \cite{boole1847logic}.

A qubit $\ket{\psi}$, on the other hand, is defined as a two-level quantum
system determined by two complex numbers $a_0,a_1\in\mathbb C$ with
$\abs{a_0}^2+\abs{a_1}^2=1$. It is commonly written as
\begin{align}
  \ket{\psi} = a_0 \ket{0} + a_1 \ket{1} = \begin{pmatrix}
    a_0 \\ a_1
  \end{pmatrix},
  \label{eq:singlequbitstate}
\end{align}
where the \emph{computational basis states} $\ket{0}$ and $\ket{1}$ represent
the digital bit states 0 and 1, respectively. Thus, a qubit is not always
either $\ket{0}$ or $\ket{1}$, but rather any complex linear combination of
both. This concept is usually called \emph{superposition} and is the first
reason why a qubit can be considered a generalized bit.

If it were only this generalization, it would be easy to imagine why the
computational model of quantum computing might have an advantage over digital
computing. However, the drawback of this computational model is that the complex
coefficients $a_0$ and $a_1$ defining the state given by
\equref{eq:singlequbitstate} cannot be observed directly. Instead, as dictated
by quantum theory, they only define probabilities $p_0=\abs{a_0}^2$
($p_1=\abs{a_1}^2$) to observe the qubit in the binary state 0 (1). Such an
observation is called \emph{measurement}. The complex coefficients are
correspondingly called \emph{probability amplitudes}. The interpretation
in the context of probability theory also requires that the amplitudes
be normalized such that $\abs{a_0}^2+\abs{a_1}^2=1$.

Thus, a constraint of the computational model of quantum computing is that,
after a measurement, the state of the qubit is reset to the observed binary
state. This mathematical constraint is the core of why algorithms for quantum
computers have been notoriously hard to find, and that only a few key algorithms
with a considerable theoretical speedup have been found
\cite{Aaronson2008limits}.

\subsection{Bloch sphere}

\begin{figure}[p]
  \centering
  \includegraphics[width=.5\linewidth]{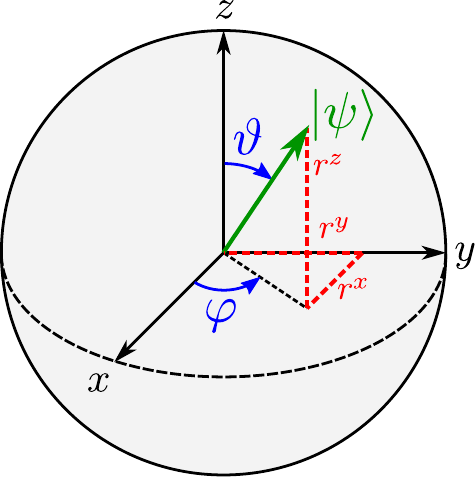}
  \caption{Bloch sphere representation of a pure single-qubit state $\ket\psi$.
  The azimuthal angle $\vartheta\in[0,\pi]$ and the polar angle $\varphi\in[0,2\pi)$
  are defined in \equref{eq:singlequbitblochangles}, and the Cartesian coordinates
  $r^x$, $r^y$, and $r^z$ are given by \equref{eq:singlequbitblochvector}}
  \label{fig:blochsphere}
\end{figure}

\begin{figure}[p]
  \centering
  \includegraphics[width=\linewidth]{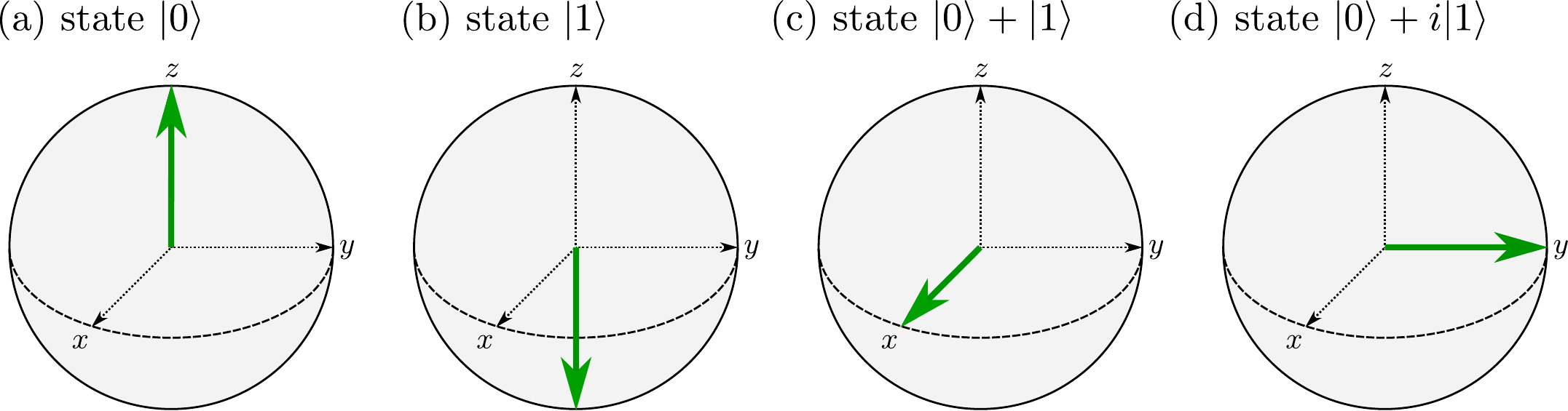}
  \caption{Bloch sphere representations of a few single-qubit states
  corresponding to eigenstates of the Pauli matrices defined in
  \equref{eq:paulimatrices}: (a) $+1$ eigenstate of $\sigma^z$, (b) $-1$
  eigenstate of $\sigma^z$, (c) $+1$ eigenstate of $\sigma^x$, and (d) $+1$
  eigenstate of $\sigma^y$. For simplicity, the normalization factor of $1/\sqrt
  2$ has been left out in the labels of (c) and (d).}
  \label{fig:blochsphereexamples}
\end{figure}

The fact that the complex coefficients $a_0$ and $a_1$ in
\equref{eq:singlequbitstate} are probability amplitudes can be used to find a
convenient parametrization of a general single-qubit state: From
$\abs{a_0}^2+\abs{a_1}^2=1$, we know that there is an angle
$\vartheta\in[0,\pi]$ such that $\abs{a_0} = \cos(\vartheta/2)$ and
$\abs{a_1}=\sin(\vartheta/2)$. Furthermore, quantum theory states that the
global phase of any state $\ket{\psi}$ is not observable, so we can set
$a_0=\cos(\vartheta/2)\ge0$ and $a_1=\exp(i\varphi)\sin(\vartheta/2)$, where
$\varphi\in[0,2\pi)$ encodes the relative phase between $a_0$ and $a_1$. The
general single-qubit state $\ket{\psi}$ given by \equref{eq:singlequbitstate}
thus becomes
\begin{align}
\ket{\psi} &= \cos\frac\vartheta 2\ket{0} + e^{i\varphi} \sin\frac\vartheta 2\ket{1}.
  \label{eq:singlequbitblochangles}
\end{align}
Due to the domain of the angles $\vartheta\in[0,\pi]$ and $\varphi\in[0,2\pi)$, the single-qubit state $\ket{\psi}$
can be visualized on the surface of a three-dimensional sphere called \emph{Bloch sphere} as shown in
\figref{fig:blochsphere}. Consequently, the representation defined by \equref{eq:singlequbitblochangles} is
called the \emph{Bloch sphere representation} and the three-dimensional vector $\vec r$ representing $\ket{\psi}$
on this sphere is called
the \emph{Bloch vector}. Its Cartesian coordinates $r^x$, $r^y$, and $r^z$ can be calculated from the expectation values
of $\ket{\psi}$ with respect to the Pauli matrices
\begin{align}
  \sigma^x &= \begin{pmatrix}
    0 & 1 \\ 1 & 0
  \end{pmatrix},&
  \sigma^y &= \begin{pmatrix}
    0 & -i \\ i & 0
  \end{pmatrix},&
  \sigma^z &= \begin{pmatrix}
    1 & 0 \\ 0 & -1
  \end{pmatrix},
  \label{eq:paulimatrices}
\end{align}
yielding
\begin{align}
  \vec r &=
  \begin{pmatrix}
    r^x \\ r^y \\ r^z
  \end{pmatrix} =
  \begin{pmatrix}
    \braket{\psi|\sigma^x|\psi} \\ \braket{\psi|\sigma^y|\psi} \\ \braket{\psi|\sigma^z|\psi}
  \end{pmatrix} =
  \begin{pmatrix}
    2\,\mathrm{Re}\,a_0^*a_1 \\ 2\,\mathrm{Im}\,a_0^*a_1 \\ \abs{a_0}^2-\abs{a_1}^2
  \end{pmatrix} =
  \begin{pmatrix}
    \sin\vartheta\cos\varphi\\
    \sin\vartheta\sin\varphi\\
    \cos\vartheta
  \end{pmatrix}.
  \label{eq:singlequbitblochvector}
\end{align}

The three Pauli matrices given in \equref{eq:paulimatrices} are unitary,
traceless, Hermitian matrices with eigenvalues $\pm 1$. The corresponding
eigenstates define the unit axes of the coordinate system of the Bloch sphere
shown in \figref{fig:blochsphere}. In particular, the eigenstates of $\sigma^z$
are the computational basis states $\ket{0}$ and $\ket{1}$, whose Bloch vectors
are given by the positive and the negative $z$ axis, respectively (see
\figref{fig:blochsphereexamples}(a) and (b)). The eigenstates of $\sigma^x$ are
denoted by $\ket{\pm} = (\ket{0} \pm \ket{1})/\sqrt 2$ with Bloch vectors lying
on the $x$ axis, and the eigenstates of $\sigma^y$ are given by $\ket{\pm
i}=(\ket{0} \pm i\ket{1})/\sqrt 2$ with Bloch vectors lying on the $y$ axis.
Both $\sigma^x$ and $\sigma^y$ eigenstates corresponding to the eigenvalue $+1$
are visualized in \figref{fig:blochsphereexamples}(c) and (d), respectively.

\subsection{Multiple qubits}

The second reason why qubits can be seen as generalized bits becomes apparent
when multiple qubits are combined. While an $n$-bit state in a digital computer
is an element of the $n$-fold Cartesian product (see \equref{eq:multiplebits}),
an $n$-qubit state in a quantum computer is an element of the $n$-fold
\emph{tensor product}. This means that an $n$-qubit state is a superposition of
all $2^n$ bit states given in \equref{eq:multiplebits}. We can therefore write
an arbitrary multi-qubit state $\ket{\psi}$ as
\begin{align}
  \ket{\psi} &= \sum\limits_{j_{0}\cdots j_{n-1}\in\{0,1\}^n} a_{j_{0}\cdots j_{n-1}} \ket{j_{0}\cdots j_{n-1}}
  = \sum\limits_{J=0}^{2^n-1} a_J \ket{J}
  = \begin{pmatrix}
    a_0 \\ \vdots \\ a_{2^n-1}
  \end{pmatrix},
  \label{eq:multiqubitstate}
\end{align}
where the integer index $J$ and its binary representation $j_0\cdots j_{n-1}$
can be used interchangeably.

Formally, a single-qubit state is an element of the two-dimensional complex
Hilbert space $\mathcal H_2 = \mathrm{span}\{\ket 0,\ket 1\}\cong\mathbb C^2$.
A multi-qubit state for $n$ qubits is then an element of the $2^n$-dimensional
complex Hilbert space $\mathcal H_{2^n}$, given by the $n$-fold tensor product
of $\mathcal H_2$,
\begin{align}
  \mathcal H_{2^n} &= \mathcal H_2^{\otimes n}
  = \mathrm{span}\{\ket 0,\ket 1\}^{\otimes n}
  = \mathrm{span}\{ \bigcup\limits_{\substack{j_{0},\ldots,j_{n-1}\\\in\{0,1\}}} \ket{j_{0}\cdots j_{n-1}}\} \nonumber\\
  &= \mathrm{span}\{\underbrace{\ket{\overbrace{0\cdots0}^{n}},\ldots,\ket{\overbrace{1\cdots1}^{n}}}_{2^n}\}
  \cong \mathbb C^{2^n}.
  \label{eq:multiqubithilbertspace}
\end{align}
To keep the notation concise, we do not write the tensor products explicitly
such that the states $\ket{j_{0}j_1\cdots j_{n-1}}$,
$\ket{j_{0}}\ket{j_1}\cdots\ket{j_{n-1}}$, and
$\ket{j_{0}}\otimes\ket{j_1}\otimes\cdots\otimes\ket{j_{n-1}}$ are understood to
be the same. Note that the basis states $\ket{j_{0}\cdots j_{n-1}}$ in the vector
space $\mathcal H_{2^n}$ correspond to the same $2^n$ bit strings $j_{0}\cdots
j_{n-1}$ comprising the space for $n$-bit states given by \equref{eq:multiplebits}.
By analogy with single-qubit states, a multi-qubit state can be an arbitrary
complex superposition of these basis states.

A property of \equref{eq:multiqubitstate} is that most of the states in the
tensor-product space $\mathcal H_{2^n}$ cannot be written as tensor products
themselves. Such states are called \emph{entangled} states and the corresponding
concept is called \emph{entanglement}. Simple examples for
entangled states in the two-qubit case $n=2$ are the so-called Bell states
\begin{subequations}
\begin{align}
  \label{eq:bellstatephi}
  \ket{\Phi^\pm} &= \frac 1 {\sqrt{2}} (\ket{00} \pm \ket{11}),\\
  \label{eq:bellstatepsi}
  \ket{\Psi^\pm} &= \frac 1 {\sqrt{2}} (\ket{01} \pm \ket{10}),
\end{align}
\end{subequations}
the last of which, $\ket{\Psi^-}$, is also known as the \emph{singlet state}.
The fact that these states are entangled can be proven by contradiction:
Assuming that there exists a tensor-product state of the form
$(a_0\ket0+a_1\ket1)(b_0\ket0+b_1\ket1)$ for any of the four Bell states
leads to contradictory equations for $a_0,a_1,b_0,b_1\in\mathbb C$.

The presence of entanglement is a consequence of the algebra (with complex
tensor-product spaces) that we use in quantum
theory to describe observations mathematically. It may seem peculiar in the
sense that two separate qubits described by $(\ket{00}+\ket{11})/\sqrt{2}$ have
a strong correlation, i.e., when we measure one qubit we seem to know the state
of the other qubit without measuring it. However, this sense of peculiarity is
rather a consequence of the way  we choose to describe the state.

For instance, suppose that we observe some process that can be described by a
probability  distribution $p(x,y)$. Then we empirically observe that one cannot
express  $p(x,y) = f(x)g(y)$ for any two functions $f$ and $g$ and therefore
call the variables  $x$ and $y$ \emph{entangled}. The existence of such
processes seems reasonable,  but does not look very peculiar in this language.
To make this example concrete  in the present context, such a probability
distribution for  an entangled state $(\ket{00}+\ket{11})/\sqrt{2}$ would be
$p(x,y)=\delta_{xy}/2$ where $x,y\in\{0,1\}$. It is obvious that $\delta_{xy}$
depends on  both $x$ and $y$ and cannot be written as a product of some $f(x)$
and $g(y)$. For a non-entangled state such as $(\ket{00}+\ket{10})/\sqrt{2}$, in
contrast, the probability distribution would be $p(x,y) = \delta_{y0}/2$, and it
can be easily expressed as a product $f(x)g(y)$ for $f(x)=1$ and
$g(y)=\delta_{y0}/2$.

So entanglement is a consequence of the fact that in quantum theory, we model
physical states using complex tensor-product spaces (see
\equref{eq:multiqubithilbertspace}). This allows us  to describe correlations
between individual components of the full space, such as the individual qubits
in \equaref{eq:bellstatephi}{eq:bellstatepsi}, which cannot be described by
states in Cartesian-product spaces (see \equref{eq:multiplebits}). For this
reason, there is no analogue of entanglement in the classical multi-bit states
used to describe digital computers.

It is worth mentioning that there exist mathematical tests for entanglement,
known as \emph{Bell tests} in the context of \emph{Bell inequalities}
\cite{Bell1964OnTheEPRParadox, Bell2004SpeakableAndUnspeakable} and
\emph{separability criteria} \cite{Horodecki1996SeparabilityEntanglementWitness,
Terhal2000bell}. If one wants to describe an experiment using a quantum
theoretical model, such a Bell test yields an answer to the question if the
observations have to be described in terms of an entangled state. Note, however,
that such a test can never prove, in a mathematical sense, that the observations
can \emph{only} be described by a quantum theoretical model. An alternative
``subquantum'' model that can describe the individual events and not only their
quantum theoretical statistics can be found in
\cite{DeRaedt2005DiscreteEventSimulationDES}
(see~\cite{Willsch2020DESQuantumWalks, DeRaedt2020EEPRB} for two particular
applications of the model).

The constraint of normalizing the states is not included in the
definition of the multi-qubit space given by \equref{eq:multiqubithilbertspace}.
The purpose of this is to keep the linearity of the algebraic structures.
Similarly, we have not made any efforts to eliminate the global phase of the
complex coefficients of the general multi-qubit state given by
\equref{eq:multiqubitstate}.  The concepts of normalization and global phase
only play a role when the complex coefficients are to be interpreted as
probability amplitudes, or when one wants to eliminate as many degrees of
freedom as possible to find concise representations of quantum states. The
latter was done to derive the Bloch sphere representation of a single-qubit
state (see \equref{eq:singlequbitblochangles} and \figref{fig:blochsphere}).

For general multi-qubit states $\ket{\psi}\in\mathcal H_{2^n}$, it is possible
to visualize the individual qubits using one Bloch sphere per qubit. However,
such a picture does not capture all the information required to describe  the
state $\ket{\psi}$, as it did for a single qubit. This can be understood from a
simple counting argument: A single-qubit state $a_0\ket{0}+a_1\ket{1}$ is
described by two complex numbers or, equivalently, four real numbers. Using  the
normalization constraint and eliminating the global phase, we were able to
reduce these four real numbers to the two angles $\vartheta$ and $\varphi$,
which can be visualized as a vector of length $1$ on the Bloch sphere. For $n$
qubits with one Bloch sphere per qubit, we would need $2n$ real numbers (or $3n$
if we do not fix the length of each Bloch vector to $1$). However, a general
$n$-qubit state such as the one given by \equref{eq:multiqubitstate} is
described by $2^{n+1}$ real numbers. Even if we subtract two for the
normalization constraint and the global phase, this is still much more than can
be visualized using $n$ spheres (see also
\cite{bengtsson2006geometryofquantumstates}).

Nevertheless, it may still be helpful to visualize each single qubit on a
separate Bloch sphere. To compute the Bloch vectors, we use the notation
\begin{align}
  \label{eq:paulimatricesmultiplequbits}
  \sigma_i^\alpha = I \otimes \cdots \otimes \sigma^\alpha \otimes \cdots \otimes I
\end{align}
for the corresponding Pauli matrices, where $I$ denotes the identity matrix on
the two-dimensional single-qubit space $\mathcal H_2$, $\alpha\in\{x,y,z\}$
labels the Pauli matrices given by \equref{eq:paulimatrices}, and
$i\in\{0,\ldots,n-1\}$ denotes the position of $\sigma^\alpha$ in this
tensor-product matrix using the same ordering as for the multi-qubit basis state
$\ket{j_{0}\cdots j_{n-1}}$. This means that applying $\sigma_i^\alpha$ to this
basis state only affects the qubit $\ket{j_i}$ such that
\begin{align}
  \sigma_i^\alpha \ket{j_{0}\cdots j_{n-1}} = \ket{j_{0}}\cdots\ket{j_{i-1}}(\sigma^\alpha\ket{j_i})\ket{j_{i+1}}\cdots\ket{j_{n-1}}.
\end{align}
Using $\sigma_i^\alpha$, we can now compute Bloch vectors $\vec r_i\in\mathbb
R^3$ for each qubit in the same manner as in \equref{eq:singlequbitblochvector}
for the general multi-qubit state $\ket{\psi}\in\mathcal H_{2^n}$ given by
\equref{eq:multiqubitstate}. A short calculation yields
\begin{align}
  \label{eq:multiqubitblochvector}
  \vec r_i &=
  \begin{pmatrix}
    r_i^x \\ r_i^y \\ r_i^z
  \end{pmatrix} =
  \begin{pmatrix}
    \braket{\psi|\sigma_i^x|\psi} \\ \braket{\psi|\sigma_i^y|\psi} \\ \braket{\psi|\sigma_i^z|\psi}
  \end{pmatrix} =
  \sum\limits_{\substack{j_{0}\cdots j_{n-1}\\\text{without $j_i$}}}
  \begin{pmatrix}
    2\,\mathrm{Re}(a_{j_{0}\cdots0\cdots j_{n-1}}^*a_{j_{0}\cdots1\cdots j_{n-1}}) \\
    2\,\mathrm{Im}(a_{j_{0}\cdots0\cdots j_{n-1}}^*a_{j_{0}\cdots1\cdots j_{n-1}}) \\
    \abs{a_{j_{0}\cdots0\cdots j_{n-1}}}^2-\abs{a_{j_{0}\cdots1\cdots j_{n-1}}}^2
  \end{pmatrix},
\end{align}
exhibiting a similar structure as \equref{eq:singlequbitblochvector}.
This expression is used for the implementation of the visualizations
discussed in \secref{sec:visualizer} and \appref{app:visualization}.

It is worth mentioning that for two-qubit states, some ideas have been  proposed
to visualize a general two-qubit state using three spheres. The requirement  is
that two of the three spheres shall correspond to the respective single-qubit
Bloch vectors computed from \equref{eq:multiqubitblochvector}. The remaining
information about the amount of entanglement in the two-qubit state is then
visualized on the third sphere. See \cite{Wie2014blochspheretwoqubits} or
\cite{rigetti2009quantumgates} for more information.

\subsection{Leakage}
\label{sec:leakage}

In practice, many physical realizations of qubits contain additional states
beyond the computational basis states $\ket{0}$ and $\ket{1}$. For
superconducting  transmon qubits, for instance, this refers to higher excited
states $\ket{2},\ket{3},\ldots$. Formally, these non-computational states are
not described by the tensor-product structure of
\equref{eq:multiqubithilbertspace}. Instead, they describe alternative states
that belong to the individual qubits. This means that the single-qubit
description from \equref{eq:singlequbitstate} needs to be extended by a direct
sum structure
\begin{align}
  \label{eq:singlequbitwithleakage}
  \ket\psi = a_0\ket 0 + a_1\ket 1 + a_2\ket 2 + \cdots \in \mathcal H_2 \oplus \mathcal H_{\mathrm{L}},
\end{align}
where $\mathcal H_{\mathrm{L}} = \mathrm{span}\{\ket 2, \ket 3, \ldots\}$.
If the state $\ket\psi$ acquires a contribution of
states from $\mathcal H_{\mathrm{L}}$, one generally speaks of \emph{leakage}.

For the multi-qubit Hilbert space $\mathcal H_{2^n}$ defined in \equref{eq:multiqubithilbertspace},
the corresponding leakage space is constructed via
\begin{align}
  \label{eq:multiqubithilbertspacewithleakage}
  (\mathcal H_2 \oplus \mathcal H_{\mathrm{L}}) \otimes \cdots \otimes (\mathcal H_2 \oplus \mathcal H_{\mathrm{L}})
  = \mathcal H_{2^n} \oplus \mathcal H_{\mathrm{Leak}},
\end{align}
where $\mathcal H_{\mathrm{Leak}}$ is spanned by each state in which at least
one part is an element of $\mathcal H_{\mathrm{L}}$.

The concept of leakage and its implications for evolutions of quantum systems
are considered further in \secref{sec:transformationssubsystemsleakage}.  Since
leakage is a particularly important issue for superconducting transmon qubits,
the concept will play an important role for the experiments studied throughout
this thesis.

\section{Quantum gates}
\label{sec:quantumgates}

Given a set of qubits as defined in the previous section, there are certain
operations that can be performed on the qubits. These operations are called
\emph{quantum gates}. They are inspired by their analogue for digital computers,
i.e., the digital logic gates.

A digital logic gate is an arbitrary function $f:\{0,1\}^n\to\{0,1\}^m$. It
takes as input a bit string of length $n$ such as the one defined  in
\equref{eq:multiplebits} and outputs another bit string of length  $m$ (not
necessarily of the same length). There is, in principle, no further limitation
for $f$.

Transferring this idea to quantum states, a quantum gate would need to be a
function acting on the space $\mathcal H_{2^n}$ defined in
\equref{eq:multiqubithilbertspace}, which is an immensely larger space. However,
there are certain restrictions for quantum gates that impose some nontrivial
limitations on these functions. In this section, we first discuss these
limitations and then define the set of elementary quantum gates that are
implemented in terms of pulses by the transmon simulator (see also
\appref{app:gateset}).

\subsection{Unitary operators}
\label{sec:unitaryoperators}

Every quantum gate is a basic operation on multi-qubit states of the form given
by \equref{eq:multiqubitstate}. A quantum gate is mathematically defined as a linear map $U:
\mathcal H_{2^n} \to \mathcal H_{2^n}$ with the important restriction that it
has to be unitary (see \appref{app:wigner} for a
review of arguments why quantum theory requires unitary linear maps). Note that
the linearity of the map $U$ implies that expressions such as
$U(\ket{\psi}+\ket{\phi})$ can be evaluated as $U\ket{\psi}+U\ket{\phi}$ for any
quantum gate $U$ and any states $\ket\psi$ and $\ket\phi$.

As we only consider finite-dimensional Hilbert spaces, we identify unitary
operators with their representation in terms of unitary matrices, i.e.,
complex invertible matrices
$U\in\mathbb C^{2^n\times 2^n}$ that satisfy $U^{-1} = U^\dagger$, where
$U^\dagger$ denotes the Hermitian conjugate (or conjugate transpose) of $U$. By
definition, such matrices conserve the norm of quantum states, so the
restriction to unitary matrices goes hand in hand with the interpretation of the
complex coefficients in \equref{eq:multiqubitstate} as probability amplitudes.

From a computational perspective, the condition of unitary square matrices
imposes severe limitations on the computational model of quantum computing. In
particular, each quantum gate has to be invertible (i.e., \emph{reversible}).
This property is not fulfilled by many of the conventional digital logic gates.
For instance, the logical \textsc{AND} gate takes two bits as input and produces
only one bit as output (namely 1 if both of its inputs are 1, and 0 otherwise).
Since it is not possible to deduce the two input bits given the output bit,
the \textsc{AND} gate is not reversible. The same applies to the logical
\textsc{OR} gate, the universal \textsc{NAND} gate, and many others. However,
all these gates can, in principle,  be made reversible by adding another output
bit. This is the idea of classical, reversible computation, which bridges the
gap to make the logical gates amenable to quantum computing (see
\cite{Fredkin1982ConservativeLogic, Bennett1973LogicalReversibility} for more
information).

\subsection{Elementary quantum gates}
\label{sec:elementaryquantumgates}

An elementary single-qubit gate is given by a three-dimensional rotation on the
Bloch sphere (see \figref{fig:blochsphere}). General single-qubit gates are
often represented as sequences of such single-qubit  rotations. This is possible
since for one qubit, the Bloch sphere yields a faithful representation of the
state, in the sense that it captures the full information contained in the
state. Furthermore, it has the advantage that single-qubit rotations directly
relate to the pulses that are used on actual quantum processors (such as the IBM
Q processors \cite{ibmquantumexperience2016}) to implement the gates
\cite{Cross2017openqasm2}. For this reason, we follow this convention and
express all single-qubit gates as a sequence of rotations on the Bloch sphere.

A single-qubit rotation on the Bloch sphere is given by $R^\alpha(\vartheta) =
\exp(-i\vartheta\sigma^\alpha/2)$, where $\alpha$ defines the axis of rotation,
$\sigma^\alpha$ denotes the respective Pauli matrix given by
\equref{eq:paulimatrices}, and $\vartheta$ is the angle of rotation.
One can compute the matrix exponential to get the closed-form expressions
\begin{subequations}
\begin{align}
  \label{eq:singlequbitrotationx}
  R^x(\vartheta) &= e^{-i\vartheta \sigma^x/2} =
  \begin{pmatrix}
    \cos(\vartheta/2) & -i\sin(\vartheta/2) \\
    -i\sin(\vartheta/2) & \cos(\vartheta/2)
  \end{pmatrix}
  ,\\
  \label{eq:singlequbitrotationy}
  R^y(\vartheta) &= e^{-i\vartheta \sigma^y/2} =
  \begin{pmatrix}
    \cos(\vartheta/2) & -\sin(\vartheta/2) \\
    \sin(\vartheta/2) & \cos(\vartheta/2)
  \end{pmatrix}
  ,\\
  \label{eq:singlequbitrotationz}
  R^z(\vartheta) &= e^{-i\vartheta \sigma^z/2} =
  \begin{pmatrix}
    \exp(-i\vartheta/2) & 0 \\
    0 & \exp(i\vartheta/2)
  \end{pmatrix}.
\end{align}
\end{subequations}

The operation $R^\alpha(\vartheta)\ket{\psi}$, where $\ket{\psi}$ is the
single-qubit state shown in \figref{fig:blochsphere}, corresponds to a rotation
of the Bloch vector $\vec r$ (see \equref{eq:singlequbitblochvector}) around the
axis $\alpha$ by an angle $\vartheta$. The sense of rotation is given by the
\emph{right-hand rule}, set by the minus sign in the exponent of
$R^\alpha(\vartheta)$.

We define three elementary combinations of $R^x(\pi/2)$ and $R^z(\vartheta)$
that are directly related to the hardware implementations of the IBM Q
processors \cite{Cross2017openqasm2} (see also
\secref{sec:optimizatingsinglequbitgate}). They are given by
\begin{subequations}
\begin{align}
  \label{eq:singlequbitU1}
  \textsc{U1}(\lambda) &= c_1 R^z(\lambda)
  ,\\
  \label{eq:singlequbitU2}
  \textsc{U2}(\phi,\lambda) &= c_2 R^z\Big(\phi+\frac\pi2\Big)\,R^x\Big(\frac\pi2\Big)\,R^z\Big(\lambda-\frac\pi2\Big)
  ,\\
  \label{eq:singlequbitU3}
  \textsc{U3}(\theta,\phi,\lambda) &= c_3 R^z(\phi+3\pi)\,R^x\Big(\frac\pi2\Big)\,R^z(\theta+\pi)\,R^x\Big(\frac\pi2\Big)\,R^z(\lambda)
  ,
\end{align}
\end{subequations}
where the complex phase factors $c_1=\exp(i\lambda/2)$ and
$c_2=c_3=\exp(i(\phi+\lambda)/2)$ are not essential for the operation of the
gates and are given for reference only.

When a single-qubit gate is applied to one of the qubits of a multi-qubit state
$\ket\psi \in \mathcal H_{2^n}$ (see \equref{eq:multiqubitstate}), the Pauli
matrices $\sigma^\alpha$ in
\equsref{eq:singlequbitrotationx}{eq:singlequbitrotationz} have to be replaced
by $\sigma_i^\alpha$ defined in \equref{eq:paulimatricesmultiplequbits}.
Accordingly, we denote the corresponding rotations by
\begin{align}
  \label{eq:singlequbitrotationmultiplequbits}
  R_i^\alpha(\vartheta) &= e^{-i\vartheta \sigma_i^\alpha/2}
  = I \otimes \cdots \otimes R^\alpha(\vartheta) \otimes \cdots \otimes I.
\end{align}
In the following chapters, we sometimes use the widespread alternative notations
$X_i^\vartheta = R_i^x(\vartheta)$, $Y_i^\vartheta = R_i^y(\vartheta)$,
$Z_i^\vartheta = R_i^z(\vartheta)$. Furthermore, we use the convention that
$\vartheta=\pi$ if the ``exponent'' $\vartheta$ is not explicitly specified.

Any of the single-qubit gates used in this work can be expressed in terms of the
single-qubit rotations defined in
\equsref{eq:singlequbitrotationx}{eq:singlequbitrotationz} or the $\textsc{U}$
gates defined in  \equsref{eq:singlequbitU1}{eq:singlequbitU3} (see
\tabref{tab:elementarygateset} in \appref{app:gateset} for a list of all gates,
their matrix representations,  and their relations to the elementary gates). Two
particularly important single-qubit gates are the $X$ gate (also known as bit
flip or \textsc{NOT} gate) and the Hadamard gate. They are defined as
\begin{align}
  X &=
  \begin{pmatrix}
    0 & 1 \\
    1 & 0
  \end{pmatrix}, &
  H &= \frac{1}{\sqrt 2}
  \begin{pmatrix}
    1 & 1 \\
    1 & -1
  \end{pmatrix}.
\end{align}
As before, we extend these single-qubit gates to multi-qubit spaces using the
notations $X_i$ and $H_i$, as done in
\equaref{eq:paulimatricesmultiplequbits}{eq:singlequbitrotationmultiplequbits}.
Note that the gate $X_i$ corresponds to the single-qubit rotation
$X_i^\vartheta$ for $\vartheta = \pi$ (up to a complex phase factor), which is
in agreement with the above convention.

In quantum computing, there is a much larger variety of single-qubit gates than
in digital computing (where the only nontrivial single-bit gate is the
\textsc{NOT} gate). Nonetheless, it turns out that almost any two-qubit gate
is sufficient to build a \emph{universal gate set} from the single-qubit
rotations \cite{Deutsch95universality, divincenzo1995twoqubitgates} (see also
\cite{DiVincenzo2000universalnearestneighbourexchangeinteraction}). In this
context, universal means that any quantum gate on $\mathcal H_{2^n}$ can be
represented as a finite sequence of gates from this set, using suitable angles
for all single-qubit rotations.

One such two-qubit gate is the controlled \textsc{NOT} (\textsc{CNOT})
gate. It is defined as
\begin{align}
  \label{eq:cnotgate}
  \textsc{CNOT} =\ \begin{blockarray}{cccc}
    \matindex{\ket{00}} & \matindex{\ket{01}} & \matindex{\ket{10}} & \matindex{\ket{11}} \\
    \begin{block}{(cccc)}
      1 & 0 & 0 & 0 \\
      0 & 1 & 0 & 0 \\
      0 & 0 & 0 & 1 \\
      0 & 0 & 1 & 0 \\
    \end{block}
  \end{blockarray}\ ,
\end{align}
such that $\textsc{CNOT} \ket{j_0}\ket{j_1}=\ket{j_0}\ket{j_0\oplus j_1}$ where
$\oplus$ denotes the \textsc{XOR} operation (or integer addition modulo $2$).
The effect of this gate is to flip the target qubit $\ket{j_1}$ if and only if
the control qubit $\ket{j_0}$ is in state $\ket{1}$. In the multi-qubit space
$\mathcal H_{2^n}$ (see \equref{eq:multiqubithilbertspace}), we use the notation
$\textsc{CNOT}_{il}$ to denote a \textsc{CNOT} gate where qubit $i$ is the
control qubit and qubit $l$ is the target qubit.

An important thing to realize is that the model of gate-based quantum computing
with elementary gates given by the single-qubit rotations in
\equsref{eq:singlequbitrotationx}{eq:singlequbitrotationz} is a model of analog
computation. Given an angle $\vartheta\in[0,2\pi)$, a high level of precision
over the controlling pulse may be necessary to implement the rotation
$R_i^\alpha(\vartheta)$ accurately. We study hardware implementations of quantum
gates (and in particular the two-qubit \textsc{CNOT} gate) in more detail in the
following chapters.

\section{Quantum circuits}
\label{sec:quantumcircuits}

For a gate-based quantum computer, an algorithm is specified in terms of a
\emph{quantum circuit}. It consists of a sequence of quantum gates as defined in
the previous section. This model is inspired by the   circuit model of digital
computation, in which an algorithm for a digital computer can, in principle,  be
decomposed into a sequence of digital logic gates.

As each quantum gate is a unitary operator, a full quantum circuit is also a
unitary operator (the set of unitary operators on a given Hilbert space forms a
group). For an $n$-qubit system described by the Hilbert space $\mathcal
H_{2^n}$ given by \equref{eq:multiqubithilbertspace}, a quantum circuit is thus
a specification of a large but sparse unitary matrix $U \in\mathbb C^{2^n\times
2^n}$. If a quantum circuit is written as a sequence of gates from a certain
gate set, this means that a (potentially large) unitary matrix is expressed as a
product of (typically small) elementary unitary matrices. In this sense, a quantum
circuit is just a decomposition of a large matrix into smaller matrices. A list
of elementary gates and their matrix representations is given in
\appref{app:gateset}.

\begin{figure}
  \centering
  \[
  \Qcircuit @C=.8em @R=.7em {
      &\lstick{\ket{j_0}}&\gate{H}&\gate{T^\dag}&\targ    &\gate{T^\dag}&\targ    &\gate{S}&\qw     &\ctrl{1}&\targ    &\ctrl{1}&\qw\\
      &\lstick{\ket{j_1}}&\qw     &\qw          &\ctrl{-1}&\qw          &\ctrl{-1}&\gate{T}&\gate{H}&\targ   &\ctrl{-1}&\targ   &\qw\\
    }
  \]
  \caption{Circuit diagram for the two-qubit QFT. The boxes denote single-qubit
  gates (see \tabref{tab:elementarygateset} in \appref{app:gateset}), and
  vertical lines denote two-qubit \textsc{CNOT} gates as defined in
  \equref{eq:cnotgate}. The solid circle in a \textsc{CNOT} gate denotes the
  control qubit, and the open circle denotes the target qubit. The matrix
  representation of this circuit is given by \equref{eq:twoqubitqft}.}
  \label{fig:twoqubitqft}
\end{figure}
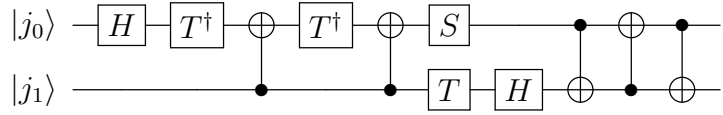

Quantum circuits are often expressed in a diagrammatic notation, where each
qubit corresponds to a horizontal line. Gates on each qubit are specified using
corresponding circuit symbols. An example for a simple circuit diagram
representing a two-qubit quantum Fourier transform (QFT) is shown in
\figref{fig:twoqubitqft}. It is taken from \cite{NielsenChuang} after rewriting
some of the gates in terms of the elementary gate set used in this
thesis.

Circuit diagrams are defined such that time increases from left to right.
Therefore, when converting between circuit diagrams and their matrix
representations, the order of the matrices has to be reversed. For instance, the
unitary matrix corresponding to the circuit shown in \figref{fig:twoqubitqft}
is computed as
\begin{align}
  U^{(\textsc{QFT2})} &= \textsc{CNOT}_{01}\,\textsc{CNOT}_{10}\,\textsc{CNOT}_{01}\,
    H_1\,T_1\,S_0\,\textsc{CNOT}_{10}\,T^\dagger_0\,\textsc{CNOT}_{10}\,T^\dagger_0\,H_0 \nonumber\\
  \label{eq:twoqubitqft}
  &= \frac 1 2
   \begin{pmatrix}
    1 & 1 & 1 & 1 \\
    1 & i & -1 & -i \\
    1 & -1 & 1 & -1 \\
    1 & -i & -1 & i
\end{pmatrix}.
\end{align}

The \textsc{QFT} can be generalized to an arbitrary number of qubits. Its
generic property is that it maps a computational basis state to a uniform
superposition over all states with relative phases dependent on the original
state. This can be seen from its definition for a basis state $\ket J\in\mathcal
H_{2^n}$ \cite{NielsenChuang},
\begin{align}
  \label{eq:multiqubitqft}
  U^{(\textsc{QFT}n)} \ket J &= \frac{1}{2^{n/2}} \sum_{K=0}^{2^n-1} e^{2\pi i J K /2^n} \ket K,
\end{align}
where we identified $J$ and its binary representation $j_{0}\cdots j_{n-1}$ as
done in \equref{eq:multiqubitstate}.

Mathematically, the QFT corresponds to a  discrete Fourier transform over the
additive group of integers modulo $2^n$. Interestingly, a universal quantum
computer using the gate set defined in \appref{app:gateset} can do a QFT in a
number of steps \emph{polynomial} in $n$, as opposed to the \emph{Fast Fourier
Transform} that requires $\mathcal O(n2^n)$ steps
\cite{Cooley1965FastFourierTransform}. This discovery forms the basis for most
of the quantum algorithms for which an exponential speedup is known.

One such algorithm is Shor's factorization algorithm \cite{shor94factoring} in
which the QFT is basically used to find the period of a suitable function  (note
that finding periods is a generic feature of any Fourier transform).
The common algebraic problem solved by most algorithms with an exponential
speedup is the \emph{hidden subgroup problem}
\cite{Jozsa2001hiddensubgroupproblem}.

The difficulty in finding these algorithms, however, is that even for an ideal
gate-based quantum computer, the complex phases in \equref{eq:multiqubitqft} are
not directly accessible  (see also the discussion at the end of
\secref{sec:singlequbits}). Furthermore, actual implementations of the QFT
require an extremely precise control over the phases of the pulses that are used
to implement the gates. These phases are related to the relative phase factors
present in \equref{eq:multiqubitqft}. We study the problem of controlling the
phases and implementing gates through pulses in more detail in the following
chapters.

\section{Quantum operations}
\label{sec:quantumoperations}

In some situations, it is expedient to consider not a single, pure quantum state
$\ket\psi\in\mathcal H_{2^n}$ representing the $n$ qubits, but a probability distribution $\{p_i\}$
over several quantum states $\ket{\psi_i}\in\mathcal H_{2^n}$. Such states are called \emph{mixed states}.
They can be represented by a density matrix
\begin{align}
  \label{eq:mixedstates}
  \rho = \sum_i p_i \ketbra{\psi_i}{\psi_i}.
\end{align}
This representation is called \emph{ensemble representation}. The fact
that $\{p_i\}$ represents a probability distribution means that each $p_i\ge0$,
and $\mathrm{Tr}\,\rho = \sum_i p_i = 1$. Further properties of $\rho$ are
$\rho=\rho^\dagger$ and that $\rho$ is a positive (semidefinite) matrix, i.e.,
$\braket{\phi|\rho|\phi} \ge 0$ for all $\ket\phi$. Note that this characterization
of density matrices is both necessary and sufficient, since each positive matrix
$\rho$ with $\mathrm{Tr}\,\rho=1$ can be written in the form of
\equref{eq:mixedstates} by taking its spectral decomposition such that $p_i\ge0$
are the eigenvalues and $\ket{\psi_i}$ are the eigenvectors of $\rho$.

The model of a gate-based quantum computer can be extended to a description
in terms of mixed states. The result of applying a quantum gate $U$, which maps a pure
quantum state $\ket\psi$ to $U\ket{\psi}$, is then described by
\begin{align}
  \label{eq:mixedstatesQuantumGate}
  \rho \mapsto U\rho U^\dagger = \sum_i p_i U\ketbra{\psi_i}{\psi_i}U^\dagger.
\end{align}

For mixed states, however, one often considers more general transformations
called \emph{quantum operations} \cite{Kraus1971}. We denote such an operation
by a map $\mathcal E$ that transforms $\rho$ according to
\begin{align}
  \label{eq:mixedstatesGeneralQuantumOperation}
  \rho\mapsto\mathcal E(\rho).
\end{align}
Typical constraints on $\mathcal E$ are that it be linear
(cf.~\appref{app:wigner}), Hermiticity-preserving ($\mathcal
E(\rho)^\dagger=\mathcal E(\rho^\dagger)$), and completely positive. The latter
means formally that if $A$ is a positive matrix on an extended Hilbert space of
arbitrary dimensionality, then also the extended map $\mathcal E\otimes\mathds1$
preserves the positivity of A \cite{Stinespring1955}. This property ensures that
a density matrix $\rho=\sum_i p_i\ketbra{\psi_i}{\psi_i}$ with non-negative
probabilities $p_i$ is mapped to another density matrix $\mathcal E(\rho)$ that
also represents non-negative probabilities. Furthermore, the ``completely'' in
completely positive ensures that this preservation of positivity also applies if
the description is extended to another system (such as extending the description
of a single qubit to two qubits, or one qubit and an environment). If $\mathcal
E$ is also trace-preserving (i.e., $\mathrm{Tr}\,\mathcal E(\rho) =
\mathrm{Tr}\,\rho$), we call the completely positive trace-preserving (CPTP) map
$\mathcal E$ \emph{quantum channel} or \emph{error channel}
(cf.~\secref{sec:errorchannels}).

\subsection{Representations of quantum operations}

There are several ways of representing a quantum operation $\mathcal E$. Since
it is a linear map, one obvious way is to write it as a matrix with elements
$\mathcal E_{(ij),(kl)} = \mathrm{Tr}\,\ketbra{i}{j}^\dagger\mathcal E(\ketbra k
l) = \braket{i|\mathcal E(\ketbra k l)|j}$. For an $n$-qubit system of dimension
$N=2^n$, this matrix has $N^2\times N^2$ complex elements. It is referred to as
the matrix representation of $\mathcal E$ in the standard basis. However, this
representation is not tuned to the particular properties of $\mathcal E$. For
instance, the property of complete positivity is not easily expressed by this
matrix.

A more useful representation is the so-called \emph{Kraus
representation}, $\mathcal E(\rho) = \sum_\alpha A_\alpha \rho B_\alpha^\dagger$, where
$A_\alpha,B_\alpha\in\mathbb C^{N\times N}$. One can show that
$\mathcal E$ is completely positive if and only if $B_\alpha = A_\alpha$
\cite{Kraus1971, Choi1975CPTPmaps}. This means that for completely positive maps $\mathcal E$,
a Kraus representation is given by
\begin{align}
  \label{eq:krausrepresentation}
  \mathcal E(\rho) = \sum_{\alpha=1}^R E_\alpha \rho E_\alpha^\dagger,
\end{align}
where $E_\alpha\in\mathbb C^{N\times N}$. The smallest number $R$ of terms in
the Kraus representation is called the \emph{Kraus rank} of $\mathcal E$.
Using the Kraus representation, one has that $\mathcal E$ is trace-preserving
if and only if $\sum_\alpha E_\alpha^\dagger E_\alpha = \mathds1$.

Another commonly used matrix representation of $\mathcal E$ is the so-called
\emph{Choi matrix} $J(\mathcal E)$. It is defined as
\begin{align}
  \label{eq:choimatrix}
  J(\mathcal E)
  = \frac 1 N \sum_{i,j=0}^{N-1} \ketbra i j\otimes\mathcal E(\ketbra i j)
  = \frac 1 N \begin{pmatrix}
    \mathcal E(\ketbra{0}{0}) & \mathcal E(\ketbra{0}{1}) & \cdots & \\
    \mathcal E(\ketbra{1}{0}) & \mathcal E(\ketbra{1}{1}) & \\
    \vdots  & & \ddots \\
      \end{pmatrix},
  % \\
  % = \frac 1 N \begin{pmatrix}
  %   \mathcal E(\ketbra{0}{0}) & \mathcal E(\ketbra{0}{1}) & \cdots & \mathcal E(\ketbra{0}{N-1}) \\
  %   \mathcal E(\ketbra{1}{0}) \\
  %   \vdots & & & \vdots \\
  %   \mathcal E(\ketbra{N-1}{0}) & & \cdots & \mathcal E(\ketbra{N-1}{N-1})
  % \end{pmatrix},
\end{align}
and it represents a matrix with $N^2\times N^2$ complex coefficients. A useful
property of the Choi matrix is that $\mathcal E$ is completely positive if and
only if $J(\mathcal E)$ is positive semidefinite \cite{Choi1975CPTPmaps}.
Furthermore, the rank of the Choi matrix (i.e., the number of non-zero singular
values) yields the minimum number $R$ of terms in the Kraus representation
given by \equref{eq:krausrepresentation}. For this reason, the Kraus rank $R$ is
also called the \emph{Choi rank} of $\mathcal E$.

The Choi matrix is often written in compact form as $J(\mathcal E) =
(\mathds1\otimes\mathcal E)(\Phi)$, where $\Phi=\ketbra\Phi\Phi$ and
$\ket\Phi=\sum_j\ket{jj}/\sqrt{N}$ is the maximally-entangled state on an
extended Hilbert space. Note also that the order of the tensor-product factors
is sometimes reversed such that $J(\mathcal E) = (\mathcal
E\otimes\mathds1)(\Phi)$ is used instead. Both definitions are equivalent, but
the matrix representation of the one given in \equref{eq:choimatrix} appears
more canonical.

Finally, there is a matrix representation that is particularly convenient for
numerical work called the \emph{Pauli transfer matrix} $G$. It is the
matrix representation of the linear map $\mathcal E$ with respect to the
Pauli basis
\begin{align}
  \label{eq:paulibasis}
  \mathcal P = \{ I, \sigma^x, \sigma^y, \sigma^z \}^{\otimes n}.
\end{align}
We denote the elements of $\mathcal P$ by $P_i$ for $i=0,\ldots,N^2-1$, where
$P_0 = I\otimes\cdots\otimes I$, $P_1 = I\otimes\cdots\otimes I \otimes
\sigma^x$,  $P_2 = I\otimes\cdots\otimes I \otimes \sigma^y$, and so on. The
basis $\mathcal P$ is orthogonal with respect to the Hilbert-Schmidt inner
product, $\mathrm{Tr}\,P_i^\dagger P_j = N \delta_{ij}$. The
corresponding normalized basis elements are given by $\widehat{P}_i =
P_i/\sqrt{N}$. In this basis, we obtain the Pauli transfer matrix of $\mathcal
E$ as
\begin{align}
  \label{eq:paulitransfermatrix}
  G_{ij} = \mathrm{Tr}\,\widehat{P}_i \mathcal E(\widehat{P}_j) = \frac 1 N \mathrm{Tr}\,P_i \mathcal E(P_j).
\end{align}
The matrix $G$ obeys the typical properties expected from matrix representations
of linear maps, namely that the columns contain the images of the basis
elements, and the composition of two maps $\mathcal E_1\circ\mathcal E_2$
corresponds to the matrix product $G_1G_2$ of their Pauli transfer matrices.
Furthermore, a quantum operation $\mathcal E$ preserves Hermiticity, so the
matrix $G$ consists only of real numbers, which makes it
convenient for numerical work. For trace-preserving and trace-decreasing quantum operations,
the matrix elements $G_{ij}$ are in the range $[-1,1]$. We also have $G_{0j} = \mathrm{Tr}\,\mathcal
E(P_j)/N$, so the first row of $G$ is given by $(1,0,\ldots,0)$ if and only if
$\mathcal E$ is trace-preserving. Another property of the Pauli transfer matrix
is that if the first column of $G$ is given by $(1,0,\ldots,0)^T$, the map
$\mathcal E$ is unital (meaning that $\mathcal E(\mathds1)=\mathds1$).
Finally, if all rows and all columns of $G$ contain exactly one non-zero entry of
magnitude $1$, the map $\mathcal E$ is known as a \emph{Clifford gate},  which
means that it maps Pauli operators to Pauli operators. Clifford gates are
useful as it can be shown that simulating quantum circuits containing only
Clifford gates is much less complex than simulating universal quantum circuits
\cite{gottesman1998gottesmanknilltheorem}.

A modern perspective including a graphical calculus for the common ways of
representing quantum operations is given in
\cite{Wood2015RepresentationQuantumOperations}. For the present work, the
(generalized) Kraus representation and the Pauli transfer matrix are the most
useful representations. The former is used in various places in
\chapref{cha:gateerrors} and \chapref{cha:fullcircuitsimulations}. The latter,
because of the convenient properties discussed above,  will prove particularly
useful for the tomography experiments studied in \secref{sec:gatesettomography}.

\subsection{Transformations of subsystems and leakage}
\label{sec:transformationssubsystemsleakage}

In \secref{sec:quantumgates}, we said that quantum theory typically considers
unitary transformations (see \appref{app:wigner} for the reasons behind this).
This notion is also contained in the formalism of quantum operations, although
it is not immediately apparent from the completely positive map $\mathcal E$
given in \equref{eq:krausrepresentation}. To illustrate this connection, we
extend the computational Hilbert space $\mathcal H_{2^n}$ given by
\equref{eq:multiqubithilbertspace} with another Hilbert space $\mathcal
H_{\mathrm{Env}}$,
\begin{align}
  \label{eq:hilbertspaceSystemEnv}
  \mathcal H = \mathcal H_{2^n} \otimes \mathcal H_{\mathrm{Env}},
\end{align}
which can be interpreted as a simple system-environment model. The idea of this
model is that the system and the environment, which are initially described
by a product state $\rho\otimes\rho_{\mathrm{Env}}$, undergo a joint unitary
transformation $U$, i.e., $\rho\otimes\rho_{\mathrm{Env}}\mapsto U (\rho\otimes\rho_{\mathrm{Env}}) U^\dagger$.
The final state of the system is then fully described by
\begin{align}
  \label{eq:krausrepresentationUnitaryExtendedSpace}
  \mathcal E(\rho) = \mathrm{Tr}_{\mathrm{Env}}\!\left(U(\rho\otimes\rho_{\mathrm{Env}})U^\dagger\right),
\end{align}
where $\mathrm{Tr}_{\mathrm{Env}}$ denotes the partial trace over the environment's degrees of freedom.
In this context, ``fully described'' means that the expectation value of each
observable $A$ on $\mathcal H_{2^n}$ is given by $\mathrm{Tr}A\mathcal E(\rho)$.

To relate \equref{eq:krausrepresentationUnitaryExtendedSpace} to the Kraus
representation given by \equref{eq:krausrepresentation}, we write the initial
state of the environment as a pure state $\rho_{\mathrm{Env}}=\ketbra{e_0}{e_0}$
(note that the space $H_{\mathrm{Env}}$ can always be chosen large enough so
that $\rho_{\mathrm{Env}}$ can be expressed as a pure state
\cite{NielsenChuang}). Choosing $E_\alpha=(\mathds1\otimes\bra{e_\alpha}) U
(\mathds1\otimes\ket{e_\alpha})$, where $\{\ket{e_\alpha}\}$ is a basis of
$\mathcal H_{\mathrm{Env}}$ completing $\ket{e_0}$, yields the Kraus
representation of $\mathcal E$ given by \equref{eq:krausrepresentation}. The
fact that the Kraus operators are given by $E_\alpha$ and $E_\alpha^\dagger$,
respectively, shows that each model of the form of
\equref{eq:krausrepresentationUnitaryExtendedSpace} is automatically completely
positive (see \cite{breuer2007openquantumsystems} for a more comprehensive
discussion).

The system-environment model can
also be extended to a description of leakage (cf.~\secref{sec:leakage})
by supplementing the Hilbert space given by \equref{eq:hilbertspaceSystemEnv}
with a direct sum for higher, non-computational states,
\begin{align}
  \label{eq:hilbertspaceSystemEnvLeakage}
  \mathcal H = (\mathcal H_{2^n} \oplus \mathcal H_{\mathrm{Leak}}) \otimes \mathcal H_{\mathrm{Env}},
\end{align}
where $\mathcal H_{\mathrm{Leak}}$ is defined in \equref{eq:multiqubithilbertspacewithleakage}.
Since $\mathcal H_{2^n}$ and $\mathcal H_{\mathrm{Leak}}$ form a direct sum and not
a direct product, one cannot trace out $\mathcal H_{\mathrm{Leak}}$ to obtain
a description of the computational subspace. Instead, one can project
the result of \equref{eq:krausrepresentationUnitaryExtendedSpace} onto $H_{2^n}$. The
corresponding quantum operation for a density matrix $\rho$ on $\mathcal H_{2^n}$ is then given by
\begin{align}
  \label{eq:krausrepresentationUnitaryExtendedSpaceProjection}
  \overline{\mathcal E}(\rho) = P\, \mathrm{Tr}_{\mathrm{Env}}\!\left(U(\rho\otimes\rho_{\mathrm{Env}})U^\dagger\right)P
  = \sum_\alpha \overline E_\alpha \rho \overline E_\alpha^\dagger,
\end{align}
where $\overline E_\alpha = P(\mathds1\otimes\bra{e_\alpha}) U (\mathds1
\otimes\ket{e_\alpha})P$, and $P$ denotes the projection onto the computational
subspace $H_{2^n}$. As before, the state $\overline{\mathcal E}(\rho)$ fully
describes the final state, in the sense that expectation values for observables
on $\mathcal H_{2^n}$ can be evaluated with $\overline{\mathcal E}(\rho)$.
Furthermore, since $\overline {\mathcal E}(\rho)$ can be written in Kraus form
(see \equref{eq:krausrepresentation}), it is automatically completely positive.
The only difference is that it may no longer be trace-preserving due to the
projection $P$.

Note that, in a description of an experiment, it may be that leakage can still
be described by trace-preserving quantum operations. This depends on how higher,
non-computational states show up in the measurement. For instance, if a
measurement reports each non-computational state as $\ket 1$, the quantum
operation would need to map each non-computational state to $\ket 1$ (which is a
non-unitary operation), but the corresponding quantum operation would be
trace-preserving. If, however, a non-computational state shows up randomly as 0
or 1, or if the measured event is classified as ``wrong'' and discarded, the
procedure would be described in terms of a trace-decreasing quantum operation.
In this case, the resulting probability distribution may need to be
renormalized, which is sometimes also be modeled by a nonlinear,
trace-preserving quantum operation of the form $\rho\mapsto\mathcal
E(\rho)/\mathrm{Tr}\,\mathcal E(\rho)$.

Finally, we remark that there are also simple quantum systems whose evolution
cannot be described by quantum operations of the form of
\equref{eq:krausrepresentation}, i.e., systems that cannot be described by
completely positive maps. In the context of
\equref{eq:krausrepresentationUnitaryExtendedSpace}, this may be the case if the
system and the environment do not start in a separable state such as
$\rho\otimes\rho_{\mathrm{Env}}$. See \cite{NielsenChuang} for a simple example
of such a system. Further characterizations of quantum systems beyond completely
positive maps are given in \cite{Carteret2008DynamicsBeyondCPmaps,
Dominy2016BeyondCPmaps}.

We utilize the formalism of quantum operations on mixed states
in \chapref{cha:gateerrors} to introduce error metrics on quantum gates and
the procedure of gate set tomography. Furthermore, practical applications of
quantum channels are given in \chapref{cha:fullcircuitsimulations} for
simple, effective error models and example circuits from the theory
of quantum fault tolerance.

%% file: chap3.tex
\chapter{Simulating superconducting transmon qubits}
\label{cha:simulation}

Over the past decades, superconducting circuits have emerged as a convenient
platform to engineer quantum mechanical systems with very few degrees of
freedom. This is remarkable in the sense that such quantum mechanical systems
are usually given by atoms or single electrons, i.e., \emph{microscopic} objects
that cannot be easily perceived. Superconducting circuits, however, are visible
to the naked eye. And although these macroscopic electrical systems are composed
of a huge number of atoms, they exhibit collective phenomena that can be
accurately described with Hamiltonians that have only a small number of charge
and phase variables. In this context, the field is commonly called
\emph{macroscopic quantum mechanics} and the systems are often referred to as
\emph{artificial atoms}. It is this property that makes superconducting circuits
ideal candidates to engineer quantum mechanical two-level systems that serve as
qubits for quantum information processors, and a huge variety of different
candidates has been studied in the literature. A review of prominent
superconducting architectures for quantum information processors and their
theoretical modeling can be found in \cite{Wendin2017SuperconductingReview}.

In this chapter, we introduce and describe the transmon simulator that is used
for most of the simulations presented in this work. \ssecref{sec:circuits}
reviews the quantum mechanical modeling of superconducting circuits with an
emphasis on the architecture used for transmon qubits. In
\secref{sec:transmonmodel}, we define the generic model Hamiltonian used to
describe the dynamics of a transmon quantum computer.
\ssecref{sec:simulationsoftware} describes the simulation packages developed for
this work, including the implementations of the numerical algorithms. The most
important transmon model systems used  for the results presented in the
remainder of this thesis are defined in \secref{sec:transmonmodelsystems}.
Finally, in \secref{sec:extractfoster}, we describe a procedure to extract
suitable model parameters from experiments or electromagnetic simulations of
the experimental devices.

\section{Superconducting circuits}
\label{sec:circuits}

A superconducting circuit is an electronic circuit in which the  circuit
elements are superconducting. This means that they conduct electricity with
practically zero resistance. Just like conventional electronic circuits, a
superconducting circuit includes basic circuit elements such as capacitors or
inductors. A circuit element of particular importance for quantum information
processors is the Josephson junction. It consists of two superconducting metals
with an insulating barrier in between. A key observation was that such a system
can be described by quantum mechanical tunneling processes of the
superconducting charge carriers, the Cooper pairs \cite{Josephson1962}.

\subsection{Quantum and classical descriptions}
\label{sec:quantumclasscical}

The quantum mechanical description of superconducting circuits emerged from a
course given by Devoret at the Les Houches School of Physics
\cite{devoret1997quantumfluctuations} and has recently been  reviewed and
updated in \cite{vooldevoret2017circuitqed}. The first step is to obtain a
classical Hamiltonian (or a Lagrangian) describing the dynamics of the
superconducting circuit. For basic circuit  elements such as capacitors or
inductors, the equations of motion determined by the Hamiltonian are the
corresponding differential equations of classical  electrodynamics (see
\cite{jackson1999classicalelectrodynamics}). Typically, the resulting
Hamiltonians are a set of harmonic oscillators with potential anharmonicities
from the Josephson junctions.

The quantum description in terms of a quantum Hamiltonian is obtained by
quantizing  the harmonic oscillators. The quantized Hamiltonian is the key
element to model dissipationless superconducting circuits for quantum
information processors. It can directly be used to obtain the dynamics of the
system as described by the TDSE given by \equref{eq:tdse1}.

For the quantum mechanical description of superconducting circuits  to be
appropriate, the following physical conditions are considered necessary
\cite{devoret1997quantumfluctuations, vooldevoret2017circuitqed}:
\begin{enumerate}
  \item The characteristic wave lengths corresponding to the oscillation
  frequencies need to be larger than the dimensions of the chip. In this case,
  the circuit is in the \emph{lumped element limit} and can be described by only
  a few collective degrees of freedom such as the charge or the flux (which
  could still follow the classical equations of motion).
  \item The temperature surrounding the system needs to be sufficiently low such
  that thermal fluctuations are much smaller than the spacing of the
  energy levels (although a high temperature does not by itself invalidate
  the quantum description).
  \item The widths of the energy levels (caused by dissipative elements such as
  resistors that induce damping and reduce quality factors) need to be much
  smaller than the spacing of the energy levels.
\end{enumerate}
For quantum harmonic oscillators, a well-known property is that the expectation
values still follow the classical equations of motion. Therefore, quantum
phenomena would only be observable in second-order expectation values (such as
the variance as a function of temperature). The presence of at least one
nonlinear circuit element (such as a Josephson junction) can make
quantum effects more directly observable
\cite{Devoret1985MacroscopicQuantumTunnelingJJ,
Martinis1987ExperimentalTestQuantumBehaviorJJ,
Clarge1988QuantumMechanicsMacroscopicPhaseJJ}.

Essentially, however, the superconducting systems under investigation are
macroscopic objects in the sense that they contain a large number of
constituents (such as aluminum atoms). Ultimately, one cannot prove that a
quantum description for a certain experiment is necessary. The reason is that
conceptually, one can never formally prove that a (potentially unknown) model
will not be able to describe the observations properly (see also
\cite{Jaynes1996ProbabilityInQuantumTheory}).

In the present case, descriptions based solely on classical electrodynamics are
occasionally explored. A nice overview concerning Josephson junctions is given
in \cite{Blackburn2016ClassicalInterpretationsQuantumJosephsonExperiments}.
Further ventures in this direction can be found in
\cite{Michielsen2005EventBasedSimulationOfUniversalQC,
Marchese2007ClassicalAnalogsRabiRamseyJosephsonJunctions,
GroenbechJensen2010EntanglementClassciallyCoupledJosephsonJunctions,
Kadin2016ProposedExperimentsTestQuantumComputing,
Blackburn2017QuantumCrossoverNotFoundJJ,
Ivakhnenko2018SimulatingQuantumPhenomenaClassicalOscillators}, the last of which
is particularly constructive. It would be interesting to investigate the
descriptive potential of these  classical models for the quantum computing
systems under investigation.

\begin{figure}
  \centering
  \includegraphics[height=.17\textheight]{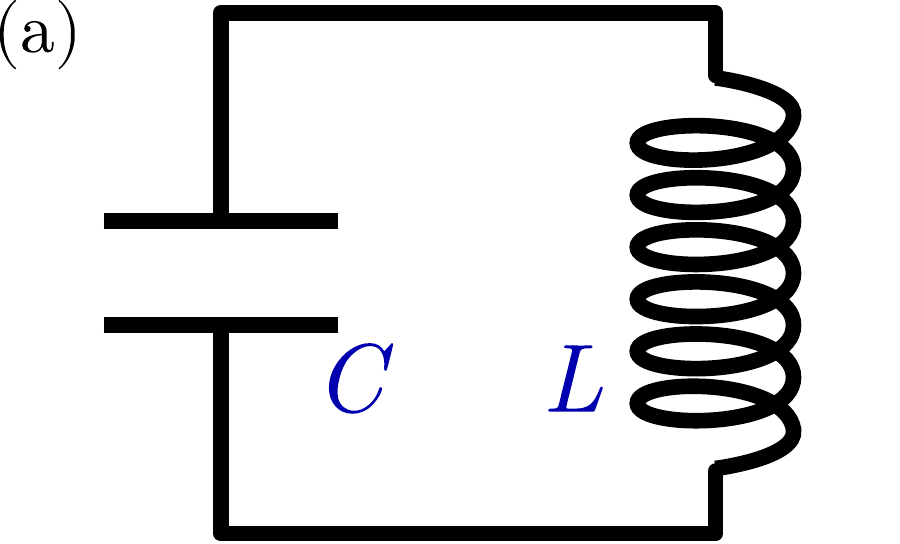}
  %\hfill
  \includegraphics[height=.17\textheight]{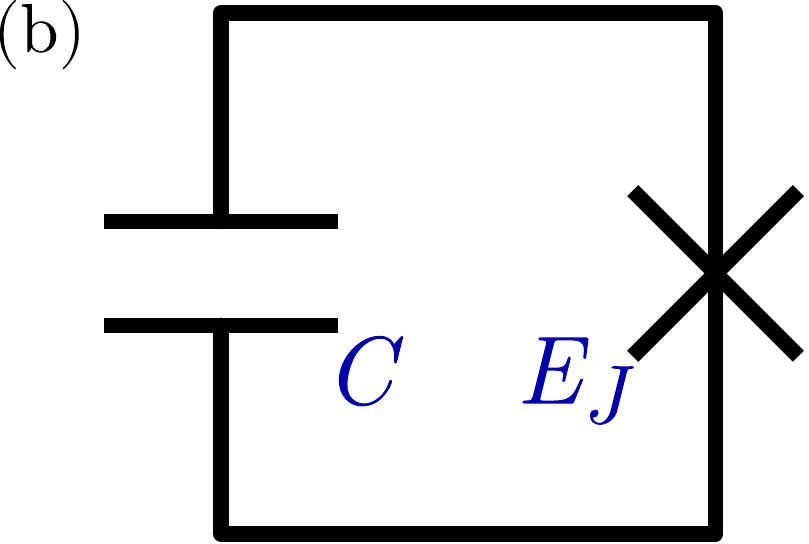}
  \caption{Lumped element circuit diagrams for (a) an LC resonator with capacitance
  $C$ and inductance $L$ and (b) a Josephson junction with capacitance $C$ and
  Josephson energy $E_J$.}
  \label{fig:lumpedelementcircuits}
\end{figure}

This project, however, is concerned with the quantum description. We show that
the discrete energy-level structure of the quantum mechanical description makes
the problem amenable to large-scale simulations.

To introduce the quantum descriptions, we consider two elementary circuit
components of a quantum information processor with transmon qubits. The first is
a simple LC resonator and the second is a Josephson junction. The lumped element
circuits for isolated, ideal versions of both are shown in
\figref{fig:lumpedelementcircuits}. We summarize the main relations; a more
detailed review can be found in \cite{vooldevoret2017circuitqed}.

\subsection{LC resonator}

The LC resonator shown in \figref{fig:lumpedelementcircuits}(a) is a typical
example for an harmonic oscillator in classical electrodynamics. Its
characteristic frequency is $\omega_{\mathrm{LC}} = 1/\sqrt{LC}$, where $C$ is the
capacitance and $L$ is the inductance. The dynamical variables are the charge
$Q$ of the capacitor and the magnetic flux $\Phi$ of the inductor. They obey the
typical differential equations for an harmonic oscillator, i.e., $\ddot Q +
\omega_{\mathrm{LC}}^2 Q = 0$ and $\ddot \Phi + \omega_{\mathrm{LC}}^2 \Phi = 0$.
The equations are Hamilton's equations of motion for the Hamiltonian
\begin{align}
  \label{eq:HamiltonianLCqphi}
  H_{\mathrm{LC}} &= \frac{Q^2}{2C} + \frac{\Phi^2}{2L}.
\end{align}

A quantum mechanical description of the system can be obtained by replacing the
variables $Q$ and $\Phi$ with operators on the Hilbert space $L^2(\mathbb R)$ of
square-integrable functions over $\mathbb R$. The operators have the spectrum
$\mathbb R$ and they obey a commutation relation $[Q,\Phi] \propto i$. They have
the same algebraic properties as  the position and momentum operators used in
the treatment of the quantum harmonic oscillator \cite{ballentine1998quantum}.
Thus the system can be diagonalized algebraically by introducing raising and
lowering operators $\hat a^\dagger$ and $\hat a$,
respectively, such that (up to a constant)
\begin{align}
  \label{eq:HamiltonianLCaadagger}
  H_{\mathrm{LC}} &= \omega_{\mathrm{LC}} \hat a^\dagger \hat a.
\end{align}
In quantum optics, the operator $\hat a^\dagger \hat a$ is called the photon
number operator and its eigenvalues $k\in\mathbb N_0$ represent the number
of photons in the corresponding monochromatic electric field of frequency
$\omega_{\mathrm{LC}}$ \cite{fox2006quantumoptics}.

\subsection{Josephson junction}

The circuit shown in \figref{fig:lumpedelementcircuits}(b) represents the
capacitance $C$ and the Josephson energy $E_J$ of a Josephson junction. The
capacitance originates directly from the
superconductor-insulator-superconductor geometry. It determines one of two
characteristic energy scales of a Josephson junction, namely the charging energy
\begin{align}
  \label{eq:JosephsonJunctionEC}
  E_C &= \frac{e^2}{2C},
\end{align}
where $e$ is the electron charge. $E_C$ can be interpreted as the electrostatic
energy required to charge the capacitor with an additional electron (although,
to be precise, the superconducting charge carriers are Cooper pairs with a
charge of $2e$, so the electrostatic energy changes in multiples of $4 E_C$).
The other characteristic energy scale is associated with the tunneling current
through the junction,
namely the Josephson energy
\begin{align}
  \label{eq:JosephsonJunctionEJ}
  E_J &= \frac{I_c}{2e},
\end{align}
where $I_c$ is the critical current representing the maximum tunneling current.
$E_J$ can be interpreted as the energy required for tunneling processes through
the insulating barrier of the Josephson junction.

A Hamiltonian for the circuit shown in
\figref{fig:lumpedelementcircuits}(b), analogous to $H_{\mathrm{LC}}$ given in
\equref{eq:HamiltonianLCqphi}, reads
\begin{align}
  \label{eq:HamiltonianJJqphi}
  H_{\mathrm{JJ}} &= \frac{Q^2}{2C} - E_J\cos(\Phi/\phi_0),
\end{align}
where $Q$ denotes the charge on the capacitor $C$, $\Phi$ stands for the
magnetic flux, and $\phi_0=\hbar/2e$ is the reduced flux quantum. The common
characterization of a Josephson junction as a weakly anharmonic oscillator can
be seen from \equref{eq:HamiltonianJJqphi} by expanding the cosine in powers of
$\Phi$ (see also \secref{sec:extractfoster}). The leading terms are then
the same as for the harmonic oscillator represented by $H_{\mathrm{LC}}$ in
\equref{eq:HamiltonianLCqphi}.

An important difference, however, is that the flux $\Phi$ in $H_{\mathrm{LC}}$
ranges from $-\infty$ to $\infty$. In $H_{\mathrm{JJ}}$, the dependence on
$\Phi$ is periodic such that restricting $\Phi/\phi_0\in[0,2\pi)$ is sufficient
to describe the system. A deeper understanding for the periodicity of $\Phi$ can
be gained by describing the superconductors on both sides of the Josephson
junction in the BCS theory of superconductivity \cite{BCS1957}. This approach
links the magnetic flux $\Phi$ directly to the phase difference of the
collective Cooper pair ground states (see \cite{Willsch2016Master,
vooldevoret2017circuitqed} for more information).

In the quantum mechanical version of $H_{\mathrm{JJ}}$, the charge $Q$ and flux
$\Phi$ are again represented by Hermitian operators. They are often made
dimensionless  such that $\hat{n} = Q/2e$ denotes the charge operator and
$\hat{\varphi} = \Phi/\phi_0$ denotes the phase operator. The Hamiltonian then
becomes
\begin{align}
  \label{eq:HamiltonianJJnphi}
  H_{\mathrm{JJ}} &=  4 E_{C} \hat n^2 - E_{J} \cos \hat\varphi.
\end{align}
Note that the spectra of the operators $\hat n$ and $\hat \varphi$ are not the
same as for the harmonic oscillator described in the previous section. In fact,
the spectrum of $\hat n$ is  $\mathbb Z$, where both positive and negative
integers are physically meaningful since $\hat n$ represents the difference in
the number of net charges on both  sides of the capacitor. Consequently, the
conjugate operator $\hat \varphi$ has  a bounded spectrum given by $[0,2\pi)$,
representing the periodic dependence on $\Phi$ in \equref{eq:HamiltonianJJqphi}.

The different spectra can lead to apparent mathematical paradoxes or
contradictions such as $1=0$ (cf.~\cite{dirac1927,
carruthers1968phaseandanglevariables,
loss1992commutationrelationsforperiodicoperators, barnett2007quantumphasebook}).
The contradictions stem from applying the operators to states that lie outside
their domain. This problem does not occur if the operator domains are
evaluated and adhered to properly (see \cite{Willsch2016Master}).

\subsection{Cooper pair box}
\label{sec:CPB}

A system known as the Cooper pair box (CPB) \cite{bouchiat1998cooperpairbox} can
be obtained by applying an external voltage bias $V_g(t)$ to the Josephson
junction shown in \figref{fig:lumpedelementcircuits}(b). The voltage bias can be
used to control the number of charges (i.e., Cooper pairs) stored on the
capacitor of the Josephson junction.  The external voltage is modeled by a
time-dependent offset to the number of charges given by $n_g(t) = C_g
V_g(t)/2e$, where $C_g$ is the capacitance of the gate through which the voltage
is applied. The Hamiltonian $H_{\mathrm{JJ}}$ given in
\equref{eq:HamiltonianJJnphi} then needs to be replaced by the time-dependent
Hamiltonian
\begin{align}
  \label{eq:HamiltonianCPBnphi}
  H_{\mathrm{CPB}} &=  4 E_{C} (\hat n - n_{g}(t))^2 - E_{J} \cos \hat\varphi.
\end{align}
An important property of the CPB is that the dynamics of the system can be
controlled externally through an electromagnetic pulse described by $n_g(t)$. In
quantum computing systems, $n_g(t)$ represents the pulses that are used to
implement quantum gates (see \chapref{cha:optimization}).

CPBs have been used to engineer qubits for quantum information processors for
almost 20 years \cite{Vion2002CPBqubitsQuantronium, blais2004circuitqed}, where
two low-energy states of the  multi-level system are used to define the qubits.
CPB qubits can be classified into two groups by means of the energy scales $E_J$
and $E_C$, namely the charge  qubit where $E_C\gg E_J$ (such that the charging
energy dominates), and the transmon qubit where $E_C \ll E_J$
\cite{koch2007transmon} (the name \emph{transmon} is derived from
\emph{transmission-line shunted plasma oscillation} and refers to a large shunt
capacitance that reduces $E_C$, see \equref{eq:JosephsonJunctionEC}).  The
transmon has turned out to be much more coherent
\cite{Paik2011HighCoherenceTransmon, Rigetti2012jaynescummingsbreaksdown} and
easier to control \cite{chowGambetta2012fidelitiesandcoherence}. It is now
employed in many experimental quantum computing platforms
\cite{NationalAcademyOfSciences2019QuantumComputing} by companies such as IBM
\cite{ibmquantumexperience2016}, Google \cite{Google2019QuantumSupremacy}, and
Rigetti Computing \cite{rigetti2017computing}.

\section{Transmon quantum computer model}
\label{sec:transmonmodel}

The topic of this thesis is the simulation of a system of transmons
and resonators by means of solving the TDSE,
\begin{align}
  i \frac{\partial}{\partial t} \ket{\Psi(t)} = H \ket{\Psi(t)},
  \label{eq:tdse3}
\end{align}
for a generic model Hamiltonian $H$. The solution to \equref{eq:tdse3}, namely
the state $\ket{\Psi(t)}$, can be used to compute any physically relevant
quantity. The model Hamiltonian $H$ includes $N_{\mathrm{Tr}}$ transmons described
as CPBs (see \equref{eq:HamiltonianCPBnphi}) to represent the qubits,
$N_{\mathrm{Res}}$ transmission-line resonators described as LC oscillators (see
\equref{eq:HamiltonianLCaadagger}), and various ways of coupling transmons and
resonators.

Throughout this work, we use units with $\hbar=1$ unless otherwise stated.
Often, time and energy are the only necessary physical quantities that appear in
this thesis. Typically, time is given in nanoseconds, and energies and
frequencies are given in gigahertz.

\subsection{Hamiltonian}

The full model Hamiltonian reads
\begin{subequations}
\begin{align}
  \label{eq:Htotal}
  H &= H_{\mathrm{Tr}} + H_{\mathrm{Res}} + H_{\mathrm{Int}},
\end{align}
where
\begin{align}
  \label{eq:HTr}
  H_{\mathrm{Tr}} &= \sum\limits_{i=0}^{N_{\mathrm{Tr}}-1} \left[ 4 E_{Ci} (\hat n_i - n_{gi}(t))^2 - E_{Ji} \cos \hat\varphi_i \right], \\
  \label{eq:HRes}
  H_{\mathrm{Res}} &= \sum\limits_{r=0}^{N_{\mathrm{Res}}-1} \Omega_r \hat a_r^\dagger\hat a_r + \sum\limits_{r=0}^{N_{\mathrm{Res}}-1} \Omega_r \epsilon_r(t)(\hat a_r + \hat a_r^\dagger), \\
  \label{eq:HIntTrRes}
  H_{\mathrm{Int}} &= \sum\limits_{r=0}^{N_{\mathrm{Res}}-1} \sum\limits_{i=0}^{N_{\mathrm{Tr}}-1} G_{ri} \hat n_i(\hat a_r + \hat a_r^\dagger) \\
  \label{eq:HIntRes}
  &+ \sum\limits_{0\le r<l<N_{\mathrm{Res}}} \lambda_{rl} (\hat a_r + \hat a_r^\dagger)(\hat a_l + \hat a_l^\dagger) \\
  \label{eq:HCC}
  &+ \sum\limits_{0\le i<j<N_{\mathrm{Tr}}} E_{Ci,Cj} \hat n_i \hat n_j.
\end{align}
\end{subequations}
The transmon Hamiltonian $H_{\mathrm{Tr}}$ given by \equref{eq:HTr} is a sum of
$N_{\mathrm{Tr}}$ CPB Hamiltonians (see \equref{eq:HamiltonianCPBnphi}). Each
transmon $i=0,\ldots,N_{\mathrm{Tr}}-1$ is defined by its charging energy $E_{Ci}$
and Josephson energy $E_{Ji}$ (see
\equaref{eq:JosephsonJunctionEC}{eq:JosephsonJunctionEJ}). $\hat n_i$ is the
number operator and $\hat\varphi_i$ is the phase operator of transmon $i$.
$n_{gi}(t)$ represents a time-dependent external pulse applied to the transmon.

The resonator Hamiltonian $H_{\mathrm{Res}}$ given by \equref{eq:HRes} is a sum of
$N_{\mathrm{Res}}$ LC resonators (see \equref{eq:HamiltonianLCaadagger}). Each
resonator $r=0,\ldots,N_{\mathrm{Res}}-1$ is defined by its frequency $\Omega_r$.
The raising and lowering operators of resonator $r$ are given by $\hat
a_r^\dagger$  and $\hat a_r$, respectively. $\epsilon_r(t)$ denotes a
time-dependent pulse  applied to resonator $r$.

The interaction between transmons and resonators is modeled by
$H_{\mathrm{Int}}$ given by \equsref{eq:HIntTrRes}{eq:HCC}. The first term given
in \equref{eq:HIntTrRes} represents a coupling between each resonator $r$ and
each transmon $i$ with coupling strength $G_{ri}$.  This coupling originates
from the capacitive interaction between the transmons and resonators (see
\cite{koch2007transmon}) since the number operator $\hat n_i$ describes the
amount of charges stored in the CPB's capacitor and $\hat a_r + \hat
a_r^\dagger$ represents the electric field of the resonator. In the simulation
model, in principle, each resonator $r$ can be coupled to each transmon $i$ such
that the matrix $G_{ri}$ is dense. However, this is hard to realize in
experiments for a system of more than a few transmons.  The architecture based
on coupling different transmons via resonators is characteristic of the
processors available on the IBM Q Experience \cite{ibmquantumexperience2016}.

The second interaction term given by \equref{eq:HIntRes} models an electric
dipole interaction between the resonators. It is typically used to describe
photonic interactions when a bath of resonators is used as a model for  an
environment \cite{koch2007transmon} (see \secref{sec:extractfoster}).

Finally, the term given by \equref{eq:HCC} describes a capacitive coupling
between the transmons,  where $E_{Ci,Cj}$ is the capacitive coupling energy
between transmon $i$ and $j$. A coupling mechanism of this form is used by
Rigetti Computing \cite{DidierRigetti2017AnalyticalParametericTransmon} (see
also \cite{ReagorRigetti2017universalparametric,
CaldwellRigetti2017parametrictransmongate}) and Google
\cite{barendsMartinis2013xmoncoherence, Neil2017GoogleBlueprintQuantumSupremacy,
Google2019QuantumSupremacy}.

The simulation model defined by \equsref{eq:Htotal}{eq:HCC} contains a large set
of parameters and time-dependent functions which can be set to arbitrary values.
However, note that in the practically relevant scenarios studied in this thesis,
only a small  subset of all parameters is non-zero. Typically, the non-zero
parameters are set to values which have been measured in corresponding
experimental setups. All energies are usually specified in gigahertz and in units of
$2\pi$ (using $\hbar=1$). Consequently, the characteristic time scale for these
systems is nanoseconds. See \tabref{tab:modelparameterscharacteristicscales} for
characteristic values of the model parameters.

\begin{table}
  \caption{Characteristic values for the parameters of the model Hamiltonian
  defined in \equsref{eq:Htotal}{eq:HCC}. The numbers are usually determined by
  the corresponding experimental setups. The complexity of the simulation grows
  with $N_{\mathrm{Tr}}$ and $N_{\mathrm{Res}}$, so these quantities are limited to
  keep the computational cost reasonable (see below). The exact numerical values
  of the energies, however, are irrelevant for the computational cost of the simulation.}
  \centering
  \label{tab:modelparameterscharacteristicscales}
  \begin{tabular}{@{}cccccccc@{}}
    \toprule
    $N_{\mathrm{Tr}}$ & $N_{\mathrm{Res}}$ & $E_{Ci}/2\pi$ & $E_{Ji}/2\pi$ & $\Omega_r/2\pi$ & $G_{ri}/2\pi$ & $\lambda_{rl}/2\pi$ & $E_{Ci,Cj}/2\pi$ \\
    \midrule
    1--20 & 1--20 & 0.1--1\,GHz & 10--15\,GHz & 4--7\,GHz & 10--100\,MHz & 1--20\,MHz & 10--100\,MHz \\
    \bottomrule
  \end{tabular}
\end{table}

\subsection{Choice of the basis}
\label{sec:choiceofbasis}

To solve the TDSE given by \equref{eq:tdse3} on a digital computer, it is
necessary to choose a basis for the state $\ket{\Psi(t)}$ such that the
solution of the TDSE corresponds to a set of complex coefficients to be
determined numerically.

For the resonators described by $H_{\mathrm{Res}}$ (see \equref{eq:HRes}), an
obvious choice for the basis vectors are the eigenstates of the photon
number operator $\hat a_r^\dagger
\hat a_r$ for resonator $r$. These are the photon number states (or Fock
states), denoted by $\ket{k_r}$ for $k_r\in\mathbb N_0$ such that
\begin{align}
  \label{eq:resonatornumberoperator}
  \hat a_r^\dagger \hat a_r = \sum\limits_{k_r} k_r \ketbra{k_r}{k_r}.
\end{align}
All other terms in $H_{\mathrm{Res}}$ are proportional to the operator
$\hat a_r + \hat a_r^\dagger$, which corresponds to the electric field in
resonator $r$. Its representation with respect to the photon number states is
\begin{align}
  \label{eq:resonatorelectricfieldoperator}
  \hat a_r + \hat a_r^\dagger = \sum\limits_{k_r} \sqrt{k_r+1} (\ketbra{k_r}{k_r+1}+\ketbra{k_r+1}{k_r})
  = \begin{pmatrix}
    0 & 1        & \\
    1 & 0        & \sqrt{2} \\
      & \sqrt{2} & 0        & \sqrt{3} \\
      &          & \sqrt{3} & 0      & \ddots \\
      &          &          & \ddots & \ddots \\
  \end{pmatrix},
\end{align}
i.e., a tridiagonal symmetric matrix with zeros on the diagonal.

For the transmons described by $H_{\mathrm{Tr}}$ (see \equref{eq:HTr}), there are
two possible choices of basis states that come to mind. One option is given by
the eigenstates of the charge number operators $\hat n_i$. They are called
\emph{charge states} and are denoted by $\ket{n_i}$ for $n_i\in\mathbb Z$ such
that
\begin{align}
  \label{eq:transmonnumberoperator}
  \hat n_i = \sum\limits_{n_i} n_i \ketbra{n_i}{n_i}.
\end{align}
The interpretation of the state $\ket{n_i}$ is that the capacitor of the
Josephson junction is charged with a net charge of $n_i$ Cooper pairs (note that
the net charge can be both positive and negative such that $n_i\in\mathbb Z$).
An advantage of this basis is that each operator in \equsref{eq:Htotal}{eq:HCC}
associated with transmon $i$ has a straightforward representation with respect
to this basis. Specifically, the operator $\cos\hat\varphi_i$ is given by
\begin{align}
  \label{eq:transmonphaseoperator}
  \cos\hat\varphi_i = \sum\limits_{n_i} \frac 1 2 (\ketbra{n_i}{n_i+1} + \ketbra{n_i+1}{n_i}).
\end{align}
It couples the charge states $\ket{n_i}$ and $\ket{n_i\pm1}$. In other words, it
describes the tunneling processes of Cooper pairs from one side of the Josephson
junction to the other, thereby changing the net  charge on the capacitor by one.
A CPB qubit simulator solving the TDSE in this basis was studied in
\cite{Willsch2016Master}.

However, for transmon simulations, a much more appropriate basis is given by the
transmon eigenstates. They are denoted by $\ket{m_i}$ for $m_i\in\mathbb N_0$
and correspond to the eigenstates of the full Josephson junction Hamiltonian
$H_{\mathrm{JJ}}$ given in \equref{eq:HamiltonianJJnphi}. This means that for each
transmon $i$, we have
\begin{align}
  \label{eq:transmoneigenstates}
  4 E_{Ci} \hat n_i^2 - E_{Ji} \cos \hat\varphi_i = \sum\limits_{m_i} E_{i,m_i}^{\mathrm{Tr}} \ketbra{m_i}{m_i},
\end{align}
where $E_{i,m_i}^{\mathrm{Tr}}$ denotes the corresponding eigenvalues (we
typically shift the eigenvalues $E_{i,m_i}^{\mathrm{Tr}}$ by the respective
ground-state energy $E_{i,0}^{\mathrm{Tr}}$ such that $E_{i,0}^{\mathrm{Tr}}=0$
for all $i$).  The ground state $\ket{m_i=0}$ and the first excited state
$\ket{m_i=1}$ are the so-called qubit states or computational states of each
transmon. Correspondingly, the energy difference $\tilde\omega_i =
E_{i,1}^{\mathrm{Tr}} - E_{i,0}^{\mathrm{Tr}}$ between the two lowest states is
called the qubit frequency. Note that in practice, it may be beneficial to use a
slightly shifted frequency $\omega_i$ for pulse control because of the presence
of other components such as resonators (see the discussion around
\equref{eq:singlequbitblochvectorTimeEvolutionRotating} or
\secref{sec:optimizatingpulseparametersResults}). More insight into the spectrum
can be gained by observing that the $\hat\varphi_i$ dependence of $\cos
\hat\varphi_i$ in leading order is $\hat\varphi_i^2$  (up to a constant). This
means that \equref{eq:transmoneigenstates} can be seen as an harmonic oscillator
with anharmonic higher-order corrections. It can be shown that the spectrum
$\{E_{i,m_i}^{\mathrm{Tr}}\}$ slightly deviates from an equidistant spectrum by
an anharmonicity $\alpha_i = E_{i,2}^{\mathrm{Tr}} - E_{i,1}^{\mathrm{Tr}} -
\tilde\omega_i\approx - E_{Ci}$ \cite{koch2007transmon}. The energy difference
between higher levels $m_i'+1$ and $m_i'$ is approximately reduced by
$m_i'|\alpha_i|$ such that $E_{i,m_i'+1}^{\mathrm{Tr}} -
E_{i,m_i'}^{\mathrm{Tr}} \approx \tilde\omega_i + \alpha m_i'$
\cite{gambetta2013controlIFF}.

To set up a simulation for \equsref{eq:Htotal}{eq:HCC} in the transmon basis
$\ket{m_i}$, we need to find a representation for the  charge operator $\hat
n_i$ given in \equref{eq:transmonnumberoperator} with respect to this basis. One
approach is given in \cite{DidierRigetti2017AnalyticalParametericTransmon},
where the authors perform a systematic perturbation theory up to
$25^{\mathrm{th}}$ order in the parameter
$\overline{\xi_i}=\sqrt{2E_{Ci}/E_{Ji}}$. For typical device parameters (see
\tabref{tab:modelparameterscharacteristicscales}), $\overline{\xi_i}$ takes
values between $0.1$ and $0.5$.

\begin{figure}[t]
  \centering
  \includegraphics[width=\textwidth]{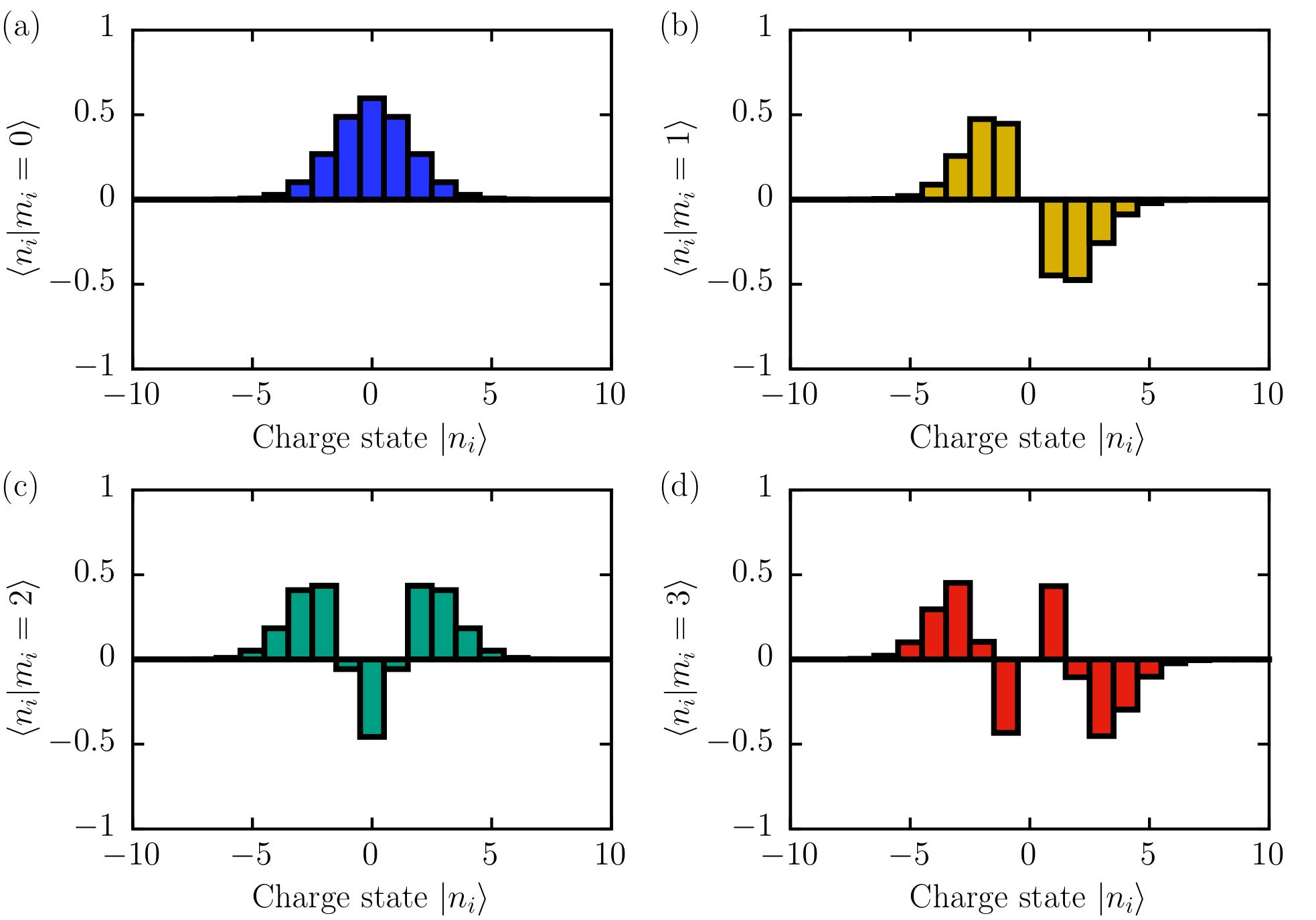}
  \caption{Characteristic distribution of the four lowest transmon eigenstates
  (a) $\ket{m_i=0}$, (b) $\ket{m_i=1}$, (c) $\ket{m_i=2}$, (d) $\ket{m_i=3}$, in
  terms of the charge basis $\ket{n_i}$. Shown are the coefficients
  $\braket{n_i|m_i}$ obtained from a numerical diagonalization of the
  tridiagonal matrix given by the left-hand side of
  \equref{eq:transmoneigenstates} for $E_{Ci} = 2\pi\times\SI{0.222}{GHz}$ and
  $E_{Ji} = 2\pi\times\SI{12.61}{GHz}$ (the KIT system,
  cf.~\tabref{tab:devicekit}). 101 charge states $n_i=-50,\ldots,50$ have been
  used for the diagonalization. Note that the sign (i.e., the global phase)  of
  $\braket{n_i|m_i}$ is irrelevant, but the relative signs of the bars  with
  even and odd parity around $n_i=0$ is characteristic.}
  \label{fig:transmonstatesinchargebasis}
\end{figure}

However, since this work is based on a computer
simulation where we have access to all numerical values of the parameters, we
take a different approach to obtain the representation of $\hat n_i$ in the
transmon basis: We construct the tridiagonal matrix for the transmon Hamiltonian
given by \equref{eq:transmoneigenstates} in terms of the charge states
$\ket{n_i}$ and use numerical diagonalization to obtain the coefficients
$\braket{n_i|m_i}$, i.e., the representation of the charge states in the
transmon  basis. The global phase of the transmon states is chosen such that the
coefficients $\braket{n_i|m_i}\in\mathbb R$. The characteristic distribution of
these coefficients is shown in \figref{fig:transmonstatesinchargebasis}.
Typically, a truncation of the tridiagonal matrix given by
\equref{eq:transmoneigenstates} to 50 charge states below and above $n_i=0$
suffices to obtain the coefficients $\braket{n_i|m_i}$ for the lowest states
$m_i=0,1,2,\ldots$ to machine precision.

Given the coefficients $\braket{n_i|m_i}$, we can obtain the matrix
representation of the charge operator $\hat n_i$ in the transmon basis. The
matrix is symmetric since $\braket{n_i|m_i}\in\mathbb R$, and its characteristic
form is
\begin{align}
  \label{eq:transmonchargeoperatortransmonbasis}
  \hat n_i = \sum\limits_{m_im_i'} n_i^{(m_i,m_i')} \ketbra{m_i}{m_i'}
  = \begin{pmatrix}
      & n_i^{(0,1)}        &  & n_i^{(0,3)}\\
    n_i^{(0,1)}  &          & n_i^{(1,2)}  & &\cdots \\
      & n_i^{(1,2)} &          & n_i^{(2,3)} \\
    n_i^{(0,3)}  &          & n_i^{(2,3)} & & \ddots \\
       & \vdots &  & \ddots
  \end{pmatrix},
\end{align}
where the coefficients $\smash{n_i^{(m_i,m_i')}} = \sum_{n_i}\smash{
n_i\braket{m_i|n_i}\braket{n_i|m_i'}}$ are non-zero only if $m_i$ and $m_i'$
differ by an odd number.  Moreover, if this odd number is large (i.e., the
coefficient $n_i^{(m_i,m_i')}$ is far away from the diagonal), the matrix
element $n_i^{(m_i,m_i')}$ tends to zero. The coefficients $n_i^{(m_i,m_i\pm1)}$
on the subdiagonal are approximately equal to those of the tridiagonal matrix
$\hat a_r+ \hat a_r^\dagger$ shown in
\equref{eq:resonatorelectricfieldoperator}. In fact, for the approximation of
the transmon as an anharmonic oscillator often found in the literature, $\hat
n_i$ is effectively replaced by the operator $-(E_{Ji}/32E_{Ci})^{1/4} (\hat b_i +
\hat b_i^\dagger)$, where $\hat b_i + \hat b_i^\dagger$ is represented by the
same tridiagonal matrix (see
\secref{sec:singletransmonresonatorsystemPerturbative}). The coefficient
$n_i^{(0,3)}$, which is dropped in this approximation, is typically smaller than
the other matrix elements by a factor of 10--50. Nevertheless, we observed that
the coefficient is still significant for an accurate simulation of the time
evolution. See \secref{sec:singletransmonresonatorsystem} and, in particular,
\figref{fig:freekitcomparetimeevolutions}  for an empirical investigation of
approximations of this type.

The total Hilbert space of $N_{\mathrm{Tr}}$ transmons and $N_{\mathrm{Res}}$
resonators is given by
\begin{align}
  \label{eq:HilbertSpaceInfinity}
  \mathcal H_{\mathrm{total}} % &= \left( \bigotimes\limits_{r=0}^{N_{\mathrm{Res}}-1} \mathrm{span}\{ \ket{k_r} : k_r\in\mathbb N_0\} \right) \otimes \left( \bigotimes\limits_{i=0}^{N_{\mathrm{Tr}}-1} \mathrm{span}\{ \ket{m_i} : m_i\in\mathbb N_0\} \right)\\
  % &= \mathrm{span}\left\{ \left( \bigotimes\limits_{r=0}^{N_{\mathrm{Res}}-1} \bigoplus\limits_{k_r\in\mathbb N_0} \ket{k_r} \right) \otimes \left( \bigotimes\limits_{i=0}^{N_{\mathrm{Tr}}-1} \bigoplus\limits_{m_i\in\mathbb N_0} \ket{m_i} \right)\right\},
  &= \left( \bigotimes\limits_{r=0}^{N_{\mathrm{Res}}-1} \bigoplus\limits_{k_r\in\mathbb N_0}\mathrm{span}\{ \ket{k_r} \} \right) \otimes \left( \bigotimes\limits_{i=0}^{N_{\mathrm{Tr}}-1} \bigoplus\limits_{m_i\in\mathbb N_0}\mathrm{span}\{ \ket{m_i} \} \right),
\end{align}
which is an instance of the system-environment model with leakage defined in
\equref{eq:hilbertspaceSystemEnvLeakage}. Although the dimension of $\mathcal
H_{\mathrm{total}}$ is infinite, the advantage of the chosen basis is that only a
small number of basis states needs to be implemented to describe the dynamics
of the system. To be precise, a transmon-resonator system initialized in the
ground  state usually requires four states from the transmon basis for each
transmon  (see \secref{sec:accuracy} below), whereas in the charge basis
$\{\ket{n_i}\}$, at least 17 states $n_i=-8,\ldots,8$ need to be taken into
account to obtain the dynamics to sufficient precision \cite{Willsch2016Master}.
A similar argument holds for the Fock states $\ket{k_r}$ used for each of the
resonators.

Therefore, for the majority of this work, we restrict the basis to four states
for each  of the $N_{\mathrm{Tr}}+N_{\mathrm{Res}}$ subsystems.  We found that  this
basis is sufficient to describe the dynamics for almost all applications.
Furthermore, it enables a large-scale high-performance implementation of the
simulation (see \secref{sec:numericalalgorithm}). The basis states are given by
the four lowest energy eigenstates $\ket{m_i}$ of each transmon and four Fock
states $\ket{k_r}$ starting at some $k^{\mathrm{offset}}_r\in\mathbb N_0$  for
each of the resonators. The offset will be set to $k^{\mathrm{offset}}_r = 0$ for
most of the simulations such that the resonator parts of the basis consist of
the four lowest energy Fock states for each resonator. Thus, the effective
Hilbert space is given by
\begin{align}
  \label{eq:HilbertSpaceTruncated}
  \mathcal H &= \mathrm{span}\{ \ket{k_0}\!\ket{k_1}\!\cdots\!\ket{k_{N_{\mathrm{Res}}-1}}\!\ket{m_0}\!\ket{m_1}\!\cdots\!\ket{m_{N_{\mathrm{Tr}}-1}} \},
  %\mathcal H &= \mathrm{span}\{ \ket{k_0k_1\cdots k_{N_{\mathrm{Res}}-1}m_0m_1\cdots m_{N_{\mathrm{Tr}}-1}} : k_r \}
\end{align}
where each $k_r \in\{ k^{\mathrm{offset}}_r, k^{\mathrm{offset}}_r+1,
k^{\mathrm{offset}}_r+2, k^{\mathrm{offset}}_r+3\}$ and each $m_i \in \{0,1,2,3\}$.
The total number of states (i.e., the dimension of the Hilbert space) is thus
$\mathrm{dim}(\mathcal H) = 4^{N_{\mathrm{Res}}+N_{\mathrm{Tr}}}$.

\section{Simulation toolkit}
\label{sec:simulationsoftware}

The software package developed for this thesis consists of several programs for
the simulation of transmon quantum computers. The central tool, \texttt{solver},
computes the solution $\ket{\Psi(t)}$ of the TDSE given in \equref{eq:tdse3}, by
which any physically relevant quantity of the system can be evaluated.
Additionally, the toolkit contains several convenience programs to either
prepare a run for \texttt{solver} or to evaluate its results. The individual
software components are:

\begin{itemize}
  \item \texttt{solver}: Compute the numerical solution $\ket{\Psi(t)}$ of the
  TDSE given by \equref{eq:tdse3} for the full model Hamiltonian $H$ given by
  \equsref{eq:Htotal}{eq:HCC}. $H$ is characterized by a set of numerical values
  for the parameters and pulse shapes for the time-dependent functions. The
  result $\ket{\Psi(t)}$ is the time evolution of a given initial state
  $\ket{\Psi(0)}$.

  \item \texttt{evaluator}: Given the solution $\ket{\Psi(t)}$ produced by
  \texttt{solver}, compute expectation values such as the Bloch vectors given in
  \equref{eq:multiqubitblochvector} or perform basis transformations to
  rotating frames. The program can also be used to evaluate the accuracy and the
  overlap between different solutions.

  \item \texttt{visualizer}: Generate 3D visualizations of the time evolution
  of the Bloch vectors computed from the solution $\ket{\Psi(t)}$ using
  real-time rendering.

  \item \texttt{optimizer}: Optimize quantum gates by finding suitable pulses
  (see \chapref{cha:optimization}). This program invokes \texttt{solver}
  repeatedly for varying pulse parameters and evaluates the success of
  implementing a certain gate by studying the time evolution produced by
  \texttt{solver}.

  \item \texttt{compiler}: Given a quantum circuit specified as a set of quantum
  gates in a certain file format such as OpenQASM \cite{Cross2017openqasm2}
  or the JUQCS instruction set defined in \cite{DeRaedt2018MassivelyParallel},
  generate the pulse shape information required for \texttt{solver}. The program
  can use the results produced by \texttt{optimizer} and can also produce
  appropriate configuration files to set up the environment for \texttt{solver}.
\end{itemize}

In this section, we give a detailed description of the algorithms underlying
\texttt{solver}, \texttt{evaluator}, and \texttt{visualizer}. Advanced
functionalities implemented by \texttt{optimizer} and \texttt{compiler} are
discussed in the \secaref{sec:optimizatingpulseparameters}{sec:compiler},
respectively.

\subsection{Numerical algorithm: \texorpdfstring{$\texttt{solver}$}{solver}}
\label{sec:numericalalgorithm}

The task of \texttt{solver} is to simulate the time evolution $\ket{\Psi(t)}$
of a given initial state $\ket{\Psi(0)}$. Formally, the time evolution can be
expressed as
\begin{align}
  \label{eq:psioftTimeEvolutionOperator}
  \ket{\Psi(t)} = \mathcal U(t,0) \ket{\Psi(0)},
\end{align}
where $\mathcal U$ denotes the unitary time-evolution operator of the system
given by
\begin{align}
  \mathcal U(t_1,t_0)&=\mathcal T \exp\!\left(-i\int_{t_0}^{t_1} H(\tilde t)\, \mathrm d\tilde t \right).
  \label{eq:totaltransmontimeevolutionoperator}
\end{align}
In this expression, $H(\tilde t)$ is the time-dependent model Hamiltonian
defined in \equsref{eq:Htotal}{eq:HCC} and $\mathcal T$ is the time-ordering
symbol (see \cite{bruus2004manybodyquantum}). The time-evolution operator
satisfies the group property $\mathcal U(t_1,t_0)=\mathcal U(t_1,t')\,\mathcal
U(t',t_0)$ such that the time evolution from $0$ to $t$ can be written as a
product of successive time evolutions with a time step $\tau$, i.e.,
\begin{align}
  \label{eq:psioftTimeEvolutionOperatorSteps}
  \ket{\Psi(t)} = \mathcal U(t,t-\tau) \cdots \mathcal U(2\tau,\tau)\,\mathcal U(\tau,0)\ket{\Psi(0)}.
\end{align}
In each step, the current state vector $\ket{\Psi(t_0)}$
is updated by computing
\begin{align}
  \ket{\Psi(t_0+\tau)} = \mathcal U(t_0+\tau,t_0) \ket{\Psi(t_0)},
  \label{eq:psioftTimeEvolutionOperatorStepSinglePropagation}
\end{align}
starting from $t_0=0$. Here, we have chosen the time  step $\tau$ to be
constant, but in principle it may be updated dynamically for each time step to
speed up the simulation, if the problem allows it. In general, the time step
$\tau$ needs to be sufficiently small such that $H(t_0)$ and $H(t_0+\tau)$ are
well approximated by $H(t_0+\tau/2)$ (this condition is affected by the time
dependence of the pulses $n_{gi}(t)$ and $\epsilon_r(t)$ in
\equaref{eq:HTr}{eq:HRes}, respectively). In this case, the time-evolution
operator $\mathcal U(t_0+\tau,t_0)$  (which is equal to
\equref{eq:totaltransmontimeevolutionoperator} for $t_1=t_0+\tau$) simplifies to
$\mathcal U(t_0+\tau,t_0)= e^{-i \tau H(t_0+\tau/2)}$. The update rule for the
time evolution given by
\equref{eq:psioftTimeEvolutionOperatorStepSinglePropagation} then becomes
\begin{align}
  \ket{\Psi(t_0+\tau)} = e^{-i \tau H(t_0+\tau/2)} \ket{\Psi(t_0)}.
  \label{eq:psioftTimeEvolutionOperatorStepSinglePropagationTimeIndependent}
\end{align}
The core of the transmon simulator implements this operation to
propagate the state vector according to
\equref{eq:psioftTimeEvolutionOperatorSteps}. Obviously, there are many
different ways to implement
\equref{eq:psioftTimeEvolutionOperatorStepSinglePropagationTimeIndependent}, but
whether a simulation for large systems is feasible or not heavily depends on the
choice of basis and the particular algorithm. A suitable choice for the
basis is the product basis of resonator eigenstates $\ket{k_r}$ and transmon
eigenstates $\ket{m_i}$ discussed in \secref{sec:choiceofbasis}.

In this basis, the solution $\ket{\Psi(t)}$ is determined by its complex
expansion coefficients $\psi_{k_0k_1\cdots m_0m_1\cdots}(t)$ defined by
\begin{align}
  \label{eq:psioftsolutioncoefficients}
  \ket{\Psi(t)} = \sum\limits_{\substack{k_0k_1\cdots\\ m_0m_1\cdots}}
  \psi_{k_0k_1\cdots m_0m_1\cdots}(t)
  \ket{k_0k_1\cdots m_0m_1\cdots }.
\end{align}
As each $k_r$ and $m_i$ in the truncated Hilbert space $\mathcal H$ (see
\equref{eq:HilbertSpaceTruncated}) can take one of four different values,
$\ket{\Psi(t)}$ is described by an array of
$4^{N_{\mathrm{Res}}+N_{\mathrm{Tr}}}$ complex coefficients. The index
$k_0k_1\cdots m_0m_1\cdots $ can thus be efficiently encoded in an integer of
$2(N_{\mathrm{Res}}+N_{\mathrm{Tr}})$ bits. We introduce the notation
\begin{align}
  \label{eq:KMindexnotation}
  \texttt{KM} =
  \underbrace{\overarrow{\texttt{00}}{k_0}\,\overarrow{\texttt{00}}{k_1}\,\cdots\,\texttt{00}}_{\texttt K}\,
  \underbrace{\overarrow{\texttt{00}}{m_0}\,\overarrow{\texttt{00}}{m_1}\,\cdots\,\texttt{00}}_{\texttt M}
  ,\:\:\ldots,\:\:
  \underbrace{\texttt{11}\,\cdots\,\texttt{11}}_{\texttt K}\,
  \underbrace{\texttt{11}\,\cdots\,\texttt{11}}_{\texttt M}
\end{align}
for this integer, where each group of two bits encodes the respective value of
$k_r \in\{
k^{\mathrm{offset}}_r,k^{\mathrm{offset}}_r+1,k^{\mathrm{offset}}_r+2,k^{\mathrm{offset}}_r+3\}$
and $m_i \in \{0,1,2,3\}$. Using this notation, the expansion in
\equref{eq:psioftsolutioncoefficients} reads
\begin{align}
  \label{eq:psioftsolutioncoefficientsKM}
  \ket{\Psi(t)} = \sum\limits_{\texttt{KM}}
  \psi_{\texttt{KM}} (t)
  \ket{\texttt{KM}}.
\end{align}
We choose unsigned 64-bit integers to represent the index $\texttt{KM}$ in the
implementations (see \appref{app:implementations}). This choice is reasonable
since the number of bits required for $N_{\mathrm{Tr}}$ transmons and
$N_{\mathrm{Res}}$ resonators is $2(N_{\mathrm{Tr}}+N_{\mathrm{Res}})$. This
means that 32-bit integers would impose the unnecessary constraint
$N_{\mathrm{Tr}}+N_{\mathrm{Res}}\le16$ even though larger simulations are
technically feasible. Integers larger than 64 bits are not required since there
exists no system today that can store more than $2^{64}$ complex numbers.

We have already studied the matrix representation with respect to
$\ket{\texttt{KM}}$ for parts of the Hamiltonian in \secref{sec:choiceofbasis}.
This knowledge can be used to derive an algorithm to implement the operation
given by
\equref{eq:psioftTimeEvolutionOperatorStepSinglePropagationTimeIndependent}.

\subsubsection{Suzuki-Trotter product-formula algorithm}

The algorithm that we use to implement
\equref{eq:psioftTimeEvolutionOperatorStepSinglePropagationTimeIndependent} is
the second-order Suzuki-Trotter product-formula algorithm
\cite{deraedt1987productformula}.  It is based on generalizations of the
Lie-Trotter formula \cite{Lie1888Transformationsgruppen,  Trotter1959Formula,
suzuki1976GeneralizedTrotterFormula, Suzuki1985ProductFormulaError}. The
algorithm belongs to a family of explicit and unconditionally stable algorithms
for linear parabolic difference equations and has been applied numerous times to
solve the TDSE, especially also in the context of quantum computing
\cite{DeRaedt2000QCE, DeRaedt2002QuantumSpinDynamics, deraedt2004computational,
Willsch2016Master, Willsch2017GateErrorAnalysis,
Willsch2018TestingFaultTolerance, WillschMadita2020PhD,
Willsch2020FluxQubitsQuantumAnnealing}. Moreover, rigorous error bounds have
been proven to assess the accuracy of the algorithm
\cite{deraedt1987productformula,
huyghebaert1990productFormulaTimeDependentErrorBounds}. Recently, the bounds
have been extended to tight error bounds for observables
\cite{WillschMadita2020PhD}. We test these bounds in
\secref{sec:accuracyperformancebenchmark}.

The first step in the derivation of the algorithm is to split the Hamiltonian
$H$ given by \equsref{eq:Htotal}{eq:HCC} at time $\tilde t = t_0+\tau/2$ into a
part $H_0$ that is diagonal with respect to $\ket{\texttt{KM}}$ and the
remaining part $W$. We obtain up to an irrelevant constant (which would only
lead to a global phase in the solution),
\begin{subequations}
  \begin{align}
    \label{eq:HtotalDecomposition}
    H &= H_0 + W , \\
    \label{eq:HtotalDecompositionH0}
    H_0 &= \sum\limits_{i=0}^{N_{\mathrm{Tr}}-1} ( 4 E_{Ci} \hat{n}_i^2 - E_{Ji} \cos \hat{\varphi}_i ) + \sum\limits_{r=0}^{N_{\mathrm{Res}}-1} \Omega_r \hat a_r^\dagger\hat a_r ,\\
    \label{eq:HtotalDecompositionW}
    W &= \sum\limits_{i=0}^{N_{\mathrm{Tr}}-1} -8 E_{Ci} n_{gi}(\tilde t) \hat{n}_i \\
      &\hphantom{=}+ \sum\limits_{r=0}^{N_{\mathrm{Res}}-1} \Omega_r \varepsilon_r(\tilde t)(\hat{a}_r+\hat{a}_r^\dagger) \\
      &\hphantom{=}+ \sum\limits_{r=0}^{N_{\mathrm{Res}}-1} \sum\limits_{i=0}^{N_{\mathrm{Tr}}-1} G_{ri} \hat{n}_i (\hat{a}_r+\hat{a}_r^\dagger) \\
      &\hphantom{=}+ \sum\limits_{0\le r<l<N_{\mathrm{Res}}} \lambda_{rl} (\hat{a}_r+\hat{a}_r^\dagger)(\hat{a}_l+\hat{a}_l^\dagger) \\
    \label{eq:HtotalDecompositionEnd}
      &\hphantom{=}+ \sum\limits_{0\le i<j<N_{\mathrm{Tr}}} E_{Ci,Cj} \hat{n}_i \hat{n}_j.
  \end{align}
\end{subequations}
The second-order product-formula decomposition for $\mathcal U(t_0+\tau,t_0)= e^{-i \tau (H_0+W)}$ is given by
\begin{align}
  \label{eq:productformulasecondorder}
  \widetilde{\mathcal U} = e^{-i \tau H_0/2} \, e^{-i \tau W} \, e^{-i \tau H_0/2}.
\end{align}
This decomposition is equal to $\mathcal U(t_0+\tau,t_0)$ up to second order in
$\tau$. It is the only approximation apart from the discretization in time, and
the error can be well controlled by the time step $\tau$ using the rigorous
bounds given in \cite{deraedt1987productformula, WillschMadita2020PhD}. To apply
the decomposition to the state vector given in
\equref{eq:psioftsolutioncoefficientsKM}, we need to derive its action on the
basis state $\ket{\texttt{KM}}$.

For the diagonal part $e^{-i \tau H_0/2}$, we make use of the spectral
representations given in
\equaref{eq:resonatornumberoperator}{eq:transmoneigenstates}, yielding
\begin{align}
  \label{eq:HtotalDecompositionH0Diagonal}
    H_0 &= \sum\limits_{\texttt{KM}} \left( k_0\Omega_0 + k_1\Omega_1 + \cdots + E_{0,m_0}^{\mathrm{Tr}} + E_{1,m_1}^{\mathrm{Tr}} + \cdots \right) \ketbra{\texttt{KM}}{\texttt{KM}}.
\end{align}
Thus we obtain the explicit expression
\begin{align}
  \label{eq:HtotalDecompositionH0DiagonalImplementation}
  e^{-i \tau H_0/2} \ket{\texttt{KM}} = \exp(-i\tau ( k_0\Omega_0 + k_1\Omega_1 + \cdots + E_{0,m_0}^{\mathrm{Tr}} + E_{1,m_1}^{\mathrm{Tr}} + \cdots )/2) \ket{\texttt{KM}},
\end{align}
which can be implemented on a computer by multiplying each coefficient
$\psi_{\texttt{KM}}$ with the corresponding phase factor. Since all
operations are independent, this operation can be easily parallelized.

The action of the non-diagonal part $e^{-i \tau W}$ on $\ket{\texttt{KM}}$ is
more complicated. One option would be to apply the  Suzuki-Trotter
product-formula decomposition again to each of the terms contained in $W$  (see
\equsref{eq:HtotalDecompositionW}{eq:HtotalDecompositionEnd}) and evaluate  the
smaller matrix exponentials explicitly. Such an approach was used in
\cite{Willsch2016Master}. However, in the present case, a non-diagonal term like
$\hat n_i\hat n_j$ in \equref{eq:HtotalDecompositionEnd} would result in
16-component updates (cf.~\equref{eq:transmonchargeoperatortransmonbasis}) of
the state vector given by \equref{eq:psioftsolutioncoefficientsKM}.
Furthermore, additional second-order decompositions  of the matrix exponential
$e^{-i \tau W}$ would introduce additional errors of order $\tau^3$ such that
the time step $\tau$ would need to be reduced.

Therefore, we follow a different route by constructing a transformation $V$ to
change to the eigenbasis of $W$ such that $\Lambda = V^\dagger W V$ is diagonal.
This means that the operator $e^{-i \tau W}$ in
\equref{eq:productformulasecondorder} can be implemented as
$V\,e^{-i\tau\Lambda}\,V^\dagger$. Using this, the decomposition used to
implement the time step
\equref{eq:psioftTimeEvolutionOperatorStepSinglePropagation} reads
\begin{align}
  \label{eq:productformulasecondorderAfterDiagW}
  \widetilde{\mathcal U} = e^{-i \tau H_0/2} \, V\,e^{-i\tau\Lambda}\,V^\dagger \, e^{-i \tau H_0/2}.
\end{align}
Since
the Hilbert space given by \equref{eq:HilbertSpaceTruncated} is a product of multiple spaces with only
four dimensions each, $V$ can be written as a tensor product of complex $4\times4$
matrices,
\begin{align}
  \label{eq:Vtensorproductof4x4matrices}
  V &= \bigotimes\limits_{r=0}^{N_{\mathrm{Res}}-1} V^{(a)}_r \bigotimes\limits_{i=0}^{N_{\mathrm{Tr}}-1} V^{(n)}_i,
\end{align}
where $V^{(a)}_r$ diagonalizes the matrix representation of $\hat a_r + \hat
a_r^\dagger$  in the Fock basis (see \equref{eq:resonatorelectricfieldoperator}
and $V^{(n)}_i$ diagonalizes the matrix representation of $\hat n_i$ in the
transmon basis (see \equref{eq:transmonchargeoperatortransmonbasis}). We obtain
these $4\times4$ matrices numerically by diagonalizing the corresponding
$4\times4$ matrix representations given in
\equref{eq:resonatorelectricfieldoperator} and
\equref{eq:transmonchargeoperatortransmonbasis}, respectively. The matrices
$V^{(a)}_r$ and $V^{(n)}_i$ only need to be computed once, i.e., before the
actual time evolution starts. Furthermore, the matrices $V^{(a)}_r$ are all
equal for different $r$, so only one of them needs to be stored in memory. In
summary, we have
\begin{subequations}
\begin{align}
  \label{eq:resonatorelectricfieldoperatorEigendecomposition}
  V^{(a)}_r\,\Lambda^{(a)}_r\,V^{(a)\dagger}_r &=
  \begin{pmatrix}
      & \sqrt{1+k^{\mathrm{offset}}_r}        & \\
    \sqrt{1+k^{\mathrm{offset}}_r} &          & \sqrt{2+k^{\mathrm{offset}}_r} \\
      & \sqrt{2+k^{\mathrm{offset}}_r} &          & \sqrt{3+k^{\mathrm{offset}}_r} \\
      &          & \sqrt{3+k^{\mathrm{offset}}_r} &
  \end{pmatrix} ,\\
  \label{eq:transmonchargeoperatortransmonbasisEigendecomposition}
   V^{(n)}_i\,\Lambda^{(n)}_i\,V^{(n)\dagger}_i &=
  \begin{pmatrix}
    & n_i^{(0,1)}        &  & n_i^{(0,3)}\\
  n_i^{(0,1)}  &          & n_i^{(1,2)}  & \\
    & n_i^{(1,2)} &          & n_i^{(2,3)} \\
  n_i^{(0,3)}  &          & n_i^{(2,3)} & \\
  \end{pmatrix} ,
\end{align}
\end{subequations}
where $n_i^{(m_i,m_i')} = \sum_{n_i} n_i\braket{m_i|n_i}\braket{n_i|m_i'}$ (see
\equref{eq:transmonchargeoperatortransmonbasis}), and the corresponding
eigenvalues are contained in the diagonal matrices $\Lambda^{(a)}_r$ and
$\Lambda^{(n)}_i$, respectively.

A proof that $V$ defined in \equref{eq:Vtensorproductof4x4matrices} diagonalizes
$W$ given by \equsref{eq:HtotalDecompositionW}{eq:HtotalDecompositionEnd} can be
obtained by explicitly computing $\Lambda = V^\dagger W V$:
\begin{align}
  \Lambda &= \sum\limits_{i=0}^{N_{\mathrm{Tr}}-1} -8 E_{Ci} n_{gi}(\tilde t) \Lambda^{(n)}_i
  + \sum\limits_{r=0}^{N_{\mathrm{Res}}-1} \Omega_r \varepsilon_r(\tilde t) \Lambda^{(a)}_r
  + \sum\limits_{r=0}^{N_{\mathrm{Res}}-1} \sum\limits_{i=0}^{N_{\mathrm{Tr}}-1} G_{ri} \,\Lambda^{(a)}_r\otimes \Lambda^{(n)}_i \nonumber\\
  &\hphantom{=}+ \sum\limits_{0\le r<l<N_{\mathrm{Res}}} \lambda_{rl} \,\Lambda^{(a)}_r\otimes \Lambda^{(a)}_l
  + \sum\limits_{0\le i<j<N_{\mathrm{Tr}}} E_{Ci,Cj} \,\Lambda^{(n)}_i\otimes\Lambda^{(n)}_j,
\end{align}
which is a direct expression for the eigenvalues of $W$. It can be used to implement the
operation $e^{-i\tau\Lambda}$ in \equref{eq:productformulasecondorderAfterDiagW} in the same
way as $e^{-i\tau H_0/2}$ in \equref{eq:HtotalDecompositionH0DiagonalImplementation}.

The only thing left for the implementation of $\widetilde{\mathcal U}$ given in
\equref{eq:productformulasecondorderAfterDiagW} is the implementation of the
basis transformation $V$. Each $4\times4$ component $V^{(a/n)}_{r/i}$ of $V$ in
\equref{eq:Vtensorproductof4x4matrices} results in four-component updates of the
coefficients $\psi_{\texttt{KM}}$ of the form
\begin{align}
  \label{eq:psioft4componentupdates}
  \begin{pmatrix}
    \psi_{*\cdots*00*\cdots*} \\
    \psi_{*\cdots*01*\cdots*} \\
    \psi_{*\cdots*10*\cdots*} \\
    \psi_{*\cdots*11*\cdots*} \\
  \end{pmatrix} &\leftarrow V^{(a/n)}_{r/i} \begin{pmatrix}
    \psi_{*\cdots*00*\cdots*} \\
    \psi_{*\cdots*01*\cdots*} \\
    \psi_{*\cdots*10*\cdots*} \\
    \psi_{*\cdots*11*\cdots*} \\
  \end{pmatrix},
\end{align}
where the notation $*\cdots*$ indicates that the $4\times4$ transformation
$V^{(a)}_r$ ($V^{(n)}_i$) needs to be done in a loop over
$\texttt{KM}=0,\ldots,4^{N_{\mathrm{Tr}}+N_{\mathrm{Res}}}-1$ where the two bits
corresponding to $k_r$ ($m_i$) are fixed (cf.~\equref{eq:KMindexnotation}).

We study three alternatives to implement this loop over $\texttt{KM}$ on a
supercomputer. There is a priori no guarantee which of the implementations
performs best on which processor. It is reasonable to focus mainly on optimizing
this part since it makes the largest contribution to the run time of the
algorithm (see \figref{fig:performancestrong} below). For this reason, we
compare the alternative implementations empirically in
\secref{sec:performancebenchmark}. A \texttt{C++} sample implementation for each
is given in Listings~\ref{code:implementation0}--\ref{code:implementation2} in
\appref{app:implementations}.

\subsubsection{Implementation 0: Complete single loop with branches}

The simplest approach consists of a complete loop over all $\texttt{KM}$
from 0 to $\mathrm{dim}(\mathcal H)-1 = 4^{N_{\mathrm{Tr}}+N_{\mathrm{Res}}}-1$. In
each iteration, we test if the two bits corresponding to the current transformation
are 0 (i.e., if $\texttt{KM} = *\cdots*00*\cdots*$). If they are, we perform the
$4\times4$ update of the coefficients $\psi_{\texttt{KM}}$. In other words,
we iterate over all $\texttt{KM}$, but only do something every fourth iteration.

This implementation might seem naive since the inner loop has four times as many
iterations as it needs. Moreover, the test $\texttt{KM} = *\cdots*00*\cdots*$
introduces branches in the code which may interrupt the sequential flow in the
instruction pipeline of the processor.

However, modern processors use branch predictors to detect patterns in the
evaluation of conditional structures \cite{Smith1981BranchPrediction,
Mittal2019BranchPrediction}.  This means that instructions which are likely to
follow the branch are already loaded into the pipeline before knowing if they
are really going to be executed. Since the branch under consideration (see lines
6 and 13 of Listing~\ref{code:implementation0}) has an easy pattern (it
evaluates to true every four iterations), it could be that branch prediction
effectively removes the overhead. Therefore, the implementation is worth
studying in more detail, also to assess the impact that optimizations on this
level can have with modern compilers and processors.

\subsubsection{Implementation 1: Reduced single loop with bitwise operations}

The next implementation explicitly reduces the number of iterations in the inner
loop by a factor of four. The price to be paid is that the actual index
\texttt{KM} needs to be computed from the reduced iteration count. This is done
by means of additional bitwise operations (see lines 8 and 18 of
Listing~\ref{code:implementation1}).

This implementation has the smallest amount of branches, at the cost of additional
computation required for the bitwise operations to obtain the index $\texttt{KM}$.
Furthermore, it requires the largest amount of code and is less readable than the
other implementations. This approach might have been the first choice for older
architectures where branches in performance-critical code directly result in
an increased run time. It is interesting to see if this intuition also holds for
modern processors.

\subsubsection{Implementation 2: Reduced nested loops}

The last implementation divides the loop over $\texttt{KM}$ into two separate,
nested loops over the higher part \texttt{K} and the lower part \texttt{M} of
the index \texttt{KM} (see \equref{eq:KMindexnotation}). A potential problem of
this implementation is that the loops themselves also introduce branches, and
the evaluation of the tests may not be as predictable as the branches in
implementation 0.

However, modern processors are well
tuned to the execution of loops with simple conditions and increments,
and separate loops over \texttt{K} and \texttt{M} are easy to parse and
parallelize. This implementation conveys the programmer's intent more clearly
and does not contain as many explicit bitwise operations as implementation 1 or
additional iterations as implementation 0.

\subsubsection{Storage of the results and the computational subspace}

The results produced by \texttt{solver} are the complex coefficients
$\psi_{\texttt{KM}}(t)$ of the state vector $\ket{\Psi(t)}$ given in
\equref{eq:psioftsolutioncoefficientsKM} at certain times $t\in\mathbb T$ during
the time evolution.
The set $\mathbb T$ usually consists of the times after each pulse
implementing a certain quantum gate. The maximum number of times in $\mathbb T$
is determined by the total duration of the time evolution divided by the time
step $\tau$. The particular set of times $\mathbb T$ at which the
coefficients are saved is often much smaller than the total number of time steps.

The coefficients $\psi_{\texttt{KM}}(t)$ are typically stored in separate text
files containing the modulus $\mathrm{abs}(\psi_{\texttt{KM}}(t))$ and the
argument $\mathrm{arg}(\psi_{\texttt{KM}}(t))$ of the complex numbers.
Additionally, the current state vector can be saved in binary format. This is
useful when the simulation is interrupted and needs to be continued at a later
point in time; for instance, if the simulation takes longer than the maximum
time that a job can allocate on a supercomputer.

Sometimes, it may not be feasible or necessary to store the coefficients
$\psi_{\texttt{KM}}(t)$ for all $4^{N_{\mathrm{Res}}+N_{\mathrm{Tr}}}$ values of the
index $\texttt{KM}$ (see \equref{eq:KMindexnotation}). It is often
sufficient to consider only the projection of $\ket{\Psi(t)}$ on
the so-called \emph{computational subspace} of the Hilbert space $\mathcal H$
given by \equref{eq:HilbertSpaceTruncated}. The projection can be formally
written in terms a projection operator defined by
\begin{align}
  \label{eq:projectionComputationalSubspace}
  P_{\mathcal H_{2^n}} \ket{\texttt{KM}} = \begin{cases}
    \ket{\texttt{KM}} & \text{if }k_0,k_1,\ldots=0\text{ and }m_0,m_1,\ldots\in\{0,1\} \\
    0 & \text{otherwise}
  \end{cases},
\end{align}
where $n=N_{\mathrm{Tr}}$. Hence $P_{\mathcal H_{2^n}}$ keeps only those states
$\ket{\texttt{KM}}$ for which the resonator part $\texttt{K}=0$ and each $m_i$
in the transmon part $\texttt{M}$ is either 0 or 1.

Since the range of the operator $P_{\mathcal H_{2^n}}$ is a subspace of
dimension $2^n$, it can be identified with the multi-qubit space $\mathcal
H_{2^n}$ defined in \equref{eq:multiqubithilbertspace}. This is the
computational subspace of $\mathcal H$. Consequently, we call all other states
$\ket{\texttt{KM}}$ for which $P_{\mathcal H_{2^n}}\ket{\texttt{KM}} = 0$
\emph{higher levels} or \emph{non-computational states} of $\mathcal H$ (see
also \secref{sec:transformationssubsystemsleakage}). This concept will play an
important role in  \chapref{cha:optimization}, when we study and optimize pulses
$n_{gi}(t)$  to implement a set of quantum gates on the computational subspace.

Note that a projection on the computational subspace may not be sufficient to
compute any arbitrary observable for the system, since transmons may
suffer from leakage or become entangled with resonators. In the latter case,
a partial trace over the  resonators' degrees of freedom would be more
appropriate (see e.g.~\equref{eq:multiqubitblochvectorTransmonTrace} below).
However, for the optimization of quantum gates, where a pulse is explicitly
optimized to render the transmon within the computational subspace, the
projection defined by \equref{eq:projectionComputationalSubspace} is
adequate.

\subsection{Evaluation of the results: \texorpdfstring{$\texttt{evaluator}$}{evaluator}}
\label{sec:evaluator}

The first step after setting up a new simulation (i.e., after specifying the
device parameters for the Hamiltonian including potential time-dependent pulses)
is to configure the simulation parameters to ensure that the results are
accurate up to a certain numerical precision. Afterwards, qubit-specific properties
such as the qubit frequency or the Bloch vectors can be calculated from the
results. These tasks are provided by \texttt{evaluator}. They typically need to
be done before pulses are optimized to implement the quantum gates.

\subsubsection{Adjusting the time step by monitoring overlap and error}

A crucial parameter of the simulation is the time step $\tau$ used to solve the
TDSE (see the discussion below \equref{eq:psioftTimeEvolutionOperatorSteps}).
Despite the existence of rigorous error bounds (see
\cite{deraedt1987productformula,WillschMadita2020PhD}), it is often crucial to
tweak $\tau$ such that the simulation produces results equal to the mathematical
solution of the TDSE (up to some desired precision), but still runs in
reasonable time on a (super)computer.  In practice, one usually starts with a
small time step and gradually increases  $\tau$ as long as the resulting state
vectors $\ket{\Psi(t)}$ (or certain desired expectation values) effectively stay
the same. For most of the simulations, we use time steps
$\tau\in\{\SI{10^{-3}}{ns}, \SI{10^{-4}}{ns}\}$.

To check whether two state vectors $\ket{\Psi^{\tau_1}(t)}$ and
$\ket{\Psi^{\tau_2}(t)}$ resulting from simulations with different time steps
$\tau_1<\tau_2$ are effectively the same, \texttt{evaluator} provides an option
to compute the respective overlap given by
\begin{align}
  \label{eq:overlapPsitau1Psitau2}
  \mathrm{overlap}(t) = \frac{\abs{\braket{\Psi^{\tau_1}(t)|\Psi^{\tau_2}(t)}}^2}{\braket{\Psi^{\tau_1}(t)|\Psi^{\tau_1}(t)}\braket{\Psi^{\tau_2}(t)|\Psi^{\tau_2}(t)}} \in[0,1].
\end{align}
This quantity has the advantage of being independent of a global phase
difference between each of the state vectors. A difference in the global phase
is typically the first observable numerical error caused by an increased time
step,  but it is irrelevant for computing physically meaningful quantities.
Although the simulation result $\ket{\Psi(t)}$ is always normalized since the
Suzuki-Trotter product-formula algorithm is unitary by definition
\cite{deraedt1987productformula}, \equref{eq:overlapPsitau1Psitau2} includes  an
explicit normalization of both state vectors in the denominator. The reason for
this is that \texttt{evaluator} can then also be applied to a projection of the
state vectors on the computational subspace (see
\equref{eq:projectionComputationalSubspace}).

A larger time step $\tau_2>\tau_1$ is sufficient for the simulation if
$\mathrm{overlap}(t)$ is 1 for all times $t\in\mathbb T$. For
convenience, \texttt{evaluator} also computes the average error between
$\ket{\Psi^{\tau_1}(t)}$ and $\ket{\Psi^{\tau_2}(t)}$ given by
\begin{align}
  \label{eq:errorPsitau1Psitau2}
  1-\frac{1}{|\mathbb T|}\sum\limits_{t\in\mathbb T}\mathrm{overlap}(t),
\end{align}
such that only a single number needs to be monitored when configuring the time step.

We present results from applying this procedure in
practice at the end of \secref{sec:accuracy}. Additionally,
we study the behavior of local and global errors with respect to rigorous
error bounds (see \figaref{fig:accuracylocalerror}{fig:accuracyglobalerror}).
However, note that a study of the error bounds requires small, undriven systems,
whereas the practical procedure outlined in this section also works for larger
systems with time-dependent Hamiltonians.

\subsubsection{Computing Bloch vectors}

Given the coefficients $\psi_{\texttt{KM}}(t)$ of the resulting state vector
$\ket{\Psi(t)}$, one could in principle compute its projection on the
computational  subspace (see \equref{eq:projectionComputationalSubspace}) and
evaluate the Bloch  vectors $\vec r_i(t)$
for each qubit according to \equref{eq:multiqubitblochvector}.

However, a characteristic problem of transmon qubits is that a significant part
of the  state may lie outside the computational subspace during the time
evolution. This  problem is known as leakage \cite{chen2016leakagemartinis,
Wood2017LeakageRB, Willsch2017GateErrorAnalysis} (see also
\secref{sec:leakage}). In this case, a simple projection as defined by
\equref{eq:projectionComputationalSubspace} may not be sufficient to compute the
observables $\vec r_i(t)= \braket{\Psi(t)|\vec \sigma_i|\Psi(t)} $. Instead, we
need to trace over all other degrees of freedom of the Hilbert space. Hence,
\equref{eq:multiqubitblochvector} becomes
\begin{align}
  \label{eq:multiqubitblochvectorTransmonTrace}
  \vec r_i(t) &=
  \sum\limits_{k_0k_1\cdots}\sum\limits_{\substack{m_0m_1\cdots\\\text{without $m_i$}}}
  \begin{pmatrix}
    2\,\mathrm{Re}(\psi_{k_0k_1\cdots m_0m_1\cdots(m_i=0)\cdots}^*(t)\psi_{k_0k_1\cdots m_0m_1\cdots(m_i=1)\cdots}(t)) \\
    2\,\mathrm{Im}(\psi_{k_0k_1\cdots m_0m_1\cdots(m_i=0)\cdots}^*(t)\psi_{k_0k_1\cdots m_0m_1\cdots(m_i=1)\cdots}(t)) \\
    \abs{\psi_{k_0k_1\cdots m_0m_1\cdots(m_i=0)\cdots}(t)}^2-\abs{\psi_{k_0k_1\cdots m_0m_1\cdots(m_i=1)\cdots}(t)}^2
  \end{pmatrix},
\end{align}
where all indices $\texttt{KM}=k_0k_1\cdots m_0m_1\cdots$ without $m_i$
enumerate all other basis states included in the simulation
(cf.~\equref{eq:HilbertSpaceTruncated}).

\subsubsection{Determining qubit frequencies}

The computational basis states for each qubit $i$ are given by the respective
lowest-energy transmon eigenstates $\ket{m_i=0}$ and $\ket{m_i=1}$. The energy
difference between these states is called the qubit transition frequency
$\omega_i$, because it corresponds to the frequency that an externally applied
pulse needs to have to drive transitions between the states. It may not be equal
to the qubit frequency $\tilde\omega_i$ obtained from diagonalization (see
the discussion below \equref{eq:transmoneigenstates}) due to the presence
of other transmons and resonators in the system.

One option to measure this frequency makes use of the fact that the  time
evolution of the state $\ket{1}$ results in a relative phase factor
$e^{-i\omega_i t}$ between the states $\ket{0}$ and $\ket{1}$. If the qubit is
prepared in the uniform superposition $\ket{+} = (\ket{0} + \ket{1})/\sqrt{2}$,
its time evolution yields a Bloch vector that rotates around the $z$ axis with a
frequency of $\omega_i$ (see \equref{eq:singlequbitblochvector}), i.e.
\begin{align}
  \vec r_i^{\,\text{(theory)}}(t) &=
  \begin{pmatrix}
    \cos(\omega_i t)\\
    -\sin(\omega_i t)\\
    0
  \end{pmatrix}.
  \label{eq:singlequbitblochvectorTimeEvolutionRotating}
\end{align}
We can thus infer a good candidate for the effective frequency of the qubit by
fitting $\omega_i$ in this equation to the time evolution of the Bloch vector of
qubit $i$ computed using \equref{eq:multiqubitblochvectorTransmonTrace}.
Specifically, this means that we first prepare the qubit in the state
$(\ket{m_i=0}+\ket{m_i=1})/\sqrt 2$, and then have it evolve freely for a
certain time $T$. The other qubits are all prepared in the state $\ket0$ (this
protocol is in agreement with the experimental procedure; see, for instance, the
red squares in \figref{fig:crosstalkExperimentResults}(b) obtained from
experiments on the \texttt{ibmqx4} processor \cite{ibmqx4}). At each time $t$,
we then compute the respective qubit's Bloch vector according to
\equref{eq:multiqubitblochvectorTransmonTrace}, and finally fit
\equref{eq:singlequbitblochvectorTimeEvolutionRotating} to the data. The squared
error for this fit is given by
\begin{align}
  \chi^2(\omega_i) = \sum_{n=1}^{N_{\mathrm{data}}}
    \left[ (r_i^x(t_n) - \cos(\omega_i t_n))^2
         + (r_i^y(t_n) + \sin(\omega_i t_n))^2 \right],
  \label{eq:qubitfrequencyfitchisq}
\end{align}
where $N_{\mathrm{data}}$ is the number of points included in the fit  (usually
much smaller than the total number of time steps required for the simulation),
$t_n$ is the $n^{\mathrm{th}}$ point in time (comprising the set $\mathbb T$
above), and $r_i^x(t_n)$ ($r_i^y(t_n)$) is the $x$ ($y$) component of $\vec
r_i(t)$ given by \equref{eq:multiqubitblochvectorTransmonTrace} at time $t_n$.
Note that it is advantageous to include the data for both ``quadratures''
$r_i^x(t_n)$ and $r_i^y(t_n)$ in the error function, instead of only fitting a
cosine function to $r_i^x(t_n)$. Otherwise the function can have additional
extrema and be harder to minimize properly.

We apply this procedure to determine frequencies since it emulates a typical
experimental procedure to infer qubit frequencies. An alternative exact
diagonalization of the full system to determine its eigenenergies is usually not
feasible for larger systems. The time $T$ is typically chosen on the order of
$\SI{1000}{ns}$ and the number of data points included in the fit is
$N_{\mathrm{data}}=10000$. The inferred qubit frequency $\omega_i$ will later
serve as an initial value for the drive frequency of the pulse optimizations to
implement quantum gates.

A technical difficulty of the procedure is that the error function
$\chi^2(\omega_i)$ given by \equref{eq:qubitfrequencyfitchisq} is a strongly
oscillating function of $\omega_i$ with many local minima and only one sharply
peaked global minimum (see for instance \figref{fig:ibm5edfrequencieschisq}
below). Therefore, standard fitting routines might have problems in locating the
right minimum.

We apply a method called \emph{Golden Section Search} which is designed to
handle the worst possible case of one-dimensional function minimization
\cite{numericalrecipes}. The method brackets the minimum by maintaining a
triplet of points $\omega_a<\omega_b<\omega_c$, and chooses the next point to be
the golden mean point (closer to $\omega_b$) within the larger segment of
$\omega_b-\omega_a$ and $\omega_c-\omega_b$. In each step, the bracketing
interval $\omega_c-\omega_a$ will be a factor of $(\sqrt{5}-1)/2\approx0.61803$
(the inverse golden section) smaller than the preceding interval. This
particular ratio stems from an optimality condition for function  minimization
similar to the bisection method for finding zeros (see \cite{numericalrecipes}
for more information).

\subsubsection{Transformation to the rotating frame}

During a free time evolution, the Bloch vector $\vec r_i(t)$ of qubit $i$
describes rotations around the $z$ axis (see
\equref{eq:singlequbitblochvectorTimeEvolutionRotating}) at the frequency
$\omega_i$ of the qubit. For transmon qubits, for which $\omega_i$ can be around
$2\pi\times\SI{5}{GHz}$, typical quantum gate implementations (such as rotations
of $\vec r_i(t)$ around the $x$ or $z$ axis) may take approximately
$\SI{80}{ns}$ (see \equref{eq:singlequbitpulseGD} in
\secref{sec:optimizatingsinglequbitgate}). Thus, $\vec r_i(t)$ performs a large
number of rotations during the time needed to apply one gate.

For this reason, it is convenient to describe the qubit in a basis rotating at
the qubit's frequency, both for the description of the pulses to implement
quantum gates and also for the purpose of visualization. This basis is commonly
called the \emph{rotating frame} (the other basis is often referred to as the
\emph{lab frame} in  this context). The rotating frame is defined as a change to
a time-dependent  basis according to $(\ket 0, \ket 1)\mapsto(\ket 0,
\exp(-i\omega_i t)\ket 1)$, effectively removing the relative phase between
the computational basis states mentioned above.

For the coefficients $\psi_{k_0k_1\cdots m_0m_1\cdots}(t)$ of
the solution $\ket{\Psi(t)}$ of the transmon simulation, this change of basis
amounts to replacing
\begin{align}
  \psi_{k_0k_1\cdots m_0m_1\cdots}(t) \mapsto e^{i t \sum_i \omega_i m_i} \psi_{k_0k_1\cdots m_0m_1\cdots}(t).
  \label{eq:rotatingframe}
\end{align}
Note that this transformation only removes the relative phase factors between
states corresponding to $m_i=0$ and $m_i=1$. It does not completely remove
relative phases of higher non-computational states $m_i>1$ since transmon
eigenenergies are not exactly equidistant (see e.g.~\cite{koch2007transmon}).
However, this is also not required as we are only interested in the
computational states when describing transmon qubits in a rotating frame.
Furthermore, the transformation \equref{eq:rotatingframe} does not affect the
probability $\vert\psi_{k_0k_1\cdots m_0m_1\cdots}(t)\vert^2$.

The geometrical effect of the rotating frame is that  the Bloch vector $\vec
r_i(t)$ given in \equref{eq:multiqubitblochvectorTransmonTrace}, after replacing
the coefficients according to \equref{eq:rotatingframe}, effectively stands
still. As shown below, this only holds on average since the influence of
higher levels and crosstalk will typically make the Bloch vectors shrink,
wiggle, or continue to rotate slowly in time (see, for  instance,
\figref{fig:crosstalkExperimentBlochSpheres}(b)). The goal of the pulse
optimizations is then to tune the pulse parameters to capture this effect such
that the result of applying a quantum gate also displaces the qubit's Bloch vector
into the desired position.

\clearpage
\subsection{Visualization of the results: \texorpdfstring{$\texttt{visualizer}$}{visualizer}}
\label{sec:visualizer}

Quantum gate implementations for transmon qubits typically consist of
time-dependent pulses whose purpose is to rotate the qubits' Bloch vectors in
their respective Bloch spheres (see \figref{fig:blochsphere}). This is a
continuous, analog operation which can be easily visualized, especially for
single-qubit gates. To engineer and assess the effect of different pulses, it is
instructive to study the time evolution of each $\vec r_i(t)$ computed from
\equref{eq:multiqubitblochvectorTransmonTrace} (potentially in a rotating frame,
see \equref{eq:rotatingframe}).

For this purpose, \texttt{visualizer} processes the coordinates of each $\vec
r_i(t)$ and renders a three-dimensional scene that can be navigated in time and
space using mouse and keyboard. The program makes  use of the high-performance
cross-platform open-source engine Irrlicht \cite{irrlicht}. It is written in
\texttt{C++} and builds on the real-time renderer that has been  developed for
the work presented in \cite{Willsch2016Master}. See \appref{app:visualization}
for some example renderings.

\section{Definition of the model systems}
\label{sec:transmonmodelsystems}

In this section, we give a definition of the most important transmon systems
used in the following chapters. Each system is characterized by its parameters
for the full model Hamiltonian given by \equsref{eq:Htotal}{eq:HCC}.
The values of the device parameters and the topology of the systems
are inspired by various transmon systems used for quantum processors
in experiments.

\subsection{Single transmon-resonator system}
\label{sec:transmonmodelkit}

The simplest system studied in this work consists of a single
transmon coupled to a single resonator. It models a system used in an experiment
that has been conducted at the Karlsruhe Institute of Technology (KIT). The
parameters are given in \tabref{tab:devicekit}. This system  is used for two
purposes in \secref{sec:singletransmonresonatorsystem} of the following chapter,
namely to study the validity of perturbative approaches and to assess the
influence of higher photon numbers  in the resonator. It is also part of
the environment model defined in the next section.

\begin{table}[p]
  \caption{Model parameters for a device operated at KIT with a single transmon
  coupled to a readout resonator \cite{DennisIoan2019}, simulated by solving the TDSE for the  model
  Hamiltonian given by \equsref{eq:Htotal}{eq:HCC}. The indices $i=0$ and
  $r=0$ have been dropped for simplicity. All energies are expressed in GHz
  ($\hbar=1$). Unspecified parameters are set to zero. An estimate for the qubit
  frequency $\omega$ and the anharmonicity $\alpha$ are obtained by
  diagonalizing the transmon Hamiltonian given in
  \equref{eq:transmoneigenstates}. They are given for reference only.}
  \centering
  \label{tab:devicekit}
  \begin{tabular}{@{}cccccccc@{}}
    \toprule
    $N_{\mathrm{Tr}}$ & $N_{\mathrm{Res}}$ & $E_{C}/2\pi$ & $E_{J}/2\pi$ & $\Omega/2\pi$ & $G/2\pi$ & $\tilde\omega/2\pi$ & $\alpha/2\pi$ \\
    \midrule
    1 & 1 & $\SI{0.222}{GHz}$ & $\SI{12.61}{GHz}$ & $\SI{5.821}{GHz}$ & $\SI{0.0349}{GHz}$ & $\SI{4.498}{GHz}$ & $\SI{-0.252}{GHz}$ \\
    \bottomrule
  \end{tabular}
\end{table}

\begin{figure}[p]
  \centering
  \includegraphics[width=.42\linewidth]{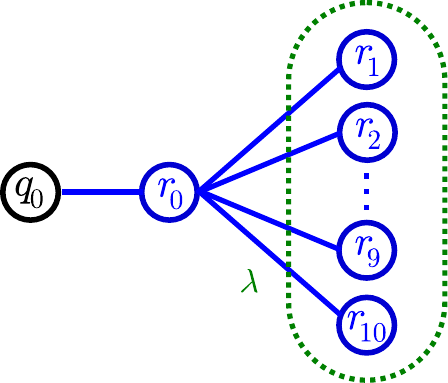}
  \caption{Setup of a system with one transmon and one resonator, coupled to a
  bath of 10 additional resonators. Such a setup is a generic model for a
  general linear superconducting environment (see \secref{sec:extractfoster}). The
  system is simulated by solving the TDSE for the model Hamiltonian given in
  \equsref{eq:Htotal}{eq:HCC} (see \tabref{tab:devicekitbath} for the model
  parameters).}
  \label{fig:kitbathtopology}
\end{figure}

\begin{table}[p]
  \caption{Model parameters for the KIT device specified in
  \tabref{tab:devicekit} coupled to a bath of 10 resonators (see
  \figref{fig:kitbathtopology}). The relation between the specified resonator
  parameters and the parameters of the model Hamiltonian given in
  \equsref{eq:Htotal}{eq:HCC} is
  $(\Omega_{r=0},\Omega_{r=1},\ldots,\Omega_{r=10}) \leftrightarrow
  (\Omega,W_{l=1},\ldots,W_{l=10})$ and $\lambda_{rl} \leftrightarrow \lambda_l$
  (for simplicity, we do not write the index $r=0$ of the central resonator).
  The notation Gaussian($\mu$,$\sigma$) means that the frequencies are drawn
  from a Gaussian distribution with mean $\Omega=2\pi\times\SI{5.821}{GHz}$ and
  standard deviation $\sigma=2\pi\times\SI{1}{GHz}$. The notation
  Uniform(0,$\lambda$) means that the coupling strengths are drawn from a
  uniform distribution between 0 and
  $\lambda\in2\pi\times\{\SI{5}{MHz},\SI{10}{MHz},\SI{20}{MHz}\}$ (see
  \secref{sec:freetransmonresonatorbathphotons}).}
  \centering
  \label{tab:devicekitbath}
  \begin{tabular}{@{}ccccc@{}}
    \toprule
    $N_{\mathrm{Tr}}$ & $N_{\mathrm{Res}}$ & $E_{C},E_{J},\Omega,G$ & $W_l/2\pi$ & $\lambda_{l}/2\pi$ \\
    \midrule
    1 & 11 & (see \tabref{tab:devicekit}) & Gaussian($\Omega$,$\sigma$) & Uniform(0,$\lambda$) \\
    \bottomrule
  \end{tabular}
\end{table}

\subsection{Transmon-resonator system coupled to a bath}
\label{sec:transmonmodelresonatorbath}

We consider the single transmon-resonator system from
\secref{sec:transmonmodelkit}, coupled to a separate bath of harmonic
oscillators. The form of the model is sketched in \figref{fig:kitbathtopology}.
Although only the resonator is explicitly connected  to the bath, the model is
completely general: as we show in \secref{sec:extractfoster}, it is uniquely
related to the Foster representation of a superconducting environment
\cite{Nigg2012BlackBoxCircuitQuantization}. This model is used extensively in
\secref{sec:freetransmonresonatorbathphotons}, where we use the bath approach to study the
transition from an isolated system to an open quantum system described by a
quantum master equation. The particular model parameters chosen for this study
are given in \tabref{tab:devicekitbath}.

\clearpage
\subsection{Two-transmon system}
\label{sec:transmonmodelibm2gst}

The smallest nontrivial system that allows for a simulation of a quantum
computer needs at least two qubits
\cite{Garainin2004SingleSpinQubitClassicalLandauLifshitz} such that the study of
two-qubit gates is possible. The parameters of such a system with two transmons
coupled by one resonator are defined in \tabref{tab:deviceibm2gst}. We use this
system frequently throughout the following chapters to analyze two-qubit gates
and small quantum algorithms. It has also been used for the results published in
\cite{Willsch2017GateErrorAnalysis}.

\begin{table}[h]
  \caption{Model parameters for a system with $N_{\mathrm{Tr}}=2$ transmons and
  $N_{\mathrm{Res}}=1$ resonator, simulated by solving the TDSE for the  model
  Hamiltonian given by \equsref{eq:Htotal}{eq:HCC}. The parameters are inspired
  by device parameters of the quantum processors available on the IBM Q
  Experience between December 2016 and September 2017
  \cite{ibmquantumexperience2016} and have been obtained using an
  electromagnetic HFSS simulation \cite{JayFirat2016} (see
  \secref{sec:extractfoster} for more information on this procedure). All
  energies are expressed in GHz ($\hbar=1$). Unspecified parameters are set to
  zero. Estimates for the qubit frequencies $\tilde\omega_i$ and the
  anharmonicities $\alpha_i$ are obtained by diagonalizing the transmon
  Hamiltonian given by \equref{eq:transmoneigenstates} (see also the actual
  frequencies given in
  \equaref{eq:statedependentfrequenciesResultsOmega0}{eq:statedependentfrequenciesResultsOmega1}).
  The resonator operates at frequency $\Omega_{r=0}/2\pi=\SI{7}{GHz}$.}
\centering
\label{tab:deviceibm2gst}
\begin{tabular}{@{}cccccc@{}}
  \toprule
  Transmon $i$ & $E_{Ci}/2\pi$ & $E_{Ji}/2\pi$ & $G_{i}/2\pi$ & $\tilde\omega_i/2\pi$ & $\alpha_i/2\pi$ \\
  \midrule
  0 & 0.301 & 13.349 & 0.07 & 5.350 & $-$0.350 \\
  1 & 0.301 & 12.292 & 0.07 & 5.120 & $-$0.353 \\
  \bottomrule
\end{tabular}
\end{table}

\subsection{Small five-transmon system}
\label{sec:transmonmodelibm5}

\begin{figure}[p]
  \centering
  \includegraphics[width=.25\linewidth]{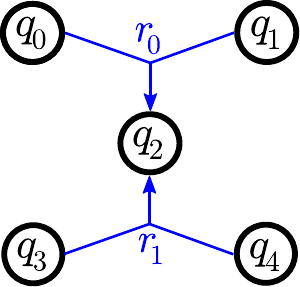}
  \caption{Setup of a system with five transmon qubits and two resonators
  inspired by the five-transmon device that was available on the IBM Q
  Experience in 2016 \cite{ibmquantumexperience2016}.  The system is described
  by the Hamiltonian given by \equsref{eq:Htotal}{eq:HCC} (see
  \tabref{tab:deviceibm51} and \tabref{tab:deviceibm52} for the model
  parameters).}
  \label{fig:ibm5topology}
\end{figure}
\begin{table}[p]
  \caption{Model parameters for a system of $N_{\mathrm{Tr}}=5$ transmons and
  $N_{\mathrm{Res}}=2$ resonators, simulated by solving the TDSE for the  model
  Hamiltonian given by \equsref{eq:Htotal}{eq:HCC}. The system is sketched in
  \figref{fig:ibm5topology}. All values are given in GHz ($\hbar=1$). The qubit
  frequencies $\omega$ have been obtained by minimizing the function given in
  \equref{eq:qubitfrequencyfitchisq} (see also
  \figref{fig:ibm5edfrequencieschisq}). The drive frequencies $f$ have been
  obtained by the single-qubit pulse-optimization procedure (see
  \secaref{sec:optimizatingsinglequbitgate}{sec:optimizatingpulseparameters}).
  The parameters of the resonators are given separately in
  \tabref{tab:deviceibm52}.}
\centering
\label{tab:deviceibm51}
\begin{tabular}{@{}cccccc@{}}
  \toprule
  & $q_0$ & $q_1$ & $q_2$ & $q_3$ & $q_4$ \\
  \midrule
  $E_C/2\pi$ & 0.301 & 0.301 & 0.301 & 0.301 & 0.301 \\
  $E_J/2\pi$ & 13.3511 & 13.1446 & 12.2942 & 12.7882 & 12.0903 \\
  $\omega/2\pi$ & 5.34732 & 5.30259 & 5.11382 & 5.22509 & 5.07094 \\
  $f$           & 5.34697 & 5.30232 & 5.11345 & 5.22506 & 5.07065 \\
  \bottomrule
\end{tabular}
\end{table}
\begin{table}[p]
  \caption{Model parameters of the resonators coupling the transmon  qubits
  specified in \tabref{tab:deviceibm51}. The parameters determine the
  resonator Hamiltonian defined in \equref{eq:HRes}. All values are given in GHz
  ($\hbar=1$).}
\centering
\label{tab:deviceibm52}
\begin{tabular}{@{}ccc@{}}
  \toprule
  & $r_0$ & $r_1$ \\
  \midrule
  $\Omega/2\pi$ & 7.01 & 6.63 \\
  $G/2\pi$ & 0.07 & 0.07 \\
  Coupled to & $q_0,q_1,q_2$ & $q_2,q_3,q_4$ \\
  \bottomrule
\end{tabular}
\end{table}

The fourth system considered in this work is a setup inspired by the
five-transmon device that was available on the IBM Q Experience in 2016
\cite{ibmquantumexperience2016} and was further characterized in
\cite{Takita2017faultTolerantStatePreparation}. It consists of five transmons
coupled by two resonators, each of which connects to three of the five
transmons. The system is sketched in \figref{fig:ibm5topology} and its model
parameters are defined in \tabref{tab:deviceibm51} and \tabref{tab:deviceibm52}.
We use this system primarily for simulations of the quantum circuit experiments
studied in \chapref{cha:fullcircuitsimulations}. In particular, we study a
circuit designed to reveal frequency shifts due to crosstalk for which we also
perform the corresponding experiment on a quantum processor (see
\secref{sec:crosstalk}).

\subsection{Large five-transmon system}
\label{sec:transmonmodelibm5ed}

\begin{figure}[p]
  \centering
  \includegraphics[width=\linewidth]{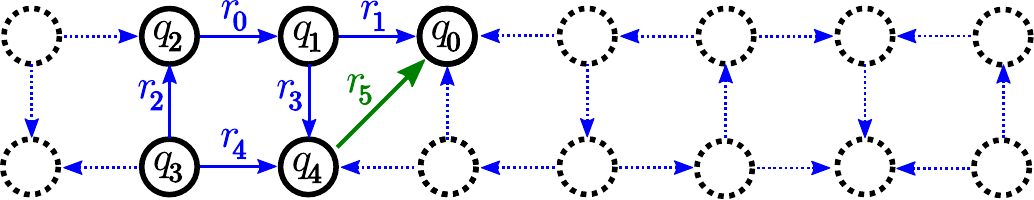}
  \caption{Setup of a system with five transmons and six resonators. The
  system is described by the Hamiltonian given by \equsref{eq:Htotal}{eq:HCC}
  (see \tabref{tab:deviceibm5ed1} and \tabref{tab:deviceibm5ed2} for the model
  parameters). It represents a subset of the 16-qubit device \texttt{ibmqx5}
  \cite{ibmquantumexperience2016} indicated by the dashed lines. An additional
  resonator $r_5$  has been added to the simulation model to enable the
  implementation of all circuits required for the fault-tolerance experiment
  considered in \secref{sec:testingfaulttolerance}.}
  \label{fig:ibm5edtopology}
\end{figure}

\begin{figure}[t]
  \centering
  \includegraphics[width=\textwidth]{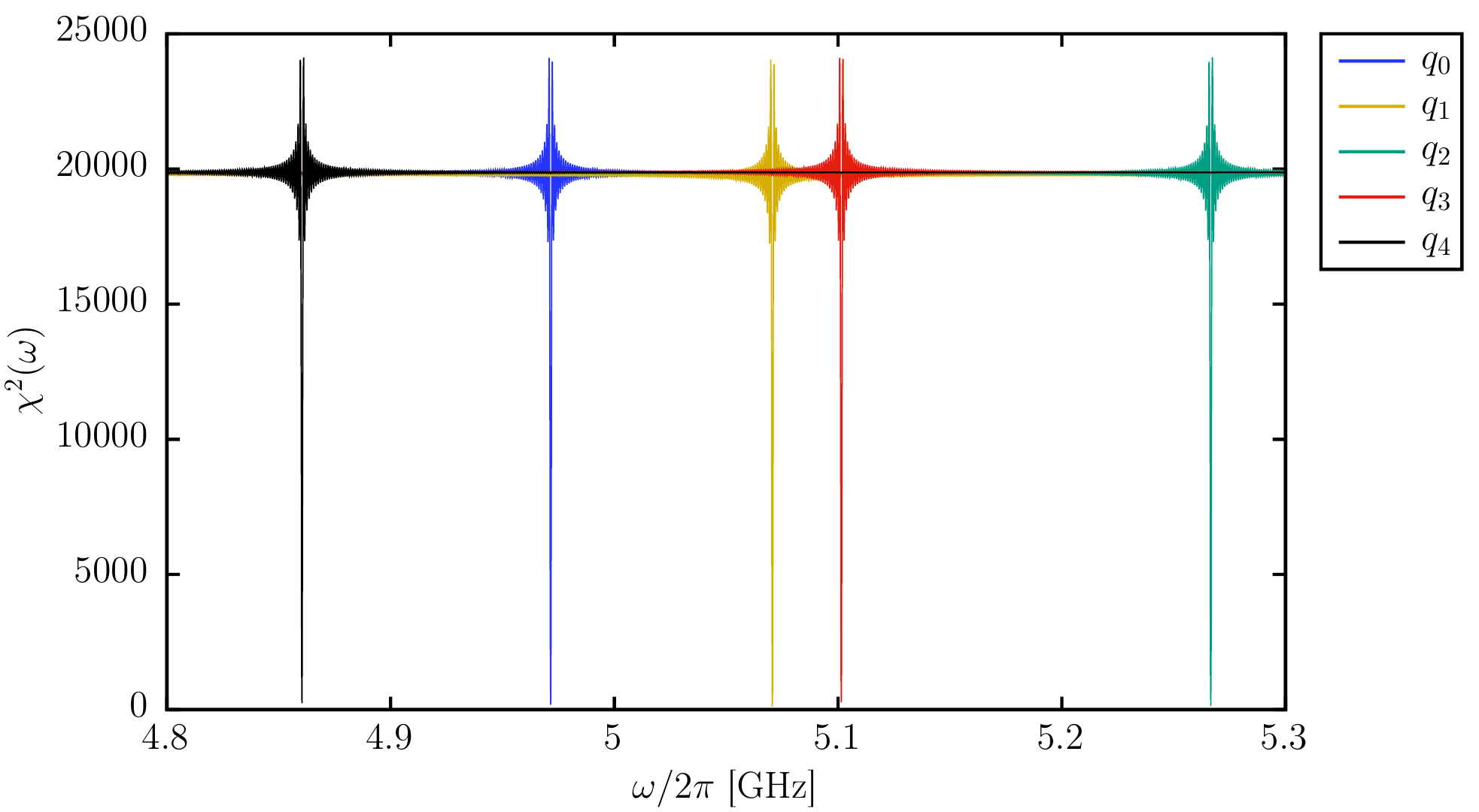}
  \caption{Squared error $\chi^2(\omega)$ defined in
  \equref{eq:qubitfrequencyfitchisq} for $N_{\mathrm{data}}=10000$ points from
  the free time evolution of the system sketched in \figref{fig:ibm5edtopology}
  up to $T=\SI{1000}{ns}$. Each line corresponds to a separate simulation in
  which the transmon $q_i$ (whose frequency is to be determined) is initialized
  in the state $\ket{+}$ and the other transmons are initialized in the state
  $\ket{0}$. The time step used for the simulations is $\tau=\SI{10^{-3}}{ns}$.
  The minima of the sharply peaked functions correspond to the qubit frequencies
  listed in \tabref{tab:deviceibm5ed1}. They are determined using the Golden
  Section Search method described above with $\omega_a/2\pi=\SI{4.8}{GHz}$ and
  $\omega_c/2\pi=\SI{5.3}{GHz}$. All simulations were performed on the
  supercomputer JURECA \cite{JURECA}.}
  \label{fig:ibm5edfrequencieschisq}
\end{figure}

The largest system used for simulations of actual quantum algorithms consists of
five transmons and six resonators (larger systems are only considered for free
evolutions and benchmarks). It is sketched in \figref{fig:ibm5edtopology} as a
subset of IBM's 16-qubit device, extended with an additional resonator. The
model parameters of the system  are given in \tabref{tab:deviceibm5ed1} and
\tabref{tab:deviceibm5ed2}. The qubit frequencies $\omega$ in
\tabref{tab:deviceibm5ed1} have been obtained by the procedure described above
(see \equref{eq:qubitfrequencyfitchisq}); the corresponding squared errors
defined for each qubit are shown in \figref{fig:ibm5edfrequencieschisq}.

\begin{table}[p]
  \caption{Model parameters for a system of $N_{\mathrm{Tr}}=5$ transmons and
  $N_{\mathrm{Res}}=6$ resonators, simulated by solving the TDSE for the model
  Hamiltonian given by \equsref{eq:Htotal}{eq:HCC}. The system is sketched in
  \figref{fig:ibm5edtopology}. All values are given in GHz ($\hbar=1$). The
  qubit frequencies $\omega$ have been obtained by minimizing the function given
  in \equref{eq:qubitfrequencyfitchisq} (see the plot in
  \figref{fig:ibm5edfrequencieschisq}). The drive frequencies $f$ have been
  obtained by the single-qubit pulse-optimization procedure (see
  \secaref{sec:optimizatingsinglequbitgate}{sec:optimizatingpulseparameters}).
  The parameters of the resonators are given separately in
  \tabref{tab:deviceibm5ed2}.}
\centering
\label{tab:deviceibm5ed1}
\begin{tabular}{@{}cccccc@{}}
  \toprule
  & $q_0$ & $q_1$ & $q_2$ & $q_3$ & $q_4$ \\
  \midrule
  $E_C/2\pi$ & 0.301 & 0.301 & 0.301 & 0.301 & 0.301 \\
  $E_J/2\pi$ & 11.6671 & 12.1273 & 13.003 & 12.2456 & 11.1943 \\
  $\omega/2\pi$ & 4.97154 & 5.07063 & 5.26657 & 5.10145 & 4.86036 \\
  $f$ & 4.97164 & 5.07043 & 5.26634 & 5.10147 & 4.86055 \\
  \bottomrule
\end{tabular}
\end{table}
\begin{table}[p]
  \caption{Model parameters of the resonators coupling the transmon  qubits
  specified in \tabref{tab:deviceibm5ed1}. The parameters determine the
  resonator Hamiltonian defined in \equref{eq:HRes}. All values are given in GHz
  ($\hbar=1$).}
\centering
\label{tab:deviceibm5ed2}
\begin{tabular}{@{}ccccccc@{}}
  \toprule
  & $r_0$ & $r_1$ & $r_2$ & $r_3$ & $r_4$ & $r_5$ \\
  \midrule
  $\Omega/2\pi$ & 6.45 & 6.25 & 6.65 & 6.65 & 6.45 & 6.85 \\
  $G/2\pi$ & 0.07 & 0.07 & 0.07 & 0.07 & 0.07 & 0.07 \\
  Coupled to & $q_1,q_2$ & $q_0,q_1$ & $q_2,q_3$ & $q_1,q_4$ & $q_3,q_4$ & $q_0,q_4$ \\
  \bottomrule
\end{tabular}
\end{table}

We mainly use the large five-transmon system to test a fault-tolerant protocol from
the  field of quantum  error correction (see
\secref{sec:testingfaulttolerance}). The same model has also been used for
the results published in \cite{Willsch2018TestingFaultTolerance}.

\clearpage
\section{Modeling electromagnetic environments}
\label{sec:extractfoster}

In this section, we describe a general method to find suitable model parameters
for simulating electromagnetic environments with the simulation method defined
in \secref{sec:transmonmodel}. The parameters can be obtained either  by
directly probing the experimental system or by performing an electromagnetic
simulation of the device.

The following construction is inspired by the black box quantization method
\cite{Nigg2012BlackBoxCircuitQuantization,
Ansari2019BlackBoxCircuitQuantization} and makes use of Foster's theorem
\cite{Foster1924Theorem}. It is also related to the method for extracting
circuit Hamiltonians described in
\cite{Bourassa2012GeneralCircuitQuantizationHamiltonian}.  Furthermore, it can
be extended to lossy electromagnetic environments to capture dissipative
dynamics and predict relaxation rates (see
\cite{Solgun2014BlackboxQuantizationExactImpedance, Solgun2015PhDThesis}).
Following a similar approach, the authors in
\cite{Solgun2019ImpedanceMicrowaveDescriptionSuperconductingQubit} demonstrate
how to extract the device parameters, including transmon-transmon and
transmon-resonator couplings, for the 16-qubit device \texttt{ibmqx5}
\cite{ibmqx5} sketched in \figref{fig:ibm5edtopology}.

To make the construction concrete, we consider in detail the
transmon-resonator-bath system defined in
\secref{sec:transmonmodelresonatorbath} for a general number $L$ of bath
resonators. In particular, we show that the system is a sufficiently general
model for a Josephson junction coupled to a linear but otherwise arbitrary
electromagnetic environment. This implies that there is no loss of generality by
modeling a bath of uncoupled harmonic oscillators that only interact with a
central resonator which is in turn coupled to a transmon. In principle, the
method can be extended to multiple transmons by following the corresponding
generalization in \cite{Nigg2012BlackBoxCircuitQuantization,
Ansari2019BlackBoxCircuitQuantization}.

\begin{figure}[t]
 \centering
 \large
 \def\svgwidth{.7\textwidth}
 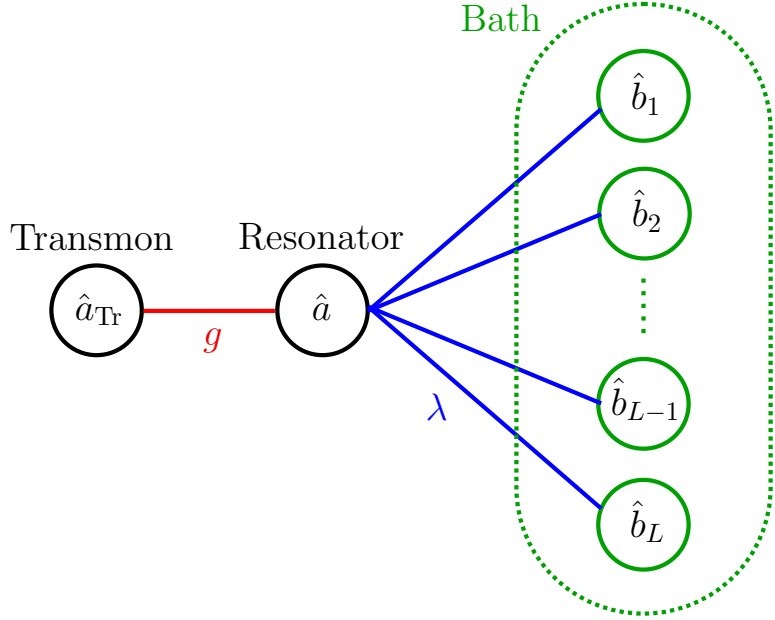
 \normalsize
 \caption{Setup of a system with one transmon and one central resonator, coupled
 to a bath of $L$ noninteracting resonators. The coupling strength between the
 central resonator  and the transmon (the bath) is $g$ ($\lambda_l$ for
 $l=1,\ldots,L$). Although only the central resonator is connected to each other
 component, we show that such a setup is sufficiently general to model a
 generic case.}
 \label{fig:extractFosterModel}
\end{figure}

The transmon-resonator-bath Hamiltonian reads (cf.~\equsref{eq:Htotal}{eq:HCC})
\begin{subequations}
\begin{align}
  \label{eq:extractFosterHTotal}
  H &= H_{\mathrm{Tr}} + H_{\mathrm{Res}} +  H_{\mathrm{Bath}}, \\
  \label{eq:extractFosterHTr}
  H_{\mathrm{Tr}} &= 4 E_{C} \hat n^2 - E_{J} \cos \hat\varphi, \\
  \label{eq:extractFosterHRes}
  H_{\mathrm{Res}} &= \Omega \hat a^\dagger\hat a + G \hat n(\hat a + \hat a^\dagger), \\
  \label{eq:extractFosterHBath}
  H_{\mathrm{Bath}} &= \sum_{l=1}^L W_l \hat b_l^\dagger\hat  b_l
  + \sum_{l=1}^L \lambda_l (\hat a+\hat a^\dagger)(\hat b_l+\hat b_l^\dagger).
\end{align}
\end{subequations}
The idea is to treat all linear contributions canonically, and later work out
the connection to the nonlinear parts (note that \emph{(non)linear} in this
context means \emph{(an)harmonic}). The only nonlinear electromagnetic
contribution to $H$ comes from the Josephson junction described by the transmon
Hamiltonian $H_{\mathrm{Tr}}$. We split the Hamiltonian $H$ into all linear and
purely nonlinear parts by expanding the cosine in $H_{\mathrm{Tr}}$,
\begin{align}
  \label{eq:extractFosterHTrLinNonlin}
  H_{\mathrm{Tr}} &= 4 E_{C} \hat n^2 + \frac{E_{J}}2 \hat\varphi^2 + H_{\mathrm{Nonlin}},
\end{align}
where the nonlinear part reads (up to a constant),
\begin{align}
  \label{eq:extractFosterHNonlin}
  H_{\mathrm{Nonlin}}
  &= E_J(1 - \cos\hat\varphi) - \frac{E_{J}}2 \hat\varphi^2
  = E_J\sum\limits_{n=2}^\infty \frac{(-1)^{n+1}}{(2n)!}\hat\varphi^{2n}
  = - \frac{E_J}{24}\hat\varphi^4 + \cdots.
\end{align}
To diagonalize the linear part of $H_{\mathrm{Tr}}$, we introduce
operators $\hat{a}_{\mathrm{Tr}}$ and $\smash{\hat{a}_{\mathrm{Tr}}^\dagger}$
such that%
\begin{subequations}
  \begin{align}
    \label{eq:extractFosterLadderOperatorsN}
    \hat{n} &= -\frac{1}{\sqrt 2}\left(\frac{E_{J}}{8E_{C}}\right)^{1/4}(\hat{a}_{\mathrm{Tr}}+\hat{a}_{\mathrm{Tr}}^\dagger), \\
    \label{eq:extractFosterLadderOperatorsPhi}
    \hat{\varphi} &= \frac{i}{\sqrt 2}\left(\frac{8E_{C}}{E_{J}}\right)^{1/4}(\hat{a}_{\mathrm{Tr}}-\hat{a}_{\mathrm{Tr}}^\dagger).
  \end{align}
\end{subequations}
Note that no approximation is required for this step (see  the discussion in
\secref{sec:singletransmonresonatorsystemPerturbative}). Using
\equaref{eq:extractFosterLadderOperatorsN}{eq:extractFosterLadderOperatorsPhi},
the linear part of the full Hamiltonian $H = H_{\mathrm{Lin}} +
H_{\mathrm{Nonlin}}$ becomes
\begin{align}
  H_{\mathrm{Lin}}
  &= \Omega_{\mathrm{Tr}}\hat{a}_{\mathrm{Tr}}^\dagger \hat{a}_{\mathrm{Tr}}
  + \Omega \hat a^\dagger\hat a
  + \sum_{l=1}^L W_l \hat b_l^\dagger\hat  b_l \nonumber\\
  &+ g (\hat{a}_{\mathrm{Tr}}+\hat{a}_{\mathrm{Tr}}^\dagger)(\hat a + \hat a^\dagger) \nonumber\\
  \label{eq:extractFosterHLin}
  &+ \sum_{l=1}^L \lambda_l (\hat a+\hat a^\dagger)(\hat b_l+\hat b_l^\dagger),
\end{align}
where $\Omega_{\mathrm{Tr}}=\sqrt{8E_CE_J}$ denotes the frequency of the linear
part of the transmon (which is different from the actual qubit frequency; see
the discussion below \equref{eq:transmoneigenstates}), and $g =
-(E_J/32E_C)^{1/4}G$. The model is sketched in \figref{fig:extractFosterModel}.

In what follows, we first review the Foster representation of an electromagnetic
environment, which gives rise to a procedure for extracting suitable model
parameters. We then construct the mapping to the parameters of the Hamiltonian
$H_{\mathrm{Lin}}$ in the rotating wave approximation (RWA). Finally, we give
the full symplectic transformation to relate the model parameters to the
Hamiltonian without the RWA.

\subsection{The Foster representation of an electromagnetic environment}
\label{sec:FosterRepresentation}

\begin{figure}[t]
  \centering
  \includegraphics[width=\textwidth]{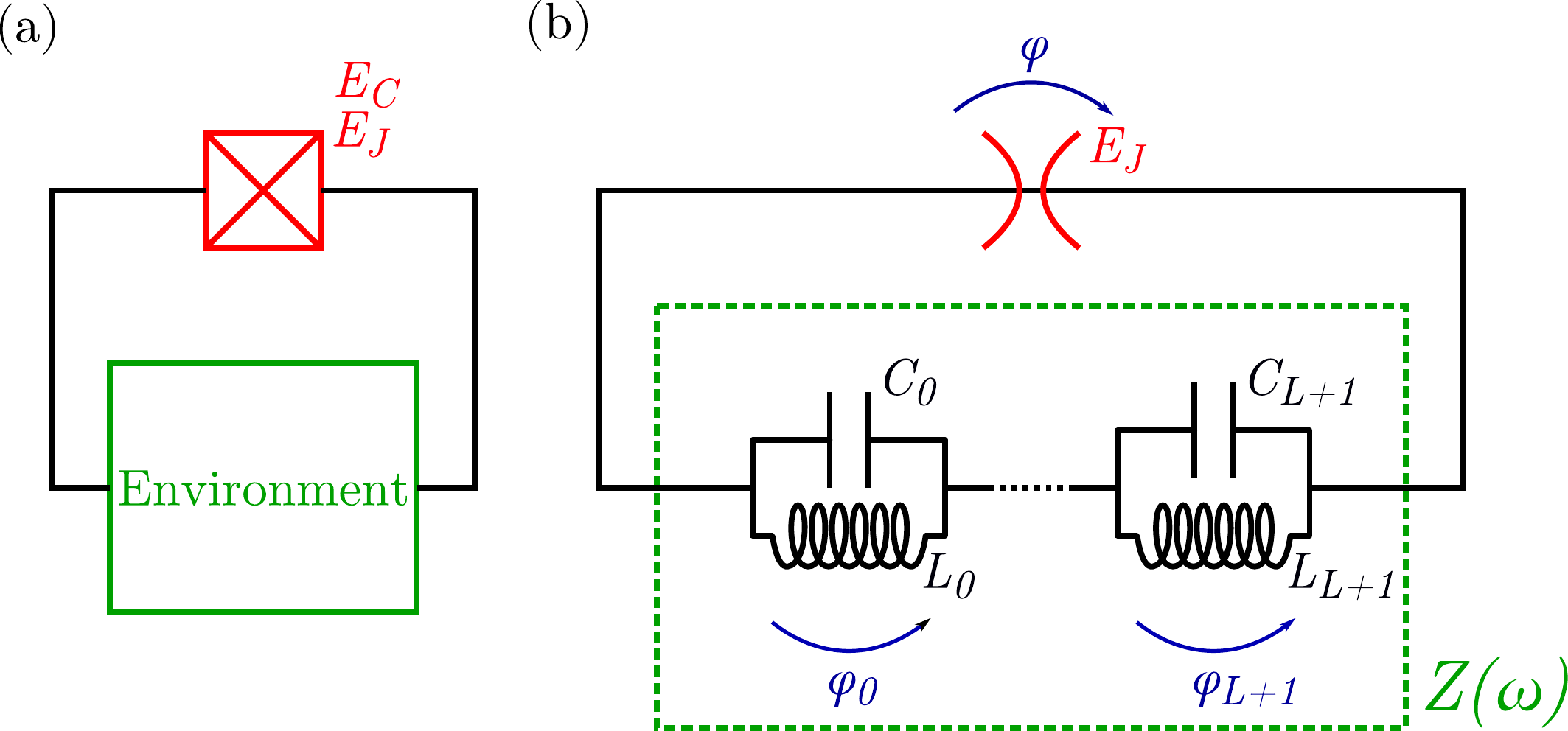}
  \caption{(a) Schematics of a transmon qubit (represented by a Josephson
  junction with  capacitive energy $E_C$ and inductive energy $E_J$) coupled to
  an electromagnetic environment. (b) The environment is characterized by its
  complex impedance $Z(\omega)$ and assumed to be linear, passive, and
  lossless. It is expressed in terms of capacitances $C_j$ and inductances $L_j$
  using the Foster representation of the first kind (see
  \equref{eq:extractFosterTheoremFirst}). Note that all linear parts of the
  Josephson junction are combined with the rest of the environment and lumped
  into $Z(\omega)$.  The only nonlinear part, represented by the red spider
  symbol, corresponds to the Hamiltonian $H_{\mathrm{Nonlin}}$ given in
  \equref{eq:extractFosterHNonlin}.}
  \label{fig:extractFosterCircuit}
\end{figure}

We consider a Josephson junction that is connected to a general, linear
electromagnetic environment (see \figref{fig:extractFosterCircuit}(a)). Such an
environment is characterized by its electrical impedance $Z(\omega)$ or,
equivalently, by its admittance $Y(\omega) = 1/Z(\omega)$. Note that for the
treatment of several Josephson junctions, the impedance $Z(\omega)$ is replaced
by an impedance matrix \cite{Nigg2012BlackBoxCircuitQuantization}.

As we consider superconducting systems, we assume in the following analysis
that the environment is lossless, i.e., no dissipative elements such
as resistors are present. This means that the impedance $Z(\omega)$ is
purely imaginary. Note, however, that  the formalism can be extended to
dissipative elements (see \cite{Solgun2014BlackboxQuantizationExactImpedance,
Solgun2015PhDThesis}).

Foster's theorem \cite{Foster1924Theorem} states that for a lossless, passive
network, $Z(\omega)$ is a complex meromorphic function of the form
\begin{align}
  \label{eq:extractFosterTheorem}
  Z(\omega) &= -i z_0
  \frac{(\overline\omega_0^2-\omega^2)\cdots(\overline\omega_L^2-\omega^2)}
  {(\omega_0^2-\omega^2)\cdots(\omega_{L+1}^2-\omega^2)},
\end{align}
where $z_0\ge0$ is a constant and $\overline\omega_j$ ($\omega_j$) are the
resonant (antiresonant) frequencies, which obey
$0\le\omega_0\le\overline\omega_0\le\cdots\le\overline\omega_L\le\omega_{L+1}$.
In particular, the function $Z(\omega)$ has poles at each $\omega_j$ (for
simplicity, we do not consider poles at zero or infinity). By writing $\omega_j =
1/\sqrt{L_jC_j}$ for suitable inductances $L_j$ and capacitances $C_j$ and using
a pole decomposition of $Z(\omega)$, Foster showed that $Z(\omega)$ can be
synthesized by the series of $LC$-oscillators  shown in
\figref{fig:extractFosterCircuit}(b). This representation is sometimes called
the \emph{Foster representation of the first kind}
\cite{felsen2009electromagneticfieldcomputation}. Explicitly, we obtain
\begin{align}
  \label{eq:extractFosterTheoremFirst}
  Z(\omega) &= \sum\limits_{j=0}^{L+1} Z_j
  = \sum\limits_{j=0}^{L+1} \frac{i\omega/C_j}{\omega_j^2-\omega^2},
\end{align}
where we used the relation $Z_j=1/(1/Z_{L_j}+1/Z_{C_j})$ for the impedance
of a parallel combination of an inductance $Z_{L_j} = i\omega L_j$
and a capacitance $Z_{C_j} = 1/i\omega C_j$.

\subsubsection{Relation to experiments}

The impedance of the environment $Z(\omega)$ can, in principle, be extracted
from current-voltage measurements. In practice, however, it is much more
convenient to obtain $Z(\omega)$ from a finite-element high-frequency structure
simulator (HFSS) to solve Maxwell's equations for a specification of the
device's geometry \cite{Nigg2012BlackBoxCircuitQuantization}.

After finding $Z(\omega)$, the next step is to determine the frequencies
$\omega_j$ corresponding to the poles of the impedance (see
\equref{eq:extractFosterTheoremFirst}) or, equivalently, the zeros of the
admittance $Y(\omega)=1/Z(\omega)$. This can be done numerically. Additionally,
we obtain the capacitances $C_j$ and inductances $L_j$ of the Foster
representation shown in \figref{fig:extractFosterCircuit}(b) using the formulas%
\begin{subequations}
\begin{align}
  \label{eq:extractFosterCapacitance}
  C_j &= \frac{\abs{\mathrm{Im}\,Y'(\omega_j)}}2,\\
  \label{eq:extractFosterInductance}
  L_j &= \frac{2}{\omega_j^2\abs{\mathrm{Im}\,Y'(\omega_j)}}.
\end{align}
\end{subequations}
These expressions can be proven by observing that the residue of $Z(\omega)$ at
$\omega=\omega_j$ is $\mathrm{Res}(Z,\omega_j) =
\lim_{\omega\to\omega_j}(\omega-\omega_j)Z(\omega)=i/2C_j$. Furthermore, since
$Z(\omega)=f(\omega)/g(\omega)$ with $f$ and $g$ holomorphic and $g(\omega_j)=0$
and $g'(\omega_j)\neq0$ (cf.~\equref{eq:extractFosterTheorem}), we have
$\mathrm{Res}(Z,\omega_j) =
\lim_{\omega\to\omega_j}(\omega-\omega_j)f(\omega)/g(\omega) =
f(\omega_j)/g'(\omega_j)$. Also,  we have
$Y'(\omega_j)=g'(\omega_j)/f(\omega_j)$ and thus
$Y'(\omega_j)=1/\mathrm{Res}(Z,\omega_j)=-2iC_j$, which yields
\equref{eq:extractFosterCapacitance}. \sequref{eq:extractFosterInductance} then
follows from $\omega_j^2=1/L_jC_j$.

Using \equaref{eq:extractFosterCapacitance}{eq:extractFosterInductance},
we also find an expression for the \emph{characteristic impedance of
a lossless resonator},
\begin{align}
  \label{eq:extractFosterCharacteristicImpedance}
  Z^{\mathrm{eff}}_j = \sqrt{\frac{L_j}{C_j}} = \frac{2}{\omega_j\abs{\mathrm{Im}\,Y'(\omega_j)}}.
\end{align}

\subsubsection{Quantization}

After extracting all fundamental modes $\omega_j$ of the environment depicted in
\figref{fig:extractFosterCircuit}(b), we can directly write down a Hamiltonian of
$L+2$ quantum harmonic oscillators for the electromagnetic environment of the system,
\begin{align}
  \label{eq:extractFosterHEnv}
  H_{\mathrm{Env}} = \sum\limits_{j=0}^{L+1} \omega_j \hat c_j^\dagger \hat c_j,
\end{align}
where $\hat c_l$ and $\hat c_l^\dagger$ are the corresponding ladder operators
(see \secref{sec:quantumclasscical} for some general remarks on quantum and
classical descriptions).

Note that the Hamiltonian in \equref{eq:extractFosterHEnv} describes the same
system as the  model Hamiltonian $H_{\mathrm{Lin}}$ given by
\equref{eq:extractFosterHLin}. In essence, $H_{\mathrm{Env}}$ is the diagonal
version of $H_{\mathrm{Lin}}$. However, it is not obvious that the model
parameters in $H_{\mathrm{Lin}}$ (i.e., the set of frequencies and coupling
coefficients) are sufficiently generic to capture all instances of
$H_{\mathrm{Env}}$. Constructing the mapping  to the parameters of
$H_{\mathrm{Lin}}$ is the purpose of this section.

Up to this point, we have a Hamiltonian $H_{\mathrm{Env}}$ to describe the
linear environment (dashed green box in \figref{fig:extractFosterCircuit}(b)),
and another Hamiltonian $H_{\mathrm{Nonlin}}$ (see
\equref{eq:extractFosterHNonlin}) for the nonlinear part represented by the
spider symbol in \figref{fig:extractFosterCircuit}(b). To make the connection
between both, we need to relate the operator $\hat\varphi\propto
(\hat{a}_{\mathrm{Tr}}-\hat{a}_{\mathrm{Tr}}^\dagger)$ given in
\equref{eq:extractFosterLadderOperatorsPhi} to the ladder operators $\hat c_j$
and $\hat c_j^\dagger$ in \equref{eq:extractFosterHEnv}. As indicated in
\figref{fig:extractFosterCircuit}(b), this relation can be made through the
phase $\hat\varphi$ using the conservation of the total magnetic flux
(cf.~\cite{Nigg2012BlackBoxCircuitQuantization}):
\begin{align}
  \label{eq:extractFosterFluxOperators}
  \hat\varphi = \sum_j \hat\varphi_j = \sum_j \frac{i}{\sqrt2} \xi_j(\hat c_j-\hat c_j^\dagger),
\end{align}
where $\xi_j = \sqrt{\vphantom{Z}\smash{Z^{\mathrm{eff}}_j}}/\phi_0$ is a
dimensionless coefficient, $Z^{\mathrm{eff}}_j$ is given by
\equref{eq:extractFosterCharacteristicImpedance}, and $\phi_0$ is the reduced
flux quantum (note that in SI units, $\xi_j = \sqrt{\hbar
\smash{Z^{\mathrm{eff}}_j}}/\phi_0$ and $\phi_0=\hbar/2e$). The convention
$\hat\varphi_j\propto i(\hat c_j-\hat c_j^\dagger)$ instead of $\hat c_j+\hat
c_j^\dagger$ (as used in \cite{Nigg2012BlackBoxCircuitQuantization}) is only for
compatibility with
\equaref{eq:extractFosterLadderOperatorsN}{eq:extractFosterLadderOperatorsPhi} and
does not affect the resulting dynamics.

\subsection{Mapping to the model Hamiltonian}

In this section, we construct the mapping from $H_{\mathrm{Env}}$
given by \equref{eq:extractFosterHEnv} to the model Hamiltonian
$H_{\mathrm{Lin}}$ given by \equref{eq:extractFosterHLin}. For simplicity,
we first consider the RWA version of the model Hamiltonian,
\begin{align}
  \label{eq:extractFosterHLinRWA}
  H_{\mathrm{Lin}}^{\mathrm{RWA}}
  &= \Omega_{\mathrm{Tr}}\hat{a}_{\mathrm{Tr}}^\dagger \hat{a}_{\mathrm{Tr}}
  + \Omega \hat a^\dagger\hat a
  + g (\hat{a}_{\mathrm{Tr}}^\dagger\hat a + \hat{a}_{\mathrm{Tr}}\hat a^\dagger)
  + \sum_{l=1}^L W_l \hat b_l^\dagger\hat  b_l
  + \sum_{l=1}^L \lambda_l (\hat a^\dagger\hat b_l+\hat a\hat b_l^\dagger),
\end{align}
and generalize the mapping to the full model Hamiltonian $H_{\mathrm{Lin}}$
afterwards.

\subsubsection{Derivation within the RWA}

First, we write both $H_{\mathrm{Lin}}^{\mathrm{RWA}}$ and $H_{\mathrm{Env}}$ in matrix form,
\begin{subequations}
\begin{align}
  \label{eq:extractFosterHLinRWAMatrix}
  H_{\mathrm{Lin}}^{\mathrm{RWA}}
  %&= \sum\limits_{j=0}^{L+1} \omega_j \hat c_j^\dagger \hat c_j
  &= \boldsymbol{\hat a}^\dagger\boldsymbol \Omega\boldsymbol{\hat a},
  &
  \boldsymbol\Omega
  &= \begin{pmatrix}
    \Omega_{\mathrm{Tr}} & g \\
    g & \Omega & \lambda_1 & \cdots & \lambda_L \\
     & \lambda_1 & W_1 \\
     & \vdots & & \ddots \\
     & \lambda_L & & & W_L \\
  \end{pmatrix},
  &
  \boldsymbol{\hat a}
  &=
  \begin{pmatrix}
    \hat{a}_{\mathrm{Tr}} \\
    \hat{a} \\
    \hat b_1 \\
    \vdots \\
    \hat b_L \\
  \end{pmatrix},\\
  \label{eq:extractFosterHEnvMatrix}
  H_{\mathrm{Env}}
  %&= \sum\limits_{j=0}^{L+1} \omega_j \hat c_j^\dagger \hat c_j
  &= \boldsymbol{\hat c}^\dagger\boldsymbol \omega\boldsymbol{\hat c},
  &
  \boldsymbol\omega
  &= \begin{pmatrix}
    \omega_0 \\
    & \ddots \\
    & & \omega_{L+1} \\
  \end{pmatrix},
  &
  \boldsymbol{\hat c}
  &=
  \begin{pmatrix}
    \hat{c}_0 \\
    \vdots \\
    \hat{c}_{L+1} \\
  \end{pmatrix}.
\end{align}
\end{subequations}
The goal is to construct a linear transformation $\boldsymbol{\hat c} = U\boldsymbol{\hat a}$,
or in components,
\begin{align}
  \label{eq:extractFosterUComponents}
  \boldsymbol{\hat c}_j = \sum_{j'} U_{jj'} \boldsymbol{\hat a}_{j'},
\end{align}
such that $H_{\mathrm{Lin}}^{\mathrm{RWA}}=H_{\mathrm{Env}}$. Here, $U$
is a matrix of dimension $(L+2)\times(L+2)$. There are three conditions that $U$ needs to meet:
\begin{subequations}
  \begin{align}
    \label{eq:extractFosterUConditionI}
    &\text{(I)} & \boldsymbol\Omega &= U^\dagger \boldsymbol\omega U, & \\
    \label{eq:extractFosterUConditionII}
    &\text{(II)} & \mathds1 &= U^\dagger U, &\\
    \label{eq:extractFosterUConditionIII}
    &\text{(III)} & \hat{a}_{\mathrm{Tr}} &\propto \sum_j \xi_j\hat c_j. &
  \end{align}
\end{subequations}
Condition (I) comes from the requirement $H_{\mathrm{Lin}}^{\mathrm{RWA}}=H_{\mathrm{Env}}$ and
yields direct expressions for the model parameters $\boldsymbol\Omega$ in terms of the results
from \secref{sec:FosterRepresentation}. Condition (II) follows from the fact that
both operators $\boldsymbol{\hat a}_j$ and $\boldsymbol{\hat c}_j$ for
$j=0,\ldots,L+1$ need to satisfy the bosonic commutation relations,
\begin{align}
  \label{eq:extractFosterUComponentsUnitary}
  \delta_{ij}
  &= [\boldsymbol{\hat c}_i,\boldsymbol{\hat c}_j^\dagger]
  = \sum_{i'j'} U_{ii'} U_{jj'}^* [\boldsymbol{\hat a}_{i'},\boldsymbol{\hat a}_{j'}^\dagger]
  = (UU^\dagger)_{ij},
\end{align}
which means that $U$ is unitary. Note that this is the point where the RWA
simplifies the construction; in general, a transformation between bosonic
operators is not necessarily unitary (see below). Finally, by condition (III),
we can make sure that the transformation is compatible with the already existing
relation between $\hat a_{\mathrm{Tr}}$  and $\hat c_j$ following from
\equaref{eq:extractFosterLadderOperatorsPhi}{eq:extractFosterFluxOperators},
namely that $(\hat{a}_{\mathrm{Tr}}-\hat{a}_{\mathrm{Tr}}^\dagger) \propto
\hat\varphi \propto \sum_j \xi_j(\hat c_j-\hat c_j^\dagger)$.

In what follows, we construct the elements of $U$. Let $\boldsymbol{u}_j$ for
$j=0,\ldots,L+1$ denote the columns of $U$, i.e.,
$U=(\boldsymbol{u}_0,\ldots,\boldsymbol{u}_{L+1})$. Although
\equref{eq:extractFosterUConditionI} states that $U$ diagonalizes
$\boldsymbol\Omega$, the construction is not straightforward, because we do not
know the elements of $\boldsymbol\Omega$ yet. Instead, we need to start from the
eigenvalues $\omega_j$ and obtain the elements of $\boldsymbol\Omega$ from
\equref{eq:extractFosterUConditionI},
\begin{align}
  \label{eq:extractFosterElementsOmega}
  \boldsymbol\Omega_{ij}=\boldsymbol u_i^\dagger\boldsymbol\omega\boldsymbol u_j.
\end{align}
Furthermore, \equref{eq:extractFosterUConditionII} requires that the vectors
$(\boldsymbol{u}_0,\ldots,\boldsymbol{u}_{L+1})$ form an orthonormal basis,
and the fact that $\boldsymbol\Omega$ in
\equref{eq:extractFosterHLinRWAMatrix} is symmetric allows us to choose
$\boldsymbol{u}_j\in\mathbb R^{L+2}$ (which makes $U$ an orthogonal matrix).

For the first vector, $\boldsymbol{u}_0$, we obtain using
\equaref{eq:extractFosterUComponents}{eq:extractFosterUComponentsUnitary},
$\boldsymbol{u}_0^\dagger \boldsymbol{\hat c} = \boldsymbol{\hat a}_0 = \hat
a_{\mathrm{Tr}}$. To satisfy condition (III) in \equref{eq:extractFosterUConditionIII}, we set
\begin{align}
  \label{eq:extractFosterU0}
  \boldsymbol{u}_0 = \frac{1}{\sqrt{\sum_j\xi_j^2}}
  \begin{pmatrix}
    \xi_0 \\
    \vdots \\
    \xi_{L+1} \\
  \end{pmatrix},
\end{align}
where $\xi_j = \sqrt{\vphantom{Z}\smash{Z^{\mathrm{eff}}_j}}/\phi_0\in\mathbb R$
is given below \equref{eq:extractFosterFluxOperators}. It is not uncommon to
find $\xi_0 \gg \xi_j$ for $j\ge1$ since the slope of the admittance
$\abs{\mathrm{Im}\,Y'(\omega_0)}$ in the denominator of
\equref{eq:extractFosterCharacteristicImpedance} is typically small for the lowest
frequency $\omega_0$ (see also Fig.~2 in the supplementary material of
\cite{Nigg2012BlackBoxCircuitQuantization}). Hence, the first component of
$\boldsymbol{u}_0$ is often the largest.
\sequref{eq:extractFosterElementsOmega} then yields the first model parameter,
\begin{align}
  \label{eq:extractFosterOmegaTR}
  \Omega_{\mathrm{Tr}}
  = \boldsymbol{u}_0^\dagger\boldsymbol\omega\boldsymbol{u}_0
  = \frac{\sum_j\xi_j^2\omega_j}{\sum_j\xi_j^2},
\end{align}
i.e., a weighted average of $\omega_j$, where the weights are given by the
effective characteristic impedances
$\xi_j^2\propto\vphantom{Z}\smash{Z^{\mathrm{eff}}_j}$. We can also determine
$E_C$ and $E_J$ at this point: By inserting $\hat a_{\mathrm{Tr}} =
\boldsymbol{u}_0^\dagger\boldsymbol{\hat c}$ and $\hat a_{\mathrm{Tr}}^\dagger =
\boldsymbol{\hat c}^\dagger \boldsymbol{u}_0$ into
\equref{eq:extractFosterLadderOperatorsPhi} and comparing with
\equref{eq:extractFosterFluxOperators}, we obtain $\sqrt{8E_C/E_J} =
\sum_j\xi_j^2$. The second equation for $E_C$ and $E_J$ comes from
\equref{eq:extractFosterOmegaTR} by using $\Omega_{\mathrm{Tr}}=\sqrt{8E_CE_J}$
(see \equref{eq:extractFosterHLin}). Combining both, we find
\begin{subequations}
\begin{align}
  \label{eq:extractFosterEC}
  E_C &= %\frac{\boldsymbol{u}_0^\dagger\boldsymbol\omega\boldsymbol{u}_0} 8 \sum_j\xi_j^2, \\
  \frac{1} 8 \sum_j\xi_j^2\omega_j, \\
  \label{eq:extractFosterEJ}
  E_J &= %\frac{\boldsymbol{u}_0^\dagger\boldsymbol\omega\boldsymbol{u}_0}{\sum_j\xi_j^2}.
  \frac{\sum_j\xi_j^2\omega_j}{(\sum_j\xi_j^2)^2}.
\end{align}
\end{subequations}

The choice of the next vector, $\boldsymbol u_1$, requires a little trick: We
want all elements $\boldsymbol\Omega_{j0}=0$ for $j\ge2$ (the first column of
$\boldsymbol\Omega$ in \equref{eq:extractFosterHLinRWAMatrix}). These elements
are given by $\boldsymbol\Omega_{j0}=\boldsymbol
u_j^\dagger\boldsymbol\omega\boldsymbol u_0$ (see
\equref{eq:extractFosterElementsOmega}). If we choose $\boldsymbol u_1$ such
that
\begin{align}
  \label{eq:extractFosterSpanCondition}
  \mathrm{span}\{\boldsymbol u_0,\boldsymbol u_1\}
  = \mathrm{span}\{\boldsymbol u_0, \boldsymbol\omega\boldsymbol u_0\},
\end{align}
we know that all $\boldsymbol u_j$ for $j\ge2$ will be orthogonal to
$\boldsymbol\omega\boldsymbol u_0$. Therefore,
$\boldsymbol\Omega_{j0}=\boldsymbol u_j^\dagger\boldsymbol\omega\boldsymbol u_0 =
0$. As $\boldsymbol\Omega_{0j} = \boldsymbol\Omega_{j0}$ (see
\equref{eq:extractFosterElementsOmega}), this choice also produces the zeros in
the first row of $\boldsymbol\Omega$ (see
\equref{eq:extractFosterHLinRWAMatrix}). A real unit vector $\boldsymbol u_1$
that accomplishes \equref{eq:extractFosterSpanCondition} and is orthogonal to
$\boldsymbol u_0$ is given by
\begin{align}
  \label{eq:extractFosterU1}
  \boldsymbol{u}_1 = \frac{\boldsymbol\omega\boldsymbol{u}_0 - \Omega_{\mathrm{Tr}}\boldsymbol{u}_0}{\abs{\boldsymbol\omega\boldsymbol{u}_0 - \Omega_{\mathrm{Tr}}\boldsymbol{u}_0}},
\end{align}
where we used that the frequencies $\omega_j$ in $\boldsymbol\omega$ are, in
general, different from $\Omega_{\mathrm{Tr}}$.
\sequref{eq:extractFosterElementsOmega} then yields the next set of model
parameters,
\begin{subequations}
  \begin{align}
    \label{eq:extractFosterOmega}
    \Omega = \boldsymbol{u}_1^\dagger\boldsymbol\omega\boldsymbol{u}_1, \\
    \label{eq:extractFosterg}
    g = \boldsymbol{u}_0^\dagger\boldsymbol\omega\boldsymbol{u}_1.
  \end{align}
\end{subequations}

The remaining columns $(\boldsymbol{u}_2,\ldots,\boldsymbol{u}_{L+1})$ need to
be orthogonal to $\mathcal S = \mathrm{span}\{\boldsymbol u_0,\boldsymbol
u_1\}$. Furthermore, they need to satisfy $\boldsymbol\Omega_{ij}=\boldsymbol
u_i^\dagger\boldsymbol\omega\boldsymbol u_j=0$ for $i,j\ge2$ with $i\neq j$
because the bottom-right  $L\times L$ block of $\boldsymbol\Omega$ in
\equref{eq:extractFosterHLinRWAMatrix} is diagonal. One way to achieve this is
to complete a set of linearly independent vectors to an orthonormal basis
$(\boldsymbol u_0, \boldsymbol u_1, \boldsymbol v_2, \ldots, \boldsymbol
v_{L+1})$ of $\mathbb R^{L+2}$ using the Gram-Schmidt procedure, followed by a
rediagonalization of $\boldsymbol\omega$ in the space spanned by the vectors
$\boldsymbol v_j$.

Another way to achieve this is to project $\boldsymbol\omega$ directly onto the
orthogonal complement $\mathcal S^\perp$ of $\mathcal S$ using the projector
\begin{align}
  P = \mathds1-\boldsymbol u_0\boldsymbol u_0^\dagger-\boldsymbol u_1\boldsymbol u_1^\dagger,
\end{align}
and to diagonalize the projected matrix $P\boldsymbol\omega P$. Since $P$
reduces the rank of $\boldsymbol\omega$ from $L+2$ to $L$, the matrix
$P\boldsymbol\omega P$ has only $L$ non-zero eigenvalues. These are exactly the
frequencies $W_l$ for $l=1,\ldots,L$ of the $L$ noninteracting oscillators of
the bath (cf.~\figref{fig:extractFosterModel}). The corresponding
eigenvectors make up the remaining columns $\boldsymbol u_{l+1}$ of $U$,
\begin{align}
  (P\boldsymbol\omega P) \boldsymbol u_{l+1} &= W_l \boldsymbol u_{l+1}.
\end{align}
Thus we find for the remaining model parameters $W_l$ and $\lambda_l$
of $H_{\mathrm{Lin}}^{\mathrm{RWA}}$ for $l=1,\ldots,L$,
\begin{subequations}
  \begin{align}
    W_l &= \boldsymbol u_{l+1}^\dagger\boldsymbol\omega\boldsymbol u_{l+1},\\
    \lambda_l &= \boldsymbol u_{l+1}^\dagger\boldsymbol\omega\boldsymbol u_1.
  \end{align}
\end{subequations}

\subsubsection{Derivation without the RWA}

Without the RWA, the model Hamiltonian $H_{\mathrm{Lin}}$ given in
\equref{eq:extractFosterHLin} also contains quadratic terms of the form
$\boldsymbol{\hat a}_j^2$ and $(\boldsymbol{\hat a}_j^\dagger)^2$ (the operators
$\boldsymbol{\hat a}$ are defined in \equref{eq:extractFosterHLinRWAMatrix}). In
this more general case, the Hamiltonians read
\begin{subequations}
  \begin{align}
    \label{eq:extractFosterHLinBogoliubovMatrix}
    H_{\mathrm{Lin}}
    &=
    \begin{pmatrix}
      \boldsymbol{\hat a}^\dagger & \boldsymbol{\hat a}
    \end{pmatrix}
    \begin{pmatrix}
      \boldsymbol\Omega/2 & \boldsymbol\Omega/2-\boldsymbol D \\
      \boldsymbol\Omega/2-\boldsymbol D & \boldsymbol\Omega/2\\
    \end{pmatrix}
    \begin{pmatrix}
       \boldsymbol{\hat a} \\
      \boldsymbol{\hat a}^\dagger \\
    \end{pmatrix}, \\
    H_{\mathrm{Env}}
    &=
    \begin{pmatrix}
      \boldsymbol{\hat c}^\dagger & \boldsymbol{\hat c}
    \end{pmatrix}
    \begin{pmatrix}
      \boldsymbol\omega/2 & \boldsymbol 0 \\
      \boldsymbol 0 & \boldsymbol\omega/2\\
    \end{pmatrix}
    \begin{pmatrix}
       \boldsymbol{\hat c} \\
      \boldsymbol{\hat c}^\dagger \\
    \end{pmatrix},
  \end{align}
\end{subequations}
where $\boldsymbol D = \mathrm{diag}(\boldsymbol\Omega/2)$, and
$\boldsymbol{\hat c}$ and $\boldsymbol\omega$ are defined in
\equref{eq:extractFosterHEnvMatrix}. This case would require a more general
transformation,
\begin{align}
  \begin{pmatrix}
     \boldsymbol{\hat c} \\
    \boldsymbol{\hat c}^\dagger \\
  \end{pmatrix}
  = T \begin{pmatrix}
     \boldsymbol{\hat a} \\
    \boldsymbol{\hat a}^\dagger \\
  \end{pmatrix},
\end{align}
that also mixes the operators in $\boldsymbol{\hat a}$ and $\boldsymbol{\hat
a}^\dagger$. The most general transformation $T$ that preserves the bosonic
commutation relations is a Bogoliubov transformation, which is also known as a
Bogoliubov-Valatin transformation as it was studied independently by Bogoliubov
and Valatin to find solutions of the BCS theory of superconductivity
\cite{Bogoliubov1958Transformation, Valatin1958BogoliubovTransformation}.
Bogoliubov transformations have to be \emph{para-unitary}, i.e.,
\begin{align}
  \label{eq:extractFosterParaUnitary}
  T^\dagger \begin{pmatrix}
    \mathds1 & \boldsymbol 0 \\
    \boldsymbol 0 & - \mathds1\\
  \end{pmatrix} T = \begin{pmatrix}
    \mathds1 & \boldsymbol 0 \\
    \boldsymbol 0 & - \mathds1\\
  \end{pmatrix}.
\end{align}
Bogoliubov transformations have been studied in great detail by Colpa
\cite{Colpa1978DiagonalizationQuadraticBosonHamiltonian}. It is not directly
obvious that a suitable para-unitary transformation producing $H_{\mathrm{Lin}} =
H_{\mathrm{Env}}$  exists, as the matrices on the diagonal and off-diagonal
blocks in \equref{eq:extractFosterHLinBogoliubovMatrix} need to be equal except
for the diagonal $\boldsymbol D$. Nonetheless, it is possible to construct a
para-unitary transformation for the present problem by following Section 5 of
\cite{Colpa1978DiagonalizationQuadraticBosonHamiltonian}.

A simpler mapping that yields $H_{\mathrm{Lin}} = H_{\mathrm{Env}}$, however,
can be obtained by working in an appropriate position-momentum representation,
i.e., by writing the bosonic operators $\boldsymbol{\hat a}$ and
$\boldsymbol{\hat c}$ as linear combinations of Hermitian operators.
Specifically, we define Hermitian operators $\boldsymbol{\hat x}$,
$\boldsymbol{\hat y}$, $\boldsymbol{\hat q}$, and $\boldsymbol{\hat p}$ such
that for $j = 0,\ldots,L+1,$
\begin{subequations}
  \begin{align}
    \label{eq:extractFosterSymplecticAXY}
    \boldsymbol{\hat a}_j &= \sqrt{\alpha_j}\,\boldsymbol{\hat x}_j + \frac{i}{\sqrt{\alpha_j}}\,\boldsymbol{\hat y}_j, \\
    \label{eq:extractFosterSymplecticCQP}
    \boldsymbol{\hat c}_j &= \sqrt{\omega_j}\,\boldsymbol{\hat q}_j + \frac{i}{\sqrt{\omega_j}}\,\boldsymbol{\hat p}_j,
  \end{align}
\end{subequations}
where $\alpha_0 = \Omega_{\mathrm{Tr}}$, $\alpha_1 = \Omega$, $\alpha_{l+1} = W_l$ for $l=1,\ldots,L$,
and $\omega_j$ are the frequencies of the linear environment (see \equref{eq:extractFosterHEnv}).
The inverse relations are given by
\begin{subequations}
  \begin{align}
    \boldsymbol{\hat x}_j &= \frac{\boldsymbol{\hat a}_j + \boldsymbol{\hat a}_j^\dagger}{2\sqrt{\alpha_j}}, & \boldsymbol{\hat y}_j &= \frac{\boldsymbol{\hat a}_j - \boldsymbol{\hat a}_j^\dagger}{2i/\sqrt{\alpha_j}}, \\
    \boldsymbol{\hat q}_j &= \frac{\boldsymbol{\hat c}_j + \boldsymbol{\hat c}_j^\dagger}{2\sqrt{\omega_j}}, & \boldsymbol{\hat p}_j &= \frac{\boldsymbol{\hat c}_j - \boldsymbol{\hat c}_j^\dagger}{2i/\sqrt{\omega_j}}.
  \end{align}
\end{subequations}
In this representation, $H_{\mathrm{Lin}}$ in \equref{eq:extractFosterHLin}
and $H_{\mathrm{Env}}$ in \equref{eq:extractFosterHEnvMatrix} amount to
\begin{subequations}
\begin{align}
  \label{eq:extractFosterHLinSymplecticMatrix}
  H_{\mathrm{Lin}}
  &=
  \begin{pmatrix}
    \boldsymbol{\hat x} & \boldsymbol{\hat y}
  \end{pmatrix}
  \begin{pmatrix}
    \boldsymbol{X} & \boldsymbol{0} \\
    \boldsymbol{0} & \mathds1
  \end{pmatrix}
  \begin{pmatrix}
    \boldsymbol{\hat x} \\
    \boldsymbol{\hat y} \\
  \end{pmatrix}, \\
  \label{eq:extractFosterHEnvSymplecticMatrix}
  H_{\mathrm{Env}}
  &=
  \begin{pmatrix}
    \boldsymbol{\hat q} & \boldsymbol{\hat p}
  \end{pmatrix}
  \begin{pmatrix}
    \boldsymbol{\omega}^2 & \boldsymbol{0} \\
    \boldsymbol{0} & \mathds1
  \end{pmatrix}
  \begin{pmatrix}
    \boldsymbol{\hat q} \\
    \boldsymbol{\hat p} \\
  \end{pmatrix},
\end{align}
\end{subequations}
where
\begin{align}
  \label{eq:extractFosterSymplecticX}
  \boldsymbol X
  &= \begin{pmatrix}
    \Omega_{\mathrm{Tr}}^2 & 2g\sqrt{\Omega_{\mathrm{Tr}}\Omega} \\
    2g\sqrt{\Omega_{\mathrm{Tr}}\Omega} & \Omega^2 & 2\lambda_1\sqrt{\Omega W_1} & \cdots & 2\lambda_L\sqrt{\Omega W_L} \\
     & 2\lambda_1\sqrt{\Omega W_1} & W_1^2 \\
     & \vdots & & \ddots \\
     & 2\lambda_L\sqrt{\Omega W_L} & & & W_L^2 \\
  \end{pmatrix},
\end{align}
and $\boldsymbol\omega^2 = \mathrm{diag}(\omega_0^2,\ldots,\omega_{L+1}^2)$.
Note that the coefficients in
\equaref{eq:extractFosterSymplecticAXY}{eq:extractFosterSymplecticCQP} were
chosen such that the bottom-right blocks in
\equaref{eq:extractFosterHLinSymplecticMatrix}{eq:extractFosterHEnvSymplecticMatrix}
are identity matrices, and the commutation relations,
\begin{align}
  \label{eq:extractFosterSymplecticCommutationRelations}
  [\boldsymbol{\hat x}_j, \boldsymbol{\hat y}_{j'}] = [\boldsymbol{\hat q}_j, \boldsymbol{\hat p}_{j'}] = \frac i 2 \delta_{jj'},
\end{align}
take the same form for each $j$ and $j'$.

To find a mapping that yields $H_{\mathrm{Lin}}=H_{\mathrm{Env}}$, we now search
for a transformation
\begin{align}
  \label{eq:extractFosterSymplecticTransformation}
  \begin{pmatrix}
     \boldsymbol{\hat q} \\
    \boldsymbol{\hat p} \\
  \end{pmatrix}
  = S \begin{pmatrix}
     \boldsymbol{\hat x} \\
    \boldsymbol{\hat y} \\
  \end{pmatrix}.
\end{align}
The transformation $S$ needs to be real (to preserve Hermiticity) and
\emph{symplectic}, i.e.
\begin{align}
  \label{eq:extractFosterSymplecticDefinition}
  S^T \begin{pmatrix}
    \boldsymbol 0 & \mathds1 \\
    - \mathds1 & \boldsymbol 0 \\
  \end{pmatrix} S = \begin{pmatrix}
    \boldsymbol 0 & \mathds1 \\
    - \mathds1 & \boldsymbol 0 \\
  \end{pmatrix},
\end{align}
to preserve the commutation relations in
\equref{eq:extractFosterSymplecticCommutationRelations}. The symplectic
condition is analogous to the para-unitary condition defined in
\equref{eq:extractFosterParaUnitary} for Bogoliubov transformations in the
ladder-operator representation (see also
\cite{Meyer2009SymplecticTransformations,
Dopico2009ParametrizationSymplecticGroup}). As an ansatz, we try
\begin{align}
  \label{eq:extractFosterSymplecticS}
  S = \begin{pmatrix}
  O & \boldsymbol 0 \\
  \boldsymbol 0 & O \\
  \end{pmatrix}
\end{align}
with an orthogonal matrix $O$, i.e., $O^TO=\mathds1$. By testing
\equref{eq:extractFosterSymplecticDefinition},  $S$ is easily seen to be
symplectic. Inserting
\equaref{eq:extractFosterSymplecticTransformation}{eq:extractFosterSymplecticS}
into \equref{eq:extractFosterHEnvSymplecticMatrix}, we see that for
$H_{\mathrm{Lin}}=H_{\mathrm{Env}}$, $O$ has to satisfy
\begin{align}
  \label{eq:extractFosterOCondition}
  \boldsymbol{X} = O^T\boldsymbol{\omega}^2O,
\end{align}
which is the analogous condition to \equref{eq:extractFosterUConditionI}. Note
that the ansatz \equref{eq:extractFosterSymplecticS} only works because the
coefficients in
\equaref{eq:extractFosterSymplecticAXY}{eq:extractFosterSymplecticCQP} are such
that the bottom-right blocks in
\equaref{eq:extractFosterHLinSymplecticMatrix}{eq:extractFosterHEnvSymplecticMatrix}
are identity matrices.

We now construct the elements of the orthogonal matrix $O$.  Let
$\boldsymbol{o}_j$ for $j=0,\ldots,L+1$ denote the columns of $O$, i.e.,
$O=(\boldsymbol{o}_0,\ldots,\boldsymbol{o}_{L+1})$. Since the matrices
$\boldsymbol{X}$ and $\boldsymbol{\omega}^2$ in
\equref{eq:extractFosterOCondition} have the same structure as
$\boldsymbol{\Omega}$ and $\boldsymbol{\omega}$ in
\equaref{eq:extractFosterHLinRWAMatrix}{eq:extractFosterHEnvMatrix}, we can use
almost the same construction for $O$ as we did for $U$ (note that the resulting
elements of $U$ were real, so $U$ was already orthogonal). The major difference
lies in the first column $\boldsymbol{o}_0$ defined by $\boldsymbol{\hat
y}_0=\boldsymbol{o}_0^T\boldsymbol{\hat p}$ since it has to be compatible with
the flux conservation \equref{eq:extractFosterFluxOperators}. This equation
reads in terms of the new coordinates
\begin{align}
  \label{eq:extractFosterSymplecticPhiP}
  \hat\varphi
  %= \sum_j \frac{i}{\sqrt2} \xi_j(\boldsymbol{\hat c}_j-\boldsymbol{\hat c}_j^\dagger)
  = -\sum_j \sqrt{\frac{2}{\omega_j}} \xi_j \boldsymbol{\hat p}_j.
\end{align}
On the other hand, we have from \equref{eq:extractFosterLadderOperatorsPhi},
\begin{align}
  \label{eq:extractFosterSymplecticPhiY}
  \hat{\varphi}
  %= \frac{i}{\sqrt 2}\left(\frac{8E_{C}}{E_{J}}\right)^{1/4}(\hat{a}_{\mathrm{Tr}}-\hat{a}_{\mathrm{Tr}}^\dagger)
  = -\sqrt{\frac{2}{\Omega_{\mathrm{Tr}}}}\left(\frac{8E_{C}}{E_{J}}\right)^{1/4} \boldsymbol{\hat y}_0.
\end{align}
Therefore, we choose the first normalized column of $O$ as (cf.~\equref{eq:extractFosterU0})
\begin{align}
  \label{eq:extractFosterSymplecticO0}
  \boldsymbol{o}_0 = \frac{1}{\sqrt{\sum_j\xi_j^2/\omega_j}}
  \begin{pmatrix}
    \xi_0/\sqrt{\omega_0} \\
    \vdots \\
    \xi_{L+1}/\sqrt{\omega_{L+1}} \\
  \end{pmatrix}.
\end{align}
The remaining construction of
$O=(\boldsymbol{o}_0,\boldsymbol{o}_1,\ldots,\boldsymbol{o}_{L+1})$  is
completely analogous to the construction of
$U=(\boldsymbol{u}_0,\boldsymbol{u}_1,\ldots,\boldsymbol{u}_{L+1})$ above: We
set (cf.~\equref{eq:extractFosterU1})
\begin{align}
  \label{eq:extractFosterSymplecticO1}
  \boldsymbol{o}_1 = \frac{\boldsymbol\omega^2\boldsymbol{o}_0 - \Omega_{\mathrm{Tr}}^2\boldsymbol{o}_0}{\abs{\boldsymbol\omega^2\boldsymbol{o}_0 - \Omega_{\mathrm{Tr}}^2\boldsymbol{o}_0}},
\end{align}
where $\Omega_{\mathrm{Tr}}^2 =
\boldsymbol{o}_0^T\boldsymbol{\omega}^2\boldsymbol{o}_0$. Then we define the
projector $P = \mathds1 - \boldsymbol{o}_0\boldsymbol{o}_0^T -
\boldsymbol{o}_1\boldsymbol{o}_1^T$, diagonalize $P\boldsymbol\omega^2P$,  and
find all non-zero eigenvalues $W_l^2$ for $l=1,\ldots,L$. The corresponding
eigenvectors $\boldsymbol{o}_{l+1}$ constitute the remaining columns of $O$.

Given $O=(\boldsymbol{o}_0,\ldots,\boldsymbol{o}_{L+1})$,
\equaref{eq:extractFosterSymplecticX}{eq:extractFosterOCondition} then yield for
the model parameters of $H_{\mathrm{Lin}}$ in \equref{eq:extractFosterHLin}:
\begin{subequations}
  \begin{align}
    \label{eq:extractFosterSymplecticOmegaTr}
    \Omega_{\mathrm{Tr}} &= \sqrt{\boldsymbol{o}_0^T\boldsymbol{\omega}^2\boldsymbol{o}_0}
    = \left(\frac{\sum_j\frac{\xi_j^2}{\omega_j}\omega_j^2}{\sum_j\frac{\xi_j^2}{\omega_j}}\right)^{1/2},\\
    \Omega &= \sqrt{\boldsymbol{o}_1^T\boldsymbol{\omega}^2\boldsymbol{o}_1},\\
    W_l &= \sqrt{\boldsymbol{o}_{l+1}^T\boldsymbol{\omega}^2\boldsymbol{o}_{l+1}},\\
    g &= \frac{\boldsymbol{o}_1^T\boldsymbol{\omega}^2\boldsymbol{o}_0}{2\sqrt{\Omega_{\mathrm{Tr}}\Omega}},\\
    \lambda_l &= \frac{\boldsymbol{o}_{l+1}^T\boldsymbol{\omega}^2\boldsymbol{o}_1}{2\sqrt{\Omega W_l}}.
  \end{align}
\end{subequations}
Finally, from \equsref{eq:extractFosterSymplecticPhiP}{eq:extractFosterSymplecticO0},
$\boldsymbol{\hat y}_0=\boldsymbol{o}_0^T\boldsymbol{\hat p}$, and
$\Omega_{\mathrm{Tr}}=\sqrt{8E_CE_J}$,  we obtain expressions
for the remaining model
parameters of the full Hamiltonian $H$ in \equref{eq:extractFosterHTr},
\begin{align}
  E_C &= \frac{1} 8 \sum_j\xi_j^2\omega_j, \\
  E_J &= \left(\sum_j\frac{\xi_j^2}{\omega_j}\right)^{-1}, \\
  G &= -\left(\frac{32E_C}{E_J}\right)^{1/4}g.
\end{align}
Note that the procedure yields all model parameters of $H$, including the
transmon energies $E_C$ and $E_J$, because all linear components have been
included in the electromagnetic environment shown in
\figref{fig:extractFosterCircuit}.

%% file: figs/transmon-resonator-bath.pdf_tex
%% Accompanies image file 'transmon-resonator-bath.pdf' (pdf, eps, ps)
\begingroup%
  \makeatletter%
  \providecommand\color[2][]{%
    \errmessage{(Inkscape) Color is used for the text in Inkscape, but the package 'color.sty' is not loaded}%
    \renewcommand\color[2][]{}%
  }%
  \providecommand\transparent[1]{%
    \errmessage{(Inkscape) Transparency is used (non-zero) for the text in Inkscape, but the package 'transparent.sty' is not loaded}%
    \renewcommand\transparent[1]{}%
  }%
  \providecommand\rotatebox[2]{#2}%
  \newcommand*\fsize{\dimexpr\f@size pt\relax}%
  \newcommand*\lineheight[1]{\fontsize{\fsize}{#1\fsize}\selectfont}%
  \ifx\svgwidth\undefined%
    \setlength{\unitlength}{150.70804866bp}%
    \ifx\svgscale\undefined%
      \relax%
    \else%
      \setlength{\unitlength}{\unitlength * \real{\svgscale}}%
    \fi%
  \else%
    \setlength{\unitlength}{\svgwidth}%
  \fi%
  \global\let\svgwidth\undefined%
  \global\let\svgscale\undefined%
  \makeatother%
  \begin{picture}(1,0.7265704)%
    \lineheight{1}%
    \setlength\tabcolsep{0pt}%
    \put(0,0){\includegraphics[width=\unitlength,page=1]{transmon-resonator-bath.pdf}}%
    \put(0.07672495,0.35064956){\color[rgb]{0,0,0}\makebox(0,0)[lt]{\lineheight{1.25}\smash{\begin{tabular}[t]{l}$\hat{a}_{\mathrm{Tr}}$\end{tabular}}}}%
    \put(0.35512675,0.34960388){\color[rgb]{0,0,0}\makebox(0,0)[lt]{\lineheight{1.25}\smash{\begin{tabular}[t]{l}$\hat{a}$\end{tabular}}}}%
    \put(0.00011664,0.43276087){\color[rgb]{0,0,0}\makebox(0,0)[lt]{\lineheight{1.25}\smash{\begin{tabular}[t]{l}Transmon\end{tabular}}}}%
    \put(0.26891306,0.43454319){\color[rgb]{0,0,0}\makebox(0,0)[lt]{\lineheight{1.25}\smash{\begin{tabular}[t]{l}Resonator\end{tabular}}}}%
    \put(0.52978238,0.69277009){\color[rgb]{0,0.62745098,0}\makebox(0,0)[lt]{\lineheight{1.25}\smash{\begin{tabular}[t]{l}Bath\end{tabular}}}}%
    \put(0.22781992,0.31993452){\color[rgb]{0.9372549,0,0}\makebox(0,0)[lt]{\lineheight{1.25}\smash{\begin{tabular}[t]{l}$g$\end{tabular}}}}%
    \put(0.48922302,0.23173764){\color[rgb]{0,0.00784314,1}\makebox(0,0)[lt]{\lineheight{1.25}\smash{\begin{tabular}[t]{l}$\lambda$\end{tabular}}}}%
    \put(0.73015468,0.59977958){\color[rgb]{0,0,0}\makebox(0,0)[lt]{\lineheight{1.25}\smash{\begin{tabular}[t]{l}$\hat{b}_{1}$\end{tabular}}}}%
    \put(0.73105278,0.46160089){\color[rgb]{0,0,0}\makebox(0,0)[lt]{\lineheight{1.25}\smash{\begin{tabular}[t]{l}$\hat{b}_{2}$\end{tabular}}}}%
    \put(0.70654663,0.23814582){\color[rgb]{0,0,0}\makebox(0,0)[lt]{\lineheight{1.25}\smash{\begin{tabular}[t]{l}$\hat{b}_{L-1}$\end{tabular}}}}%
    \put(0.729025,0.09347285){\color[rgb]{0,0,0}\makebox(0,0)[lt]{\lineheight{1.25}\smash{\begin{tabular}[t]{l}$\hat{b}_{L}$\end{tabular}}}}%
  \end{picture}%
\endgroup%

%% file: chap4.tex
\chapter{Free time evolution}
\label{cha:freeevolution}

A straightforward application of the simulation framework presented in the
previous chapter is the free (undriven) evolution of various transmon-resonator
systems. This is interesting for three reasons: First, it does not require the
specification of pulses such that the systems can be easily scaled up and
benchmarked with respect to accuracy and performance. Second, small free systems
are simple enough such that one can compare to perturbative results often used
in analytical  works. And third, an undriven evolution may be easier to study in
laboratory experiments.

A free time evolution mathematically means that in the model Hamiltonian
given by \equsref{eq:Htotal}{eq:HCC}, the time-dependent fields $n_{gi}(t)$ and
$\epsilon_r(t)$ are zero. The model Hamiltonian
considered in this chapter can therefore be reduced to
\begin{subequations}
\begin{align}
  \label{eq:Htotalfree}
  H^{\mathrm{free}} &= H_{\mathrm{Tr}}^{\mathrm{free}} + H_{\mathrm{Res}}^{\mathrm{free}}, \\
  \label{eq:HTrfree}
  H_{\mathrm{Tr}}^{\mathrm{free}} &= \sum\limits_{i=0}^{N_{\mathrm{Tr}}-1} \left[ 4 E_{Ci} \hat n_i^2 - E_{Ji} \cos \hat\varphi_i \right], \\
  H_{\mathrm{Res}}^{\mathrm{free}} &= \sum\limits_{r=0}^{N_{\mathrm{Res}}-1} \Omega_r \hat a_r^\dagger\hat a_r + \sum\limits_{r=0}^{N_{\mathrm{Res}}-1} \sum\limits_{i=0}^{N_{\mathrm{Tr}}-1} G_{ri} \hat n_i(\hat a_r + \hat a_r^\dagger) \nonumber\\
  \label{eq:HResfree}
  &+ \sum\limits_{0\le r<l<N_{\mathrm{Res}}} \lambda_{rl} (\hat a_r + \hat a_r^\dagger)(\hat a_l + \hat a_l^\dagger).
\end{align}
\end{subequations}
We use this Hamiltonian to test the implementation of the main part of the
simulation algorithm (see \secref{sec:simulationsoftware}). This includes
accuracy and performance benchmarks discussed in
\secref{sec:accuracyperformancebenchmark}, where we use small systems to compare
the simulation results to exact diagonalization, and larger systems to assess
weak and strong scaling performance on parallel supercomputers. In
\secref{sec:singletransmonresonatorsystem}, we simulate a single
transmon-resonator system to analyze the accuracy of known analytical results
based on perturbation theory. \ssecref{sec:freetransmonresonatorbathphotons} is
centered around a larger system of resonators that play the role of a bath as a
model for an open quantum system.  The results are compared to those obtained
from a Lindblad master equation. In particular, we consider the time evolution
of a transmon as a function of the number of photons  in its readout resonator.
A corresponding experiment has been conducted at the Karlsruhe Institute of
Technology (KIT), and we use the same device parameters to define the particular
instance of the model Hamiltonian $H^{\mathrm{free}}$ (see \tabref{tab:devicekit}
and \tabref{tab:devicekitbath}). Finally, in
\secref{sec:statedependentfrequenciesfree}, we study the free evolution of two
coupled transmon qubits for different initial states,  by which we characterize
state-dependent frequency shifts caused by the resonator-mediated exchange
interaction.

\section{Accuracy and performance benchmarks}
\label{sec:accuracyperformancebenchmark}

In this section, we study the accuracy and the performance of the simulation
algorithm described in \secref{sec:numericalalgorithm}. For the former, the
number of transmons $N_{\mathrm{Tr}}$ and resonators $N_{\mathrm{Res}}$ in the
system is kept small to compare the results  to exact diagonalization, which is
infeasible if the size of the Hilbert space is  too large. For the latter, we
study increasing system sizes with a focus on the scaling of the algorithm. All
simulations were performed on the supercomputer JURECA \cite{JURECA}.

\subsection{Accuracy}
\label{sec:accuracy}

A mandatory step for any implementation of a numerical algorithm is to verify
its correctness. In this context, correctness concerns two points: First, the
algorithm should be stable with respect to different initial conditions (i.e.,
different initial states for the TDSE in \equref{eq:tdse3}). This is guaranteed
as the Suzuki-Trotter product-formula algorithm is unconditionally stable, since
all updates of the state $\ket{\Psi(t)}$ in the decomposition are unitary by
construction (see
\equaref{eq:totaltransmontimeevolutionoperator}{eq:productformulasecondorderAfterDiagW}).
Second, the result of the algorithm has to agree, up to a certain controllable
precision, with the exact solution of the mathematical problem. Specifically,
this means that the resulting coefficients of the solution $\ket{\Psi(t)}$ have
to agree with the mathematical solution of the TDSE in \equref{eq:tdse3}.
Verifying this is the purpose of the present section.

The systems used to study the accuracy of the simulation are small
such that they can still be diagonalized exactly: a single transmon-resonator
device produced at KIT (see \secref{sec:transmonmodelkit}), and a device with two
transmons coupled by a single resonator inspired by one of the early processors
available on the IBM Q Experience (see \secref{sec:transmonmodelibm2gst}).

For the Suzuki-Trotter product-formula algorithm, the numerical error can be
controlled by the time step $\tau$ used to integrate the TDSE. Specifically, we
test the following local and global error bounds for the second-order
product-formula algorithm:
\begin{subequations}
\begin{align}
  \label{eq:accuracyErrorGlobalPhase}
  \|\Psi - \overline{\Psi}\| &\le \mathrm{const}\cdot N_{\tau}\cdot\tau^3, \\
  \label{eq:accuracyErrorOverlap}
  1-\abs{\braket{\Psi|\overline{\Psi}}}^2 &\le \mathrm{const}\cdot N_{\tau}^2\cdot\tau^6,
\end{align}
\end{subequations}
where $\Psi$ represents the coefficients of the state vector after $N_{\tau}$
time steps of size $\tau$, and $\overline{\Psi}$ represents the exact solution
at time $N_{\tau}\tau$. \sequref{eq:accuracyErrorGlobalPhase} was proven in
\cite{deraedt1987productformula,
huyghebaert1990productFormulaTimeDependentErrorBounds}, along with an expression
for the constant in terms of commutators of $H_0$ and $W$ given in
\equsref{eq:HtotalDecomposition}{eq:HtotalDecompositionEnd}. To prove
\equref{eq:accuracyErrorOverlap},  note that
$1-\abs{\braket{\Psi|\overline{\Psi}}}^2\le
1-(\mathrm{Re}\braket{\Psi|\overline{\Psi}})^2\le
2(1-\mathrm{Re}\braket{\Psi|\overline{\Psi}})=\|\Psi-\overline{\Psi}\|^2$.

The local error is obtained by performing a single time step ($N_\tau=1$) for
different values of $\tau$. The global error, on the other hand, is obtained by
keeping $\tau$ fixed and performing $N_\tau\tau$ as a function of $N_\tau$. Note
that a nice property of the algorithm is that the bound on the global error in
\equref{eq:accuracyErrorGlobalPhase} only grows linearly with the number of
time steps $N_\tau$.

\begin{figure}
  \centering
  \includegraphics[width=\textwidth]{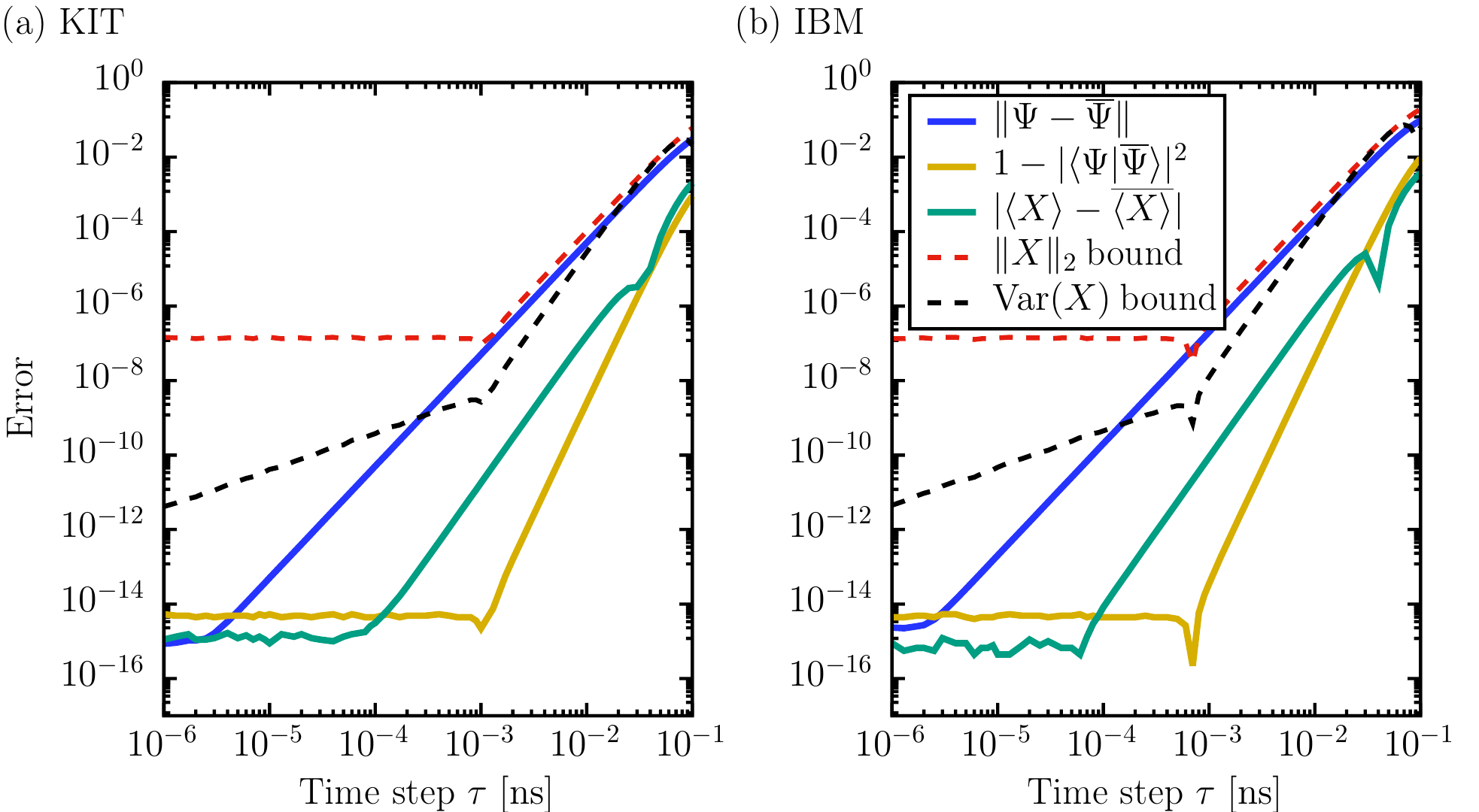}
  \caption{Local error between the result of the simulation $\ket{\Psi}$ and the
  exact result $\ket{\overline\Psi}$ after a single time
  step $\tau$. The purpose of this local error plot is to verify the scaling laws given in
  \equaref{eq:accuracyErrorGlobalPhase}{eq:accuracyErrorOverlap} as a function
  of the time step $\tau$ for (a) the KIT system defined in \secref{sec:transmonmodelkit}, and
  (b) the IBM system defined in \secref{sec:transmonmodelibm2gst}. Additionally, the
  plots show errors for the observable $X$ given by
  \equref{eq:accuracyObservablesDefinition} and the respective bounds given by
  \equaref{eq:accuracyObservableBoundInf}{eq:accuracyObservableBoundVar}.}
  \label{fig:accuracylocalerror}
\end{figure}

Both local and global errors are shown in
\figaref{fig:accuracylocalerror}{fig:accuracyglobalerror}, respectively. The
initial state of the simulations is set to $\ket+$ ($\ket{++}$) for the KIT
(IBM) system. This means that we study a worst-case scenario because of the
fast, observable rotations performed by the Bloch vectors (see
\equref{eq:singlequbitblochvectorTimeEvolutionRotating}). We obtain the mathematical
solution $\overline{\Psi}$ by exact diagonalization of the full system,
which can be done up to machine precision \cite{Demmel2008AccuracyLapackEigensolvers}.

For the local error shown in \figref{fig:accuracylocalerror}, we used four
states for each transmon and each resonator to obtain $\overline{\Psi}$ in order
to verify the power-law scaling in
\equaref{eq:accuracyErrorGlobalPhase}{eq:accuracyErrorOverlap} as a function of
$\tau$.  \sfigref{fig:accuracylocalerror} shows that the scaling laws are
satisfied, as soon as the errors leave the range of machine precision around
$10^{-15}$ where the numbers are practically zero.

\begin{figure}
  \centering
  \includegraphics[width=\textwidth]{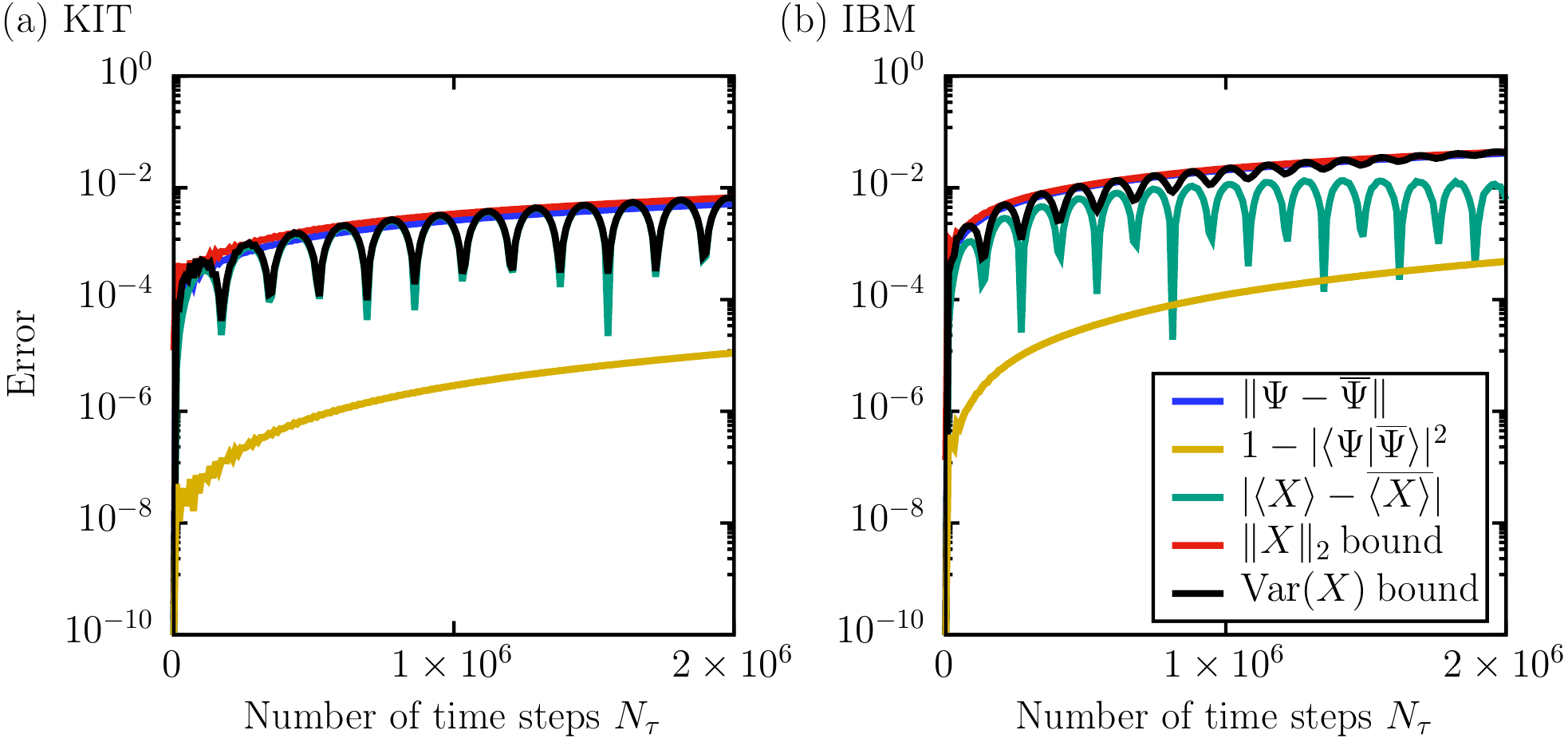}
  \caption{Global error between the result of the simulation $\ket{\Psi}$ and
  the exact result $\ket{\overline\Psi}$. (a) the KIT system defined in
  \secref{sec:transmonmodelkit}, (b) the IBM system defined in
  \secref{sec:transmonmodelibm2gst}.  Additionally, the plots show errors for the
  observable $X$ given in \equref{eq:accuracyObservablesDefinition} and the
  respective bounds given in
  \equaref{eq:accuracyObservableBoundInf}{eq:accuracyObservableBoundVar}. The
  total time evolution corresponds to $\SI{2}{\micro s}$ using a time step of
  $\tau = \SI{10^{-3}}{ns}$.}
  \label{fig:accuracyglobalerror}
\end{figure}

 For the global error shown in \figref{fig:accuracyglobalerror}, we used 10
 states to obtain $\overline{\Psi}$ such that $\overline{\Psi}$
 is equal to the mathematical result (up to machine precision). For this reason,
 \figref{fig:accuracyglobalerror} also includes the error made by truncating the
 Hilbert space given in \equref{eq:HilbertSpaceTruncated}. We see that even for
 the largest time step used in this work ($\tau = \SI{10^{-3}}{ns}$), the errors
 are sufficiently small.

\subsubsection{Accuracy of observables}

Often, one is not interested in the full state vector itself (especially in
cases where the system is so large that it is not practical to store the
coefficients on disk anymore). Rather, the goal of the simulation is to produce
the expectation value $\langle X\rangle=\braket{\Psi|X|\Psi}$ of an observable
$X$. In practice, an expectation value may often be more accurate than suggested
by the error bounds given in
\equaref{eq:accuracyErrorGlobalPhase}{eq:accuracyErrorOverlap}. A reason for
this is that the error $\|\Psi-\overline\Psi\|$ is often dominated by a
difference in the global phase, which does not lead to errors in expectation
values.

Therefore, we additionally consider the actual error on the
observable $|\langle X\rangle - \langle\overline X\rangle|$, where
$\langle\overline X\rangle = \braket{\overline\Psi|X|\overline\Psi}$. This
quantity gives a measure of the discriminability of different states. For any
Hermitian operator $X$, one can prove the following bounds (see
\appref{app:errorobservable} and
\cite{WillschMadita2020PhD} for more information):
\begin{subequations}
\begin{align}
  \label{eq:accuracyObservableBoundInf}
  \abs{\langle X\rangle - \langle\overline X\rangle} &\le 2 \sqrt \Delta\, \|X\|_2,\\
  \label{eq:accuracyObservableBoundVar}
  %\abs{\langle X\rangle - \langle\overline X\rangle} &\le
  %  2 \sqrt{\mathrm{Var}_{\Psi}(X)} \sqrt{1-\abs{\braket{\Psi|\overline{\Psi}}}^2}
  %  \left(\abs{\braket{\Psi|\overline{\Psi}}} + \sqrt{2}\sqrt{1-\abs{\braket{\Psi|\overline{\Psi}}}^2}\right),
  %\\
  \abs{\langle X\rangle - \langle\overline X\rangle} &\le
  2  \sqrt \Delta
  \sqrt{\mathrm{Var}_{\Psi}(X)}\,\abs{\braket{\Psi|\overline{\Psi}}}
  + 2 \Delta\,\|X\|_2,
\end{align}
\end{subequations}
where $\Delta=1-\abs{\braket{\Psi|\overline{\Psi}}}^2$ is the
\emph{distinguishability} between $\ket{\Psi}$ and $\ket{\overline{\Psi}}$,
$\|X\|_2$ denotes the spectral norm (largest singular value) of $X$, and
$\mathrm{Var}_{\Psi}(X)=\langle X^2\rangle-\langle X\rangle^2$ is the variance
of $X$ with respect to the state $\ket\Psi$. Note that both bounds are general
bounds for how well different quantum states can be distinguished. However, the
second bound may be preferable because it depends, to leading order in $\sqrt
\Delta$, only on the variance of the observable $X$ in the respective state,
instead of the state-independent operator norm $\|X\|_2$. Both bounds can be
related to the time step $\tau$ by using $\Delta\propto\tau^6$ (see
\equref{eq:accuracyErrorOverlap}), without knowledge of the exact result
$\ket{\overline\Psi}$. Furthermore, they can be estimated by evaluating the
expressions in \cite{deraedt1987productformula,
huyghebaert1990productFormulaTimeDependentErrorBounds}  for the constant
prefactor in \equref{eq:accuracyErrorOverlap}.

As an application, we study the observables
\begin{align}
  \label{eq:accuracyObservablesDefinition}
  X = \begin{cases}
    \sigma_0^x & \text{(KIT system)} \\
    \sigma_0^x\sigma_1^x & \text{(IBM system)} \\
  \end{cases}.
\end{align}
In both cases, we have $\|X\|_2~=1$ and $\mathrm{Var}_{\Psi}(X)=1-\langle
X\rangle^2$. Since the initial state of the simulation is set to $\ket+$
($\ket{++}$) for the KIT (IBM) system, the observables $X$ in
\equref{eq:accuracyObservablesDefinition} measure the  fast rotations performed
by the Bloch vectors (see
\equref{eq:singlequbitblochvectorTimeEvolutionRotating}). Therefore, we expect
errors to surface quickly.

Indeed,
\figref{fig:accuracylocalerror} shows that the local error of $\abs{\langle
X\rangle - \langle\overline X\rangle}$ increases earlier than
$1-\abs{\braket{\Psi|\overline{\Psi}}}^2$, but is still a factor of 10--100
smaller than the bounds given in
\equaref{eq:accuracyObservableBoundInf}{eq:accuracyObservableBoundVar}.
Interestingly, though, the $\mathrm{Var}_{\Psi}(X)$ bound is tight for the global error
shown in \figref{fig:accuracyglobalerror}(a), even after two million time steps
of size $\tau=\SI{10^{-3}}{ns}$ (corresponding to $\SI{2}{\micro s}$). For the
slightly larger systems studied in \figref{fig:accuracyglobalerror}(b), we see
that the global error of $\abs{\langle X\rangle - \langle\overline X\rangle}$
saturates after $\SI{1}{\micro s}$ while the respective bounds increase.

Finally, we evaluate the error defined by \equref{eq:errorPsitau1Psitau2}
between results obtained with two different time steps $\tau_1=\SI{10^{-4}}{ns}$
and $\tau_2=\SI{10^{-3}}{ns}$. We consider the overlap in
\equref{eq:overlapPsitau1Psitau2} between $\ket{\Psi^{\tau_1}(t)}$ and
$\ket{\Psi^{\tau_2}(t)}$ after every $\SI{10}{ns}$ for a total time of
$\SI{2}{\micro s}$. This means that the set $\mathbb T$ in
\equref{eq:errorPsitau1Psitau2} contains $200$ items. For the KIT system, we
obtain an error of $3.34\times10^{-8}$. The IBM system yields an error of
$9.50\times10^{-7}$. Hence, we conclude that a time  step of $\SI{10^{-3}}{ns}$
is sufficient for both systems. In contrast to the rigorous error bounds studied
above, this procedure  to determine the time step does not require knowledge of
the exact result $\ket{\overline\Psi}$, so it can also be used for much larger
systems with time-dependent Hamiltonians.

\subsection{Performance}
\label{sec:performancebenchmark}

Having verified the accuracy of the simulation method, the next step is to
assess the performance of the transmon simulator. This requires a simulation of
larger systems such that most of the time spent goes into updating the state
vector $\ket{\Psi}$ for each time step of size $\tau$. As discussed in
\secref{sec:simulationsoftware} (see
\equaref{eq:psioftTimeEvolutionOperatorStepSinglePropagationTimeIndependent}
{eq:productformulasecondorderAfterDiagW}), this update is given by
\begin{align}
  \label{eq:psioftTimeEvolutionUpdateRule}
  \ket{\Psi} \leftarrow \underbrace{e^{-i \tau H_0/2} \, V\,e^{-i\tau\Lambda}\,V^\dagger \, e^{-i \tau H_0/2}}_{\widetilde{\mathcal U}} \ket{\Psi}.
\end{align}
The transformation associated with the operator $V$ is the most complicated part
(see  \equref{eq:Vtensorproductof4x4matrices}; the other operators are
diagonal). Therefore, we expect the bottleneck to be in this transformation. We
compare three different implementations of this transformation (see
\appref{app:implementations}).

Performance benchmarks are usually obtained by measuring the run time as a
function of the system size $N_{\mathrm{Tr}}+N_{\mathrm{Res}}$ and the number of
parallel threads $N_{\mathrm{Threads}}$. We consider two different system
configurations. For configuration (1), we vary $N_{\mathrm{Tr}}$ while keeping
$N_{\mathrm{Res}}=1$ fixed, and for configuration (2), we vary both
$N_{\mathrm{Tr}}=N_{\mathrm{Res}}$. The model parameters are given in
\tabref{tab:devicebenchmark}.  For each of the following benchmarks, we perform
$1000$ time steps of size $\tau=\SI{10^{-3}}{ns}$. All simulations were
performed on the JURECA cluster \cite{JURECA} at the J\"ulich Supercomputing
Centre.

For a given system size $N_{\mathrm{Tr}}+N_{\mathrm{Res}}$, the memory needed to
store the state vector $\ket{\Psi}$ is
\begin{align}
  \label{eq:performanceMemory}
  \mathrm{dim}(\mathcal H)\times\SI{16}{bytes} = \SI{4^{N_{\mathrm{Tr}}+N_{\mathrm{Res}}+2}}{bytes},
\end{align}
where the size of the Hilbert space is $\mathrm{dim}(\mathcal H) =
4^{N_{\mathrm{Tr}}+N_{\mathrm{Res}}}$ (see \equref{eq:HilbertSpaceTruncated}) and
the factor of $16$ is due to the use of complex double-precision floating point
numbers. Therefore, we expect the computational work to grow exponentially with
the system size $N_{\mathrm{Tr}}+N_{\mathrm{Res}}$, and the goal is to work
against this by increasing the number of threads.

\begin{table}
  \caption{Model parameters for performance benchmarks for various system sizes
  $N_{\mathrm{Tr}}+N_{\mathrm{Res}}$ between $1$ and $16$
  (cf.~\equsref{eq:Htotalfree}{eq:HResfree}). For configuration (1),
  $N_{\mathrm{Tr}}$ is variable and $N_{\mathrm{Res}}=1$. For configuration (2),
  $N_{\mathrm{Tr}}=N_{\mathrm{Res}}$ are both variable. The symbols $u_i^I$,
  $u_r^I$, and $u_{ri}^I$ denote uniform random numbers drawn from the interval
  $I$, where $i=0,\ldots,N_{\mathrm{Tr}}-1$ and $r=0,\ldots,N_{\mathrm{Res}}-1$. The
  parameters are inspired by the IBM device specified in
  \tabref{tab:deviceibm2gst}. All energies are expressed in GHz ($\hbar=1$).
  Unspecified parameters are set to zero.}
\centering
\label{tab:devicebenchmark}
\begin{tabular}{@{}lcccc@{}}
  \toprule
  \multicolumn{1}{c}{Configuration} & $E_{Ci}/2\pi$ & $E_{Ji}/2\pi$ & $\Omega_r/2\pi$ & $G_{ri}/2\pi$\\
  \midrule
  (1) $N_{\mathrm{Tr}}$ variable, $N_{\mathrm{Res}}=1$ & 0.301 & $10+u_i^{[0,5)}$ & $5.5+u_r^{[0,3)}$ & $0.055+u_{ri}^{[0,0.03)}$ \\
  (2) $N_{\mathrm{Tr}}=N_{\mathrm{Res}}$ variable & 0.301 & $10+u_i^{[0,5)}$ & $5.5+u_r^{[0,2)}$ & $0.055+u_{ri}^{[0,0.02)}$ \\
  \bottomrule
\end{tabular}
\end{table}

\begin{figure}[p]
  \centering
  \includegraphics[width=\textwidth]{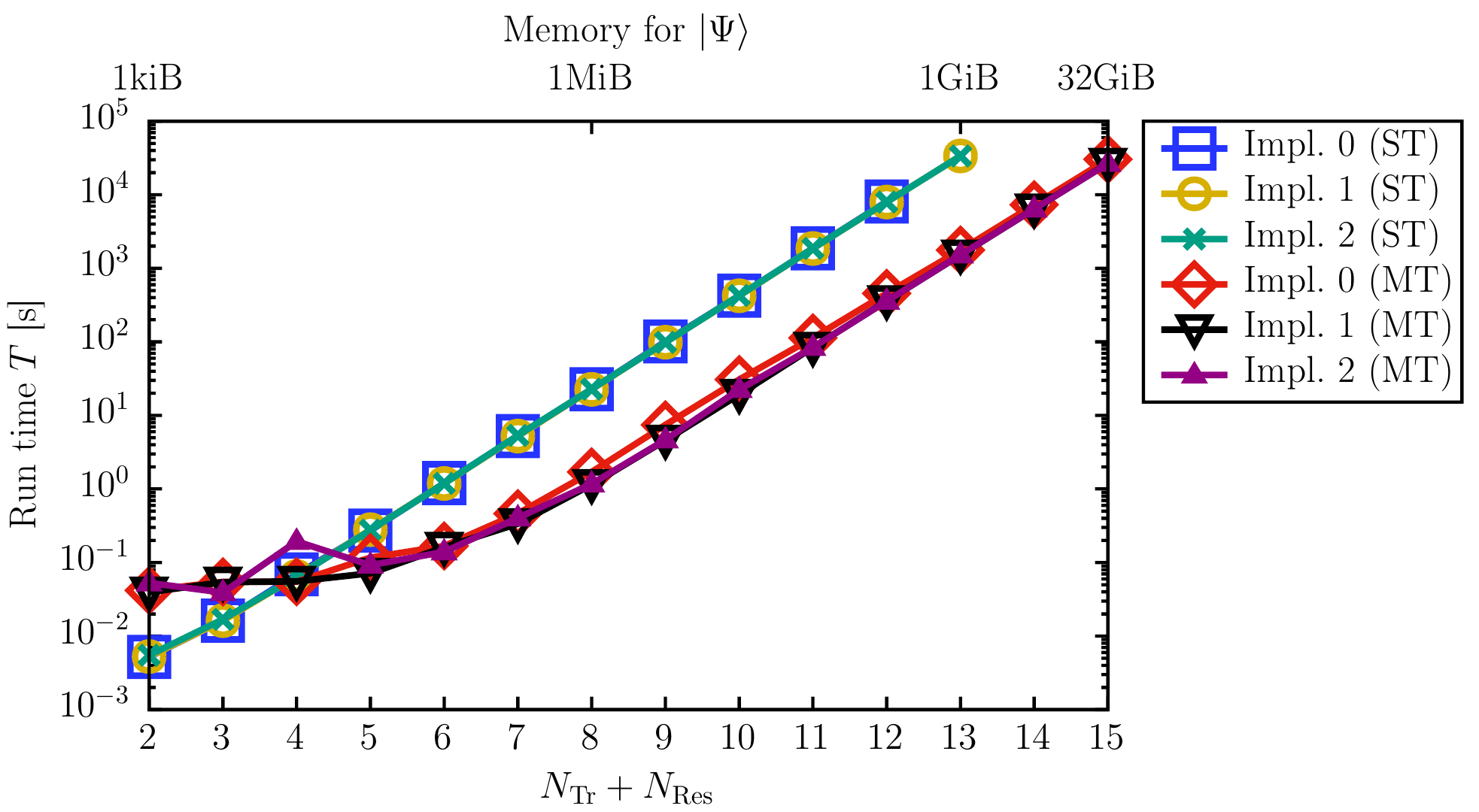}
  \caption{Run time $T$ as a function of the system size
  $N_{\mathrm{Tr}}+N_{\mathrm{Res}}$, for all three implementations (see
  \appref{app:implementations}) in the single-threaded case (ST) using one core
  and the multi-threaded case (MT) using 48 threads on 24 cores. The memory
  required to store the state vector $\ket{\Psi}$ is linked to the system size
  via \equref{eq:performanceMemory}. Lines are drawn to guide the eye.}
  \label{fig:performanceruntime}
\end{figure}

As a simple check, we measure the run time $T$ as a function of the system size,
for each of the three implementations (see \appref{app:implementations}) in both
single-threaded and multi-threaded runs. The result for configuration (1) is
shown in \figref{fig:performanceruntime}. The run times for configuration (2)
are almost the same (data not shown). We see that in the single-threaded case,
the run time $T$ grows as expected already for very small systems. In the
multi-threaded case, the scaling only starts at system sizes around $7$. This
effect is due to the overhead of managing 48 threads, compared to the relatively
small computational work required for $N_{\mathrm{Tr}}+N_{\mathrm{Res}}<7$. At
this stage, one cannot see an essential difference between the three
implementations yet.

\begin{figure}[p]
  \centering
  \includegraphics[width=\textwidth]{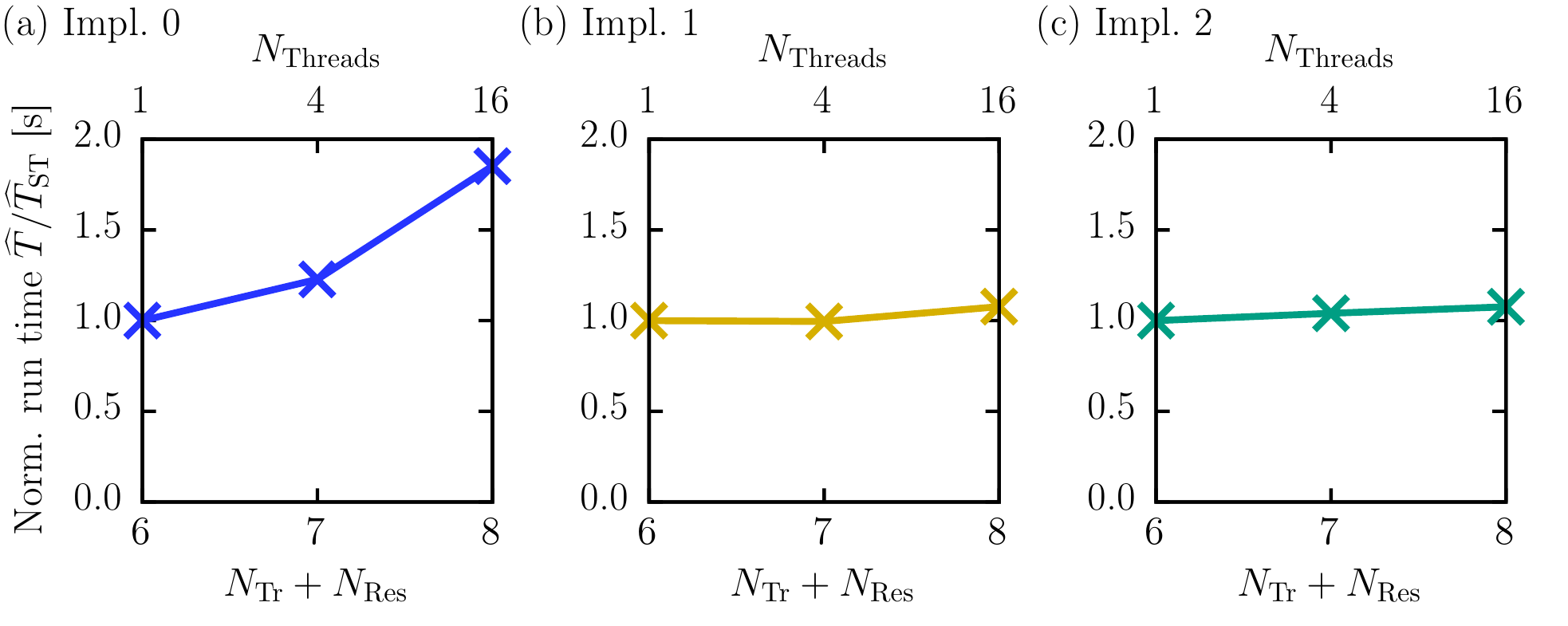}
  \caption{Weak scaling performance of the normalized run time $\widehat
  T/\widehat T_{\mathrm{ST}}$ using (a) implementation 0 where $\widehat
  T_{\mathrm{ST}}=\SI{1.202}{s}/6$, (b) implementation 1 where $\widehat
  T_{\mathrm{ST}}=\SI{1.188}{s}/6$, (c) implementation 2 where $\widehat
  T_{\mathrm{ST}}=\SI{1.188}{s}/6$. Lines are
  drawn to guide the eye.}
  \label{fig:performanceweak}
\end{figure}

\subsubsection{Weak and strong scaling}

The performance of a parallel implementation is usually evaluated using two
scaling metrics, namely the \emph{weak scaling} performance and the \emph{strong
scaling} performance. The weak scaling performance is obtained by simultaneously
increasing the system size $N_{\mathrm{Tr}}+N_{\mathrm{Res}}$ and the number of
threads $N_{\mathrm{Threads}}$, while keeping the computational work per thread
constant. For the present algorithm, increasing the system size
$N_{\mathrm{Tr}}+N_{\mathrm{Res}}$ by 1 increases the size of the state vector
$\ket\Psi$ by a factor of 4 (see \equref{eq:performanceMemory}). However, the
total computational work is increased by \emph{more than} a factor of 4. This
effect is easily seen in
Listings~\ref{code:implementation0}--\ref{code:implementation2} in
\appref{app:implementations}: the number of operations in the  inner loop grows
by a factor of 4 and, additionally, the number of iterations in one of the outer
loops grows by 1 (the reason is that increasing
$N_{\mathrm{Tr}}+N_{\mathrm{Res}}$ adds additional four-component updates  to
the transformation $V$ given in \equref{eq:Vtensorproductof4x4matrices}).
Therefore, the total computational work is proportional to
$(N_{\mathrm{Tr}}+N_{\mathrm{Res}}) 4^{N_{\mathrm{Tr}}+N_{\mathrm{Res}}}$. Since
the number of threads $N_{\mathrm{Threads}}$ can only be increased by a factor
of 4, we consider the rescaled run time  $\widehat T =
T/(N_{\mathrm{Tr}}+N_{\mathrm{Res}})$. We normalize this quantity by the time
$\widehat T_{\mathrm{ST}}$ obtained for the single-threaded case at
$N_{\mathrm{Tr}}+N_{\mathrm{Res}}=6$.

The weak scaling performance is shown in \figref{fig:performanceweak}(a)--(c)
for each of the three implementations, using configuration (1) since the system
size in configuration (2) can only be increased in steps of 2
(cf.~\tabref{tab:devicebenchmark}). We find that implementations 1 and 2 show
almost ideal scaling behavior while implementation 0 does not scale as
favorably. This means that relying on branch predictors, as done for
implementation 0, can hamper parallel scalability. In particular, it is
the additional non-loop branches present in implementation 0
(cf.~Listing~\ref{code:implementation0})
that destroy the scalability. In \figref{fig:performanceweak}, the difference
between the three implementations is much more evident than in
\figref{fig:performanceruntime}, where the fact that implementation 0 is slower
can only be seen in the slightly higher red diamonds and the missing blue square
at $N_{\mathrm{Tr}}+N_{\mathrm{Res}}=13$ (in this case, the computation did not
finish within the allotted time of 24 hours).

The strong scaling performance is obtained by increasing the number of threads
$N_{\mathrm{Threads}}$ while keeping the computational work (i.e., the system
size) fixed. We choose a moderate system of size $10$ from configuration (2)
(see \tabref{tab:devicebenchmark}) with $N_{\mathrm{Tr}}=5$ transmons and
$N_{\mathrm{Res}}=5$ resonators. The number of threads
$N_{\mathrm{Threads}}$ is varied from $1$ to $48$ in steps of $1$.

\begin{figure}[p]
  \centering
  \includegraphics[width=0.75\textwidth]{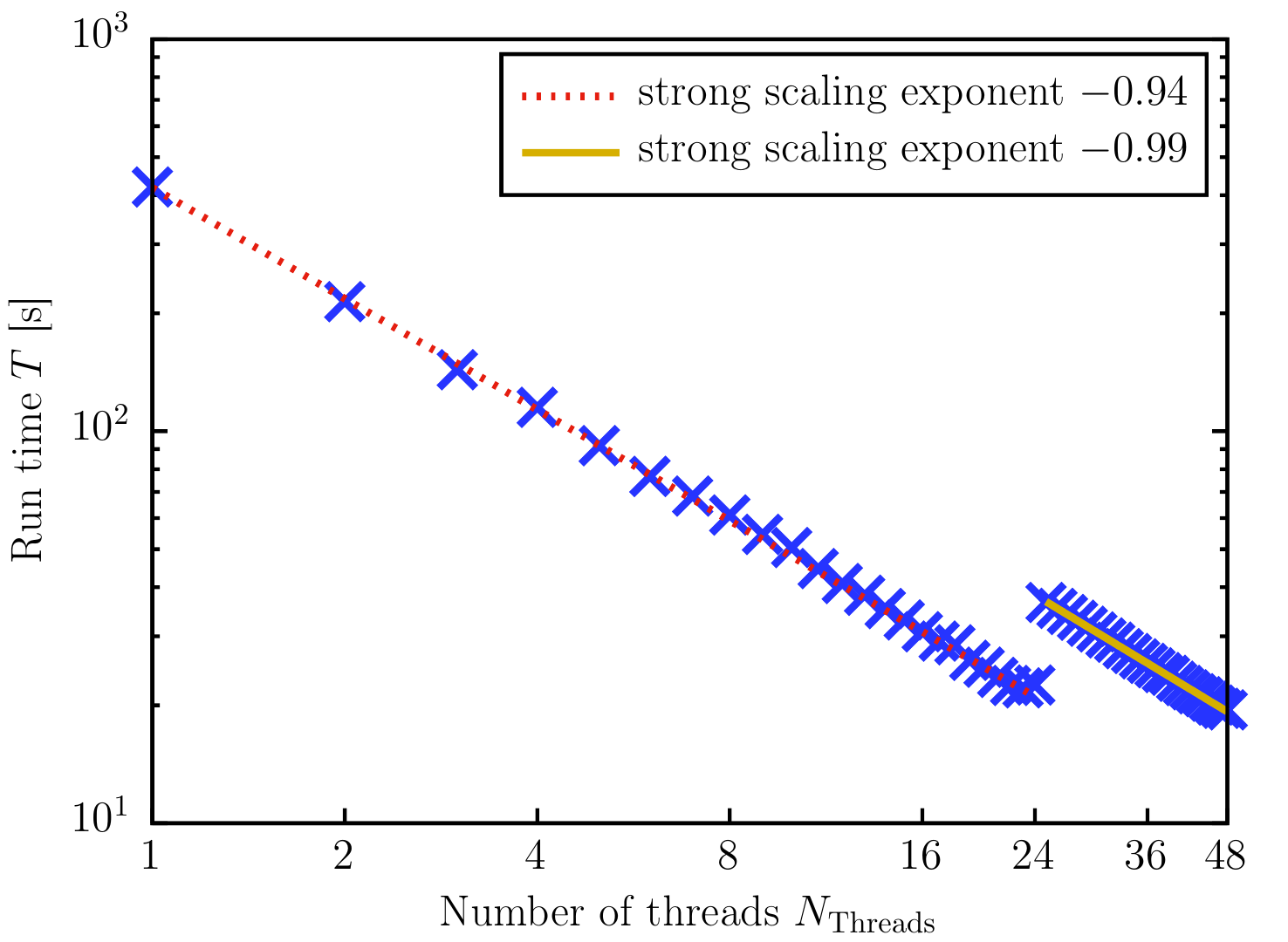}
  \caption{Strong scaling performance for a system of $N_{\mathrm{Tr}}=5$
  transmons and $N_{\mathrm{Res}}=5$ resonators. Blue crosses show the run time
  $T$ as a function of the number of threads $N_{\mathrm{Threads}}$.  Lines
  indicate a fit of the function $f(x)=ax^b$ to the data, where $b$ is the
  strong scaling exponent (ideally $-1$) and $a$ is a constant. The line is
  dotted in the domain $1\le N_{\mathrm{Threads}}\le24$ without hyper-threading,
  and solid in the domain $25\le N_{\mathrm{Threads}}\le48$ with
  hyper-threading. }
  \label{fig:performancestrong}
\end{figure}

The strong scaling performance is shown in \figref{fig:performancestrong} for
implementation 2 (the results for implementation 1 are almost identical). A node
on JURECA has 24 physical cores which can each process two threads using
hyper-threading (also called simultaneous multithreading) \cite{JURECA}.
Therefore, we find two separate domains, namely the domain $1\le
N_{\mathrm{Threads}}\le24$ where hyper-threading is off and the domain $25\le
N_{\mathrm{Threads}}\le48$ where hyper-threading is on. In each domain, a fit of
the function $f(x)=ax^b$ to the observed run times yields an almost ideal strong
scaling exponent of $b\approx-1$. Although  the run time decreases at the
hyper-threading threshold from $N_{\mathrm{Threads}}=24$ to
$N_{\mathrm{Threads}}=25$, we find that the simulation performs best when the
full capacity of the node with $N_{\mathrm{Threads}}=48$ is used.

\begin{figure}[p]
  \centering
  \includegraphics[width=\textwidth]{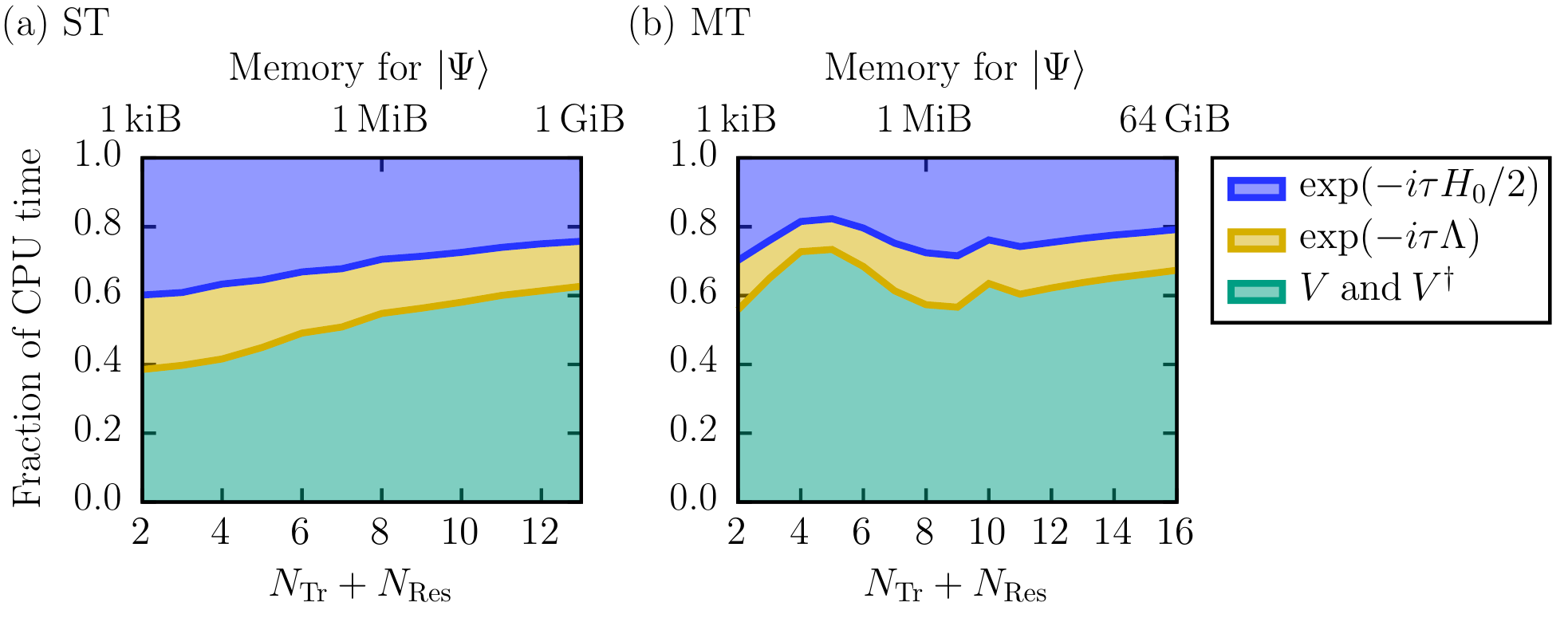}
  \caption{Breakdown of the time required for the different transformations for
  the update rule given in \equref{eq:psioftTimeEvolutionUpdateRule};  (a)
  single-threaded case using one core, (b) multi-threaded case using 48 threads
  on 24 cores. The run-time fractions are shown for both $\exp(-i \tau H_0/2)$
  transformations (blue), for the $\exp(-i\tau\Lambda)$ transformation (yellow),
  and for both $V$ and $V^\dagger$ (green). The memory required
  to store the state vector $\ket{\Psi}$ (top axis) is linked to the system size
  (bottom axis) via \equref{eq:performanceMemory}.}
  \label{fig:performanceprofiling}
\end{figure}

To analyze which part of the update rule given in
\equref{eq:psioftTimeEvolutionUpdateRule} takes the longest fraction of the run
time, we perform a breakdown of the computational work.  For this purpose, we
measure the run time for each of the transformations included in
$\widetilde{\mathcal U}$ as a function of  the system size
$N_{\mathrm{Tr}}+N_{\mathrm{Res}}$. For each transformation, we compute the
median of $10$ run times, measured every $100$ time steps. The result is shown
in \figref{fig:performanceprofiling} for both the single-threaded and  the
multi-threaded case,  using configuration (1) and  implementation 2. As
expected, the bottleneck for larger systems is the transformation $V$. The
irregularities for smaller system sizes $N_{\mathrm{Tr}}+N_{\mathrm{Res}}<7$ in
the multi-threaded case shown in \figref{fig:performanceprofiling}(b) again
reflect the above-mentioned overhead required for managing too many threads for
smaller systems.

We conclude that, among the three implementations for $V$
(cf.~\appref{app:implementations}), implementations 1 and 2 are preferable since
they are faster and show almost ideal weak and strong scaling performance. We
use implementation 2 for the simulation work presented in the remainder of this
thesis (see also the discussion at the end of \secref{sec:numericalalgorithm}).

\clearpage
\section{Single transmon-resonator system}
\label{sec:singletransmonresonatorsystem}

In this section, we study the elementary case of a single transmon coupled to a
single resonator. Such a system is small enough to make detailed analytical
investigations possible. It has often been considered in the literature (e.g.,
in the initial proposals of the circuit QED architecture
\cite{blais2004circuitqed} and the transmon \cite{koch2007transmon}). Also, the
system can be related to the famous Jaynes-Cummings model
\cite{JaynesCummings1963}. Therefore, this simple setup is an ideal candidate to
relate results from the transmon simulator to previous analytical work.

The Hamiltonian of the single transmon-resonator system is given by
\begin{align}
  \label{eq:Htotalfreesingle}
  H_{\mathrm{Single}} &= 4 E_{C} \hat n^2 - E_{J} \cos \hat\varphi
  + \Omega \hat a^\dagger\hat a + G \hat n(\hat a + \hat a^\dagger),
\end{align}
and corresponds to $H^{\mathrm{free}}$ given in
\equsref{eq:Htotalfree}{eq:HResfree} for $N_{\mathrm{Tr}}=N_{\mathrm{Res}}=1$. We
use the model parameters of a corresponding system manufactured at KIT
(see \tabref{tab:devicekit}). We first give a brief summary of some analytical
results for this model and then check their range of validity with the
simulation.

\subsection{Overview of known perturbative results}
\label{sec:singletransmonresonatorsystemPerturbative}

The Hamiltonian $H_{\mathrm{Single}}$ given in \equref{eq:HilbertSpaceInfinity}
expressed in the transmon basis $\{\ket{m}\}$ reads
\begin{align}
  \label{eq:HtotalfreesingleTransmonBasis}
  H_{\mathrm{Single}} &= \sum\limits_{m} E_{m}^{\mathrm{Tr}} \ketbra{m}{m}
  + \Omega \hat a^\dagger\hat a
  + \sum\limits_{mm'} G n^{(m,m')} \ketbra{m}{m'}(\hat a + \hat a^\dagger),
\end{align}
where $E_{m}^{\mathrm{Tr}}$ denotes the transmon eigenenergies (see
\equref{eq:transmoneigenstates}),  and $n^{(m,m')}$ denotes the matrix elements
of the charge operator $\hat n$ in  the transmon basis (see
\equref{eq:transmonchargeoperatortransmonbasis}). In this representation,
$H_{\mathrm{Single}}$ resembles the Jaynes-Cummings  model
\cite{JaynesCummings1963}, which is why it is often called \emph{generalized
Jaynes-Cummings Hamiltonian} \cite{koch2007transmon}.

Several approximations to this Hamiltonian are frequently found in the
literature. The first step in most of them is an approximation of the transmon
as an anharmonic oscillator (AO), based on the observation that the energy
levels $E_{m}^{\mathrm{Tr}}$ are almost equidistant. To apply this
approximation, we introduce operators $\hat{a}_{\mathrm{Tr}}$ and
$\hat{a}_{\mathrm{Tr}}^\dagger$
(cf.~\equaref{eq:extractFosterLadderOperatorsN}{eq:extractFosterLadderOperatorsPhi})
such that
\begin{subequations}
  \begin{align}
    \label{eq:transmonLadderOperatorsN}
    \hat{n} &= -\frac{1}{\sqrt 2}\left(\frac{E_{J}}{8E_{C}}\right)^{1/4}(\hat{a}_{\mathrm{Tr}}+\hat{a}_{\mathrm{Tr}}^\dagger), \\
    \label{eq:transmonLadderOperatorsPhi}
    \hat{\varphi} &= \frac{i}{\sqrt 2}\left(\frac{8E_{C}}{E_{J}}\right)^{1/4}(\hat{a}_{\mathrm{Tr}}-\hat{a}_{\mathrm{Tr}}^\dagger),
  \end{align}
\end{subequations}
with $[\hat{a}_{\mathrm{Tr}},\hat{a}_{\mathrm{Tr}}^\dagger]=1$. The operators
$\hat{n}$ and $\hat{\varphi}$ can always be written in this way because they are
Hermitian and obey $[\hat{n},\hat{\varphi}]=i$ (note that, strictly  speaking,
the commutator $[\hat n,\hat\varphi]=i$ is only well defined on the domain of
periodic functions; see Appendix B in \cite{Willsch2016Master}). In the
literature, the phase factors $-1$ and $i$ are sometimes put in different places
(cf.~\cite{devoret1997quantumfluctuations, koch2007transmon,
gambetta2013controlIFF}), but this does not change the resulting time
evolutions.

For the AO approximation, the operators $\hat{a}_{\mathrm{Tr}}$ and
$\hat{a}_{\mathrm{Tr}}^\dagger$ are effectively replaced by \emph{ladder
operators} $\hat b = \sum_m\sqrt{m+1}\ketbra{m}{m+1}$ and $\hat b^\dagger =
\sum_m\sqrt{m+1}\ketbra{m+1}{m}$, respectively (see \cite{koch2007transmon}).
This means that the matrix representation  of $\hat n$ in the transmon basis
(see \equref{eq:transmonchargeoperatortransmonbasis}) is approximated by a
tridiagonal matrix.

We first study the accuracy for the case in which the AO approximation is only
done for the interaction term $G \hat n(\hat a + \hat a^\dagger)$, but in
combination with the rotating wave approximation (RWA), $(\hat b + \hat
b^\dagger)(\hat a + \hat a^\dagger)\approx\hat b\hat a^\dagger + \hat b^\dagger
\hat a$. We define (cf.~\equref{eq:HtotalfreesingleTransmonBasis})
\begin{align}
  \label{eq:HtotalfreesingleAOIntRWA}
  H_{\mathrm{Single}}^{\mathrm{AOIntRWA}} &= \sum\limits_{m} E_{m}^{\mathrm{Tr}} \ketbra{m}{m}
  + \Omega \hat a^\dagger\hat a
  + g (\hat b\hat a^\dagger + \hat b^\dagger\hat a),
\end{align}
where
\begin{align}
  \label{eq:transmonresonatorexchangecoupling}
  g&=-\left(\frac{E_J}{32E_C}\right)^{1/4}G
\end{align}
is the rescaled transmon-resonator coupling.

Since $g$ is small compared to the other energy scales
(cf.~\tabref{tab:devicekit}), the Hamiltonian in
\equref{eq:HtotalfreesingleAOIntRWA} is often diagonalized in first-order
perturbation theory (PT) for the eigenstates $\ket{km}$ of the diagonal part
(note that $\hat a^\dagger\hat a=\sum_k k\ketbra k k$ is also diagonal in this
basis). One obtains \cite{koch2007transmon}
\begin{align}
  \label{eq:transmonDressedStates}
  \overline{\ket{km}} = \frac{1}{\mathcal N_{km}}\left(\ket{km}
  + \frac{g}{\Delta_{m}}\hat a\hat b^\dagger\ket{km}
  - \frac{g}{\Delta_{m-1}}\hat a^\dagger\hat b\ket{km} \right),
\end{align}
where $\Delta_{m}=E_{m+1}^{\mathrm{Tr}}-E_{m}^{\mathrm{Tr}}-\Omega$, and the
factor $\mathcal N_{km}$ is chosen such that $\overline{\ket{km}}$ is
normalized. The states $\overline{\ket{km}}$ are often called \emph{dressed}
states since the transmon states are dressed by a small photonic component
\cite{fox2006quantumoptics} (see also
\cite{Pommerening2020WhatIsMeasuredMultiQubitChip}). Using
\equref{eq:transmonDressedStates}, we define the second approximation under
investigation by
\begin{align}
  \label{eq:HtotalfreesingleAOIntPT}
  H_{\mathrm{Single}}^{\mathrm{AOIntPT}} &= \sum\limits_{km} (E_{m}^{\mathrm{Tr}} + k\Omega) \overline{\ketbra{km}{km}},
\end{align}
where the eigenvalues $E_{m}^{\mathrm{Tr}} + k\Omega$ are the same as  for the
original Hamiltonian given in \equref{eq:HtotalfreesingleTransmonBasis}, because
they are not affected by the off-diagonal interaction operator in first-order
PT. Note that the same result can be obtained using the Schrieffer-Wolff
transformation \cite{schriefferwolff1966, BravyiDivincenzo2011schriefferwolff}
(see \cite{richer2013perturbative} for the calculation).

A third alternative is to extend the AO approximation (which was only
done for the interaction before) to the full transmon
(cf.~\cite{gambetta2013controlIFF}),
\begin{align}
  \label{eq:HtotalfreesingleAOFull}
  H_{\mathrm{Single}}^{\mathrm{AOFull}} &= \bar\omega\hat b^\dagger\hat b + \frac{\bar\alpha}2 \hat b^\dagger\hat b (\hat b^\dagger\hat b -1)
  + \Omega \hat a^\dagger\hat a
  + g (\hat b + \hat b^\dagger)(\hat a + \hat a^\dagger),
\end{align}
where $\bar\omega=\sqrt{8 E_C E_J}-E_C$ and $\bar\alpha=-E_C$ are the resulting
approximations to the qubit frequency and the anharmonicity, respectively (see
also the discussion below \equref{eq:transmoneigenstates}). In this form, the
transmon's character as an AO is evident: Using $\hat b^\dagger\hat b = \sum_m m
\ketbra m m$, \equref{eq:HtotalfreesingleAOFull} states that the eigenvalues
$E_m^{\mathrm{Tr}}$ are approximated by $\bar\omega m + \bar\alpha m(m-1)/2$
such  that successive eigenvalues differ by
$E_{m+1}^{\mathrm{Tr}}-E_m^{\mathrm{Tr}}\approx\bar\omega +\bar\alpha m$.  Note
that unlike \equref{eq:HtotalfreesingleAOIntRWA}, the RWA has not been used in
\equref{eq:HtotalfreesingleAOFull}.

Finally, we consider the two-level approximation (TLA) of the transmon. This is
the crudest approximation since it retains only the two lowest-energy
eigenstates $\ket{m=0}$ and $\ket{m=1}$ of \equref{eq:HtotalfreesingleAOFull}.
Expressing the ladder operators in terms of Pauli matrices
$\sigma^z=-\ketbra00+\ketbra11$ and $\sigma^x=\ketbra01+\ketbra10$, we obtain
(up to a constant)
\begin{align}
  \label{eq:HtotalfreesingleTLA}
  H_{\mathrm{Single}}^{\mathrm{TLA}} &= -\frac{\bar\omega}2 \sigma^z
  + \Omega \hat a^\dagger\hat a
  + g \sigma^x(\hat a + \hat a^\dagger),
\end{align}
which resembles the Jaynes-Cummings model of an atom coupled to an electric
field (see \cite{JaynesCummings1963,gerry2005introductoryquantumoptics}), except
that the RWA has not been used. The reason for not using the RWA is that all
deviations from the exact time evolution can then be attributed to the TLA.

In summary, we list all of these approximations in terms of the transmon states
$\{\ket{m}\}$ or the dressed transmon-resonator states
$\{\overline{\ket{km}}\}$, respectively:
\begin{subequations}
\begin{align}
  \label{eq:HtotalfreesingleAOIntRWATransmonBasis}
  H_{\mathrm{Single}}^{\mathrm{AOIntRWA}} &= \sum\limits_{m} E_{m}^{\mathrm{Tr}} \ketbra{m}{m}
  + \Omega \hat a^\dagger\hat a \nonumber\\
  &\quad+ \sum\limits_{m} g \sqrt{m+1} (\ketbra{m}{m+1}\hat a^\dagger + \ketbra{m+1}{m}\hat a),\\
  \label{eq:HtotalfreesingleAOIntPTTransmonBasis}
  H_{\mathrm{Single}}^{\mathrm{AOIntPT}} &= \sum\limits_{km} (E_{m}^{\mathrm{Tr}} + k\Omega) \overline{\ketbra{km}{km}},\\
  \label{eq:HtotalfreesingleAOFullTransmonBasis}
  H_{\mathrm{Single}}^{\mathrm{AOFull}} &= \sum\limits_{m} (\bar\omega m + \frac{\bar\alpha}{2}m(m-1)) \ketbra{m}{m}
  + \Omega \hat a^\dagger\hat a \nonumber\\
  &\quad+ \sum\limits_{m} g \sqrt{m+1} (\ketbra{m}{m+1} + \ketbra{m+1}{m})(\hat a + \hat a^\dagger),\\
  \label{eq:HtotalfreesingleTLATransmonBasis}
  H_{\mathrm{Single}}^{\mathrm{TLA}} &= \bar\omega \ketbra{m=1}{m=1}
  + \Omega \hat a^\dagger\hat a \nonumber\\
  &\quad+ g (\ketbra{m=0}{m=1}+\ketbra{m=1}{m=0})(\hat a + \hat a^\dagger).
\end{align}
\end{subequations}

\subsection{Comparison to simulation results}

\begin{figure}[p]
  \centering
  \includegraphics[width=\textwidth]{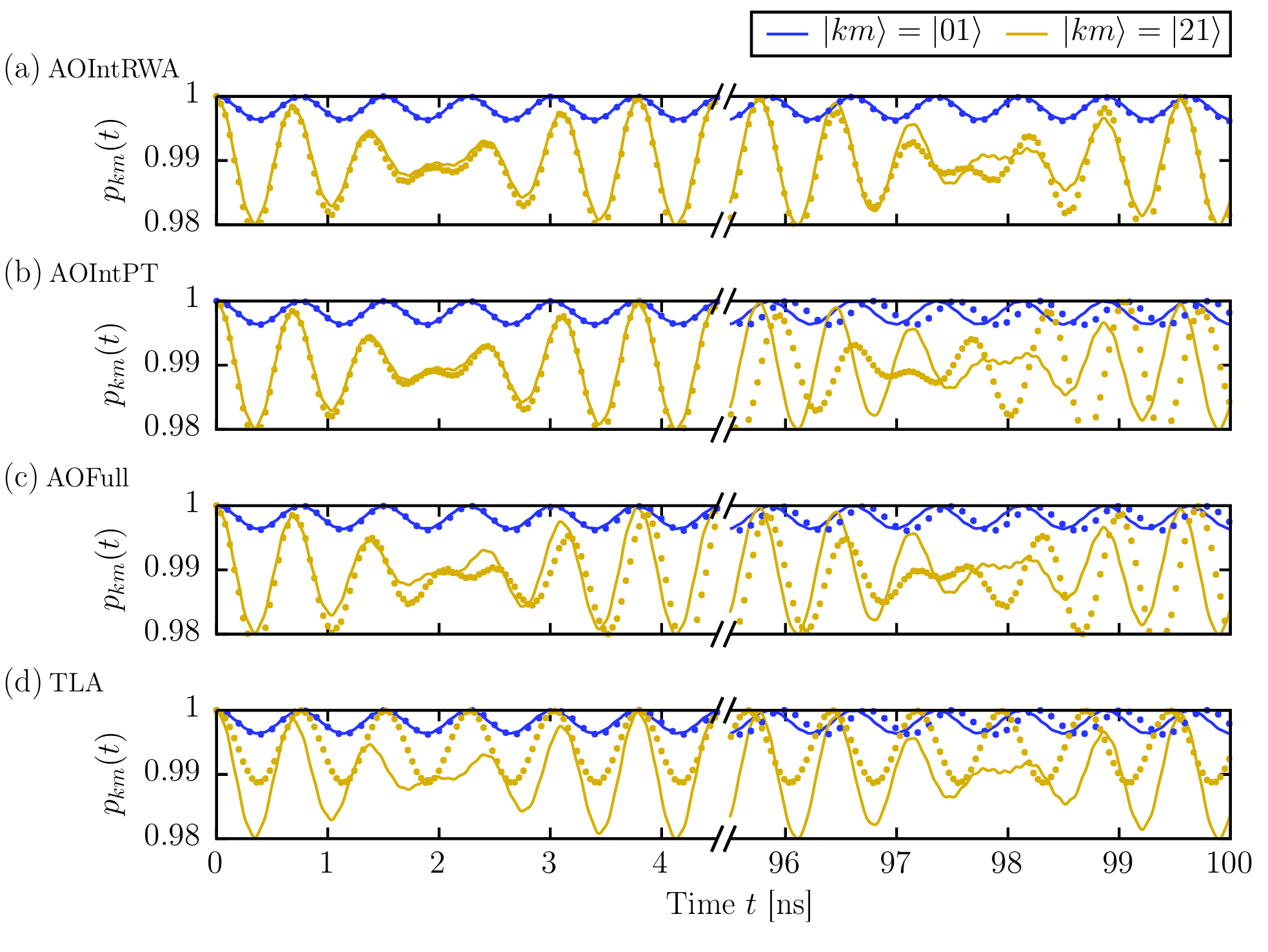}
  \caption{Time evolution of two different initial states $\ket{km}=\ket{01},\ket{21}$ to compare the exact
  solution (solid lines) and various common approximations (dots):
  (a) using an anharmonic oscillator for the interaction term combined with the RWA (see \equref{eq:HtotalfreesingleAOIntRWA}),
  (b) the same but in first-order PT (see \equref{eq:HtotalfreesingleAOIntPT}),
  (c) using an anharmonic oscillator for the whole transmon (see \equref{eq:HtotalfreesingleAOFull}),
  (d) using a two-level system to describe the transmon (see \equref{eq:HtotalfreesingleTLA}).
  Shown is the probability $p_{km}(t)$ to measure the system in the state $\ket{km}$
  (see \equref{eq:freekitcomparetimeevolutionspkm}).
  The exact time evolution corresponds to the Hamiltonian $H_{\mathrm{Single}}$
  given in \equref{eq:Htotalfreesingle}, and the approximations (a)--(d) correspond to the
  Hamiltonians given in
  \equsref{eq:HtotalfreesingleAOIntRWATransmonBasis}{eq:HtotalfreesingleTLATransmonBasis}}
  \label{fig:freekitcomparetimeevolutions}
\end{figure}

We measure the success of the different approximations given in
\equsref{eq:HtotalfreesingleAOIntRWATransmonBasis}{eq:HtotalfreesingleTLATransmonBasis}
by how well they predict the time evolution. For every Hamiltonian $\widetilde
H\in\smash{\{H_{\mathrm{Single}}^{\mathrm{AOIntRWA}},}$
$\smash{H_{\mathrm{Single}}^{\mathrm{AOIntPT}},}$
$\smash{H_{\mathrm{Single}}^{\mathrm{AOFull}},}$ $\smash{H_{\mathrm{Single}}^{\mathrm{TLA}}\}}$
in \equsref{eq:HtotalfreesingleAOIntRWATransmonBasis}
{eq:HtotalfreesingleTLATransmonBasis}, we obtain the time evolution by numerical
diagonalization. This is done by diagonalizing the matrix representation of
$\widetilde H$ with respect to the joint transmon-resonator basis $\{\ket{km}\}$
for $k,m\in\{0,\ldots,9\}$. We then use the resulting eigenvalues and
eigenvectors of $\widetilde H$ to compute the time-evolution operator
$\widetilde U(t) = \exp(-i\widetilde H t)$. Note that this procedure is only
possible since the matrices are small enough and the Hamiltonians do not contain
time-dependent terms.

We compute the probability $p_{km}(t)$ for $0\le t \le\SI{100}{ns}$ to
measure the system in the state $\ket{km}$ if it has been initialized in
$\ket{km}$ at $t=0$. Using the time-evolution operator $\widetilde U(t)$, this
probability is given by%
\begin{align}
  \label{eq:freekitcomparetimeevolutionspkm}
  p_{km}(t) = \abs{\braket{km|\widetilde U(t)|km}}^2.
\end{align}

The reference time evolution is the exact result governed by the Hamiltonian
given in \equref{eq:Htotalfreesingle}. It is obtained using  the transmon
simulator defined in \secref{sec:simulationsoftware} (note that the error for
this system can be made arbitrarily small by scaling the time step $\tau$; see
\figref{fig:accuracylocalerror}(a)). Explicitly, using the notation from
\equref{eq:psioftsolutioncoefficients}, we set
$\psi_{k_0m_0}(0)=\delta_{k_0k}\delta_{m_0m}$, perform the time evolution, and
compute $p_{km}(t)=\abs{\psi_{km}(t)}^2$.

The result is shown in \figref{fig:freekitcomparetimeevolutions} for the initial
states $\ket{km}=\ket{01},\ket{21}$, i.e., the transmon is always initialized in
the excited qubit state $\ket{m=1}$ and the resonator is populated with either
$k=0$  or $k=2$ photons. We see that if the AO approximation is only used  for
the interaction, both the short-term and the long-term evolution are described
reasonably well (see \figref{fig:freekitcomparetimeevolutions}(a)). We checked
that in this case, the result is the same whether the RWA is used for this term
or not. The perturbative result shown in
\figref{fig:freekitcomparetimeevolutions}(b) can describe the short-term
evolution equally well, but acquires a shift for longer time evolutions. The
same applies if the whole transmon is approximated as an  anharmonic oscillator
(see \figref{fig:freekitcomparetimeevolutions}(c)). However, in this case a
slight drift for the $k=2$ case is also already observable at $t=\SI{4}{ns}$.
Still, in all approximations for which the transmon is described by more than
two levels, the average amplitude of the oscillations is correct. As shown in
\figref{fig:freekitcomparetimeevolutions}(d), this is not true anymore for the
two-level approximation. It suffers both from a drift and also from a reduced
amplitude in the $k=2$ case. The strength of the interaction for $k\neq0$ is
thus underestimated.

We can conclude from this that higher transmon states play an important role in
mediating the interaction between the transmon and the resonator. We find that
none of the approximations are suitable for the optimization of quantum gate
pulses which take $\mathcal O(\SI{100}{ns})$. The reason is that all
approximations develop a drift at this time scale, and drifts are related to
inaccurate relative phases which are required to be very precise in order to
implement e.g. the \textsc{CNOT} gate.  As shown in
\secref{sec:statedependentfrequenciesfree}, however, there still exist
nontrivial cases in which even a two-level approximation describes the time
evolution rather well.

\section{Transmon-resonator system coupled to a bath}
\label{sec:freetransmonresonatorbathphotons}

In practice, it is almost impossible to keep readout resonators completely void
of photons (see, for instance, \cite{Suri2015ResonatorPhotonOccupancy,
Bultink2016ActiveResonatorResetPhotons}). This means, in the context of this
work, that it is difficult to prepare a resonator exactly in the state
$\ket{k=0}$ (cf.~\equref{eq:psioftsolutioncoefficients}). An interesting
question is how much a transmon qubit is affected by the presence of photons in
the resonator.

We study this question using three complimentary approaches. First, we consider
an isolated transmon-resonator system as defined in
\secref{sec:transmonmodelkit}. Next, we study its coupling to a heat bath using
the model defined in \secref{sec:transmonmodelresonatorbath} (this model  is the
same that can be used to describe electromagnetic environments, see
\secref{sec:extractfoster}).  These first two approaches are based on solving
the TDSE given in \equref{eq:tdse3} using the transmon simulator described in
\secref{sec:simulationsoftware}. For the third approach, we use a quantum master
equation approach. This allows us to address the transition from a closed system
over the system-bath model to the effective evolution described by a quantum
master equation.

\subsection{Simulation models}
\label{sec:freetransmonresonatorbathphotonsSimulationModels}

The central system considered in this section is the single transmon-resonator
system defined in \secref{sec:transmonmodelkit}. We supplement the
system with a bath of $L=10$ resonators to model an open quantum system. The
bath Hamiltonian essentially represents a collection of harmonic oscillators,
i.e., a boson bath. Thus, the model Hamiltonian for the TDSE is given by $H =
H_{\mathrm{Single}} + H_{\mathrm{Bath}}$, where
\begin{subequations}
  \begin{align}
    \label{eq:HtotalfreeBathSingle}
    H_{\mathrm{Single}} &= 4 E_{C} \hat n^2 - E_{J} \cos \hat\varphi
    + \Omega \hat a^\dagger\hat a + G \hat n(\hat a + \hat a^\dagger), \\
    \label{eq:HtotalfreeBathBath}
    H_{\mathrm{Bath}} &= \sum_{l=1}^L W_l \hat b_l^\dagger\hat  b_l
     + \sum_{l=1}^L \lambda_l (\hat a+\hat a^\dagger)(\hat b_l+\hat b_l^\dagger).
  \end{align}
\end{subequations}
For clarity, we use the symbols $\hat b_l$ to distinguish the bath resonators
from  the central resonator. Thus, the mapping to the model Hamiltonian
$H_{\mathrm{Res}}^{\mathrm{free}}$ given in \equref{eq:HResfree} is $\hat
a_{r=0}\leftrightarrow\hat a$ and $\hat a_{l}\leftrightarrow\hat b_l$ for
$l=1,\ldots,L$. The specification of the bath parameters and their relation to
the parameters of the full model Hamiltonian used in the simulation framework
(see \equsref{eq:Htotal}{eq:HCC}) is given in \tabref{tab:devicekitbath}. The
topology of the system is sketched in \figref{fig:kitbathtopology}.

\subsubsection{Isolated system}

The first approach focuses on the isolated transmon-resonator system.
As in \secref{sec:singletransmonresonatorsystem}, we solve the TDSE
\begin{align}
  \label{eq:photonsApproach1}
  i \frac{\partial}{\partial t} \ket{\Psi(t)} &= H_{\mathrm{Single}} \ket{\Psi(t)},
\end{align}
where $H_{\mathrm{Single}}$ is given in \equref{eq:HtotalfreeBathSingle},
corresponding to $H^{\mathrm{free}}$ given in
\equsref{eq:Htotalfree}{eq:HResfree} for $N_{\mathrm{Tr}}=N_{\mathrm{Res}}=1$. We
use the model parameters of a transmon system manufactured at KIT
(see \tabref{tab:devicekit}).

The system's initial state is given by $\ket{\Psi(0)}=\ket{k,m=0}$ such that the
transmon is initialized in its ground state and the resonator is populated with
$k$ photons. We intentionally do not initialize the system in an eigenstate of
the full Hamiltonian (such as the dressed states given in
\equref{eq:transmonDressedStates}). The reason for this is that in an
eigenstate, the time evolution would be trivial for each $k$. Instead, the goal
is to assess the impact of $k$ additional photons on the transmon-resonator
interaction and its consequences for the time evolution of the transmon system.

We analyze the free time evolution for various $k$ between $0$ and $180$. At
each time $t$, we evaluate the probability $p_{m\neq0}(t)$ to find the transmon
in a higher excited state $\ket{m}$ with $m\neq0$. This probability is given by
\begin{align}
  \label{eq:freekitkprobabilitymnot0Psi}
  p_{m\neq0}(t) = 1 - \bra{\Psi(t)}(\ketbra{m=0}{m=0})\ket{\Psi(t)}, %\sum_{k_0} \abs{\psi_{k_00}(t)}^2,
\end{align}
where $\ket{\Psi(t)}$ is given in \equref{eq:psioftsolutioncoefficients}.
Since the Hilbert space given in \equref{eq:HilbertSpaceTruncated} includes
only four states for the resonator, we make use of the parameter
$k^{\mathrm{offset}}$ to study larger values of $k$. Specifically, we set
$k^{\mathrm{offset}}=k-2$ (for $k\ge2$) such that two Fock states below
$\ket{k}$ and one Fock state above $\ket{k}$ are taken into account. We have
verified by exact diagonalization that with this choice, the quantities studied
for this system are accurate up to three significant digits.

\subsubsection{System with bath}

For the second approach, we use the transmon simulator to solve the TDSE
\begin{align}
  \label{eq:photonsApproach2}
  i\frac{\partial}{\partial t} \ket{\Psi(t)} &= (H_{\mathrm{Single}}+H_{\mathrm{Bath}}) \ket{\Psi(t)},
\end{align}
where $H_{\mathrm{Single}}$ and $H_{\mathrm{Bath}}$ are given in
\equaref{eq:HtotalfreeBathSingle}{eq:HtotalfreeBathBath}, respectively. We
consider various coupling strengths $\lambda_l$ that are chosen uniformly from
$[0,\lambda]$, where
$\lambda\in2\pi\times\{\SI{5}{MHz},\SI{10}{MHz},\SI{20}{MHz}\}$. Furthermore,
the frequencies $W_l$ of the bath modes are chosen randomly from a Gaussian
distribution centered around $\Omega$ (see \tabref{tab:devicekitbath}).  The
choice of random bath parameters is motivated by the observation that for such
large, generic models, randomness in the bath parameters is required to model
generic effects \cite{Jin2013BathNeedsRandomness}. Note that in principle, this
particular system-bath setup is a sufficiently general model for a
superconducting environment (see \secref{sec:extractfoster}, where we also
describe a procedure to extract the bath parameters $W_l$ and $\lambda_l$ from
electromagnetic HFSS simulations).

We solve \equref{eq:photonsApproach2} for an initial state given by a
product of the state $\ket{k,m=0}$ (as before) and the zero temperature ground
state of the bath. Note that the state vectors
$\ket{\Psi(t)}$ in \equaref{eq:photonsApproach1}{eq:photonsApproach2} are
defined on different Hilbert spaces. Specifically, the Hilbert space for
\equref{eq:photonsApproach1} includes one transmon and one resonator, whereas
the one for \equref{eq:photonsApproach2} contains 11 resonators and is thus much
larger (cf.~\equref{eq:psioftsolutioncoefficients}). In the case $\lambda=0$,
however, the solution of \equref{eq:photonsApproach2} coincides with the
solution of \equref{eq:photonsApproach1} after projection onto the smaller
Hilbert space. All simulations for the approach given by
\equref{eq:photonsApproach2} were performed on the supercomputers JURECA
\cite{JURECA} and JUWELS \cite{JUWELS}.

As before, we study the effect of a larger number of photons $k$ in the system's
resonator by evaluating \equref{eq:freekitkprobabilitymnot0Psi}. As opposed to
the isolated case, the photons now directly couple to the environment. This
means that for a larger number of photons $k$, energy can leak out of the
resonator into the environment. Furthermore, energy exchange between the
transmon and the environment can be virtually mediated by the resonator.
Therefore, a reasonable expectation would be that the average transmon
excitation $\langle p_{m\neq0}(t)\rangle$ is reduced.

We remark that simulating \equref{eq:photonsApproach2} also allows for a study
of the theory used to describe the Purcell effect for transmons, which starts
from a bath model similar to \equref{eq:HtotalfreeBathBath} to derive the
transmon relaxation rate (see \cite{koch2007transmon}).

\subsubsection{Master equation}

As a third approach, we consider a Lindblad master equation (also known as GKLS
equation) \cite{Gorini1976CPmapsLindblad,Lindblad1976} for the system's
reduced density matrix $\rho(t)$,
\begin{align}
  \label{eq:masterequation}
    \frac{\partial}{\partial t} \rho(t) &= -i[H_{\mathrm{Single}}, \rho(t)] + \sum_{\gamma}\kappa_\gamma \mathcal D[A_\gamma](\rho(t)),
  % \frac{\partial}{\partial t}\rho(t) &= -i[H_{\mathrm{Single}}, \rho(t)]
  % + \frac\kappa 2 (2 \hat a\rho(t) \hat a^\dagger
  % - \hat a^\dagger \hat a \rho(t)
  % - \rho(t) \hat a^\dagger \hat a),
\end{align}
where the dissipator $\mathcal D[A_\gamma]$ is defined by $\mathcal
D[A_\gamma](B)=A_\gamma B A_\gamma^\dagger-(A_\gamma^\dagger A_\gamma B + B
A_\gamma^\dagger A_\gamma)/2$ for some operator $B$. This type of master
equation is the most general form for the generator of the quantum dynamical
semigroup \cite{breuer2007openquantumsystems}. The associated time evolution is
automatically CPTP (cf.~\secref{sec:quantumoperations}). In this context, the
reference to the algebraic structure of a semigroup is due to the fact that an
inverse time evolution is not required \cite{Kossakowski1972DynamicalSemigroup}
(unitary time evolutions, on the other hand, form a group because they are
reversible). A comprehensive historical account of the events that led to
\equref{eq:masterequation}, including a survey of the results, is given in
\cite{Chruscinksi2017HistoryLindbladEquation}.

For a single dissipator $\mathcal D[\hat a]$, the master equation in \equref{eq:masterequation} reads
\begin{align}
  \label{eq:photonsApproach3}
  \frac{\partial}{\partial t} \rho(t) &= -i[H_{\mathrm{Single}}, \rho(t)]
  + \frac{\kappa}2 \left(2\hat a \rho(t)\hat a^\dagger-\hat a^\dagger\hat a\rho(t) - \rho(t)\hat a^\dagger \hat a\right),
\end{align}
where $\kappa=2\pi\times\SI{2.7}{MHz}$ denotes the photon loss rate obtained
from  a corresponding experiment at KIT \cite{DennisIoan2019}. In addition to
the single dissipator $\mathcal D[\hat a]$, one frequently considers effects of
the form $\mathcal D[\hat a^\dagger]$ for environment-induced photon excitation,
or corresponding versions for qubit relaxation, excitation, and dephasing (see
e.g.~\cite{Suri2015ResonatorPhotonOccupancy}). The motivation to consider
$\mathcal D[\hat a]$ the most dominant dissipator is that the readout resonator
is directly coupled to the ``outside world'' (i.e, the transmission line, whose
temperature may be much higher than the effective temperature of the transmons).
Therefore, photons in the resonator can easily leak out and take energy away
from the system. This is especially true if the number of photons $k$ in the
resonator is large.

In the case of the master equation, the solution is given by the system's
density matrix $\rho(t)$. Therefore, the probability from
\equref{eq:freekitkprobabilitymnot0Psi} to find the transmon in a higher excited
state $\ket m$ with $m\neq0$
is given by
\begin{align}
  \label{eq:freekitkprobabilitymnot0Rho}
  p_{m\neq0}(t) = 1-\braket{m=0|\rho(t)|m=0}.
\end{align}

We emphasize that the bath approach in \equref{eq:photonsApproach2} is more
general than the other two: The solution reduces to the one of
\equref{eq:photonsApproach1} if the bath coupling strength $\lambda=0$, and a
Markovian master equation of the form of \equref{eq:photonsApproach3} for the
reduced density matrix
$\rho(t)=\mathrm{Tr}_{\mathrm{Bath}}\,\ketbra{\Psi(t)}{\Psi(t)}$ can describe  the
more complicated dynamics generated by a system-bath model only under certain
conditions \cite{breuer2007openquantumsystems} (see
\cite{zhao2016masterequation, DeRaedt2017relaxation} for detailed investigations
of this point).

\subsection{Results}

\begin{figure}[p]
  \centering
  \includegraphics[width=\textwidth]{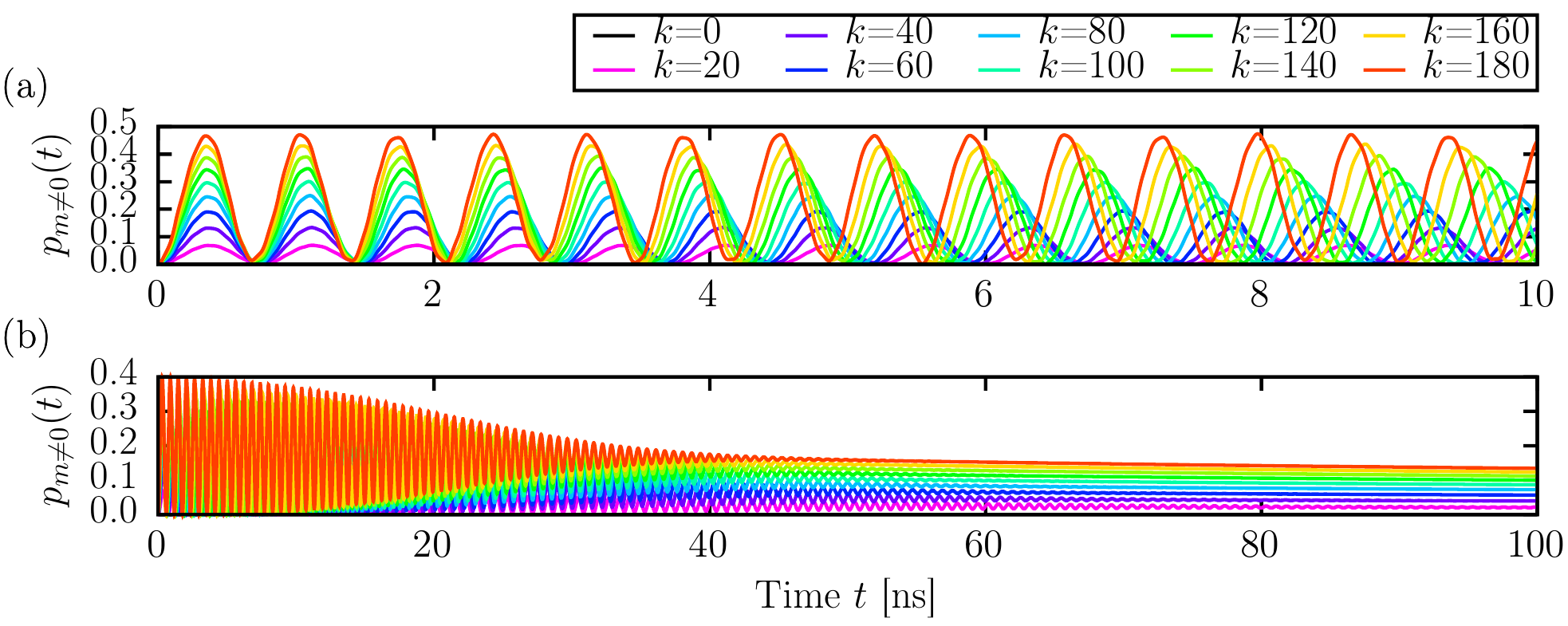}
  \caption{Time evolution of the probability $p_{m\neq0}(t)$ to find the
  transmon in any excited state $\ket{m}$ for $m\neq0$. $p_{m\neq0}(t)$ is obtained
  by solving (a) the TDSE given in \equref{eq:photonsApproach1} (see
  \equref{eq:freekitkprobabilitymnot0Psi}), (b) the master equation given
  in \equref{eq:photonsApproach3} (see \equref{eq:freekitkprobabilitymnot0Rho}).
  Different colors correspond to a different number of photons $k$ in the resonator.
  }
  \label{fig:freekitktimeevolution}
\end{figure}

\begin{figure}[p]
  \centering
  \includegraphics[width=\textwidth]{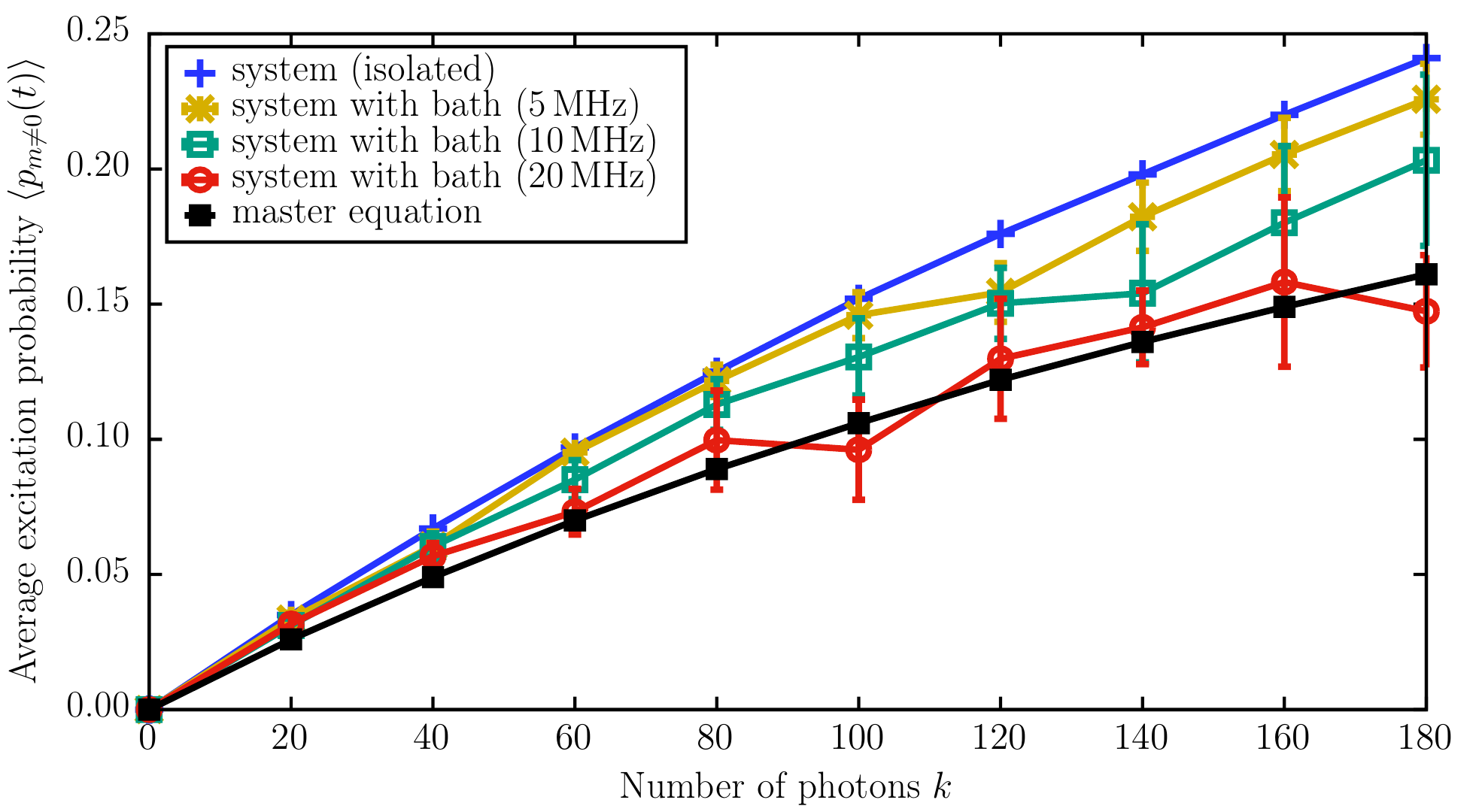}
  \caption{Average probability to find the transmon in any excited state
  $\ket{m}$ for $m\neq0$ as a function of the number of photons $k$ in the
  resonator, using the three approaches described in
  \secref{sec:freetransmonresonatorbathphotonsSimulationModels}. Each point
  corresponds to the time average $\langle p_{m\neq0}(t)\rangle$ given  in
  \equref{eq:photonsAverageExcitation}. For the isolated system and the master
  equation, the corresponding probabilities $p_{m\neq0}(t)$ are shown in
  \figref{fig:freekitktimeevolution}. For the bath simulations, each point
  (error bar) is the mean (standard deviation) of 10 independent results for
  $\langle p_{m\neq0}(t)\rangle$, each of which uses different, random bath
  parameters that are distributed as specified in \tabref{tab:devicekitbath}.
  }
  \label{fig:freekitkaverage}
\end{figure}

The time evolution of $p_{m\neq0}(t)$ for $k=0,20,\ldots,180$ photons
is shown in \figref{fig:freekitktimeevolution}. For the isolated system,
\figref{fig:freekitktimeevolution}(a) shows oscillations of the transmon between
$\ket{m=0}$
and higher excited states. For $k=180$ (red curve), this oscillation is
quite strong such that after $\SI{0.33}{ns}$, $p_{m\neq0}(t)$ already reaches
$46\%$. A closer inspection (data not shown) yields that at this point in time,
the probability to find $\ket{m=1}$ ($\ket{m=2}$) is $37\%$ ($9\%$). Hence, an
increased number of photons can also excite the qubit to higher,
non-computational  states. \sfigref{fig:freekitktimeevolution} also shows that
with increasing $k$, the frequency of the oscillation between $\ket{m=0}$ and
higher states grows.

In \figref{fig:freekitktimeevolution}(b), we show the corresponding probability
$p_{m\neq0}(t)$ as described by the Lindblad master equation given in
\equref{eq:photonsApproach3}. For the first $\SI{10}{ns}$, the time evolution is
almost equal to the isolated transmon-resonator case shown in
\figref{fig:freekitktimeevolution}(a). After that, the photon loss modeled by
the dissipator $\mathcal D[\hat a]$  becomes observable, and the oscillations of
$p_{m\neq0}(t)$ decay. Note that the state at the end of the depicted time
evolution is not the steady state,  since the photon-loss mechanism in this
model would continue to take energy from the system until the resonator is
completely depleted of photons.

We average the probability $p_{m\neq0}(t)$ shown in
\figref{fig:freekitktimeevolution} over a period $0\le t\le T$ to obtain the
average excitation probability
\begin{align}
  \label{eq:photonsAverageExcitation}
  \langle p_{m\neq0}(t)\rangle = \frac 1 T \int\limits_0^T p_{m\neq0}(t)\,\mathrm{d}t,
\end{align}
for each number of photons $k=0,20,\ldots,180$ and each of the three
approaches  introduced above.  For the TDSE-based approaches given by
\equaref{eq:photonsApproach1}{eq:photonsApproach2}, we take $T=\SI{20}{ns}$ such
that enough oscillations contribute to the average. For the master-equation
approach given by \equref{eq:photonsApproach3}, we take $T=\SI{100}{ns}$,
because this is roughly the time scale of the measurement process in the
corresponding experiment \cite{DennisIoan2019}. The result is shown in
\figref{fig:freekitkaverage}.

For the bath simulations (yellow, green, and red curves in
\figref{fig:freekitkaverage}), we additionally average $\langle
p_{m\neq0}(t)\rangle$ over ten independent runs using different bath parameters.
Since for each $k$, we also have random parameters, the averages give a clear
indication of the generic trend. However, we also see fluctuations (represented
by error bars) for different baths. See below for an analysis of this point.

Generically, we see that the average excitation probability increases with the
number of photons $k$. The overall effect is most pronounced for the isolated
system (blue line in \figref{fig:freekitkaverage}). This makes sense since
without an environment, energy can only be exchanged between the transmon and
the resonator (see \equref{eq:HtotalfreeBathSingle}). Qualitatively, this
interaction is proportional to $\hat a + \hat a^\dagger \sim \sqrt{k+1}$
(cf.~\equref{eq:resonatorelectricfieldoperator}). Therefore, a larger number of
photons $k$ can lead to a higher excitation of the transmon.

In the system-bath models, however, energy from the resonator and (in second
order)  from the transmon can dissipate into the environment. Therefore, we see
in  \figref{fig:freekitkaverage} that the transmon excitation decreases. In
particular, for increasing system-bath couplings $\lambda$, the results approach
the purely dissipative situation modeled by the Lindblad master equation (black
line). Note that in general, TDSE dynamics of a system coupled to a bath can
exhibit much more complicated, non-Markovian behavior that is incompatible
with a Lindblad master equation \cite{DeRaedt2017relaxation}.

\subsubsection{Bath fluctuations}

The remarkably nice transition from an isolated system (TDSE) to a dissipative
system (master equation) by gradually increasing the system-bath coupling
$\lambda$ cannot be observed in every simulation. This
is the reason for the fluctuations represented by the error bars in
\figref{fig:freekitkaverage}, especially for stronger system-bath couplings (see
the red line corresponding to $\lambda=2\pi\times\SI{20}{MHz}$).

\begin{figure}[t]
  \centering
  \includegraphics[width=\textwidth]{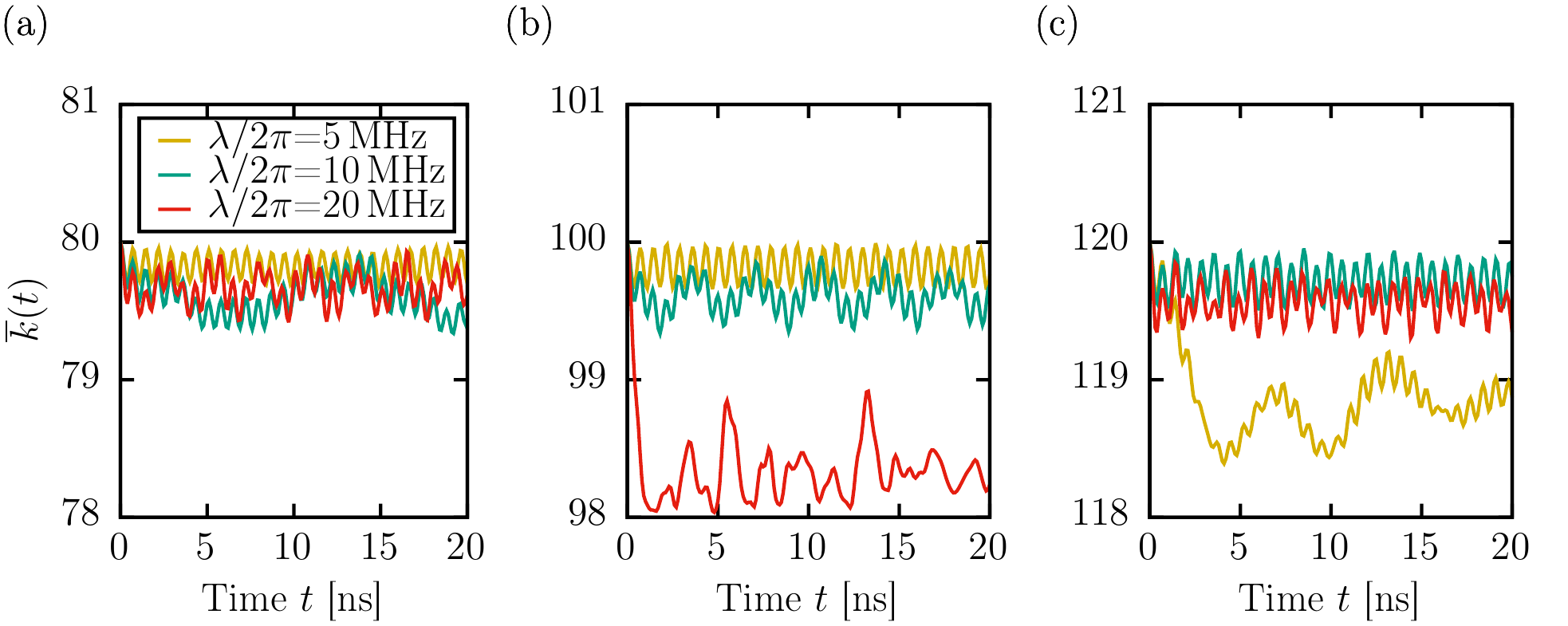}
  \caption{Time evolution of the average number of photons $\overline k(t)$ (see
  \equref{eq:photonsNumberOperatorAverage}) in the resonator in the presence
  of the bath described by \equref{eq:HtotalfreeBathBath}. The model is
  schematically shown in \figref{fig:kitbathtopology}. For each initial number
  of photons $\overline k(0)=k$ with (a) $k=80$, (b) $k=100$, and (c) $k=120$,
  we show one time evolution from the 10 independent runs that yield a point on
  the yellow, green, and red lines in \figref{fig:freekitkaverage}. The same
  colors yellow, green, and red are used to represent different coupling strengths
  $\lambda$.}
  \label{fig:freekitkaverageevolution}
\end{figure}

A closer look at the data shows that most bath configurations follow the
average, but a few particular configurations cause a dip in the excitation
probability. This dip is observable in the red line at $k=100$ and $k=180$
in \figref{fig:freekitkaverage}. To understand the reason for this, we evaluate
the time evolution of the average photon number in the resonator,
\begin{align}
  \label{eq:photonsNumberOperatorAverage}
  \overline{k}(t) = \bra{\Psi(t)}\hat a^\dagger\hat a\ket{\Psi(t)},
\end{align}
where $\ket{\Psi(t)}$ is the solution of the TDSE given by \equref{eq:photonsApproach2}.

\sfigref{fig:freekitkaverageevolution}(a) shows a representative result for
$\overline{k}(t)$ for each $\lambda$, taken from one of the ten runs for each
$\lambda$ corresponding to $k=80$ in \figref{fig:freekitkaverage}.  We see that
over the course of the time evolution, $\overline{k}(t)$ stays  nicely within
the four simulated Fock states (cf.~\equref{eq:HilbertSpaceTruncated}).

However, the red line in \figref{fig:freekitkaverageevolution}(b) shows an
extreme case for $\lambda=2\pi\times\SI{20}{MHz}$, in which the average photon
number $\overline{k}(t)$ immediately drops and hits the computational boundary
at $k=98$. This case corresponds to a particular configuration of the bath
that causes a dip in the corresponding red line in
\figref{fig:freekitkaverage} at $k=100$. A similar situation is depicted in
\figref{fig:freekitkaverageevolution}(c) for $\lambda=2\pi\times\SI{5}{MHz}$
(yellow line). This is the cause of the slightly less pronounced dip in the
corresponding yellow line at $k=120$ in \figref{fig:freekitkaverage}.

To understand this immediate drop in the photon number, we investigated the
corresponding spectral properties of the bath. In many cases, there is a certain
bath mode with a frequency $W_l$ close to the resonator frequency $\Omega$ and a
particularly strong coupling $\lambda_l$ (see \equref{eq:HtotalfreeBathBath}).
It seems reasonable that such a resonant condition leads to a special situation.
However, this explanation does not hold for each instance that exhibits this
behavior, and it is complicated to recognize the resonant pathway in all cases.
This points out an opportunity to improve the model and suggests
an interesting venue for further research.

\subsection{Additional ways to improve the models}
\label{sec:bathadditionalaspects}

For the results presented above, we initialized the resonator
in a Fock state $\ket{k}$ with $k$ photons. A more adequate model of the
experimental situation would be to consider a coherent state
\cite{Glauber1963CoherentStates, fox2006quantumoptics}, given by
\begin{align}
  \label{eq:coherentstate}
  \ket{\alpha} &= \sum_{k'} e^{-|\alpha|^2/2}\frac{\alpha^{k'}}{{k'}!} \ket{k'}.
\end{align}
For such a state, $\alpha=\sqrt{k}$ would represent the electromagnetic field in
the resonator with an average photon number $k$. We ran simulations including up
to 300 Fock states (data not shown) and found that the average excitation
probability $\langle p_{m\neq0}(t)\rangle$ shown in \figref{fig:freekitkaverage}
is only marginally reduced (from $24\%$ to $22\%$ at the maximum for
$k=180$). For the master equation, the effect is even weaker, with a
decrease by less than $0.1\%$. The time evolution $p_{m\neq0}(t)$, however, is
more interesting, with sharply peaked oscillations at a 5--10 times higher
frequency on top of the curves shown in \figref{fig:freekitktimeevolution}.

Future work will go into extending the model such that larger baths with
additional  Fock states can be simulated. This would allow a precise
classification of the bath configurations that lead to the immediate drop in the
photon number illustrated in \figref{fig:freekitkaverageevolution}(b) for
$\lambda=2\pi\times\SI{20}{MHz}$ (red line). Furthermore, it enables an
initialization of the bath in a thermal state at a certain temperature $T\neq0$
such that finite temperature effects can be studied.

From a statistical physics point of view, the transition from the isolated
system over the system-bath model to the master equation shown in
\figref{fig:freekitkaverage} could be scrutinized. In this respect, it would
be interesting to study a more general type of master equation
including photon excitations $\mathcal D[\hat a^\dagger]$ and additional
dissipators for the transmon itself, which are often used to describe
experimental observations (see e.g.~\cite{Suri2015ResonatorPhotonOccupancy}).

In the context of modeling experiments, it would be an exciting idea to use,
instead of random bath parameters, the frequencies $W_l$ and couplings
$\lambda_l$ representing the superconducting environment of the particular
device. A procedure to extract these parameters from experiments or
electromagnetic HFSS simulations of the device is described in
\secref{sec:extractfoster}. We plan to continue in this direction for a new
sample manufactured at KIT \cite{DennisIoan2019}.

Finally, an interesting conceptual question is to what extent the generic
features are specific to the bosonic bath considered in this section. For
instance, an alternative model for an environment would be a spin bath to model
a system of two-level defects in materials \cite{Mueller2009RelaxationTLS,
Mueller2019UnderstandingTLSinAmorphousSolids, WillschMadita2020PhD,
Willsch2020FluxQubitsQuantumAnnealing}. Similarly, one could replace the bosonic
bath with a fermionic bath of superconducting (Bogoliubov) quasiparticles
\cite{Kivelson1990QuasiparticlesFermions}, which are also considered to be
sources of relaxation, decoherence, and electromagnetic dissipation
\cite{Catelani2011QuasiparticleRelaxation,
Catelani2012QuasiparticlesDecoherence, Pop2014QuasiparticlesDissipation}.

\clearpage
\section{Effective \texorpdfstring{$ZZ$}{ZZ} interaction for coupled transmons}
\label{sec:statedependentfrequenciesfree}

We consider a pair of transmon qubits coupled by a resonator. This way of
coupling transmon qubits is the primary architecture studied in this work. The
effective transmon coupling mediated by a resonator has been frequently studied
in the literature (see for instance \cite{blais2004circuitqed,
majer2007coupling, liparaoanu2008capacitive, gambetta2013controlIFF,
richer2013perturbative, billangeon2015longitudinal,
Willsch2017GateErrorAnalysis, Ku2020SuppressionZZinCSFQ}). Analytical calculations often use perturbation
theory to obtain dominant effective couplings such as $\sigma_0^x\sigma_1^x$ or
$(\sigma_0^x\sigma_1^x+\sigma_0^y\sigma_0^y)/2$. In this section, we study a
much weaker coupling of the type $\sigma_0^z\sigma_1^z$. Albeit very small, this
coupling is still relevant for experiments; we will later construct a quantum circuit by
which its effects can be directly observed in the IBM Q processors (see
\secref{sec:crosstalk}). Note that it is also possible to obtain this coupling
from a complete microwave description of the system
\cite{Solgun2019ImpedanceMicrowaveDescriptionSuperconductingQubit}.

A coupling of the type $\sigma_0^z\sigma_1^z$, also known as longitudinal coupling,
makes the frequency of one qubit depend on the state of the other qubit.
To see this, consider a Hamiltonian describing two qubits (Q0, Q1) of the form
\begin{align}
  \label{eq:statedependentfrequenciesTwoQubitHamiltonian}
  H_{ZZ} &= -\frac{\omega_0'}2 \sigma_0^z
   -\frac{\omega_1'}2 \sigma_1^z
   +J \sigma_0^z \sigma_1^z.
\end{align}
When Q1 is in state $\ket 0$, the frequency of Q0 is $\omega_0'-2J$ (given by
the difference between the eigenvalues of $\ket{10}$ and $\ket{00}$). However,
when Q1 is in state $\ket 1$, the frequency of Q0 is $\omega_0'+2J$. This means
that the frequency of Q0 depends on the state of Q1. In the same manner, the
frequency of Q1 depends on the state of Q0.

To study this effect for a pair of transmon qubits, we simulate a system of
$N_{\mathrm{Tr}}=2$ transmons coupled by $N_{\mathrm{Res}}=1$ resonator (see
\tabref{tab:deviceibm2gst} for the model parameters). We determine the frequency
of Q0 for three different initial states of Q1, namely $\ket{0},\ket{+}$, and
$\ket{1}$. To determine the frequency, we make use of the procedure described in
\secref{sec:evaluator}. Thus, we initialize Q0 in the state $\ket{+}$ and obtain
the frequency from the time evolution of its Bloch vector $\vec r_0(t)$ (see
\equaref{eq:singlequbitblochvectorTimeEvolutionRotating}{eq:qubitfrequencyfitchisq}).
The resonator is always initialized in its ground state $\ket{k=0}$.

\begin{figure}[t]
  \centering
  \includegraphics[width=\textwidth]{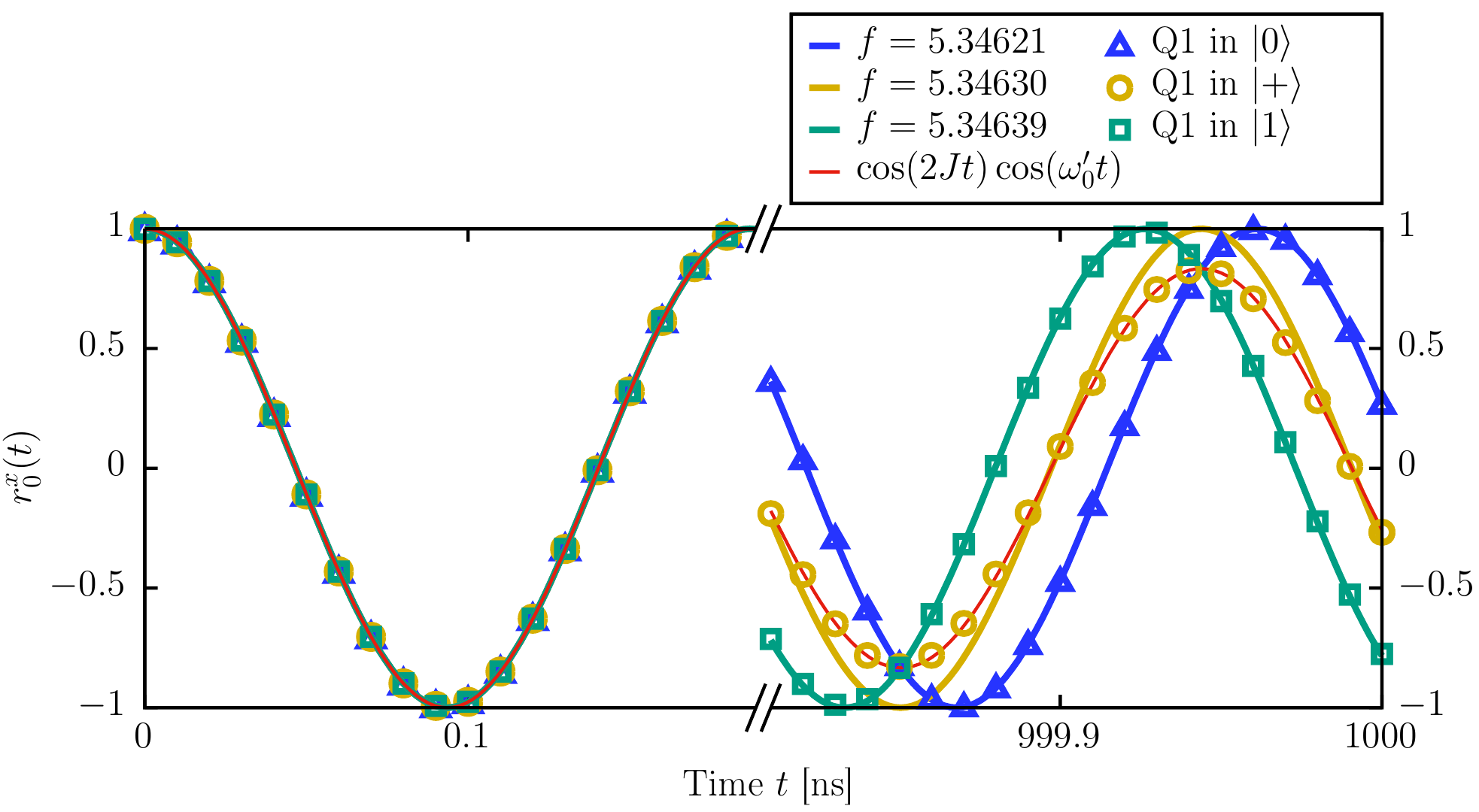}
  \caption{Time evolution of the $x$ component of the Bloch vector $\vec r_0(t)$
  of Q0 (cf.~\equref{eq:multiqubitblochvectorTransmonTrace}), for different
  initial states of Q1 (indicated by blue, yellow, and green colors). Points
  represent simulation results. Blue, yellow, and green lines correspond to the
  functions $\cos(2\pi f t)$, where the frequency $f$ is given in GHz in the
  legend. The red line corresponds to the function $\cos(2Jt)\cos(\omega_0't)$
  (see \equref{eq:statedependentfrequenciesTwoQubitHamiltonianTimeEvoltionR0X}),
  where $J$ and $\omega_0'$ are given in
  \equaref{eq:statedependentfrequenciesResultsOmega0}{eq:statedependentfrequenciesResultsJ},
  respectively. The difference in the frequencies is only observable after a
  long time evolution (right panel of the plot).}
  \label{fig:statedependentfrequenciesfree}
\end{figure}

\sfigref{fig:statedependentfrequenciesfree} shows a plot of the time evolution
of the Bloch vector's $x$ component $r_0^x(t)$ for the three different initial
states of Q1, along with the frequencies determined by the procedure described
in \secref{sec:evaluator}. As expected, the frequency of Q0 depends on the state
of Q1. Since the difference in frequency is on the order of
$\Delta f\approx\SI{0.0001}{GHz}$, we need to simulate the  time evolution
up to $\SI{1}{\micro s}$  (corresponding to $\SI{1}{\micro s}/\tau=10^6$ time
steps) to observe the difference in $f$.

As can be seen, the frequencies used for the cosine functions (lines) describe
the simulation results (points) very accurately. However, on closer inspection,
we see that the amplitude of the oscillation corresponding to the case where  Q1
is in state $\ket+$ (yellow circles) decreases over time, which is not described
accurately by $\cos(2\pi f t)$. The reason for this is that the magnitude of the
Bloch vector $\vec r_0(t)$ becomes smaller which corresponds to entanglement
building up in the state.

Interestingly, the effect is described correctly by the effective
Hamiltonian $H_{ZZ}$ given by
\equref{eq:statedependentfrequenciesTwoQubitHamiltonian}. To see this, we
compute the corresponding time evolution under $H_{ZZ}$,
\begin{align}
  e^{-itH_{ZZ}}\ket{++} &= \frac 1 2 \Big(
  e^{it(\omega_0'+\omega_1'-2J)/2} \ket{00}
  + e^{it(\omega_0'-\omega_1'+2J)/2} \ket{01} \nonumber\\
  \label{eq:statedependentfrequenciesTwoQubitHamiltonianTimeEvoltion++}
  &\quad+ e^{it(-\omega_0'+\omega_1'+2J)/2} \ket{10}
  + e^{it(-\omega_0'-\omega_1'-2J)/2} \ket{11}
  \Big).
\end{align}
Evaluating the expectation value $r_0^x(t)=\langle\sigma_0^x\rangle$ for this state
yields
\begin{align}
  \label{eq:statedependentfrequenciesTwoQubitHamiltonianTimeEvoltionR0X}
  r_0^x(t) &= \cos(2Jt)\cos(\omega_0't).
\end{align}
The parameters $\omega_0'$, $\omega_1'$, and $J$ can be obtained from the time
evolution of the four states $(\ket{+0},\ket{+1},\ket{0+},\ket{1+})$. Using the
same procedure as before, we extract the corresponding four frequencies
$f_i^{\pm}$ from the data for $\vec r_i(t)$ (see
\equaref{eq:singlequbitblochvectorTimeEvolutionRotating}{eq:qubitfrequencyfitchisq}).
For instance, the frequency $f_0^-$ ($f_0^+$) obtained from the time evolution
of $\ket{+0}$ ($\ket{+1}$) corresponds to the blue (green) line in
\figref{fig:statedependentfrequenciesfree}. The frequencies $f_i^{\pm}$ are
related to the parameters $\omega_0'$, $\omega_1'$, and $J$ via $2\pi
f_i^{\pm}=\omega_i'\pm2J$ (see
\equref{eq:statedependentfrequenciesTwoQubitHamiltonian}). Therefore, we have
$\omega_i'=2\pi(f_i^++f_i^-)/2$ and $J=2\pi(f_i^+-f_i^-)/4$ (which is the same
for both $i=0$ and $i=1$ up to $\SI{10^{-10}}{GHz}$), which evaluates to
\begin{subequations}
  \begin{align}
    \label{eq:statedependentfrequenciesResultsOmega0}
    \omega_0' &= 2\pi\times\SI{5.346300}{GHz},\\
    \label{eq:statedependentfrequenciesResultsOmega1}
    \omega_1' &= 2\pi\times\SI{5.116707}{GHz},\\
    \label{eq:statedependentfrequenciesResultsJ}
    %J &= 2\pi\times\SI{0.0000465558}{GHz}
    J &= 2\pi\times\SI{46.6}{kHz}.
  \end{align}
\end{subequations}
The frequencies $\omega_i'$ are shifted with respect to the individual qubit
frequencies $\tilde\omega_i$ given in \tabref{tab:deviceibm2gst} due to the
presence of the resonator. Note the large number of significant digits that is
required to resolve the value of $J$.

The function $\cos(2Jt)\cos(\omega_0't)$ given in
\equref{eq:statedependentfrequenciesTwoQubitHamiltonianTimeEvoltionR0X} is shown
as a red line in \figref{fig:statedependentfrequenciesfree}. We see that it
describes the decrease in amplitude very accurately, despite the extremely small
value of $J$. In \secref{sec:gatesettomography}, we will reproduce the
same value for $J$ (see \equref{eq:GSTResultsZZinteractionJ}) using a much more
sophisticated procedure called \emph{gate set tomography} (GST).

It is worth mentioning that the effective Hamiltonian $H_{ZZ}$ given in
\equref{eq:statedependentfrequenciesTwoQubitHamiltonian} can also be derived from the
original transmon-resonator Hamiltonian by doing a perturbative
diagonalization.  Such a calculation is given in \cite{gambetta2013controlIFF}
and, using a more recent technique, in
\cite{Magesan2018CrossResonanceGateEffectiveHamiltonians}. The starting
point of the calculation is a two-transmon version of the anharmonic oscillator
Hamiltonian $H_{\mathrm{Single}}^{\mathrm{AOIntRWA}}$ given in \equref{eq:HtotalfreesingleAOIntRWA}.
Furthermore, in \cite{billangeon2015longitudinal}, a derivation starting from a two-qubit
version of $H_{\mathrm{Single}}^{\mathrm{TLA}}$ given in \equref{eq:HtotalfreesingleTLA}
is presented, without resorting to the RWA.
Both calculations yield the correct type of longitudinal $ZZ$
interaction.
Furthermore, the order of magnitude of the frequency corrections is
right: The difference between $\omega_i'$ given in
\equaref{eq:statedependentfrequenciesResultsOmega0}
{eq:statedependentfrequenciesResultsOmega1} and the original frequencies
$\tilde\omega_i$ given in \tabref{tab:deviceibm2gst} is approximately equal to
the Lamb shift $-g_i^2/(\Omega-\tilde\omega_i)\in\{2\pi\times\SI{-0.0035}{GHz},
2\pi\times\SI{-0.0029}{GHz}\}$, where $g_i=-(E_{Ji}/32E_{Ci})^{1/4}G_i$.
However, the respective values for the longitudinal coupling strength $J$ given in
\equref{eq:statedependentfrequenciesResultsJ} are different:
\begin{subequations}
  \begin{align}
    \label{eq:statedependentfrequenciesResultsJinPTAOIntRWA}
    J^{\mathrm{AOIntRWA}} &\approx 2\pi\times\SI{240}{kHz},\\
    \label{eq:statedependentfrequenciesResultsJinPTTLA}
    J^{\mathrm{TLA}} &\approx 2\pi\times\SI{5}{kHz}.
  \end{align}
\end{subequations}
The interaction strength $J^{\mathrm{AOIntRWA}}$ is stronger because it refers to
the eigenbasis obtained after the perturbative diagonalization. The reason for
$J^{\mathrm{TLA}}$ being too small is that the TLA discards contributions from
higher levels. Thus, higher transmon states play an important role in mediating
the exchange interaction between transmon qubits.  Independent of these deviations,
however, all perturbative calculations yield the important scaling law $J\propto
G^4$, showing that the resonator-mediated exchange interaction depends
sensitively on the respective transmon-resonator couplings.

Although the magnitude of the coupling is extremely small (see
\equref{eq:statedependentfrequenciesResultsJ}), one can find circuits by which
its consequences, namely the state-dependent frequencies, can be directly
observed in the IBM Q processors (see \secref{sec:crosstalk}). Furthermore, the
effective evolution described by $H_{ZZ}$ given in
\equref{eq:statedependentfrequenciesTwoQubitHamiltonian}, including the same
value for $J$ given in \equref{eq:statedependentfrequenciesResultsJ}, is also
found, without using prior information, by the black box model of GST (see
\equref{eq:GSTResultsZZinteractionJ} in  \secref{sec:gatesettomography}).

\section{Conclusions}

In this chapter, we studied undriven time evolutions. After verifying known
error bounds for the transmon simulator (see
\equaref{eq:accuracyErrorGlobalPhase} {eq:accuracyErrorOverlap}), we found that
new error bounds for observables \cite{WillschMadita2020PhD} (see
\equaref{eq:accuracyObservableBoundInf}{eq:accuracyObservableBoundVar}) are
indeed tight. In detailed benchmarks, we observed that the simulation algorithm
exhibits nearly ideal weak and strong scaling behavior (see
\figsref{fig:performanceruntime}{fig:performanceprofiling}) on the supercomputer
JURECA \cite{JURECA} for implementations 1 and 2
(cf.~\appref{app:implementations}). We compared the time evolution produced by
the transmon simulator to known perturbative results, and found that the
perturbative results qualitatively predict the time evolution properly, but all
develop a drift after a short time (see
\figref{fig:freekitcomparetimeevolutions}), which makes them unsuitable for
accurate pulse optimization. By coupling an isolated system to a bosonic heat
bath, we observed that system-environment models based on the solution of the
TDSE can, under certain conditions, be effectively described  by the CPTP
dynamics generated by a Lindblad master equation (see
\figref{fig:freekitkaverage}). Finally, we characterized the resonator-mediated
exchange interaction between coupled transmons. Although the magnitude of this
interaction is extremely small (see
\equref{eq:statedependentfrequenciesResultsJ}), we note that  it is accurately
reproducible using a black box model (see
\secref{sec:gatesettomographyRunning}). Furthermore, its consequences such as
the state-dependent frequency shifts of neighboring qubits shown in
\figref{fig:statedependentfrequenciesfree}, can be directly observed in the IBM
Q processors (see \secref{sec:crosstalk}).

%% file: chap5.tex
\chapter{Optimizing pulses for quantum gates}
\label{cha:optimization}

For gate-based quantum computers, a quantum gate is implemented by a certain
external action on the system. A natural way of interacting with a
superconducting system is to apply an electromagnetic pulse. In the transmon
systems considered in this thesis, such a pulse is a microwave voltage pulse,
applied to each qubit through its respective transmission line (see
\secref{sec:CPB}). A voltage pulse is modeled by the external time-dependent
functions $n_{gi}(t)$ in the Hamiltonian given by \equsref{eq:Htotal}{eq:HCC}.
We consider a generic sum of microwave voltage pulses
\begin{align}
  n_{gi}(t) = \sum\limits_{j}^{} \Omega_{ij}(t) \cos( 2\pi f_{ij} t - \gamma_{ij} ),
  \label{eq:genericvoltagepulses}
\end{align}
where $\Omega_{ij}(t)$ is the envelope of pulse $j$ on qubit $i$, $f_{ij}$ is
the corresponding drive frequency, and $\gamma_{ij}$ is an offset phase.
The generic expression for microwave pulses given by
\equref{eq:genericvoltagepulses} is motivated by the form of the microwave signals
used in typical experiments to implement quantum gates (see \cite{McKay2016VZgate} for more information).

To implement a particular quantum gate, the pulses $n_{gi}(t)$ given by
\equref{eq:genericvoltagepulses} must be chosen such that the time evolution of
the full system corresponds to the unitary transformation representing the
desired quantum gate. In other words, if $U:\mathcal H_{2^n}\to\mathcal H_{2^n}$
is the unitary operator corresponding to the desired quantum gate (see
\secref{sec:quantumgates}), the functions $n_{gi}(t)$ for $0\le t < T$ need to
be chosen such that the time-evolution operator of the full system $\mathcal
U(T,0)$ (see \equref{eq:totaltransmontimeevolutionoperator}) implements $U$
(potentially in a certain rotating frame, meaning that the columns of $\mathcal
U(T,0)$ are transformed according to \equref{eq:rotatingframe}).

The challenge, however, is that the time-evolution operator $\mathcal U(T,0)$ acts on the
much larger Hilbert space $\mathcal H$ of all transmons and
resonators given by \equref{eq:HilbertSpaceTruncated}. Therefore, the
operators $U$ and $\mathcal U(T,0)$ can only be equal after projecting
$\mathcal U(T,0)$ on the smaller computational subspace $\mathcal H_{2^n}$, yielding
\begin{align}
  M = P_{\mathcal H_{2^n}} \mathcal U(T,0) P_{\mathcal H_{2^n}},
  \label{eq:MprojectionUtotal}
\end{align}
where $P_{\mathcal H_{2^n}}$ denotes the projection operator defined in
\equref{eq:projectionComputationalSubspace}. In almost all practical cases, the
projected matrix $M$ is not unitary anymore, so it is, strictly speaking,
impossible to make $M$ equal to the desired quantum gate $U$. Intuitively, this
means that non-computational states affect the time evolution of the total
system, a particular problem for transmon qubits known as \emph{leakage}
\cite{chen2016leakagemartinis, Willsch2017GateErrorAnalysis, Wood2017LeakageRB}
(see also \secref{sec:leakage}).
The best one can hope for is to find a pulse resulting in a transformation $M$
that approximates the desired quantum gate $U$ as closely as possible. The
important question is whether such a fundamentally imperfect implementation is
sufficient in practice.

Therefore, the aim of this chapter is to develop an optimization scheme for a
set of pulse parameters for \equref{eq:genericvoltagepulses} to implement the
closest approximation to $U$. The particular set of parameters depends on the
kind of quantum gate to be optimized. In
\secaref{sec:optimizatingsinglequbitgate}{sec:optimizatingtwoqubitgate}, we
specify this set of parameters for single- and two-qubit gates, respectively. In
\secref{sec:optimizatingpulseparameters}, we describe the optimization procedure
and present results for some of the model systems used for this work.
\ssecref{sec:compiler} gives an example of the compilation process to translate
a quantum circuit into a sequence of pulses. Finally, we give a brief overview
of alternative optimization techniques in
\secref{sec:alterativegateoptimizationtechniques}. Applying the optimized pulses
to actual quantum circuits and comparing their performance to experimental
implementations is the topic of the following chapters.

\section{Single-qubit pulses}
\label{sec:optimizatingsinglequbitgate}

Applying a pulse of the form of \equref{eq:genericvoltagepulses} has the effect
that the qubit represented by transmon $i$ is rotated around its Bloch sphere.
Specifically, within the RWA, one can show \cite{gambetta2013controlIFF} that a pulse of the form
\begin{align}
  \label{eq:singlequbitpulseexample}
  \Omega(t)\cos(2\pi f t - \gamma),
\end{align}
on transmon $i$, where $f=\omega_i/2\pi$ is given by the qubit frequency,
corresponds to a rotation by an angle
\begin{align}
  \label{eq:singlequbitpulseangle}
  \vartheta = b_i \int\limits_0^T \Omega(t)\,\mathrm dt,
\end{align}
where $T$ is the duration of the pulse, and
\begin{align}
  \label{eq:singlequbitAmplitudeToEnergy}
  b_i = 8 E_{Ci}
  \left(\frac{E_{Ji}}{32E_{Ci}}\right)^{1/4}
\end{align}
is the energy scale of the dimensionless amplitudes. In other words, the area
under the envelope $\Omega(t)$ determines the angle of rotation. The axis of
rotation is defined by the phase $\gamma$ in
\equref{eq:singlequbitpulseexample}. In particular, $\gamma=0$ ($\gamma=\pi/2$)
corresponds to the $x$ ($y$) axis. Furthermore, one can show that choosing
$f\neq\omega_i/2\pi$ results in additional rotations around the $z$ axis. See
\cite{gambetta2013controlIFF,Willsch2016Master} for more information on these
properties.

For the above relations, the coefficients of the state vector have to be
expressed in the rotating frame (see \equref{eq:rotatingframe}). Note that,
although the choice of frame does not affect the result of the final
measurement, it matters when we want to interpret or visualize the coefficients
of the intermediate state vector $\ket{\Psi(t)}$.

As discussed in \secref{sec:elementaryquantumgates}, every single-qubit gate can
be expressed in terms of rotations on the Bloch sphere. We implement the
particular set of single-qubit rotations $X^{\pi/2}=R^x(\pi/2)$ (a $\pi/2$
rotation of the qubit around the $x$ axis) and $Z^{\vartheta}=R^z(\vartheta)$
(an arbitrary rotation by an angle $\vartheta$ around the $z$ axis). These gates
represent the elementary building blocks  of the $\textsc{U1},\textsc{U2}$, and
$\textsc{U3}$ gates given by \equsref{eq:singlequbitU1}{eq:singlequbitU3}, which
in turn can be used to express all standard single-qubit gates (see
\tabref{tab:elementarygateset} in \appref{app:gateset}). This choice of
elementary single-qubit gates is the same that was made for the IBM Q processors
\cite{Cross2017openqasm2}.

\subsection{The VZ gate}
\label{sec:singlequbitVZgate}

In principle, the information given above is sufficient to find candidates for
pulses to implement both $X^{\pi/2}$ and $Z^{\vartheta}$. However, one can
simplify the hardware implementation further by using the concept of a virtual Z
gate (VZ gate). This concept is common practice in the transmon architecture under
investigation \cite{McKay2016VZgate}. Therefore, we also implement this concept
in the transmon simulator.

The VZ gate is based on the fact that the phase $\gamma$ in
\equref{eq:singlequbitpulseexample} defines the axis of rotation in the $xy$
plane. In other words, $\gamma$ determines the frame of reference in which the
qubit is defined. Therefore, whenever the next gate in a sequence of gates is
$Z^{\vartheta}$ or $\textsc{U1}(\lambda)$, instead of applying a pulse, we
rotate our personal frame of reference. This affects the phases $\{\gamma\}$ of
\emph{all the following pulses} according to a given rule.

For single-qubit pulses, this rule corresponds to an exchange of operations
according to the scheme
\begin{align}
  \label{eq:singlequbitpulseVZgatescheme}
  \mathrm{pulse}^{(m)}(\gamma)\cdots \mathrm{pulse}^{(1)}(\gamma)\, Z^{\vartheta} \ket\Psi
  = Z^{\vartheta}\, \mathrm{pulse}^{(m)}(\gamma-\vartheta)\cdots \mathrm{pulse}^{(1)}(\gamma-\vartheta) \ket\Psi.
\end{align}
The advantage of this scheme is that no time on the hardware is required to
implement the family of rotations $Z^{\vartheta}$. This is the reason why the VZ
gate is called a \emph{virtual} gate.

However, the downside is that during the time evolution, we have to keep track
of the VZ phases $\gamma_i$ for each transmon $i$. Furthermore, for every
elementary quantum gate, we need to define how it commutes with $Z^{\vartheta}$
and how $\vartheta$ affects the phases of the underlying pulses. For the
single-qubit pulse to implement $X^{\pi/2}$  and the more complicated two-qubit
pulses, this rule is given in the  following sections (see
\equref{eq:singlequbitpulseGDruleVZ} and \figref{fig:twoqubitpulseruleVZ},
respectively).

\subsection{The GD pulse}
\label{sec:singlequbitGDpulse}

To implement $X^{\pi/2}$ with a VZ phase parameter $\gamma$, we use a
microwave pulse of the form
\begin{align}
  \label{eq:singlequbitpulseGDnodrag}
  \Omega_G(t)\cos(2\pi f t-\gamma),
\end{align}
where the envelope $\Omega_G(t)$ is a Gaussian defined as
\begin{align}
  \Omega_{\mathrm{G}}(t) &=
  \Omega_X\frac{\exp\!\left(-\frac{(t-T_X/2)^2}{2\sigma^2}\right) - \exp\!\left(-\frac{T_X^2}{8\sigma^2}\right)}
  {1 - \exp\!\left(-\frac{T_X^2}{8\sigma^2}\right)}.
  \label{eq:gaussianpulse}
\end{align}
Here, $\Omega_X$ is the amplitude that needs to be chosen such that the angle
$\vartheta$ given by \equref{eq:singlequbitpulseangle} equals $\pi/2$, $T_X$ is
the duration  of the pulse (typically around \SI{80}{ns}), and $\sigma=T_X/4$
characterizes the width of the Gaussian. Note that the Gaussian is shifted
vertically such that $\Omega_{\mathrm{G}}(0)=\Omega_{\mathrm{G}}(T)=0$.

For transmon qubits, simple Gaussian pulses like the one given in
\equref{eq:singlequbitpulseGDnodrag} can drive the qubit out of the
computational subspace such that higher levels $\ket m$ with $m\ge 2$ are
excited. To mitigate this effect, a technique known as DRAG has become standard
\cite{motzoi2009drag, chow2010dragexperiment, gambetta2010dragtheory,
Theis2018DRAGstatusAfter10years}. There are several alternatives of implementing
DRAG (see \cite{gambetta2013controlIFF,Willsch2016Master} for a comparison), but
the common concept is that a shifted microwave pulse with an amplitude
given by the derivative $\dot\Omega_{\mathrm{G}}(t)$ of the Gaussian in
\equref{eq:gaussianpulse} is added to \equref{eq:singlequbitpulseGDnodrag}. The
prefactor of this term is the so-called DRAG coefficient $\beta_X$.
We also implement this concept for the transmon simulator.

The single-qubit pulse to implement $X^{\pi/2}$ with DRAG
correction is given by
\begin{align}
  \mathrm{GD}^{\pi/2}(\gamma) :\quad
    \Omega_{\mathrm{G}}(t)\cos(2\pi f t-\gamma)
    + \beta_X\dot\Omega_{\mathrm{G}}(t)\cos(2\pi ft-(\gamma+\pi/2)),
  \label{eq:singlequbitpulseGD}
\end{align}
and is a characterized by four pulse parameters $(f,T_X,\Omega_X,\beta_X)$.
These parameters are tuned in the pulse optimization procedure discussed in
\secref{sec:optimizatingpulseparameters}. Initial values for the optimization
are either given by theory or taken from experiments. Specifically,  the drive
frequency $f$ is initialized to the qubit frequency $f_i=\omega_i/2\pi$
(determined  using the procedure described in \secref{sec:evaluator}); the time
$T_X$ is typically kept fixed at around \SI{80}{ns} (given by the corresponding processor, see e.g.~\cite{ibmqx5}); the
drive strength $\Omega_X$ is determined from \equref{eq:singlequbitpulseangle};
and the DRAG coefficient $\beta_X$ is set to $-1/2\alpha$, where $\alpha$ is the
anharmonicity (see the text below \equref{eq:transmoneigenstates}). The phase
$\gamma$ in \equref{eq:singlequbitpulseGD} is used to implement VZ gates
according to the rule
\begin{align}
  \label{eq:singlequbitpulseGDruleVZ}
  \mathrm{GD}^{\pi/2}(\gamma)\,Z^{\vartheta}\ket{\Psi}
  = Z^{\vartheta}\,\mathrm{GD}^{\pi/2}(\gamma-\vartheta)\ket{\Psi}.
\end{align}

By analogy with the notation used for single-qubit gates in multi-qubit systems
(see \equref{eq:singlequbitrotationmultiplequbits}), we denote a $\mathrm{GD}$
pulse on qubit $i$ by $\mathrm{GD}_i^{\pi/2}(\gamma)$. In addition to
$\mathrm{GD}_i^{\pi/2}(\gamma)$, we also define a pulse
$\mathrm{GD}_i^{\pi}(\gamma)$ that is supposed to implement a full $X_i^{\pi}$
rotation (i.e., a bit flip). This pulse is used mainly as a building block for the
two-qubit pulses defined in the next section; a single-qubit $X_i^{\pi}$ gate is
typically implemented as $\textsc{U3}_i(\pi,0,\pi)$ (see also \equref{eq:singlequbitU3}),
i.e., in terms of two $\mathrm{GD}_i^{\pi/2}(\gamma)$ pulses as done for the
IBM Q processors \cite{Cross2017openqasm2}.

Technically, a pulse for the $X^\pi$ rotation differs
from the $X^{\pi/2}$ rotation in that the drive amplitude $\Omega_X$ is twice as
large, since the angle of rotation given by \equref{eq:singlequbitpulseangle} is
directly proportional to $\Omega_X$. However, the other parameters may also come
out differently in the parameter-optimization process. In general, it is
nontrivial to predict the best set of parameters $(f,T_X,\Omega_X,\beta_X)$  for
a full transmon-resonator system.

\subsection{The zero pulse}
\label{sec:singlequbitzeropulse}

We define a particular pulse representing an undriven
time evolution for a time $T$, denoted by $\mathrm{zero}(T)$. It is used as
an explicit expression for an identity gate. Typically, $T=T_X$ is the time
used for single-qubit gates (see above).

In the transmon computer model defined in \secref{sec:transmonmodel}, a
zero pulse on qubit $i$ corresponds to setting $n_{gi}(t)=0$. Usually, it is not
necessary to specify the zero pulse explicitly, since $n_{gi}(t)=0$ whenever no
pulse is specified. However, it can be used to configure free time evolutions as
studied in \chapref{cha:freeevolution}, or to have the simulation run freely for
some time after the last actual pulse has been applied.

Note that for some quantum computing systems, an identity gate is implemented by
explicit pulses with zero net effect. For instance, a second-order dynamical
decoupling sequence of the form $X^{\pi}Y^{\pi}X^{\pi}Y^{\pi}$
\cite{Khodjasteh2009DynamicalDecoupling} was found to improve the performance of
identity gates in a recent trapped-ion qubit system
\cite{BlumeKohout2017DemoRigorousThresholdGST}. However, we choose to implement
the identity gate in terms of an undriven time evolution, which is also done for
the IBM Q processors \cite{ibmquantumexperience2016, Cross2017openqasm2}.
Additionally, this implementation allow us to study and understand the emerging
effects in coupled transmon systems (see
\secaref{sec:statedependentfrequenciesfree}{sec:crosstalk},
and also the gate set tomography experiments discussed in
\secref{sec:gatesettomographyRunning}).

\section{Two-qubit pulses}
\label{sec:optimizatingtwoqubitgate}

The universal two-qubit gate used for the quantum computer simulations presented
in this thesis is the \textsc{CNOT} gate defined by \equref{eq:cnotgate}. A
prominent pulse to implement the \textsc{CNOT} gate is the cross-resonance (CR)
pulse \cite{rigetti2010CR, chow2011CR, Groot2012selectivedarkening}. Its idea
was first proposed in \cite{paraoanu2008CRchargequbits}, and the knowledge and
methods about how to use it for transmon quantum computers have continuously
improved over the past years \cite{Corcoles2013processverification,
sheldon2016procedure, Takita2017faultTolerantStatePreparation,
Tripathi2019CrossResonanceGate,
Magesan2018CrossResonanceGateEffectiveHamiltonians,
Malekakhlagh2020FirstPrinciplesCrossResonance}.

\subsection{CNOT gates based on the CR effect}
\label{sec:cnotbasedonCReffect}

\begin{figure}
  \centering
  \includegraphics[width=\textwidth]{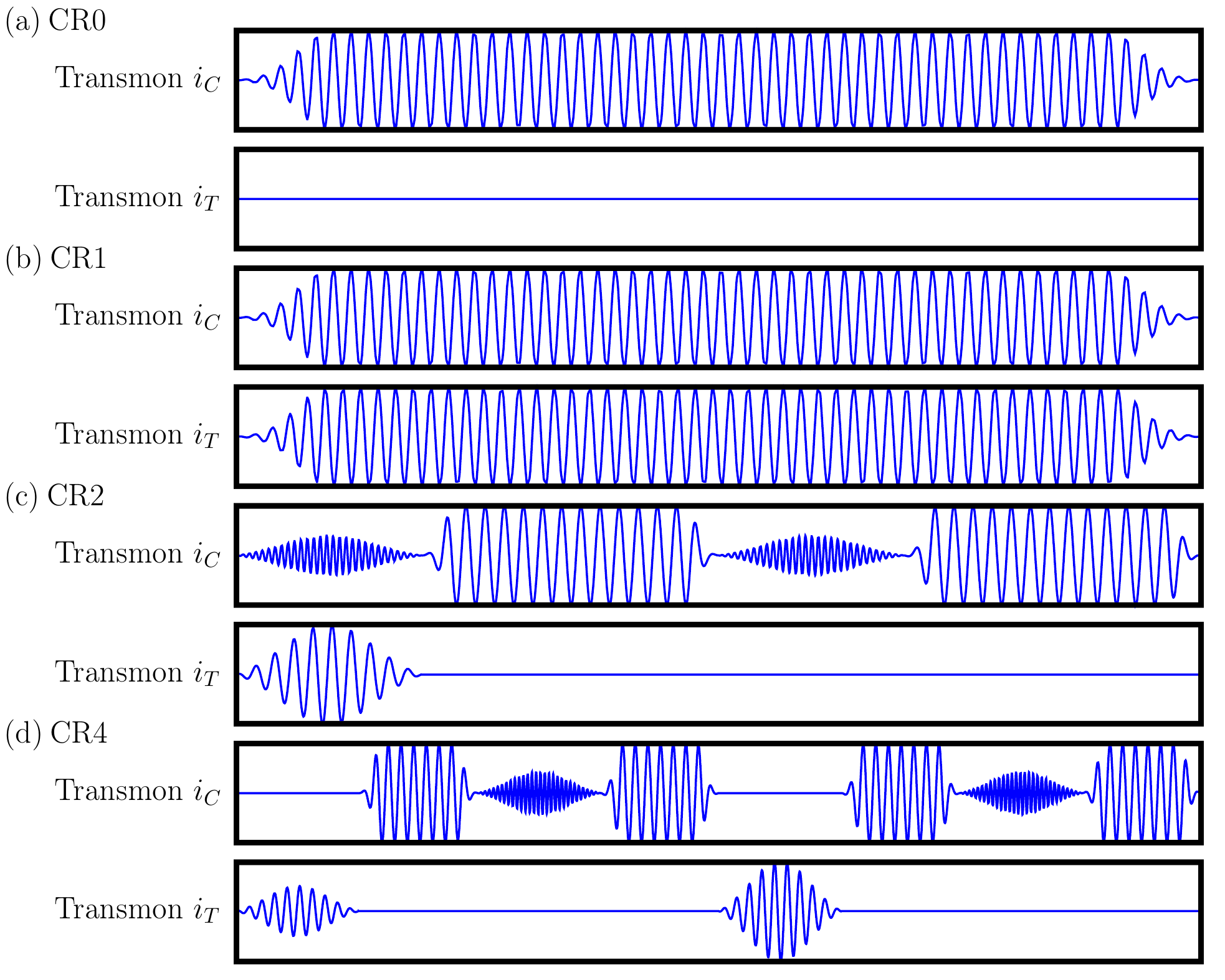}
  \caption{CR pulse sequences as a function of time, corresponding to
  $n_{gi}(t)$ given by \equref{eq:genericvoltagepulses} where $i=i_C$ $(i=i_T)$
  denotes the control (target) qubit, (a) basic CR0 pulse on the control
  qubit at the frequency of the target qubit (see \equref{eq:twoqubitpulseCR0});
  (b)--(d) three different realizations CR1, CR2, and CR4 of a \textsc{CNOT} gate using
  combinations of CR pulses (see \equref{eq:twoqubitpulseCR0}) and single-qubit
  GD pulses (see \equref{eq:singlequbitpulseGD}). Gaussians represent the GD
  pulses and implement $X^{\pi/2}$ and $X^{\pi}$ rotations. Flat-topped
  Gaussians represent the CR pulses. The CR1 gate consists only of flat-topped
  Gaussian pulses at the target frequency. The CR2 gate is an echoed CR gate
  containing two additional $X^{\pi}$ pulses on the control qubit and one
  $X^{\pi/2}$ pulse on the target qubit. The CR4 gate is a four-pulse echoed CR
  gate that contains an additional $X^{\pi}$ pulse on the target qubit. See
  \figref{fig:twoqubitpulseruleVZ} for the full pulse specifications.}
  \label{fig:crossresonancepulses}
\end{figure}

We consider a \textsc{CNOT} gate between a control qubit $i_C$ and a target
qubit  $i_T$. The basic CR pulse is defined as a microwave pulse applied to the
control qubit at the resonance frequency $f_{i_T}=\omega_{i_T}/2\pi$ of the
target qubit. Since the frequencies of adjacent transmon qubits typically differ
by $\SI{100-300}{MHz}$, the CR pulse is a slightly off-resonant pulse on the
control qubit. For the time-dependent pulses $n_{gi}(t)$ given by
\equref{eq:genericvoltagepulses}, we define the basic CR pulse (denoted by
$\mathrm{CR0}$) as a flat-topped Gaussian microwave pulse
\begin{align}
  \label{eq:twoqubitpulseCR0}
  \mathrm{CR0}(\gamma):\quad \Omega_{GF}(t)\cos(2\pi f_{i_T} t-\gamma),
\end{align}
where $\gamma$ is the VZ phase (cf.~\secref{sec:singlequbitVZgate}), and the
envelope $\Omega_{GF}(t)$ of the CR pulse is a flat-topped Gaussian. The latter
is formally defined as
\begin{align}
  \label{eq:gaussianflattoppulse}
  \Omega_{GF}(t) = \begin{cases}
    \Omega_G(t)\left\vert\Big(\substack{\Omega_X\\T_X\\\sigma}\Big)\mapsto\Big(\substack{\Omega_{\mathrm{CR}}\\2T_{\mathrm{rise}}\\T_{\mathrm{rise}}/3}\Big)\right.
      & (0\le t \le T_{\mathrm{rise}}) \\
    \Omega_{\mathrm{CR}}
      & (T_{\mathrm{rise}}\le t \le T_{\mathrm{rise}}+T_{\mathrm{CR}}) \\
    \Omega_G(t-T_{\mathrm{CR}})\left\vert\Big(\substack{\Omega_X\\T_X\\\sigma}\Big)\mapsto\Big(\substack{\Omega_{\mathrm{CR}}\\2T_{\mathrm{rise}}\\T_{\mathrm{rise}}/3}\Big)\right.
      & (T_{\mathrm{rise}}+T_{\mathrm{CR}}\le t \le 2T_{\mathrm{rise}}+T_{\mathrm{CR}}) \\
\end{cases},
\end{align}
where $\Omega_{\mathrm{CR}}$ is the amplitude of the CR pulse, $T_{\mathrm{CR}}$
is the time of the flat part in the middle of the pulse, and
$T_{\mathrm{rise}}=\SI{15}{ns}$ is the time of the Gaussian rise at the
beginning and the Gaussian fall at the end of the pulse. The total duration of
the CR pulse is thus $T_{\mathrm{tot}}=T_{\mathrm{CR}}+\SI{30}{ns}$. The
parameters for the Gaussian $\Omega_G(t)$ given by \equref{eq:gaussianpulse}
have to be replaced as indicated in \equref{eq:gaussianflattoppulse}. The basic
CR0 pulse given by \equref{eq:twoqubitpulseCR0} is schematically shown in
\figref{fig:crossresonancepulses}(a). It works as a building block for the
pulse sequences defined below to implement the \textsc{CNOT} gate.

\begin{figure}
  \centering
  \captionsetup[subfigure]{position=top,textfont=normalfont,singlelinecheck=off,justification=raggedright,labelformat=parens}
  \subfloat[\label{sfig:CR00}CR0 $\ket{00}$]{
    \includegraphics[width=.22\textwidth]{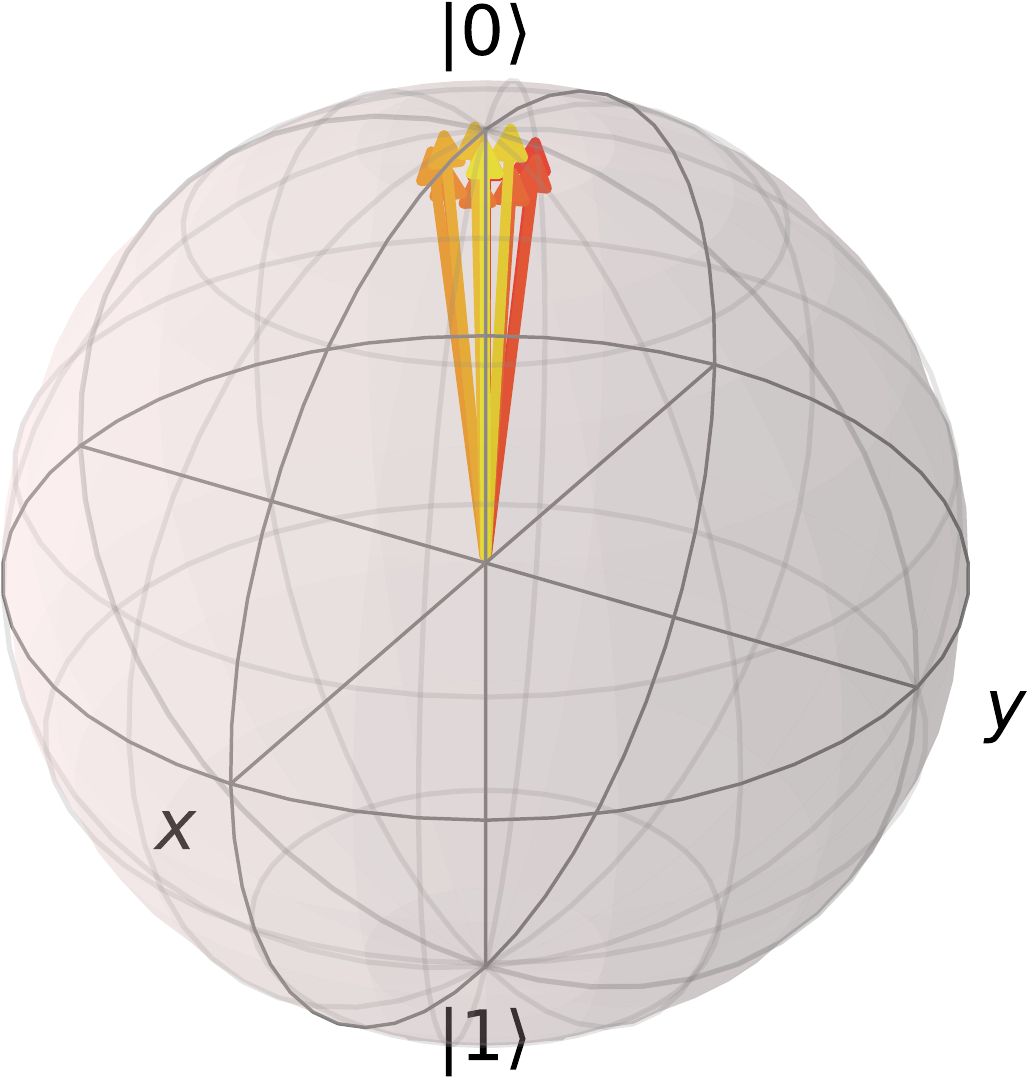}
    \includegraphics[width=.22\textwidth]{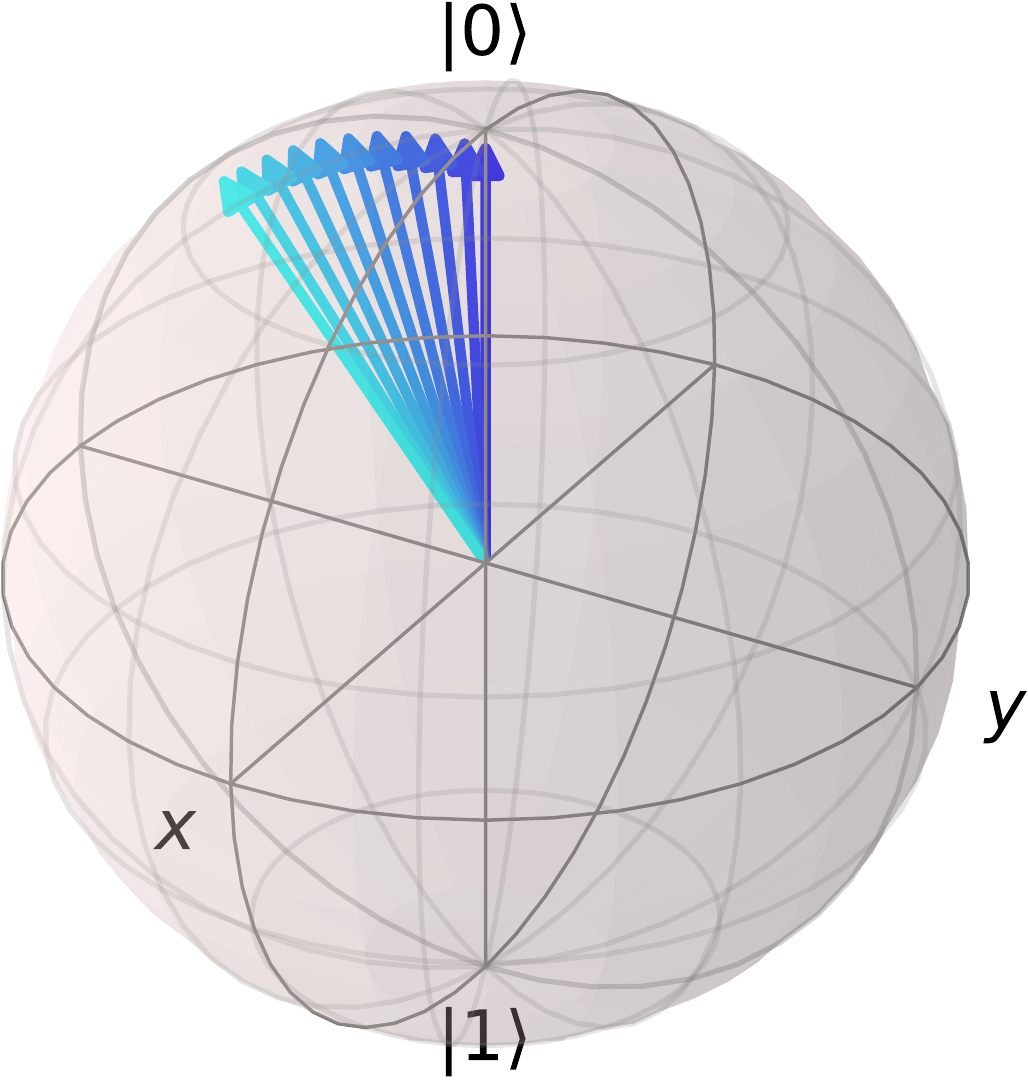}
  }\hfill
  \subfloat[\label{sfig:CR01}CR0 $\ket{10}$]{
    \includegraphics[width=.22\textwidth]{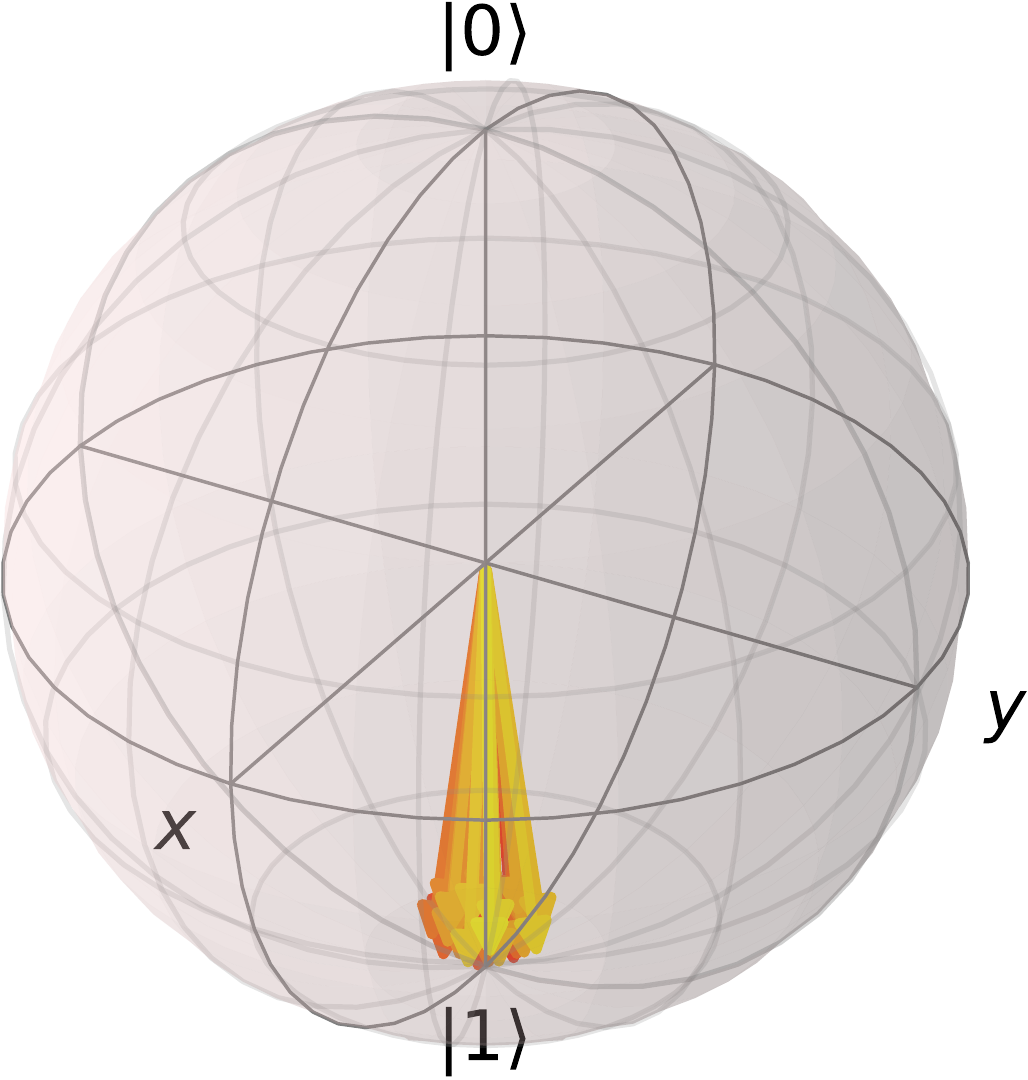}
    \includegraphics[width=.22\textwidth]{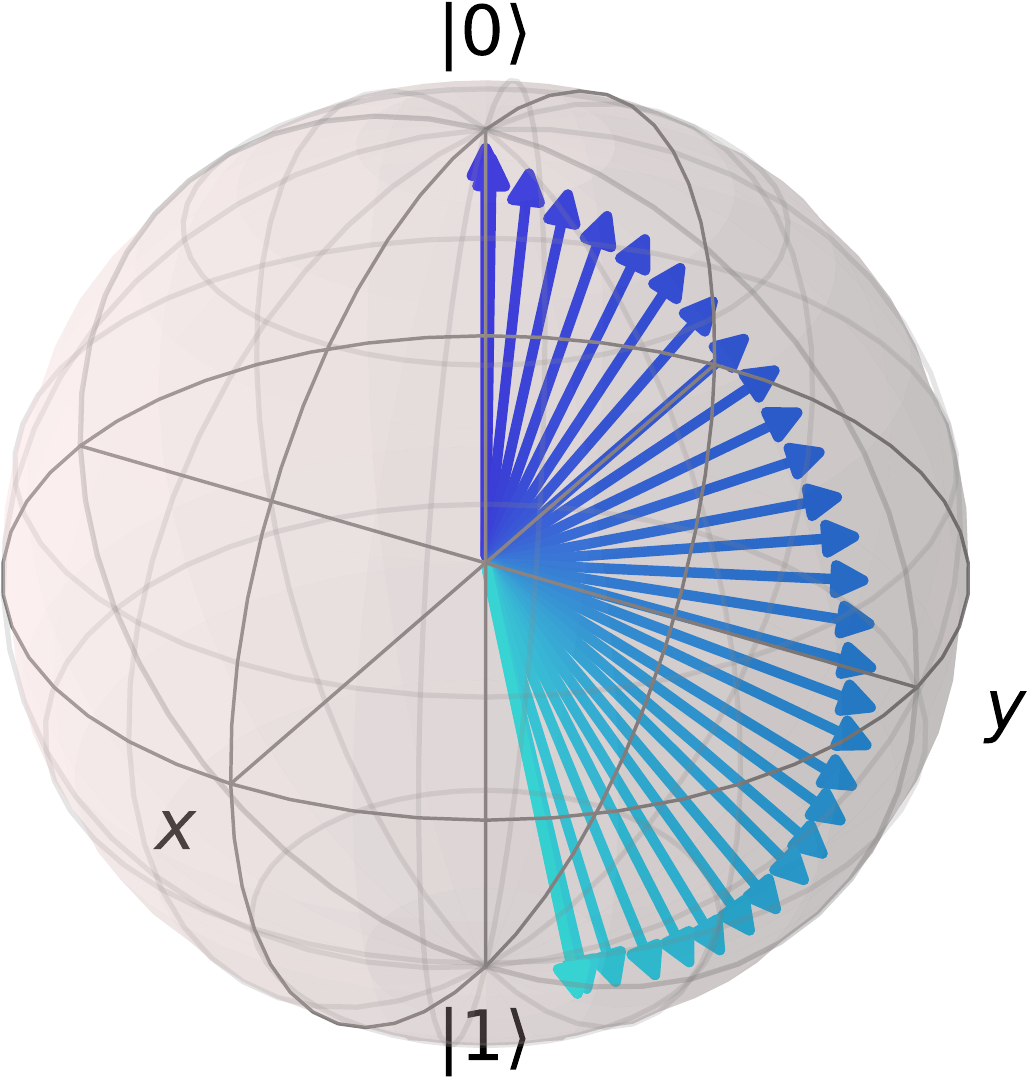}
  }\\
  \subfloat[\label{sfig:CR10}CR1 $\ket{00}$]{
    \includegraphics[width=.22\textwidth]{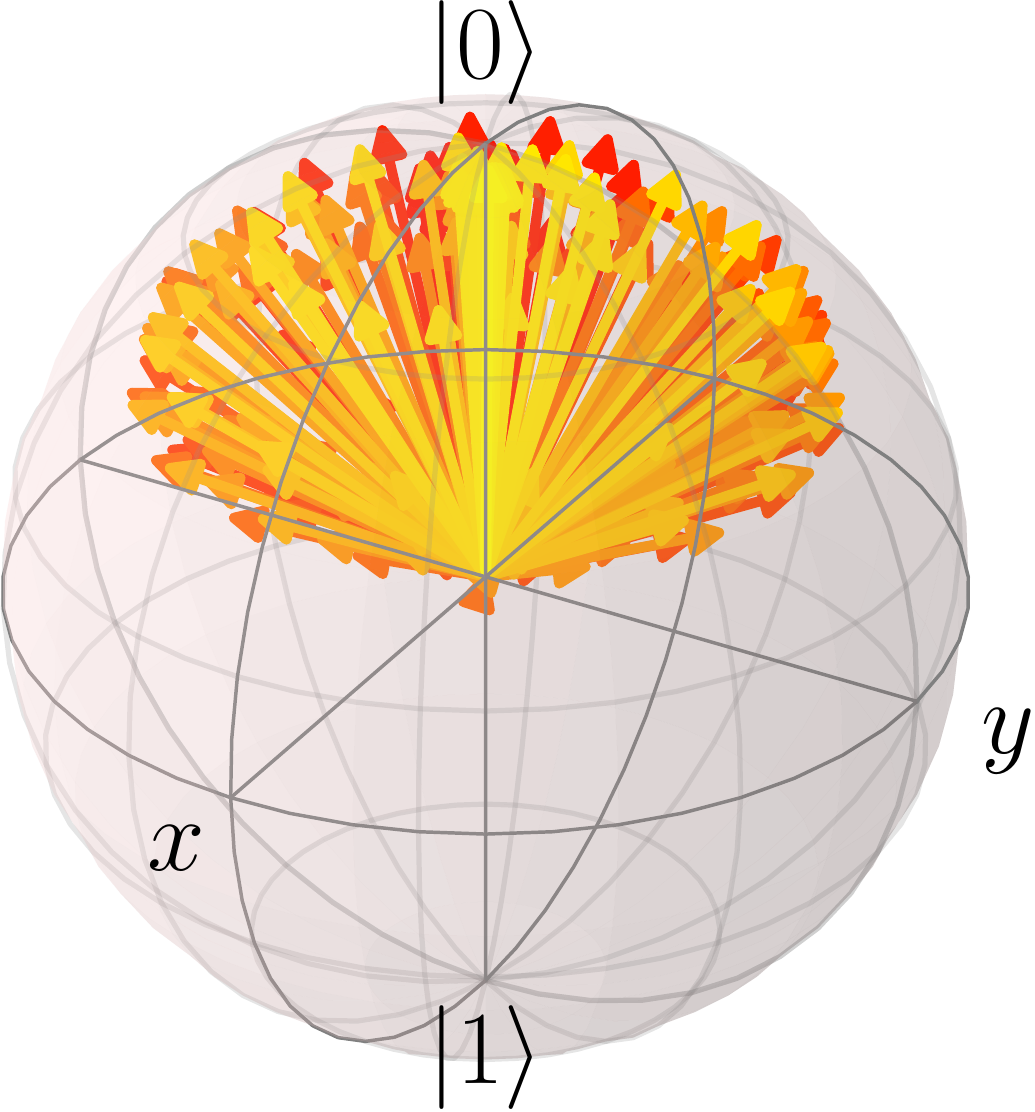}
    \includegraphics[width=.22\textwidth]{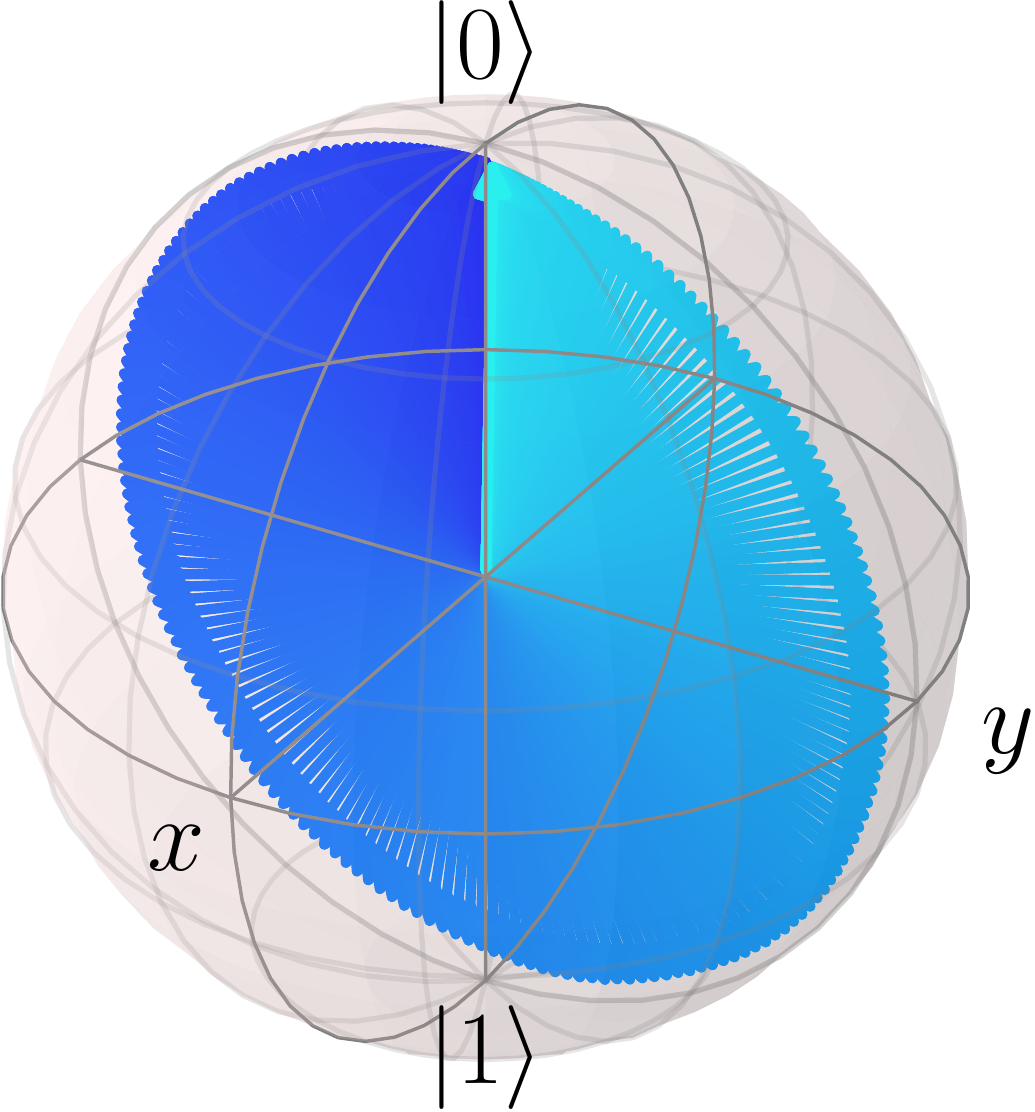}
  }\hfill
  \subfloat[\label{sfig:CR11}CR1 $\ket{10}$]{
    \includegraphics[width=.22\textwidth]{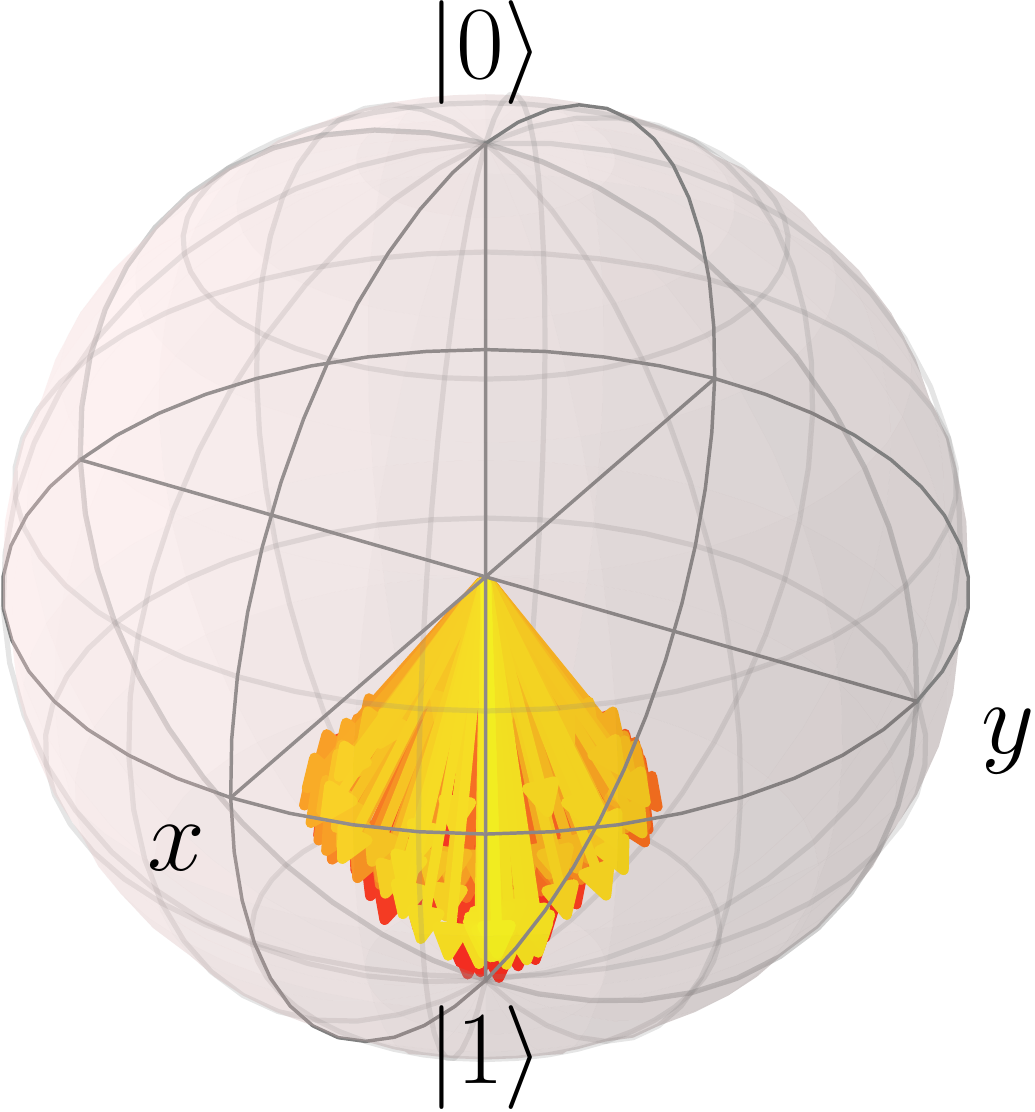}
    \includegraphics[width=.22\textwidth]{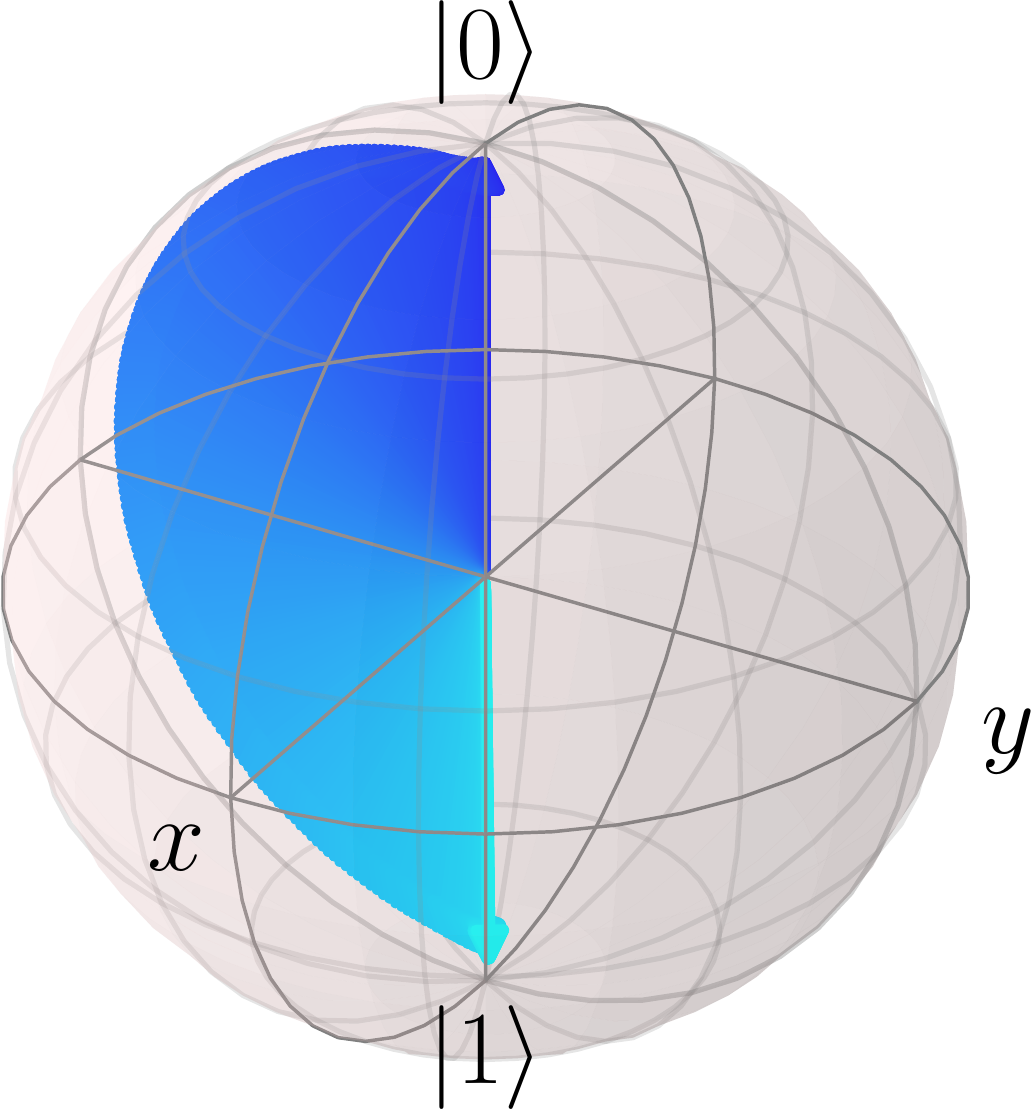}
  }\\
  \subfloat[\label{sfig:CR20}CR2 $\ket{00}$]{
    \includegraphics[width=.22\textwidth]{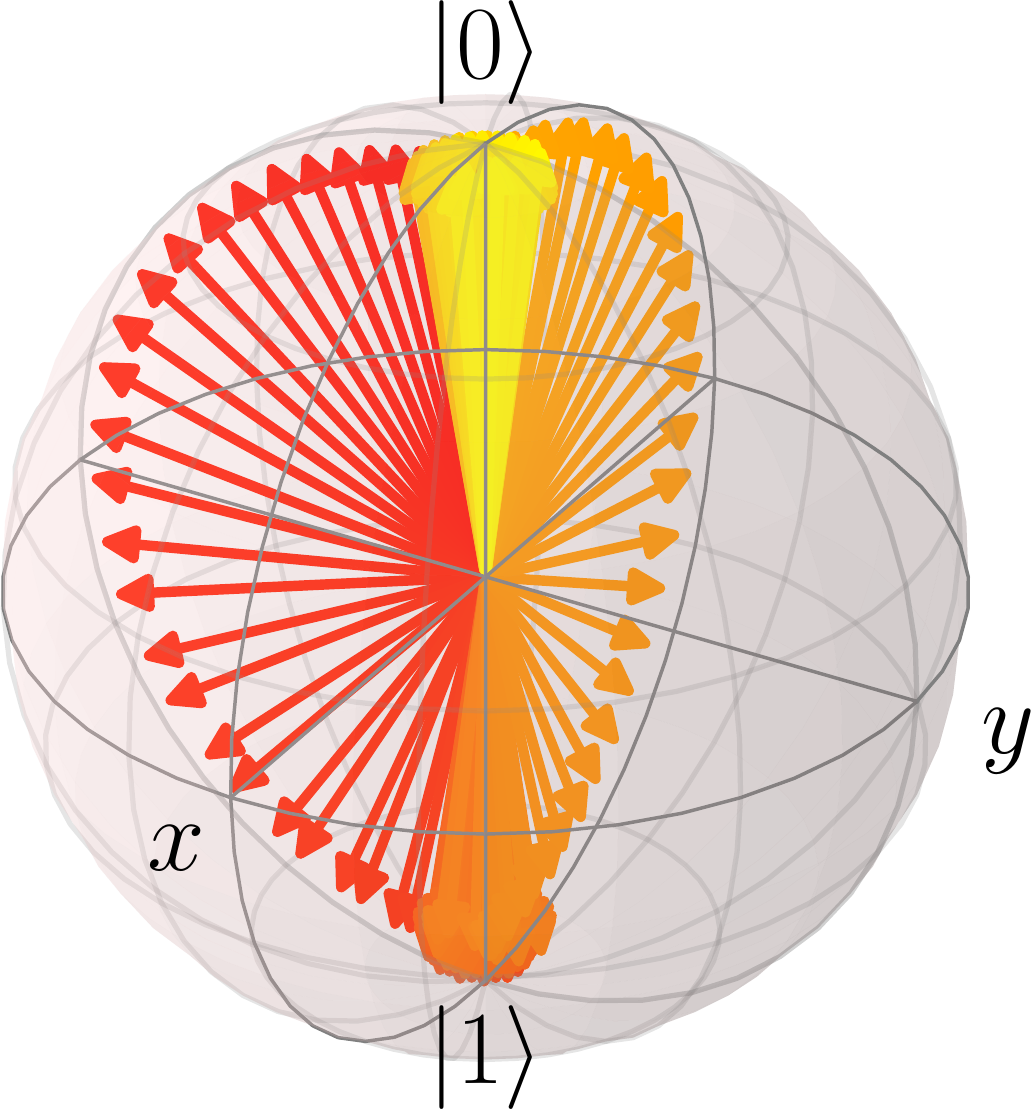}
    \includegraphics[width=.22\textwidth]{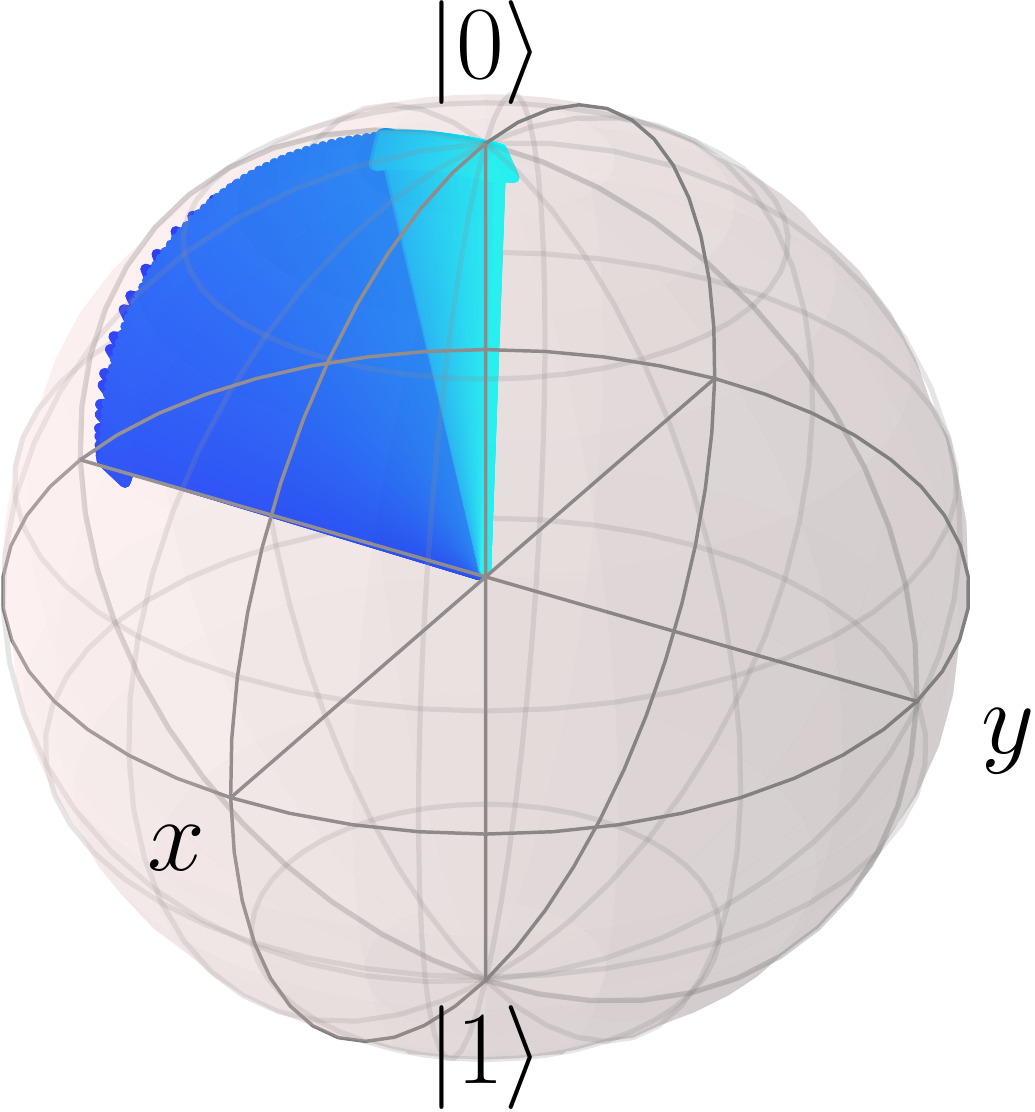}
  }\hfill
  \subfloat[\label{sfig:CR21}CR2 $\ket{10}$]{
    \includegraphics[width=.22\textwidth]{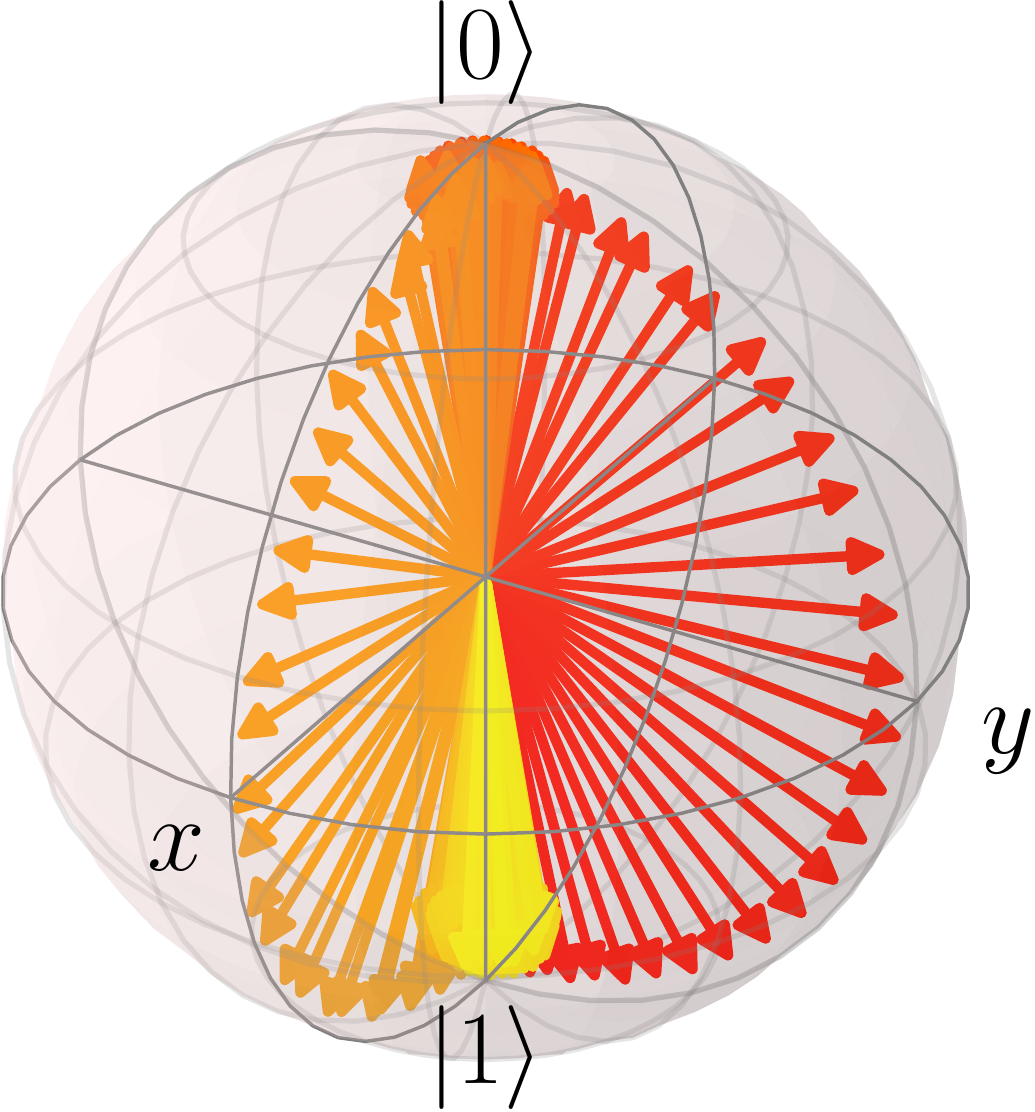}
    \includegraphics[width=.22\textwidth]{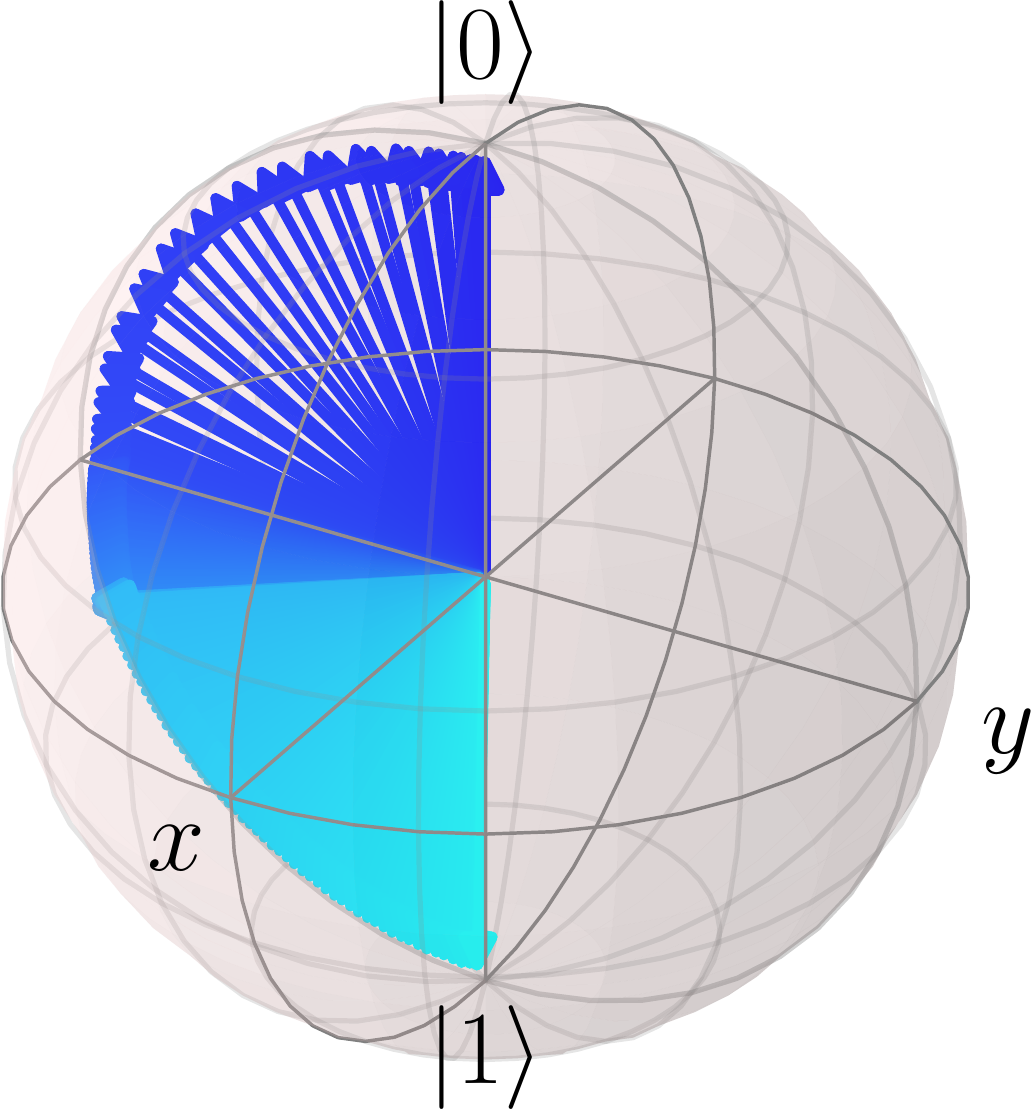}
  }\\
  \subfloat[\label{sfig:CR40}CR4 $\ket{00}$]{
    \includegraphics[width=.22\textwidth]{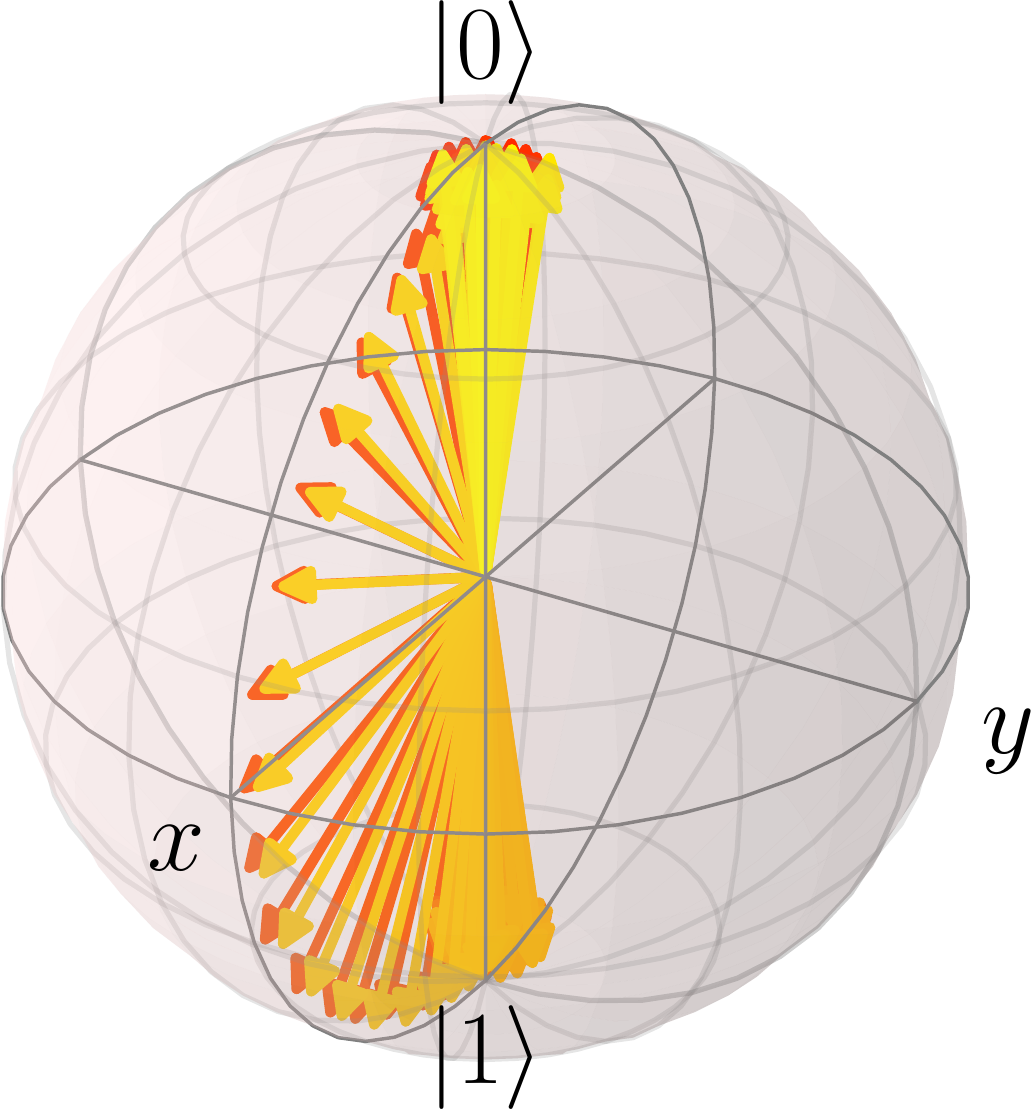}
    \includegraphics[width=.22\textwidth]{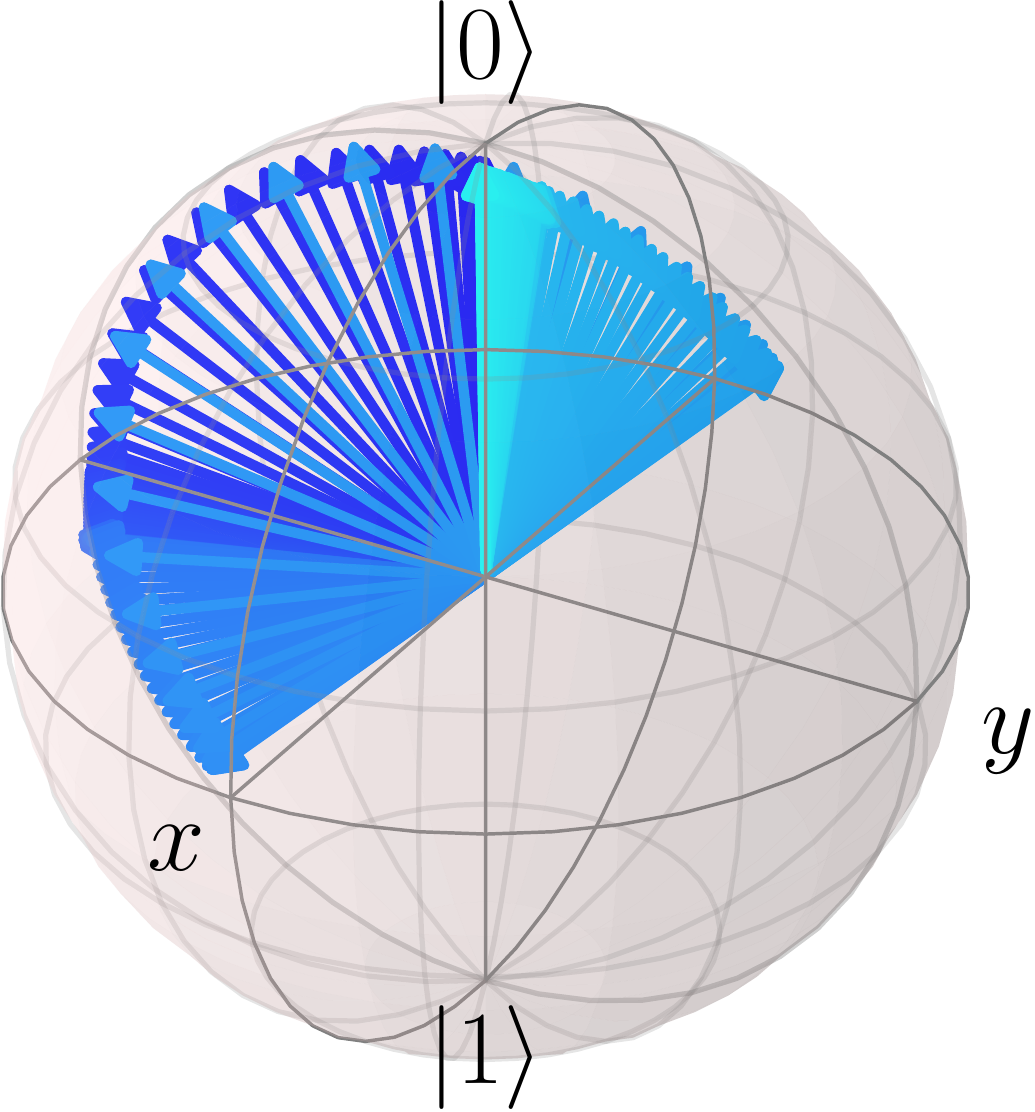}
  }\hfill
  \subfloat[\label{sfig:CR41}CR4 $\ket{10}$]{
    \includegraphics[width=.22\textwidth]{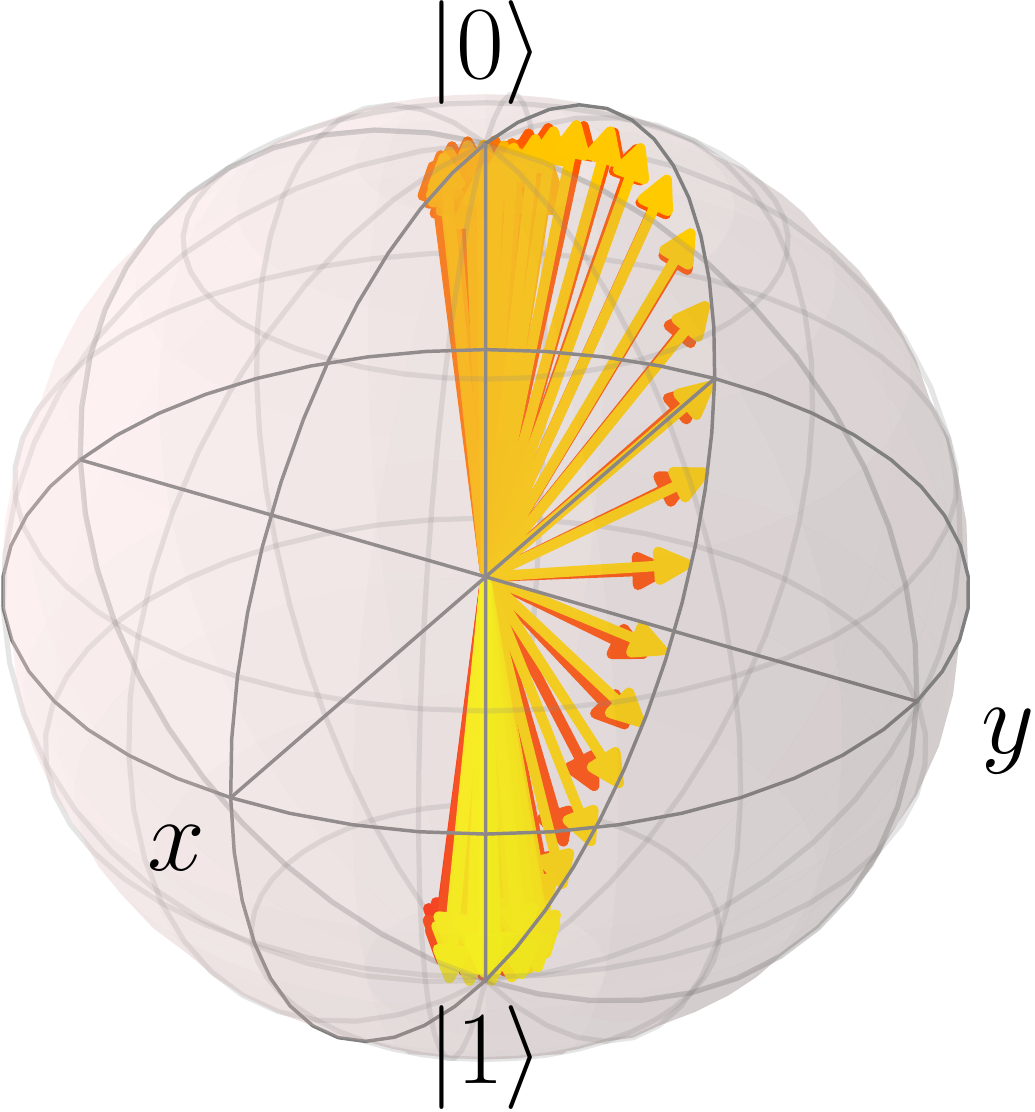}
    \includegraphics[width=.22\textwidth]{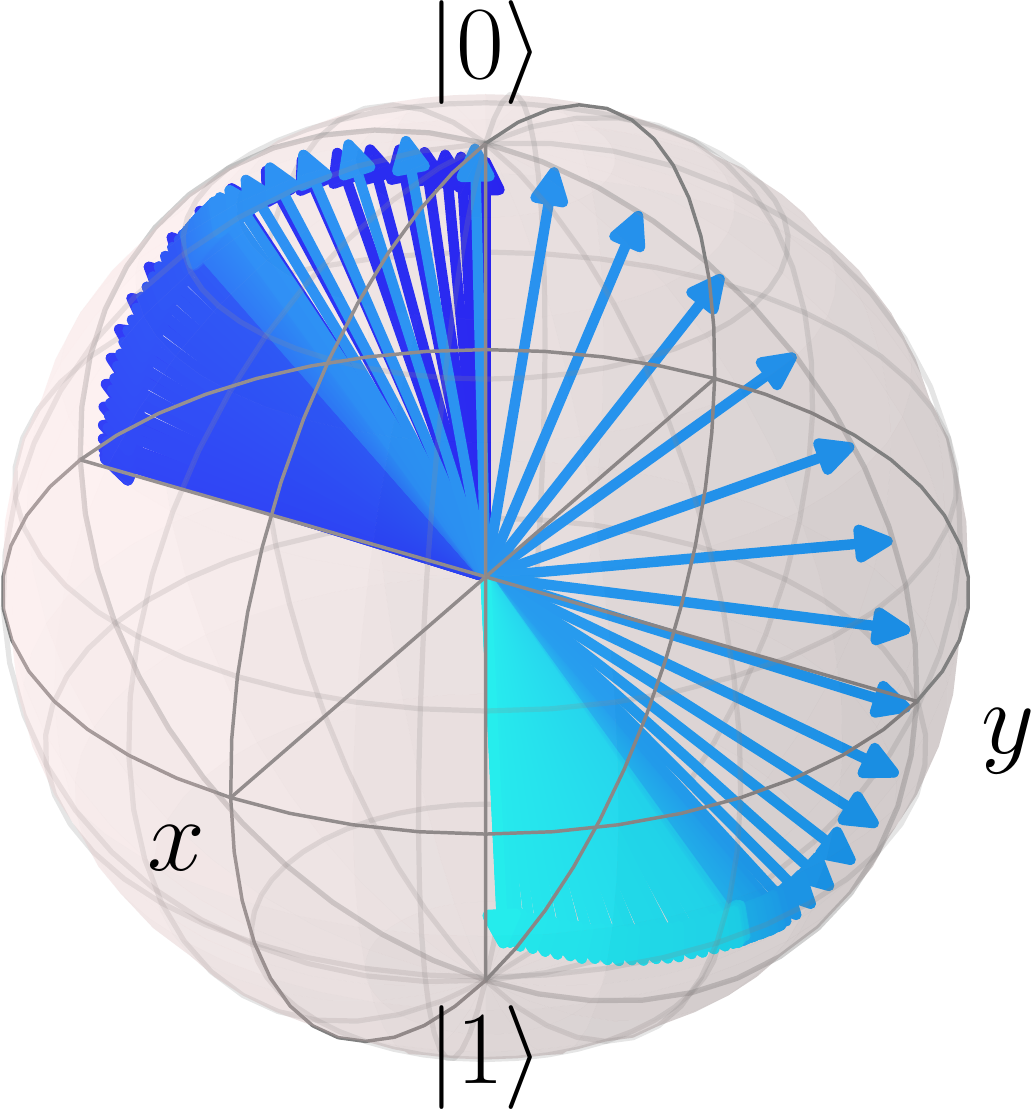}
  }\null
  \caption{Bloch-sphere representation of the time evolutions of two transmons
  under the application of the CR pulses shown in
  \figref{fig:crossresonancepulses}; (a) and (b) basic CR0 pulse with
  $\Omega_{\mathrm{CR}}=0.01$ and $T_{\mathrm{CR}}=\SI{270}{ns}$; (c)--(h) CR1,
  CR2, and CR4 pulse implementing a $\textsc{CNOT}$ gate. The time $t$ is
  encoded in the color of the arrows, i.e., from red to yellow (blue to cyan)
  for the control (target) qubit. The pulse parameters for (c)--(h)
  result from the optimization  procedure described in
  \secref{sec:optimizatingpulseparameters} and are listed in
  \appref{app:pulseparameters}. The model parameters of the simulated transmon
  system are given in \tabref{tab:deviceibm2gst}. The Bloch vectors are computed
  by \equref{eq:multiqubitblochvectorTransmonTrace} (not renormalized) in a frame
  rotating at the frequencies given in
  \tabref{tab:deviceibm2gstPulseParametersGD}. The data has been visualized
  with \texttt{QuTiP} \cite{Johansson2012Qutip, Johansson2013Qutip2}.}
  \label{fig:twoqubitpulseblochevolution}
\end{figure}

The effect of the basic CR0 pulse can be illustrated in
terms of the effective two-qubit Hamiltonian
\begin{align}
  \label{eq:twoqubitpulseEffectiveHamiltonian}
  H_{\mathrm{eff}} = \frac{J_{\mathit{IX}}}2 \sigma_{i_T}^x + \frac{J_{\mathit{ZX}}}2 \sigma_{i_C}^z \sigma_{i_T}^x + \frac{J_{\mathit{ZI}}}2 \sigma_{i_C}^z ,
\end{align}
where $i_C$ is the index of the control qubit (to which the pulse is applied),
$i_T$ is the index of the target qubit (which determines the frequency of the pulse),
and the coefficients $J_{\mathit{IX}}$, $J_{\mathit{ZX}}$, and $J_{\mathit{ZI}}$
represent the strength of the two-qubit terms. Both
$J_{\mathit{IX}}$ and $J_{\mathit{ZX}}$ are approximately proportional to
$\Omega_{\mathrm{CR}}$ (see below). After an application of the CR0 pulse, the
implemented transformation is approximately given by
\begin{align}
  \label{eq:twoqubitpulseCR0ImplementedMatrix}
  \exp(-i H_{\mathrm{eff}}T_{\mathrm{tot}}) \propto\ \begin{blockarray}{cc}
    \matindex{\text{Control in }\ket 0} & \matindex{\text{Control in }\ket 1} \\
    \begin{block}{(cc)}
      R^x((J_{\mathit{IX}}+J_{\mathit{ZX}})T_{\mathrm{tot}}) & \boldsymbol0 \\
      \boldsymbol0 & R^x((J_{\mathit{IX}}-J_{\mathit{ZX}})T_{\mathrm{tot}})e^{i\eta}  \\
    \end{block}
  \end{blockarray}\ ,
\end{align}
where $R^x(\vartheta)$ is the matrix of the single-qubit $x$ rotation defined in
\equref{eq:singlequbitrotationx} and $e^{i\eta}$ is a phase factor. The
operation expressed by \equref{eq:twoqubitpulseCR0ImplementedMatrix} is an $x$
rotation of the target qubit by an angle
$\vartheta=(J_{\mathit{IX}}+J_{\mathit{ZX}})T_{\mathrm{tot}}$
($\vartheta=(J_{\mathit{IX}}-J_{\mathit{ZX}})T_{\mathrm{tot}}$) if the control
qubit is in state $\ket 0$ ($\ket 1$). See
\figref{fig:twoqubitpulseblochevolution}(a) and (b) for a Bloch-sphere
visualization of the time evolution, computed from the simulation of the
two-transmon system defined in  \secref{sec:transmonmodelibm2gst} under the
application of a CR0 pulse. In each plot, the number of arrows per time plotted
is constant, so larger spacings between successive arrows represent a faster
relative time evolution. Furthermore, the vectors are not renormalized, so
the fact that the magnitude of all vectors is almost one means that the states
evolve as almost unentangled states if the initial state is a computational
basis state.

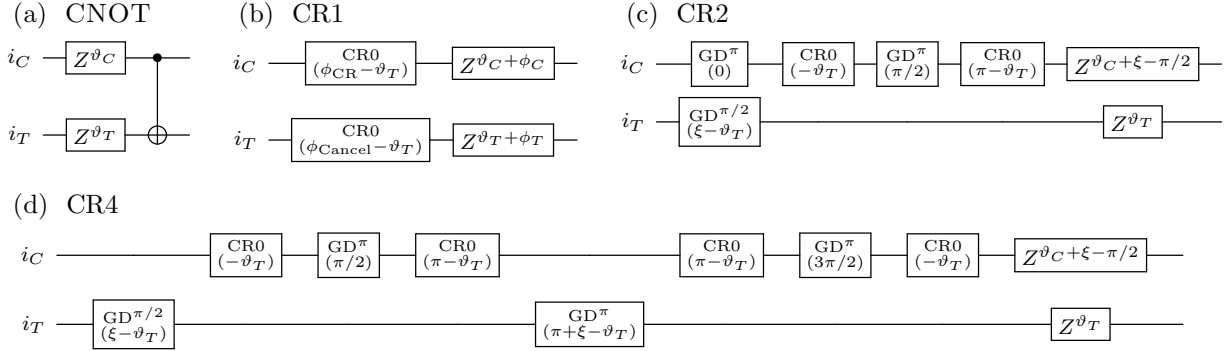
\begin{figure}
  \scriptsize
  \centering
  \captionsetup[subfigure]{position=top,textfont=normalfont,singlelinecheck=off,justification=raggedright,labelformat=parens}
  \subfloat[\label{sfig:CNOT} \textsc{CNOT}]{
    $
      \Qcircuit @C=1em @R=2em {
       & \lstick{i_C} & \gate{Z^{\vartheta_C}} & \ctrl{1} & \qw \\
       & \lstick{i_T} & \gate{Z^{\vartheta_T}} & \targ & \qw \\
      }
    $
  }\hfill
  \subfloat[\label{sfig:CR1} CR1]{
    $
      \Qcircuit @C=1em @R=1.5em {
	       & \lstick{i_C} & \gate{\substack{\mathrm{CR0}\\(\phi_{\mathrm{CR}}-\vartheta_T)}} & \gate{Z^{\vartheta_C+\phi_C}} & \qw \\
         & \lstick{i_T} & \gate{\substack{\mathrm{CR0}\\(\phi_{\mathrm{Cancel}}-\vartheta_T)}} & \gate{Z^{\vartheta_T+\phi_T}} & \qw \\
      }
    $
  }\hfill
  \subfloat[\label{sfig:CR2} CR2]{
    $
      \Qcircuit @C=1em @R=.5em {
  	   & \lstick{i_C} & \gate{\substack{\mathrm{GD}^{\pi}\\(0)}}                 & \gate{\substack{\mathrm{CR0}\\(-\vartheta_T)}} & \gate{\substack{\mathrm{GD}^{\pi}\\(\pi/2)}} & \gate{\substack{\mathrm{CR0}\\(\pi-\vartheta_T)}} & \gate{Z^{\vartheta_C+\xi-\pi/2}} & \qw \\
	     & \lstick{i_T} & \gate{\substack{\mathrm{GD}^{\pi/2}\\(\xi-\vartheta_T)}} &                                            \qw &                                          \qw &                                               \qw & \gate{Z^{\vartheta_T}}           & \qw \\
      }
    $
  }\\
  \subfloat[\label{sfig:CR4} CR4]{
    $
      \Qcircuit @C=1.6em @R=1em {
	      & \lstick{i_C} & \qw & \gate{\substack{\mathrm{CR0}\\(-\vartheta_T)}} & \gate{\substack{\mathrm{GD}^{\pi}\\(\pi/2)}} & \gate{\substack{\mathrm{CR0}\\(\pi-\vartheta_T)}} & \qw & \gate{\substack{\mathrm{CR0}\\(\pi-\vartheta_T)}} & \gate{\substack{\mathrm{GD}^{\pi}\\(3\pi/2)}} & \gate{\substack{\mathrm{CR0}\\(-\vartheta_T)}} & \gate{Z^{\vartheta_C+\xi-\pi/2}} & \qw \\
	      & \lstick{i_T} & \gate{\substack{\mathrm{GD}^{\pi/2}\\(\xi-\vartheta_T)}} & \qw & \qw & \qw & \gate{\substack{\mathrm{GD}^{\pi}\\(\pi+\xi-\vartheta_T)}} & \qw & \qw & \qw & \gate{Z^{\vartheta_T}} & \qw \\
      }
    $
  }\hfill\null
  \caption{Specifications of the pulse sequences shown in
  \figref{fig:crossresonancepulses}(b)--(d) to implement the \textsc{CNOT} gate;
  (a) elementary \textsc{CNOT} gate with preceding local $z$ rotations; (b)--(d)
  CR1, CR2, and CR4 schemes to implement the \textsc{CNOT} gate with local $z$
  rotations, where the $z$ gates have been propagated to the end of
  the sequence to make the pulse scheme compatible with the VZ gate. Note that
  the VZ phases of the control qubit are unaffected by $\vartheta_C$ because of
  \equref{eq:cnotgatecommuteswithZoncontrol}. The single-qubit GD pulse is
  defined in \equref{eq:singlequbitpulseGD} and the basic CR0 pulse is defined
  in \equref{eq:twoqubitpulseCR0}.}
  \label{fig:twoqubitpulseruleVZ}
\end{figure}

Note that the effective Hamiltonian given by
\equref{eq:twoqubitpulseEffectiveHamiltonian}  does not accurately model the
intermediate time evolution during the application of the pulse including the
finite rise and fall time. For instance, the control qubit  does not stand still
during the evolution, as indicated by the red arrows in
\figref{fig:twoqubitpulseblochevolution}.  However, the result of the pulse
application is described well enough by
\equref{eq:twoqubitpulseCR0ImplementedMatrix}: an $x$ rotation of the target
qubit, for which the angle depends on the state of the control qubit.  Since
this operation maps the target qubit to different states depending on whether
the control qubit is in state $\ket 0$ or $\ket 1$, it is an entangling
operation and can be used to assemble a \textsc{CNOT} gate.

We consider three candidates, denoted by CR1, CR2, and CR4, to implement a
\textsc{CNOT} gate based on the elementary CR0 pulse. They are schematically
plotted in \figref{fig:crossresonancepulses}(b)--(d). For each pulse, the time
evolution of a two-transmon system is shown in
\figref{fig:twoqubitpulseblochevolution}(c)--(h). While the target qubit (blue)
can be easily seen to end up in the proper state, the time evolution of the
control qubit (red) can be much more complicated.

\subsubsection{The CR1 pulse}

As a first candidate to implement the \textsc{CNOT} gate, we consider the
CR1 pulse shown in \figref{fig:crossresonancepulses}(b). This pulse sequence was
introduced in \cite{Willsch2017GateErrorAnalysis} and a demonstration of
how to implement it on an IBM Q processor by means of \emph{Qiskit Pulse} is
given in \cite{Alexander2020QiskitPulse}.

The CR1 pulse consists of two simultaneous elementary CR0 pulses:
one on the control qubit and one on the target qubit. Both pulses have the
frequency $f_{i_T}$ of the target qubit. The amplitude of the second pulse is denoted by
$\Omega_{\mathrm{Cancel}}$ and is typically smaller than the control pulse
amplitude $\Omega_{\mathrm{CR}}$. Since
the second pulse on the target qubit is a resonant driving, its effect is a simple $x$
rotation whose angle is determined by \equref{eq:singlequbitpulseangle}.
Therefore, $\Omega_{\mathrm{Cancel}}$ directly changes the strength
$J_{\mathit{IX}}$ of the unconditional $x$ rotation given in
\equref{eq:twoqubitpulseEffectiveHamiltonian}. This behavior is also confirmed
by the simulation results discussed in the next section (see
\figref{fig:crossresonanceamplitudesscan}(c) and (d)).

We can use this fact in the following way: The elementary CR0 pulse with amplitude
$\Omega_{\mathrm{CR}}$ on the control qubit sets the magnitude of
$J_{\mathit{IX}}$ and  $J_{\mathit{ZX}}$. By changing $\Omega_{\mathrm{Cancel}}$
on the target qubit, we can adjust $J_{\mathit{IX}}$ independently such that
\begin{subequations}
\begin{align}
  \label{eq:twoqubitpulseCR1equation1}
  (J_{\mathit{IX}}+J_{\mathit{ZX}})T_{\mathrm{tot}}\ \mathrm{mod}\ 2\pi &= 0,\\
  \label{eq:twoqubitpulseCR1equation2}
  (J_{\mathit{IX}}-J_{\mathit{ZX}})T_{\mathrm{tot}}\ \mathrm{mod}\ 2\pi &= \pi,
\end{align}
\end{subequations}
and the implemented transformation in
\equref{eq:twoqubitpulseCR0ImplementedMatrix} becomes proportional to a
$\textsc{CNOT}$ gate (up to local $z$ rotations). The effect of
\equref{eq:twoqubitpulseCR1equation1} can be seen in
\figref{fig:twoqubitpulseblochevolution}(c), where the target qubit undergoes a
full $2\pi$ rotation and remains effectively unchanged. Similarly, the effect of
\equref{eq:twoqubitpulseCR1equation2} is shown in
\figref{fig:twoqubitpulseblochevolution}(d), where the target qubit undergoes  a
$\pi$ rotation such that it ends up in the $\ket{1}$ state.

There may still be spurious local $z$ rotations. One is represented by the phase
factor  $e^{i\eta}$ in \equref{eq:twoqubitpulseCR0ImplementedMatrix}, which stems
from the coefficient $J_{\mathit{ZI}}$. Because of such phase errors, the
operation on the computational subspace actually amounts to
\begin{align}
  \label{eq:twoqubitpulseCR1ImplementedMatrix}
  \begin{pmatrix}
e^{i\chi_1} \\ & e^{i\chi_2} \\ & & & e^{i\chi_3} \\ & & e^{i\chi_4}
  \end{pmatrix}.
\end{align}
However, the phase errors $\chi_i$ can be corrected with phase shifts
$\phi_{\mathrm{CR}}$ and $\phi_{\mathrm{Cancel}}$ in the basic CR0 pulses
defined in \equref{eq:twoqubitpulseCR0}, followed by local $z$ rotations
$Z_{i_C}^{\phi_C}\otimes Z_{i_T}^{\phi_T}$.

In summary, the $\textsc{CNOT}_{i_Ci_T}$ implementation using the CR1 pulse is
\begin{align}
  \label{eq:twoqubitpulseCR1}
  \textsc{CNOT}_{i_Ci_T} = (Z_{i_C}^{\phi_C}\otimes Z_{i_T}^{\phi_T}) (\mathrm{CR0}_{i_C}(\phi_{\mathrm{CR}})\otimes \mathrm{CR0}_{i_T}(\phi_{\mathrm{Cancel}})).
\end{align}
The CR1 gate depends on $7$ parameters $(f_{i_T},T_{\mathrm{CR}}, \Omega_{\mathrm{CR}},
\Omega_{\mathrm{Cancel}}, \phi_{\mathrm{CR}}, \phi_{\mathrm{Cancel}}, \phi_{C},
\phi_{T})$ to be optimized in the optimization procedure. In principle, one
phase parameter could be eliminated; however, we found that keeping an additional
phase parameter helps in mitigating phase errors caused by other components of
the system. The full pulse sequence including
VZ phases is specified in \figref{fig:twoqubitpulseruleVZ}(b).

\subsubsection{The CR2 pulse}

The CR2 pulse implements the \textsc{CNOT} gate using a two-pulse echo scheme.
The idea has been analyzed in \cite{Corcoles2013processverification} and further
specified in the supplementary material of
\cite{Takita2017faultTolerantStatePreparation}. It is also currently used for
the processors on the IBM Q Experience \cite{ibmquantumexperience2016} (see the
IBM Q backend specifications). The sequence of pulses is schematically shown in
\figref{fig:crossresonancepulses}(c).

The idea of the echo scheme is that the CR0 pulse is split into two parts with
opposite amplitudes $\Omega_{\mathrm{CR}}$. Both parts are defined to have the
duration $T_{\mathrm{CR}}+\SI{30}{ns}$. Between these two parts, the control qubit
is inverted using a $\mathrm{GD}^{\pi}$ pulse. In this way, the
$J_{\mathit{IX}}$ component in  \equref{eq:twoqubitpulseEffectiveHamiltonian} is
canceled, whereas the  desired $J_{\mathit{ZX}}$ component is doubled. Besides
canceling the  $J_{\mathit{IX}}$ component, this scheme also addresses the
$J_{\mathit{ZI}}$  component in \equref{eq:twoqubitpulseEffectiveHamiltonian}
and the residual longitudinal interaction of the form given in
\equref{eq:statedependentfrequenciesTwoQubitHamiltonian}.

The amplitude $\Omega_{\mathrm{CR}}$ and the time $T_{\mathrm{CR}}$ of each of
the two CR0 parts are chosen such that
\begin{align}
  \label{eq:twoqubitpulseCR2equation}
  J_{\mathit{ZX}}T_{\mathrm{tot}} &= \frac \pi 4.
\end{align}
This means that the combined effect of the two CR0 pulses and the intermediate
$\mathrm{GD}^{\pi}$ pulse is a $\pi/2$ rotation  of the target qubit in one
direction if the control qubit is in state $\ket 0$, and in the other direction
if the control qubit is in state $\ket 1$. The \textsc{CNOT} gate is completed
with an additional $\mathrm{GD}^{\pi}$ pulse  on the control qubit (whose VZ
phase can be used to take care of $J_{\mathit{ZI}}$) and a $\mathrm{GD}^{\pi/2}$
pulse on the target qubit. These pulses have been moved to the beginning of the
pulse sequence shown in \figref{fig:crossresonancepulses}(c).

The time evolution generated by the CR2 pulse sequence is shown in
\figref{fig:twoqubitpulseblochevolution}(e) and (f): First, the target qubit is
rotated by $\pi/2$ to the negative $y$ axis. Then the first CR0 pulse rotates it
either back towards $\ket 0$ (\figref{fig:twoqubitpulseblochevolution}(e)) or
further on towards $\ket 1$ (\figref{fig:twoqubitpulseblochevolution}(f)). The
second CR0 pulse finishes this rotation. Note also that the angular velocity of
the target qubit is different for both CR0 parts (indicated by the abrupt
transition from blue to cyan). This is due to the different magnitudes of
$\abs{J_{\mathit{IX}}\pm J_{\mathit{ZX}}}$.

In practice, we use a slightly more sophisticated relation to obtain good
initial values from \equref{eq:twoqubitpulseCR2equation} for the time and the
amplitude of the CR pulses. The idea is to account for the finite rise and fall
of $\Omega_{GF}(t)$ given by \equref{eq:gaussianflattoppulse} by integrating
over time. The area under the envelope then yields the angle of rotation,
analogous to the single-qubit result in \equref{eq:singlequbitpulseangle}. The
relation reads
\begin{align}
  \label{eq:twoqubitpulseCR2equationMoreSophisticated}
  \int\limits_0^{T_{\mathrm{CR}}+\SI{30}{ns}} J^{\mathrm{theory}}_{\mathit{ZX}}\vert_{\Omega_{\mathrm{CR}}\mapsto\Omega_{GF}(t)}\,\mathrm dt &= \mathrm{sign}(J^{\mathrm{theory}}_{\mathit{ZX}})\frac \pi 4,
\end{align}
where $J^{\mathrm{theory}}_{\mathit{ZX}}$ is given by the first term of the
perturbative result given in \equref{eq:twoqubitpulseCRtheoryZX} below.
Typically, we set $\Omega_{\mathrm{CR}}=0.01$ and solve
\equref{eq:twoqubitpulseCR2equationMoreSophisticated} for $T_{\mathrm{CR}}$ to
get initial values for the optimization.

The set of parameters specifying the CR2 pulse is $(f_{i_C},f_{i_T},T_{\mathrm{CR}},
\Omega_{\mathrm{CR}})$. Implicitly, the CR2 pulse also depends on the parameters
$(T_X^\pi,\Omega_X^\pi,\beta_X^\pi)_C$ of the $\mathrm{GD}^\pi$ pulses on the
control qubit, as well as the parameters
$(T_X^{\pi/2},\Omega_X^{\pi/2},\beta_X^{\pi/2})_T$ of the $\mathrm{GD}^{\pi/2}$
pulse on the target qubit. Additionally, a phase shift $\xi=\pi$ is required if
$J_{\mathit{ZX}}<0$ for  $\Omega_{\mathrm{CR}}>0$. This can happen when
$\omega_{i_C}<\omega_{i_T}$ or $\omega_{i_C}>\omega_{i_T}+\abs{\alpha_{i_C}}$
(see \equref{eq:twoqubitpulseCRtheoryZX} or Fig.~5.12 in
\cite{Willsch2016Master} for more information). The full pulse sequence is
specified in \figref{fig:twoqubitpulseruleVZ}(c) and has a total
duration of $2(T_{\mathrm{CR}}+\SI{30}{ns})+(T_X^\pi)_C+\max\{(T_X^\pi)_C,(T_X^{\pi/2})_T\}$.

\subsubsection{The CR4 pulse}

The CR4 pulse splits the basic CR0 pulse into four parts and contains another
set of GD pulses to echo out additional phase errors. It has been used in
\cite{takita2016demonstration} and further refined in
\cite{Takita2017faultTolerantStatePreparation} (see the corresponding
supplementary material). The full pulse is sketched in
\figref{fig:crossresonancepulses}(d).

The time evolution of a two-transmon system under the application of a CR4 pulse
is shown in \figref{fig:twoqubitpulseblochevolution}(g) and (h). We see that, as
in CR2, the target qubit undergoes an initial $\pi/2$ rotation followed by
two CR0 pulses. Then, a $\pi$ pulse moves it to the opposite side of the Bloch
sphere. Another set of two CR0 pulses finally rotates it back to $\ket 0$ if the
control qubit is in state $\ket 0$
(see \figref{fig:twoqubitpulseblochevolution}(g)), or to $\ket 1$ if the
control qubit is in state $\ket 1$
(see \figref{fig:twoqubitpulseblochevolution}(h)).

Similar to the CR2 pulse, a CR4 pulse is determined by the amplitude
$\Omega_{\mathrm{CR}}$ and the time $T_{\mathrm{CR}}$ for each of the four CR0
parts. Implicitly, the CR4 pulse also depends on the parameters of all
intermediate GD pulses. The full pulse sequence is specified in
\figref{fig:twoqubitpulseruleVZ}(d).

\subsubsection{Implementation of the VZ gate in CR pulses}

All \textsc{CNOT} pulses need to be compatible with the VZ gate introduced in
\secref{sec:singlequbitVZgate}. This means that we have to define how the pulses
commute with Z gates in the spirit of \equref{eq:singlequbitpulseVZgatescheme}.
Since the CR$n$ pulses are two-qubit pulses, the scheme has to take into account
the two phases $\vartheta_C$ and $\vartheta_T$ of the control and the target
qubit, respectively:
\begin{align}
  \label{eq:twoqubitpulseVZgatescheme}
  \textsc{CNOT}_{i_Ci_T}\, (Z_{i_C}^{\vartheta_C}\otimes Z_{i_T}^{\vartheta_T}) \ket\Psi
  = (Z_{i_C}^{\vartheta_C'}\otimes Z_{i_T}^{\vartheta_T'})\, \mathrm{CR}n(\{\gamma'\}) \ket\Psi.
\end{align}
The VZ phases $\{\gamma'\}$ of the intermediate pulses and the resulting phases
$\vartheta_C'$ and $\vartheta_T'$ are summarized in
\figref{fig:twoqubitpulseruleVZ}(b)--(d).

Note that only the target phase $\vartheta_T$ has an influence on the VZ phases
of the control qubit. The reason for this is that
\begin{align}
  \label{eq:cnotgatecommuteswithZoncontrol}
  \textsc{CNOT}_{i_Ci_T}\,Z^{\vartheta_C}_{i_C} &= Z^{\vartheta_C}_{i_C}\,\textsc{CNOT}_{i_Ci_T},
\end{align}
meaning that the \textsc{CNOT} gate commutes with $z$ rotations on the control qubit.

\subsection{Analysis of \texorpdfstring{$IX$}{IX} and \texorpdfstring{$ZX$}{ZX} interactions}

The effective interaction strengths $J_{\mathit{IX}}$ and $J_{\mathit{ZX}}$ can
be both estimated analytically or extracted from simulations (such as the one
shown in \figref{fig:twoqubitpulseblochevolution}(a) and (b)). The procedure for
the  simulation emulates the procedure used in experiments
\cite{sheldon2016procedure}. In this section, we explore both routes and
systematically compare the results.

\begin{figure}
  \centering
  \includegraphics[width=\textwidth]{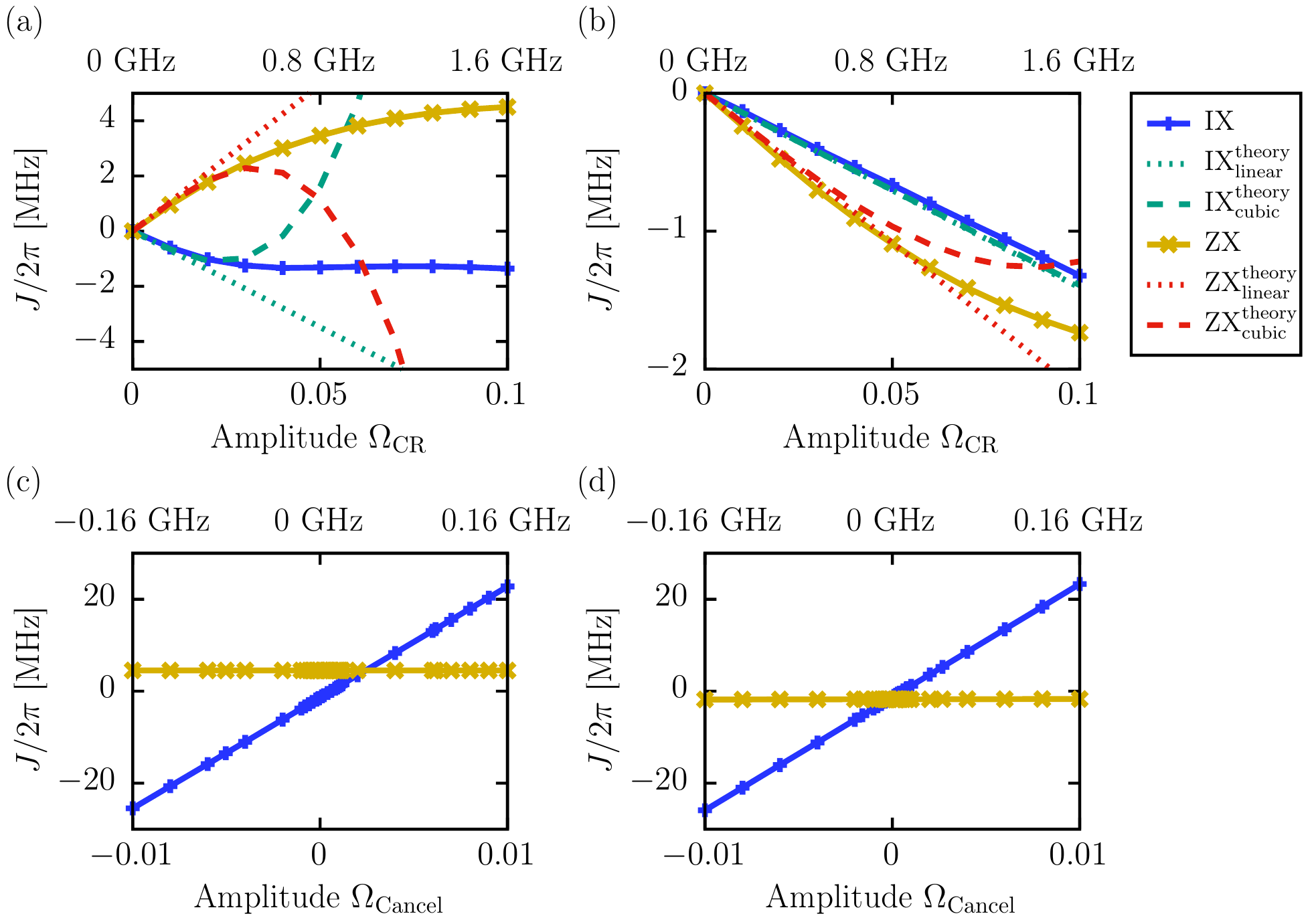}
  \caption{Interaction strengths $J_{\mathit{IX}}$ and $J_{\mathit{ZX}}$ (see
  \equref{eq:twoqubitpulseEffectiveHamiltonian}) as a function of the CR drive
  amplitudes $\Omega_{\mathrm{CR}}$ and $\Omega_{\mathrm{Cancel}}$. (a) and (b)
  represent applications of a CR0 pulse (see
  \figref{fig:crossresonancepulses}(a)) with amplitude $\Omega_{\mathrm{CR}}$ on
  the control qubit; (c) and (d) represent applications of a CR1 pulse (see
  \figref{fig:crossresonancepulses}(b)) with $\Omega_{\mathrm{CR}}=0.1$ for the
  control qubit fixed and $\Omega_{\mathrm{Cancel}}$ for the target qubit
  variable. Each CR pulse has a total duration of
  $T_{\mathrm{CR}}+\SI{30}{ns}=\SI{500}{ns}$. For (a) and (c), the control qubit
  is $i_C=0$; for (b) and (d), the control qubit is $i_C=1$. The dimensionless
  amplitudes can be converted to the strength of the drive by multiplying them
  with the conversion factor given in \equref{eq:singlequbitAmplitudeToEnergy}
  (shown on top of the figures). Each point in the figures results from two
  simulations of the system defined in \secref{sec:transmonmodelibm2gst}: one
  where the control qubit is in state $\ket 0$ and one where the control qubit
  is in state $\ket 1$.  The theory predictions given by
  \equaref{eq:twoqubitpulseCRtheoryIX}{eq:twoqubitpulseCRtheoryZX} are shown as
  dotted (dashed) lines for the linear (cubic) approximations.}
  \label{fig:crossresonanceamplitudesscan}
\end{figure}

Analytic expressions for $J_{\mathit{IX}}$ and $J_{\mathit{ZX}}$ can be derived
perturbatively. See \cite{Magesan2018CrossResonanceGateEffectiveHamiltonians}
for an extensive perturbative calculation, yielding analytic expressions up to
third order in the drive strength $\Omega_{\mathrm{CR}}$ for the coefficients
$J_{\mathit{IX}}$ and $J_{\mathit{ZX}}$. The results are
\begin{subequations}
\begin{align}
  \label{eq:twoqubitpulseCRtheoryIX}
  J_{\mathit{IX}}^{\mathrm{theory}} &=
  -\frac{J_{\mathrm{xch}}}{\alpha_{i_C}+\Delta} b_{i_C}\Omega_{\mathrm{CR}}
  +\frac{J_{\mathrm{xch}}\alpha_{i_C}\Delta}
  {(\alpha_{i_C}+\Delta)^3(\alpha_{i_C}+2\Delta)(3\alpha_{i_C}+2\Delta)}
  (b_{i_C}\Omega_{\mathrm{CR}})^3
  ,
  \\
  \label{eq:twoqubitpulseCRtheoryZX}
  J_{\mathit{ZX}}^{\mathrm{theory}} &=
  -\frac{J_{\mathrm{xch}}\alpha_{i_C}}{\Delta(\alpha_{i_C}+\Delta)} b_{i_C}\Omega_{\mathrm{CR}}
  +\frac{J_{\mathrm{xch}}\alpha_{i_C}^2(3\alpha_{i_C}^3+11\alpha_{i_C}^2\Delta+15\alpha_{i_C}\Delta^2+9\Delta^3)}
  {2\Delta^3(\alpha_{i_C}+\Delta)^3(\alpha_{i_C}+2\Delta)(3\alpha_{i_C}+2\Delta)}
  (b_{i_C}\Omega_{\mathrm{CR}})^3
  ,
\end{align}
\end{subequations}
where $\Delta=\omega_{i_C}-\omega_{i_T}$ is the difference between the
qubit frequencies, $\alpha_{i_C}$ is the anharmonicity of the control qubit,
$b_{i_C}$ is the conversion factor between energies and dimensionless amplitudes
(see \equref{eq:singlequbitAmplitudeToEnergy}), and $J_{\mathrm{xch}}$ is the
effective transmon-exchange coupling. The latter can be approximated as
\begin{align}
  \label{eq:twoqubitexchangecoupling}
  J_{\mathrm{xch}}= \frac{g_{i_C}g_{i_T}(\omega_{i_C}+\omega_{i_T}-2\Omega)}
  {2(\omega_{i_C}-\Omega)(\omega_{i_T}-\Omega)},
\end{align}
where $g_{i_C}$ and $g_{i_T}$ are the rescaled transmon-resonator couplings
given by \equref{eq:transmonresonatorexchangecoupling}, and $\Omega$ is the
frequency of the resonator. Both the linear terms and the cubic corrections
in \equaref{eq:twoqubitpulseCRtheoryIX}{eq:twoqubitpulseCRtheoryZX} are
shown as dotted and dashed lines, respectively, in \figref{fig:crossresonanceamplitudesscan}(a)
and (b). We remark that \equaref{eq:twoqubitpulseCRtheoryIX}{eq:twoqubitpulseCRtheoryZX} can also be
obtained as a special case in an extensive theoretical study using the
energy-basis representation of a transmon \cite{Malekakhlagh2020FirstPrinciplesCrossResonance}.

From the simulations, we extract $J_{\mathit{IX}}$ and $J_{\mathit{ZX}}$ by
measuring the oscillations of the target qubit conditional on the control qubit
being in state $\ket{0}$ and state $\ket{1}$. This means that for each amplitude
$\Omega_{\mathrm{CR}}$, we simulate two time evolutions for both states of the
control qubit. The resulting evolutions of the target qubit are described by
\equref{eq:twoqubitpulseCR0ImplementedMatrix} and visualized in
\figref{fig:twoqubitpulseblochevolution}(a) and (b). The procedure to obtain
$J_{\mathit{IX}}$ and $J_{\mathit{ZX}}$ from the data is described in
\cite{sheldon2016procedure} (see also Fig.~5.12 in
\cite{Willsch2016Master}).

As shown in \figref{fig:crossresonanceamplitudesscan}(a) and (b), the linear
terms in \equaref{eq:twoqubitpulseCRtheoryIX}{eq:twoqubitpulseCRtheoryZX}
correctly describe the regime of weak driving $\Omega_{\mathrm{CR}}\lesssim0.3$.
Although the cubic correction to the theoretical results properly captures the
sign of the curvature, it diverges quickly from the numerical results. The fact
that sometimes the numerical result is not exactly between the linear and cubic
theory predictions for $0\le\Omega_{\mathrm{CR}}\le0.05$ may be due to the
approximation made for the exchange coupling in
\equref{eq:twoqubitexchangecoupling}.

For fixed $\Omega_{\mathrm{CR}}=0.01$, we additionally apply a CR0 pulse at the
target frequency on the target qubit. Technically, this combination of pulses
corresponds to the CR1 pulse scheme sketched in
\figref{fig:crossresonancepulses}(b). The amplitude $\Omega_{\mathrm{Cancel}}$
of the second pulse is varied  between $-0.01$ and $0.01$. The same analysis as
before is used to obtain the coefficients  $J_{\mathrm{IX}}$ and
$J_{\mathrm{ZX}}$. The result is plotted in
\figref{fig:crossresonanceamplitudesscan}(c) and (d). As can be seen, the
additional  pulse on the target qubit linearly displaces $J_{\mathrm{IX}}$ and
does not affect $J_{\mathrm{ZX}}$. This property has been used to satisfy the
CR1 pulse conditions given by
\equaref{eq:twoqubitpulseCR1equation1}{eq:twoqubitpulseCR1equation2}.

\section{Optimization of pulse parameters}
\label{sec:optimizatingpulseparameters}

The functionality discussed in this section is part of the \texttt{optimizer}
module of the software toolkit developed for this thesis
(cf.~\secref{sec:simulationsoftware}). Its task is to optimize a  set of pulse
parameters $\vec x$ to implement a quantum gate $U$. This is done by first
generating the appropriate pulse information  for the time-dependent functions
$n_{gi}(t)$ in \equsref{eq:Htotal}{eq:HCC}, and then invoking \texttt{solver}
with this pulse information for different initial states from the computational
basis. From the results, \texttt{optimizer} infers new pulse parameters $\vec x'$
and invokes \texttt{solver} again. Eventually, the procedure converges to a set
of pulse parameters that can then be used by \texttt{compiler} to translate
arbitrary quantum circuits into pulse information. In what follows, we outline
the procedure to determine the pulse parameters for a particular quantum
gate $U$.

Given a certain voltage pulse $n_{gi}(t)$ of the form of
\equref{eq:genericvoltagepulses}, the goal is to implement a quantum gate $U$ on
the computational subspace. We denote the actual transformation of the
computational subspace  after the application of the pulse by the matrix $M$
defined in \equref{eq:MprojectionUtotal}.

The matrix $M$ depends on the set of parameters $\vec x$ defining the
particular pulse. For the single- and two-qubit pulses of interest, these
parameters have been specified in the previous sections. They typically
consist of times, amplitudes, and phases. The goal is to optimize the parameters
$\vec x$, starting from some initial values suggested by theory, with
the objective to make $M$ as close as possible to $U$.

Quantitatively, we measure closeness between $M$ and $U$ in terms of the
matrix distance
\begin{align}
  \Delta(M,U) &= \left\|M - z U\right\|_F^2,
  \label{eq:matrixdistanceobjective}
\end{align}
which is induced by the Frobenius norm $\|A\|_F^2~= \sum_{ij} \abs{A_{ij}}^2$.
The phase factor $z=\pm\sqrt{\mathrm{Tr}(MU^\dagger)/\mathrm{Tr}(MU^\dagger)^*}$
is chosen such that the difference due to the global phases of $M$ and $U$ is
minimal, since quantum gates are considered equivalent if they only differ by a
global phase (the two candidates for $z$ can be derived by minimizing
\equref{eq:matrixdistanceobjective} w.r.t.~$z=e^{i\zeta}$ for $\zeta\in\mathbb R$). In
principle, one could use other, more sophisticated quantities to measure
closeness between quantum gates. Obvious examples include the common gate
metrics studied in \secref{sec:gatemetrics}, e.g., the fidelity or the diamond
distance. However, we found that for practical purposes, the choice of the
distance function does not significantly change the quality of the resulting
quantum gates. Furthermore, \equref{eq:matrixdistanceobjective} is numerically
well suited for optimization and yields a reasonably fast convergence.

We construct $M$ by initializing the system in each of the computational basis
states at $t=0$ and simulating its time evolution $\ket{\Psi(t)}$ under the
application of the pulse for $0\le t\le T$. Each final state vector
$\ket{\Psi(T)}$ is transformed to the rotating frame (see  \equref{eq:rotatingframe}) and projected onto the computational
subspace to obtain the columns of $M$.

The size of the matrix $M$ is, in principle, determined by the computational
subspace on which the quantum gate $U$ shall be implemented. Specifically, this
means that $M$ and $U$ are $2\times2$ complex matrices for single-qubit gates,
and $4\times4$ complex matrices for two-qubit gates. However, if numerically
feasible, we sometimes optimize $M$ on the whole computational subspace, as done
for the transmon-resonator system studied in
\cite{Willsch2017GateErrorAnalysis}.

For a set of pulse parameters $\vec x$, the evaluation of the objective
function $\Delta(M,U)$ given by \equref{eq:matrixdistanceobjective} is a
complicated procedure that involves several simulations of the time evolutions
of a joint transmon-resonator system. It is therefore nontrivial to find
suitable gradients of $\Delta(M,U)$ with respect to $\vec x$. Fortunately,
there is a  multidimensional, gradient-free algorithm well suited for the optimization of
a few parameters in the case where the most complicated step is the evaluation
of the objective function: the Nelder--Mead algorithm
\cite{NelderMead1965,numericalrecipes}.

\subsection{The Nelder--Mead algorithm}
\label{sec:neldermead}

Many minimization algorithms in multiple dimensions are based on the evaluation
or estimation of gradients of the objective function. This either requires
analytic expressions for the function's reaction to changes in the  parameters,
or repeated function evaluations to trace changes in the function values back to
changes in the parameters. Furthermore, typical multidimensional minimization
algorithms base their computational strategy on well-known minimization
algorithms in a single dimension. Popular examples of these are quasi-Newton
methods such as BFGS \cite{numericalrecipes} or L-BFGS-B
\cite{Zhu1997LBFGSBalgorithm, Morales2011LBFGSBalgorithmImprovement}.

The Nelder--Mead algorithm (also known as the downhill simplex method) is based
on a completely different approach that does not require the evaluation of
gradients. It is an entirely self-contained, direct search method based on
geometrical heuristics. In principle, it can also be applied to  constrained or
discrete optimization problems \cite{Luersen2004NelderMeadConstrained,
Audet2018NelderMeadConstrained}. Due to its conciseness, it is often used as a
first step to produce reasonable results and may become the method of choice if
only a few parameters need to be optimized and the evaluation of the objective
function is rather complicated. This is the case for the present work, where
less than 10 pulse parameters need to be optimized but the evaluation of the
objective function may take several minutes on a supercomputer (e.g.~for systems
with more than 10 transmons or resonators,
cf.~\figref{fig:performanceruntime}).

\begin{figure}
  \centering
  \def\svgwidth{\textwidth}
  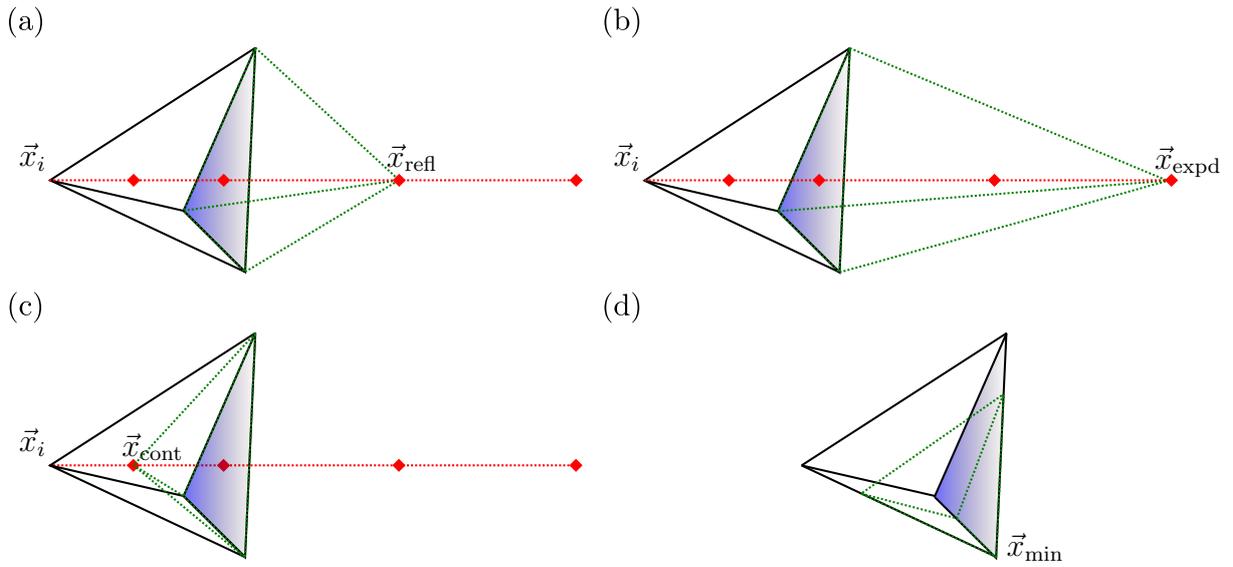
  \caption{Illustration of the four central operations of the Nelder--Mead
  algorithm in $N=3$ dimensions. The simplex is defined by $N+1=4$ points. Shown
  are the basic Nelder--Mead operations which (a) reflect, (b) expand, (c) contract, or (d)
  shrink the simplex. The solid, black simplex  represents the initial simplex.
  The dashed, green simplex represents the  simplex after the corresponding
  operation.}
  \label{fig:neldermeadsteps}
\end{figure}

To find a minimum of a function $F(\vec x)$ in $N$ real dimensions (i.e., $\vec
x \in\mathbb R^N$), the Nelder--Mead method maintains a set of $N+1$ points
$S=(\vec x_0,\ldots,\vec x_N)$. Geometrically, $S$ defines a simplex in $N$
dimensions. At each step in the minimization, the ``highest'' point $\vec x_i$,
for which $F_i = F(\vec x_i) =\max_j F(\vec x_j)$, is reflected along a line
through the opposite face of the simplex (i.e., through the centroid of the face
spanned by all the other points $\vec x_j$ for $j\neq i$, see
\figref{fig:neldermeadsteps}(a)).

After this reflection, four things can happen:
\begin{enumerate}
  \item If the new point $\vec x_{\mathrm{refl}}$ is better than at least one
  $\vec x_j$ for $j\neq i$, but not better than the current minimum, it is
  taken as the new $\vec x_i$.

  \item If $\vec x_{\mathrm{refl}}$ is even better than the current minimum, it is
  expanded further along the line to $\vec x_{\mathrm{expd}}$ (see
  \figref{fig:neldermeadsteps}(b)). The better point of $\vec x_{\mathrm{refl}}$ and
  $\vec x_{\mathrm{expd}}$ is taken as the new $\vec x_i$.

  \item Otherwise, if $\vec x_{\mathrm{refl}}$ is worse than all $\vec x_j$ for
  $j\neq i$, it is contracted back along the line to $\vec x_{\mathrm{cont}}$ (see
  \figref{fig:neldermeadsteps}(c)). If this point is
  better than the previous maximum $\vec x_i$, it is taken as the new $\vec x_i$.

  \item Only if $\vec x_{\mathrm{cont}}$ is still not better than the previous
  maximum $\vec x_i$, the simplex is shrunk towards the current minimum $\vec
  x_{\mathrm{min}}$ as shown in \figref{fig:neldermeadsteps}(d) (this will be
  the exception).
\end{enumerate}
Most of the time, the algorithm will either take the reflected point $\vec
x_{\mathrm{refl}}$ or  the contracted point $\vec x_{\mathrm{cont}}$
(cf.~Figs.~\ref{fig:optimizationxpih}(b) and \ref{fig:optimizationcnot}(b)
below). See \cite{numericalrecipes} for an example implementation or
\cite{Willsch2019NelderMead} for a modular implementation of the algorithm.

As with many optimization algorithms, the choice of the initial parameters is
crucial. Therefore, it is a good idea to use initial values from theory (such as
those outlined in
\secaref{sec:optimizatingsinglequbitgate}{sec:optimizatingtwoqubitgate}) or from
an initial scan  of parameters. Each point of the simplex is then initialized to
\begin{subequations}
  \begin{align}
    \label{eq:neldermeadInitialSimplex0}
    \vec x_0 &= \vec x_{\mathrm{init}},\\
    \label{eq:neldermeadInitialSimplex1}
    \vec x_1 &= \vec x_{\mathrm{init}} + \delta \vec x\cdot (1\,0\,\cdots\,0\,0)^T,\\
    &\makesamesize[r]{=}{\vdots}\\
    \label{eq:neldermeadInitialSimplexN}
    \vec x_{N} &= \vec x_{\mathrm{init}} + \delta \vec x\cdot (0\,0\,\cdots\,0\,1)^T,
  \end{align}
\end{subequations}
where $\vec x_{\mathrm{init}}$ is the initial set of parameters, and $\delta
\vec x$ is a set of characteristic scales for each parameter.

We terminate the optimization when the fractional range of all function values
$\vec F = (F(\vec x_0),\ldots,F(\vec x_N))$ of the simplex becomes smaller than a certain tolerance $\varepsilon$
\cite{numericalrecipes}. Quantitatively, this means that
\begin{align}
  \label{eq:neldermeadDelta}
  \delta = 2\frac{\abs{\max\vec F-\min\vec F}}{\abs{\max\vec F} + \abs{\min\vec F} + \texttt{tiny}} < \varepsilon,
\end{align}
where $\texttt{tiny}=10^{-10}$ is used to prevent division by zero. Typically,
we choose $\varepsilon=10^{-4}$ to make sure that the optimization does not run
indefinitely because of limited resolution in $\Delta(M,U)$ due to the finite
time step $\tau$.

In general, no multidimensional minimization algorithm can guarantee to find the
global  optimum; but a local minimum may still produce close-to-optimal results.
For the Nelder--Mead algorithm, the stability of the solutions can be tested by
restarting the algorithm from the previous minimum with a new set of
characteristic  scales $\delta \vec x$ that may be chosen a factor of $10$
smaller than the  previous characteristic scales. This step may also help to
escape from local minima that are not yet close to optimal.

\subsection{Optimization results}
\label{sec:optimizatingpulseparametersResults}

We use the \texttt{optimizer} module (cf.~\secref{sec:simulationsoftware}) to
optimize quantum gate pulses for the two-transmon system defined in
\secref{sec:transmonmodelibm2gst}, the small five-transmon system defined in
\secref{sec:transmonmodelibm5}, and the large five-transmon system defined in
\secref{sec:transmonmodelibm5ed}. The pulses are used for the quantum circuit
simulations described in the following chapters.  All resulting pulse parameters
are summarized in \appref{app:pulseparameters}.

\begin{figure}
  \centering
  \includegraphics[width=\textwidth]{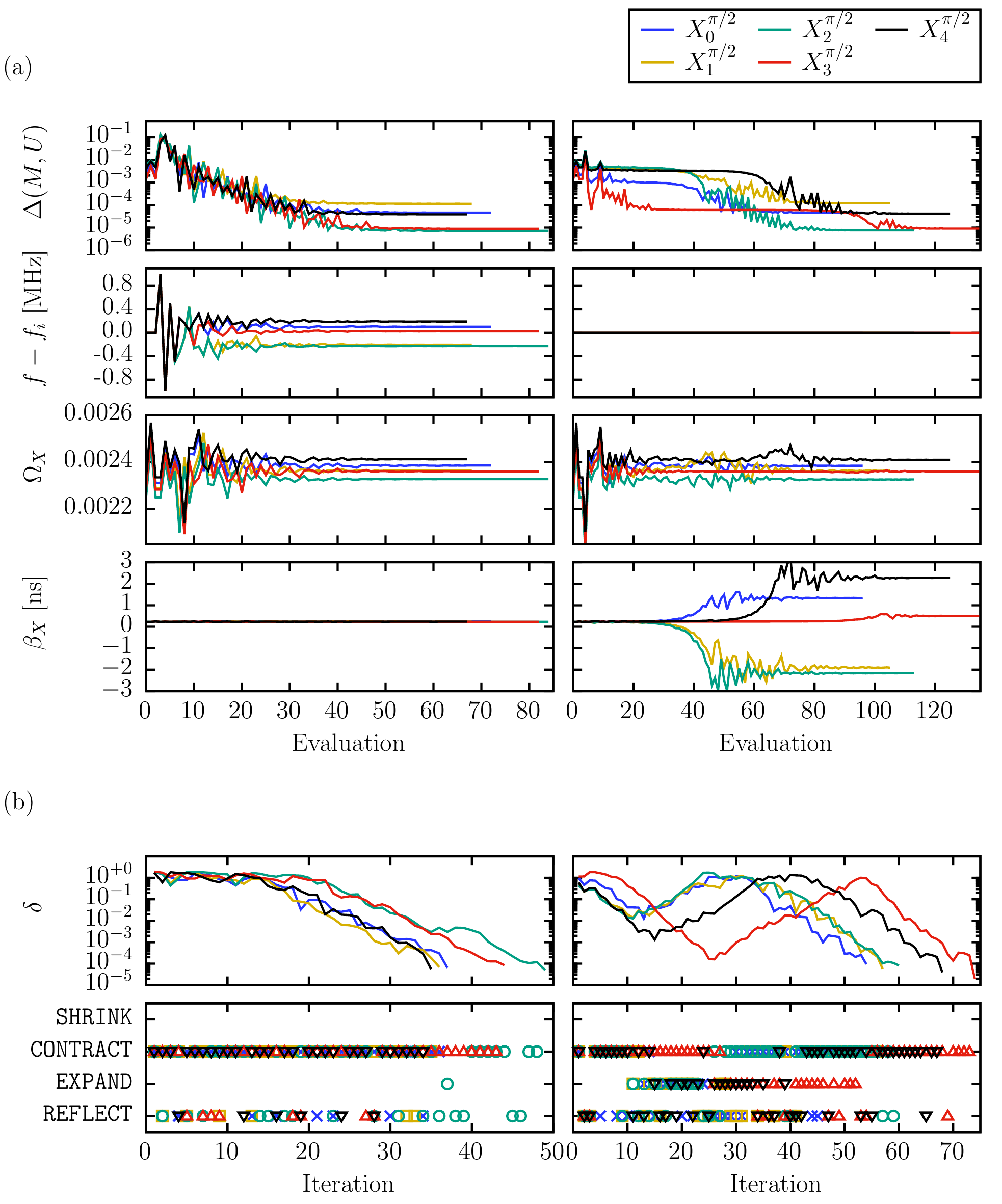}
  \caption{Optimization of single-qubit GD pulses for a system with five
  transmons and six resonators with frequency tuning (left panels) and without (right
  panels); (a) values of the pulse parameters specified in
  \equref{eq:singlequbitpulseGD} at each evaluation of
  $\Delta(M,U)$ given by \equref{eq:matrixdistanceobjective}; (b) convergence
  criterion $\delta$ given by \equref{eq:neldermeadDelta} and Nelder--Mead
  operation (see \figref{fig:neldermeadsteps}) at each iteration of the
  optimization.  The system is sketched in \figref{fig:ibm5edtopology} (see also
  \tabref{tab:deviceibm5ed1} and \tabref{tab:deviceibm5ed2}). The resulting
  parameters are given in \tabref{tab:deviceibm5edPulseParametersGD}.
  All optimizations were performed on the supercomputer JURECA \cite{JURECA}.}
  \label{fig:optimizationxpih}
\end{figure}

\begin{figure}
  \centering
  \includegraphics[width=\textwidth]{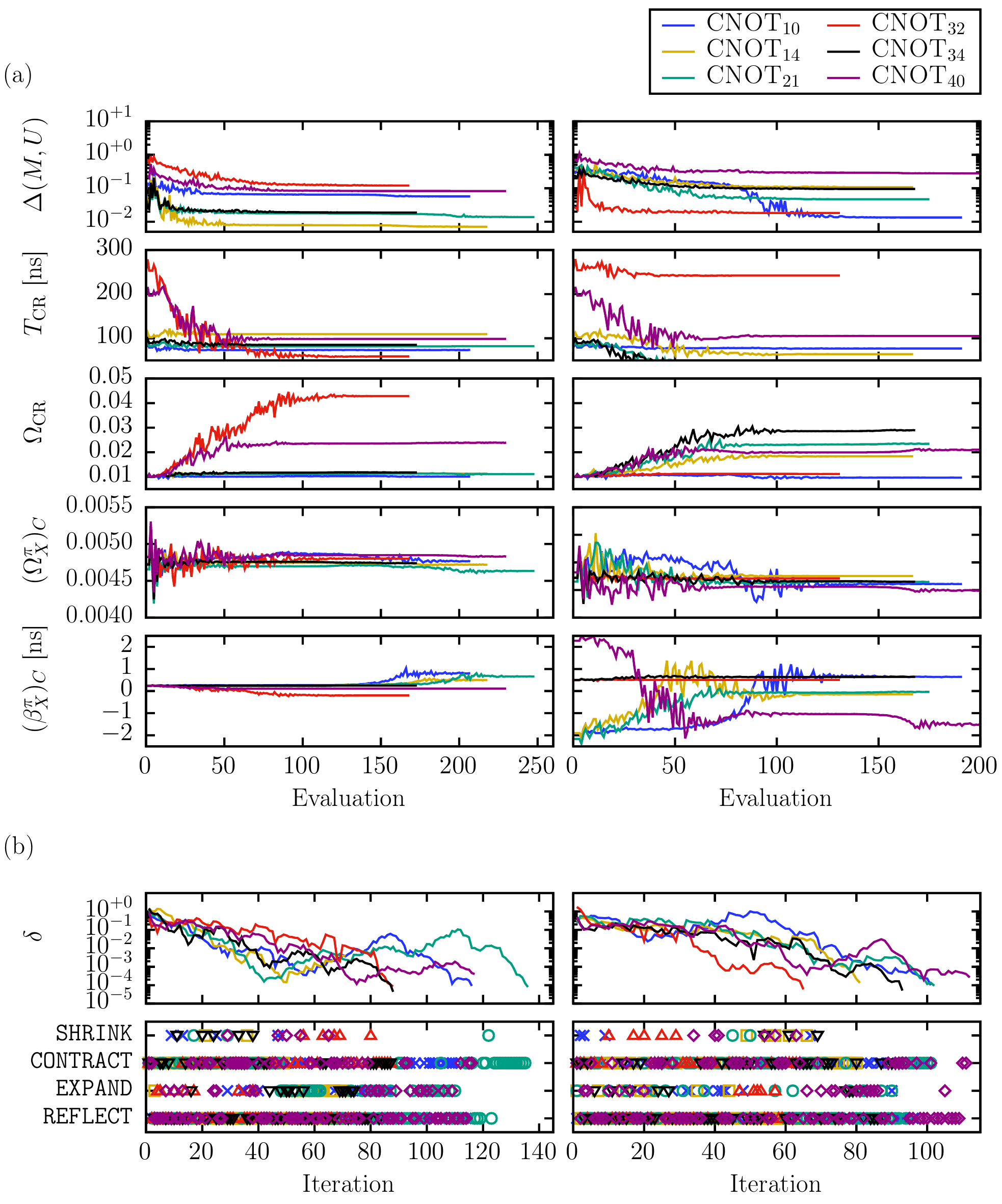}
  \caption{The same as in \figref{fig:optimizationxpih} but for the optimization
  of the two-qubit CR2 pulses specified in \secref{sec:cnotbasedonCReffect},
  using the GD pulses with frequency tuning (left panels) and without (right
  panels). The resulting parameters are given in
  \tabref{tab:deviceibm5edPulseParametersCR2} in \appref{app:pulseparameters}.
  All optimizations were performed on the supercomputer JURECA \cite{JURECA}.}
  \label{fig:optimizationcnot}
\end{figure}

As an example, we present results for the optimization process of gate pulses
for the large five-transmon system (see \secref{sec:transmonmodelibm5ed}) used
for the fault-tolerance experiments discussed in
\secref{sec:testingfaulttolerance}. We  consider two sets of gate pulses that
are optimized separately. For the first set, labeled \emph{with frequency
tuning} (or \texttt{withf} in the tables given in \appref{app:pulseparameters}),
the drive frequency $f$ of the single-qubit pulses defined by
\equref{eq:singlequbitpulseGD} is included in the set of parameters to optimize.
The idea is that slightly off-resonant driving (i.e.~$f\neq f_i$) can mitigate
phase errors since it induces rotations about the $z$ axis
\cite{gambetta2013controlIFF}. The second gate set, labeled \emph{without
frequency tuning}, keeps the frequencies $f$ fixed at the qubit frequencies
$f_i$. The optimization process of the first (second) set is shown on the left
(right) panels in \figaref{fig:optimizationxpih}{fig:optimizationcnot}. Note
that the relation between the two different scales on the $x$ axis, i.e., the
number of objective function evaluations and  the number of Nelder--Mead
iterations, is not one-to-one since some Nelder--Mead operations require several
function evaluations.

\sfigref{fig:optimizationxpih} illustrates the optimization process of the
single-qubit GD pulses. As discussed in
\secref{sec:optimizatingsinglequbitgate}, the GD pulses are used to implement
$X_i^{\pi/2}=R_i^x(\pi/2)$ (the target matrix is defined in
\equref{eq:singlequbitrotationmultiplequbits}). At the first evaluation,
\figref{fig:optimizationxpih}(a) shows that the initial values mentioned below
\equref{eq:singlequbitpulseGD} already yield reasonable candidates, for which
the matrix distance $\Delta(M,R^x(\pi/2))$ defined by
\equref{eq:matrixdistanceobjective} is between $10^{-2}$ and $10^{-3}$. The
following three evaluations correspond to the initialization of the simplex
according to \equsref{eq:neldermeadInitialSimplex0}
{eq:neldermeadInitialSimplexN}. After that, the optimization begins its search
through the parameter space. As shown in \figref{fig:optimizationxpih}(b), the
minimization with frequency tuning (left panels) converges after less than 50
iterations for each transmon $i$. We see that in most iterations, the simplex
is contracted or reflected (cf.~\figref{fig:neldermeadsteps}). Note that
the DRAG coefficients $\beta_X$ do not change significantly for the gate set
with frequency tuning.

For the gate set without frequency tuning, i.e., when the drive frequency
$f=f_i$ is fixed, this behavior is completely different (see the right panels of
\figref{fig:optimizationxpih}): When the convergence criterion $\delta$ reaches
a local minimum after approximately 20--30 iterations, the DRAG coefficients
$\beta_X$ start to diverge significantly from their initial values. This means
that the optimization without frequency tuning tries to compensate for the
inability to change $f$ by drastically changing $\beta_X$. From
\equref{eq:singlequbitpulseGD}, we see that $\beta_X$ controls a pulse whose
phase is shifted by $\pi/2$. Thus it can be exploited to trigger effective
rotations about the $z$ axis. In this way, changing $\beta_X$ can have the same
effect as the slightly off-resonant driving included in the optimizations with
frequency tuning. Note also that the simplex is frequently expanded during the
drastic increase of $\abs{\beta_X}$ (cf.~\figref{fig:optimizationxpih}(b)).

Eventually, the optimizations with and without frequency tuning converge to
pulse parameters that implement the $X^{\pi/2}$ gate with a similar matrix
distance. To make further statements about  the quality of the different
single-qubit gate sets, we study the corresponding  gate metrics in
\secref{sec:gatemetricsresults}.

\sfigref{fig:optimizationcnot} shows the optimization of two-qubit CR2 pulses to
implement \textsc{CNOT} gates between each pair of qubits with a corresponding
resonator (cf.~\figref{fig:ibm5edtopology}). The optimization process looks
similar to the single-qubit case shown in \figref{fig:optimizationxpih} except
that more parameters need to be optimized. Furthermore, the time evolution due
to the CR2 pulse is more complicated
(cf.~\figref{fig:twoqubitpulseblochevolution}(e) and (f)). As shown in
\figref{fig:optimizationcnot}(b), this is reflected by  the fact that the
convergence criterion $\delta$ goes a more convoluted path,  on which the
Nelder--Mead optimization occasionally needs to do the \texttt{SHRINK}
operation illustrated in \figref{fig:neldermeadsteps}(d).

For some gates, such as $\textsc{CNOT}_{34}$ with frequency tuning, the distance
$\Delta(M,\textsc{CNOT})$ shown in \figref{fig:optimizationcnot}(a) (black line on
the left panels) does not turn out much better than the initial point. This
suggests that in this case, the theory outlined in
\secref{sec:optimizatingtwoqubitgate} already provides good pulse parameters.
However, for $\textsc{CNOT}_{32}$ and $\textsc{CNOT}_{40}$ with frequency tuning
(red and purple lines on the left panels), the optimization finds that making
the pulses much shorter (by reducing $T_{\mathrm{CR}}$)  and stronger (by
increasing $\Omega_{\mathrm{CR}}$) yields better gates. For the gates without
frequency tuning shown on the right panels of  \figref{fig:optimizationcnot}(a),
we again see the effect observed for the single-qubit pulses, namely that the
DRAG coefficients $(\beta_X^\pi)_C$ are modified drastically by the optimization
process to compensate for the fixed drive frequency.

As with the single-qubit gates, we study the quality of the resulting two-qubit
gates in more detail by analyzing various gate error rates in
\secref{sec:gatemetricsresults}. The metrics of the particular two-qubit gates
discussed here are presented in \tabref{tab:ibm5edgatemetrics}.

The result of the \texttt{optimizer} module is a set of parameters defining
each elementary quantum gate pulse for a system. This serves as input
for the \texttt{compiler} module discussed in the following section.

\section{Compiling quantum circuits}
\label{sec:compiler}

The \texttt{compiler} module is the last component of the simulation toolkit
introduced in \secref{sec:simulationsoftware}. It takes as input a specification
of the elementary pulses (as produced by \texttt{optimizer}) and a quantum
circuit to compile. The circuit can be in one of several file formats such as
OpenQASM \cite{Cross2017openqasm2} or the JUQCS instruction set
\cite{DeRaedt2018MassivelyParallel}. The output is a sequence of pulse
information describing the time-dependent functions in the model Hamiltonian
given by \equsref{eq:Htotal}{eq:HCC}. This is used by the transmon simulator to
perform the simulation.

In this section, we discuss an example of the compilation process. The model
system is the two-transmon system defined in \secref{sec:transmonmodelibm2gst}. We
consider the compilation of the simple circuit shown in
\figref{fig:compilerexamplecircuit}. It contains one single-qubit gate and one
two-qubit gate and creates the  maximally entangled state
$(\ket{00}+\ket{11})/\sqrt 2$.

To compile the circuit into a sequence of pulses, \texttt{compiler} needs a
specification of the elementary pulse parameters produced by \texttt{optimizer}.
For the system of interest, the input file is given in
Listing~\ref{code:compilergatepulses}. It represents a subset of the full set
of pulse parameters given in \tabref{tab:deviceibm2gstPulseParametersGD} and \tabref{tab:deviceibm2gstPulseParametersCR}
in \appref{app:pulseparameters}.
The parameters of the single-qubit GD pulses are defined by
\equref{eq:singlequbitpulseGD}. The parameters of the two-qubit CR2 pulse are
defined in \secref{sec:singlequbitGDpulse} (see the text below
\equref{eq:twoqubitpulseCR2equationMoreSophisticated}).

The first step to translate a circuit into a sequence of pulses is
to express all single-qubit gates in terms of the elementary
single-qubit \textsc{U} gates defined in
\equsref{eq:singlequbitU1}{eq:singlequbitU3}.
In the present case, this means that we need to write the gate \texttt{-Y} as
\begin{align}
  \label{eq:compilerRewriteYGateAsU}
  \Qcircuit @C=0em @R=.7em {& \gate{\texttt{-Y}} & \qw}\,= \textsc{U2}(0,0) = Z^{\pi/2}\,X^{\pi/2}\,Z^{-\pi/2}.
\end{align}
The first gate acting on the qubits is $Z^{-\pi/2}$. Since $Z^\vartheta$
gates are implemented by means of the VZ gate defined in
\secref{sec:singlequbitVZgate}, it only affects the VZ phase of the following
pulses. The next gate is $X^{\pi/2}$, which is implemented by $\mathrm{GD}^{\pi/2}(0)$.
According to \equref{eq:singlequbitpulseGDruleVZ}, we have
\begin{align}
  \label{eq:compilerexamplesingleYgate}
  Z^{\pi/2}\,\mathrm{GD}^{\pi/2}(0)\,Z^{-\pi/2}\ket\Psi
  = Z^{\pi/2}\,Z^{-\pi/2}\,\mathrm{GD}^{\pi/2}(\pi/2)\ket\Psi
  = \mathrm{GD}^{\pi/2}(\pi/2)\ket\Psi,
\end{align}
so the first pulse resulting from the compilation is
$\mathrm{GD}^{\pi/2}(\pi/2)$. It is specified in line 2 of
Listing~\ref{code:compilerpulses}: a pulse for the first $\SI{83}{ns}$ with
phase $\pi/2$ and Gaussian envelope $\Omega_G(t)$ (see
\equref{eq:gaussianpulse}) defined by its duration \texttt{T}, amplitude
\texttt{A}, and width \texttt{sigma}.

\begin{figure}[t]
  \centering
  \[
    \Qcircuit @C=.8em @R=.7em {
      &\lstick{\ket{0}}&\gate{\texttt{-Y}}&\ctrl{1}&\qw\\
      &\lstick{\ket{0}}&\qw               &\targ   &\qw\\
    }
  \]
  \caption{Circuit diagram for an example of the process performed  by the
  \texttt{compiler} module. See \tabref{tab:elementarygateset} in
  \appref{app:gateset} for a definition of the circuit elements.}
  \label{fig:compilerexamplecircuit}
\end{figure}

\begin{lstfloat}[t]
  \caption{Optimized pulse parameters for the system defined in \secref{sec:transmonmodelibm2gst}}
\begin{lstlisting}[style=myconfigstyle,label=code:compilergatepulses]
#name    type f      TX OmegaX   betaX
xpih-0   GD   5.3463 83 0.002221 0.2309
xpih-1   GD   5.1167 83 0.002269 0.2891

#name    type fiC    fiT    TCR      OmegaCR TX OmegaXpiC betaXpiC OmegaXpihT betaXpihT
cnot-0-1 CR2  5.3463 5.1167 102.9746 0.01111 83 0.004444  0.2193   0.002269   0.2891
\end{lstlisting}
\end{lstfloat}

\begin{lstfloat}[b]
  \caption{Compiled pulse information for the circuit in \figref{fig:compilerexamplecircuit}}
\begin{lstlisting}[style=myconfigstyle,label=code:compilerpulses]
#i  Tstart    Tend      f       phase    pulse      T        A          sigma (Trise)
0   0         83        5.3463  1.57080  gauss      83       0.0022212  20.75
0   0         83        5.3463  3.14159  gaussdot   83       0.0005129  20.75
0   83        166       5.3463  0        gauss      83       0.0044440  20.75
0   83        166       5.3463  1.57080  gaussdot   83       0.0009744  20.75
1   83        166       5.1167  0        gauss      83       0.0022686  20.75
1   83        166       5.1167  1.57080  gaussdot   83       0.0006558  20.75
0   166       298.975   5.1167  0        gaussflat  132.975  0.0111083  5      15
0   298.975   381.975   5.3463  1.57080  gauss      83       0.0044440  20.75
0   298.975   381.975   5.3463  3.14159  gaussdot   83       0.0009744  20.75
0   381.975   514.950   5.1167  3.14159  gaussflat  132.975  0.0111083  5      15
\end{lstlisting}
\end{lstfloat}

From the second term in the definition of the GD pulse in
\equref{eq:singlequbitpulseGD},  we obtain line 3 of
Listing~\ref{code:compilerpulses}: a DRAG pulse at the same time as the first
pulse, characterized by a phase of $\pi$, the identifier \texttt{gaussdot}, and
a much smaller amplitude \texttt{A} than the main Gaussian pulse. Since the $Z$ gates in
\equref{eq:compilerexamplesingleYgate} cancel, no VZ phase is carried over to
the following pulses.

The next gate in \figref{fig:compilerexamplecircuit} is a two-qubit
\textsc{CNOT} gate which is implemented in terms of a CR2 pulse shown in
\figref{fig:crossresonancepulses}(c). As specified in
\figref{fig:twoqubitpulseruleVZ}(c), this requires a $\mathrm{GD}^{\pi}(0)$
pulse on qubit 0 (defined in lines 4 and 5) and a $\mathrm{GD}^{\pi/2}(0)$ pulse
on qubit 1 (defined in lines 6 and 7). Since no VZ phase was carried over from
the previous pulses, we have $\vartheta_C = \vartheta_T = 0$. In line 8, we find
the first flat-topped Gaussian with phase $0$ and a duration of
$T_{\mathrm{CR}}+2T_{\mathrm{rise}}$ with $T_{\mathrm{rise}}=\SI{15}{ns}$. After
another $\mathrm{GD}^{\pi}(0)$ pulse on qubit 0,  the CR2 pulse is finalized by
the second flat-topped Gaussian defined in line 11. The echo scheme implemented
by the CR2 pulse is reflected by the phase difference of $\pi$ between line 8
and line 11. This is the full information needed to describe the CR2 pulse
visualized in \figref{fig:twoqubitpulseblochevolution}(e) and (f).

As the last line of Listing~\ref{code:compilerpulses} shows, the full pulse
sequence  takes $\SI{514.950}{ns}$. This is the time for which \texttt{solver}
needs to compute the time evolution to simulate the circuit shown in
\figref{fig:compilerexamplecircuit}.

\section{Alternative gate optimization techniques}
\label{sec:alterativegateoptimizationtechniques}

The goal of obtaining the best pulse parameters to implement a desired quantum
gate can be addressed by many different optimization techniques. In
\secref{sec:neldermead}, we have discussed why a Nelder--Mead optimization is a
reasonable approach for the present case, in which the evaluation of the
objective function by simulating the time evolution is the most expensive step.

However, given the vast number of optimization and machine learning
techniques, one may pursue other approaches in the hope of finding better pulse
parameters or completely new pulse shapes. Such efforts belong to the field
of quantum control and are often addressed by applying the techniques directly
to the experiment. Especially the popularity of deep reinforcement learning
(DeepRL) \cite{Goodfellow2016DeepLearning}, which has stemmed from the recent
successes in playing Atari  games \cite{Mnih2015DeepRLDQNAtari} or board games
such as Go \cite{Silver2016DeepRLAlphaGo}, may inspire efforts to apply the
framework to quantum control. For this reason, a few research groups have
recently implemented strategies to apply DeepRL to control optimization
\cite{Niu2019DeepRLQuantumControlPulsesPiecewiseConstant,
An2019DeepRLQuantumGateControl} (see also
\cite{Palittapongarnpim2017QuantumControl}). It would be interesting to
study such methods to obtain completely new pulses, and to test them
using both the transmon simulator and the real processor by means of the
recently released OpenPulse interface \cite{McKay2018OpenPulse}.

For the purpose of the present work, however, we choose not to optimize the
pulses further. One reason is that the pulse parameters found by the
Nelder--Mead method are in the same range that is used in experiments, so the
resulting pulses resemble the experimental approach. In fact, as we shall see in
the following chapter, the pulses often exhibit a much better performance than
their experimental equivalents. Hence we leave the endeavor of studying more
optimization methods for pulses in simulated transmon quantum computers for
future work.

\section{Conclusions}
\label{sec:chapter5conclusions}

In this chapter, we studied how time-dependent pulses can be used and optimized
to implement quantum gates. Two results obtained in this analysis are
noteworthy: First, the third-order perturbative results for the CR interaction
strengths given by
\equaref{eq:twoqubitpulseCRtheoryIX}{eq:twoqubitpulseCRtheoryZX} are useful for
obtaining initial pulse parameters, but the predictions quickly diverge from the
actual interaction strengths for larger CR drive amplitudes (see
\figref{fig:crossresonanceamplitudesscan}). Secondly, when the drive frequencies
are not tuned during the pulse optimization, the DRAG coefficients $\beta_X$ can
be (mis)used to compensate for phase errors such that equally good quantum gates
are still possible (cf.~the fourth row in
\figref{fig:optimizationxpih}(a)).

However, it is important to realize that, because of the presence of higher
transmon states in the time evolution of the full system, the projection of the
full time-evolution operator on the computational subspace (see
\equref{eq:MprojectionUtotal}) is inherently non-unitary.  Therefore, even in
theory, it is impossible to realize perfect quantum gates on the computational
subspace (unless certain resonance conditions are met exactly, which is
practically impossible). This implies that bare quantum gate implementations are
always faulty. The question is whether the associated error rates can still be
made small enough to allow for a reasonable operation of the devices. This
question is studied in the following chapters.

%% file: figs/neldermead.pdf_tex
%% Accompanies image file 'neldermead.pdf' (pdf, eps, ps)
\begingroup%
  \makeatletter%
  \providecommand\color[2][]{%
    \errmessage{(Inkscape) Color is used for the text in Inkscape, but the package 'color.sty' is not loaded}%
    \renewcommand\color[2][]{}%
  }%
  \providecommand\transparent[1]{%
    \errmessage{(Inkscape) Transparency is used (non-zero) for the text in Inkscape, but the package 'transparent.sty' is not loaded}%
    \renewcommand\transparent[1]{}%
  }%
  \providecommand\rotatebox[2]{#2}%
  \newcommand*\fsize{\dimexpr\f@size pt\relax}%
  \newcommand*\lineheight[1]{\fontsize{\fsize}{#1\fsize}\selectfont}%
  \ifx\svgwidth\undefined%
    \setlength{\unitlength}{444.16487566bp}%
    \ifx\svgscale\undefined%
      \relax%
    \else%
      \setlength{\unitlength}{\unitlength * \real{\svgscale}}%
    \fi%
  \else%
    \setlength{\unitlength}{\svgwidth}%
  \fi%
  \global\let\svgwidth\undefined%
  \global\let\svgscale\undefined%
  \makeatother%
  \begin{picture}(1,0.45786233)%
    \lineheight{1}%
    \setlength\tabcolsep{0pt}%
    \put(0,0){\includegraphics[width=\unitlength,page=1]{neldermead.pdf}}%
    \put(0.82065439,0.00324329){\color[rgb]{0,0,0}\makebox(0,0)[lt]{\lineheight{1.25}\smash{\begin{tabular}[t]{l}$\vec x_{\mathrm{min}}$\end{tabular}}}}%
    \put(0,0){\includegraphics[width=\unitlength,page=2]{neldermead.pdf}}%
    \put(0.00844272,0.32615441){\color[rgb]{0,0,0}\makebox(0,0)[lt]{\lineheight{1.25}\smash{\begin{tabular}[t]{l}$\vec x_i$\end{tabular}}}}%
    \put(0.3107087,0.32575863){\color[rgb]{0,0,0}\makebox(0,0)[lt]{\lineheight{1.25}\smash{\begin{tabular}[t]{l}$\vec x_{\mathrm{refl}}$\end{tabular}}}}%
    \put(0,0){\includegraphics[width=\unitlength,page=3]{neldermead.pdf}}%
    \put(0.49728257,0.32581606){\color[rgb]{0,0,0}\makebox(0,0)[lt]{\lineheight{1.25}\smash{\begin{tabular}[t]{l}$\vec x_i$\end{tabular}}}}%
    \put(0.94223109,0.32407006){\color[rgb]{0,0,0}\makebox(0,0)[lt]{\lineheight{1.25}\smash{\begin{tabular}[t]{l}$\vec x_{\mathrm{expd}}$\end{tabular}}}}%
    \put(0,0){\includegraphics[width=\unitlength,page=4]{neldermead.pdf}}%
    \put(0.00844322,0.08975551){\color[rgb]{0,0,0}\makebox(0,0)[lt]{\lineheight{1.25}\smash{\begin{tabular}[t]{l}$\vec x_i$\end{tabular}}}}%
    \put(0.0928842,0.08598287){\color[rgb]{0,0,0}\makebox(0,0)[lt]{\lineheight{1.25}\smash{\begin{tabular}[t]{l}$\vec x_{\mathrm{cont}}$\end{tabular}}}}%
    \put(0,0){\includegraphics[width=\unitlength,page=5]{neldermead.pdf}}%
  \end{picture}%
\endgroup%

%% file: chap6.tex
\chapter{Errors in quantum gates}
\label{cha:gateerrors}

The goal of this chapter is to characterize the optimized quantum gate pulses
and compare their performance to experiments. In general, neither the pulses
used in experiments  nor the optimized pulses used for the transmon simulator
can implement perfect quantum gates (see \secref{sec:chapter5conclusions}).  The
advantage of the simulator, though, is that we have  full access to the complex
coefficients of the quantum state. This means  that a much deeper analysis of
the intrinsic errors can be performed.

Note that all errors observed in the simulation are inherently part of the
unitary time evolution of the full transmon-resonator system. However, this does
not mean that they have to be unitary maps on the computational subspace (such
as systematic over- or underrotations). Often, unitary maps on a large system
are not unitary on a smaller part of the system (see also
\secref{sec:quantumoperations}).

We study quantum gate errors in four complementary ways:
\begin{enumerate}
  \item by evaluating common gate metrics such as the fidelity
  \cite{nielsen2002gatefidelity}, the diamond  distance
  \cite{kitaev1997diamondnorm}, and the unitarity \cite{Wallman2015unitarity};
  \item by analyzing repeated gate applications;
  \item by performing gate set tomography using the black box model of a quantum
  computer;
  \item by assessing the performance of actual quantum algorithms.
\end{enumerate}
Results for (a) and (b) are presented in
\secaref{sec:gatemetrics}{sec:repeatedgates}, respectively. These findings can
be directly compared to experiments on the IBM Q Experience
\cite{ibmquantumexperience2016}. \ssecref{sec:gatesettomography} contains
results for (c), i.e., an extensive gate set tomography of the simulated
two-transmon device. Real quantum algorithms (d) are partially tested in this
chapter (such as the quantum Fourier transform  in \secref{sec:repeatedgates} or
the circuits for gate set tomography in \secref{sec:gatesettomography}), but
this is mainly the topic of the following chapter. Some of the results presented
in \secaref{sec:gatemetrics}{sec:repeatedgates} of this chapter have been
published in \cite{Willsch2017GateErrorAnalysis}.

\clearpage
\section{Evaluation of gate metrics}
\label{sec:gatemetrics}

The performance of an implementation for a certain quantum gate is often
measured in terms of various gate metrics. Such metrics are meant to be single
numbers representing the degree of success or failure to which the quantum gate
has been implemented. Some of these, such as the diamond distance, are
sufficient in the mathematical sense, meaning that a value of 0 implies a
perfect implementation of the gate. Of course, in an experiment, it is not
possible to unambiguously prove a perfect implementation, simply because a
particular quantity cannot be estimated with zero error in practice.
Nevertheless, a good value may still inspire confidence in the underlying
implementation.

Gate metrics are defined in terms of quantum operations, i.e., completely
positive linear maps on the space of density matrices on the computational
subspace $\mathcal H_{2^n}$ (see \secref{sec:quantumoperations}). We define two
particular quantum operations,
\begin{subequations}
  \begin{align}
    \label{eq:gatemetricsGideal}
    \mathcal G_{id}(\ketbra\psi\psi) &= U\ketbra\psi\psi U^\dagger,\\
    \label{eq:gatemetricsGactual}
    \mathcal G_{ac}(\ketbra\psi\psi) &= M\ketbra\psi\psi M^\dagger.
  \end{align}
\end{subequations}
Here, $\mathcal G_{id}$ denotes the \emph{ideal} quantum operation, where $U$ is the
intended unitary quantum gate (cf.~\secref{sec:quantumgates}), and $\mathcal
G_{ac}$ denotes the \emph{actual} operation performed on the computational
subspace. The matrix $M$ in \equref{eq:gatemetricsGactual} is the transformation
of the computational subspace obtained after applying the quantum gate pulse to
the full system. Formally, $M$ is given by the projection in
\equref{eq:MprojectionUtotal},  i.e., $M=P_{\mathcal H_{2^n}} \mathcal U(T,0)
P_{\mathcal H_{2^n}}$, where $\mathcal U(T,0)$ is the time evolution operator
resulting from the simulation (potentially expressed in a rotating frame according to \equref{eq:rotatingframe}), and $P_{\mathcal H_{2^n}}$ is the projector onto
the computational subspace. Note that $n$ denotes the number of qubits involved
in the quantum gate $U$, which may be smaller than the total number of qubits.

We construct the matrix $M$ in \equref{eq:gatemetricsGactual} using the same
procedure that was used for the optimization of the pulses (see
\secref{sec:optimizatingpulseparameters}). Specifically, for single-qubit gates,
we have  $M\in\mathbb C^{2\times2}$, so constructing $M$ requires two
simulations of the time evolutions under the particular pulse (one for the
initial state $\ket 0$ and one for the initial state $\ket 1$). Similarly,
constructing $M$ for a two-qubit gate requires four simulations.

Because of the projection in \equref{eq:MprojectionUtotal}, $M$ is generally not
a unitary matrix. Nevertheless, the quantum operations defined in
\equaref{eq:gatemetricsGideal}{eq:gatemetricsGactual} are completely positive
maps. This can be seen since $\braket{\phi|(\mathcal
G_{ac}\otimes\mathds1)(A)|\phi} = \braket{\phi'|A|\phi'}\ge0$ for all positive
operators $A$ (using $\ket{\phi'}=(M\otimes\mathds1)^\dagger\ket{\phi}$). We
remark that an alternative definition using $M^{-1}$ instead of $M^\dagger$ in
\equref{eq:gatemetricsGactual} would not preserve Hermiticity, so the computed
gate metrics would not be real numbers.

\subsection{Average gate fidelity}

The average gate fidelity between the quantum operations defined in
\equaref{eq:gatemetricsGideal}{eq:gatemetricsGactual} is given by
 \cite{horodecki1999fidelity, nielsen2002gatefidelity}
\begin{align}
  \label{eq:gatemetricsaveragegatefidelity}
  F_{\mathrm{avg}}
  &= \int \mathrm{d}\!\ket{\psi} \bra{\psi} \mathcal G_{ac}(\mathcal G_{id}^{-1}(\ketbra\psi\psi)) \ket{\psi}
  ,
  %= \int \mathrm{d}\!\ket{\psi} \abs{\bra{\psi} MU^\dagger\ket{\psi}}^2,
\end{align}
where the integral is taken over normalized pure states. In practice,
\equref{eq:gatemetricsaveragegatefidelity} means that the overlap of a state
with itself after applying $\mathcal G_{ac}$ and $\mathcal G_{id}^{-1}$ is
averaged over random pure states from the unit sphere
\cite{bengtsson2006geometryofquantumstates}, yielding an average measure of
agreement between the ideal and the actual gate operation.

Note that some authors \cite{Bongioanni2010TomographyTraceDecreasingMaps,
Sekino2014QuantumInformationDivision} use a different formula for the fidelity
in the case of trace-decreasing quantum operations. These formulas are effectively based
on trace-preserving extensions of the quantum operation. However, we decide to use
\equref{eq:gatemetricsaveragegatefidelity} also for trace-decreasing quantum operations.
The reason is that, if a quantum operation does not preserve the trace, it suffers from
leakage, which should be reflected accordingly by a reduced fidelity. Otherwise,
we could have quantum operations with leakage out of the computational subspace that still
attain fidelities unreasonably close to unity.

In practice, one could evaluate the average gate fidelity in
\equref{eq:gatemetricsaveragegatefidelity} by sampling the integrand
$\bra{\psi}\mathcal G_{ac}(\mathcal G_{id}^{-1}(\ketbra\psi\psi)) \ket{\psi} =
\abs{\bra{\psi} MU^\dagger\ket{\psi}}^2$ for a sufficiently large number of
random states $\ket\psi$ \cite{bengtsson2006geometryofquantumstates}.
This procedure is used for the
unitarity defined below.

For the fidelity, however, we can use an alternative closed-form expression proven in
\appref{app:prooffidelity},
\begin{align}
  \label{eq:gatemetricsaveragegatefidelityEvaluation}
  F_{\mathrm{avg}}
  &= \frac{\abs{\mathrm{Tr}(MU^\dagger)}^2 + \mathrm{Tr}(M^\dagger M)}{N(N+1)},
\end{align}
where $N=2^n$ denotes the dimension of the computational subspace. It can either
be derived by generalizing a well-known relation between the average gate
fidelity  and the entanglement fidelity \cite{horodecki1999fidelity,
nielsen2002gatefidelity} to the case of non-trace-preserving quantum operations (see
\secref{sec:prooffidelityalgebraic}), or by evaluating the integral in
\equref{eq:gatemetricsaveragegatefidelity} directly (see
\secref{sec:prooffidelityanalytic}). \sequref{eq:gatemetricsaveragegatefidelityEvaluation} can also be found in
\cite{Pedersen2007FidelityGeneralQuantumOperations}.

In experiments, the average gate fidelity is typically estimated using
randomized benchmarking (RB) \cite{emerson2005randomunitaryoperatorstwirling,
Knill2008randomizedbenchmarking, Magesan2012RBandDiamondNorm}. However, such
results must be treated with caution as it has been shown that RB cannot measure
$F_{\mathrm{avg}}$ \cite{Qi2019RBdoesnotmeasurefidelity}, but typically produces
numbers that overestimate the performance of the gates
\cite{proctor2017RandomizedBenchmarking, Lin2019GateMetricsGaugeFreedom}.

\subsection{Diamond distance}
\label{sec:diamonddistance}

The diamond distance is a mathematical construct that measures the difference
between quantum operations. It was introduced in \cite{kitaev1997diamondnorm,
aharonov2008thresholdtheorem} and is the relevant quantity for many
threshold theorems in the theory of fault-tolerant quantum computation
\cite{Terhal2005ftqcForLocalNonmarkovianNoise, aliferis2006extendedrectangles,
aliferis2007FTQCwithLeakage, aharonov2008thresholdtheorem,
ng2009FTQCversusGaussianNoise, Sanders2016ThresholdTheorem}.  Unfortunately,
evaluating the diamond distance in practice is nontrivial
\cite{hendrickx2010DiamondNormNPhard, watrous2018theoryofQI}. Furthermore, the
relation to more accessible quantities such as the average gate fidelity given
by \equref{eq:gatemetricsaveragegatefidelity} is not straightforward
\cite{Wallman2014RBwithConfidence, Sanders2016ThresholdTheorem}. Therefore,
we look at these issues in more detail.

\subsubsection{Definition of the diamond distance}

The diamond distance between the two quantum operations $\mathcal G_{id}$ and $\mathcal G_{ac}$ is defined as
\begin{align}
  \label{eq:gatemetricsdiamonddistance}
  \eta_\Diamond &= \frac{1}{2} \left\|\mathcal G_{ac}\circ\mathcal G_{id}^{-1} - \mathds1\right\|_\Diamond,
\end{align}
where $\|\cdot\|_\Diamond$ denotes the diamond norm. For a general superoperator
$\mathcal T$, the diamond norm is defined as \cite{kitaev1997diamondnorm, Aharonov1998DiamondNorm}
\begin{align}
  \label{eq:gatemetricsdiamondnorm}
  \|\mathcal T\|_\Diamond\,
  = \sup_{X\neq0} \frac{\|(\mathcal T\otimes\mathds1)(X)\|_{\mathrm{Tr}}}{\|X\|_{\mathrm{Tr}}},
  %= \sup_{\|X\|_{\mathrm{Tr}}=1} \|(\mathcal T\otimes\mathds1)(X)\|_\mathrm{Tr},
\end{align}
where $\|\cdot\|_{\mathrm{Tr}}$ denotes the trace norm
defined as $\|X\|_{\mathrm{Tr}}~=\mathrm{Tr}\sqrt{X^\dagger X}$, i.e., the sum
of the singular values of $X$. The identity operator $\mathds1$ in \equref{eq:gatemetricsdiamondnorm}
acts on a space that is at least as large as the space of matrices that $\mathcal T$
acts on. Interestingly, one can show that it does not need to be larger \cite{Gilchrist2005fidelities,
Johnston2009ComputingStabilizedNormsQC}. This means that the supremum can be computed by
extending the Hilbert space with another Hilbert space of the same dimensionality.

The fact that the diamond norm is the same even if the Hilbert space is
extended by a much larger space is a stability property that is sometimes
expressed in an alternative definition of the diamond norm (see e.g.~\cite{Sanders2016ThresholdTheorem}),
\begin{align}
  \label{eq:gatemetricsdiamondnormSupSup}
  \left\|\mathcal T\right\|_\Diamond =
  \sup_{\mathcal H'}
  \sup_{\rho\in\mathrm{dens}(\mathcal H\otimes\mathcal H')}
  \left\| (\mathcal T\otimes\mathds1)(\rho)\right\|_{\mathrm{Tr}},
\end{align}
where $\mathcal H'$ is an arbitrary
ancillary Hilbert space and $\mathrm{dens}(\mathcal H\otimes\mathcal H')$ denotes
the set of all density matrices  on the joint Hilbert space $\mathcal
H\otimes\mathcal H'$.
\sequref{eq:gatemetricsdiamondnormSupSup} means that the diamond distance
$\eta_\Diamond$ corresponds to the worst-case error, since the trace
norm is maximized over all ancillary Hilbert spaces $\mathcal H'$ added to
$\mathcal H$ and any input density matrix $\rho$ on this joint space. The diamond
distance is thus a very strong measure, in the sense that a small value for
this quantity is not easy to achieve for implementations of quantum gates.
Note that the diamond distance can be interpreted as
the distance between ideal and actual probability distributions,
since the trace norm is directly related to the total variation
distance \cite{Gilchrist2005fidelities, Sanders2016ThresholdTheorem}.

We note that in the mathematical literature, the diamond norm is also known as the
completely bounded trace norm \cite{paulsen2003completelyboundedmapsdiamond}.
Many properties of this norm in relation to its use in quantum information
can be found in \cite{watrous2018theoryofQI}.

\subsubsection{Computation of the diamond distance}

We consider two ways to compute the diamond distance given by
\equref{eq:gatemetricsdiamonddistance}. The first method is based on a direct
evaluation of the definition given in \equref{eq:gatemetricsdiamondnorm}, so it
includes a maximization problem. The second method makes use of a minimization
algorithm presented in \cite{Johnston2009ComputingStabilizedNormsQC}. If both
methods (i.e., the minimization and the maximization) yield the same numeric
quantity up to a reasonable number of digits, we have estimated the true diamond
distance with sufficient accuracy.

Historically, the first deep connection between the evaluation of the diamond
distance between two quantum channels and a convex optimization problem was made
in \cite{Gilchrist2005fidelities}. Apart from the methods we pursue in this
work, similar algorithms have been mentioned in
\cite{Zarikian2006AlternatingProjectionAlgorithmDiamondNorm,
BenAroya2010ComplexityApproximatingDiamondNorm}, and a class of algorithms based
on semidefinite programming has been given in
\cite{Watrous2009diamondnormAlgorithm, Watrous2012diamondnormAlgorithm,
watrous2018theoryofQI}, including proofs of their efficiency.

\subsubsection{Maximization algorithm}

The definition of the diamond norm given by \equref{eq:gatemetricsdiamondnorm}
is valid for a general linear map $\mathcal T$. However, in almost all practical
situations, $\mathcal T$ is at least Hermiticity-preserving, i.e.,
$\mathcal T(\rho)^\dagger = \mathcal T(\rho^\dagger)$.
This is also true in the present case, in which
\begin{align}
  \label{eq:gatemetricsdiamondnormT}
  \mathcal T(\rho) = \mathcal G_{ac}(\mathcal G_{id}(\rho)) - \rho = W \rho W^\dagger - \rho,
\end{align}
where $W=MU^\dagger$
(cf.~\equaref{eq:gatemetricsGideal}{eq:gatemetricsGactual}). In this case, the
supremum in \equref{eq:gatemetricsdiamondnorm} is attained by a pure state
$X=\ketbra x x$ with $\ket x\in\mathcal H_{2^n}\otimes\mathcal H_{2^n}$
\cite{watrous2018theoryofQI}. Since $\|X\|_{\mathrm{Tr}}~=\braket{x|x}=1$,
\equref{eq:gatemetricsdiamonddistance} becomes
\begin{align}
  \label{eq:gatemetricsdiamondnormPureState}
  \eta_\Diamond = \frac 1 2 \max_{\ket{x}} \left\|(W\otimes\mathds1)\ketbra x x (W^\dagger\otimes\mathds1)-\ketbra x x\right\|_{\mathrm{Tr}}.
\end{align}
This expression requires the evaluation of the trace norm of a rank-2 matrix of
the form $R=\alpha\ketbra v v - \beta\ketbra x x$, where $\alpha,\beta\ge0$ and
$\ket v$  is a normalized pure state. For the trace norm $\|R\|_{\mathrm{Tr}}$ of
such a matrix, one can derive a closed-form expression:
\begin{align}
  \label{eq:gatemetricsdiamondnormTraceDistanceRank2}
  \|\alpha\ketbra v v - \beta\ketbra x x \|_{\mathrm{Tr}}\: = \sqrt{(\alpha+\beta)^2 - 4\alpha\beta\abs{\braket{v|x}}^2}.
\end{align}
This can be shown by noting that a rank-2 matrix has at most two non-zero
singular values. Since the rank-2 matrix $R$ is Hermitian (and thus normal), its
singular values are the absolute values of its eigenvalues $\mu_{\pm}$, for
which a short calculation yields
\begin{align}
  \label{eq:gatemetricsdiamondnormTraceDistanceRank2Eigenvalues}
  \mu_{\pm} = \frac{\alpha-\beta}2\pm\frac 1 2\sqrt{(\alpha+\beta)^2-4\alpha\beta\abs{\braket{v|x}}^2}.
\end{align}
As both eigenvalues $\mu_{\pm}$ have opposite signs (meaning $\abs{\mu_\pm} =
\pm\mu_\pm$ since $\alpha,\beta>0$ and $0\le\abs{\braket{v|x}}\le1$), we obtain the trace norm
$\|R\|_{\mathrm{Tr}}\:\,=\abs{\mu_+}+\abs{\mu_-}=\mu_+-\mu_-$.

To apply
this result to \equref{eq:gatemetricsdiamondnormPureState}, we set
$\alpha=\braket{x|W^\dagger W\otimes\mathds1|x}\in[0,1]$, $\beta=1$, $\ket
v=(W\otimes\mathds1)\ket{x}/\sqrt\alpha$ and
$\abs{\braket{v|x}}^2=\abs{\braket{x|W\otimes\mathds1|x}}^2/\alpha$. Thus we
obtain
\begin{align}
  \label{eq:gatemetricsdiamondnormPureStateEvaluated}
  \eta_\Diamond = \frac 1 2 \max_{\ket{x}} \sqrt{(\braket{x|W^\dagger W\otimes\mathds1|x}+1)^2-4\abs{\braket{x|W\otimes\mathds1|x}}^2}.
\end{align}
For the one- and two-qubit quantum operations of interest,
this result describes a quadratic optimization problem that can
directly be solved on a computer \cite{boyd2004ConvexOptimization}.

Furthermore, we show in \appref{app:diamonddistanceunitary} that if both quantum
operations $\mathcal G_{ac}=M\cdot M^\dagger$ and $\mathcal
G_{id}=U\cdot U^\dagger$ are unitary, we can obtain an explicit result
from \equref{eq:gatemetricsdiamondnormPureStateEvaluated}. This yields an
elementary proof of the statements given in \cite{Aharonov1998DiamondNorm} and
\cite{Johnston2009ComputingStabilizedNormsQC}. In most situations considered in
the present work, however, $M$ is not exactly unitary due to leakage out of the
computational subspace (see below).

\clearpage
\subsubsection{Minimization algorithm}

For the second algorithm to compute the diamond norm $\|\mathcal T\|_\Diamond$,
we use a slightly modified version of the minimization algorithm  presented in
\cite{Johnston2009ComputingStabilizedNormsQC}. The algorithm is based on
minimizing over all generalized Kraus representations
\begin{align}
  \label{eq:gatemetricsGeneralizedChoiKrausRepresentation}
  \mathcal T(\rho) = \sum_l A_l \rho B_l,
\end{align}
where $A_l$ and $B_l$ are generalized Kraus operators of $\mathcal T$
(cf.~\equref{eq:krausrepresentation} in \secref{sec:quantumoperations}).
We have
\begin{align}
  \label{eq:gatemetricsdiamonddistanceQuantityToMinimize}
  \|\mathcal T\|_\Diamond\: = \inf_{A_l,B_l}
  \left\{\left\|\sum_l A_l^\dagger A_l\right\|_2^{1/2} \left\|\sum_l B_l B_l^\dagger\right\|_2^{1/2}\right\},
\end{align}
where $\|\cdot\|_2$ denotes the spectral norm (largest singular value). Note
that  this expression differs from the one given in
\cite{Johnston2009ComputingStabilizedNormsQC} by the position of the Hermitian
conjugate. The reason for this is that the completely bounded (spectral) norm computed in
\cite{Johnston2009ComputingStabilizedNormsQC} and the diamond norm computed here
are related to each other by replacing $A_l\mapsto A_l^\dagger$ and $B_l\mapsto
B_l^\dagger$.

To compute $\eta_\Diamond$ given by \equref{eq:gatemetricsdiamonddistance}, we
set $\mathcal T = \mathcal G_{ac}\circ\mathcal G_{id}^{-1} - \mathds1$.  We
obtain a generalized Kraus representation of $\mathcal T$ by inserting
\equaref{eq:gatemetricsGideal}{eq:gatemetricsGactual}:
\begin{align}
  \label{eq:gatemetricsdiamonddistanceA}
  \mathcal T(\rho)
  = \mathcal G_{ac}(\mathcal G_{id}^{-1}(\rho)) - \rho
  = MU^\dagger \rho UM^\dagger - \rho,
\end{align}
where we can identify $A_1=MU^\dagger, A_2 = \mathds1, B_1=UM^\dagger$, and
$B_2=-\mathds1$. We assume that both $(A_1, A_2)$ and $(B_1, B_2)$ are linearly
independent, since otherwise we would have $MU^\dagger=\gamma\mathds1$ for some $\gamma$ such that
we could directly obtain $\eta_\Diamond=\abs{1-\abs{\gamma}^2}/2$ (which is
consistent with \equref{eq:gatemetricsdiamondnormPureStateEvaluated} for $W=\gamma\mathds1$).

All
other generalized Kraus representations of $\mathcal T$  are related to
$(A_1,A_2)$ and $(B_1,B_2)$ by an invertible complex $2\times2$ matrix
$S$. Thus, the infimum over $A_l$ and $B_l$ in
\equref{eq:gatemetricsdiamonddistanceQuantityToMinimize} can be written as an
infimum over all invertible $S\in\mathbb C^{2\times2}$ after replacing
$(A_1,A_2)^T\mapsto S^{-1}(A_1, A_2)^T$ and $(B_1,B_2)\mapsto (B_1,B_2)S$.
Combining this with \equref{eq:gatemetricsdiamonddistanceQuantityToMinimize}, we
obtain
\begin{align}
  \label{eq:gatemetricsdiamonddistanceComputation}
  \eta_\Diamond &= \frac1 2 \inf_S \left\{ \left\| \begin{pmatrix}
  UM^\dagger, & \mathds1
\end{pmatrix} S^{-\dagger} S^{-1} \begin{pmatrix}
  MU^\dagger \\
  \mathds1
\end{pmatrix}\right\|_2^{1/2} \left\| \begin{pmatrix}
  UM^{\dagger}, & -\mathds1
\end{pmatrix} S S^\dagger \begin{pmatrix}
  MU^\dagger \\
  -\mathds1
\end{pmatrix}\right\|_2^{1/2} \right\},
\end{align}
where the infimum is taken over all invertible $S\in\mathbb C^{2\times2}$.

In practice, we evaluate \equref{eq:gatemetricsdiamonddistanceComputation} by
sampling over $10^4$ random $S$ and then applying the minimization method
described in \secref{sec:neldermead} to the eight real coefficients of the
best $S$ sampled. We verified that this procedure,
together with the maximization method discussed above,
produces reliable values for $\eta_\Diamond$ in all relevant cases.

\clearpage
\subsubsection{Relation to the fidelity}

The relation between the average gate fidelity given by
\equref{eq:gatemetricsaveragegatefidelity} and the diamond distance given by
\equref{eq:gatemetricsdiamonddistance} is not obvious. As discussed in
\cite{Sanders2016ThresholdTheorem}, this has led some groups to make partly
unjustified claims about reaching fault-tolerance thresholds
\cite{chowGambetta2012fidelitiesandcoherence,  barends2014superconducting}. As a
consequence, the connection between fidelity and diamond distance has been
further studied in the literature \cite{Wallman2014RBwithConfidence,
Wallman2015ErrorRates, Kueng2016ComparingExperimentsToThreshold}. In these
references, two bounds relating both quantities have been proven. We have
\begin{align}
  \label{eq:gatemetricsdiamonddistanceBounds}
  \eta_\Diamond^{\mathrm{Pauli}}\le\eta_\Diamond\le\eta_\Diamond^{\mathrm{ub}},
\end{align}
where, for trace-preserving quantum operations,
\begin{subequations}
\begin{align}
  \label{eq:gatemetricsdiamonddistanceBoundPauliTP}
  \eta_\Diamond^{\mathrm{Pauli}} &= \frac{N+1}{N}(1-F_{\mathrm{avg}}),\\
  \label{eq:gatemetricsdiamonddistanceBoundUpper}
  \eta_\Diamond^{\mathrm{ub}} &= \sqrt{N(N+1)(1-F_{\mathrm{avg}})}.
\end{align}
\end{subequations}
For trace-decreasing quantum operations, however, $\eta_\Diamond$ can actually
be lower than $\eta_\Diamond^{\mathrm{Pauli}}$.  Therefore, we prove a new lower
bound, which also holds for trace-decreasing quantum operations, in
\appref{app:proofdiamondnormbound}. Applying the result in
\equref{eq:diamondnormBoundResult} to $\mathcal E = MU^\dagger\cdot UM^\dagger$
yields
\begin{align}
  \label{eq:gatemetricsdiamonddistanceBoundPauliLower}
  \eta_\Diamond^{\mathrm{lb}} &= \eta_\Diamond^{\mathrm{Pauli}}
  - \frac {N+2} {2N}\left(1-\frac{\mathrm{Tr}\,M^\dagger M} N\right),
\end{align}
which is a proper lower bound on $\eta_\Diamond$. Note that this bound coincides with
\equref{eq:gatemetricsdiamonddistanceBoundPauliTP} in the trace-preserving case
$\mathrm{Tr}\,M^\dagger M= N$.

The lower bound $\eta_\Diamond^{\mathrm{Pauli}}$ given by
\equref{eq:gatemetricsdiamonddistanceBoundPauliTP} is saturated if the actual
operation $\mathcal G_{ac}$ can be represented as a Pauli channel
\cite{Sanders2016ThresholdTheorem}. This means that $\mathcal G_{ac}$ has a
Kraus representation in which each Kraus operator is proportional to a tensor
product of  Pauli matrices. Such quantum operations are typically considered to
represent errors that are easy to correct. We shall see that for the
simulations considered in this work, we often find
$\eta_\Diamond\gg\eta_\Diamond^{\mathrm{Pauli}}$
(cf.~\tabref{tab:ibm2gstgatemetrics}).

The upper bound $\eta_\Diamond^{\mathrm{ub}}$ given by
\equref{eq:gatemetricsdiamonddistanceBoundUpper} grows exponentially with the
number of qubits, which is the reason for the fact that impressively high
fidelities do not generally imply a sufficiently small diamond distance
\cite{Sanders2016ThresholdTheorem}. Furthermore, the authors in
\cite{Sanders2016ThresholdTheorem} have shown that the upper bound is
asymptotically tight, which means there cannot be any bound on
$\eta_\Diamond$ in terms of $N$ and $F_{\mathrm{avg}}$ that scales
better than exponentially in the number of qubits.

\subsection{Unitarity}

Because of the projection required to obtain the transformation $M$ of the
computational subspace under a certain pulse, the actual operation $\mathcal G_{ac}$
given by \equref{eq:gatemetricsGactual} is often not unitary. This means that
$M^\dagger\neq M^{-1}$, implying that the quantum operation is not trace-preserving such
that we typically have $\mathrm{Tr}(\mathcal G_{ac}(\rho))<\mathrm{Tr}(\rho)$.
This is the mathematical manifestation of the fact that the systems under investigation
suffer from \emph{leakage} into non-computational states (see also the discussion
below \equref{eq:MprojectionUtotal}).

To quantify the effects of leakage for a particular quantum gate pulse,  a
quantity called \emph{unitarity} has been introduced \cite{Wallman2015unitarity,
wallman2017leakage}. It is defined as
\begin{align}
  \label{eq:gatemetricsunitarity}
  u &= \frac{N}{N-1} \int \mathrm{d}\!\ket{\psi}\,
  \mathrm{Tr}\!\left[ \mathcal G_{ac}'(\ketbra\psi\psi)^\dagger \mathcal G_{ac}'(\ketbra\psi\psi) \right],
\end{align}
where
\begin{align}
  \label{eq:gatemetricsUnitarityGacPrime}
  \mathcal G_{ac}'(\ketbra\psi\psi) = \mathcal G_{ac}(\ketbra\psi\psi-\mathds1/N) - \mathrm{Tr}\!\left[\frac{\mathcal G_{ac}(\ketbra\psi\psi-\mathds1/N)}{\sqrt{N}}\right]\mathds1.
\end{align}
The rationale behind this definition is that \equref{eq:gatemetricsunitarity}
corresponds to the average purity $\mathrm{Tr}(\rho^\dagger\rho)$ where $\rho$
is the output of the quantum operation acting on a pure state. The reason that the
quantum operation is $\mathcal G_{ac}'$ instead of $\mathcal G_{ac}$ is that otherwise,
one could define explicitly trace-decreasing or non-unital quantum operations with $u=1$
(see \cite{Wallman2015unitarity}). As for the average gate fidelity given by
\equref{eq:gatemetricsaveragegatefidelity}, the integral is taken over random
states from the unit sphere. Since we have no closed-form expression in this
case, we evaluate $u$ by sampling over $10^5$ random states. A typical procedure
to generate random states is given by setting
$\ket\psi=\sum_j(a_j+ib_j)\ket{j}$, where the coefficients $a_j$ and $b_j$ are
first drawn from a normal distribution and subsequently normalized
\cite{bengtsson2006geometryofquantumstates}.

As the definition of the unitarity $u$ given by \equref{eq:gatemetricsunitarity}
involves only the actual quantum operation $\mathcal G_{ac}$ and not the ideal
gate $\mathcal G_{id}$, the value $u=1$ does not imply equivalence between the
two operations. In that sense, it differs from other metrics such as the diamond
distance introduced in \secref{sec:diamonddistance}. Nevertheless, the unitarity
is a useful quantity to measure how incoherent the errors appear on the
computational subspace. Therefore, it is an interesting metric to study in the
present case, in which all errors are actually systematic and coherent on the
total transmon-resonator Hilbert space.

\subsection{Results}
\label{sec:gatemetricsresults}

We evaluate the average gate fidelity $F_{\mathrm{avg}}$ given by
\equref{eq:gatemetricsaveragegatefidelity}, the diamond distance $\eta_\Diamond$
given by \equref{eq:gatemetricsdiamonddistance}, and the unitarity $u$ given by
\equref{eq:gatemetricsunitarity} for all three transmon systems for which we
have optimized quantum gate pulses
(cf.~\secref{sec:optimizatingpulseparametersResults}). For the two-transmon
system, we additionally evaluate the bounds given in
\equsref{eq:gatemetricsdiamonddistanceBoundPauliTP}{eq:gatemetricsdiamonddistanceBoundPauliLower}.

\subsubsection{Two-transmon system}

The gate metrics for the two-transmon system defined in
\secref{sec:transmonmodelibm2gst} are given in \tabref{tab:ibm2gstgatemetrics}.
We see that the overall performance of the gates is reasonably good, in that the
fidelity $F_{\mathrm{avg}}$ and the unitarity $u$ are close to one, and the
diamond distance $\eta_\Diamond$ is close to zero. The fidelities are in the
same range as those reported in experiments for similar pulse schemes
\cite{kelly2014optimalcontrolrb, barends2014superconducting,
sheldon2015singlequbitfidelities, sheldon2016procedure, gambetta2015building}.
In fact, top fidelities reported in experiments are sometimes better; this is an
artifact of the RB procedure used to measure the fidelities: As shown in
\cite{proctor2017RandomizedBenchmarking}, the number reported by RB often
overestimates the performance of the gate. Indeed, we shall see below (cf., for
instance, \figref{fig:repeatedgatescnot}(a)) that in practice, the performance
of the pulses optimized for the transmon simulator is often much better.

For all gates (including the single-qubit gates), we have optimized the
transformation $M$ of the entire, four-dimensional computational subspace (see
\equref{eq:MprojectionUtotal}). In this respect, the two-transmon case differs
from the optimizations for the five-transmon systems discussed below, for which
only the relevant part of the computational subspace is considered.

\begin{table}
  \renewcommand{\arraystretch}{0.7}
  \caption{Gate metrics for the two-transmon system defined in \secref{sec:transmonmodelibm2gst}.
  The pulse types are defined in \secref{sec:optimizatingsinglequbitgate} and \secref{sec:optimizatingtwoqubitgate}
  (see \appref{app:pulseparameters} for the optimized pulse parameters).
  The label in the third column refers to the internal name used for the compiler
  (see Listing~\ref{code:compilergatepulses} in \secref{sec:compiler}).
  $\Delta$ is the distance objective given by \equref{eq:matrixdistanceobjective}.
  The reported gate metrics are the average gate fidelity $F_{\mathrm{avg}}$ (see \equref{eq:gatemetricsaveragegatefidelity}),
  the diamond distance $\eta_\Diamond$ (see \equref{eq:gatemetricsdiamonddistance}),
  and the unitarity $u$ (see \equref{eq:gatemetricsunitarity}).
  The bounds $\eta_\Diamond^{\mathrm{lb}}$, $\eta_\Diamond^{\mathrm{Pauli}}$, and $\eta_\Diamond^{\mathrm{ub}}$
  are given by \equsref{eq:gatemetricsdiamonddistanceBoundPauliTP}{eq:gatemetricsdiamonddistanceBoundPauliLower},
  respectively.}
  \centering
  \label{tab:ibm2gstgatemetrics}
\begin{tabular}{@{}lllccccccc@{}}
  \toprule
  Gate & Pulse & Label & $\Delta$ & $F_{\mathrm{avg}}$ & $\eta_\Diamond$ & $\eta_\Diamond^{\mathrm{lb}}$ & $\eta_\Diamond^{\mathrm{Pauli}}$ & $\eta_\Diamond^{\mathrm{ub}}$ & $u$\\
  \midrule
  $X^{\pi/2}_0$        & $\mathrm{GD}^{\pi/2}$ & \texttt{xpih-0}   & $2.2\e{-3}$ &      0.9946 &      0.027 & 0.003 &   0.007 &      0.33 &      0.990 \\
  $X^{\pi/2}_1$        & $\mathrm{GD}^{\pi/2}$ & \texttt{xpih-1}   & $2.3\e{-3}$ &      0.9942 &      0.028 & 0.003 &   0.007 &      0.34 &      0.989 \\
  $X^{\pi}_0$          & $\mathrm{GD}^{\pi}$   & \texttt{xpi-0}    & $1.3\e{-3}$ &      0.9949 &      0.020 & 0.003 &   0.006 &      0.32 &      0.990 \\
  $X^{\pi}_1$          & $\mathrm{GD}^{\pi}$   & \texttt{xpi-1}    & $1.5\e{-3}$ &      0.9943 &      0.023 & 0.003 &   0.007 &      0.34 &      0.989 \\
  \midrule
  $\textsc{CNOT}_{01}$ & CR1                   & \texttt{cnot-0-1} & $1.3\e{-3}$ &      0.9842 &      0.029 & 0.008 &   0.020 &      0.56 &      0.969 \\
  $\textsc{CNOT}_{10}$ & CR1                   & \texttt{cnot-1-0} & $2.3\e{-3}$ &      0.9951 &      0.033 & 0.003 &   0.006 &      0.31 &      0.991 \\
  \midrule
  $\textsc{CNOT}_{01}$ & CR2                   & \texttt{cnot-0-1} & $6.1\e{-3}$ &      0.9943 &      0.048 & 0.004 &   0.007 &      0.34 &      0.991 \\
  $\textsc{CNOT}_{10}$ & CR2                   & \texttt{cnot-1-0} & $5.6\e{-3}$ &      0.9947 &      0.048 & 0.003 &   0.007 &      0.32 &      0.992 \\
  \midrule
  $\textsc{CNOT}_{01}$ & CR4                   & \texttt{cnot-0-1} & $5.4\e{-3}$ &      0.9934 &      0.049 & 0.004 &   0.008 &      0.36 &      0.989 \\
  $\textsc{CNOT}_{10}$ & CR4                   & \texttt{cnot-1-0} & $4.5\e{-3}$ &      0.9946 &      0.044 & 0.003 &   0.007 &      0.33 &      0.991 \\
  \bottomrule
\end{tabular}
\end{table}

Note that a large fidelity does not always
correspond to a small diamond distance. Especially for the \textsc{CNOT} gates, we
observe similar fidelities as for the single-qubit gates but an
almost twice as large diamond distance (see also \cite{Sanders2016ThresholdTheorem}).
An extreme case is the CR1-type $\textsc{CNOT}_{01}$: The fidelity
$F_{\mathrm{avg}}=0.9842$
is much worse than for the other gates, whereas the diamond distance
$\eta_\Diamond=0.029$
has the best value found for all two-qubit gates.

A decrease in fidelity often corresponds to a decrease in unitarity. From
this, we conclude that leakage due to non-computational states in the transmons
and the resonator is the dominant
source of error for the pulses optimized for this system, even though the
pulse-shaping techniques DRAG \cite{motzoi2009drag} and VZ corrections \cite{McKay2016VZgate}
have been used.

The diamond distances given in \tabref{tab:ibm2gstgatemetrics} are always within
the bounds given in \equref{eq:gatemetricsdiamonddistanceBounds}. In most cases, we
also find $\eta_\Diamond\gg\eta_\Diamond^{\mathrm{Pauli}}$, suggesting that the
systematic errors included in the pulses are inherently different from the
simple Pauli-type errors (see \cite{Sanders2016ThresholdTheorem}).

\subsubsection{Small five-transmon system}

The gate metrics of the elementary pulses for the small five-transmon system
defined in \secref{sec:transmonmodelibm5} are given in
\tabref{tab:ibm5gatemetrics}. For the pulse optimizations, the matrices $M$ and
$U$ used in \equref{eq:matrixdistanceobjective} are $2\times2$ matrices for all
single-qubit gates $X_i^{\pi/2}$, and $4\times4$ matrices for all two-qubit
gates $\textsc{CNOT}_{ij}$. This is different from the two-transmon results
given in \tabref{tab:ibm2gstgatemetrics}, for which always the full
four-dimensional computational subspace was considered. Although this  does not
really affect the quality of the optimized pulses, it can be seen in the
objective function $\Delta$ used for the optimization (see
\tabref{tab:ibm5gatemetrics}), because it typically differs by about two orders
of magnitude between single-qubit and two-qubit pulses.

\begin{table}
  \renewcommand{\arraystretch}{0.7}
  \caption{Gate metrics for the small five-transmon system defined in \secref{sec:transmonmodelibm5}.
  The pulse types are defined in \secref{sec:optimizatingtwoqubitgate}
  (see \appref{app:pulseparameters} for the optimized pulse parameters).
  $\Delta$ is the distance objective given by \equref{eq:matrixdistanceobjective},
  $F_{\mathrm{avg}}$ is the average gate fidelity given by \equref{eq:gatemetricsaveragegatefidelity},
  $\eta_\Diamond$ is the diamond distance given by \equref{eq:gatemetricsdiamonddistance},
  and $u$ is the unitarity given by \equref{eq:gatemetricsunitarity}.}
  \centering
  \label{tab:ibm5gatemetrics}
\begin{tabular}{@{}lllcccc@{}}
  \toprule
  Gate & Pulse & Label & $\Delta$ & $F_{\mathrm{avg}}$ & $\eta_\Diamond$ & $u$ \\
  \midrule
  $X^{\pi/2}_0$ & $\mathrm{GD}^{\pi/2}$ & \texttt{xpih-0-withf} & $1.30\e{-4}$ & 0.9879 & 0.011 & 0.9759 \\
  $X^{\pi/2}_1$ & $\mathrm{GD}^{\pi/2}$ & \texttt{xpih-1-withf} & $1.28\e{-4}$ & 0.9879 & 0.011 & 0.9758 \\
  $X^{\pi/2}_2$ & $\mathrm{GD}^{\pi/2}$ & \texttt{xpih-2-withf} & $2.20\e{-4}$ & 0.9837 & 0.015 & 0.9675 \\
  $X^{\pi/2}_3$ & $\mathrm{GD}^{\pi/2}$ & \texttt{xpih-3-withf} & $7.76\e{-6}$ & 0.9965 & 0.003 & 0.9930 \\
  $X^{\pi/2}_4$ & $\mathrm{GD}^{\pi/2}$ & \texttt{xpih-4-withf} & $2.71\e{-4}$ & 0.9828 & 0.016 & 0.9658 \\
  \midrule
  $\textsc{CNOT}_{02}$ & CR2 & \texttt{cnot-0-2-withf} & $5.04\e{-2}$ & 0.9691 & 0.136 & 0.9584 \\
  $\textsc{CNOT}_{12}$ & CR2 & \texttt{cnot-1-2-withf} & $2.92\e{-2}$ & 0.9708 & 0.107 & 0.9534 \\
  $\textsc{CNOT}_{32}$ & CR2 & \texttt{cnot-3-2-withf} & $1.46\e{-2}$ & 0.9848 & 0.082 & 0.9754 \\
  $\textsc{CNOT}_{42}$ & CR2 & \texttt{cnot-4-2-withf} & $6.63\e{-2}$ & 0.9774 & 0.177 & 0.9812 \\
  \bottomrule
\end{tabular}
\end{table}

Another difference to the two-transmon system is that also the drive frequency
$f$ has been optimized (cf.~\secref{sec:optimizatingpulseparametersResults}).
This is indicated by the label \texttt{withf} in \tabref{tab:ibm5gatemetrics}.
The reason is that we found that when keeping the drive frequency $f$ fixed,
especially the two-qubit gates yield diamond distances approximately twice as
large and fidelities as low as $0.94$ (data not shown).

The resulting gate metrics in \tabref{tab:ibm5gatemetrics} are slightly worse
than those obtained for the two-transmon system (see
\tabref{tab:ibm2gstgatemetrics}). This is reasonable since every transmon is
coupled to at least one additional transmon through a resonator (see the
topology of the system in \figref{fig:ibm5topology}). This means that a much
larger number of states are present in the joint time evolution of the system
and affect the gate operation. In particular, the \textsc{CNOT} gates, for which
the target qubit is coupled to three additional qubits (typically called
``spectator'' qubits \cite{Takita2017faultTolerantStatePreparation}), suffer
from this aspect in that they yield diamond distances larger by a factor of 2--4
compared to the two-transmon system.

A particularly noteworthy result is the single-qubit gate $X_3^{\pi/2}$, which
attains by far the best values for all gate metrics, also in comparison to the
two-transmon results shown in \tabref{tab:ibm2gstgatemetrics}. However, when
looking at an actual application (see the fourth Bloch vector visualized in
\figref{fig:visualization} in \appref{app:visualization}), the performance of
$X^{\pi/2}_1$ is better despite much worse gate metrics.

\subsubsection{Large five-transmon system}

\begin{table}
  \renewcommand{\arraystretch}{0.7}
  \caption{The same as in \tabref{tab:ibm5gatemetrics}
  for the large five-transmon system defined in \secref{sec:transmonmodelibm5ed}.
  As indicated by the label in the third column,
  all gate pulses have been optimized both with and without frequency tuning (cf.~\secref{sec:optimizatingpulseparametersResults}).}
  \centering
  \label{tab:ibm5edgatemetrics}
\begin{tabular}{@{}lllcccc@{}}
  \toprule
  Gate & Pulse & Label & $\Delta$ & $F_{\mathrm{avg}}$ & $\eta_\Diamond$ & $u$ \\
  \midrule
  $X^{\pi/2}_0$ & $\mathrm{GD}^{\pi/2}$ & \texttt{xpih-0} & $4.60\e{-5}$ & 0.9930 & 0.007 & 0.9860 \\
  $X^{\pi/2}_1$ & $\mathrm{GD}^{\pi/2}$ & \texttt{xpih-1} & $1.19\e{-4}$ & 0.9884 & 0.011 & 0.9770 \\
  $X^{\pi/2}_2$ & $\mathrm{GD}^{\pi/2}$ & \texttt{xpih-2} & $7.52\e{-6}$ & 0.9962 & 0.002 & 0.9925 \\
  $X^{\pi/2}_3$ & $\mathrm{GD}^{\pi/2}$ & \texttt{xpih-3} & $8.99\e{-6}$ & 0.9965 & 0.003 & 0.9930 \\
  $X^{\pi/2}_4$ & $\mathrm{GD}^{\pi/2}$ & \texttt{xpih-4} & $4.17\e{-5}$ & 0.9934 & 0.006 & 0.9868 \\
  \midrule
  $X^{\pi/2}_0$ & $\mathrm{GD}^{\pi/2}$ & \texttt{xpih-0-withf} & $4.59\e{-5}$ & 0.9930 & 0.007 & 0.9860 \\
  $X^{\pi/2}_1$ & $\mathrm{GD}^{\pi/2}$ & \texttt{xpih-1-withf} & $1.14\e{-4}$ & 0.9887 & 0.011 & 0.9774 \\
  $X^{\pi/2}_2$ & $\mathrm{GD}^{\pi/2}$ & \texttt{xpih-2-withf} & $7.20\e{-6}$ & 0.9963 & 0.002 & 0.9927 \\
  $X^{\pi/2}_3$ & $\mathrm{GD}^{\pi/2}$ & \texttt{xpih-3-withf} & $8.85\e{-6}$ & 0.9965 & 0.003 & 0.9930 \\
  $X^{\pi/2}_4$ & $\mathrm{GD}^{\pi/2}$ & \texttt{xpih-4-withf} & $3.87\e{-5}$ & 0.9936 & 0.006 & 0.9873 \\
  \midrule
  $\textsc{CNOT}_{10}$ & CR2 & \texttt{cnot-1-0} & $1.34\e{-2}$ & 0.9852 & 0.071 & 0.9758 \\
  $\textsc{CNOT}_{14}$ & CR2 & \texttt{cnot-1-4} & $1.08\e{-1}$ & 0.9621 & 0.177 & 0.9668 \\
  $\textsc{CNOT}_{21}$ & CR2 & \texttt{cnot-2-1} & $4.68\e{-2}$ & 0.9714 & 0.119 & 0.9615 \\
  $\textsc{CNOT}_{32}$ & CR2 & \texttt{cnot-3-2} & $1.83\e{-2}$ & 0.9852 & 0.088 & 0.9777 \\
  $\textsc{CNOT}_{34}$ & CR2 & \texttt{cnot-3-4} & $9.54\e{-2}$ & 0.9671 & 0.179 & 0.9720 \\
  $\textsc{CNOT}_{40}$ & CR2 & \texttt{cnot-4-0} & $2.78\e{-1}$ & 0.9347 & 0.284 & 0.9783 \\
  \midrule
  $\textsc{CNOT}_{10}$ & CR2 & \texttt{cnot-1-0-withf} & $5.70\e{-2}$ & 0.9751 & 0.149 & 0.9728 \\
  $\textsc{CNOT}_{14}$ & CR2 & \texttt{cnot-1-4-withf} & $7.13\e{-3}$ & 0.9841 & 0.056 & 0.9712 \\
  $\textsc{CNOT}_{21}$ & CR2 & \texttt{cnot-2-1-withf} & $1.38\e{-2}$ & 0.9806 & 0.081 & 0.9668 \\
  $\textsc{CNOT}_{32}$ & CR2 & \texttt{cnot-3-2-withf} & $1.21\e{-1}$ & 0.9644 & 0.207 & 0.9764 \\
  $\textsc{CNOT}_{34}$ & CR2 & \texttt{cnot-3-4-withf} & $1.88\e{-2}$ & 0.9832 & 0.090 & 0.9740 \\
  $\textsc{CNOT}_{40}$ & CR2 & \texttt{cnot-4-0-withf} & $8.27\e{-2}$ & 0.9739 & 0.168 & 0.9806 \\
  \bottomrule
\end{tabular}
\end{table}

The gate metrics of the elementary pulses for the large five-transmon system
defined in \secref{sec:transmonmodelibm5ed} are given in
\tabref{tab:ibm5edgatemetrics}. As for the small five-qubit system, the optimized
matrix $M$ is a $2\times2$ matrix for all single-qubit gates $X_i^{\pi/2}$, and
a $4\times4$ matrix for all two-qubit gates $\textsc{CNOT}_{ij}$.

However, in this case, we consider both sets of gate pulses, i.e., with and
without frequency tuning. The reason for this is that the difference in the gate
metrics of the \textsc{CNOT} gates with and without frequency tuning is not as
pronounced as for the small five-qubit system. Furthermore, the single-qubit
gates show almost no difference, so it is reasonable to compare their
performance in actual quantum circuits instead.

Compared to the results for the small five-qubit system presented in
\tabref{tab:ibm5gatemetrics}, the single-qubit gates emerge considerably better
from the optimization. Only the $X_1^{\pi/2}$ gate appears less promising.
Therefore, although the parameter space is quite large, the gate metrics alone
seem to suggest that the topology of the larger system shown in
\figref{fig:ibm5edtopology} (where only two transmons are connected by a
resonator)  is better suited for single-qubit gates than the topology of the
smaller system depicted in \figref{fig:ibm5topology}.

The two-qubit gate metrics reported in \tabref{tab:ibm5gatemetrics} show much
larger fluctuations. Based solely on these metrics, there is no clear winner
between the gates with and without frequency tuning. This makes sense regarding
the much more complicated implementation of the two-qubit CR2 gates (see
\figref{fig:crossresonancepulses} in \secref{sec:optimizatingtwoqubitgate}).

The pulse parameters of the gates with and without frequency tuning come out
very differently: As can be seen in \tabref{tab:deviceibm5edPulseParametersGD}
in \appref{app:pulseparameters}, for the gates without frequency tuning, the
optimization often finds exotic values for the DRAG coefficients $\beta_X$. As
the resulting gate metrics given in \tabref{tab:ibm5edgatemetrics} are almost
the same, though, it seems that the DRAG coefficients are effectively used to
compensate for incorrect drive frequencies. This observation agrees with the
conclusions from the previous chapter (see \secref{sec:chapter5conclusions}).

We remark that in practical applications, it turns out that the gate set with
frequency tuning comes closer to experimental results and also shows better
performance on average (see \cite{Willsch2018TestingFaultTolerance}).

\section{Repeated gate applications}
\label{sec:repeatedgates}

The simplest practical test of the performance of quantum gates is to apply them
repeatedly. Such an experiment yields a first quantitative assessment about how
the accumulation of errors materializes in actual quantum algorithms. We first
consider the evolution of the diamond distance under repeated pulse
applications, and then analyze the evolution of observable error rates in both
simulations and experiments.

\subsection{Evolution of the diamond distance}
\label{sec:evolutiondiamonddistance}

\begin{figure}
  \centering
  \includegraphics[width=\textwidth]{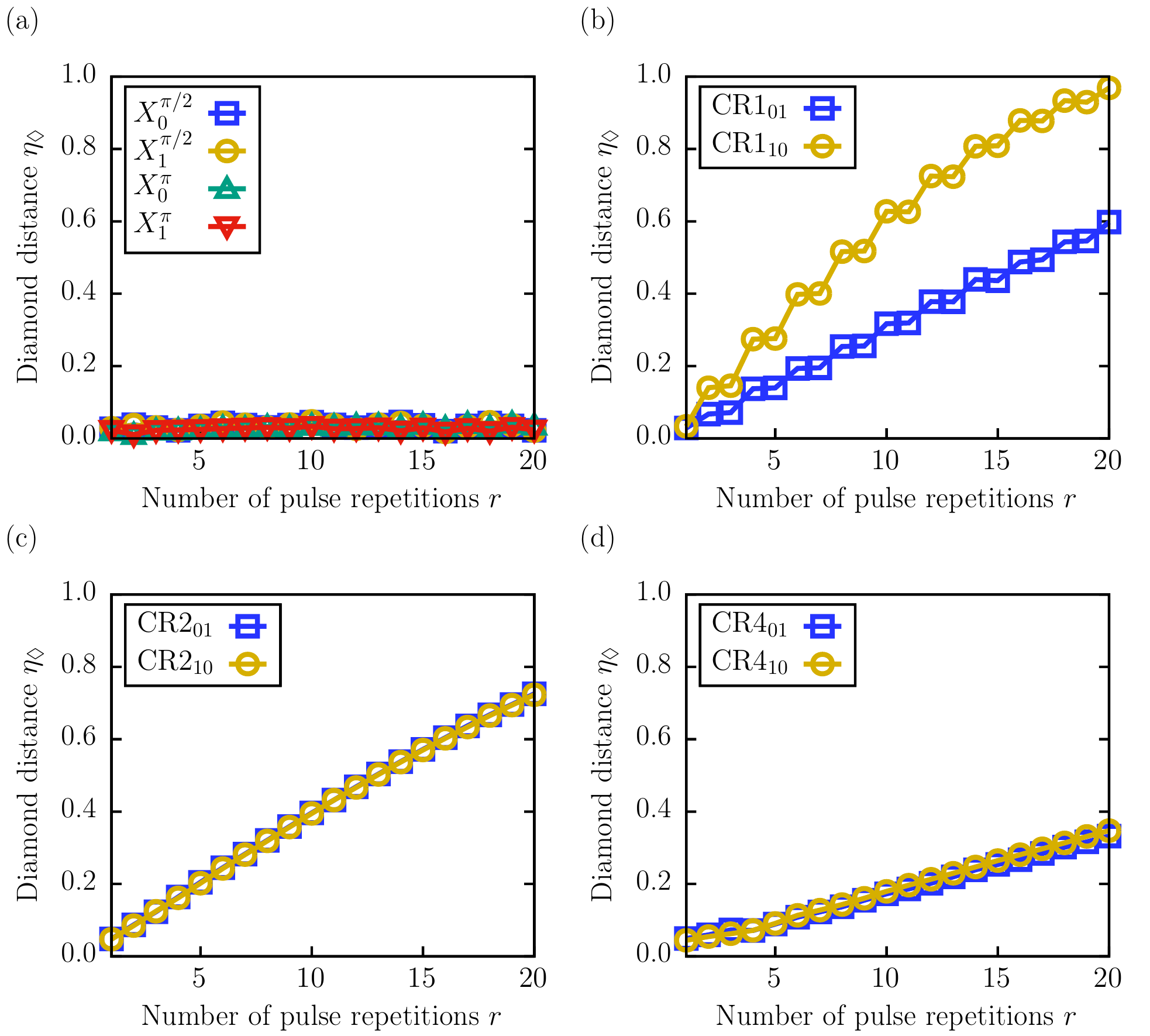}
  \caption{Error rate after repeated application of (a) single-qubit $X$ gates
  and (b)--(d) two-qubit \textsc{CNOT} gates (the indicated CR pulse sequences
  are given in \figref{fig:crossresonancepulses}). Shown is the evolution of
  the diamond distance $\eta_\Diamond$  (see \secref{sec:diamonddistance})
  between the ideal operation $\mathcal G_{id}$ given by
  \equref{eq:repeatedgateapplicationsGid}, corresponding to  $r$ applications of
  the respective quantum gate $U$, and the actual operation $\mathcal
  G_{ac}$ given by \equref{eq:repeatedgateapplicationsGac}. The simulated system
  is the two-transmon system defined in  \secref{sec:transmonmodelibm2gst}. The
  corresponding gate metrics are given in \tabref{tab:ibm2gstgatemetrics}.}
  \label{fig:repeatedgateserrorrates}
\end{figure}

For the first experiment on repeated gate applications, we use the two-transmon
system because many errors are most clearly understood when reduced to the
smallest reproducible case. Specifically, we test $r=1,\ldots,20$ repetitions of
gate pulses from the elementary gate set optimized for the two-transmon system.
This involves the single-qubit $X^{\pi/2}$ and $X^{\pi}$ rotations corresponding
to the elementary $\mathrm{GD}$ pulses (see
\secref{sec:optimizatingsinglequbitgate}), and the two-qubit $\textsc{CNOT}$
gates corresponding to the CR1, CR2, and CR4 pulses
(\secref{sec:optimizatingtwoqubitgate}).

\begin{table}
  \renewcommand{\arraystretch}{1.0}
  \caption{Comparison of the CR2 and CR4 pulses
  (cf.~\figref{fig:crossresonancepulses}) for a single $\textsc{CNOT}_{01}$
  gate, 20 successive $\textsc{CNOT}_{01}$ gates, and four successive QFT
  applications (one QFT contains five \textsc{CNOT} pulses and two GD pulses;
  see \figref{fig:twoqubitqft} for the circuit and \secref{sec:compiler} for how
  gates are mapped to pulses). The numbers reported are the diamond distances
  defined in \secref{sec:diamonddistance}, but the same qualitative trends are
  true for  the other gate metrics (data not shown). The results for
  $\textsc{CNOT}^1$ are taken from \tabref{tab:ibm2gstgatemetrics}. The
  results for $\textsc{CNOT}^{1}$ and $\textsc{CNOT}^{20}$ can also be seen as
  blue squares in
  \figref{fig:repeatedgateserrorrates}(c) and (d).}
  \centering
  \label{tab:repeatedgatesqft}
\begin{tabular}{@{}cccc@{}}
  \toprule
  Pulse & $\textsc{CNOT}^1$ & $\textsc{CNOT}^{20}$ & $\mathrm{QFT}^4$ \\
  \midrule
  CR2 & 0.048 & 0.73 & 0.27 \\
  CR4 & 0.049 & 0.33 & 0.32 \\
  \bottomrule
\end{tabular}
\end{table}

For each gate pulse, we run the two-transmon simulation with the corresponding
pulses  applied 20 times. Given the duration $T$ of a certain pulse
(cf.~\tabref{tab:deviceibm2gstPulseParametersGD} and
\tabref{tab:deviceibm2gstPulseParametersCR} in \appref{app:pulseparameters}),
this means that the time evolution is simulated for $0\le t\le 20T$ for each
initial state from the computational basis
$(\ket{00},\ket{01},\ket{10},\ket{11})$. At each time $t=rT$ after $r$ pulse
applications, we extract the transformation matrix $M(rT)=P_{\mathcal H_{2^n}}
\mathcal U(rT,0) P_{\mathcal H_{2^n}}$ of the computational subspace according
to \equref{eq:MprojectionUtotal} (see also
\secref{sec:optimizatingpulseparameters}).

If the implementation of a particular quantum gate $U$ was perfect, the
transformation $M(rT)$ would be equivalent to $U^r$. In this section,
we measure the error rate between $M(rT)$ and $U^r$ in terms of the diamond
distance $\eta_\Diamond$ defined in \equref{eq:gatemetricsdiamonddistance},
where the ideal operation and the actual operation are given by
\begin{subequations}
  \begin{align}
    \label{eq:repeatedgateapplicationsGid}
    \mathcal G_{id}(\rho)&=U^r\,\rho\,(U^\dagger)^r,\\
    \label{eq:repeatedgateapplicationsGac}
    \mathcal G_{ac}(\rho)&=M(rT)\,\rho\,M(rT)^\dagger.
  \end{align}
\end{subequations}
The results are presented in \figref{fig:repeatedgateserrorrates}. We note that
the same qualitative results for each of the curves can be seen for the other
gate metrics studied in \secref{sec:gatemetrics} (data not shown).

The performance of the single-qubit gates shown in
\figref{fig:repeatedgateserrorrates}(a) is reasonably good. The error rate of
around 2\% for $r=1$ (which corresponds to the results for $\eta_\Diamond$ given in
\tabref{tab:ibm2gstgatemetrics}) does not grow after repeated gate applications.

This is different for the two-qubit gates shown in
\figref{fig:repeatedgateserrorrates}(b)--(d), for which the error rate after
repeated applications grows approximately linearly with $r$ for all pulse
schemes. For the CR1-type \textsc{CNOT}, we also observe a different performance
if control and target qubit are interchanged. This error seems to have been
canceled by the echo schemes used for the CR2 and CR4 pulses.

Comparing \figref{fig:repeatedgateserrorrates}(b) and (c), we see that the
CR1-type gate $\textsc{CNOT}_{01}$ (blue squares in
\figref{fig:repeatedgateserrorrates}(b)) achieves the smallest error rate after
repeated applications. Interestingly, this gate pulse has the worst results for
the average gate fidelity and the unitarity (see
\tabref{tab:ibm2gstgatemetrics}). This means that the gate metrics do not
reflect the actual performance of the pulses in repeated applications.

The fact that we find the best performance for the CR4 pulse scheme (see
\figref{fig:repeatedgateserrorrates}(d)) agrees with experimental observations
\cite{Takita2017faultTolerantStatePreparation}. Note in particular that this is
in contrast to the results suggested by the gate metrics in
\tabref{tab:ibm2gstgatemetrics}. For instance, the CR4-type gate
$\textsc{CNOT}_{01}$  has the worst diamond distance $\eta_\Diamond=0.049$ after
a single pulse  application, but performs best of all in repeated applications
(blue squares in \figref{fig:repeatedgateserrorrates}(d)). Still, this
observation can change again when looking at the performance of the gates in a
practical quantum algorithm: In \tabref{tab:repeatedgatesqft}, we see that the
same CR4 pulse performs worse than CR2 when used 20 times in a quantum Fourier
transform (QFT). Therefore, the gate metrics can also not reliably predict the
performance in practical applications.

Finally, it is worth mentioning that the transformation $M(rT)$ after $r>1$
pulse applications is always closer to the ideal result $U^r$ than the product
$M(T)^r$ (data not shown). This means that the pulses have been tuned to the
full time evolution of the transmon-resonator system including non-computational
states; in a sense, the pulses are capable of using the more
complicated dynamics on the larger space for improved results on the
computational subspace. It also means that the effect of higher levels on the
time evolution cannot be neglected. This is in agreement with the findings
reported in \figref{fig:freekitcomparetimeevolutions}(d), which show that the
two-level approximation cannot sufficiently describe the dynamics of full
transmon systems. However, as we show in \secref{sec:gatesettomography}, there
exist maps on the two-qubit reduced density matrix that can reliably describe
the dynamics under repeated pulse applications.

\subsection{Relation to experiments}

\begin{figure}
  \centering
  \includegraphics[width=\textwidth]{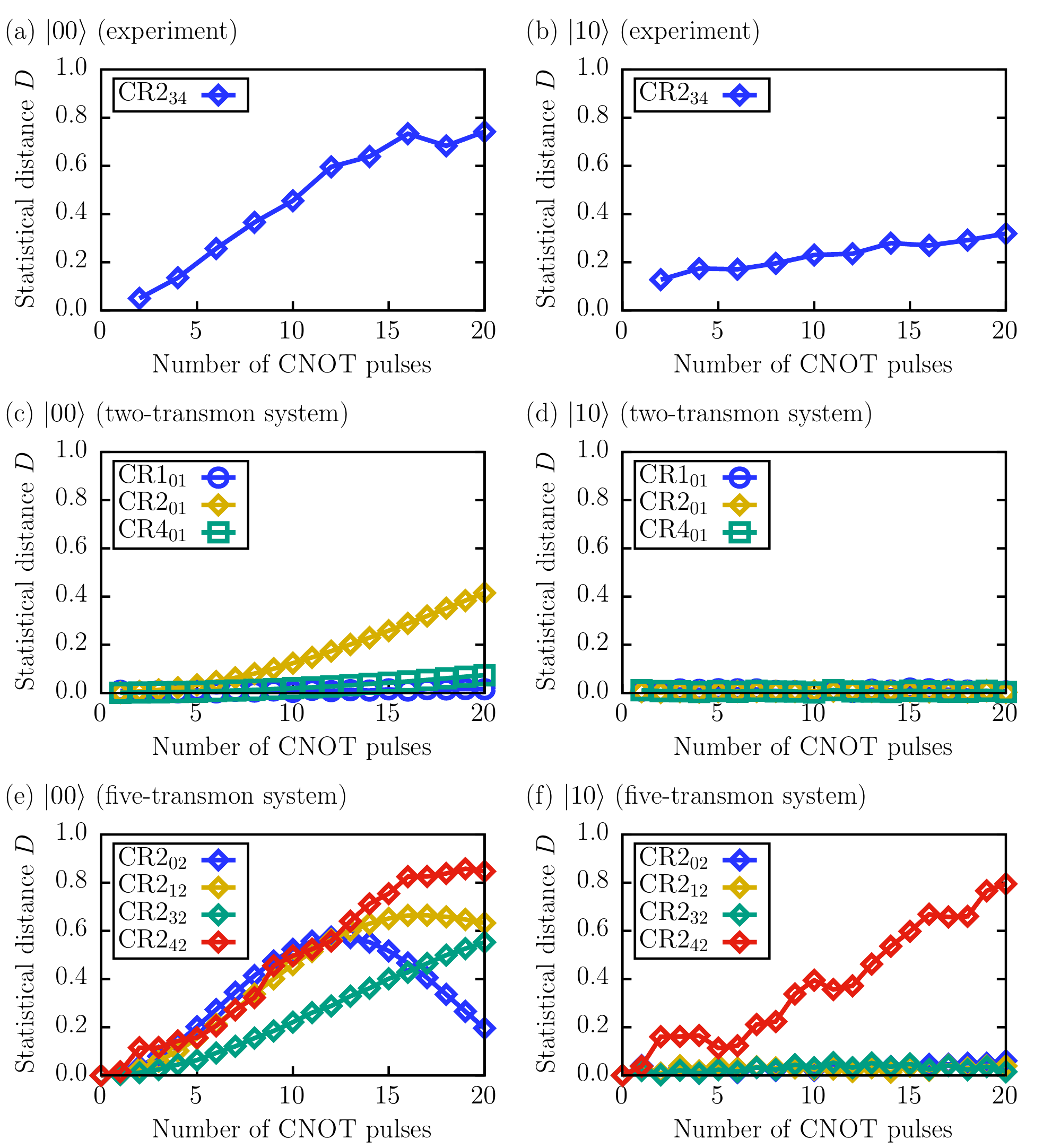}
  \caption{Evolution of the statistical distance $D$ given by
  \equref{eq:statisticaldistance}, measured after repeated applications of the
  two-qubit $\textsc{CNOT}$ gate. (a) and (b) show results from an IBM Q
  processor (see text), (c) and (d) have been obtained from the two-transmon simulation
  defined in \secref{sec:transmonmodelibm2gst}, and (e) and (f) have been
  obtained from the five-transmon simulation defined in
  \secref{sec:transmonmodelibm5}. The gate metrics for the simulated systems are
  given in \tabref{tab:ibm2gstgatemetrics} and \tabref{tab:ibm5gatemetrics},
  respectively. The left (right) panels correspond to the initial state
  $\ket{00}$ ($\ket{10}$). Different markers correspond to different
  cross-resonance pulses (see \figref{fig:crossresonancepulses}). Different
  colors correspond to pulses between different qubits. All simulations of the
  five-transmon system were performed on the supercomputer JURECA
  \cite{JURECA}.}
  \label{fig:repeatedgatescnot}
\end{figure}

The goal of this section is to relate the gate metrics obtained in
\secref{sec:gatemetricsresults} and their evolution studied in
\secref{sec:evolutiondiamonddistance} to the performance of a real device. When
executing a quantum circuit on a real device with two qubits, the result is a
two-bit string $j_0j_1$. To relate the results to the ideal, theoretical
probabilities $\smash{p^{(\mathrm{id})}_{j_0j_1}}$, one has to execute the same
circuit repeatedly and estimate the experimental relative frequencies
$\smash{p^{(\mathrm{exp})}_{j_0j_1}}$ of each bit string $j_0j_1$ by sampling.
The number of samples is called \emph{shots}. For the simple experiments studied
in this section and  also the more complicated circuits in
\chapref{cha:fullcircuitsimulations},  we always consider the maximum number of
8192 shots.

To compare two probability distributions $p_{j_0j_1}$ and $\tilde p_{j_0j_1}$,
we use the statistical distance
\begin{align}
  \label{eq:statisticaldistance}
  D = \frac 1 2 \sum_{J} \abs{p_J - \tilde p_J},
\end{align}
where the sum is over all two-bit strings $J=j_0j_1$. In what follows, the
distribution $p_J$ is always the result $\smash{p^{(\mathrm{id})}_{j_0j_1}}$ expected
from an ideal, gate-based quantum computer as introduced in
\chapref{cha:quantumcomputing}. The other distribution, $\tilde p_J$, is either
given by relative frequencies $\smash{p^{(\mathrm{exp})}_{j_0j_1}}$ measured in an
experiment, or probabilities
$\smash{p^{(\mathrm{sim})}_{j_0j_1}=\abs{\braket{m_0=j_0,m_1=j_1|\Psi}}^2}$ obtained
from the state vector $\ket{\Psi}$ (see \equref{eq:psioftsolutioncoefficients})
produced by the transmon simulator.

This distance measure $D$ given by \equref{eq:statisticaldistance} is also
known as the \emph{total variation distance}. It can be interpreted as the
minimum fraction of samples that must be altered to achieve  $\tilde p_J \to
p_J$. Furthermore, the total variation distance induces the diamond  distance
$\eta_\Diamond$ defined in \secref{sec:diamonddistance} on the space  of quantum
channels. See \cite{Sanders2016ThresholdTheorem} for more information.

We perform 20 repetitions of the \textsc{CNOT} gate on the input states
$\ket{00}$ and $\ket{10}$ using both a real device and also different pulse
schemes on the simulated systems. For the experiment on the real processor, we
used qubits Q3 and Q4 of the five-qubit processor that was available on the IBM
Q Experience \cite{ibmquantumexperience2016} on August 17, 2017. The
\textsc{CNOT} gate was implemented by means of the CR2 pulse defined in
\figref{fig:crossresonancepulses}(c) (see also
\figref{fig:twoqubitpulseruleVZ}(c)). The RB error rates reported by the
processor at the time of execution were $0.0376$ for the \textsc{CNOT} gate,
$0.0031$ and $0.0016$ for the single-qubit gates, and $0.033$ and $0.06$ for the
readout errors.

The results of these conceptually simple experiments are shown in
\figref{fig:repeatedgatescnot}(a) and (b). We see that on the real processor,
the \textsc{CNOT} gate on the state $\ket{00}$ yields a reasonable performance
if applied twice, but degrades rapidly for more repetitions. For the initial
state $\ket{10}$ shown in \figref{fig:repeatedgatescnot}(b), we see a larger
offset of about 0.2 if applied twice; however, the actual error increases only
very slightly over repeated applications of the gate.

For the simulations, we first use the small two-transmon system defined in
\secref{sec:transmonmodelibm2gst} to compare the different pulse implementations
CR1, CR2, and CR4 of the \textsc{CNOT} gate (see
\figref{fig:crossresonancepulses}). The results shown in
\figref{fig:repeatedgatescnot}(c) and (d) demonstrate that only the two-pulse
echoed \textsc{CNOT} gate, CR2, shows a performance similar to the experiment.
This is reasonable as this is the same pulse scheme that is used for the
experiment. However, the offset of about 0.2 for the initial state $\ket{10}$
(see \figref{fig:repeatedgatescnot}(b)) cannot be observed in the simulation.
Although it can be modeled by including a simple measurement error of $0.2$ (in
the spirit of \equref{eq:errordetectionMeasurementError} in
\secref{sec:testingfaulttolerance}), it is more likely due to environmental
effects that are not included in the two-transmon simulation.

Note especially that the other pulse schemes, CR1 and CR4, perform much better
than CR2 (see \figref{fig:repeatedgatescnot}(c)). A similar trend  has been
observed in experiments with CR1 \cite{Alexander2020QiskitPulse} and  CR4
\cite{Takita2017faultTolerantStatePreparation}. This is not at all reflected by
the gate metrics shown in \tabref{tab:ibm2gstgatemetrics}. In particular, the
fidelity $F_{\mathrm{avg}}$ is nearly the same for all pulses, despite the
strong difference in the actual gate performance. In fact, the only gate with a
comparably bad fidelity, $\mathrm{CR1}_{01}$, performs unexpectedly well (see
\figref{fig:repeatedgatescnot}(c) and (d)). Also the diamond distances shown in
\figref{fig:repeatedgateserrorrates} do not suggest that. We note that, since
the diamond distance is related to the worst-case statistical distance
\cite{Sanders2016ThresholdTheorem}, it is possible that for another input state,
the statistical distances would show a comparable increase for repeated pulses.

In \figref{fig:repeatedgatescnot}(e) and (f), we present simulation results for
all \textsc{CNOT} gates on the small five-qubit system defined in
\secref{sec:transmonmodelibm5}. Again, the errors are not properly captured by
the corresponding gate metrics given in \tabref{tab:ibm5gatemetrics}. Especially
the performance of the pulse $\mathrm{CR2}_{42}$ applied to the input state
$\ket{10}$ (the red line in \figref{fig:repeatedgatescnot}(f)) degrades quickly,
although this pulse scored the best unitarity and the second-best fidelity of
all two-qubit gates (see \tabref{tab:ibm5gatemetrics}). In this case, only the
diamond distance suggests the bad performance.

Apart from this, we observe that the performance of the simulated transmon
system presented in \figref{fig:repeatedgatescnot}(e) comes much closer to the
result of the experiment shown in \figref{fig:repeatedgatescnot}(a).  Note that
the agreement can only be qualitative because the simulated five-transmon system
corresponds to a different device that does not have a \textsc{CNOT} gate
between transmons 3 and 4 (cf.~\secref{sec:transmonmodelibm5} and
\tabref{tab:deviceibm5PulseParametersCR2}).  Still, the observation that the
five-transmon simulation comes  much closer to the experiment than the
two-transmon simulation  gives positive evidence for the hypothesis that the
additional features included in the five-transmon simulation, i.e., leakage and
crosstalk due to additional transmons and resonators,  can capture most of the
errors observed in the experiment.

\section{Gate set tomography}
\label{sec:gatesettomography}

\begin{figure}
  \centering
  \includegraphics[width=.5\textwidth]{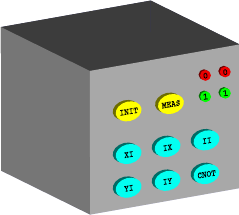}
  \caption{Black box model of a quantum computer. For GST, a quantum computer is
  considered as a black box with only digital input and output, agnostic to the
  physical details of the implementation. The black box provides buttons for
  initialization and measurement (yellow), and several gate operations (cyan;
  see \tabref{tab:GSTPulseImplementations} for the meaning of the labels). On each
  measurement, the black box produces binary output in the form of light signals
  indicating a bit string. In GST, the best quantum theoretical two-level
  description is fitted to the frequency of bit strings produced by the black
  box.}
  \label{fig:blackboxmodel}
\end{figure}

The results presented in the previous sections demonstrate that common gate
metrics such as the average fidelity or the diamond norm cannot reliably predict
the performance of a sequence of quantum gates
(cf.~\cite{Iyer2017smallQCneededforFT}). It has also been recognized in the
literature that the RB-number, i.e., the quantity produced by randomized
benchmarking \cite{proctor2017RandomizedBenchmarking}, suffers from the same
problems \cite{BlumeKohout2017DemoRigorousThresholdGST}. The obvious conclusion
is that a quantum gate is too complicated to be characterized by a single
number.

Therefore, in this section, we explore a much more sophisticated approach to
characterize implementations of quantum gates called \emph{gate set tomography}
(GST) \cite{BlumeKohout2013GST, Greenbaum2015GST}.  In short, the idea of GST is
to treat a quantum device as a black box that takes only digital input and
produces only digital output. One then fits the best quantum theoretical
description to experiments performed with the black box. An example black box
model for a two-qubit device is shown in \figref{fig:blackboxmodel}. A recent
application to a quantum processor based on ion traps is presented in
\cite{BlumeKohout2017DemoRigorousThresholdGST}, for which an open-source
implementation of GST called \texttt{pyGSTi} \cite{Nielsen2018pyGSTi0944} has
been developed. This package has also been used for the GST experiments
presented in this thesis.

Our goal is to test the GST procedure for output generated by the transmon
simulation developed for this work (see \secref{sec:simulationsoftware}). In
particular, the objective  is to study whether the resulting quantum theoretical
description has a better predictive power than the common gate metrics. As shown
below, the answer to this question is affirmative: The two-level description
produced by GST can reliably predict the performance of gates when used
repeatedly in quantum algorithms.

\subsection{The idea of GST}

In quantum theory, we describe the general state of a system in terms of a
density matrix $\rho$. The evolution of the state is represented
by a linear map $\rho\mapsto \mathcal G(\rho)$, where $\mathcal G$ is typically
a quantum operation as introduced in \secref{sec:quantumoperations}. After the
evolution, the system is measured and the probability to obtain outcome $J$ is
given by
\begin{align}
  \label{eq:GSTBornRule}
  p_J = \mathrm{Tr}\,E_J \mathcal G(\rho).
\end{align}
Here, the set of operators $\{E_J\}$ represents the measurement. The minimum
requirements on $\{E_J\}$ are that the $E_J$ are positive semidefinite and
$\sum_J E_J = \mathds1$ such that the $p_J$ are valid probabilities. In the
literature,  a set $\{E_J\}$ with these properties is called POVM
\cite{NielsenChuang}  and \equref{eq:GSTBornRule} is a generalization of
the Born rule (see also \cite{Fuchs2001QuantumFoundationsInformation,
Fuchs2002QuantumFoundationsInformation}).

\subsubsection{The problem of tomography}

The central idea of quantum tomography is to construct, from experimentally
observed relative frequencies $p_J$, an element of the quantum theoretical
description given by \equref{eq:GSTBornRule}. If the element to reconstruct is
the density matrix $\rho$, one speaks of \emph{quantum state tomography}
\cite{Smithey1993QuantumStateTomography, Leibfried1996QuantumStateTomography}.
If the objective is to find a representation of the measurement $\{E_J\}$, the
procedure is known as \emph{quantum measurement tomography}
\cite{Luis1999QuantumMeasurementTomography}. And lastly, characterizing the map
$\mathcal G$ is called \emph{quantum process tomography}
\cite{Chuang1997QuantumProcessTomography, Poyatos1997QuantumProcessTomography}.

All these kinds of tomography have a fundamental flaw, namely that they require
the other elements of \equref{eq:GSTBornRule} to be postulated. For instance,
quantum process tomography requires the system to be prepared in various known
states $\rho$ and measured with various known operators $E_J$, which are
typically implemented using the very same gates $\mathcal G$ that process tomography
tries to characterize \cite{BlumeKohout2013GST}. In that sense, all three kinds
of tomography are self-referential and circular. Note that this problem also
affects more recent tomography proposals \cite{Xin2017QST,
Helsen2019SpectralQuantumTomography}.

\subsubsection{The solution provided by GST}

The idea of GST is to solve this problem by self-consistently estimating the
state $\rho$, the process $\mathcal G$, and the measurement $\{E_J\}$ of
\equref{eq:GSTBornRule} at the same time, using only experimentally observed
frequencies $p_J$. The philosophy is that all elements of the description in
\equref{eq:GSTBornRule} should be accessible from the observed data. If the
system is prepared in a certain state $\rho$, the data should reveal it;
similarly, if the measurement is in a different basis, the data should suggest
it; etc. The central viewpoint is that obtaining enough data by playing around
with the black box should be sufficient to predict its future behavior
\cite{BlumeKohout2013GST}.

To frame this idea, the system is treated as a black box with very limited
control. Such a black box for a two-qubit quantum computer is shown in
\figref{fig:blackboxmodel}. It has several buttons for interaction and two sets
of lights signaling the output, but the actual physical implementation is
irrelevant. The mapping of the buttons to the mathematical description in
\equref{eq:GSTBornRule} is as follows. The yellow \texttt{INIT} button is to be
described by the mathematical object $\rho$. Each cyan button is to be described
by a map in the set $\{\mathcal G_{\texttt{XI}}, \mathcal G_{\texttt{YI}},
\mathcal G_{\texttt{IX}}, \mathcal G_{\texttt{IY}}, \mathcal G_{\texttt{II}},
\mathcal G_{\texttt{CNOT}}\}$ (the target operation indicated by the indices is
given in \tabref{tab:GSTPulseImplementations} below). The process $\mathcal G$ is
then an arbitrary sequence of elements in this set, corresponding to the
sequence of buttons pressed. Finally, the yellow \texttt{MEAS} button makes the
system produce an event
$J\in\{00,01,10,11\}$, which is to be
described by $\{E_J\}$. An experiment in this model takes a simple form:
\begin{enumerate}[label=\arabic*.]
  \item Press the \texttt{INIT} button to initialize the system.
  \item Press a sequence of gate buttons from the set $\{\texttt{XI}, \texttt{YI},
  \texttt{IX}, \texttt{IY}, \texttt{II},\texttt{CNOT}\}$.
  \item Press the \texttt{MEAS} button and record the outcome.
\end{enumerate}
For each experiment corresponding to the same sequence of buttons pressed,
the relative frequencies $p_J$ are computed from the outcomes. The best
description in terms of $\rho$, $\mathcal G$, and $E_J$ is then obtained from the
data.

It is important to recognize that only the frequencies $p_J$ of each event $J$
from the data are used. All other information about the individual events need
to be discarded. This approach is typical of all quantum theoretical models, in
the sense that quantum theory cannot model the individual events
\cite{DeRaedt2019Separation} or long-time correlations between individual events
\cite{Willsch2020LongTimeCorrelations}.

\subsubsection{Mathematical framework}

To fit a mathematical description in terms of $E_J$, $\mathcal G$, and $\rho$ to
the measured relative frequencies $p_J$, it is useful to introduce a vector
representation of \equref{eq:GSTBornRule}. We obtain such a representation by
expanding the $N\times N$ matrices $E_J$ and $\rho$ in the normalized Pauli
basis $\mathcal P$ defined in \equref{eq:paulibasis},
\begin{subequations}
\begin{align}
  \label{eq:GSTVectorizedRho}
  \rho &= \sum_i \rho_i \widehat{P}_i,\\
  \label{eq:GSTVectorizedEJ}
  E_J &= \sum_i e_{Ji} \widehat{P}_i,
\end{align}
\end{subequations}
where $\rho_i=\mathrm{Tr}\,\widehat{P}_i\rho$ and
$e_{Ji}=\mathrm{Tr}\,\widehat{P}_iE_J$.  We arrange the coefficients
$\{\rho_i\}$ and $\{e_{Ji}\}$ in $N^2$-dimensional vectors denoted by
$\sket{\rho}$ and $\sket{E_J}$, respectively.

In the literature, the vector space corresponding to this vector representation
of matrices is sometimes called Hilbert-Schmidt space. The inner product on
this space is given by the Hilbert-Schmidt inner product $\sbraket{X}{Y} =
\mathrm{Tr}\,X^\dagger Y$ for complex matrices $X,Y$.

Inserting \equaref{eq:GSTVectorizedRho}{eq:GSTVectorizedEJ} into
\equref{eq:GSTBornRule} yields
\begin{align}
  \label{eq:GSTBornRuleVectorized}
  p_J = \sum_{ij} e_{Ji} \left(\mathrm{Tr}\,\widehat{P}_i\mathcal G(\widehat{P}_{j})\right) \rho_j
  = \sbra{E_J} G \sket{\rho},
\end{align}
where the matrix $G$ representing the map $\mathcal G$ is
the Pauli transfer matrix defined in \equref{eq:paulitransfermatrix}. We use
a similar notation to distinguish between the maps $\{\mathcal G_{\texttt{XI}},\ldots\}$
and their matrix representations $\{G_{\texttt{XI}},\ldots\}$.

Due to the use of the Pauli basis, all coefficients in the vectors $\sket\rho$
and $\sket{E_J}$ and the matrices $G$ are real. Still, the total number of
coefficients for two qubits ($N=4$) is $16+4\times16+6\times16^2 = 1616$ (16 for
$\rho$, 16 for each $E_J$, and $16^2$ for each of the gate matrices
$G_{\texttt{XI}},  G_{\texttt{YI}},  G_{\texttt{IX}}, G_{\texttt{IY}},
G_{\texttt{II}}$, and $G_{\texttt{CNOT}}$). This means that a considerable
number of experiments needs to be run to obtain enough relative frequencies $p_J$
to determine the coefficients accurately.

These experiments are not chosen arbitrarily. Rather, they are constructed
systematically to yield a sufficient amount of information about the
coefficients, while still targeting operations such as single-qubit rotations
that are commonly used in current quantum information processors
\cite{BlumeKohout2017DemoRigorousThresholdGST}. This construction works as
follows: The description of data from an experiment, which corresponds to a
sequence $(s_1,\ldots,s_L)$ of pressed cyan buttons in
\figref{fig:blackboxmodel}, has the structure
\begin{align}
  \label{eq:GSTFullFiducialsGerms}
  p_J^{(s_1,\ldots,s_L)} = \sbra{E_J}
  \underbrace{G_{s_L}\cdots G_{s_*}}_{F^{(M)}}
  \underbrace{G_{s_*}\cdots G_{s_*}}_{g^l}
  \underbrace{G_{s_*}\cdots G_{s_1}}_{F^{(P)}}
  \sket{\rho},
\end{align}
where $F^{(M)}$ is called a measurement fiducial, $g^l$ is called a germ power,
and $F^{(P)}$ is called a preparation fiducial. The fiducials are
chosen to prepare an \emph{informationally complete} set of states (for instance, the
eigenstates of the Pauli matrices, four of which are shown  in
\figref{fig:blochsphereexamples}(a)--(d)). The germs $g$, raised to
logarithmically spaced integer powers such as $l\in\{1,2,4,8,16,32\}$, are chosen to amplify
certain systematic errors expected from the gates. All germs and fiducials are
sequences of the six gate matrices $G_{\texttt{XI}}, G_{\texttt{YI}},
G_{\texttt{IX}}, G_{\texttt{IY}},  G_{\texttt{II}}$, and $G_{\texttt{CNOT}}$.
They explicitly include empty sequences such that also ``bare''  experiments
corresponding to $\sbraket{E_J\vert F_i^{(M)}F_j^{(P)}}{\rho}$ and, in
particular, $\sbraket{E_J}{\rho}$ are represented by the data. See
\cite{BlumeKohout2013GST, BlumeKohout2017DemoRigorousThresholdGST} for more
information on these design choices.

When describing experimentally obtained data $\{p_J^{(s_1,\ldots,s_L)}\}$, the
mathematical objects contained in \equref{eq:GSTFullFiducialsGerms} can only be
determined up to an invertible matrix $M$. The reason for this is that the
$p_J^{(s_1,\ldots,s_L)}$ stay the same if all objects in
\equref{eq:GSTFullFiducialsGerms} are replaced according to
$\sbra{E_J}~\mapsto\sbra{E_J}M^{-1}$, $G\mapsto MGM^{-1}$, and
$\sket\rho\mapsto M\sket\rho$. This fundamental freedom in GST is called the
\emph{GST gauge}.

There are several ways to fix a certain gauge transformation $M$. For instance,
if one requires the description of the state $\rho$ to be independent of the
operations that are performed on it (as typically required by consistency in
quantum theoretical descriptions, see \cite{DeRaedt2019Separation}), one could
choose a gauge $M$ that always maps the fitted $\rho$ to some fixed
representation. In general, it can be problematic to fix a certain gauge and
interpret the resulting, gauge dependent error rates (see also
\cite{Lin2019GateMetricsGaugeFreedom}). For the following analysis, however, the
goal is to predict the $p_J^{(s_1,\ldots,s_L)}$, so the choice of the gauge is
irrelevant. Therefore, we use the default gauge optimization performed by
\texttt{pyGSTi}.

The actual self-consistent fitting procedure involves several steps. The first
step is to construct a matrix $\texttt{AB}$, where $\texttt{AB}_{J+Ni,j}$
contains  the relative frequency corresponding to $\sbraket{E_J\vert
F_i^{(M)}F_j^{(P)}}{\rho}$, i.e., obtained from the ``bare'' experiments. The
(pseudo-)inverse of this matrix yields initial estimates for $\rho$, $E_J$, and
$G$ (this step requires the fiducials to be be informationally complete).  This
part of the procedure is called \emph{linear inversion GST} and is described in
detail in \cite{BlumeKohout2013GST, Greenbaum2015GST}. A difference
between the single-qubit GST described in these references and the two-qubit GST
studied here is that all data for
$J\in\{00,01,10,11\}$ is included
in $\texttt{AB}$. The full iterative procedure to refine the initial estimates
self-consistently is explained in
\cite{BlumeKohout2017DemoRigorousThresholdGST} (see Fig.~1 of this work).

We remark that, because of the vast number of parameters to be estimated in GST,
it would be interesting to explore techniques from deep learning to improve the
fitting procedure. These techniques have specifically been invented to fit
statistical models with millions of parameters to observed data
\cite{Goodfellow2016DeepLearning}.

\subsection{Running GST}
\label{sec:gatesettomographyRunning}

\begin{table}
  \caption{Summary of the pulses used to implement the gates corresponding to
  the buttons of the black box shown in \figref{fig:blackboxmodel} by means of
  the two-transmon simulation model defined in
  \secref{sec:transmonmodelibm2gst}.
  The general single-qubit GD pulses are defined in \secref{sec:singlequbitGDpulse}.
  The implementation of $Y^{\pi/2}=\texttt{-Y}$ in particular is derived in \equaref{eq:compilerRewriteYGateAsU}{eq:compilerexamplesingleYgate}.
  The identity gate is implemented as an undriven time evolution for $T_X=\SI{83}{ns}$,
  i.e., the duration of the single-qubit pulses (see \secref{sec:singlequbitzeropulse}).
  The two-qubit \textsc{CNOT} gate is implemented using the CR2 pulse (see \secref{sec:optimizatingtwoqubitgate}).
  All pulse parameters are listed in \appref{app:pulseparameters}, and the
  target gates are defined in \tabref{tab:elementarygateset}.
  Gate metrics for the elementary pulses are given in \tabref{tab:ibm2gstgatemetrics}.
  }
  \centering
  \label{tab:GSTPulseImplementations}
  \begin{tabular}{@{}clcc@{}}
    \toprule
    Button & Pulse & Target gate & Description \\
    \midrule
    \texttt{XI}   & $\mathrm{GD}^{\pi/2}_0(0)$     & $X^{\pi/2}_0$        & $\pi/2$ rotation of qubit $0$ about the $x$ axis \\
    \texttt{YI}   & $\mathrm{GD}^{\pi/2}_0(\pi/2)$ & $Y^{\pi/2}_0$        & $\pi/2$ rotation of qubit $0$ about the $y$ axis \\
    \texttt{IX}   & $\mathrm{GD}^{\pi/2}_1(0)$     & $X^{\pi/2}_1$        & $\pi/2$ rotation of qubit $1$ about the $x$ axis \\
    \texttt{IY}   & $\mathrm{GD}^{\pi/2}_1(\pi/2)$ & $Y^{\pi/2}_1$        & $\pi/2$ rotation of qubit $1$ about the $y$ axis \\
    \texttt{II}   & $\mathrm{zero}(T_X)$           & I                    & identity gate \\
    \texttt{CNOT} & CR2 & $\textsc{CNOT}_{01}$ & \textsc{CNOT} between qubit $0$ (control) and $1$ (target) \\
    \bottomrule
  \end{tabular}
\end{table}

We apply GST to the two-transmon simulation model defined in
\secref{sec:transmonmodelibm2gst}. This requires 58990 quantum circuits (see
below) to be compiled into pulses (cf.~\secref{sec:compiler}) and simulated by
solving the TDSE. From the final state vector $\ket\Psi$, only the relative
frequencies for the two-qubit states $\ket{00},\ket{01},\ket{10},$ and
$\ket{11}$ are extracted. In this way, the simulation represents the
implementation that is hidden inside the black box shown in
\figref{fig:blackboxmodel}. The particular pulses used to implement the cyan
buttons are summarized in \tabref{tab:GSTPulseImplementations}.

We use the software package \texttt{pyGSTi} \cite{Nielsen2018pyGSTi0944} to
produce the input (the quantum circuits) to the black box  and to analyze the
output (the relative frequencies $p_J$). The list of simulated quantum circuits
comprises 58990 gate sequences with a maximum of 38 gates per sequence.
\texttt{pyGSTi} respects the structure of \equref{eq:GSTFullFiducialsGerms}, using 16
preparation fiducials $F^{(P)}$, 11 measurement fiducials $F^{(M)}$, and 89
germs $g$ raised to the power $l\in\{1,2,4,8,16,32\}$. Since the output $p_J$ is passed to
\texttt{pyGSTi} in terms of actual integer counts for each event
$J\in\{00,01,10,11\}$, we compute counts for 1000 samples. This number reflects
the expected accuracy of the product-formula algorithm for $\tau=\SI{10^{-3}}{ns}$
(cf.~\figref{fig:accuracyglobalerror}(b)). Furthermore, \texttt{pyGSTi} offers
several modes for estimating the objects $\rho$, $\mathcal G$, and $E_J$ in
\equref{eq:GSTBornRule}:
\begin{enumerate}[labelwidth=*,labelindent=1.5cm]
  \item[{Full}:] Fully unconstrained maps,
  \item[{TP}:] Constrain the maps to be trace-preserving quantum operations (cf.~\secref{sec:quantumoperations}),
  \item[{CPTP}:] Constrain the maps to be CPTP quantum channels (cf.~\secref{sec:quantumoperations}).
\end{enumerate}
In this section, we report the CPTP estimates, because they proved to be the
most reliable estimates when trying to predict the performance of
quantum gate circuits (see below). The estimated initial
density matrix reads
\begin{align}
  \label{eq:GSTResultCPTPEstimateRho}
  \rho = \ketbra{00}{00} + .009(\ketbra{00}{01}+\ketbra{01}{00}) + \mathcal O(10^{-3}),
\end{align}
and the estimated operators describing the measurement are
\begin{subequations}
  \begin{align}
  \label{eq:GSTResultCPTPEstimateE00}
  E_{00} &= \ketbra{00}{00} + .009(\ketbra{00}{01}+\ketbra{01}{00}) + \mathcal O(10^{-3}),\\
  \label{eq:GSTResultCPTPEstimateE01}
  E_{01} &= \ketbra{01}{01} - .009(\ketbra{00}{01}+\ketbra{01}{00}) + \mathcal O(10^{-3}),\\
  \label{eq:GSTResultCPTPEstimateE10}
  E_{10} &= \ketbra{10}{10} + \mathcal O(10^{-3}),\\
  \label{eq:GSTResultCPTPEstimateE11}
  E_{11} &= \ketbra{11}{11} + \mathcal O(10^{-3}).
  \end{align}
\end{subequations}
Up to the small off-diagonal component in $\rho$, $E_{00}$, and $E_{01}$, these
estimates match the expected result given by the first term. The small deviation
may be a consequence of the gauge transformation found by
\texttt{pyGSTi}, since we still have  $\mathrm{Tr}\,E_{00}\rho=1+\mathcal
O(10^{-4})$ and $\mathrm{Tr}\,E_{01}\rho=0+\mathcal O(10^{-4})$.

\subsubsection{Visualization of the Pauli transfer matrices}

\begin{figure}
  \centering
  \captionsetup[subfigure]{position=top,textfont=normalfont,singlelinecheck=off,justification=raggedright,labelformat=parens}
  \subfloat[\label{sfig:gstGxi}$G_{\texttt{XI}}$]{
    \includegraphics[width=.42\textwidth]{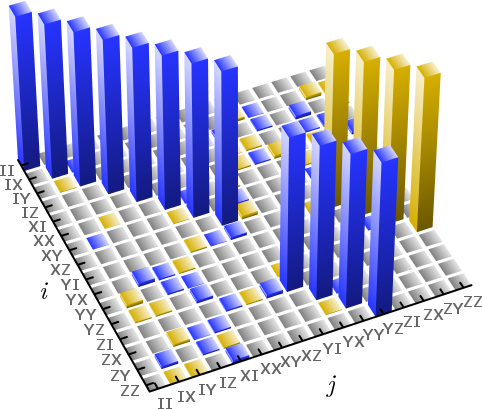}
  }\hfill
  \subfloat[\label{sfig:gstGyi}$G_{\texttt{YI}}$]{
    \includegraphics[width=.42\textwidth]{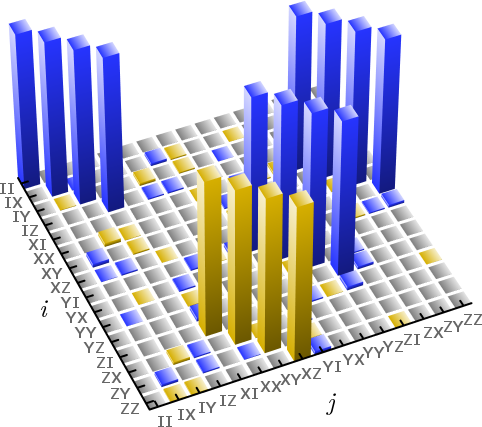}
  }\\
  \subfloat[\label{sfig:gstGix}$G_{\texttt{IX}}$]{
    \includegraphics[width=.42\textwidth]{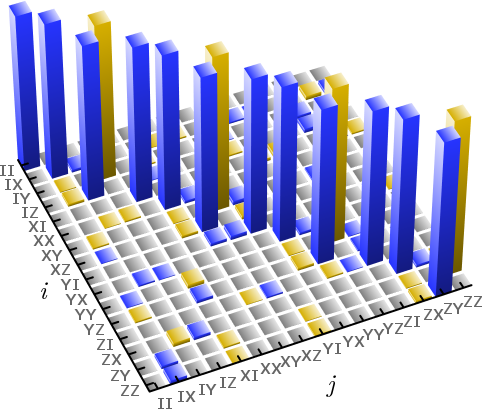}
  }\hfill
  \subfloat[\label{sfig:gstGiy}$G_{\texttt{IY}}$]{
    \includegraphics[width=.42\textwidth]{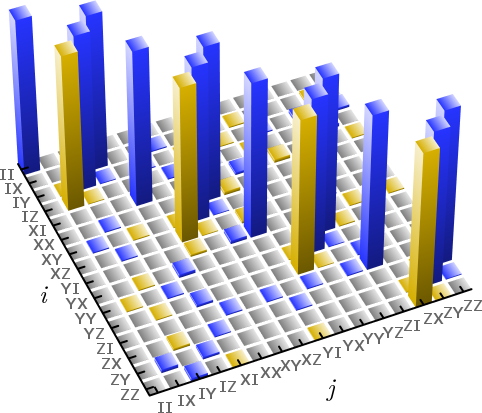}
  }\\
  \subfloat[\label{sfig:gstGii}$G_{\texttt{II}}$]{
    \includegraphics[width=.42\textwidth]{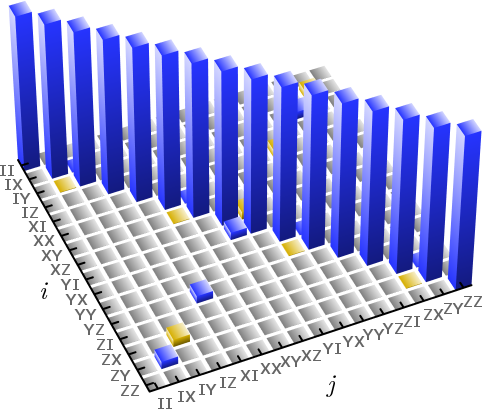}
  }\hfill
  \subfloat[\label{sfig:gstGcnot}$G_{\texttt{CNOT}}$]{
    \includegraphics[width=.42\textwidth]{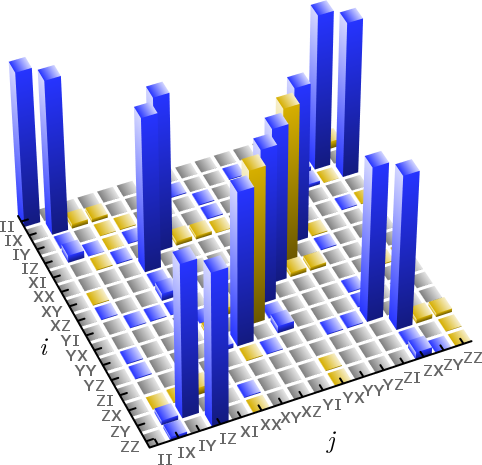}
  }\null
  \caption{Estimated gates resulting from GST experiments on a two-transmon
  simulation of 58990 circuits implemented by pulses. Shown are the Pauli
  transfer matrices $G_s$ (see \equref{eq:paulitransfermatrix}), where the subscript
  $s$ indicates the button of the black box model shown in
  \figref{fig:blackboxmodel}. Blue (yellow) bars indicate positive (negative)
  values. Gray areas indicate absolute values below $10^{-3}$. All large bars
  represent values close to $\pm1$.}
  \label{fig:gstptms}
\end{figure}

In \figref{fig:gstptms}, we show the resulting estimates for the gate processes
$\mathcal G$ in terms of their Pauli transfer matrices $G$ (see
\equref{eq:paulitransfermatrix}), which consist of $16\times16$ real numbers in
the range $[-1,1]$. One thing to note is that for all matrices $G$, the first
row $i=0$ and the first column $j=0$ (both corresponding to the axis label
$\texttt{II}$) contain only one non-zero entry at $G_{00}=1$. This reflects the
fact that the estimated maps are both trace-preserving and unital.

For each gate, the expected target gate would be represented by the same image
with all small blue and yellow bars replaced by gray areas such that only one
large bar occurs in each row and each column in \figref{fig:gstptms}. This
reflects the fact that the target gates are Clifford gates. The result shows
that the operations performed by the pulses are best described by non-Clifford
gates.

The largest deviations (i.e., the largest bars that should have been gray areas)
can be observed for the identity gate $G_{II}$ shown in  \figref{fig:gstptms}(e)
and the $\textsc{CNOT}$ gate shown in \figref{fig:gstptms}(f). For the identity
gate, the deviations occur systematically on the antidiagonal. This effect is
analyzed in the following section. It corresponds to an intrinsic error of the
form $\sigma_0^z\otimes\sigma_1^z$. A related behavior can also be observed in
experiments on the IBM Q processors (see \secref{sec:crosstalk}).

\subsubsection{Axis-angle decompositions of the estimated gates}

To gain further insight into the GST results, we apply a decomposition
algorithm to express the gate maps $\mathcal G$ in terms of a Hamiltonian
generator,
\begin{align}
  \label{eq:GSTGateDecomposition}
  \mathcal G(\rho) \approx e^{-iH} \rho\, e^{iH}.
\end{align}
Details on the decomposition algorithm are explained in
\appref{app:gatedecomposition}. A similar algorithm is implemented by the
\texttt{pyGSTi} package \cite{Nielsen2018pyGSTi0944}. Alternative approaches to
find effective Hamiltonians have been studied in  \cite{richer2013perturbative,
Willsch2016Master, WillschMadita2020PhD}.

We consider a Hamiltonian $H$ expressed in the Pauli basis (see \equref{eq:paulibasis}),
\begin{align}
  \label{eq:GSTGateHamiltonian}
  H = \sum_{k=0}^{d-1} h_k P_k/2,
\end{align}
where $d=N^2=4^n$ and $h_k\in\mathbb R$. The reason for this parametrization is that
given $h_k$, the action of $e^{-iH}$ can be interpreted as a rotation
characterized by an angle $\varphi$ and an axis $\widehat h$ according to
\begin{subequations}
  \begin{align}
    \label{eq:GSTHamiltonianAngle}
    \varphi &= \sqrt{\sum_k h_k^2},\\
    \label{eq:GSTHamiltonianAxis}
    \widehat h_k &= \frac{h_k}{\sqrt{\sum_k h_k^2}}.
  \end{align}
\end{subequations}
For instance, an ideal $\pi/2$ rotation about the $x$ axis for qubit $1$
($X_1^{\pi/2}$) corresponds to $\varphi=\pi/2$ and $\widehat h_k = \delta_{k1}$
since $P_1=I\otimes\sigma^x$ (cf.~\equref{eq:paulibasis}).

%\pdfpageattr {/Group << /S /Transparency /I true /CS /DeviceRGB>>} % fix the transparency acroread-9 simulate overprinting issue caused by these figures with /Transparency (moved to header.tex)
\begin{table}
  \caption{Axis-angle decompositions and errors of the estimated gates $\mathcal G$.
  The decompositions are obtained from the Hamiltonian given by \equref{eq:GSTGateHamiltonian},
  from which the angle $\varphi$ and the axis $\widehat h$ are extracted via
  \equaref{eq:GSTHamiltonianAngle}{eq:GSTHamiltonianAxis}.
  The notation ``$x\texttt{m}y$'' means $x\times10^{-y}$.
  Blue (yellow) colors highlight significant positive (negative) coefficients.
  The decomposition error $\gamma$ measures the error of the approximation
  given by \equref{eq:GSTGateDecomposition} (see \equref{eq:GateDecompositionHamiltonianProjectionError} for the precise definition).
  The target gates are defined in \tabref{tab:elementarygateset}.
  The target error $\eta_\Diamond$ is the diamond distance between the full
  estimated map $\mathcal G$ (not only its axis-angle decomposition) and the target gate
  (cf.~\secref{sec:diamonddistance}).
  }
  \centering
  \label{tab:gstgatedecompositions}
  \begin{tabular}{@{}ccccccc@{}}
    \toprule
    Gate & Angle $\varphi$ & Axis $\widehat h$ & Error $\gamma$ & Target & Target error $\eta_\Diamond$ \\
    \midrule
    $\mathcal G_{\texttt{XI}}$   & $0.5001\pi$ & \raisebox{-.5\height}{\includegraphics[width=4cm,height=3.3cm,trim={.65cm 0 0 .65cm},clip]{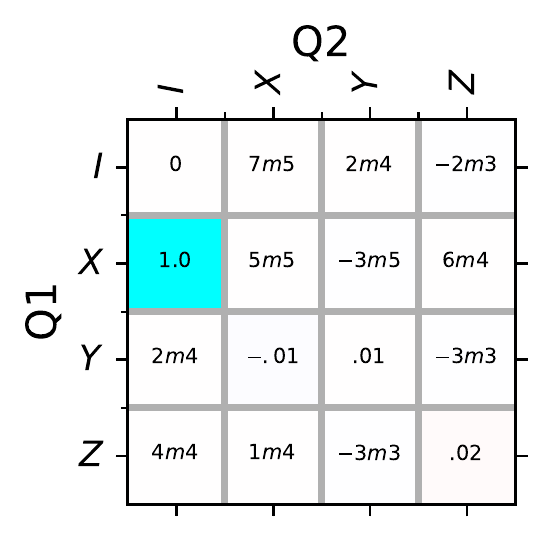}} & $2.8\e{-5}$ & $X^{\pi/2}_0$        & $0.023$ \\ % $0.097$
    $\mathcal G_{\texttt{YI}}$   & $0.5002\pi$ & \raisebox{-.5\height}{\includegraphics[width=4cm,height=3cm,trim={.65cm 0 0 1cm},clip]{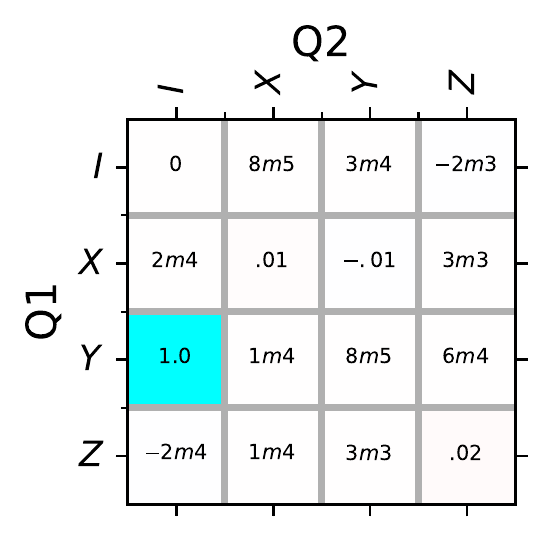}}     & $2.0\e{-5}$ & $Y^{\pi/2}_0$        & $0.023$ \\ % $0.098$
    $\mathcal G_{\texttt{IX}}$   & $0.5001\pi$ & \raisebox{-.5\height}{\includegraphics[width=4cm,height=3cm,trim={.65cm 0 0 1cm},clip]{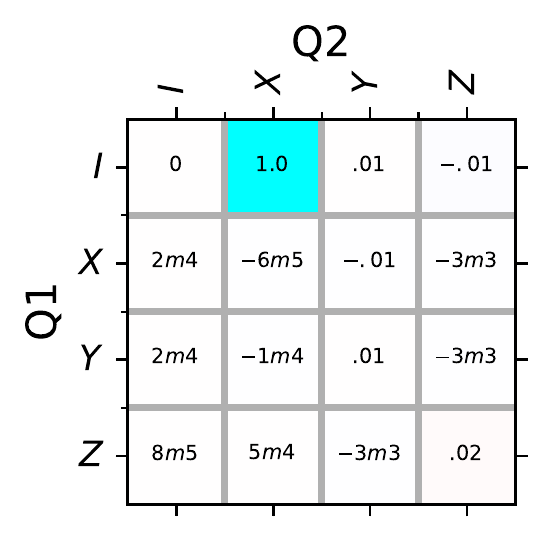}}     & $2.1\e{-4}$ & $X^{\pi/2}_1$        & $0.029$ \\ % $0.117$
    $\mathcal G_{\texttt{IY}}$   & $0.5001\pi$ & \raisebox{-.5\height}{\includegraphics[width=4cm,height=3cm,trim={.65cm 0 0 1cm},clip]{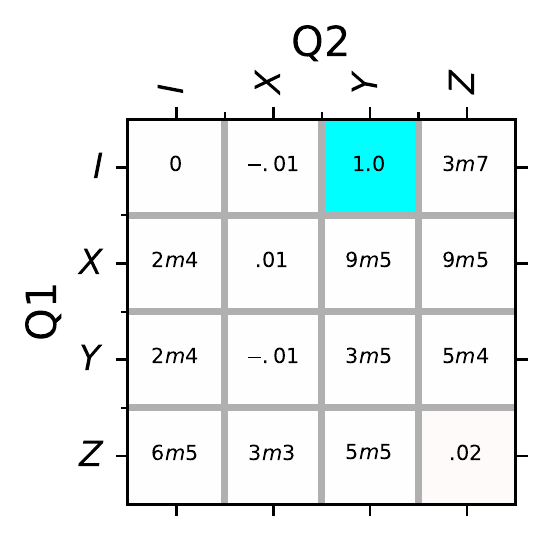}}     & $1.2\e{-5}$ & $Y^{\pi/2}_1$        & $0.022$ \\ % $0.092$
    $\mathcal G_{\texttt{II}}$   & $0.0155\pi$ & \raisebox{-.5\height}{\includegraphics[width=4cm,height=3cm,trim={.65cm 0 0 1cm},clip]{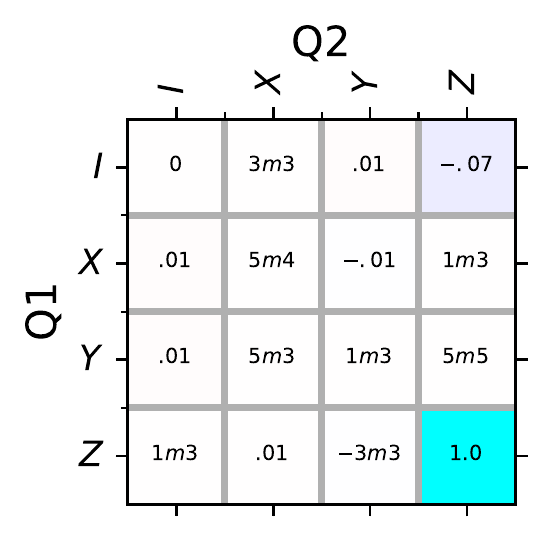}}     & $2.4\e{-5}$ & $I$                  & $0.026$ \\ % $0.138$
    $\mathcal G_{\texttt{CNOT}}$ & $0.8655\pi$ & \raisebox{-.5\height}{\includegraphics[width=4cm,height=3cm,trim={.65cm 0 0 1cm},clip]{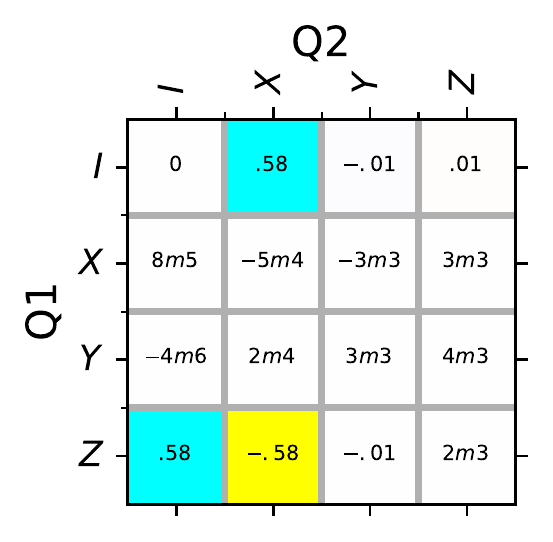}}   & $4.8\e{-2}$ & $\textsc{CNOT}_{01}$ & $0.041$ \\ % $0.176$
    \bottomrule
  \end{tabular}
\end{table}

The results of the axis-angle decomposition are given in
\tabref{tab:gstgatedecompositions}. For the single-qubit gates $\mathcal
G_{\texttt{XI}}$, $\mathcal G_{\texttt{YI}}$, $\mathcal G_{\texttt{IX}}$, and
$\mathcal G_{\texttt{IY}}$, GST almost perfectly reproduces the $\pi/2$
rotations that the GD pulses have been designed to implement. Furthermore, the
decomposition errors $\gamma$ are reasonable small, and the computed diamond
distances $\eta_\Diamond$ agree with the gate metrics reported in
\tabref{tab:ibm2gstgatemetrics}.

The most interesting results are given by the decompositions of the identity
gate $\mathcal G_{\texttt{II}}$ and the \textsc{CNOT} gate. For the identity
gate, GST extracted, without prior knowledge, the exact same type
of $\sigma_0^z\otimes\sigma_1^z$ interaction studied in
\figref{fig:statedependentfrequenciesfree}. From the parameters
given in \tabref{tab:gstgatedecompositions}, we can also compute the
interaction strength of the effective Hamiltonian $H_{ZZ}$ introduced in
\equref{eq:statedependentfrequenciesTwoQubitHamiltonian},
\begin{align}
  \label{eq:GSTResultsZZinteractionJ}
  J^{\mathrm{GST}} &= \frac{\varphi \widehat h_{15}}{2 T_{X}} = 2\pi\times\SI{46.6}{kHz},
\end{align}
where $T_X=\SI{83}{ns}$ is the duration of the identity gate, and  $\widehat
h_{15}=0.9973$ is the $ZZ$ entry of the axis $\widehat h$. This result perfectly
matches the result given by \equref{eq:statedependentfrequenciesResultsJ}. Note
that this $ZZ$ effect is still qualitatively compatible with $\mathcal
G_{\texttt{II}}$ implementing an identity gate since the angle $\varphi=0.0155\pi$
is close to 0. In \secref{sec:crosstalk}, we construct a circuit to observe
related crosstalk effects in an IBM Q processor.

For the estimated \textsc{CNOT} gate $\mathcal G_{\texttt{CNOT}}$, the
axis-angle decomposition reported in \tabref{tab:gstgatedecompositions}
yields an effective Hamiltonian of the form
\begin{align}
  \label{eq:GSTResultsCNOT}
  \frac 1 2 0.865\pi(.58[\sigma_0^z + \sigma_1^x - \sigma_0^z \sigma_1^x] - .01 (\sigma_1^y-\sigma_1^z+\sigma_0^z\sigma_1^y) + \mathcal O(10^{-3})).
\end{align}
This Hamiltonian agrees very well with the effective Hamiltonian expected from
the CR pulse (see \equref{eq:twoqubitpulseEffectiveHamiltonian}). Furthermore,
the next-order terms of the form $IY$, $IZ$, and $ZY$ resemble the contributions
observed experimentally in \cite{sheldon2016procedure} (see also Fig.~5.12 in
\cite{Willsch2016Master}). It is remarkable that GST reliably resolves all these
effects given only the measured relative frequencies $p_J$, without any prior knowledge about the transmon dynamics such as the CR pulse used internally to implement the gate.

However, one should also be careful not to put too much trust in the
decomposition of $\mathcal G_{\texttt{CNOT}}$, since in this case, the
decomposition error $\gamma$ in \tabref{tab:gstgatedecompositions} is larger
than for the other gates. One reason for this could be that a minimal Kraus
representation of $\mathcal G_{\texttt{CNOT}}(\rho)$
(cf.~\equref{eq:krausrepresentation}) needs more than one term such that
\equref{eq:GSTGateDecomposition} is not easily achievable. However, we examined
the corresponding Kraus rank by studying the singular values of the Choi matrix
$J(\mathcal G_{\texttt{CNOT}})$ (see \equref{eq:choimatrix}) and found that all
but one singular value are smaller than $3\e{-4}$. Instead, what happens in this
case is that the corresponding Lindblad operator $\mathcal L=\log\,\mathcal
G_{\texttt{CNOT}}$ does not exactly have the form required to write it in terms
of a Hamiltonian (see \equref{eq:GateDecompositionHamiltonianPTMExplicit} in
\appref{app:gatedecomposition} for more information).

Intuitively, this decomposition error expresses the lingering effect of
non-computational levels on the result after the pulse application. This
additional information, which is properly captured by the GST result visualized
in \figref{fig:gstptms}(f),  is essential to reliably predict the effect of many
repeated pulse applications. This is demonstrated in the following section.

\subsection{Predicting repeated pulse applications}
\label{sec:gatesettomographyPredictability}

To test the predictive power of GST, as opposed to that of the gate metrics
studied in \secref{sec:repeatedgates}, we simulate $r=1,\ldots,1000$ repeated gate
pulses. We compute the statistical distance $D$ given by
\equref{eq:statisticaldistance}, where $p_J$ is the distribution for an ideal,
gate-based quantum computer $\smash{p^{(\mathrm{id})}_J}$, and $\tilde p_J$ is either
given by $\smash{p^{(\mathrm{sim})}_{j_0j_1}=\abs{\braket{m_0=j_0,m_1=j_1|\Psi}}^2}$, obtained
from the transmon simulation model defined in \secref{sec:transmonmodel}, or by a GST
prediction $p^{(\mathrm{GST})}_J$.

The distribution predicted by GST is computed through \equref{eq:GSTBornRule}.
In vector form, this means that
\begin{align}
  \label{eq:GSTPrediction}
  p^{(\mathrm{GST})}_J =
  \sbra{E_J}
  (G_s)^r G_{\mathrm{(prep)}}
  \sket{\rho},
\end{align}
where $\sket{\rho}$ is the vectorized initial density matrix (see \equref{eq:GSTVectorizedRho}),
$G_{\mathrm{(prep)}}$ denotes the gates used to prepare a certain initial state,
$(G_s)^r$ represents $r$ repetitions of a certain gate $s\in\{\texttt{XI}, \texttt{YI},
\texttt{IX}, \texttt{IY}, \texttt{II},\texttt{CNOT}\}$,
and $\sbra{E_J}$ is a vectorized measurement operator (see \equref{eq:GSTVectorizedEJ}).

Note that also the initial state preparation $G_{\mathrm{(prep)}}$ explicitly
uses the GST estimated gates $G_s$, because the purpose of GST is to
self-consistently describe all preparation, gate, and measurement processes.
We test four particular initial states corresponding to the following gate sequences:
\begin{subequations}
\begin{align}
  \label{eq:GSTInitialStatePrep00}
  \ket{00} &: & G_{\mathrm{(prep)}}\sket\rho &= \sket\rho, & \\
  \ket{10} &: & G_{\mathrm{(prep)}}\sket\rho &= G_{\texttt{XI}}G_{\texttt{XI}}\sket\rho, \\
  \ket{\mathrm{++}} &: & G_{\mathrm{(prep)}}\sket\rho &= G_{\texttt{YI}}G_{\texttt{IY}}\sket\rho, \\
  \label{eq:GSTInitialStatePrepPsi}
  \ket{\Psi^-} &: & G_{\mathrm{(prep)}}\sket\rho &= G_{\texttt{CNOT}}G_{\texttt{YI}}G_{\texttt{XI}}G_{\texttt{XI}}G_{\texttt{IX}}G_{\texttt{IX}}\sket\rho,
\end{align}
\end{subequations}
where $\ket{\Psi^-}\propto\ket{01} - \ket{10}$ is the singlet state,
and $G_s$ denotes the Pauli transfer matrix of the gate map $\mathcal G_s$
(see \equref{eq:paulitransfermatrix}). For the transmon simulation, the relation
between gates and the pulses used to implement them is given in
\tabref{tab:GSTPulseImplementations}.

We test two particular GST predictions. The first prediction is based on the
CPTP estimate, for which the preparation $\rho$ and the measurements $E_J$ are
given by \equsref{eq:GSTResultCPTPEstimateRho}{eq:GSTResultCPTPEstimateE11}, and
the gates $G_s$ correspond to the Pauli transfer matrices shown in
\figref{fig:gstptms}.  The second prediction uses, for each gate $G_s$, its
corresponding decomposition in terms of an effective Hamiltonian given in
\tabref{tab:gstgatedecompositions} (see also \appref{app:gatedecomposition}).
Besides the CPTP estimate, we also analyzed the ``Full'' estimate and the TP
estimate introduced above. However,  the CPTP estimates proved to be
the most reliable (data not shown).

\begin{figure}
  \centering
  \includegraphics[width=\textwidth]{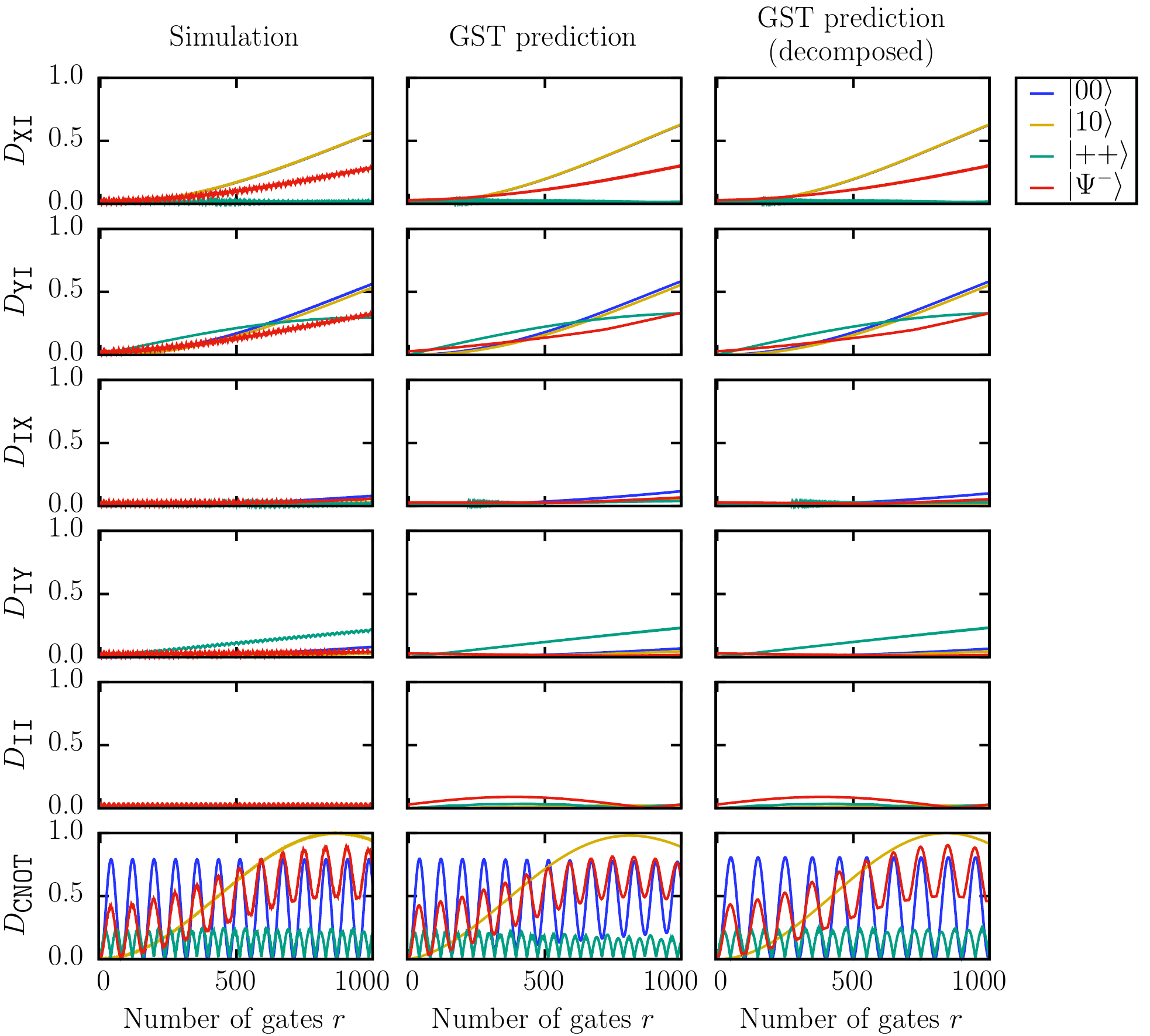}
  \caption{Test of the predictive power of GST for up to 1000 gates implemented
  through pulses. Shown is the statistical distance $D_s$ to the ideal result
  (cf.~\equref{eq:statisticaldistance}), where the subscript $s$ indicates the
  gate (see \tabref{tab:GSTPulseImplementations}), corresponding to a certain
  button of the black box model shown in \figref{fig:blackboxmodel}. The left
  column shows simulation results for the transmon model described in
  \secref{sec:transmonmodelibm2gst}. The middle column shows the GST prediction
  computed through \equref{eq:GSTPrediction}. The right column shows the same
  except that not the full estimates $\mathcal G_s$, but their Hamiltonian
  axis-angle decompositions are used (see \tabref{tab:gstgatedecompositions}).
  Different colors indicate different initial states that are also prepared
  using the estimated gates, as stated in
  \equsref{eq:GSTInitialStatePrep00}{eq:GSTInitialStatePrepPsi}. For all but
  $D_{\texttt{CNOT}}$, only every second data point is plotted (otherwise, there
  would be small oscillations on some of the lines that would render them too
  thick to be distinguishable). Note that the first 20 points on the blue and
  yellow lines in the bottom-left panel correspond to the yellow lines in
  \figref{fig:repeatedgatescnot}(c) and (d).}
  \label{fig:gstpredictability}
\end{figure}

The results of the simulation, the GST prediction, and the decomposed GST
prediction are shown in \figref{fig:gstpredictability}. The overall agreement
between the simulation and the GST prediction up to $r=500$ repetitions is
remarkable, especially since only  sequences of up to 38 gates of the form of
\equref{eq:GSTFullFiducialsGerms}  have been used to run GST. Evidently, the
predictive power slightly decreases for $r=1000$ pulse applications, but it
remains sufficient. In general, it is far superior to the predictive power of
the gate metrics, which in some cases even fail to predict the performance for
two successive pulse applications (cf.~\figref{fig:repeatedgateserrorrates}).

There are tiny oscillations in the simulation results that are most apparent on
the red lines in \figref{fig:gstpredictability}, corresponding to results
for the initial state $\ket{\Psi^-}$. These oscillations are
smoothed out in the corresponding GST predictions. Furthermore, small deviations
in amplitude between simulation and GST prediction can be seen for the identity
gate and the \textsc{CNOT} gate (last two rows in \figref{fig:gstpredictability}).
These effects stem from entanglement with non-computational states during the time
evolution which are not completely captured by the GST model. They are most pronounced
for the state $\ket{\Psi^-}$ because of the $\textsc{CNOT}$ used in the state
preparation (cf.~\equref{eq:GSTInitialStatePrepPsi}). Still, the overall performance
is sufficiently well described by the GST results.

The right column in \figref{fig:gstpredictability} might suggest that a similar
conclusion is appropriate for the predictive power of the  effective
Hamiltonians given in \tabref{tab:gstgatedecompositions}, obtained from the
axis-angle decompositions. Indeed, for all but the \textsc{CNOT} gate, the
corresponding decomposition error $\gamma$ is quite low. However, for the
\textsc{CNOT} gate, we see that the periods  of the blue, green, and red lines
in the bottom-right panel of \figref{fig:gstpredictability} are wrong. This
means that already for $r=20$  repetitions, the prediction can be wrong.
Specifically, for the initial  state $\ket{00}$ (blue line), we have
$D_{\texttt{CNOT}}\approx0.41$ for the simulation and the GST prediction, while
$D_{\texttt{CNOT}}\approx0.28$ for the decomposed GST prediction. Hence the
decomposition given by \equref{eq:GSTGateDecomposition} is not suitable to
obtain  reliable predictions for the two-qubit gate.

From the GST results, another interesting conclusion about the accuracy of the
simulation algorithm can be drawn. For $D_{\texttt{CNOT}}$ (the last row in
\figref{fig:gstpredictability}), $r=1000$ repetitions correspond to a simulation
of the time evolution over $\SI{433}{\micro s}$. Using a time step  of
$\tau=\SI{10^{-3}}{ns}$ for the product-formula algorithm
(cf.~\equref{eq:productformulasecondorderAfterDiagW}), this corresponds to more
than $4\e8$ time steps. GST, in contrast, uses only data from at most 38 gate
pulses, corresponding to  a time evolution for at most $\SI{14}{\micro s}$.
The agreement between simulation results and the GST prediction also after such
a long time evolution gives confidence in the accuracy of the product-formula
algorithm in actual applications. This was not obvious from the error analysis
of the simulated system (see \figref{fig:accuracyglobalerror}(b)).

In summary, GST and the underlying black box philosophy provide a reliable model
to effectively describe and predict quantum gate applications in complicated
two-transmon systems. It solves the problem of circular reasoning present in
many alternative proposals of tomography by self-consistently fitting
descriptions of the preparation, the gates, and the measurement to the data. Of
course, one may argue where the state preparation procedure actually starts (see
\cite{ballentine1998quantum}), and  to what extent repeated pulses really
generate the same time evolution or if non-Markovian effects dominate the
system. Furthermore, GST is not scalable to arbitrarily sized qubit systems due
to the exponential growth of the mathematical objects required in the
description. However, efforts at applying the idea of GST to systems with more
than two qubits are being explored \cite{Nielsen2017EfficientGSTMultipleQubits,
Song2019GSTforUniversalErrorMitigation, Govia2019BootstrappingQPT}. And most
importantly, GST shows that it is possible to obtain reliable descriptions for
gate-based quantum computers by only pushing well-chosen sequences of buttons on
a black box and analyzing the digital output.

\section{Conclusions}
\label{sec:chapter6conclusions}

The purpose of this chapter was to characterize the performance of quantum gate
pulses in transmon systems. We first studied common gate metrics such as the
average gate fidelity given by \equref{eq:gatemetricsaveragegatefidelity}, the
diamond distance given by \equref{eq:gatemetricsdiamonddistance}, and the
unitarity given by \equref{eq:gatemetricsunitarity}. Two byproducts of the
analysis were (1) an explicit expression for the fidelity proven in
\appref{app:prooffidelity} and (2)  a new lower bound on the diamond distance
proven in \appref{app:proofdiamondnormbound}; both of which apply, unlike
previous results, also to trace-decreasing quantum operations.

From the evaluated gate metrics, we concluded that the errors are systematic
and inherently different from simple Pauli-type errors (compare $\eta_\Diamond$
and $\eta_\Diamond^{\mathrm{Pauli}}$ in \tabref{tab:ibm2gstgatemetrics}).
Furthermore, we found that a large  part of the reduction in fidelity can be
attributed to non-unitary evolutions of (and in particular, leakage out of) the
computational subspace (compare $F_{\mathrm{avg}}$ and $u$ in
\tabref{tab:ibm5gatemetrics} and \tabref{tab:ibm5edgatemetrics}). Regarding  the
non-unitarity, repeated applications showed that the optimized gates have been
tuned to the participation of higher transmon states in the time evolution  (see
the final remark in \secref{sec:evolutiondiamonddistance}).

Although the gate metrics help in characterizing the errors, we found that none
of the metrics are suitable for predicting the performance in repeated pulse
applications. For instance, the pulse $\mathrm{CR4}_{01}$ has the worst diamond
distance (see \tabref{tab:ibm2gstgatemetrics}), but performs best in repeated
applications (see \figref{fig:repeatedgateserrorrates}(d)). Also, the gate
$X_3^{\pi/2}$ has by far the best gate metrics of all optimized transmon pulses
(see \tabref{tab:ibm5gatemetrics}), but the resulting Bloch vector
is still more tilted than for $X_1^{\pi/2}$ after repeated applications (see
\figref{fig:visualization} in \appref{app:visualization}). The
conclusion is that the gate metrics cannot reliably predict the performance in
practical applications, and that quantum gate pulses are generally much
too complicated to be characterized by a single number.

By evaluating error rates for experiments on an IBM Q processor, we found that
the \textsc{CNOT} gate performs much worse when the control qubit is in state
$\ket0$. In the simulation, this systematic error could also be observed for the
same CR2 pulse that was used in the experiment (see
\figref{fig:repeatedgatescnot}(c)). The optimized CR1 and CR4 pulses shown in
\figref{fig:crossresonancepulses}, however, did not suffer from the same
systematic error. Furthermore, we found that only by simulating a larger system
with five transmons and two resonators, we could observe similar deviations from
the ideal result in both simulation and experiment (see
\figref{fig:repeatedgatescnot}(e)). This implies that the crosstalk between
transmons, which is inherently part of the full transmon system's time
evolution, is responsible for most of the errors (see also the explicit
crosstalk experiment in \secref{sec:crosstalk}).

\clearpage

Finally, by performing an extensive GST, we found that the resulting CPTP
estimates provide a much more accurate, effective discrete description of the
evolution generated by the optimized gate pulses. The estimates were obtained
only from the relative frequencies observed in a black box model of the  system.
They showed that the implemented quantum gates, unlike the intended target
gates, are best described  by non-Clifford operations (see
\figref{fig:gstptms}). We found that GST was capable of reproducing the exact
same $ZZ$ interaction during the identity gate (see
\equref{eq:GSTResultsZZinteractionJ}) that  was also found in
\secref{sec:statedependentfrequenciesfree}. Secondly, it accurately reproduced
the effective evolution expected from the CR pulses without prior knowledge  of
the theory (see \equref{eq:GSTResultsCNOT}). And lastly, the GST results
exhibited an exceptional predictive power for up to 1000 pulse applications (see
\figref{fig:gstpredictability}) that is far superior to the conventional gate
metrics.

%% file: chap7.tex
\chapter{Selected quantum circuit experiments}
\label{cha:fullcircuitsimulations}

In this chapter, we combine the results from the previous chapters, i.e.,
the transmon simulator developed in \chapref{cha:simulation} and benchmarked in
\chapref{cha:freeevolution}, the optimized gate pulses described in
\chapref{cha:optimization}, and the actual quantum gates characterized in
\chapref{cha:gateerrors}, and apply them to several selected classes of quantum
circuits.

In principle, some more or less complicated circuits have already been simulated
for the results of the previous chapters (e.g., the gate set tomography
results in \secref{sec:gatesettomography} required 58990
quantum circuits with time evolutions up to $\SI{433}{\micro s}$).

Unlike the previous chapters, however, the present chapter focuses entirely on
relating the simulation results to experiments on quantum processors such as the
\texttt{ibmqx4} \cite{ibmqx4} or the \texttt{ibmqx5} \cite{ibmqx5} which are
available on the IBM Q Experience \cite{ibmquantumexperience2016}.

In particular, we consider:
\begin{enumerate}[leftmargin=1.6cm]
  \item an observation of crosstalk and the induced systematic errors;
  \item a characterization of the singlet state $(\ket{01}-\ket{10})/\sqrt{2}$;
  \item a test of a protocol from the theory of quantum fault tolerance.
\end{enumerate}
For each of these experiments, we run the corresponding quantum circuits on a
quantum processor  and compare the results to  data produced by the transmon
simulator. The experiment corresponding to (a) consists of a particular family
of circuits, inspired by the systematic effects observed in the simulation (see
\secref{sec:statedependentfrequenciesfree}). For this experiment, we obtain
almost perfect agreement between the simulator and an IBM Q processor (see
\secref{sec:crosstalk}). The results for (b) extend previous work published in
\cite{Michielsen2017BenchmarkingQC, Willsch2017GateErrorAnalysis}, where the
agreement between experiment and simulation was only qualitative.  We consider
several alternatives such as modified pulse  parameters or effective error
models for the environment (see \secref{sec:singletstate}). Finally, in
\secref{sec:testingfaulttolerance}, we implement a protocol from the theory of
quantum error correction and fault tolerance. We find that the protocol provides
a systematic way to improve the results in both the simulation and the real
processor, suggesting that the dominant errors are of the same nature. Some of
these results have previously been published in
\cite{Willsch2018TestingFaultTolerance}.

\section{Crosstalk experiments}
\label{sec:crosstalk}

In typical quantum computer experiments, individual qubits tend to interact even
if no gate is performed between them. This kind of \emph{crosstalk} is an
always-on coupling that is inherently part of the free evolution of the quantum
system. In this section, we examine such crosstalk using both the five-transmon
model defined in \secref{sec:transmonmodelibm5} and the five-qubit
processor \texttt{ibmqx4} of the IBM Q Experience
\cite{ibmqx4}. We study a particular circuit designed to
amplify the interaction and compare results from the simulation with data
obtained from the experiment.

Crosstalk effects during the free evolution of a two-transmon system have been
studied in \secref{sec:statedependentfrequenciesfree}. In this simple case, the
interaction resulted in state-dependent frequency shifts (see
\figref{fig:statedependentfrequenciesfree}). They could be described in terms of
an effective $ZZ$ interaction of the form of
\equref{eq:statedependentfrequenciesTwoQubitHamiltonian}. In
\secref{sec:gatesettomography}, we obtained the same effective two-level
description from the black box model of GST (see
\equref{eq:GSTResultsZZinteractionJ}).

However, the five-transmon system defined in \secref{sec:transmonmodelibm5} has
many additional states in the transmons and the resonators that take part in the
full time evolution. Furthermore, the interaction topology between the qubits is
significantly more complicated  (see \figref{fig:ibm5topology}). For this
reason, it may be hard to derive  a similarly simple, effective model
analytically, and GST is not easily doable anymore. Therefore, it seems a good
opportunity to compare the simulation results to an experiment based on the same
architecture. Obviously, this requires a certain family of quantum circuits to
characterize the effect.

\subsection{Circuit and simulation results}

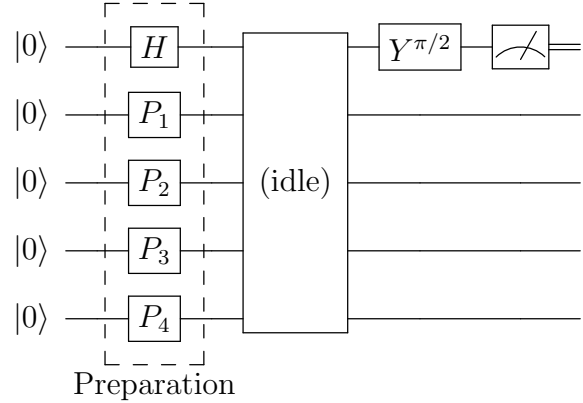
\begin{figure}
  \centering
  \[
    \Qcircuit @C=1em @R=.7em @!R {
      \lstick{\ket{0}}&\qw&\gate{H}  &\qw&\multigate{4}{\text{(idle)}}&\gate{Y^{\pi/2}}&\meter&\cw\\
      \lstick{\ket{0}}&\qw&\gate{P_1}&\qw&       \ghost{\text{(idle)}}&\qw               &\qw   &\qw\\
      \lstick{\ket{0}}&\qw&\gate{P_2}&\qw&       \ghost{\text{(idle)}}&\qw               &\qw   &\qw\\
      \lstick{\ket{0}}&\qw&\gate{P_3}&\qw&       \ghost{\text{(idle)}}&\qw               &\qw   &\qw\\
      \lstick{\ket{0}}&\qw&\gate{P_4}&\qw&       \ghost{\text{(idle)}}&\qw               &\qw   &\qw\gategroup{1}{3}{5}{3}{1.5em}{--}\\
                      &   &\mbox{Preparation}
    }
  \]
  \caption{Circuit diagram for crosstalk experiments on a five-transmon system.
  Qubit 0 is prepared in the state $\ket+$ using the $H$ gate. Qubits 1 to 4 are
  prepared in one of the states $\{\ket0,\ket+,\ket1\}$, corresponding to the
  gates $P_i\in\{I,H,X\}$. The ``idle'' gate denotes a free time evolution for a
  time $T_{\mathrm{idle}}$, without the application of any pulses or any
  externally generated interaction between the qubits. The $Y^{\pi/2}$ gate
  rotates the state $\ket+$ to the state $\ket1$. Therefore, ideally, the
  measurement of qubit 0 at the end should always return 1, corresponding to a
  Bloch vector $\vec r_0 = (0,0,-1)^T$ (see \equref{eq:crosstalkCircuitIdeal}).
  The standard gates used in this circuit are defined in
  \tabref{tab:elementarygateset} in \appref{app:gateset}.}
  \label{fig:crosstalkCircuit}
\end{figure}

In the free evolution studied in \secref{sec:statedependentfrequenciesfree}, the
resonator-mediated interaction caused the uniform superposition $\ket{+}=(\ket 0 +
\ket 1)/\sqrt 2$ of a qubit to evolve differently depending on the state of its
neighboring qubit. This is also the reason that the interaction could be
quantified accurately in the black box model of GST (see
\secref{sec:gatesettomography}). Therefore, we study a family of quantum
circuits that prepare one qubit in the state $\ket +$ and characterize its free
evolution as a function of the state of the other qubits. The general form of
the circuits is given in \figref{fig:crosstalkCircuit}.

Conceptually, the circuit is very simple. The first part of the circuit prepares
the system in a product state, where  the state of qubit 0 is given by $\ket+$.
In the simulation, this is implemented using the gate
$H=\textsc{U2}(0,\pi)=Z^{\pi/2}\,X^{\pi/2}\,Z^{\pi/2}$, which is compiled into
a single-qubit GD pulse according to
\equaref{eq:singlequbitpulseGD}{eq:singlequbitpulseGDruleVZ}. The other qubits
are prepared in one of the states $\{\ket0,\ket+,\ket1\}$, using one of the
gates $P_i\in\{I,H,X\}$. In the simulation, these gates are  implemented using
the pulse $\mathrm{zero}(T_X)$ (see \secref{sec:singlequbitzeropulse}), one GD
pulse, or two GD pulses, respectively (cf.~\tabref{tab:elementarygateset} and
\secref{sec:singlequbitGDpulse}). The idle gate is implemented by a free time
evolution for a duration $T_{\mathrm{idle}}$, corresponding to the pulse
$\mathrm{zero}(T_{\mathrm{idle}})$ (see \secref{sec:singlequbitzeropulse}).
Finally, the gate $Y^{\pi/2}$ on qubit 0 is implemented as one GD
pulse as described explicitly in \secref{sec:compiler}.

\begin{figure}
  \centering
  \captionsetup[subfigure]{position=top,textfont=normalfont,singlelinecheck=off,justification=raggedright,labelformat=parens}
  \subfloat[\label{sfig:crosstalka}$Y_0^{\pi/2}\,H_0H_1H_2H_3H_4\,\ket{00000}$]{
    \includegraphics[width=\textwidth]{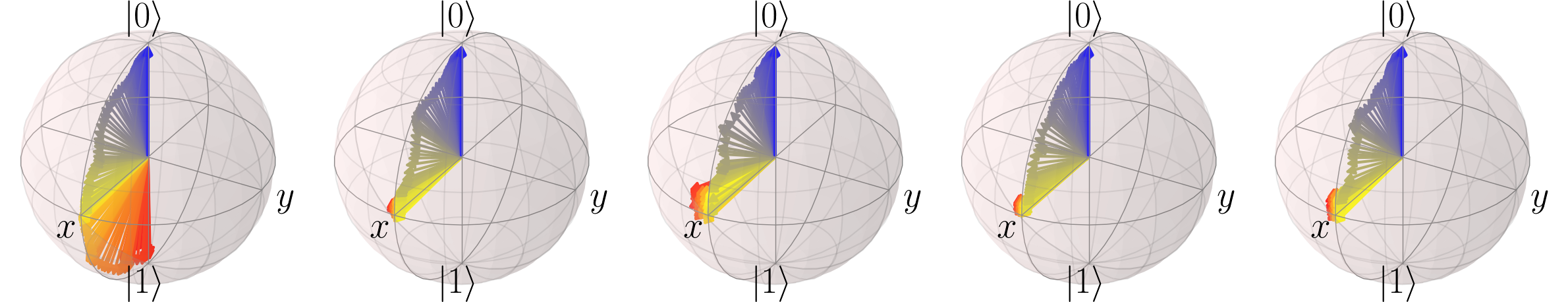}
  }\\
  \subfloat[\label{sfig:crosstalkb}$Y_0^{\pi/2}\,\text{(idle)}\,H_0H_1H_2H_3H_4\,\ket{00000}$]{
    \includegraphics[width=\textwidth]{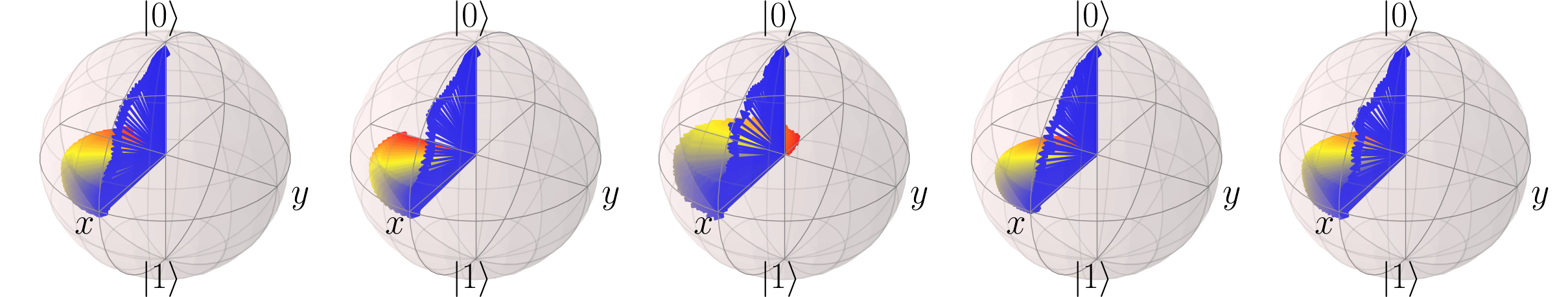}
  }\\
  \subfloat[\label{sfig:crosstalkc}$Y_0^{\pi/2}\,H_0X_1X_2X_3X_4\,\ket{00000}$]{
    \includegraphics[width=\textwidth]{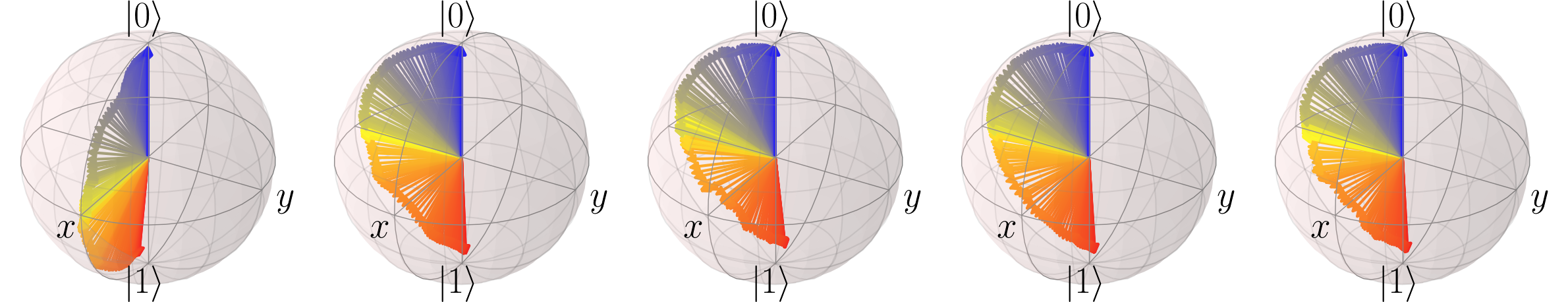}
  }\\
  \subfloat[\label{sfig:crosstalkd}$Y_0^{\pi/2}\,\text{(idle)}\,H_0X_1X_2X_3X_4\,\ket{00000}$]{
    \includegraphics[width=\textwidth]{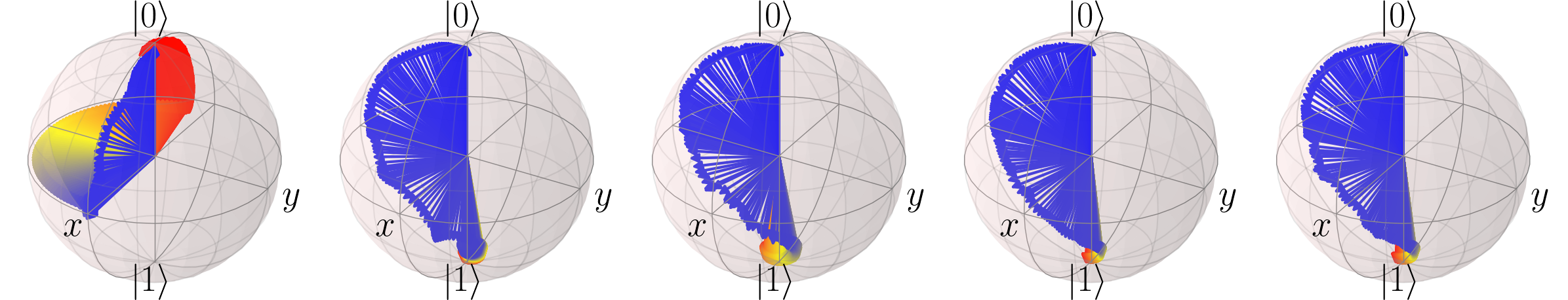}
  }\null
  \caption{Bloch-sphere representation of the time evolution of five transmon
  qubits during the application of the pulses corresponding to the gate circuit
  given in \figref{fig:crosstalkCircuit} with
  (a) $P_i=H$, $T_{\mathrm{idle}}=\SI{0}{ns}$,
  (b) $P_i=H$, $T_{\mathrm{idle}}=\SI{1600}{ns}$,
  (c) $P_i=X$, $T_{\mathrm{idle}}=\SI{0}{ns}$,
  (d) $P_i=X$, $T_{\mathrm{idle}}=\SI{1600}{ns}$.
  The time $t$ is encoded in the color of the arrows (from blue to red). The
  model parameters of the simulated transmon system are given in
  \tabref{tab:deviceibm51} and \tabref{tab:deviceibm52}. The Bloch vectors $\vec
  r_i(t)$ at time $t$ are computed according to
  \equref{eq:multiqubitblochvectorTransmonTrace} in a frame rotating at the
  frequencies $f_i$ given in \tabref{tab:deviceibm51}. The simulation results have
  been visualized with \texttt{QuTiP}
  \cite{Johansson2012Qutip, Johansson2013Qutip2}.}
  \label{fig:crosstalkExperimentBlochSpheres}
\end{figure}

Ideally, the idle gate would not affect any of the qubits. Therefore,
the ideal circuit outcome is given by
\begin{align}
  \label{eq:crosstalkCircuitIdeal}
  \ket{00000}
  \stackrel{\mathrm{Prep.}}{\mapsto} \ket{+{q_1}{q_2}{q_3}{q_4}}
  \stackrel{\mathrm{Idle}}{\mapsto} \ket{+{q_1}{q_2}{q_3}{q_4}}
  \stackrel{Y^{\pi/2}_0}{\mapsto} \ket{1{q_1}{q_2}{q_3}{q_4}}
  \stackrel{\mathrm{Meas.}}{\mapsto} 1,
\end{align}
where $\ket{q_i}=P_i\ket{0}\in\{\ket0,\ket+,\ket1\}$. This corresponds to
a Bloch vector $\vec r_0$ that is aligned with the negative $z$ axis, i.e.,
$r_0^z=-1$.

However, crosstalk between the transmons during the idle gate is expected to
change the state of qubit 0 in a way that depends on the state of the other
qubits. We simulate the five-transmon system by solving the TDSE given by
\equref{eq:tdse3} with the time step $\tau=\SI{10^{-3}}{ns}$ as described in
\secref{sec:simulationsoftware}. A few representative examples of the evolution
of the qubits' Bloch vectors are shown in
\figref{fig:crosstalkExperimentBlochSpheres}.

In \figref{fig:crosstalkExperimentBlochSpheres}(a), all qubits are rotated to
the positive $x$ axis using an $H$ gate. The idle gate is absent such that the
final $Y^{\pi/2}$ gate rotates qubit 0 to the negative $z$ axis as expected. In
this case, crosstalk between the transmons only makes the Bloch vectors wiggle
slightly, which can be seen in the not perfectly straight motion of the arrows.

When
the duration of the idle gate is set to $T_{\mathrm{idle}}=\SI{1600}{ns}$,
however, each Bloch vector keeps on rotating clockwise about the $z$ axis
(see \figref{fig:crosstalkExperimentBlochSpheres}(b)). Furthermore, the magnitude
of the Bloch vectors shrinks, which represents entanglement forming between the
qubits. Since the Bloch vector of qubit 0 ends up near the negative $y$ axis,
the final $Y^{\pi/2}$ gate does not rotate the qubit, resulting in a 50\% error
of the final measurement.

In \figref{fig:crosstalkExperimentBlochSpheres}(c) and (d), we see that the
crosstalk effect on qubit 0 is even stronger if qubits 1 to 4 are prepared in the state
$\ket 1$. The same idle time $T_{\mathrm{idle}}=\SI{1600}{ns}$ renders the Bloch
vector of qubit 0 on the negative $x$ axis. The final $Y^{\pi/2}$ gate then
rotates the qubit back to $\ket0$ (see
\figref{fig:crosstalkExperimentBlochSpheres}(d)), while it should have ended up
in state $\ket1$. Using this effect, one could,  in principle, engineer a
circuit with 100\% error rate. Note that this effect  qualitatively agrees with
the increased frequency observed in \figref{fig:statedependentfrequenciesfree},
since if the qubit's frequency is larger than that of the rotating frame,
the Bloch vector performs a clockwise rotation about the $z$ axis. In the
next section, we study to what extent the same effect can be observed in
a real five-transmon processor.

\subsection{Comparison with experiments on the IBM Q Experience}

When executing the circuit in \figref{fig:crosstalkCircuit} on a real five-qubit
processor, the result is given by the relative frequencies
$p_{0/1}^{\mathrm{(exp)}}$, corresponding to the relative number of events that
the measurement of qubit 0 returned 0 or 1.  From the frequencies, we compute
the $z$ component of the qubit's Bloch vector according to
\equref{eq:singlequbitblochvector} such that $r_0^z =
\smash{p_0^{\mathrm{(exp)}} - p_1^{\mathrm{(exp)}}} =
2\smash{p_0^{\mathrm{(exp)}}} - 1$. As the ideal result for each instance of the
circuit in \figref{fig:crosstalkCircuit} is $\ket1$ (i.e., $r_0^z=-1$), the
error rate in terms of the statistical distance $D$ (see
\equref{eq:statisticaldistance}) is given by $p_0^{\mathrm{(exp)}}$. In terms of
the coordinates of the Bloch vector $\vec r_0$, we have
\begin{align}
  \label{eq:crosstalkStatisticalDistanceBlochZ}
  D=\frac{1+r_0^z}2.
\end{align}
In the simulation, we compute the Bloch vector $\vec r_0$ through
\equref{eq:multiqubitblochvectorTransmonTrace}. Since this formula includes a
projection onto the computational subspace,  using
\equref{eq:crosstalkStatisticalDistanceBlochZ} for the error effectively yields
$D=p_0^{\mathrm{(sim)}}+p_{\ge2}^{\mathrm{(sim)}}/2$. Conceptually, this means
that a measurement of the transmon in a higher, non-computational state is
interpreted  as a 50\% chance to find the qubit in state $\ket0$ or $\ket1$.
Another reasonable alternative would be to always count a higher state as
$\ket1$. In practice, however, this conceptual choice only has a negligible
effect on the result, because $p_{\ge2}^{\mathrm{(sim)}}<0.002$ in all cases
under investigation.

\begin{figure}
  \centering
  \includegraphics[width=\textwidth]{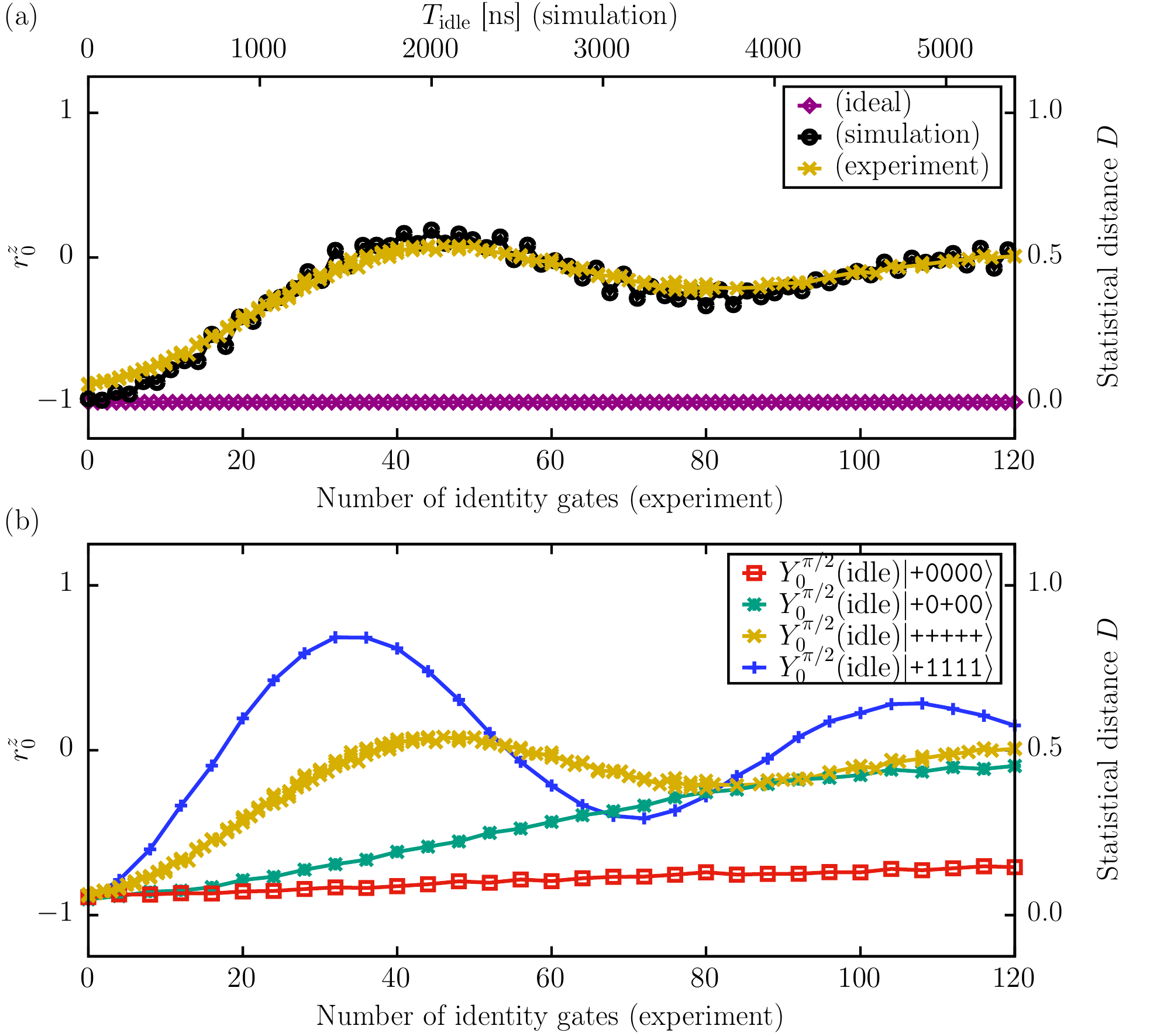}
  \caption{Results of the crosstalk experiments based on the circuit given in \figref{fig:crosstalkCircuit};
  (a) ideal, simulation, and experimental results for the circuit $Y^{\pi/2}_0\mathrm{(idle)}\ket{+\!+\!+\!+\!+}$;
  (b) experimental results for the same type of circuit with different state preparations $P_i$ as indicated by the legend.
  Shown is the $z$ component of the Bloch vector $\vec r_0=\langle\vec\sigma_0\rangle$ of qubit 0.
  The right axis shows the error rate in terms of the statistical distance $D$
  defined in \equref{eq:statisticaldistance}, which is linked to $r_0^z$ via \equref{eq:crosstalkStatisticalDistanceBlochZ}.
  In the experiment, the ``idle'' part is implemented by a certain number of identity
  gates given on the bottom axis.
  In the simulation, the ``idle'' part is implemented by the pulse
  $\mathrm{zero}(T_{\mathrm{idle}})$ (see \secref{sec:singlequbitzeropulse}),
  where $T_{\mathrm{idle}}$ is shown on the top axis.
  The simulation results in (a) at $T_{\mathrm{idle}}=\SI{0}{ns}$
  ($T_{\mathrm{idle}}=\SI{1600}{ns}$) correspond to the time evolutions
  visualized in \figref{fig:crosstalkExperimentBlochSpheres}(a) (\figref{fig:crosstalkExperimentBlochSpheres}(b)).}
  \label{fig:crosstalkExperimentResults}
\end{figure}

We use the five-qubit processor \texttt{ibmqx4} on the IBM Q Experience
\cite{ibmqx4} to execute the circuit given in
\figref{fig:crosstalkCircuit}. In the experiment, the idle gate is implemented
by $n_{\mathrm{idle}}=0,\ldots,120$ identity gates. All experiments were run
between February 21, 2018 and March 19, 2018 with 8192 shots.

In \figref{fig:crosstalkExperimentResults}(a), we show results from both
simulation and experiment for the circuit
$Y_0^{\pi/2}\mathrm{(idle)}\ket{+\!+\!+\!+\!+}$, corresponding to the
Bloch-vector evolutions in \figref{fig:crosstalkExperimentBlochSpheres}(a) and
(b). The agreement between simulation and experiment is remarkable. Both are
equally far away from the ideal result $r_0^z=-1$. The only observable
difference between simulation and experimental results for $r_0^z$ is the
slightly reduced amplitude and the tiny oscillations between successive circuit
simulations. These are apparently not resolved by the experiment and appear to
be smoothed out. This points out a limitation in the degree to which the device
could be engineered to implement the laws of quantum theory on a macroscopic
scale. The most obvious explanation would be that the deviation is caused by
influences from the environment that are not included in the simulation model
(cf.~also \figref{fig:singletcircuitwitherrorchannels}, where a similarly
reduced amplitude can be described in terms of environmental effects). Further
investigation by including the qubits' readout resonators or an effective
environment in the simulation (in the spirit of \secref{sec:extractfoster})
could shed light on this aspect. Nevertheless, the otherwise excellent agreement
seems to suggest that for this particular experiment, the most dominant sources
of error are the crosstalk effects that are inherently included in the unitary
evolution of the simulated transmon model defined in \secref{sec:transmonmodel}.

Comparing the time scale used in the simulation (top axis in
\figref{fig:crosstalkExperimentResults}(a)) with the number of identity gates
used in the experiment (bottom axis in
\figref{fig:crosstalkExperimentResults}(a)), we find that a single identity gate
corresponds to $T_{\mathrm{idle}}=\SI{45}{ns}$. According to the calibration
data obtained from the processor, the single-qubit gate duration at the time of
the experiment was \SI{83.3}{ns} with a buffer of \SI{6.7}{ns} between gates. A
difference in the exact time scales is to be expected because the device
parameters of the simulated transmon system (see \figref{fig:ibm5topology} and
\tabref{tab:deviceibm51}) do not exactly match those of \texttt{ibmqx4}
\cite{ibmqx4}. Furthermore, the time scale depends sensitively on the
transmon-resonator couplings $G$ given in \tabref{tab:deviceibm52},  in the
sense that the effective longitudinal interaction $J$ given by
\equref{eq:statedependentfrequenciesTwoQubitHamiltonian} is proportional to
$G^4$ \cite{billangeon2015longitudinal}. In the experiment, $G$ is not measured
but estimated from simulations \cite{JayFirat2016}. The fact that the transmon
simulator and the IBM Q processors still agree so well suggests that this
effect is independent of the exact values of the parameters of the full
Hamiltonian given by \equsref{eq:Htotal}{eq:HCC}.

The agreement between simulation and experiment shown in
\figref{fig:crosstalkExperimentResults}(a) suggests that also other trends seen
in the simulation are observable in the  experiment. For instance, in
\figref{fig:crosstalkExperimentBlochSpheres}(d), we found that the
crosstalk-induced rotation during the idle gate grows in speed when all other
transmons are prepared in state $\ket1$. To test this hypothesis, we  execute
the circuit given in \figref{fig:crosstalkCircuit} for three additional state
preparations on the processor \texttt{ibmqx4} \cite{ibmqx4}. The results are
shown in \figref{fig:crosstalkExperimentResults}(b).

Indeed, we see that with all other qubits in state $\ket1$, the deviation from
the ideal result, indicated by the deviation of the blue plusses from $r_0^z=-1$,
becomes even stronger. Furthermore, if only the
central qubit is prepared in the state $\ket+$ (green crosses), the impact is not as pronounced.
This also demonstrates that crosstalk effects go beyond the
simple two-qubit picture studied in \secref{sec:statedependentfrequenciesfree}.

Finally, with all other qubits in state $\ket0$, the Bloch vector of qubit 0
effectively stands still during the idle gate (red squares in
\figref{fig:crosstalkExperimentResults}(b)).  As this is the same situation that
has been chosen to determine qubit frequencies in \secref{sec:evaluator} (see
the text below \equref{eq:singlequbitblochvectorTimeEvolutionRotating}), it
shows that our method is in agreement with the experiment.

\section{Characterization of the singlet state}
\label{sec:singletstate}

In this section, we consider a conceptually simple experiment that is commonly
used to test whether a system exhibits quantum behavior, in the sense that
it produces correlations described by an entangled state. Specifically, we
consider the singlet state
\begin{align}
  \label{eq:singletstate}
  \ket{\Psi^-} &= \frac 1 {\sqrt{2}} (\ket{01} - \ket{10}),
\end{align}
which is one of the maximally entangled Bell states defined in
\equaref{eq:bellstatephi}{eq:bellstatepsi}. A property of this particular Bell
state is that independent measurements of both qubits always yield opposite
results. This property is called \emph{perfectly anti-correlated}. From
\equref{eq:singletstate}, we see that this holds when measuring in the
computational basis, i.e., when measuring the observables $\sigma_0^z$ and
$\sigma_1^z$: whenever one qubit is found in state $\ket0$ (with eigenvalue $+1$
since $\sigma^z\ket0=+\ket0$), the other qubit is found in state $\ket1$ (with
eigenvalue $-1$ since $\sigma^z\ket1=-\ket1$). One can show \cite{NielsenChuang}
that this property holds for any other pair of observables
$\vec\sigma_0\cdot\vec v$ and $\vec\sigma_1\cdot\vec v$, where $\vec v\in\mathbb
R^3$ is the measurement  direction with $\|\vec v\|\,=1$, and $\vec\sigma_i
=(\sigma_i^x, \sigma_i^y, \sigma_i^z)$ is the vector of Pauli matrices on qubit
$i$. Moreover, if $\vec v$ is replaced by $\vec a$ for qubit 0 and $\vec b$ for
qubit 1, a short calculation yields
\begin{align}
  \label{eq:singletExpectationCorrelation}
  \braket{\Psi^-|(\vec\sigma_0\cdot\vec a)(\vec\sigma_1\cdot\vec b)|\Psi^-}
  = -\vec a\cdot\vec b = -\cos\vartheta,
\end{align}
where $\vartheta=\sphericalangle(\vec a,\vec b)$ is the angle between the
measurement directions $\vec a$ and $\vec b$. Thus, $\vartheta$ determines the
degree of correlation expected in the measurements of both qubits. Furthermore,
we have
\begin{subequations}
\begin{align}
  \label{eq:singletExpectation0}
  \braket{\Psi^-|\vec\sigma_0\cdot\vec a|\Psi^-} = 0,\\
  \label{eq:singletExpectation1}
  \braket{\Psi^-|\vec\sigma_1\cdot\vec b|\Psi^-} = 0.
\end{align}
\end{subequations}
This means that, when measuring only one of the two observables, a measurement
is expected to produce an equal number of $+1$'s and $-1$'s.

\subsection{Experiment}

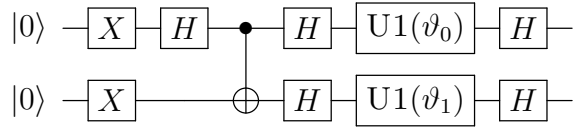
\begin{figure}
  \centering
  \[
    \Qcircuit @C=.8em @R=.7em {
      &\lstick{\ket{0}}&\gate{X}&\gate{H}&\ctrl{1}&\gate{H}&\gate{\textsc{U1}(\vartheta_0)}&\gate{H}&\qw\\
      &\lstick{\ket{0}}&\gate{X}&\qw     &\targ   &\gate{H}&\gate{\textsc{U1}(\vartheta_1)}&\gate{H}&\qw\\
    }
  \]
  \caption{Circuit diagram for experiments on the singlet state $\ket{\Psi^-}$
  given by \equref{eq:singletstate}. The first four gates prepare the state. The
  last six gates are used to implement a measurement of the state in the
  eigenbasis of $(\vec\sigma_0\cdot\vec a)(\vec\sigma_1\cdot\vec b)$ as a
  function of $\vartheta_0$ and $\vartheta_1$
  (cf.~\equref{eq:singletMeasurementBasisChange}). All circuit elements
  are defined in \tabref{tab:elementarygateset} in
  \appref{app:gateset}.}
  \label{fig:singletcircuit}
\end{figure}

We perform an in-depth characterization of the singlet state by studying the
expectation values given in
\equsref{eq:singletExpectationCorrelation}{eq:singletExpectation1} for various
measurement directions $\vec a$ and $\vec b$. Specifically, we choose $\vec a =
(0,\sin\vartheta_0,\cos\vartheta_0)^T$ and $\vec b =
(0,\sin\vartheta_1,\cos\vartheta_1)^T$ such that $-\vec a\cdot\vec b =
-\cos(\vartheta_1-\vartheta_0)$. The quantum gate circuit to implement this
experiment is given in \figref{fig:singletcircuit}. It consists of a few gates
used to prepare the singlet state (using the fact that
$\ket{\Psi^-}=\textsc{CNOT}_{01}\,H_0\,X_0X_1\ket{00}$), and a set of gates to
implement the measurement. For the latter, note that the gate sequence
$H\textsc{U1}(\vartheta_0)H$ can be used to transform between the computational
basis (i.e.~the eigenbasis of $\sigma^z$) and the eigenbasis of
$\vec\sigma\cdot\vec a$ since
\begin{align}
  \label{eq:singletMeasurementBasisChange}
  (H\textsc{U1}(\vartheta_0)H)^\dagger\,\sigma^z\,(H\textsc{U1}(\vartheta_0)H)
    = \sin\vartheta_0\,\sigma^y+\cos\vartheta_0\,\sigma^z = \vec\sigma\cdot\vec a.
\end{align}
A similar relation holds for $\vartheta_1$ and $\vec b$.
One can compute the ideal probability distribution of
the final state of the circuit shown in \figref{fig:singletcircuit} as
\begin{align}
  p(j_0j_1\vert\vartheta_0\vartheta_1)
  &= \abs{\braket{j_0j_1|(H\textsc{U1}(\vartheta_0)H)\otimes (H\textsc{U1}(\vartheta_1)H)|\Psi^-}}^2 \nonumber\\
  \label{eq:singletProgramGatesResult}
  &= \frac{1-(-1)^{j_0+j_1}\cos(\vartheta_1-\vartheta_0)}{4}.
  %= \begin{cases}
  %  \abs{e^{i\vartheta_1} - e^{i\vartheta_0}}^2/8 & (j_0j_1=00,11)\\
  %  \abs{e^{i\vartheta_1} + e^{i\vartheta_0}}^2/8 & (j_0j_1=01,10)\\
  %\end{cases}
\end{align}
To verify that this distribution complies with
\equsref{eq:singletExpectationCorrelation}{eq:singletExpectation1}, one can
compute the expectation values as
\begin{subequations}
  \begin{align}
    \label{eq:singletProgramExpectationCorrelation}
    E_{01}(\vartheta_0,\vartheta_1) &= \sum_{j_0j_1} (-1)^{j_0}(-1)^{j_1} p(j_0j_1\vert\vartheta_0\vartheta_1) = -\cos(\vartheta_1-\vartheta_0),\\
    \label{eq:singletProgramExpectation0}
    E_0(\vartheta_0,\vartheta_1) &=\sum_{j_0j_1} (-1)^{j_0} p(j_0j_1\vert\vartheta_0\vartheta_1) = 0,\\
    \label{eq:singletProgramExpectation1}
    E_1(\vartheta_0,\vartheta_1) &=\sum_{j_0j_1} (-1)^{j_1} p(j_0j_1\vert\vartheta_0\vartheta_1) = 0,
  \end{align}
\end{subequations}
where we used the fact that $j_0$ ($j_1$) labels the eigenvalue $(-1)^{j_0}$
($(-1)^{j_1}$) of $\vec\sigma_0\cdot\vec a$ ($\vec\sigma_1\cdot\vec b$).

When using devices such as an IBM Q processor for this experiment, one does not obtain a
probability distribution $p(j_0j_1\vert\vartheta_0\vartheta_1)$ directly.
Instead, the device produces a number of samples (usually 8192) for each setting
of $\vartheta_0$ and $\vartheta_1$. The resulting frequencies for each bit
string,
\begin{align}
  \label{eq:singletFrequencies}
  f(j_0j_1) = \frac{\text{\# of samples $j_0j_1$}}{8192},
\end{align}
are then used as estimators for the probabilities
$p(j_0j_1\vert\vartheta_0\vartheta_1)$. We denote the corresponding estimates
for the expectation values given in
\equsref{eq:singletProgramExpectationCorrelation}{eq:singletProgramExpectation1}
by $F_{01}(\vartheta_0,\vartheta_1)$, $F_{0}(\vartheta_0,\vartheta_1)$, and
$F_{1}(\vartheta_0,\vartheta_1)$, respectively. The goal of the experiment is to
see how well $F_{01},F_0$, and $F_1$ agree with the theoretical result
$E_{01},E_0$, and $E_1$.

Some aspects of the experiment have already been studied.  In
\cite{Michielsen2017BenchmarkingQC}, we performed the experiment on an IBM Q
processor as part of a collection of benchmarking circuits (the results are
plotted as hollow circles in \figaref{fig:singletfitsim}{fig:singletfitchannel}
below). In \cite{Willsch2017GateErrorAnalysis}, we simulated the experiment
using the optimized pulses discussed in \chapref{cha:optimization} (the
corresponding results are shown in \figref{fig:singletfitsim}(a) and (b)). The
particular pulse parameters are given in
\tabref{tab:deviceibm2gstPulseParametersGD} and
\tabref{tab:deviceibm2gstPulseParametersCR} in \appref{app:pulseparameters}.

As can be seen in \figref{fig:singletfitsim}(a) and (b), there is quite some
difference in the results, in the sense that the IBM Q processor produces rather
large, systematic deviations while the simulation performs reasonably well in
comparison with the ideal result (dashed black lines). Note that the gate
metrics given in \tabref{tab:ibm2gstgatemetrics} are almost equal to the error
rates of the processor, so the deviations between simulation and experiment are
not captured by the gate metrics (cf.~also the conclusions in
\secref{sec:chapter6conclusions}).

\begin{figure}[p]
  \centering
  \includegraphics[width=\textwidth]{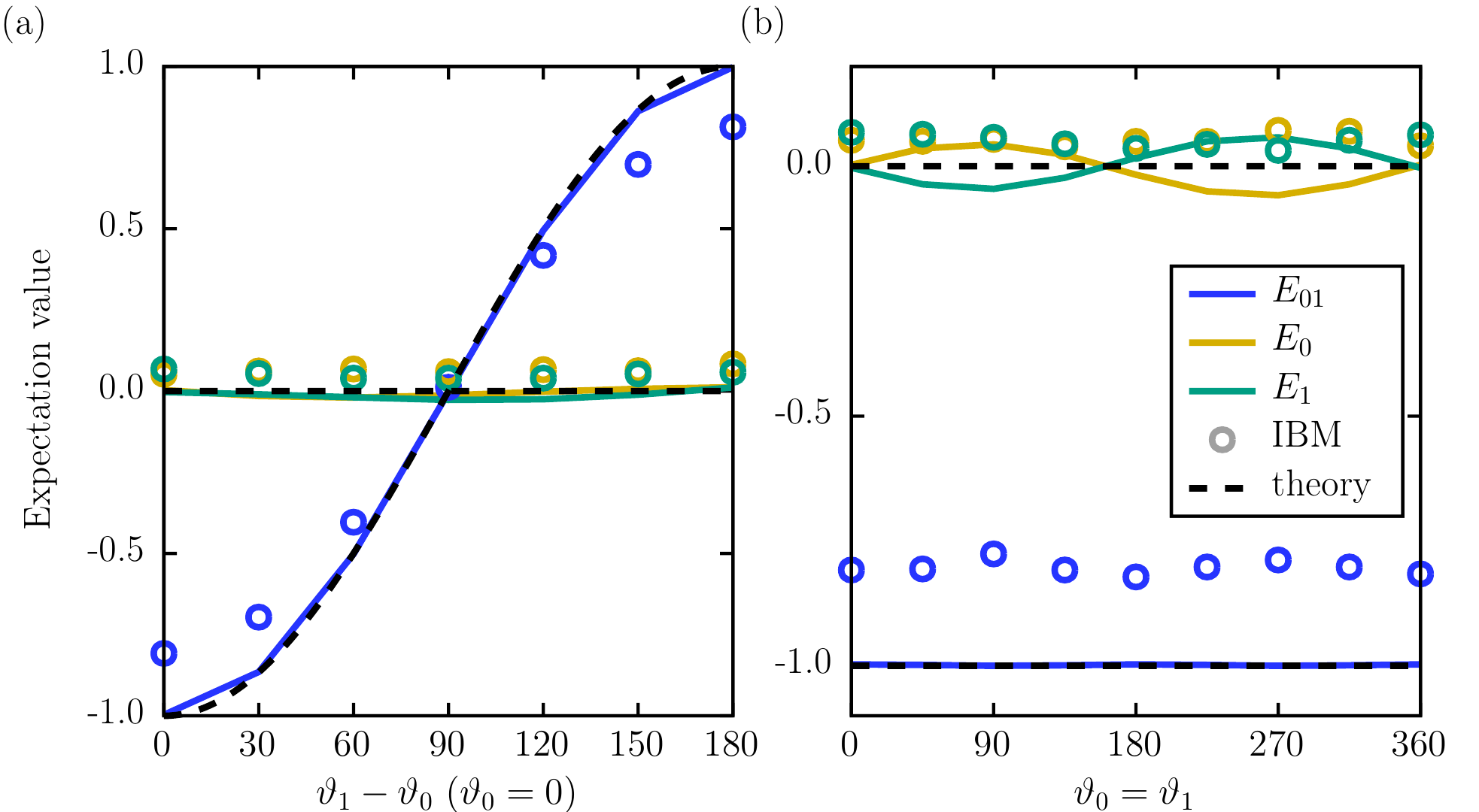}
  \includegraphics[width=\textwidth]{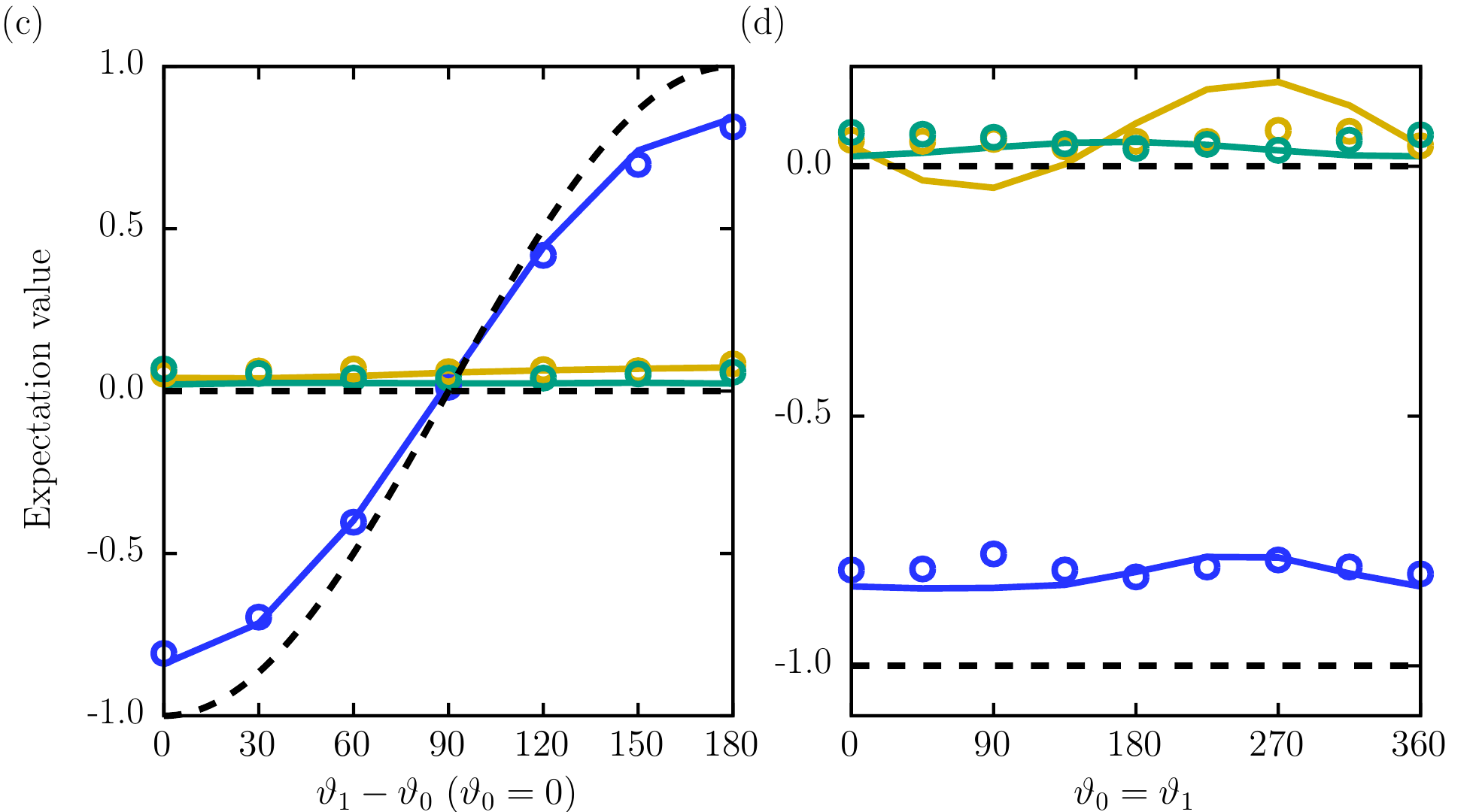}
  \caption{Results for the singlet-state characterization using (a),(b) the
  original optimized pulse parameters given in \appref{app:pulseparameters} and
  (c),(d) the modified pulse parameters that have been fitted to the
  experimental results (see
  \tabref{tab:deviceibm2gstPulseParametersAfterSingletFit}). Lines show the
  expectation values $E_{01}$, $E_0$, and $E_1$ defined in
  \equsref{eq:singletProgramExpectationCorrelation}{eq:singletProgramExpectation1},
  computed using the probabilities $p(j_0j_1\vert\vartheta_0\vartheta_1)$ from
  the two-transmon simulation (cf.~\secref{sec:transmonmodelibm2gst}). In all
  figures, hollow circles represent the expectation values $F_{01}$, $F_0$, and $F_1$
  obtained from experiments on the IBM Q processor that was available on
  February 16, 2017 (see \cite{Michielsen2017BenchmarkingQC}), computed using
  the relative frequencies $f(j_0j_1\vert\vartheta_0\vartheta_1)$ given by
  \equref{eq:singletFrequencies}. Dashed lines
  represent the theoretical result, computed from the ideal probability
  distribution given by \equref{eq:singletProgramGatesResult}.}
  \label{fig:singletfitsim}
\end{figure}

\begin{table}[t]
  \caption{Pulse parameters used for the simulation results shown in
  \figref{fig:singletfitsim}. The original values represent the optimized pulses
  discussed in \chapref{cha:optimization} (the corresponding results are shown
  in \figref{fig:singletfitsim}(a) and (b)). They are taken from the pulse
  parameters given in \tabref{tab:deviceibm2gstPulseParametersGD} and
  \tabref{tab:deviceibm2gstPulseParametersCR} in \appref{app:pulseparameters}.
  The fitted values result from minimizing the objective function given by
  \equref{eq:singletRMSobjectivefunction} (the corresponding results are shown
  in \figref{fig:singletfitsim}(c) and (d)). The duration $T_X$ of the
  single-qubit GD pulses is the same for all gates. The two-qubit \textsc{CNOT}
  pulse is based on the CR2 scheme (see \figref{fig:crossresonancepulses}(c)).}
  \centering
  \label{tab:deviceibm2gstPulseParametersAfterSingletFit}
\begin{tabular}{@{}llrr@{}}
  \toprule
  Pulse name & Parameter & Original value & Fitted value \\
  \midrule
  \texttt{xpih-0} & $f_0\,[\mathrm{GHz}]$                 &            5.3463   &          5.34647  \\
                  & $(\Omega_X^{\pi/2})_0$                &  0.002221           &       0.001980  \\
                  & $(\beta_X^{\pi/2})_0\,[\mathrm{ns}]$  &    0.2309           &         0.2584  \\
  \midrule
  \texttt{xpih-1} & $f_1\,[\mathrm{GHz}]$                 &            5.1167   &          5.11691  \\
                  & $(\Omega_X^{\pi/2})_1$                &  0.002269           &       0.002296  \\
                  & $(\beta_X^{\pi/2})_0\,[\mathrm{ns}]$  &    0.2891           &         0.4698  \\
  \midrule
  \texttt{cnot-0-1} & $T_{\mathrm{CR}}\,[\mathrm{ns}]$      &     102.9746   &          102.9720  \\
                    & $\Omega_{\mathrm{CR}}$                &   0.01111      &       0.00979  \\
                    & $(\Omega_X^\pi)_C$                    &  0.004444      &       0.003415  \\
                    & $(\beta_X^\pi)_C\,[\mathrm{ns}]$      &    0.2193      &         0.2239 \\
                    & $(\Omega_X^{\pi/2})_T$                &  0.002269      &       0.002032  \\
                    & $(\beta_X^{\pi/2})_T\,[\mathrm{ns}]$ &    0.2891       &         0.2742 \\
  \midrule
  (for all pulses) & $T_X\,[\mathrm{ns}]$                  &                83   &          82.893  \\
  \bottomrule
\end{tabular}
\end{table}

\subsubsection{Adjusting the pulse parameters}

Since the gates implemented by the optimized pulses for the two-transmon
simulation model perform much better, an interesting question is whether the
errors can be reproduced solely from the unitary control errors caused by the
time-dependent pulses. To address this question, we try to fit the pulse
parameters such that simulation results and experimental results match. If this
is possible, it would suggest that the errors for this experiment are largely
due to the pulses chosen to implement the gates. Otherwise, it would point at
external sources of error such as effects due to the environment or the
additional transmons and resonators on the processor (see below).

To fit the pulse parameters, we utilize the  Nelder--Mead optimization method
introduced in \secref{sec:neldermead}. In every iteration, the singlet
characterization is evaluated using the current pulse parameters. The objective
function used for the optimization is given by the root mean square
\begin{align}
  \label{eq:singletRMSobjectivefunction}
  \sqrt{\frac{1}{3N_{\vartheta}}\sum\limits_{\vartheta_0,\vartheta_1}
  \left((E_{01}(\vartheta_0,\vartheta_1)-F_{01}(\vartheta_0,\vartheta_1))^2
  + \sum\limits_{l=0,1}(E_{l}(\vartheta_0,\vartheta_1)-F_{l}(\vartheta_0,\vartheta_1))^2
  \right)},
\end{align}
where $N_{\vartheta}=16$ is the number of configurations
($\vartheta_0,\vartheta_1$) used in the experiment. The 13 fitted pulse
parameters  are shown in
\tabref{tab:deviceibm2gstPulseParametersAfterSingletFit}. Most of the resulting
parameters differ by less than 10\% from their respective initial values.

The performance of the singlet characterization using the fitted pulse
parameters  is shown in \figref{fig:singletfitsim}(c) and (d). We see that
applying the new pulses yields much better agreement between simulation and
experimental results. The expectation values $E_{01}$, $E_0$, and $E_1$ deviate
equally strongly from the theory. This is especially true for the case shown in
\figref{fig:singletfitsim}(c). However, the large oscillation of $E_0$ in
\figref{fig:singletfitsim}(d) (yellow line) is not present in the experimental
results (yellow circles); instead, the experimental results rather indicate a
constant offset.

We can understand the cause of this large oscillation in the simulation results
by investigating the corresponding pulses. As shown in
\figref{fig:singletcircuit}, the parameters $\vartheta_0$ and $\vartheta_1$
occur in the \textsc{U1} gates. These gates are implemented by changing the
phase of the following pulses  for the $H$ gates. Specifically, they change the
phase $\gamma_{ij}\in[0,2\pi)$ of the microwave pulses defined in
\equref{eq:genericvoltagepulses} (see \secref{sec:singlequbitVZgate}). This
change would leave the state of the system invariant if the prepared state was
the singlet state given by \equref{eq:singletstate}. However,  as the pulse
parameters have been modified, the logical conclusion is that a slightly
different state that is sensitive to these phases has been prepared by the
pulses (cf.~also the state vector in \cite{Michielsen2017BenchmarkingQC}
obtained from the data produced by the IBM Q processor).

In the experimental data, it is not the oscillation, but rather  the almost
constant offset of $E_0$ and $E_1$ from 0 that stands out. This suggests that
another kind of error may be the dominant cause for the deviation from the
ideal singlet expectation values.

\subsection{Effective error model}
\label{sec:errorchannels}

An obvious alternative source of errors in the device is the environment which
is not included in the two-transmon system. To address this hypothesis, a simple
method to include effective environment-induced errors is provided by the theory
of quantum fault tolerance. In this formalism, errors in the system are modeled
by an error channel $\mathcal E(\rho)$, which is a linear map on the system's
density matrix $\rho$, i.e.
\begin{align}
  \label{eq:singletKrausOperatorEvolution}
  \rho \mapsto \mathcal E(\rho) = \sum\limits_\alpha M_\alpha \rho M_\alpha^\dagger,
\end{align}
where $M_\alpha$ are linear (Kraus) operators defining the error channel
(cf.~the definition of general quantum operations in \secref{sec:quantumoperations}).
The description is expressed in
terms of  $\rho$ instead of pure states $\ket{\psi}$ such that
non-unitary  maps are supported and damping can be modeled. Note that the
evolution of a smaller system coupled to a larger system can always be written
in the form of \equref{eq:singletKrausOperatorEvolution} if the  evolution of
the joint system is unitary and the initial state of the joint system is
a product state (see \equref{eq:krausrepresentationUnitaryExtendedSpace}).

In this section, we study separate error
channels for each qubit. The simplest type of error channels $\mathcal E$ considered in the theory of quantum
fault tolerance are the so-called depolarizing channel and the amplitude damping
channel \cite{NielsenChuang}.

\subsubsection{Depolarizing channel}

The depolarizing channel $\mathcal E_{\mathrm{dep}}$ is defined as
\begin{align}
  \label{eq:singletDepolarizingChannel}
  \mathcal E_{\mathrm{dep}}(\rho) = (1-p_x-p_y-p_z)\rho + p_x\sigma^x\rho\sigma^x + p_y\sigma^y\rho\sigma^y + p_z\sigma^z\rho\sigma^z,
\end{align}
where $p_x,p_y,p_z\in[0,1]$ and $p_x+p_y+p_z<1$.
The interpretation is that at each application of $\mathcal E_{\mathrm{dep}}$, a bit flip
happens with probability $p_x$, a phase flip happens with probability $p_z$, and
a joint bit and phase flip happens with probability $p_y$. A depolarizing
channel is called symmetric if $p_x= p_y=p_z$. It is worth mentioning that a
symmetric depolarizing channel for an arbitrary number of qubits $n$ can be
written as
\begin{align}
  \label{eq:singletDepolarizingChannelSymmetricMixture}
  \mathcal E_{\mathrm{dep}}^{(\mathrm{sym},n)}(\rho) = F\rho + (1-F)\frac I {2^n},
\end{align}
where the parameter $F\in[0,1]$ can be interpreted as a fidelity. This channel
gives rise to a simple mixture of a uniform distribution and the quantum state
$\rho$, which was also considered as a model for the quantum supremacy
experiment \cite{Google2019QuantumSupremacy}.

\subsubsection{Amplitude damping channel}

The amplitude damping channel $\mathcal E_{\mathrm{amp}}$ is defined by
the following set of Kraus operators in the representation given by
\equref{eq:singletKrausOperatorEvolution}:
\begin{subequations}
  \begin{align}
    \label{eq:singletAmplitudeDampingChannel1}
    M_0 &= \sqrt{p} \begin{pmatrix}
      1 & 0 \\
      0 & \sqrt{1-\gamma}\\
    \end{pmatrix},&
    M_1 &= \sqrt{p} \begin{pmatrix}
      0 & \sqrt{\gamma} \\
      0 & 0\\
    \end{pmatrix},\\
    \label{eq:singletAmplitudeDampingChannel2}
    M_2 &= \sqrt{1-p} \begin{pmatrix}
      \sqrt{1-\gamma} & 0 \\
      0 & 1\\
    \end{pmatrix},&
    M_3 &= \sqrt{1-p} \begin{pmatrix}
      0 & 0 \\
      \sqrt{\gamma} & 0\\
    \end{pmatrix},
  \end{align}
\end{subequations}
where $p,\gamma\in[0,1]$. If $p\neq1$, the channel is sometimes called
\emph{generalized} amplitude damping channel as it can also excite the qubit. In
general, the parameter $p$ determines if the energy exchange with the
environment rather causes a decay ($p\approx1$) or an excitation ($p\approx0$)
of the qubit, and the parameter  $\gamma$ is the corresponding rate (per unit
time).

\subsubsection{Application of the effective error model}

\begin{figure}[p]
  \centering
  \[
    \Qcircuit @C=.8em @R=.7em {
      &\lstick{\ket{0}}&\gate{X}&\measure{\mathcal E_{\mathrm{dep}}}&\gate{H}&\measure{\mathcal E_{\mathrm{dep}}}&\ctrl{1}&\measure{\mathcal E_{\mathrm{dep}}}&\gate{H}&\measure{\mathcal E_{\mathrm{dep}}}&\gate{\textsc{U1}(\vartheta_0)}&\gate{H}&\measure{\mathcal E_{\mathrm{dep}}}&\measure{\mathcal E_{\mathrm{amp}}}&\qw\\
      &\lstick{\ket{0}}&\gate{X}&\measure{\mathcal E_{\mathrm{dep}}}&\qw     &\measure{\mathcal E_{\mathrm{dep}}}&\targ   &\measure{\mathcal E_{\mathrm{dep}}}&\gate{H}&\measure{\mathcal E_{\mathrm{dep}}}&\gate{\textsc{U1}(\vartheta_1)}&\gate{H}&\measure{\mathcal E_{\mathrm{dep}}}&\measure{\mathcal E_{\mathrm{amp}}}&\qw\\
    }
  \]
  \caption{Circuit diagram for experiments on the singlet state, extended by
  effective error channels to test if they can model the experimentally observed
  deviation from the theoretical result (see \figref{fig:singletcircuit}). The
  depolarizing channel $\mathcal E_{\mathrm{dep}}$ is defined in
  \equref{eq:singletDepolarizingChannel} and is inserted after every step  that
  would correspond to a pulse in an implementation. The amplitude damping
  channel $\mathcal E_{\mathrm{amp}}$  is defined in
  \equaref{eq:singletAmplitudeDampingChannel1}{eq:singletAmplitudeDampingChannel2}.}
  \label{fig:singletcircuitwitherrorchannels}
\end{figure}
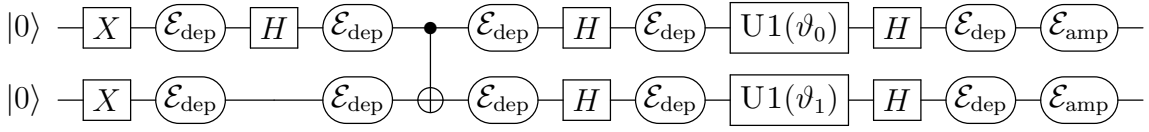

We extend the circuit given in \figref{fig:singletcircuit} with a depolarizing
channel after every step that would correspond to a new pulse, and an amplitude
damping channel at the end of the circuit. A motivation to have the latter only
at the end is given by the fact that energy exchange with the environment is
most likely to occur during the measurement process, where information leaves
the system and enters the environment
(cf.~\cite{Jacobs2014QuantumMeasurementTheory, girvin2014circuitqed}). However,
we also experimented with other approaches that produced only slightly worse
results. The new circuit to model effects from the environment in this simple
way is shown in \figref{fig:singletcircuitwitherrorchannels}.

\begin{figure}[p]
  \centering
  \includegraphics[width=\textwidth]{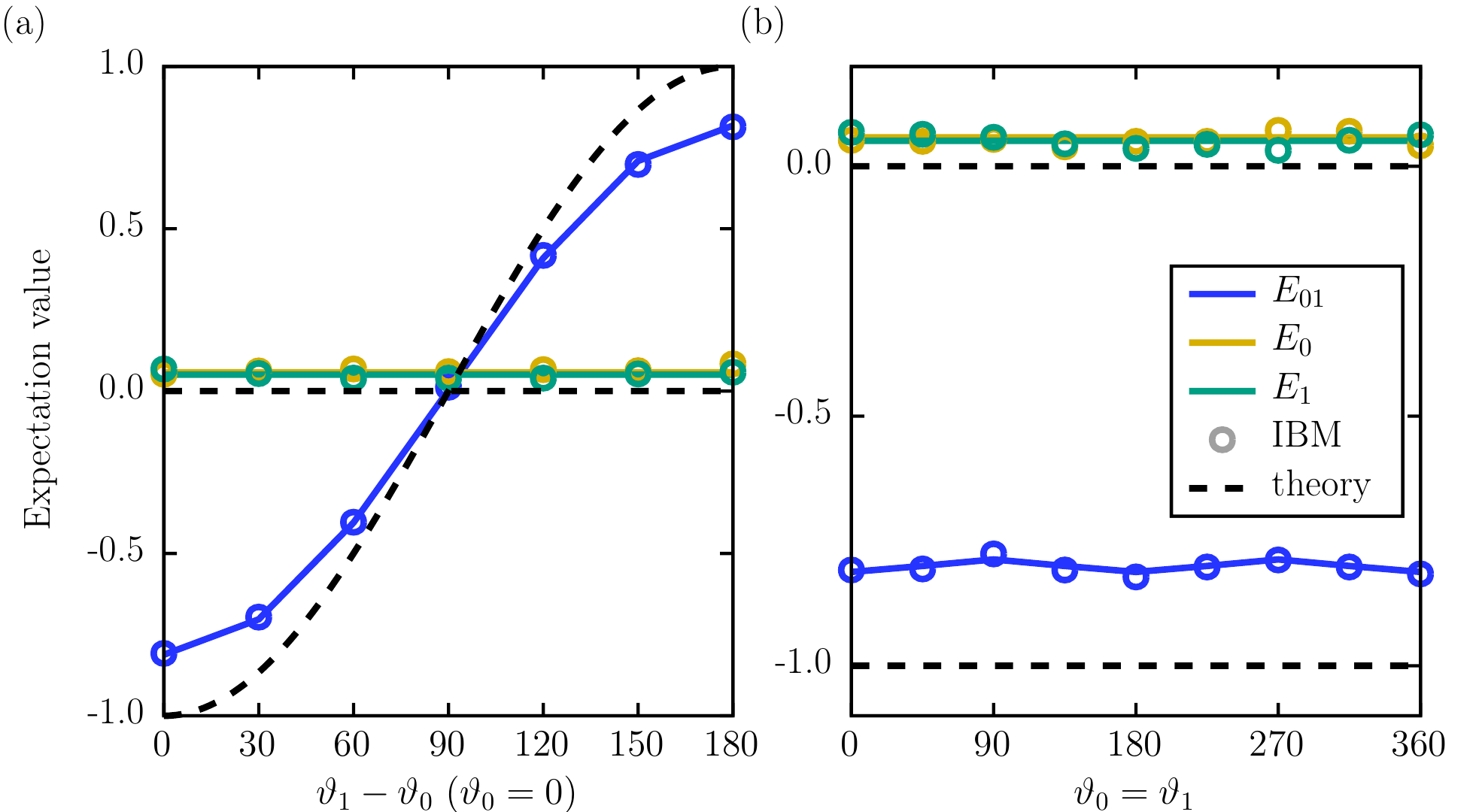}
  \caption{Same as \figref{fig:singletfitsim}, except that $E_{01}$, $E_{0}$,
  and $E_{1}$ are not obtained from the transmon simulation, but from an ideal
  quantum computer simulator such as JUQCS \cite{DeRaedt2018MassivelyParallel,
  Willsch2020BenchmarkingWithJUQCS} combined with the error channels defined in
  \equsref{eq:singletDepolarizingChannel}{eq:singletAmplitudeDampingChannel2}.
  The corresponding circuit is shown in
  \figref{fig:singletcircuitwitherrorchannels}. The parameters of the error
  channels are given in \tabref{tab:singletErrorChannelParameters}.}
  \label{fig:singletfitchannel}
\end{figure}

\begin{table}[p]
  \caption{Resulting parameters $(p_x, p_y, p_z)$ of the depolarizing channel $\mathcal E_{\mathrm{dep}}$ defined in
  \equref{eq:singletDepolarizingChannel} and $(p, \gamma)$ of the amplitude damping channel $\mathcal
  E_{\mathrm{amp}}$ defined in
  \equaref{eq:singletAmplitudeDampingChannel1}{eq:singletAmplitudeDampingChannel2}.
  The channels are added to the circuit as shown in \figref{fig:singletcircuitwitherrorchannels}
  with the same parameters at each step, but different parameters for different qubits.}
\centering
\label{tab:singletErrorChannelParameters}
\begin{tabular}{@{}cccccc@{}}
  \toprule
  Qubit & $p_x$ & $p_y$ & $p_z$ & $p$ & $\gamma$\\
  \midrule
  0 & 0.007 & 0.001 & 0.000 & 0.818 & 0.092 \\
  1 & 0.001 & 0.001 & 0.003 & 0.980 & 0.053 \\
  \bottomrule
\end{tabular}
\end{table}

The result of fitting the error-channel parameters $p_x,p_y,p_z,p$, and $\gamma$
for each qubit  to the experimentally observed data is shown in
\figref{fig:singletfitchannel}. We find that the effective model can describe
the observations very well. In particular, it includes the systematic zig-zag
pattern observed for $E_{01}$ in the case $\vartheta_0=\vartheta_1$ shown in
\figref{fig:singletfitchannel}(b). We remark that a single bit-flip channel with
$p_x\neq0$ for qubit 0 suffices to model this behavior as well as the reduced
amplitude of $E_{01}$ shown in \figref{fig:singletfitchannel}(a). However, only
the amplitude damping channel can model the constant offset of $E_0$ and $E_1$
from 0.

The channel parameters resulting from the fit are given in
\tabref{tab:singletErrorChannelParameters}. They can be interpreted in the way
that the control qubit 0 is more prone to bit-flip errors (0.7\%), while the
target qubit 1 is rather susceptible to phase errors (0.3\%). This is reasonable
since for the target qubit, the phase is the most sensitive quantity for the
operation  of the \textsc{CNOT} gate because a phase shift on the target qubit
affects the pulses on both qubits (see the occurrence of $\vartheta_T$ for both
qubits in the echoed CR scheme in \figref{fig:twoqubitpulseruleVZ}(c)). We also
see that the interaction with the environment mainly causes qubit relaxation
($p\approx1$). The corresponding decay rates $\gamma=0.092$ ($\gamma=0.053$) for
the control (target) qubit are roughly compatible with the qubit relaxation
times $T_1$ of the order of $\SI{40}{\micro s}$: for a single application of
$\mathcal E_{\mathrm{amp}}$ over the course of the circuit in
\figref{fig:singletcircuitwitherrorchannels}, whose execution time
$T_{\mathrm{exec}}$ is a few microseconds, we have $T_{\mathrm{exec}}/\gamma
\approx T_1$. The fact that the decay rate $\gamma$ of the control qubit is
larger than that of the target qubit makes sense as the control qubit of the
\textsc{CNOT} gate is driven much more strongly
(cf.~\figref{fig:crossresonancepulses}(c)), so it is more susceptible to energy
exchange with the environment.

In conclusion, there are errors in the device that can be very well described by
simple error channels, such as the depolarizing channel and the amplitude
damping channel. As such simple error channels belong to the error model that is
addressed by the theory of quantum error correction, this suggests that error
correction could work reasonably well for the transmon architecture (see the
following section). However, note also that there are more difficult, correlated
errors in the device that are very well described by the five-transmon
simulation model (see \figref{fig:crosstalkExperimentResults}(a)). It would be
interesting to see whether the five-transmon model can also describe the
experiments studied in this section better than the two-transmon model. Initial
evidence for this idea is given by the results shown in
\figref{fig:repeatedgatescnot}. We leave a detailed analysis of the singlet-state
characterization with the five-transmon model for future work.

\section{Testing quantum fault tolerance}
\label{sec:testingfaulttolerance}

Both simulations and experiments presented in the previous sections suggest that
controlling gate-based quantum computers to high accuracy is considerably
difficult. For the simulations, this is true even though the model studied in
this work is inherently quantum mechanical (see \secref{sec:transmonmodel}). The
same observation is supported by many other experiments implementing the
gate-based quantum computer model \cite{sheldon2015singlequbitfidelities,
gambetta2015building, Neil2017GoogleBlueprintQuantumSupremacy,
Michielsen2017BenchmarkingQC, Google2019QuantumSupremacy}.

The most prominent, long-term solution to this problem is proposed by the theory
of quantum error correction and fault tolerance \cite{Shor1996FaultTolerantQC,
divincenzo1996fivequbitcodeimplementation, Gottesman1998TheoryFTQC,
Campbell2017RoadsTowardsFTQC}. The basic idea is that additional \emph{physical}
qubits are used to encode a smaller number of so-called \emph{logical} qubits.
The logical qubits are designed to be tolerant to errors within certain
mathematical models.

Simple models consider discrete, uncorrelated errors such as spontaneous bit or
phase  flips (see \equref{eq:singletDepolarizingChannel}), whereas more
sophisticated  models consider non-Markovian errors in a general Hamiltonian
framework \cite{Terhal2005ftqcForLocalNonmarkovianNoise,
aliferis2006extendedrectangles, aliferis2007FTQCwithLeakage,
aharonov2008thresholdtheorem, ng2009FTQCversusGaussianNoise}. In these models,
so-called \emph{threshold theorems} are derived. They state that, if the error
rates are below a certain threshold, arbitrarily long quantum computation is
possible by using a suitable fault-tolerant protocol. As with any mathematical
model, however, it is a priori unclear if its predictions hold in practice.
Specifically, it is unclear whether such a fault-tolerant protocol
systematically improves the results in an actual experiment.

To address this question, we test a full fault-tolerant protocol explicitly
designed for small experiments \cite{Gottesman2016quantumfaulttolerance}, using
both the five-transmon, six-resonator model defined in
\secref{sec:transmonmodelibm5ed}, and the 16-qubit processor \texttt{ibmqx5} on
the IBM Q Experience \cite{ibmqx5}. The protocol is based on the four-qubit code
\cite{Leung1997fourqubitcode, Vaidman1996fourqubitcode,
Grassl1997fourqubitcode}, which has recently also been studied in other experiments
\cite{Linke2016FTIonTrapQubits, Vuillot2017ErrorDetectionIBM,
Takita2017faultTolerantStatePreparation,
HarperFlammia2018FaultToleranceInTheIBMQ}. We find that the protocol
systematically improves the results in the presence of the inherent control and
measurement errors of the studied transmon systems. Part of the work presented in
this section has been published in
\cite{Willsch2018TestingFaultTolerance}.

\subsection{Fault-tolerant protocol}

A fault-tolerant protocol considers the full procedure of (1) preparing a
certain  initial state, (2) applying a few quantum gates to the state, and (3)
measuring the result to obtain a distribution of bit strings. For steps (1) and
(2), the protocol provides a \emph{circuit encoding} to translate  each
\emph{bare circuit} (or \emph{logical circuit}) into an \emph{encoded circuit}
that requires a larger number of physical qubits. Here, the term ``circuit''
explicitly includes the gates used to prepare the initial state. For step (3),
the fault-tolerant protocol states how the measured bit strings, obtained by
measuring the larger number of physical qubits, are to be interpreted to obtain
a distribution for the logical qubits. In particular, we consider bare and
encoded circuits for two logical qubits.

\subsubsection{Definition of the circuit encoding}

The protocol under investigation encodes two logical qubits in four physical
qubits ${q_1}{q_2}{q_3}{q_4}$ and an additional so-called \emph{ancillary} qubit
${q_0}$. The code can detect an arbitrary single-qubit error
\cite{Gottesman2016quantumfaulttolerance}. Such a code is typically expressed
using the notation $[[4,2,2]]$, where a code of the form
$[[n_{\mathrm{phy}},n_{\mathrm{log}},d]]$ means that the code uses
$n_{\mathrm{phy}}$ physical qubits to encode the state of $n_{\mathrm{log}}$
logical qubits, and the distance $d$ includes information about the number of
errors that the code can detect or correct \cite{NielsenChuang}.

The logical two-qubit states of the $[[4,2,2]]$ code are defined as
\begin{subequations}
  \begin{align}
    \label{eq:errordetection00L}
    \overline{\ket{00}} &= (\ket{0000}+\ket{1111})/\sqrt{2},\\
    \overline{\ket{01}} &= (\ket{1100}+\ket{0011})/\sqrt{2} ,\\
    \overline{\ket{10}} &= (\ket{1010}+\ket{0101})/\sqrt{2} ,\\
    \label{eq:errordetection11L}
    \overline{\ket{11}} &= (\ket{0110}+\ket{1001})/\sqrt{2}.
  \end{align}
\end{subequations}
By linear combination, one can derive the encoded versions of other
logical two-qubit states, e.g.,
\begin{subequations}
  \begin{align}
    \label{eq:errordetection0+L}
    \overline{\ket{0+}} &= (\overline{\ket{00}} + \overline{\ket{01}})/\sqrt2 = (\ket{0000}+\ket{1100}+\ket{0011}+\ket{1111})/2 ,\\
    \label{eq:errordetectionPhi+L}
    \overline{\ket{\Phi^+}} &= (\overline{\ket{00}} + \overline{\ket{11}})/\sqrt2 = (\ket{0000}+\ket{0110}+\ket{1001}+\ket{1111})/2.
  \end{align}
\end{subequations}

\begin{table}[p]
  \caption{List of the three initial states considered in this experiment. For
  each state, we list the bare and encoded versions (see
  \equsref{eq:errordetection00L}{eq:errordetectionPhi+L}) of their preparation
  circuits. The preparation of the encoded version of $\ket{00}$ requires an
  additional ancilla qubit. Elementary gates are defined in
  \tabref{tab:elementarygateset} in \appref{app:gateset}.}
  \centering
  \label{tab:errordetectionstates}
  \footnotesize
  \begin{tabular}{@{}lll@{}}
  \toprule
  State & \multicolumn{1}{c}{Bare version} & \multicolumn{1}{c}{Encoded version} \\
  \midrule
  $\ket{00}$
  & \:\:\qquad
  \Qcircuit @C=1.5em @R=1.5em @!R {
  \lstick{{q_3}\ket{0}}&\qw\\
  \lstick{{q_4}\ket{0}}&\qw\\&}
  & \:\:\qquad
  \Qcircuit @C=.5em @R=.2em @!R {
  \lstick{{q_0}\ket{0}}&\qw     &\qw      &\qw      &\qw      &\targ    &\targ    &\meter&\cw\\
  \lstick{{q_1}\ket{0}}&\qw     &\qw      &\targ    &\qw      &\ctrl{-1}&\qw      &\qw   &\qw\\
  \lstick{{q_2}\ket{0}}&\qw     &\targ    &\ctrl{-1}&\qw      &\qw      &\qw      &\qw   &\qw\\
  \lstick{{q_3}\ket{0}}&\gate{H}&\ctrl{-1}&\qw      &\ctrl{1} &\qw      &\qw      &\qw   &\qw\\
  \lstick{{q_4}\ket{0}}&\qw     &\qw      &\qw      &\targ    &\qw      &\ctrl{-4}&\qw   &\qw\\&}
  \\
  \midrule
  $\ket{0+}$
  & \:\:\qquad
  \Qcircuit @C=.5em @R=.5em @!R {
  \lstick{{q_3}\ket{0}}&\qw     &\qw\\
  \lstick{{q_4}\ket{0}}&\gate{H}&\qw\\&}
  & \:\:\qquad
  \Qcircuit @C=.5em @R=.2em @!R {
  \lstick{{q_1}\ket{0}}&\qw     &\targ    &\qw      &\qw      &\qw\\
  \lstick{{q_2}\ket{0}}&\gate{H}&\ctrl{-1}&\qw      &\qw      &\qw\\
  \lstick{{q_3}\ket{0}}&\qw     &\qw      &\gate{H} &\ctrl{1} &\qw\\
  \lstick{{q_4}\ket{0}}&\qw     &\qw      &\qw      &\targ    &\qw\\&}
  \\
  \midrule
  $\ket{\Phi^+}$
  & \:\:\qquad
  \Qcircuit @C=.5em @R=.5em @!R {
  \lstick{{q_3}\ket{0}}&\gate{H}&\ctrl{1} &\qw\\
  \lstick{{q_4}\ket{0}}&\qw     &\targ    &\qw\\&}
  & \:\:\qquad
  \Qcircuit @C=.5em @R=.2em @!R {
  \lstick{{q_1}\ket{0}}&\qw     &\qw      &\gate{H} &\ctrl{3} &\qw\\
  \lstick{{q_2}\ket{0}}&\qw     &\targ    &\qw      &\qw      &\qw\\
  \lstick{{q_3}\ket{0}}&\gate{H}&\ctrl{-1}&\qw      &\qw      &\qw\\
  \lstick{{q_4}\ket{0}}&\qw     &\qw      &\qw      &\targ    &\qw\\&}
  \\
  \bottomrule
\end{tabular}
\end{table}
\begin{table}[p]
  \caption{List of the bare and encoded gate elements used to construct
  circuits to test the fault-tolerant protocol (see \equsref{eq:errordetectionX1L}{eq:errordetectionCZL}).
  Elementary gates are defined in \tabref{tab:elementarygateset} in \appref{app:gateset}.}
  \centering
  \label{tab:errordetectiongates}
  \footnotesize
  \begin{minipage}{0.4\textwidth}
    \begin{tabular}{@{}lll@{}}
      \toprule
      Gate & \multicolumn{1}{c}{Bare version} & \multicolumn{1}{c}{Encoded version} \\
      \midrule
      X1
      & \qquad
      \Qcircuit @C=.5em @R=.5em @!R {
      \lstick{{q_3}}&\gate{X} &\qw\\
      \lstick{{q_4}}&\qw      &\qw\\&}
      & \qquad
      \Qcircuit @C=.5em @R=.2em @!R {
      \lstick{{q_1}}&\gate{X} &\qw\\
      \lstick{{q_2}}&\qw      &\qw\\
      \lstick{{q_3}}&\gate{X} &\qw\\
      \lstick{{q_4}}&\qw      &\qw\\&}
      \\
      \midrule
      X2
      & \qquad
      \Qcircuit @C=.5em @R=.5em @!R {
      \lstick{{q_3}}&\qw      &\qw\\
      \lstick{{q_4}}&\gate{X} &\qw\\&}
      & \:\:\qquad
      \Qcircuit @C=.5em @R=.2em @!R {
      \lstick{{q_1}}&\gate{X} &\qw\\
      \lstick{{q_2}}&\gate{X} &\qw\\
      \lstick{{q_3}}&\qw      &\qw\\
      \lstick{{q_4}}&\qw      &\qw\\&}
      \\
      \midrule
      Z1
      & \:\:\qquad
      \Qcircuit @C=.5em @R=.5em @!R {
      \lstick{{q_3}}&\gate{Z} &\qw\\
      \lstick{{q_4}}&\qw      &\qw\\&}
      & \:\:\qquad
      \Qcircuit @C=.5em @R=.2em @!R {
      \lstick{{q_1}}&\gate{Z} &\qw\\
      \lstick{{q_2}}&\gate{Z} &\qw\\
      \lstick{{q_3}}&\qw      &\qw\\
      \lstick{{q_4}}&\qw      &\qw\\&}
      \\
      \bottomrule
    \end{tabular}
  \end{minipage}
  %\hfill
  \begin{minipage}{0.56\textwidth}
    \begin{tabular}{@{}lll@{}}
      \toprule
      Gate & \multicolumn{1}{c}{Bare version} & \multicolumn{1}{c}{Encoded version} \\
      \midrule
      Z2
      & \:\:\qquad
      \Qcircuit @C=.5em @R=.5em @!R {
      \lstick{{q_3}}&\qw      &\qw\\
      \lstick{{q_4}}&\gate{Z} &\qw\\&}
      & \:\:\qquad
      \Qcircuit @C=.5em @R=.2em @!R {
      \lstick{{q_1}}&\gate{Z} &\qw\\
      \lstick{{q_2}}&\qw      &\qw\\
      \lstick{{q_3}}&\gate{Z} &\qw\\
      \lstick{{q_4}}&\qw      &\qw\\&}
      \\
      \midrule
      HHS
      & \:\:\qquad
      \Qcircuit @C=.5em @R=.5em @!R {
      \lstick{{q_3}}&\gate{H} &\ctrl{1} &\gate{H} &\ctrl{1} &\gate{H} &\ctrl{1} &\qw\\
      \lstick{{q_4}}&\gate{H} &\targ    &\gate{H} &\targ    &\gate{H} &\targ    &\qw\\&}
      & \:\:\qquad
      \Qcircuit @C=.5em @R=.2em @!R {
      \lstick{{q_1}}&\gate{H} &\qw\\
      \lstick{{q_2}}&\gate{H} &\qw\\
      \lstick{{q_3}}&\gate{H} &\qw\\
      \lstick{{q_4}}&\gate{H} &\qw\\&}
      \\
      \midrule
      CZ
      & \:\:\qquad
      \Qcircuit @C=.5em @R=.5em @!R {
      \lstick{{q_3}}&\qw      &\ctrl{1} &\qw      &\qw\\
      \lstick{{q_4}}&\gate{H} &\targ    &\gate{H} &\qw\\&}
      & \:\:\qquad
      \Qcircuit @C=.5em @R=.2em @!R {
      \lstick{{q_1}}&\gate{S} &\qw      &\qw\\
      \lstick{{q_2}}&\gate{S} &\gate{Z} &\qw\\
      \lstick{{q_3}}&\gate{S} &\gate{Z} &\qw\\
      \lstick{{q_4}}&\gate{S} &\qw      &\qw\\&}
      \\
      \bottomrule
    \end{tabular}
  \end{minipage}
\end{table}

Using the definition of the logical two-qubit basis states given by
\equsref{eq:errordetection00L}{eq:errordetection11L}, we can derive expressions
for logical gates in this code. For instance, a logical bit flip on qubit 1,
denoted by $\overline{\mathrm X1}$, has to map $\overline{\ket{0j}}$ to
$\overline{\ket{1j}}$ and $\overline{\ket{1j}}$ to $\overline{\ket{0j}}$ for
$j=0,1$. This is accomplished by flipping the physical qubits $q_1$ and $q_3$ in
\equsref{eq:errordetection00L}{eq:errordetection11L}. Therefore, we have
$\overline{\mathrm X1}=X_1X_3$. In the same way, we find $\overline{\mathrm
X2}=X_1 X_2$. Similar expressions can be derived for the logical phase flips,
namely $\overline{\mathrm Z1}=Z_1Z_2$ and $\overline{\mathrm Z2}=Z_1Z_3$.

In addition to these single-qubit gates, we consider two particular two-qubit
gates. The first is a Hadamard gate on both logical qubits followed by swapping
the qubits. This gate is denoted by HHS. On a bare two-qubit state, such a
transformation can be implemented by the gate sequence
$\mathrm{HHS}=\textsc{CNOT}_{12}\textsc{CNOT}_{21}\textsc{CNOT}_{12}H_1H_2$. In
the code space defined by \equsref{eq:errordetection00L}{eq:errordetection11L},
this is accomplished by $\overline{\mathrm{HHS}}=H_1H_2H_3H_4$. Finally, we
consider the controlled-phase gate CZ. This is an entangling gate with the
matrix representation $\mathrm{diag}(1,1,1,-1)$. A bare implementation of this
gate is given by $\mathrm{CZ}=H_2\textsc{CNOT}_{12}H_2$. On the encoded states
defined by \equsref{eq:errordetection00L}{eq:errordetection11L}, we reach the
same effect with the gate sequence $\overline{\mathrm{CZ}} = Z_2 Z_3 S_1 S_2 S_3
S_4$. In summary, we consider six logical gates whose encoded versions
are given by
\begin{subequations}
  \begin{align}
    \label{eq:errordetectionX1L}
    \overline{\mathrm{X1}} &= X_1 X_3,\\
    \overline{\mathrm{X2}} &= X_1 X_2,\\
    \overline{\mathrm{Z1}} &= Z_1 Z_2,\\
    \overline{\mathrm{Z2}} &= Z_1 Z_3,\\
    \overline{\mathrm{HHS}} &= H_1 H_2 H_3 H_4,\\
     \label{eq:errordetectionCZL}
    \overline{\mathrm{CZ}} &= Z_2 Z_3 S_1 S_2 S_3 S_4,
  \end{align}
\end{subequations}
where the definitions of the elementary gates on the right-hand side
are given in \tabref{tab:elementarygateset} in \appref{app:gateset}.

The goal of this section is to compare bare versions of a two-qubit circuit with
their corresponding encoded versions. We consider circuits composed of one of
the initial states $\ket{00}$, $\ket{0+}$, and $\ket{\Phi^+}$, and an arbitrary
combination of gates from the gate set
$\{\mathrm{X1},\mathrm{X2},\mathrm{Z1},\mathrm{Z2},\mathrm{HHS},\mathrm{CZ}\}$.
Note that this gate set is not universal, in the sense that not all quantum
algorithms can be encoded. The requirement of universality was dropped in favor
of having a full fault-tolerant protocol (including the state preparation) that
is applicable to small experiments \cite{Gottesman2016quantumfaulttolerance}.

In \tabref{tab:errordetectionstates} and \tabref{tab:errordetectiongates}, we
list all gate sequences used to construct both bare and encoded versions of the
circuits. The labels $q_i$ refer to the qubit labels that are used in the
experiment below. Note that all circuits need to be expressed in terms of the
gates that are supported by the transmon architecture. Therefore, the bare
version of the HHS gate contains three \textsc{CNOT} gates (see
\tabref{tab:errordetectiongates}). Similarly, the encoded version of the state
preparation of $\ket{00}$ requires five \textsc{CNOT} gates and one additional
ancillary qubit (see \tabref{tab:errordetectionstates}).

The motivation for the ancillary qubit is given in
\cite{Gottesman2016quantumfaulttolerance} and is based on the simple model
of discrete, uncorrelated single-qubit errors such as spontaneous bit or phase
flips. An error within this model can be detected by the ancilla qubit
being in state $\ket1$ or the resulting four-qubit state having an odd
parity (i.e., an odd number of 1's), since each four-qubit state in \equsref{eq:errordetection00L}{eq:errordetection11L}
has an even parity.

\subsubsection{Evaluation}

We consider a two-qubit circuit made from an initial state given in
\tabref{tab:errordetectionstates} and a sequence of gates from
\tabref{tab:errordetectiongates}. In what follows, we define
$p^{(\mathrm{id})}_{j_0j_1}$ to be the probability distribution of
two-bit strings $j_0j_1$ as computed by an ideal gate-based quantum
computer (see \chapref{cha:quantumcomputing}). This distribution is
compared to the distributions $p^{(\mathrm{bare})}_{j_0j_1}$
and $p^{(\mathrm{enc})}_{j_0j_1}$, which are obtained from the evaluation of the
bare and encoded version of the circuit as follows.

The evaluation of the bare version of the circuit using the two qubits
${q_3}{q_4}$ is straightforward: In the experiment, we execute the circuit a
certain number of times (typically 8192) and count the number $n_{j_0j_1}$ of
measured bit strings $j_0j_1$. We then obtain a distribution of relative
frequencies $p^{(\mathrm{bare})}_{j_0j_1}=n_{j_0j_1}/8192$. In the simulation,
we directly obtain the probabilities
$\smash{p^{(\mathrm{bare})}_{j_0j_1}=\abs{\braket{m_3=j_0,m_4=j_1|\Psi}}^2}$ from
the state vector $\ket{\Psi}$ given by \equref{eq:psioftsolutioncoefficients}.

For the encoded version of the circuit, the distribution
$\smash{p_{j_0j_1}^{(\mathrm{enc})}}$ is constructed by evaluating the data as
dictated by the fault-tolerant protocol: Since the encoded circuit involves five
physical qubits, a measurement of all qubits at the end of the circuit produces
five bits. If the bit corresponding to the ancilla ${q_0}$ is 1, or if the
four-bit string corresponding to ${q_1}{q_2}{q_3}{q_4}$ has an odd parity (i.e.,
one or three 1's), the run is discarded. Otherwise, we can map the resulting
four-bit string to a logical two-qubit state $\overline{\ket{j_0j_1}}$ using
\equsref{eq:errordetection00L}{eq:errordetection11L}. By counting all these
mapped two-bit strings, we generate the frequency distribution
$p_{j_0j_1}^{(\mathrm{enc})}$.

The step of systematically discarding some of the measured bit strings is called
\emph{postselection} procedure. The corresponding ratio of bit strings that are
not discarded is called the postselection ratio $r$. Note that the essential
idea of fault-tolerant protocols based on postselection is that the
postselection procedure is systematic. In other words, it is not required to
know the ideal, theoretical result of the quantum circuit in order to perform
the postselection procedure. Instead, the protocol provides a fixed set of rules
(such as ``$q_0$ is 1'' or ``$q_1q_2q_3q_4$ has an odd parity'')
that can be checked for the measured bit string to see whether it should be
discarded. See \tabref{tab:errordetectionfullExamineFault} below for an
example application of the postselection procedure.

To compare the performance of the bare and encoded circuits, we
compare the resulting distributions $p^{(\mathrm{bare/enc})}_{j_0j_1}$
to the ideal distribution $p^{(\mathrm{id})}_{j_0j_1}$ by means of the
statistical distance (cf.~\equref{eq:statisticaldistance}),
\begin{align}
  \label{eq:errordetectionDbareenc}
  D_{\text{bare/enc}} &= \frac{1}{2} \sum\limits_{j_0j_1}^{}
  \left|p_{j_0j_1}^{(\mathrm{bare/enc})} - p_{j_0j_1}^{(\mathrm{id})} \right|.
\end{align}
To distinguish between simulation and experimental results, we use the notations
$\smash{D_{\mathrm{bare/enc}}^{(\mathrm{sim})}}$ and
$\smash{D_{\mathrm{bare/enc}}^{(\mathrm{exp})}}$, respectively. Note that from the
transmon simulator, we can directly obtain the probabilities
$\smash{p^{(\mathrm{bare/enc})}_{j_0j_1}}$ from the distribution defined by the state
vector $\ket{\Psi}$, so the intermediate step of sampling and counting the
outcomes is omitted.

\subsection{Test systems and circuits}

To test the fault-tolerant protocol, the performance of bare and encoded
circuits needs to be compared for a representative set of circuits. We generate
such a representative set by applying the procedure defined in
\cite{Gottesman2016quantumfaulttolerance} using the maximum logical circuit
length $T=10$, the repetition parameter $\mathrm{RP}=6$, and the periodicity
$P=3$. We obtain a total of $465$ circuits, composed of $155$ logical gate
sequences from \tabref{tab:errordetectiongates} for each of the three initial
states given in \tabref{tab:errordetectionstates}. A list of all circuits is
given in Listing~\ref{code:fullcircuits}, sorted by the total time required for
the simulation of all underlying quantum gate pulses.

\begin{lstfloat}
  \caption{Definition of all $465=3\times155$ circuits used to test the
  fault-tolerant protocol, generated according to the procedure specified in
  \cite{Gottesman2016quantumfaulttolerance}. Each of the $155$ lines consists of
  an ID (red), a sequence of logical gates from \tabref{tab:errordetectiongates}
  (blue), and a placeholder for the three initial states
  $\texttt{|i>}\in\{\ket{00},\ket{0+},\ket{\Phi^+} \}$ (see
  \tabref{tab:errordetectionstates}). The circuits are sorted in ascending order
  by the total time $T_{\mathrm{bare}}^{\ket{00}}\in[0,\SI{10}{\micro s}]$
  required for simulating the time evolution of all pulses for the bare version
  corresponding to the initial state $\ket{00}$ (cf. \secref{sec:compiler}).}
\begin{lstlisting}[style=fullcircuitstyle,label=code:fullcircuits]
 0 |i>                                 52 X2 Z1 X2 Z1 X2 Z1 X2 Z1 |i>        104 Z2 X1 Z2 X2 CZ HHS CZ |i>
 1 Z2 |i>                              53 Z1 Z1 X2 X2 Z1 Z1 X2 X2 |i>        105 HHS CZ X2 CZ |i>
 2 Z2 Z2 |i>                           54 X2 X1 CZ X1 |i>                    106 X1 Z2 HHS CZ CZ |i>
 3 Z2 Z2 Z2 |i>                        55 X1 X1 X1 X1 X1 |i>                 107 HHS X2 Z2 CZ CZ |i>
 4 Z2 Z2 Z2 Z2 |i>                     56 X2 Z1 X2 Z1 X2 Z1 X2 Z1 X2 Z1 |i>  108 Z2 CZ X1 HHS X1 CZ |i>
 5 Z2 Z2 Z2 Z2 Z2 |i>                  57 Z2 CZ X2 X2 X1 Z1 |i>              109 Z2 X1 Z1 Z1 CZ Z1 X2 Z1 HHS CZ |i>
 6 Z2 Z2 Z2 Z2 Z2 Z2 |i>               58 Z1 X2 Z2 CZ X2 X1 X1 |i>           110 CZ CZ CZ CZ CZ |i>
 7 Z2 Z2 Z2 Z2 Z2 Z2 Z2 |i>            59 CZ X1 X2 Z1 Z1 X1 X1 Z1 Z1 Z2 |i>  111 X2 CZ HHS X2 CZ Z2 |i>
 8 Z2 Z2 Z2 Z2 Z2 Z2 Z2 Z2 |i>         60 X1 X1 X1 X1 X1 X1 |i>              112 X2 X1 CZ HHS CZ Z1 Z1 X1 X2 Z2 |i>
 9 Z2 Z2 Z2 Z2 Z2 Z2 Z2 Z2 Z2 |i>      61 X2 X2 X2 X2 X2 X2 |i>              113 HHS HHS X2 |i>
10 Z2 Z2 Z2 Z2 Z2 Z2 Z2 Z2 Z2 Z2 |i>   62 CZ CZ |i>                          114 Z1 HHS HHS X2 |i>
11 Z1 Z2 |i>                           63 X1 X1 X1 X1 X1 X1 X1 |i>           115 X1 HHS X1 HHS |i>
12 Z1 Z2 Z1 Z2 |i>                     64 CZ CZ X1 |i>                       116 X2 Z2 HHS X2 Z2 HHS |i>
13 Z1 Z2 Z1 Z2 Z1 Z2 |i>               65 X1 CZ Z2 X1 CZ Z2 |i>              117 X2 HHS Z2 Z2 X2 HHS Z2 Z2 |i>
14 Z1 Z2 Z1 Z2 Z1 Z2 Z1 Z2 |i>         66 CZ Z2 Z1 Z2 X1 CZ X2 |i>           118 Z1 Z2 Z1 HHS X1 Z1 Z2 Z1 HHS X1 |i>
15 Z1 Z2 Z1 Z2 Z1 Z2 Z1 Z2 Z1 Z2 |i>   67 CZ CZ X2 |i>                       119 Z1 Z2 X2 Z1 HHS Z1 Z2 X2 Z1 HHS |i>
16 Z2 Z1 |i>                           68 X1 X1 X1 X1 X1 X1 X1 X1 |i>        120 CZ CZ CZ CZ CZ CZ |i>
17 Z2 Z1 Z2 Z1 |i>                     69 X2 X2 X2 X2 X2 X2 X2 X2 |i>        121 CZ CZ X1 CZ CZ X1 CZ CZ X1 |i>
18 Z2 Z1 Z2 Z1 Z2 Z1 |i>               70 HHS |i>                            122 HHS X1 Z2 X1 Z2 HHS X1 Z2 X1 Z2 |i>
19 Z2 Z1 Z2 Z1 Z2 Z1 Z2 Z1 |i>         71 HHS Z1 |i>                         123 Z2 Z1 HHS HHS X2 X1 Z2 CZ |i>
20 Z2 Z1 Z2 Z1 Z2 Z1 Z2 Z1 Z2 Z1 |i>   72 Z1 Z2 CZ CZ X1 X1 X2 X2 |i>        124 CZ X1 Z2 HHS X2 X1 Z1 Z1 HHS |i>
21 Z1 |i>                              73 X1 X1 X1 X1 X1 X1 X1 X1 X1 |i>     125 CZ CZ X2 CZ CZ X2 CZ CZ X2 |i>
22 X2 Z1 |i>                           74 X1 HHS |i>                         126 X1 X1 X1 HHS Z2 X1 X1 X1 HHS Z2 |i>
23 X1 |i>                              75 X2 Z2 HHS |i>                      127 CZ CZ CZ CZ CZ CZ CZ |i>
24 Z1 Z1 X1 |i>                        76 X2 HHS Z2 Z2 |i>                   128 HHS CZ HHS CZ |i>
25 Z2 Z1 X2 Z2 |i>                     77 Z1 Z2 Z1 HHS X1 |i>                129 Z1 CZ Z1 Z2 HHS Z1 CZ Z1 Z2 HHS |i>
26 Z2 X1 Z2 X2 |i>                     78 Z1 Z2 X2 Z1 HHS |i>                130 HHS CZ X2 HHS X1 X1 Z1 X1 X2 |i>
27 Z2 X2 Z1 |i>                        79 HHS X1 Z2 Z2 |i>                   131 HHS CZ HHS X2 CZ |i>
28 X2 |i>                              80 Z2 Z2 Z2 X2 Z1 CZ CZ Z2 X2 X2 |i>  132 HHS HHS HHS Z1 |i>
29 X1 X1 |i>                           81 X1 X1 X1 X1 X1 X1 X1 X1 X1 X1 |i>  133 X2 Z2 Z1 HHS Z1 HHS HHS |i>
30 X2 X2 |i>                           82 X2 X2 X2 X2 X2 X2 X2 X2 X2 X2 |i>  134 CZ HHS X1 Z2 X2 X2 X2 Z2 HHS CZ |i>
31 X2 Z1 X2 Z1 |i>                     83 CZ CZ CZ |i>                       135 CZ CZ CZ CZ CZ CZ CZ CZ |i>
32 Z1 Z1 X1 Z1 Z1 X1 |i>               84 HHS X2 X2 Z2 Z2 X1 X1 |i>          136 X1 HHS X1 HHS X1 HHS |i>
33 Z1 X2 X2 |i>                        85 X1 CZ Z2 X1 CZ Z2 X1 CZ Z2 |i>     137 X2 Z2 HHS X2 Z2 HHS X2 Z2 HHS |i>
34 Z2 Z1 X2 Z2 Z2 Z1 X2 Z2 |i>         86 HHS X1 Z2 X1 Z2 |i>                138 Z1 X2 HHS CZ X1 HHS CZ CZ X1 X1 |i>
35 Z2 X1 Z2 X2 Z2 X1 Z2 X2 |i>         87 Z1 Z2 X1 Z1 HHS X2 |i>             139 CZ CZ CZ CZ CZ CZ CZ CZ CZ |i>
36 Z1 Z1 X2 X2 |i>                     88 HHS X1 X2 X1 X2 Z2 Z1 X1 |i>       140 X1 X1 HHS Z2 HHS HHS X2 Z2 CZ |i>
37 X1 X2 X1 |i>                        89 X1 X1 X1 HHS Z2 |i>                141 X1 CZ HHS CZ HHS Z1 CZ CZ X2 |i>
38 Z1 X2 Z1 X2 X1 Z1 |i>               90 HHS CZ |i>                         142 X1 Z2 HHS CZ CZ X1 Z2 HHS CZ CZ |i>
39 X1 X1 X1 X2 X2 Z2 |i>               91 Z1 CZ Z1 Z2 HHS |i>                143 CZ Z2 HHS Z2 HHS CZ Z2 HHS |i>
40 X2 Z1 X2 Z1 X2 Z1 |i>               92 HHS Z2 CZ Z1 |i>                   144 HHS CZ X2 CZ HHS CZ X2 CZ |i>
41 Z1 Z1 X1 Z1 Z1 X1 Z1 Z1 X1 |i>      93 X2 Z2 HHS CZ |i>                   145 CZ CZ CZ CZ CZ CZ CZ CZ CZ CZ |i>
42 X1 X1 X1 |i>                        94 Z1 HHS CZ X2 Z2 |i>                146 HHS CZ HHS CZ HHS CZ |i>
43 X1 X2 X1 X2 X1 |i>                  95 X2 Z2 Z1 HHS CZ |i>                147 HHS HHS X2 HHS HHS X2 |i>
44 CZ |i>                              96 CZ X2 CZ CZ X2 X2 Z2 Z2 |i>        148 Z1 HHS HHS X2 Z1 HHS HHS X2 |i>
45 CZ Z2 |i>                           97 CZ CZ CZ CZ |i>                    149 X2 CZ HHS HHS HHS Z2 CZ CZ Z1 |i>
46 Z2 Z1 CZ |i>                        98 Z1 X2 Z1 HHS CZ Z2 Z2 X2 |i>       150 X1 HHS X1 HHS X1 HHS X1 HHS |i>
47 CZ Z2 Z1 Z2 |i>                     99 Z2 HHS CZ X2 X2 Z1 Z2 X1 X1 |i>    151 HHS CZ HHS CZ HHS CZ HHS CZ |i>
48 X1 CZ Z2 |i>                       100 CZ CZ X1 CZ CZ X1 |i>              152 X1 HHS X1 HHS X1 HHS X1 HHS X1 HHS |i>
49 Z2 Z1 X1 Z2 CZ |i>                 101 X2 HHS Z1 X1 X2 CZ X2 |i>          153 HHS HHS X2 HHS HHS X2 HHS HHS X2 |i>
50 X1 X1 X1 X1 |i>                    102 CZ CZ HHS |i>                      154 HHS CZ HHS CZ HHS CZ HHS CZ HHS CZ |i>
51 X2 X2 X2 X2 |i>                    103 CZ CZ X2 CZ CZ X2 |i>
\end{lstlisting}
\end{lstfloat}

We test the fault-tolerant protocol using both the transmon simulator
and a real quantum processor on the IBM Q Experience
\cite{ibmquantumexperience2016}. For the simulation, we use the large
five-transmon system defined in \secref{sec:transmonmodelibm5ed}. The topology
is sketched in \figref{fig:ibm5edtopology}. The qubit labels
${q_0}{q_1}{q_2}{q_3}{q_4}$ given in this figure are the same that are used to
define the circuit components in \tabref{tab:errordetectionstates} and
\tabref{tab:errordetectiongates}. Note that the resonator $r_5$ from ${q_4}$ to
${q_0}$ has been added to the simulation model to also permit the implementation
of the state preparation of the encoded state $\overline{\ket{00}}$ given in
\tabref{tab:errordetectionstates}. All simulations were performed on the
supercomputers JURECA \cite{JURECA} and JUWELS \cite{JUWELS}.

For the real quantum processor, we use five qubits from the 16-qubit processor
\texttt{ibmqx5} \cite{ibmqx5}. The qubit mapping with respect to the  circuit
elements shown in \tabref{tab:errordetectionstates} and
\tabref{tab:errordetectiongates} is ${q_0}{q_1}{q_2}{q_3}{q_4} \mapsto
Q_4Q_3Q_2Q_{15}Q_{14}$, where ${q_0}=Q_4$ is the ancillary qubit. This
five-qubit subset of the 16-qubit device is also indicated in the topology graph
of the simulation model in \figref{fig:ibm5edtopology}. Note that the resonator
$r_5$ in this figure does not exist in the real device. Therefore, the encoded
version of the initial state $\ket{00}$ in \tabref{tab:errordetectionstates}
cannot be performed on the real device. This means that only $5/6$ of all
experiments defined below can be executed on the processor (the 6 coming from 3
initial states times 2 circuit versions). The idea is that, if simulation and
experiment  agree for these $5/6$ of all experiments, an extrapolation of the
simulation results for the remaining $1/6$ of the experiments may give an
estimate for the performance on a potentially new quantum processor on which the
additional connection between the two qubits would exist.

\subsection{Results}

\begin{figure}
  \centering
  \includegraphics[width=\textwidth]{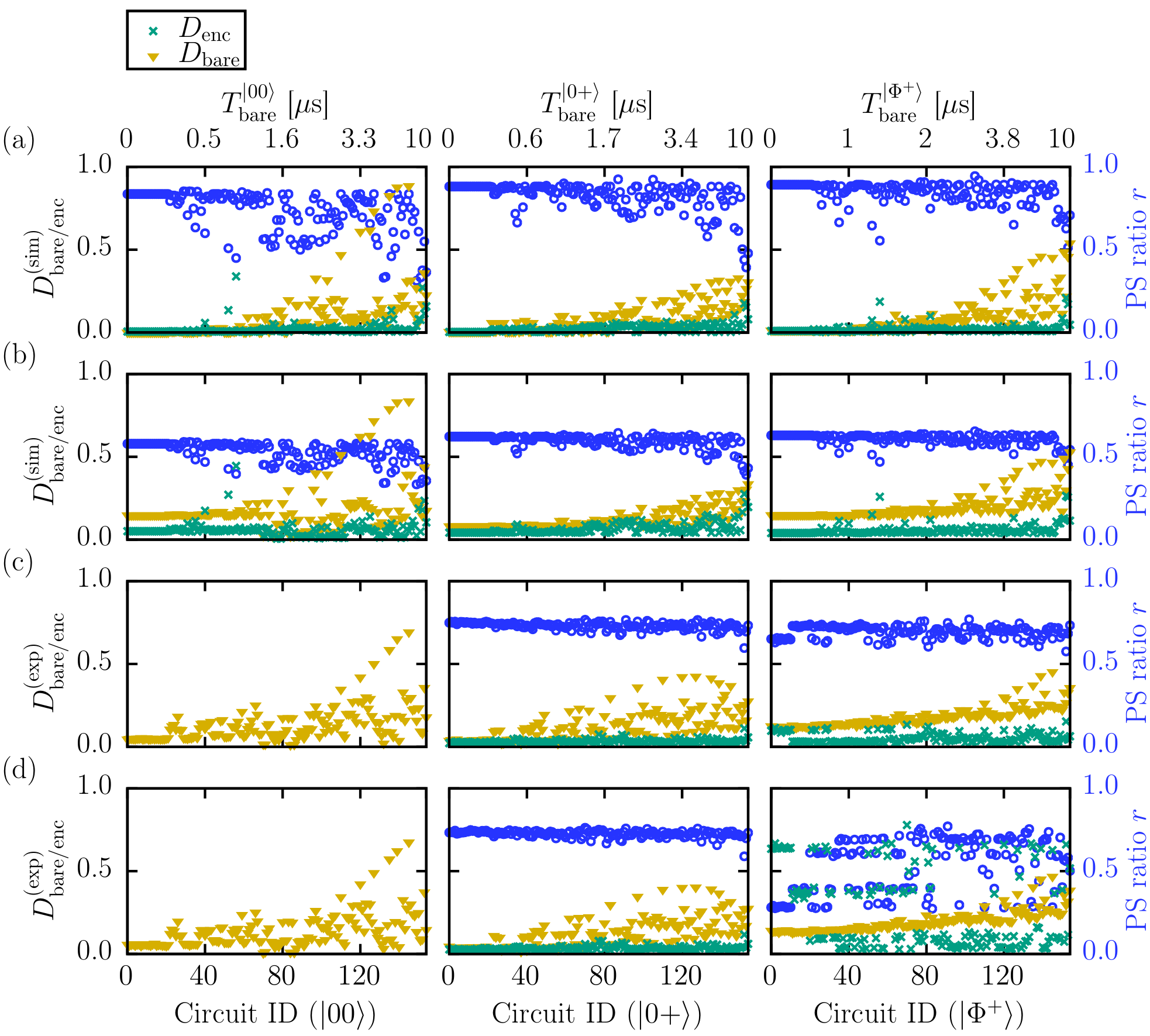}
  \caption{Test of the fault-tolerant protocol using (a) the transmon simulation
  model defined in \secref{sec:transmonmodelibm5ed}; (b) the same model with an
  additional measurement error of $\epsilon=0.08$
  (cf.~\equref{eq:errordetectionMeasurementError}); (c) the \texttt{ibmqx5}
  \cite{ibmqx5} on April 19, 2018; (d) the \texttt{ibmqx5} on April 20, 2018.
  Shown are the statistical distances $D_{\mathrm{bare}}$ for the bare circuits
  (yellow triangles) and $D_{\mathrm{enc}}$ for the encoded circuits (green
  crosses) as defined in \equref{eq:errordetectionDbareenc}, as well as the
  corresponding postselection ratios $r$ (blue circles). The three panels in
  each row correspond to the three initial states in
  \tabref{tab:errordetectionstates} (the left panels in (c) and (d) only contain
  data for the bare circuits because the encoded circuits for the initial state
  $\ket{00}$ cannot be run on the \texttt{ibmqx5}). The circuit IDs on the
  bottom axis are defined in Listing~\ref{code:fullcircuits}. They are sorted by
  the total time $T_{\mathrm{bare}}^{\ket{00}}$ required to simulate all pulses
  of the bare circuits for the initial state $\ket{00}$ (some representative
  time scales are indicated on the top axis).
  }
  \label{fig:errordetectionfull}
\end{figure}

For each of the 465 circuits given in Listing~\ref{code:fullcircuits}, we
execute the corresponding bare and encoded versions. This defines a total of 930
experiments that are both simulated and run on the \texttt{ibmqx5} processor. We
evaluate the statistical distances $D_{\mathrm{bare/enc}}$ (see
\equref{eq:errordetectionDbareenc}) between the obtained distributions and the
ideal probability distributions. All results are summarized in
\figref{fig:errordetectionfull}.

\sfigref{fig:errordetectionfull}(a) shows the statistical distances
$D_{\mathrm{bare/enc}}^{(\mathrm{sim})}$ obtained from the transmon simulation
model specified in \secref{sec:transmonmodelibm5ed} using the pulses with
frequency tuning defined in \tabref{tab:deviceibm5edPulseParametersGD} and
\tabref{tab:deviceibm5edPulseParametersCR2} (results for pulses without
frequency tuning are given in \cite{Willsch2018TestingFaultTolerance}). As
indicated  in this figure, the longest time evolution takes approximately
$\SI{10}{\micro s}$. We see that for most of the 465 circuits, the encoded
version performs better than the bare version, especially for longer
time evolutions.

However, a particular exception stands out: In the left panel of
\figref{fig:errordetectionfull}(a), three encoded circuits (shown as green
crosses) with circuit IDs $(40, 52, 56)$ and pulse durations
$T_{\mathrm{bare}}^{\ket{00}} \in (\SI{0.48}{\micro s}, \SI{0.64}{\micro s},
\SI{0.80}{\micro s})$ have a significantly higher statistical distance than the
corresponding bare circuits. Listing~\ref{code:fullcircuits} shows that these
circuits correspond to $(\mathrm{X2\,Z1})^w\ket{00}$ where $w\in(3,4,5)$. For
the bare version, such a circuit basically resembles the repeated application of
$X^\pi$ pulses studied in \secref{sec:repeatedgates}. The reason for this is
that the $Z_1$ gate needs no pulse but only affects the VZ phase of qubit 1
(cf.~\secref{sec:singlequbitVZgate}). Therefore, from the results shown in
\figref{fig:repeatedgateserrorrates}(a), we can expect a reasonably good
performance for the bare version. The encoded version, however, requires the
complicated state preparation of $\overline{\ket{00}}$ (see
\tabref{tab:errordetectionstates}). Furthermore, each
$\overline{\mathrm{X2}}=X_1X_2$ gate requires two $X^\pi$ pulses (see
\tabref{tab:errordetectionstates}), whose phases $\gamma$ change by $\pi$
because of the intermediate $\overline{\mathrm{Z1}}=Z_1Z_2$ gates
(cf.~\secref{sec:singlequbitVZgate}). The fact that we do not see this effect
for the encoded circuit $\overline{\mathrm{X2}}^4\,\overline{\ket{00}}$ with ID
51 implies that it is indeed the different phases that cause the deviation.
Apparently, this leads to a significant phase error for the encoded circuit that
cannot be detected by postselection anymore. The most severe case with ID 56 can
also be seen for the initial state $\overline{\ket{\Phi^+}}$ in the right panel
of \figref{fig:repeatedgateserrorrates}(a). Still, it is remarkable that the
encoded circuits perform so much better in all the other cases, even
though the state preparations in \tabref{tab:errordetectionstates} (especially
that for $\overline{\ket{00}}$) are much more involved.

In the simulations, we obtain the distributions $p_{j_0j_1}^{(\mathrm{bare/enc})}$
for \equref{eq:errordetectionDbareenc} directly from the state vector $\ket{\Psi}$
given by \equref{eq:psioftsolutioncoefficients}. In particular, this means that
no measurement error is included in the simulations.
To understand the effect of measurement errors on the fault-tolerant scheme,
we consider the simple model that for each qubit,
a 0 (1) is mistakenly counted as a 1 (0) with probability $\epsilon$.
For a distribution $p_{J}$ of $n$-bit strings $J=j_0j_1\cdots j_{n-1}$,
this model is implemented by the transformation
\begin{align}
  \label{eq:errordetectionMeasurementError}
  p_J \mapsto \sum\limits_{J'=0}^{2^n-1} p_{J'} \epsilon^{\Delta(J,J')}
  (1-\epsilon)^{n-\Delta(J,J')},
\end{align}
where $\Delta(J,J')$ is the Hamming distance between $J$ and $J'$, i.e., the
minimum number of bits that must be flipped to achieve $J=J'$.  We have $n=2$ in
the bare case and $n=5$ in the encoded case. Technically, a measurement error
according to \equref{eq:errordetectionMeasurementError} can be interpreted as a
depolarizing channel (see~\equref{eq:singletDepolarizingChannel}) immediately
before the measurement.

\sfigref{fig:errordetectionfull}(b) shows the effect of adding a measurement
error of $\epsilon=0.08$ to the simulation results from
\figref{fig:errordetectionfull}(a). The overall effect is that the statistical
distances $D_{\text{bare/enc}}^{(\mathrm{sim})}$ increase, but those of the bare
circuits increase more than those of the encoded circuits. Thus, measurement
errors of this sort can be detected and mitigated using the fault-tolerant
scheme, which is also reflected by an overall decrease in the postselection
ratios in \figref{fig:errordetectionfull}(b). However, note that the particular
family of encoded circuits $(\overline{\mathrm{X2}}\,\overline{\mathrm{Z1}})^w$
discussed above still stands out by performing worse than the corresponding
unencoded circuits.

In \figref{fig:errordetectionfull}(c), we present results obtained from
running the fault-tolerance test on the
\texttt{ibmqx5} \cite{ibmqx5} on April 19, 2018. For the bare versions, all
465 circuits given in Listing~\ref{code:fullcircuits} can be executed.
For the encoded versions, 155 out of the 465 circuits (corresponding to the
initial state $\overline{\ket{00}}$) cannot be executed, since the resonator
$r_5$ in \figref{fig:ibm5edtopology} does not exist in the device. Therefore, the
left panel of \figref{fig:errordetectionfull}(c) only shows results for the bare
circuits.

We see that using the fault-tolerant protocol systematically improves the
results. Furthermore, we find good qualitative agreement between experiment and
simulation. Extrapolating the simulation results suggests that encoded
circuits for the initial state $\overline{\ket{00}}$ would perform similarly
well if the device were extended by the appropriate connection.

\begin{table}
  \caption{Evaluation of the data for bit strings produced by the 16-qubit
  processor \texttt{ibmqx5} \cite{ibmqx5} for the encoded circuit
  $\mathrm{Z2}^4\,\ket{\Phi^+}$. The circuit has ID 4 in
  Listing~\ref{code:fullcircuits} and the result corresponds to one of the
  leftmost green crosses on the right panels in
  \figref{fig:errordetectionfull}(c) and (d).  Shown are the 20 most frequent
  outcomes for the experiment on April 19 (left) and April 20 (right), sorted by
  relative frequency. To demonstrate the postselection procedure, discarded bit
  strings are highlighted in {\color{red}red}, with the reason given in the third
  column. Bit strings that are not discarded are highlighted in {\color{blue}blue}. The
  qubit mapping on the device is ${q_0}{q_1}{q_2}{q_3}{q_4} \mapsto
  Q_4Q_3Q_2Q_{15}Q_{14}$, where ${q_0}=Q_4$ is the ancillary qubit (see also
  \figref{fig:ibm5edtopology}). As this data reveals, a hardware fault
  apparently caused a bit flip of qubit $Q_2$ or $Q_{15}$ with very high
  probability on April 20 such that the four most frequent outcomes are
  erroneously discarded.}
  \centering
  \label{tab:errordetectionfullExamineFault}
\begin{tabularx}{\linewidth}{@{}cYc|cYc@{}}
  \toprule
  \multicolumn{3}{c|}{April 19, 2018} & \multicolumn{3}{c}{April 20, 2018} \\
  \midrule
  Outcome & Frequency & Counted as & Outcome & Frequency & Counted as \\
  \midrule
  \textcolor{blue}{\texttt{0 0000}} & 0.163 & $\overline{\ket{00}}$  & \textcolor{red}{\texttt{0 0100}}  & 0.167 & (odd parity)        \\
  \textcolor{blue}{\texttt{0 1001}} & 0.161 & $\overline{\ket{11}}$  & \textcolor{red}{\texttt{0 1011}}  & 0.166 & (odd parity)        \\
  \textcolor{blue}{\texttt{0 0110}} & 0.130 & $\overline{\ket{11}}$  & \textcolor{red}{\texttt{0 0010}}  & 0.160 & (odd parity)        \\
  \textcolor{blue}{\texttt{0 1111}} & 0.123 & $\overline{\ket{00}}$  & \textcolor{red}{\texttt{0 1101}}  & 0.153 & (odd parity)        \\
  \textcolor{red}{\texttt{0 0010}}  & 0.061 & (odd parity)           & \textcolor{blue}{\texttt{0 1100}} & 0.054 & $\overline{\ket{01}}$ \\
  \textcolor{red}{\texttt{0 1011}}  & 0.053 & (odd parity)           & \textcolor{blue}{\texttt{0 1010}} & 0.051 & $\overline{\ket{10}}$ \\
  \textcolor{red}{\texttt{0 0100}}  & 0.050 & (odd parity)           & \textcolor{blue}{\texttt{0 0011}} & 0.043 & $\overline{\ket{01}}$ \\
  \textcolor{red}{\texttt{0 1101}}  & 0.049 & (odd parity)           & \textcolor{blue}{\texttt{0 0000}} & 0.043 & $\overline{\ket{00}}$ \\
  \textcolor{red}{\texttt{0 1000}}  & 0.037 & (odd parity)           & \textcolor{blue}{\texttt{0 0101}} & 0.038 & $\overline{\ket{10}}$ \\
  \textcolor{red}{\texttt{0 0001}}  & 0.025 & (odd parity)           & \textcolor{blue}{\texttt{0 1001}} & 0.037 & $\overline{\ket{11}}$ \\
  \textcolor{red}{\texttt{0 1110}}  & 0.023 & (odd parity)           & \textcolor{red}{\texttt{0 1000}}  & 0.014 & (odd parity)        \\
  \textcolor{blue}{\texttt{0 1010}} & 0.021 & $\overline{\ket{10}}$  & \textcolor{blue}{\texttt{0 0110}} & 0.011 & $\overline{\ket{11}}$ \\
  \textcolor{blue}{\texttt{0 1100}} & 0.019 & $\overline{\ket{01}}$  & \textcolor{red}{\texttt{0 0001}}  & 0.010 & (odd parity)        \\
  \textcolor{red}{\texttt{0 0111}}  & 0.018 & (odd parity)           & \textcolor{red}{\texttt{1 0010}}  & 0.009 & (wrong ancilla)       \\
  \textcolor{blue}{\texttt{0 0011}} & 0.012 & $\overline{\ket{01}}$  & \textcolor{blue}{\texttt{0 1111}} & 0.009 & $\overline{\ket{00}}$ \\
  \textcolor{blue}{\texttt{0 0101}} & 0.012 & $\overline{\ket{10}}$  & \textcolor{red}{\texttt{1 0100}}  & 0.006 & (wrong ancilla)       \\
  \textcolor{red}{\texttt{1 0000}}  & 0.008 & (wrong ancilla)        & \textcolor{red}{\texttt{1 1011}}  & 0.006 & (wrong ancilla)       \\
  \textcolor{red}{\texttt{1 1001}}  & 0.006 & (wrong ancilla)        & \textcolor{red}{\texttt{1 1101}}  & 0.005 & (wrong ancilla)       \\
  \textcolor{red}{\texttt{1 1111}}  & 0.006 & (wrong ancilla)        & \textcolor{red}{\texttt{0 1110}}  & 0.004 & (odd parity)        \\
  \textcolor{red}{\texttt{1 0110}}  & 0.006 & (wrong ancilla)        & \textcolor{red}{\texttt{0 0111}}  & 0.002 & (odd parity)        \\
  \bottomrule
\end{tabularx}
\end{table}

We repeated the experiments on the \texttt{ibmqx5} several times. Most of the time,
the results were of the type shown in \figref{fig:errordetectionfull}(c).
However, the fault-tolerance test was not successful every time we ran the
experiment. One such result from April 20, 2018 is shown in
\figref{fig:errordetectionfull}(d). As one can see, many of the encoded
circuits for the initial state $\overline{\ket{\Phi^+}}$ (right panel of \figref{fig:errordetectionfull}(d))
have unusually
high statistical distances and low postselection ratios.
We examined the corresponding calibration parameters and found that
qubits $Q_{14}$ and $Q_{15}$ had higher readout errors around 12\%. Typically,
the reported readout errors are between 4\% and 10\%. However, this
alone cannot explain the high error rates observed in the experiment.

Therefore, we take a closer look at the data.
\stabref{tab:errordetectionfullExamineFault} shows results obtained for the
encoded circuit $\mathrm{Z2}^4\,\ket{\Phi^+}$ corresponding to ID 4 in
Listing~\ref{code:fullcircuits}. The table shows the  most frequent five-bit
strings obtained on April 19 (left) and April 20 (right). After evaluating the
data using the postselection procedure illustrated in the table, we obtain
the frequency distributions $\smash{p_{j_0j_1}^{(\mathrm{enc})}}$, which are used to
evaluate the statistical distance $\smash{D_{\mathrm{enc}}^{(\mathrm{exp})}}$ given by
\equref{eq:errordetectionDbareenc}. On April 19, we obtained
$\smash{D_{\mathrm{enc}}^{(\mathrm{exp})}}\approx0.10$ (shown as one of the first green
crosses in the  right panel of \figref{fig:errordetectionfull}(c)). On April 20,
however, we found $\smash{D_{\mathrm{enc}}^{(\mathrm{exp})}}\approx0.65$, which corresponds to the much
worse result for ID 4 in the right panel of
\figref{fig:errordetectionfull}(d).

\stabref{tab:errordetectionfullExamineFault} reveals the reason for this: on
April 20, the four most frequent bit strings were discarded due to an odd
parity. In other words, they cannot be assigned to one of the logical basis
states given by \equsref{eq:errordetection00L}{eq:errordetection11L}. If one of
the central two bits corresponding to qubit $Q_2$ and $Q_{15}$ were flipped,
however, the result would be almost the same as on April 19. It seems that a
hardware fault caused a bit flip of qubit $Q_2$ or $Q_{15}$ with very high
probability on April 20. Apparently, this systematic error cannot be corrected
by the fault-tolerant protocol. We remark that, although reproducible, this
problem occurred only in a minority of all runs on the IBM Q processor.

In conclusion, we find that the fault-tolerant protocol provides a systematic
procedure to improve the results by encoding a logical state redundantly in a
larger number of physical qubits. This is remarkable because the errors in the
transmon simulation model and in the real device are not at all guaranteed to be
of the simple type assumed in the design of the protocol (see, for instance,
\figref{fig:crosstalkExperimentResults}(a)). Especially, the long circuit used
to encode the initial state $\ket{00}$ (see \tabref{tab:errordetectionstates})
might have seemed unlikely to improve the results. However, we also see that all
encoded gate elements used in the scheme do not require two-qubit gates (see
\tabref{tab:errordetectiongates}). Instead, these more error-prone gates
(cf.~\figaref{fig:repeatedgateserrorrates}{fig:repeatedgatescnot}) occur only in
the initial state preparation of a circuit. This design feature may provide an
alternative explanation why the particular protocol under investigation can
improve the results. Furthermore, in \cite{Willsch2018TestingFaultTolerance}, we
also studied the performance of the fault-tolerant protocol in the presence
of an environment (see also the model studied in
\secref{sec:freetransmonresonatorbathphotons}). These results suggest that the
scheme does not provide significant improvements when the errors are dominated
by decoherence, and a related conclusion was drawn for existing stabilizer
codes in \cite{Naus2018consequencesForErrorCorrection}.  Nevertheless, as long
as the errors in quantum information processors are dominated by the inherent
control and measurement errors in transmon systems, our results suggest that the
performance of a quantum computer can be systematically improved by using an
appropriate fault-tolerant protocol.

\section{Conclusions}

The purpose of the first experiment studied in this chapter was to observe
crosstalk effects, predicted by the transmon simulator, in an IBM Q processor.
We designed a class of circuits for this purpose (see
\figref{fig:crosstalkCircuit}), inspired by the state-dependent  frequency
shifts observed in previous simulations (see
\figref{fig:statedependentfrequenciesfree} and the fifth row in
\tabref{tab:gstgatedecompositions}). The five-transmon simulation results and
experimental results from the \texttt{ibmqx4} \cite{ibmqx4} showed almost
perfect agreement for time evolutions up to several microseconds  (see
\figref{fig:crosstalkExperimentResults}). We also observed that these effects
are generic for the transmon architecture in the sense that they did not depend
on the exact values of the device parameters. Moreover, the simulation results
suggested that the effects can be systematically reduced or enhanced, which  was
also confirmed by the experiment (see
\figref{fig:crosstalkExperimentResults}(b)).  This observation suggests that the
correlated crosstalk errors inherently included in the full dynamics of the
simulated transmon system (see \figref{fig:crosstalkExperimentBlochSpheres}) are
a very good model for the errors in the real processor.

For a set of experiments on the singlet state, we addressed the question if the
errors previously observed for a five-transmon processor
\cite{Michielsen2017BenchmarkingQC, Willsch2017GateErrorAnalysis} can be
described purely in terms of miscalibrated pulses for a simulated two-transmon
system. To a certain extent, this was possible (see \figref{fig:singletfitsim}),
but the two-transmon model does not seem to be capable of describing all
deviations  from the ideal result.  It would be interesting to see if a more
extensive analysis using the five-transmon simulation is more appropriate, as
indicated by previous results given in
\figaref{fig:repeatedgatescnot}{fig:crosstalkExperimentResults}. However, in
this case,  we found that simple error channels such as a depolarizing channel
and an amplitude damping channel offer a much simpler way of describing the
observed results (see \figref{fig:singletfitchannel}).

Finally, we tested a full protocol from the theory of quantum fault tolerance,
motivated by the observation that simple error channels could describe the
errors seen in the singlet-state experiment
(cf.~\figref{fig:singletfitchannel}). The extensive test comprised a total of
930 quantum circuits that were both  simulated using the large five-transmon
system defined in \secref{sec:transmonmodelibm5ed} and run on the processor
\texttt{ibmqx5} \cite{ibmqx5}. An analysis of the experimental results revealed
a systematic hardware fault in the processor during some runs (see
\tabref{tab:errordetectionfullExamineFault}). Most of the time, however, we
observed that the fault-tolerant protocol provides a systematic procedure  to
improve the results. As this was true for both simulations and experiments (see
\figref{fig:errordetectionfull}(a)--(c)), we conclude that the fault-tolerant
protocol systematically improves the quantum computer's performance if the
errors are due to the  intrinsic control and measurement errors present in
transmon systems.

Given the general qualitative and often also quantitative agreement  between
transmon simulation and experiment, we can conclude that the IBM Q processors
have been engineered very carefully to implement the quantum theoretical model
of superconducting transmon systems. It is interesting to see if further
development of the processors can also bring them sufficiently close to an
implementation of the computational model of an ideal gate-based quantum
computer.

%% file: conclusion.tex
\chapter{Discussion and conclusion}
\label{sec:conclusion}

The goal of this project was to develop a transmon simulator that utilizes  the
resources of digital supercomputers to study the emerging technology of transmon
quantum computers. NISQ devices of this architecture are currently built by
several companies such as IBM \cite{ibmquantumexperience2016}, Google
\cite{Google2019QuantumSupremacy}, and Rigetti Computing
\cite{rigetti2017computing}. We designed and implemented a simulation algorithm
that computes the real-time dynamics of a system of transmons and couplers by
solving the time-dependent Schr\"odinger equation (TDSE) for a generic model
Hamiltonian representing the quantum computing hardware.

The model features an arbitrary number of transmons and resonators, as well as
time-dependent pulses used to implement quantum gates on the qubits.
Furthermore, we described a way to simulate electromagnetic environments with
the model, together with a systematic procedure to extract suitable model
parameters from experiments or electromagnetic solvers. The numerical algorithm
used to solve the TDSE is unconditionally stable
\cite{deraedt1987productformula} and can be used to  obtain the dynamics for
several hundred microseconds on a sub-picosecond scale. In principle, the size
of the model is only limited by the available computational resources
on the supercomputer. In this work, we presented results for the simulation of
up to 16 transmons and resonators, described by more than four billion complex
coefficients.

We used simulations of free time evolutions to benchmark the simulation
algorithm and found excellent weak and strong scaling on the
supercomputer JURECA \cite{JURECA}. By studying the algorithm's accuracy, we
demonstrated that recently proven error bounds for the product-formula algorithm
\cite{WillschMadita2020PhD} are tight.

The simulation approach inherently includes effects beyond the ideal gate-based
quantum computer model, such as leakage to higher, non-computational states,
crosstalk between the transmons, entanglement between transmons  and resonators,
and control errors through imperfect pulses applied to the system. All of these
effects are known to be limiting factors in current transmon architectures
\cite{Wood2017LeakageRB, Theis2018DRAGstatusAfter10years,
Google2019QuantumSupremacy}.

Additionally, we performed experiments on real transmon devices available on the
IBM Q Experience \cite{ibmquantumexperience2016}. As the size of these NISQ
devices is in the range of what can be simulated on the supercomputers JURECA
\cite{JURECA} and JUWELS \cite{JUWELS}, it was possible to relate the simulation
results directly to data obtained from real quantum processors.

\subsubsection{Relation to perturbative results, master equations, and experiments}

We investigated known perturbative results for the model and found that, while
they provide a simple, effective description of the system, they develop a drift
in time that makes them unsuitable for pulse optimizations. By simulating the
dynamics of a transmon system coupled to a bath of harmonic oscillators, we
observed a transition from TDSE-based approaches to a Lindblad master equation,
with the result that the latter can be an adequate, effective description under
certain conditions. Finally, we characterized the resonator-mediated exchange
interaction between transmons and found that it is well described by an
appropriate $ZZ$ interaction on a two-qubit subspace. We demonstrated that the
strength  of this interaction can be accurately determined from simulated time
evolutions or experiments such as gate set tomography.

\subsubsection{Characterizing and predicting the performance of optimized quantum gates}

By studying an optimization procedure for the pulses used to implement quantum
gates, we found that the Nelder--Mead method is a suitable candidate to obtain
pulse parameters that have error rates of the same magnitude as those reported
in experiments. In fact, the pulses used for the transmon simulator were often
found to perform better in actual applications. This was especially true for two
variants (CR1 and CR4) of the echoed cross-resonance pulse (CR2) that is
routinely used to implement the two-qubit CNOT gate
\cite{ibmquantumexperience2016}.

We proved two statements to relate the average gate fidelity
\cite{nielsen2002gatefidelity} to more sophisticated gate metrics such as the
diamond distance \cite{kitaev1997diamondnorm}. Unlike previous results, the
relations also apply to trace-decreasing quantum operations, which are  relevant
for transmon systems where leakage is an important limitation. We found that the
gate metrics provide useful information about the amount of leakage and the
accuracy of a single application of a quantum gate pulse.

However, none of these gate metrics was found to be suitable for predicting  the
performance of repeated pulse applications in actual quantum algorithms. We
observed several cases with poor gate metrics and exceptional performance,  as
well as almost ideal gate metrics but bad performance in practical applications
(see also \cite{Willsch2017GateErrorAnalysis,
McKay2019ThreeQubitRBGateMetrics}). Only the diamond distance turned out to
provide an upper bound on the observable statistical distance, and even in this
case, the diamond distance also needed to be evaluated for repeated pulse
applications.

As an alternative to the common gate metrics, we extensively studied the
approach of gate set tomography to characterize quantum gates.  Gate set
tomography requires only the observable relative frequencies from experiments,
and we found that especially the completely positive trace-preserving estimates
have an exceptional predictive power for the performance of quantum gates in
actual applications.

\clearpage
\subsubsection{Crosstalk, effective error models, and quantum fault tolerance}

We studied a family of quantum circuits designed to probe crosstalk between
transmon qubits. We found that the time evolution predicted by the transmon
simulator was in excellent agreement with experimental results obtained from the
processor \texttt{ibmqx4} \cite{ibmqx4}, suggesting that the simulation model
provides a very good description of the errors present in the device.

For a class of quantum circuits designed to characterize the singlet state, we
found that the deviations from the ideal result are most likely not caused by
miscalibrated pulses. However, the deviations could be well described in terms
of simple error channels from the theory of quantum fault tolerance.

This motivated us to perform an extensive test of a fault-tolerant protocol
\cite{Gottesman2016quantumfaulttolerance}. Although the protocol cannot correct
decoherence errors \cite{Willsch2018TestingFaultTolerance}, both simulations and
experiments on the \texttt{ibmqx5} \cite{ibmqx5} suggest that the protocol
systematically improves the quantum computer's performance in the presence of
control and measurement errors characteristic of the transmon architecture.

\subsubsection{Outlook}

Many scenarios studied in this work reflect the empirical observation that
engineering accurate gate-based quantum computers is remarkably difficult. It is
important that detailed simulations such as those presented in this thesis are
continuously carried out to understand limitations and find potential ways to
overcome systematic errors. There are many interesting paths along which the
present work can be continued:

\begin{itemize}
  \item \textbf{Environment simulations:} In its current form, the model
  Hamiltonian defined in \secref{sec:transmonmodel} supports a bath of harmonic
  oscillators. This is a bosonic bath that can be used to model electromagnetic
  environments using  the procedure described in \secref{sec:extractfoster}. It
  would be interesting to extend the formalism to multiple transmons by
  following \cite{Nigg2012BlackBoxCircuitQuantization,
  Ansari2019BlackBoxCircuitQuantization}. Also, the bath simulations can be
  extended as described in \secref{sec:bathadditionalaspects}. In particular, it
  would  be interesting to analyze the consequences of replacing the bosonic
  bath with a fermionic bath or a spin bath as studied in
  \cite{Willsch2018TestingFaultTolerance, WillschMadita2020PhD,
  Willsch2020FluxQubitsQuantumAnnealing}.

  \item \textbf{Pulse optimization:} The problem of finding optimal pulses to
  implement quantum gates  is an area of active research. With the success of
  advanced machine learning techniques such as deep reinforcement learning (see
  \secref{sec:alterativegateoptimizationtechniques}), it would be interesting to
  find and investigate new pulses using these approaches as done in
  \cite{Niu2019DeepRLQuantumControlPulsesPiecewiseConstant,
  An2019DeepRLQuantumGateControl}. Also, IBM has recently introduced OpenPulse
  \cite{McKay2018OpenPulse}, by which new pulses studied with the transmon
  simulator (such as the CR variants in \figref{fig:crossresonancepulses}) can
  directly be executed on the device. It would be compelling to use this
  interface to compare pulses for the transmon simulator directly with
  experiments on the real device.
  \clearpage

  \item \textbf{Modeling experiments:} For some of the experiments studied in
  this project, minor differences between simulation and experiment were still
  observable. Future work could go into understanding these differences in
  detail. For instance, the tiny oscillations on the black curve in
  \figref{fig:crosstalkExperimentResults}(a) appear to be smoothed out in the
  experiment, which may be understood by adding an environment to the
  simulation in the sense of \secref{sec:extractfoster} or
  \figref{fig:freekitkaverage}. Also, it would be interesting to see if a
  five-transmon simulation for the singlet experiment can describe the
  deviations better than the two-transmon simulation (see
  \figref{fig:singletfitsim}). A completely different approach would be to study
  classical models of quantum information devices (see
  \cite{Blackburn2016ClassicalInterpretationsQuantumJosephsonExperiments,
  Ivakhnenko2018SimulatingQuantumPhenomenaClassicalOscillators}) and to
  understand in which respect their descriptive power differs from the quantum
  theoretical model used in this work.

  \item \textbf{Extending the simulation algorithm:} A promising direction would
  be to extend  the simulation model with time-dependent magnetic fluxes to
  include  flux-tunable transmons. This would require modifications to the
  numerical algorithm described in \secref{sec:simulationsoftware} and enable
  the simulation of transmon systems controlled by high-speed flux lines
  \cite{Roth2017OpensuperQExchangetypegate} as pursued by the European FET
  Flagship project OpenSuperQ \cite{opensuperq}. Research efforts in this
  direction are currently underway.
\end{itemize}

%% file: app.tex
\begin{appendices}

\numberwithin{figure}{section}
\numberwithin{table}{section}
\numberwithin{equation}{section}
\numberwithin{lstfloat}{section}
\renewcommand{\thesection}{\Alph{section}}
\renewcommand{\thefigure}{\thesection.\arabic{figure}}
\renewcommand{\thetable}{\thesection.\arabic{table}}
\renewcommand{\thelstfloat}{\thesection.\arabic{lstfloat}}
\renewcommand*{\sectionformat}{\makebox[0pt][l]{\raisebox{1.5em}{Appendix \thesection}}}
\renewcommand*{\sectionmarkformat}{Appendix \thesection:\enskip}

\section{Visualization of quantum gate implementations}
\label{app:visualization}

\begin{figure}
  \centering
  \includegraphics[width=\linewidth]{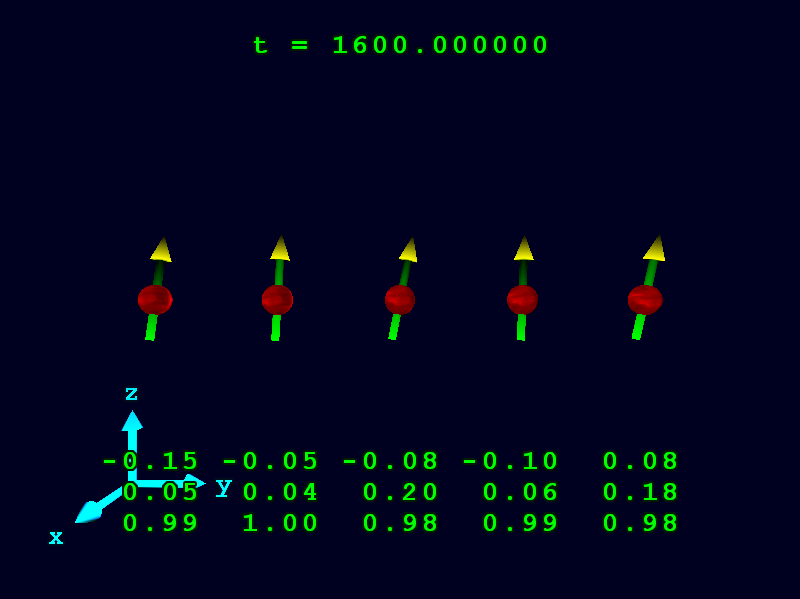}
  \caption{Example of the scene rendered by \texttt{visualizer}. The screenshot
  shows the Bloch vectors $\vec r_i(t)$ (see
  \equref{eq:multiqubitblochvectorTransmonTrace}) of five transmon qubits after
  the successive application of two $X$ gates (implemented as four $X^{\pi/2}$
  gates) on each of the qubits. The simulated system is the small five-transmon
  system defined in \secref{sec:transmonmodelibm5}. The corresponding gate
  metrics are given in \tabref{tab:ibm5gatemetrics}. The Bloch vectors
  corresponding to qubit $i=0,\ldots,4$ are shown from left to right.  As the
  pulse for an $X^{\pi/2}$ gate for this system takes $\SI{80}{ns}$ (see
  \tabref{tab:deviceibm5PulseParametersGD}), the depicted final time is
  $t=\SI{1600}{ns}$. Note that the Bloch vectors at the end of the time
  evolution are not perfectly straight. In particular, the second and the fourth
  Bloch vector are similarly tilted even though the corresponding gate metrics
  (see $X_1^{\pi/2}$ and $X_3^{\pi/2}$ in \tabref{tab:ibm5gatemetrics}) are of
  very different quality.}
  \label{fig:visualization}
\end{figure}

In \figref{fig:visualization}, we show a screenshot of the scene rendered by
\texttt{visualizer} (see \secref{sec:visualizer}) using the \texttt{C++} engine
Irrlicht \cite{irrlicht}. The scene is generated from time-evolution data of the
five-qubit transmon system used to study repeated gate applications in
\secref{sec:repeatedgates}. It shows the final state after applying two
successive $X$ gates on each  of the qubits. Although the gate sequence should
technically compose an identity operation, one can see that the Bloch vectors
are not perfectly straight, despite excellent gate metrics of the $X$ gate
pulses (especially for $X_3^{\pi/2}$, see \tabref{tab:ibm5gatemetrics}).

\begin{sidewaysfigure}
  \centering
  \includegraphics[width=\columnwidth]{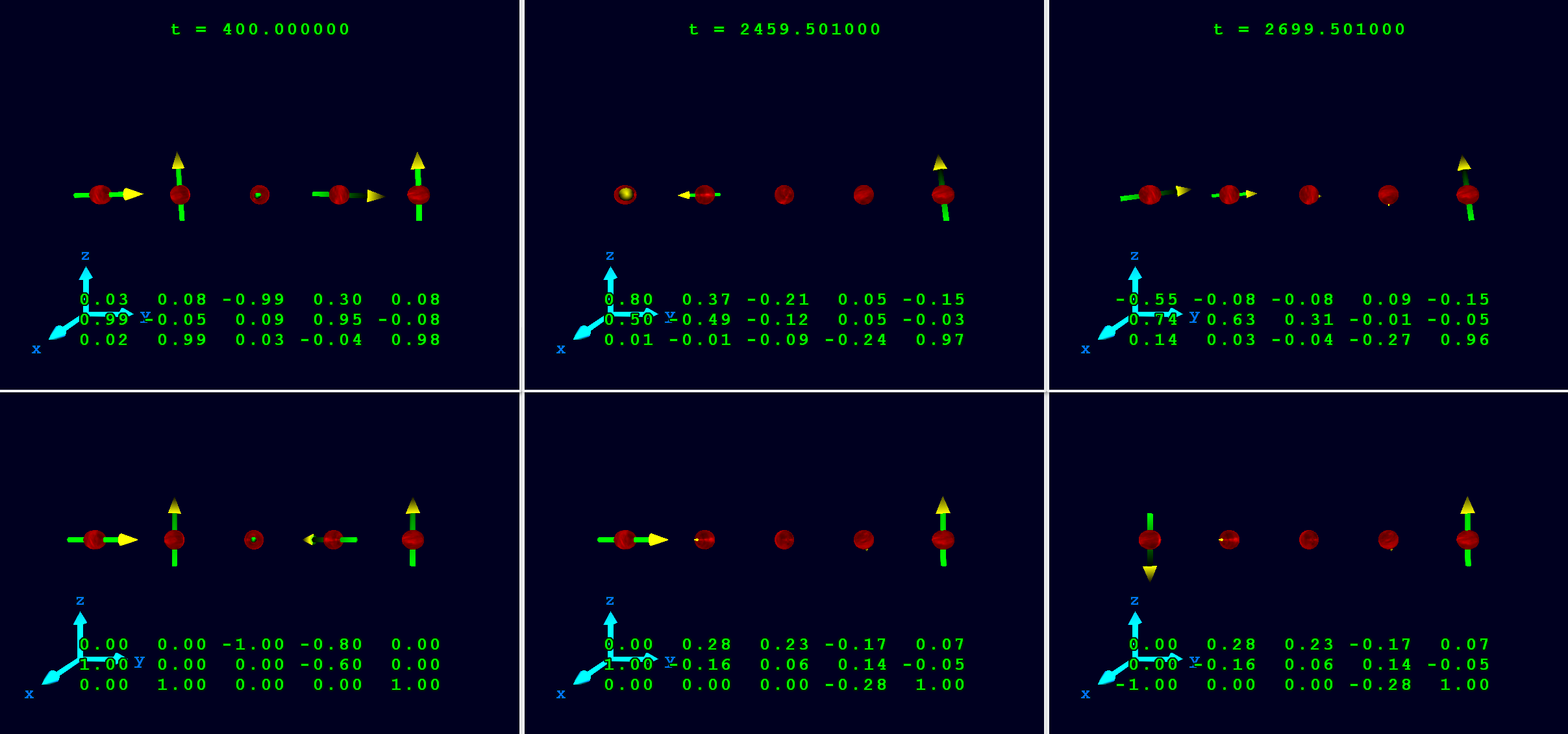}
  \caption{Selection of scenes rendered by \texttt{visualizer}. The screenshot
  shows the Bloch vectors $\vec r_i(t)$ (see
  \equref{eq:multiqubitblochvectorTransmonTrace}) of five qubits at the
  indicated times $t$ during the application of a random sequence of quantum
  gates. The top three images correspond to the time evolution of a five-qubit
  transmon system where the gates are implemented by pulses. The bottom three
  images show the exact Bloch vectors predicted by the ideal pen-and-paper
  quantum computer model described in \chapref{cha:quantumcomputing} and
  simulated using JUQCS \cite{DeRaedt2018MassivelyParallel,
  Willsch2020BenchmarkingWithJUQCS}. The rotating frames are different such that
  different directions of the arrows in the $xy$ plane may not indicate an
  error. The projection on the $z$ axis, however, indicates errors such that
  measuring the first qubit shown in the two images on the right would
  lead to an error of over 50\%.}
  \label{fig:visualizationrandom}
\end{sidewaysfigure}

\sfigref{fig:visualizationrandom} shows a selection of visualizations resulting
from the simulation of a quantum circuit with random gates from the  standard
gate set (cf.~\appref{app:gateset}) at three different points in time using a
five-transmon simulation (first row) and the ideal quantum computer simulator
JUQCS \cite{DeRaedt2018MassivelyParallel, Willsch2020BenchmarkingWithJUQCS}
(second row). The particular circuit reveals how crosstalk results in
state-dependent frequencies, inducing phase errors that can dramatically
increase the error in the system's output distribution in certain cases
(see also \figref{fig:crosstalkExperimentResults}).

\clearpage
\section{Elementary gate set used for the simulation}
\label{app:gateset}

In \tabref{tab:elementarygateset}, we summarize the set of single-qubit gates
and their representations in terms of the elementary single-qubit gates
$\textsc{U1}$, $\textsc{U2}$, and $\textsc{U3}$ defined in
\equsref{eq:singlequbitU1}{eq:singlequbitU3}.
The two-qubit gate used to make the gate set universal
\cite{DiVincenzo2000universalnearestneighbourexchangeinteraction} is the
\textsc{CNOT} gate defined in \equref{eq:cnotgate}.

Since the topic of this thesis is the simulation of transmon qubit systems,  and
the first platform offering access to transmon qubit processors was the  IBM Q
Experience \cite{ibmquantumexperience2016}, the gate set implemented for the
simulation is inspired by the gates available on the IBM Q Experience
\cite{Cross2017openqasm2}. Additionally, the gate set includes gates implemented
by JUQCS, which was used to compare the simulated gates to their equivalent on
an ideal universal quantum computer (see \cite{DeRaedt2007MassivelyParallel,
DeRaedt2018MassivelyParallel,
Willsch2020BenchmarkingWithJUQCS} for more information on JUQCS). At the time of
writing, JUQCS was used to simulate some of the larger circuits for Google's
quantum supremacy experiment \cite{Google2019QuantumSupremacy}, as well as
universal quantum circuits with up to 48 qubits, setting the largest simulation
of a universal gate-based quantum computer to date
\cite{DeRaedt2018MassivelyParallel}. In this context, \emph{universal} means
that arbitrary circuits can be simulated. If, however, only a specific subset of
circuits with a small number of entangling gates need to be simulated, a much
larger number of qubits is possible
\cite{pednault2017breaking49qubitbarrier, boixo2017lowdepthcircuitgraphs,
Chen2018ClassicalSimulation, Chen201864qubits,
MarkovBoixo2018QuantumSupremacySimulationWrapup,
VillalongaBoixo2019EstablishingQuantumSupremacyFrontierSummit,
Willsch2020BenchmarkingWithJUQCS}.

\begin{table}
  \caption{Summary of all single-qubit gates used in this work, including their circuit
  symbols and matrix representations. The gates listed in the first column
  and the alternative representations listed in the second column
  are only equivalent up to a global phase. The relations to the \textsc{U}
  gates defined in \equsref{eq:singlequbitU1}{eq:singlequbitU3}
  are given in the
  last column if the gates are implemented in this way. For any gate $G$,
  there is also the inverse operation $G^\dagger$ given by the Hermitian
  conjugate of the corresponding matrix.}
  \centering
  \label{tab:elementarygateset}
  %\begin{tabularx}{\linewidth}{@{}ccccc@{}}
  \begin{tabular}{@{}ccccc@{}}
    \toprule
    Gate & Alternatives & Symbol & Matrix & In \textsc{U} gates \\
    \midrule
    \addlinespace
    $I$ & Identity, Idle &
    $\Qcircuit @C=1em @R=.7em {& \gate{I} & \qw}$ &
    $\begin{pmatrix}
      1 & 0 \\
      0 & 1
    \end{pmatrix}$ & --- \\ %& --- \\
    \addlinespace
    $\textsc{U1}(\lambda)$ & \textsc{U1} gate &
    $\Qcircuit @C=1em @R=.7em {& \gate{\textsc{U1}(\lambda)} & \qw}$ &
    $\begin{pmatrix}
      1 & 0 \\
      0 & e^{i\lambda}
    \end{pmatrix}$ & --- \\ %& see \equref{eq:singlequbitU1}\\ %$c_1 R^z(\lambda)$ \\
    \addlinespace
    $\textsc{U2}(\phi,\lambda)$ & \textsc{U2} gate &
    $\Qcircuit @C=1em @R=.7em {& \gate{\textsc{U2}(\phi,\lambda)} & \qw}$ &
    $\dfrac{1}{\sqrt{2}}
    \begin{pmatrix}
      1 & -e^{i\lambda} \\
      e^{i\phi} & e^{i(\phi+\lambda)}
    \end{pmatrix}$ & --- \\ % & see \equref{eq:singlequbitU2}\\ %$c_2 R^z(\phi+\frac\pi2) R^x(\frac\pi2) R^z(\lambda-\frac\pi2)$ \\
    \addlinespace
    $\textsc{U3}(\theta,\phi,\lambda)$ & \textsc{U3} gate &
    $\Qcircuit @C=1em @R=.7em {& \gate{\textsc{U3}(\theta,\phi,\lambda)} & \qw}$ &
    $\begin{pmatrix}
      \cos\frac\theta2 & -e^{i\lambda} \sin\frac\theta2 \\
      e^{i\phi}\sin\frac\theta2 & e^{i(\phi+\lambda)}\cos\frac\theta2
    \end{pmatrix}$ & --- \\ % & see \equref{eq:singlequbitU3}\\ %$\substack{c_3 R^z(\phi+3\pi) R^x(\frac\pi2) \\R^z(\theta+\pi) R^x(\frac\pi2) R^z(\lambda)}$ \\
    \addlinespace
    $X$ & Bit flip &
    $\Qcircuit @C=1em @R=.7em {& \gate{X} & \qw}$ &
    $\begin{pmatrix}
      0 & 1 \\
      1 & 0
    \end{pmatrix}$ & $\textsc{U3}(\pi,0,\pi)$ \\ %& $R^x(\pi)$ \\
    \addlinespace
    $Y$ & Bit\&Phase flip &
    $\Qcircuit @C=1em @R=.7em {& \gate{Y} & \qw}$ &
    $\begin{pmatrix}
      0 & -i \\
      i & 0
    \end{pmatrix}$ & $\textsc{U3}(\pi,\frac\pi2,\frac\pi2)$ \\
    \addlinespace
    $Z$ & Phase flip &
    $\Qcircuit @C=1em @R=.7em {& \gate{Z} & \qw}$ &
    $\begin{pmatrix}
      1 & 0 \\
      0 & -1
    \end{pmatrix}$ & $\textsc{U1}(\pi)$ \\
    \addlinespace
    $H$ & Hadamard &
    $\Qcircuit @C=1em @R=.7em {& \gate{H} & \qw}$ &
    $\dfrac{1}{\sqrt{2}} \begin{pmatrix}
      1 & 1 \\
      1 & -1
    \end{pmatrix}$ & $\textsc{U2}(0,\pi)$ \\ %& $R^x(\pi)$ \\
    \addlinespace
    $S$ & $Z^{\pi/2}$, $R^z(\frac{\pi}{2})$  &
    $\Qcircuit @C=1em @R=.7em {& \gate{S} & \qw}$ &
    $\begin{pmatrix}
      1 & 0 \\
      0 & i
    \end{pmatrix}$ & $\textsc{U1}(\frac\pi2)$ \\
    \addlinespace
    %$S^\dagger$ & Phase flip &
    %$\Qcircuit @C=1em @R=.7em {& \gate{Z} & \qw}$ &
    %$\begin{pmatrix}
    %  1 & 0 \\
    %  0 & -1
    %\end{pmatrix}$ & $\textsc{U1}(\pi)$ \\
    %\addlinespace
    $T$ & $Z^{\pi/4}$, $R^z(\frac{\pi}{4})$  &
    $\Qcircuit @C=1em @R=.7em {& \gate{T} & \qw}$ &
    $\begin{pmatrix}
      1 & 0 \\
      0 & e^{i\pi/4}
    \end{pmatrix}$ & $\textsc{U1}(\frac\pi4)$ \\
    \addlinespace
    %$T^\dagger$ & Phase flip &
    %$\Qcircuit @C=1em @R=.7em {& \gate{Z} & \qw}$ &
    %$\begin{pmatrix}
    %  1 & 0 \\
    %  0 & -1
    %\end{pmatrix}$ & $\textsc{U1}(\pi)$ \\
    %\addlinespace
    $\texttt{+X}$ & $X^{-\pi/2}$, $R^x(-\frac{\pi}{2})$ &
    $\Qcircuit @C=1em @R=.7em {& \gate{\texttt{+X}} & \qw}$ &
    $\dfrac{1}{\sqrt{2}}
    \begin{pmatrix}
      1 & i \\
      i & 1
    \end{pmatrix}$ & $\textsc{U2}(\frac\pi2,-\frac\pi2)$ \\
    \addlinespace
    $\texttt{-X}$ &  $X^{\pi/2}$, $R^x(\frac{\pi}{2})$ &
    $\Qcircuit @C=1em @R=.7em {& \gate{\texttt{-X}} & \qw}$ &
    $\dfrac{1}{\sqrt{2}}
    \begin{pmatrix}
      1 & -i \\
      -i & 1
    \end{pmatrix}$ & $\textsc{U2}(-\frac\pi2,\frac\pi2)$ \\
    $\texttt{+Y}$ &  $Y^{-\pi/2}$, $R^y(-\frac{\pi}{2})$ &
    $\Qcircuit @C=1em @R=.7em {& \gate{\texttt{+Y}} & \qw}$ &
    $\dfrac{1}{\sqrt{2}}
    \begin{pmatrix}
      1 & 1 \\
      -1 & 1
    \end{pmatrix}$ & $\textsc{U2}(-\pi,\pi)$ \\
    \addlinespace
    $\texttt{-Y}$ &  $Y^{\pi/2}$, $R^y(\frac{\pi}{2})$ &
    $\Qcircuit @C=1em @R=.7em {& \gate{\texttt{-Y}} & \qw}$ &
    $\dfrac{1}{\sqrt{2}}
    \begin{pmatrix}
      1 & -1 \\
      1 & 1
    \end{pmatrix}$ & $\textsc{U2}(0,0)$ \\
    \addlinespace
    $\texttt{R}(k)$ & Phase gate &
    $\Qcircuit @C=1em @R=.7em {& \gate{\texttt{R}(k)} & \qw}$ &
    $\begin{pmatrix}
      1 & 0 \\
      0 & e^{2\pi i/2^k}
    \end{pmatrix}$ & $\textsc{U1}(\frac{2\pi}{2^k})$ \\
    \addlinespace
    \bottomrule
  \end{tabular}
\end{table}

\clearpage
\section{The reason for linear and unitary transformations in quantum theory}
\label{app:wigner}

The computational model of a gate-based quantum computer described in
\chapref{cha:quantumcomputing} is largely based on linear and unitary
transformations to describe transitions  between states. In fact, the use of
unitary matrices  can be seen as the characteristic difference between quantum
computers and digital or probabilistic machines (which would use Boolean or
stochastic matrices, respectively) \cite{Bernstein97quantumcomplexityturing}. An
interesting question is therefore: Where did the use of unitary transformations
in quantum theory come from?

To address this question, we review some of the mathematical arguments that have
been given during the development of quantum theory for the use of linear and
unitary transformations.  Note that none of these arguments can prove that Nature has
to be described by linear equations and unitary maps; they only illustrate why
humans have developed quantum theory on the basis of linear and unitary
transformations.

Historically, the question why quantum theory should be linear and unitary was
first  posed when the TDSE given by \equref{eq:tdse1} was introduced by
Schr\"odinger \cite{Schroedinger1926a,Schroedinger1926b}. As the emergence of
his equation was considered ``ad hoc'' by some researchers, they started a
search for arguments why transformations of a quantum state $\ket{\Psi}$ should
be linear and unitary. The first such argument was given by Wigner
\cite{wigner1931grouptheory,wigner1959grouptheory} and later formulated more
rigorously by Lomont and Mendelson \cite{Lomont1963WignersTheorem} and Bargmann
\cite{Bargmann1964WignersTheorem}. It has become widely known as \emph{Wigner's
theorem}.

\subsection{Wigner's theorem}

The argument behind Wigner's theorem starts from the assumption that in any experiment,
one can only ever observe transition probabilities of the form
$\abs{\braket{\phi|\psi}}^2$ between quantum states $\ket{\phi}$ and $\ket{\psi}$.
Since the laws of Nature are believed to be invariant under space-time symmetry operations,
the observable transition probabilities shall be conserved.
Mathematically, this means that any transformation $U$ between quantum states shall conserve the absolute value of the
inner product between complex vectors, i.e.,
\begin{align}
  \forall \ket{\phi},\ket{\psi} :\quad \abs{\braket{\phi|\psi}} = \abs{\braket{\phi'|\psi'}},
  \label{eq:conservationoftransitionprobabilities}
\end{align}
where $\ket{\phi'} = U(\ket{\phi})$ and $\ket{\psi'} = U(\ket{\psi})$.
Wigner's theorem essentially proves that among all conceivable transformations $U$,
only \emph{linear and unitary} or \emph{antilinear and antiunitary} operators
are compatible with \equref{eq:conservationoftransitionprobabilities}.
The hardest part of the proof is to show that $U$ must be either linear or antilinear.

Another equivalent formulation of Wigner's theorem often seen in the earlier literature
is stated in terms of equivalence classes $\boldsymbol{\psi} := \{c\ket{\psi}:c\in\mathbb C,\abs c=1\}$ (so-called \emph{rays})
\cite{Lomont1963WignersTheorem,Bargmann1964WignersTheorem}.
The motivation for this is the idea that two complex vectors $\ket{\psi}$ and $\ket{\psi'}$ describe the same physical state
if they differ only by a complex phase. Therefore, only the equivalence class $\boldsymbol{\psi}$
is understood to represent the actual physical state.
Wigner's condition then takes the form
\begin{align}
  \forall \boldsymbol{\phi},\boldsymbol{\psi} :\quad (\boldsymbol\phi,\boldsymbol\psi) = (\mathbf T(\boldsymbol{\phi}),\mathbf T(\boldsymbol{\psi})),
  \label{eq:conservationoftransitionprobabilitiesrays}
\end{align}
where $\mathbf T$ denotes the given symmetry transformation.
The square of the inner product $(\boldsymbol{\psi},\boldsymbol{\phi})$ in this space can
directly be interpreted as the transition probability between $\boldsymbol{\psi}$ and $\boldsymbol{\phi}$.
To obtain the more abstract formulation given in \equref{eq:conservationoftransitionprobabilities},
the ray expressions given in \equref{eq:conservationoftransitionprobabilitiesrays}
are defined as $\mathcal T(\boldsymbol\psi) := U(\ket\psi)$
and $(\boldsymbol\psi,\boldsymbol\phi):=\abs{\braket{\psi|\phi}}$,
where the complex vectors $\ket\psi\in\boldsymbol{\psi}$ and $\ket\phi\in\boldsymbol{\phi}$
are the representatives chosen for the computation.
This ray formulation of Wigner's theorem comes at the cost of having to prove that the ray expressions
are well-defined, in the sense that the results do not depend on the representatives.

There is also a more recent formulation of Wigner's condition that can be seen as combining
the mathematical convenience of \equref{eq:conservationoftransitionprobabilities} and the
interpretational character of \equref{eq:conservationoftransitionprobabilitiesrays}.
It is based on the density matrix representations of the pure states $\rho_\psi = \ketbra \psi \psi$
and $\rho_\phi = \ketbra \phi \phi$.
These objects are rank-1 projectors and have the advantage that they do not entail the ambiguity of
a complex phase factor in representing the same physical state.
In this formulation, Wigner's condition reads
\begin{align}
  \forall  \rho_\psi,\rho_\phi :\quad  \mathrm{Tr}\,\rho_\psi\rho_\phi =  \mathrm{Tr}\,f(\rho_\psi)f(\rho_\phi),
  \label{eq:conservationoftransitionprobabilitiesdensitymatrices}
\end{align}
where the symmetry transformation is denoted by the function $f$ (cf.~the
generalized Born rule in \equref{eq:GSTBornRule}). In this form, Wigner's
theorem has been extended to non-bijective transformations \cite{Geher2015WignerAnotherProof}. Instead of
\emph{linear and unitary} or \emph{antilinear and antiunitary} transformations,
it then states that the function $f$ has to be implemented by a \emph{linear or
antilinear isometry} $W$ such that $f(\rho_\psi) = W\rho_\psi W^\dagger$.

\subsubsection{Sketch of the proof}

In what follows, we briefly sketch the main ideas of the proof given by Geh\'er
\cite{Geher2015WignerAnotherProof}. To keep it simple, we consider bijective
transformations and focus on the formulation of the condition stated in
\equref{eq:conservationoftransitionprobabilities}.

Let $\{\ket{i}\}$ denote an orthonormal basis (ONB) of the Hilbert space $\mathcal H$. Note that, in the context of
quantum computing, we almost always deal with finite-dimensional Hilbert spaces such that
$i\in\{1,\ldots,N\}$ where $N=\mathrm{dim}\,\mathcal H$ (see below for more information on the infinite-dimensional case).

The proof first defines a set of $N$ vectors $\{\ket{i'}\}$ as the images of
this basis under $U$, i.e., $\ket{i'} := U(\ket i)$. One then constructs a
linear or antilinear map $\widetilde U:\mathcal H\to\mathcal H$ from these
images such that $\widetilde U\ket i := \ket{i'}$. Using
\equref{eq:conservationoftransitionprobabilities}, one can show that
$\{\ket{i'}\}$ is also an ONB of $\mathcal H$. By studying the action of $U$ on
the states $\ket i - \ket{i+1}$ and $\ket i + e^{i\pi/2}\ket{i+1}$, one can show
that the given transformation $U$ and the constructed $\widetilde U$ coincide,
and that there are only two options, namely the linear, unitary and the
antilinear, antiunitary case. It is instructive to study both cases and their
relevance in quantum theory separately.

\subsubsection{The linear, unitary case}
In the linear case, we have
\begin{align}
  U(a\ket{\phi} + b\ket{\psi}) = a U(\ket{\phi}) + b U(\ket{\psi})
\end{align}
for all states $\ket\phi$ and $\ket\psi$ and all $a,b\in\mathbb C$.
Given that $U$ is linear, it is quite straightforward to prove that $U$ has to be unitary.
We give a simple proof of this statement.

Since $U$ is linear, it can be represented by a matrix with matrix elements
\begin{align}
  \braket{i|U^\dagger U|j} &= \abs{\braket{i|U^\dagger U|j}} e^{i\varphi_{ij}} = \abs{\braket{i|j}} e^{i\varphi_{ij}} = \delta_{ij} e^{i\varphi_{ij}},
  \label{eq:Umustbeunitary}
\end{align}
where $\varphi_{ij}\in[0,2\pi)$ and we have used
\equref{eq:conservationoftransitionprobabilities} to eliminate $U^\dagger U$. In
the case $i\neq j$, we have $\braket{i|U^\dagger U|j} = 0$, so the phase
$\varphi_{ij}$ is irrelevant. In the case $i=j$, the left-hand side of
\equref{eq:Umustbeunitary} reads $\braket{i|U^\dagger U|i} = \norm{U\ket{i}}^2
>0$, so the phase factor $e^{i\varphi_{ii}}=1$. Hence we have
$\braket{i|U^\dagger U|j} = \delta_{ij}$ for all $i,j$. A similar argument
yields $\braket{i|UU^\dagger|j} = \delta_{ij}$, so we have $U^\dagger U =
UU^\dagger = \mathds1$, which by definition means that $U$ is unitary.

Given that transformations of the physical state $\ket{\Psi}$ can be implemented
by unitary operators $U$, i.e., $\ket{\Psi}\mapsto U\ket{\Psi}$, a derivation of
the TDSE given in \equref{eq:tdse1} is very simple: Each unitary operator can be
expressed as the exponential $\exp(K)$ of a skew-Hermitian operator $K$. Writing
$K=-i H t$ where $H$ is a Hermitian operator, we have $\ket{\Psi}\mapsto \exp(-i
H t)\ket{\Psi}$, or equivalently, \equref{eq:tdse1}.

\subsubsection{The antilinear, antiunitary case}
An antilinear operator is defined by the relation
\begin{align}
  U(a\ket{\phi} + b\ket{\psi}) = a^* U(\ket{\phi}) + b^* U(\ket{\psi})
\end{align}
for all states $\ket\phi$ and $\ket\psi$ and all $a,b\in\mathbb C$.
There are two things to keep in mind about antilinear operators:
First, the conventional Dirac notation may cause problems,
in the sense that in general $(\bra{i}U)\ket{j}\neq\bra{i}(U\ket{j})$.
This means that the common expression $\braket{i|U|j}$ must be read with caution.
Second, the \emph{complex conjugation} operator, which is the simplest example of an antilinear operator,
is dependent on the basis in terms of which it is defined. In particular,
two antilinear operators $U$ and $V$ defined by
$U(\sum_j a_j \ket{j})=\sum_j a_j^* \ket{j}$
and
$V(\sum_j a_j \ket{\tilde j})=\sum_j a_j^* \ket{\tilde j}$
with a slightly different basis $\ket{\tilde j}=i\ket j$ are not the same.

Wigner's theorem states that in the antilinear case,
the operator $U$ also needs to be antiunitary. This means that, in terms of the
notation used in \equref{eq:conservationoftransitionprobabilities}, the operator
$U$ has to satisfy $\braket{\phi'|\psi'}=\braket{\phi|\psi}^*$.

In general, antilinear operators are far less common in quantum theory than linear operators \cite{ballentine1998quantum}.
One reason for this is that we often work with transformations that depend on
a continuous parameter $t$ and shall satisfy the group relation $U(t_1)U(t_2)=U(t_1+t_2)$.
Since the product of two antilinear operators is always linear, such a transformation
cannot be implemented by an antilinear operator.

\subsubsection{Additional literature}

In the past decades,
many alternative proofs of Wigner's theorem have been published
(see e.g.~\cite{Gyory2004WignerHistory,Geher2015WignerAnotherProof,Barvinek2017WignerProof}).
They all explore different routes for the proof in the general case of an infinite-dimensional Hilbert space $\mathcal H$,
where the mathematical apparatus of functional analysis is utilized \cite{vonneumann1955,gustafson2003mathematicalconcepts}.
Some of the proofs require \emph{separability} of $\mathcal H$ \cite{Simon2008WignerTheoremNewProofs}
(i.e., the basis $\{\ket{i}\}$ is still countable such that $i\in\mathbb N$)
or differentiability of the transformation \cite{Mouchet2013WignerAlternativeProofDerivatives}.
Geh\'er's more recent proof can deal with non-separable Hilbert spaces
and non-bijective transformations \cite{Geher2015WignerAnotherProof}.
In addition to this, generalizations of the proof to several other algebraic structures have been considered
\cite{Molnar1998WignersTheoremAlgebraic,Geher2017WignerGrassmann}.
The diversity of mathematical arguments to prove Wigner's theorem is interesting.
However, all these proofs start from the conservation of the absolute value of inner products expressed in
\equref{eq:conservationoftransitionprobabilities}, meaning that
the observable transition probabilities shall be conserved.
They provide no conceptual alternative to approach the question where unitary
transformations in quantum theory come from. Therefore, we now look at some
approaches that do provide conceptual alternatives.

\subsection{Alternative approaches}

In this section, we list some alternative approaches to the question about
the use of linearity and unitarity in quantum theory. Each of these provides
separate insights into the topic.

\subsubsection{The physically compelling approach}

An approach with a conceptual alternative to Wigner's theorem was explored by
Jordan \cite{Jordan1962LinearQuantumDynamics, Jordan2009LinearQM}. His goal was
to derive Wigner's original condition in
\equref{eq:conservationoftransitionprobabilities} and the density matrix
formulation in \equref{eq:conservationoftransitionprobabilitiesdensitymatrices}
from a physically more compelling starting point. One such starting point is the
condition that the state of the system does not depend on anything outside the
system, but still allows for a description as part of a larger system. Jordan
shows that this condition implies that transformations between quantum states
must be linear \cite{Jordan2006LinearQuantumDynamics}. Once this is established,
one can derive Wigner's condition given by
\equref{eq:conservationoftransitionprobabilities}
\cite{Jordan1962LinearQuantumDynamics}. The same arguments given above then
yield the unitarity of the transformations and, correspondingly, the TDSE given
by \equref{eq:tdse1}.

\subsubsection{The necessity approach}

A deeper insight into the use of unitary operators was presented by Land\'e in
1969 \cite{Lande1969QuantumFactAndFiction3}, who explored the question ``Why do
the probabilities interfere by way of a matrix product law for the probability
amplitudes?'' He shows that transition probabilities $P(A\to B)$ and $P(B\to C)$
can be related, in a triangular form, to a transition probability $P(A\to
C)$ only by means of unitary matrices. There is no mysterious, fundamental
requirement for unitarity. Rather, the use of unitary operators in quantum
theory is a necessity due to the physicist's ambition to express such a triangular
relation in a convenient mathematical form.

\subsubsection{The consistency approach}

An interesting, fundamentally different approach has been described by  Caticha
\cite{caticha1998consistencyamplitudesprobability}. He demonstrates that  the
\emph{only consistent way} to manipulate probability amplitudes is by means of
linear, unitary transformations. The idea goes back to Cox's work on the
consistent use of probabilities and their connection to logic
\cite{Cox1946probability, Cox1961ProbableInference}. The fundamental character
of probability theory as extended logic
has been comprehensively presented by Jaynes \cite{jaynes2003probability}.

\subsubsection{The information-based approach}

A more recent approach advocated by Fuchs is based on the concept of
\emph{information}. \cite{Fuchs2001QuantumFoundationsInformation,
Fuchs2002QuantumFoundationsInformation}. The premise is that a
quantum state $\ket\Psi$ does not represent an objective entity that exists in
Nature. Rather, $\ket\Psi$ and the probability amplitudes  that it contains are
only the concise representation of our subjective information about Nature. The
subjectivity, in particular, implies that quantum theory does not need
interpretations \cite{Fuchs2000QuantumTheoryNeedsNoInterpretation}, and also
that the time evolution of a state vector $\ket\Psi$ does not represent the real
time evolution of a physical system, but rather the evolution of our personal
state of knowledge about the system
\cite{Fuchs2000QuantumTheoryNeedsNoInterpretation}. This point of view is
inspired by the Bayesian interpretation of probabilities (see also
\cite{jaynes2003probability}). As soon as probability amplitudes are seen as
representing information, their consistent evolution under linear and unitary
transformations follows \cite{Hardy2001QuantumTheoryFiveAxioms,
Schack2003QuantumTheoryFourAxioms}. In fact, in a simplified context, some argue
that the difference between probability theory and quantum theory is that, for
the former, the 1-norm of vectors is conserved (so they are transformed by
stochastic matrices), while for the latter, the 2-norm is conserved, so they are
transformed by unitary matrices \cite{Aaronson2013QuantumComputingDemocritus}.

\subsubsection{The logical inference approach}

A very conclusive approach is the logical inference (LI) approach
\cite{DeRaedt2014LIQuantumTheory, DeRaedt2015LIreviewQuantumTheory}. It starts
from the actual data that is obtained in experiments, i.e.,  individually
observed events. LI then tries to infer the \emph{most robust description} of
this data, given that the experiments are reproducible. This approach first
yields a nonlinear optimization problem for the description of the data. By
reformulating this optimization problem, one can then derive the equations of
quantum theory, such as the TDSE given by \equref{eq:tdse1}
\cite{DeRaedt2014LIQuantumTheory}, the Klein-Gordon equation
\cite{Donker2016LIKleinGordon},  the Pauli equation
\cite{DeRaedt2015LIPauliEquation},  or other well-known equations for quantum
mechanical key experiments \cite{deraedt2015LISternGerlachEPRB}. In the LI
approach, the linear and unitary character of the equations follows from a
reformulation of the optimization problem. Similar to Caticha's approach
mentioned above, the ideas behind LI are based on inductive reasoning in
the presence of uncertainty \cite{Cox1946probability, Cox1961ProbableInference,
tribus1969rationaldescriptions, jaynes2003probability}.

\clearpage
\subsection{General remarks}

It is worth mentioning that also nonlinear and non-unitary transformations are
sometimes used in the context of quantum theory. Nonlinear expressions for the
state vector $\ket\Psi$ are used as an effective tool to address complex
systems. For instance, in density functional theory, a linear multi-particle
Schr\"odinger equation is effectively replaced by a nonlinear single-particle
Schr\"odinger equation \cite{Engel2011DensityFunctionalTheory}. A simple example
for a non-unitary map is the evolution of a subsystem, which  is typically
described by a quantum operation (see \secref{sec:quantumoperations}). Also, the
measurement process in quantum theory is typically described in terms of
projections which are non-unitary maps (see, for instance, the description of
POVMs in the context of \equref{eq:GSTBornRule}).

Regarding the mathematical arguments presented above, one should keep in mind
that there can never be a way to formally prove the need of linear and unitary
transformations in descriptions of Nature. The only thing that we can prove is
that within a certain theory such as quantum theory, under certain assumptions,
the mathematical objects of the theory evolve under linear
and unitary transformations. But in
principle, any experiment might inspire a new theory capable of better
describing the observed data without requiring the concept of linear and unitary
operators.

\clearpage
\section{Implementations of the four-component transformations \texorpdfstring{$V$}{V} and \texorpdfstring{$V^\dagger$}{Vdagger}}
\label{app:implementations}

This appendix includes \texttt{C++} code examples for each of the three
alternative implementations of the four-component transformations discussed in
\secref{sec:numericalalgorithm} and benchmarked in
\secref{sec:performancebenchmark}. They constitute the $V$ and $V^\dagger$
operations in the second-order Suzuki-Trotter product-formula
\equref{eq:psioftTimeEvolutionUpdateRule}. $V$ and $V^\dagger$ consist of a
tensor product of $4\times4$ matrices (see
\equref{eq:Vtensorproductof4x4matrices}), which require four-component updates
of the form of \equref{eq:psioft4componentupdates} for all $0\le i <
N_{\mathrm{Tr}}$ and $0\le r < N_{\mathrm{Res}}$. The central loops of the three
different  implementations for the transformation $V$ are given in
Listings~\ref{code:implementation0}--\ref{code:implementation2}. A description
of the variables and function names used in the code listings is given in
\tabref{tab:codelistingsVariablesFunctions}.

\begin{lstfloat}
  \caption{Implementation 0: Complete single loop with branches}
\begin{lstlisting}[style=mycodestyle,label=code:implementation0]
uint64_t inc = 1;  // inc = 0b 00..00 01 00..00
for(int i = NTr-1; i >= 0; --i, inc <<= 2) {
    uint64_t mask = 0b11 * inc;  // mask = 0b 00..00 11 00..00
    #pragma omp for
    for(uint64_t KM = 0; KM < dim; ++KM)
        if((KM & mask) == 0)
            mul4x4(&Vn[16*i], &psi[KM], inc);
}
for(int r = NRes-1; r >= 0; --r, inc <<= 2) {
    uint64_t mask = 0b11 * inc;  // mask = 0b 00..00 11 00..00
    #pragma omp for
    for(uint64_t KM = 0; KM < dim; ++KM)
        if((KM & mask) == 0)
            mul4x4(&Va[16*r], &psi[KM], inc);
}
\end{lstlisting}
\end{lstfloat}

\begin{lstfloat}
  \caption{Implementation 1: Reduced single loop with bitwise operations}
\begin{lstlisting}[style=mycodestyle,label=code:implementation1]
uint64_t inc = 1;  // inc = 0b 00..00 01 00..00
for(int i = NTr-1; i >= 0; --i, inc <<= 2) {
    uint64_t rmask = inc - 1;          // rmask = 0b 00..00 00 11..11
    uint64_t lmask = dim - 1 - rmask;  // lmask = 0b 11..11 11 00..00
    #pragma omp for
    for(uint64_t KMred = 0; KMred < dim/4; ++KMred) {
        // KM = 0b **..** 00 **..**
        uint64_t KM = ((KMred & lmask) << 2) | (KMred & rmask);
        mul4x4(&Vn[16*i], &psi[KM], inc);
    }
}
for(int r = NRes-1; r >= 0; --r, inc <<= 2) {
    uint64_t rmask = inc - 1;          // rmask = 0b 00..00 00 11..11
    uint64_t lmask = dim - 1 - rmask;  // lmask = 0b 11..11 11 00..00
    #pragma omp for
    for(uint64_t KMred = 0; KMred < dim/4; ++KMred) {
        // KM = 0b **..** 00 **..**
        uint64_t KM = ((KMred & lmask) << 2) | (KMred & rmask);
        mul4x4(&Va[16*r], &psi[KM], inc);
    }
}
\end{lstlisting}
\end{lstfloat}

\begin{lstfloat}
  \caption{Implementation 2: Reduced nested loops}
\begin{lstlisting}[style=mycodestyle,label=code:implementation2]
uint64_t inc = 1;  // inc = 0b 00..00 01 00..00
for(int i = NTr-1; i >= 0; --i, inc <<= 2) {
    uint64_t incl = inc << 2;  // incl = 0b 00..01 00 00..00
    #pragma omp for collapse(2)
    for(uint64_t K = 0; K < dim; K += incl)  // K = 0b **..** 00 00..00
        for(uint64_t M = 0; M < inc; ++M)    // M = 0b 00..00 00 **..**
            mul4x4(&Vn[16*i], &psi[K | M], inc);
}
for(int r = NRes-1; r >= 0; --r, inc <<= 2) {
    uint64_t incl = inc << 2; // incl = 0b 00..01 00 00..00
    #pragma omp for collapse(2)
    for(uint64_t K = 0; K < dim; K += incl)  // K = 0b **..** 00 00..00
        for(uint64_t M = 0; M < inc; ++M)    // M = 0b 00..00 00 **..**
            mul4x4(&Va[16*r], &psi[K | M], inc);
}
\end{lstlisting}
\end{lstfloat}

\begin{table}
  \renewcommand{\arraystretch}{1.1}
  \caption{Description of the identifiers used in Listings~\ref{code:implementation0}--\ref{code:implementation2}.}
  \centering
  \label{tab:codelistingsVariablesFunctions}
  \begin{tabularx}{\linewidth}{@{}lX@{}}
    \toprule
    Identifier & Description \\
    \midrule
    %\texttt{uint64_t} & Unsigned 64-bit integer type \\
    \texttt{NTr} & Number of transmons $N_{\mathrm{Tr}}$ \\
    \texttt{NRes} & Number of resonators $N_{\mathrm{Res}}$ \\
    \texttt{dim} & Dimension of the Hilbert space
    ($\mathrm{dim}(\mathcal H) = 4^{N_{\mathrm{Tr}}+N_{\mathrm{Res}}}$, see \equref{eq:HilbertSpaceTruncated})\\
    \texttt{psi} & Coefficients $\psi_{\texttt{KM}}(t)$ of the state vector $\Psi(t)$ (see \equref{eq:psioftsolutioncoefficientsKM})\\
    \texttt{Vn} & Matrix $V^{(n)}_i$ including the eigenstates of the charge operator in the transmon basis (see \equref{eq:transmonchargeoperatortransmonbasisEigendecomposition})\\
    \texttt{Va} & Matrix $V^{(a)}_r$ including the eigenstates of the operator $\hat a_r+\hat a_r^\dagger$ in the Fock basis (see \equref{eq:resonatorelectricfieldoperatorEigendecomposition})\\
    \texttt{inc} & Integer indicating the current position of the $4\times4$ transformation
    given in \equref{eq:psioft4componentupdates}. It is left-shifted by two bits after
    each iteration of the outer loops. \\
    \texttt{i} & Transmon index of the current $4\times4$ transformation, corresponding to the two bits indicated by \texttt{inc} \\
    \texttt{r} & Resonator index of the current $4\times4$ transformation, corresponding to the two bits indicated by \texttt{inc} \\
    \texttt{K}, \texttt{M}, \texttt{KM} & (Parts of) the index \texttt{KM} given by \equref{eq:KMindexnotation} \\
    \texttt{KMred} & Reduced index consisting of \texttt{KM} without the two bits indicated by \texttt{inc} \\
    \texttt{mask} & Bit mask with the two bits indicated by \texttt{inc} set \\
    \texttt{rmask} & Bit mask for the bits on the right of the position indicated by \texttt{inc} \\
    \texttt{lmask} & Bit mask for the bits on the left of the position indicated by \texttt{inc} (including the two bits in the middle) \\
    \texttt{incl} & Increment for the \texttt{K} part of the index \texttt{KM} given by \equref{eq:KMindexnotation} \\
    \texttt{mul4x4} & A vectorized complex $4\times4$ transformation implementing
    \equref{eq:psioft4componentupdates}. The inner loops including this operation
    are independent and can be parallelized. This has been indicated in the
    code listings using a simple OpenMP directive. \\
    \bottomrule
  \end{tabularx}
\end{table}

\clearpage
\section{Error bounds for observables}
\label{app:errorobservable}

The accuracy of the product-formula algorithm (see
\secref{sec:numericalalgorithm}) can be controlled  by means of rigorous error
bounds for the solution of the TDSE \cite{deraedt1987productformula,
huyghebaert1990productFormulaTimeDependentErrorBounds} (see
\equaref{eq:accuracyErrorGlobalPhase}{eq:accuracyErrorOverlap}).  However, the
bounds apply to the full state vector, so they may be impractical if we are only
interested in the expectation value of a certain observable.

Therefore, we tested two general error bounds for expectation
values of observables in \secref{sec:accuracy} (see
\equaref{eq:accuracyObservableBoundInf}{eq:accuracyObservableBoundVar}).
The bounds are given by
\begin{subequations}
\begin{align}
  \label{eq:accuracyObservableBoundInfApp}
  \abs{\braket{\psi|A|\psi} - \braket{\phi|A|\phi}} &\le 2 \sqrt \Delta\, \|A\|_2,\\
  \label{eq:accuracyObservableBoundVarApp}
  \abs{\braket{\psi|A|\psi} - \braket{\phi|A|\phi}} &\le
  2  \sqrt \Delta
  \sqrt{\vphantom{\Delta}\smash{\mathrm{Var}_{\psi}(A)}}\,\abs{\braket{\psi|\phi}}
  + 2 \Delta\,\|A\|_2,
\end{align}
\end{subequations}
where $A$ is an observable (i.e., a Hermitian operator), $\ket\psi$ and
$\ket\phi$ are pure states, $\Delta=1-\abs{\braket{\psi|\phi}}^2$ is the
distinguishability between $\ket{\psi}$ and $\ket{\phi}$, $\|A\|_2$ denotes the
spectral norm (largest singular value) of $A$, and
$\mathrm{Var}_{\psi}(A)=\langle A^2\rangle-\langle A\rangle^2$ is the variance
of $A$ with respect to the state $\ket\psi$.

The second bound was shown to be tight (see
\figref{fig:accuracyglobalerror}(a)), and a general proof for its validity is
given in \cite{WillschMadita2020PhD}. For the first bound, we give a short and
elementary proof in this appendix.

First, note that for any Hermitian operator $B$ with eigenvectors $\ket b$,
eigenvalues $b$, and singular values $\abs b$, we have
\begin{align}
  \label{eq:hoelder}
  \abs{\mathrm{Tr}\,BA}
  = \bigg\vert\sum_b b \underbrace{\braket{b|A|b}}_{\le\|A\|_2}\bigg\vert
  %\le \underbrace{\sum_b \abs b}_{\|B\|_{\mathrm{Tr}}}\|A\|_2,
  \le \bigg(\sum_b \abs b\bigg) \|A\|_2\:
  = \|B\|_{\mathrm{Tr}} \|A\|_2,
\end{align}
where $\|B\|_{\mathrm{Tr}}~=\sum_b\abs b$ is the trace norm of $B$. We remark
that \equref{eq:hoelder} is a special case of H\"older's inequality for Schatten
norms, which states that $\abs{\mathrm{Tr}\,X^\dagger Y} \le
\norm{\sigma(X)}_p\norm{\sigma(Y)}_{p*}$, where $X$ and $Y$ are operators,
$\sigma(X)$ and $\sigma(Y)$ denote vectors of their respective singular
values, $\norm{v}_p = (\sum_i \abs{v_i}^p)^{1/p}$ is the $p$-norm of a vector
$v$, and $p,p^*\in[1,\infty]$ are chosen so that $1/p+1/p^*=1$
\cite{watrous2018theoryofQI}. In this notation, $p=1$ and $p^*=\infty$
correspond to  the trace norm and the spectral norm, respectively.

Applying this result to \equref{eq:accuracyObservableBoundInfApp}, we have
\begin{align}
  \abs{\braket{\psi|A|\psi} - \braket{\phi|A|\phi}}
  = \abs{\mathrm{Tr}[(\ketbra\psi\psi-\ketbra\phi\phi)A]}
  \le \|(\ketbra\psi\psi-\ketbra\phi\phi)\|_{\mathrm{Tr}} \|A\|_2.
\end{align}
To find the trace norm of $\ketbra\psi\psi-\ketbra\phi\phi$, we make use of
\equref{eq:gatemetricsdiamondnormTraceDistanceRank2} for $\alpha=\beta=1$ (the
proof is given in \secref{sec:diamonddistance}):
\begin{align}
  \|(\ketbra\psi\psi-\ketbra\phi\phi)\|_{\mathrm{Tr}}~=2\sqrt{1-\abs{\braket{\psi|\phi}}^2}.
\end{align}
Recognizing the distinguishability $\Delta=1-\abs{\braket{\psi|\phi}}^2$,
we obtain
\begin{align}
  \abs{\braket{\psi|A|\psi} - \braket{\phi|A|\phi}} &\le 2 \sqrt \Delta\, \|A\|_2,
\end{align}
which proves \equref{eq:accuracyObservableBoundInfApp}.

\clearpage
\section{Pulse parameters for quantum gates}
\label{app:pulseparameters}

For reference, we list all pulse parameters used for the various multi-transmon
systems under investigation (see \secref{sec:transmonmodelsystems}). The pulse
parameters are the result of the pulse optimization procedure discussed in
\chapref{cha:optimization} (see
\secaref{sec:optimizatingsinglequbitgate}{sec:optimizatingtwoqubitgate} for the
particular meaning of the single-qubit and two-qubit pulse parameters).
The performance of the individual gates listed in this appendix
is discussed in detail in \chapref{cha:gateerrors}.

\begin{table}[h]
  \footnotesize
  \caption{Parameters of the optimized single-qubit GD pulses defined in \equref{eq:singlequbitpulseGD}
  for the two-transmon system (see \secref{sec:transmonmodelibm2gst}).
  For the compilation process (cf.~\secref{sec:compiler}), these parameters are
  specified in the form of Listing~\ref{code:compilergatepulses}.}
  \centering
  \label{tab:deviceibm2gstPulseParametersGD}
\begin{tabular}{@{}ccccc@{}}
  \toprule
  Pulse name & $f\,[\mathrm{GHz}]$ & $T_X\,[\mathrm{ns}]$ & $\Omega_X$ & $\beta_X\,[\mathrm{ns}]$ \\
  \midrule
  \texttt{xpih-0} & 5.3463 & 83 & 0.002221 & 0.2309 \\
  \texttt{xpih-1} & 5.1167 & 83 & 0.002269 & 0.2891 \\
  \texttt{xpi-0}  & 5.3463 & 83 & 0.004444 & 0.2193 \\
  \texttt{xpi-1}  & 5.1167 & 83 & 0.004538 & 0.2239 \\
  \bottomrule
\end{tabular}
\end{table}

\begin{table}[h]
  \footnotesize
  \caption{Parameters of the optimized two-qubit pulses defined in \secref{sec:optimizatingtwoqubitgate}
  for the two-transmon system (see \secref{sec:transmonmodelibm2gst}).
  We tested three kinds of CR pulses (cf.~\figref{fig:crossresonancepulses}):
  CR1 (see the text below \equref{eq:twoqubitpulseCR1}),
  CR2 (see the text below \equref{eq:twoqubitpulseCR2equationMoreSophisticated}),
  and CR4.
  The pulse name indicates the control qubit $i_C$ and the target qubit $i_T$
  in the form \texttt{cnot-$i_C$-$i_T$}.
  The duration of all single-qubit GD pulses included in these schemes is
  $(T_X^\pi)_C=(T_X^{\pi/2})_T=\SI{83}{ns}$.
  For the compilation process (cf.~\secref{sec:compiler}), these parameters are
  specified in the form of Listing~\ref{code:compilergatepulses}.}
  \centering
  \label{tab:deviceibm2gstPulseParametersCR}
  \medskip
\begin{tabular}{@{}ccccccccc@{}}
  \toprule
  \textbf{CR1}\\
  Pulse name & $f_{i_T}\,[\mathrm{GHz}]$ & $T_{\mathrm{CR}}\,[\mathrm{ns}]$ & $\Omega_{\mathrm{CR}}$ & $\Omega_{\mathrm{Cancel}}$ & $\phi_{\mathrm{CR}}$ & $\phi_{\mathrm{Cancel}}$ & $\phi_{C}$ & $\phi_{T}$ \\
  \midrule
  \texttt{cnot-0-1} & 5.1166 &  41.865 & 0.0793 &  0.00618 &  0.54 & 0.00 & -2.10 & 0.04 \\
  \texttt{cnot-1-0} & 5.3464 & 128.193 & 0.0940 & -0.00162 & -2.89 & 1.72 &  3.25 & 1.40 \\
  \bottomrule
  \toprule
  \textbf{CR2}\\
  Pulse name & $f_{i_C}\,[\mathrm{GHz}]$ & $f_{i_T}\,[\mathrm{GHz}]$ & $T_{\mathrm{CR}}\,[\mathrm{ns}]$ & $\Omega_{\mathrm{CR}}$ & $(\Omega_X^\pi)_C$ & $(\beta_X^\pi)_C\,[\mathrm{ns}]$ & $(\Omega_X^{\pi/2})_T$ & $(\beta_X^{\pi/2})_T\,[\mathrm{ns}]$ \\
  \midrule
  \texttt{cnot-0-1} & 5.3463 & 5.1167 & 102.9746 & 0.01111 & 0.004444 & 0.2193 & 0.002269 & 0.2891 \\
  \texttt{cnot-1-0} & 5.1167 & 5.3463 &  71.5580 & 0.07058 & 0.004538 & 0.2239 & 0.002221 & 0.2309 \\
  \bottomrule
  \toprule
  \textbf{CR4}\\
  Pulse name & $f_{i_C}\,[\mathrm{GHz}]$ & $f_{i_T}\,[\mathrm{GHz}]$ & $T_{\mathrm{CR}}\,[\mathrm{ns}]$ & $\Omega_{\mathrm{CR}}$ & \multicolumn{4}{c}{GD pulse parameters from \tabref{tab:deviceibm2gstPulseParametersGD}} \\
  \midrule
  \texttt{cnot-0-1} & 5.3463 & 5.1167 & 50.2385 & 0.01018 & \multicolumn{4}{c}{\multirow{2}{*}{$(\Omega_X^\pi,\beta_X^\pi)_C$ and $(\Omega_X^{\pi/2},\beta_X^{\pi/2},\Omega_X^\pi,\beta_X^\pi)_T$}} \\
  \texttt{cnot-1-0} & 5.1167 & 5.3463 & 30.1557 & 0.06934  \\
  \bottomrule
\end{tabular}
\end{table}

\begin{table}
  \footnotesize
  \caption{Parameters of the single-qubit GD pulses defined in \equref{eq:singlequbitpulseGD}
  for the small five-transmon system (see \secref{sec:transmonmodelibm5}).
  The parameters have been obtained from the pulse
  optimization procedure described in \secref{sec:optimizatingpulseparameters}.
  The label \texttt{withf} indicates pulses with frequency tuning such that the tuned frequency $f$ may differ slightly from the qubit frequency (cf.~\tabref{tab:deviceibm51}).}
  \centering
  \label{tab:deviceibm5PulseParametersGD}
\begin{tabular}{@{}ccccc@{}}
  \toprule
  Pulse name & $f\,[\mathrm{GHz}]$ & $T_X\,[\mathrm{ns}]$ & $\Omega_X$ & $\beta_X\,[\mathrm{ns}]$ \\
  \midrule
  \texttt{xpih-0-withf} & 5.34697 & 80 & 0.00231 & 0.246 \\
  \texttt{xpih-1-withf} & 5.30232 & 80 & 0.00232 & 0.220 \\
  \texttt{xpih-2-withf} & 5.11345 & 80 & 0.00236 & 0.218 \\
  \texttt{xpih-3-withf} & 5.22506 & 80 & 0.00233 & 0.232 \\
  \texttt{xpih-4-withf} & 5.07065 & 80 & 0.00237 & 0.228 \\
  \bottomrule
\end{tabular}
\end{table}

\begin{table}
  \footnotesize
  \caption{Parameters of the two-qubit CR2 pulses defined in \secref{sec:optimizatingtwoqubitgate}
  (see the text below \equref{eq:twoqubitpulseCR2equationMoreSophisticated})
  for the small five-transmon system (see \secref{sec:transmonmodelibm5}).
  The parameters have been obtained from the pulse
  optimization procedure described in \secref{sec:optimizatingpulseparameters}.
  The pulse name indicates the control qubit $i_C$ and the target qubit $i_T$
  in the form \texttt{cnot-$i_C$-$i_T$}.
  The label \texttt{withf} indicates pulses with frequency tuning such that the tuned frequencies $f_{i_C}$ and $f_{i_T}$
  may differ slightly from the qubit frequencies (cf.~\tabref{tab:deviceibm51}).
  The duration of all single-qubit GD pulses included in the CR2 pulse is always
  $(T_X^\pi)_C=(T_X^{\pi/2})_T=\SI{80}{ns}$.}
  \label{tab:deviceibm5PulseParametersCR2}
\begin{tabular}{@{}ccccccccc@{}}
  \toprule
  Pulse name & $f_{i_C}\,[\mathrm{GHz}]$ & $f_{i_T}\,[\mathrm{GHz}]$ & $T_{\mathrm{CR}}\,[\mathrm{ns}]$ & $\Omega_{\mathrm{CR}}$ & $(\Omega_X^\pi)_C$ & $(\beta_X^\pi)_C\,[\mathrm{ns}]$ & $(\Omega_X^{\pi/2})_T$ & $(\beta_X^{\pi/2})_T\,[\mathrm{ns}]$ \\
  \midrule
  \texttt{cnot-0-2-withf} & 5.34697 & 5.11345 & 100.341 & 0.0113 & 0.00463 & 0.250 & 0.00236 & 0.218 \\
  \texttt{cnot-1-2-withf} & 5.30232 & 5.11345 & 121.308 & 0.0103 & 0.00465 & 0.230 & 0.00236 & 0.218 \\
  \texttt{cnot-3-2-withf} & 5.22506 & 5.11345 & 88.442  & 0.0114 & 0.00469 & 0.240 & 0.00236 & 0.218 \\
  \texttt{cnot-4-2-withf} & 5.07065 & 5.11345 & 48.632  & 0.0114 & 0.00502 & 0.223 & 0.00236 & 0.218 \\
  \bottomrule
\end{tabular}
\end{table}

\begin{table}
  \footnotesize
  \caption{Parameters of the single-qubit GD pulses defined in \equref{eq:singlequbitpulseGD}
  for the large five-transmon system (see \secref{sec:transmonmodelibm5ed}).
  The parameters have been obtained from the pulse
  optimization procedure described in \secref{sec:optimizatingpulseparameters}.
  The corresponding optimization process is visualized in \figref{fig:optimizationxpih}.
  The label \texttt{withf} indicates that the drive frequency $f$ has also been
  optimized such that it may differ slightly from the qubit frequency (cf.~\tabref{tab:deviceibm5ed1}).}
  \centering
  \label{tab:deviceibm5edPulseParametersGD}
\begin{tabular}{@{}ccccc@{}}
  \toprule
  Pulse name & $f\,[\mathrm{GHz}]$ & $T_X\,[\mathrm{ns}]$ & $\Omega_X$ & $\beta_X\,[\mathrm{ns}]$ \\
  \midrule
  \texttt{xpih-0} & 4.97154 & 80 & 0.00238 & 1.335 \\
  \texttt{xpih-1} & 5.07063 & 80 & 0.00236 & -1.904 \\
  \texttt{xpih-2} & 5.26657 & 80 & 0.00233 & -2.165 \\
  \texttt{xpih-3} & 5.10145 & 80 & 0.00236 & 0.498 \\
  \texttt{xpih-4} & 4.86036 & 80 & 0.00241 & 2.276 \\
  \midrule
  \texttt{xpih-0-withf} & 4.97164 & 80 & 0.00239 & 0.239 \\
  \texttt{xpih-1-withf} & 5.07043 & 80 & 0.00236 & 0.238 \\
  \texttt{xpih-2-withf} & 5.26634 & 80 & 0.00233 & 0.229 \\
  \texttt{xpih-3-withf} & 5.10147 & 80 & 0.00236 & 0.232 \\
  \texttt{xpih-4-withf} & 4.86055 & 80 & 0.00241 & 0.236 \\
  \bottomrule
\end{tabular}
\end{table}

\begin{table}
  \footnotesize
  \caption{Parameters of the two-qubit CR2 pulses defined in \secref{sec:optimizatingtwoqubitgate}
  (see the text below \equref{eq:twoqubitpulseCR2equationMoreSophisticated})
  for the large five-transmon system (see \secref{sec:transmonmodelibm5ed}).
  The parameters have been obtained from the pulse
  optimization procedure described in \secref{sec:optimizatingpulseparameters}.
  The corresponding optimization process is visualized in \figref{fig:optimizationcnot}.
  The pulse name indicates the control qubit $i_C$ and the target qubit $i_T$
  in the form \texttt{cnot-$i_C$-$i_T$}.
  The label \texttt{withf} indicates pulses with frequency tuning such that the tuned frequencies $f_{i_C}$ and $f_{i_T}$
  may differ slightly from the qubit frequencies (cf.~\tabref{tab:deviceibm5ed1}).
  The duration of all single-qubit GD pulses included in the CR2 pulse is always
  $(T_X^\pi)_C=(T_X^{\pi/2})_T=\SI{80}{ns}$.}
  \label{tab:deviceibm5edPulseParametersCR2}
\begin{tabular}{@{}ccccccccc@{}}
  \toprule
  Pulse name & $f_{i_C}\,[\mathrm{GHz}]$ & $f_{i_T}\,[\mathrm{GHz}]$ & $T_{\mathrm{CR}}\,[\mathrm{ns}]$ & $\Omega_{\mathrm{CR}}$ & $(\Omega_X^\pi)_C$ & $(\beta_X^\pi)_C\,[\mathrm{ns}]$ & $(\Omega_X^{\pi/2})_T$ & $(\beta_X^{\pi/2})_T\,[\mathrm{ns}]$ \\
  \midrule
  \texttt{cnot-1-0} & 5.07063 & 4.97154 & 76.955  & 0.0097 & 0.00461 & 0.640 & 0.00238 & 1.335 \\
  \texttt{cnot-1-4} & 5.07063 & 4.86036 & 64.161  & 0.0183 & 0.00476 & -0.148 & 0.00241 & 2.276 \\
  \texttt{cnot-2-1} & 5.26657 & 5.07063 & 33.398  & 0.0235 & 0.00465 & -0.036 & 0.00236 & -1.904 \\
  \texttt{cnot-3-2} & 5.10145 & 5.26657 & 242.064 & 0.0111 & 0.00471 & 0.508 & 0.00233 & -2.165 \\
  \texttt{cnot-3-4} & 5.10145 & 4.86036 & 33.247  & 0.0290 & 0.00465 & 0.640 & 0.00241 & 2.276 \\
  \texttt{cnot-4-0} & 4.86036 & 4.97154 & 105.151 & 0.0210 & 0.00449 & -1.511 & 0.00238 & 1.335 \\
  \midrule
  \texttt{cnot-1-0-withf} & 5.07043 & 4.97164 & 73.538  & 0.0101 & 0.00477 & 0.798 & 0.00239 & 0.239 \\
  \texttt{cnot-1-4-withf} & 5.07043 & 4.86055 & 109.439 & 0.0114 & 0.00472 & 0.502 & 0.00241 & 0.236 \\
  \texttt{cnot-2-1-withf} & 5.26634 & 5.07043 & 82.077  & 0.0111 & 0.00463 & 0.661 & 0.00236 & 0.238 \\
  \texttt{cnot-3-2-withf} & 5.10147 & 5.26634 & 58.763  & 0.0429 & 0.00480 & -0.198 & 0.00233 & 0.229 \\
  \texttt{cnot-3-4-withf} & 5.10147 & 4.86055 & 85.294  & 0.0118 & 0.00474 & 0.247 & 0.00241 & 0.236 \\
  \texttt{cnot-4-0-withf} & 4.86055 & 4.97164 & 98.599  & 0.0239 & 0.00483 & 0.115 & 0.00239 & 0.239 \\
  \bottomrule
\end{tabular}
\end{table}

\clearpage
\section{Average fidelity of trace-decreasing quantum operations}
\label{app:prooffidelity}

In this appendix, we provide two separate proofs for a generalized version of an
explicit relation between the average fidelity $F_{\mathrm{avg}}$ (defined
in terms of an integral over random states) and the entanglement fidelity
$F_{\mathrm{ent}}$ (typically accessible in closed-form),
\begin{align}
  \label{eq:fidelityAvgEntRelation}
  F_{\mathrm{avg}}(\mathcal E) = \frac{d F_{\mathrm{ent}}(\mathcal E) + \mathrm{Tr}\,\mathcal E(\mathds1/d)}{d+1}.
\end{align}
Here, $d$ is the dimension of the Hilbert space (typically $2^n$ for an
$n$-qubit system), and $\mathcal E(\rho)=\sum_\alpha E_\alpha\rho
E_\alpha^\dagger$  is a completely positive map which is not necessarily
trace-preserving, i.e., $\sum_\alpha E_\alpha^\dagger E_\alpha \le \mathds1$ (cf.~\secref{sec:quantumoperations}). In the
special case that $\mathcal E$ is trace-preserving, we have $\mathrm{Tr}\,\mathcal E(\mathds1/d) = 1$ such
that \equref{eq:fidelityAvgEntRelation} reduces to the well-known
expression given in \cite{horodecki1999fidelity}.

After stating some preliminary definitions to settle the notation, we give
both an algebraic proof using methods from quantum information theory and
an elementary, analytic proof by direct calculation.

\subsection{Preliminaries}

The \emph{fidelity} between two quantum states $\rho$ and $\sigma$ is defined as
\cite{Jozsa1994fidelity} (see also \cite{Fuchs1995PhD})
\begin{align}
  \label{eq:fidelityMixedStates}
  F(\rho,\sigma) = \|\sqrt\rho\sqrt\sigma\|_{\mathrm{Tr}}^2\,
  = \left(\mathrm{Tr}\sqrt{\sqrt\rho\sigma\sqrt\rho}\right)^2,
\end{align}
where we used the definition of the trace norm
$\|X\|_{\mathrm{Tr}}~=\mathrm{Tr}\sqrt{X^\dagger X}$ (sum of the singular
values) of $X$. If one of the states is pure, e.g.~$\rho=\ketbra\psi\psi$, the
fidelity simplifies to the overlap $\braket{\psi|\sigma|\psi}$.

The \emph{average fidelity} of a quantum operation $\mathcal E$ is defined by averaging
the fidelity $F(\ketbra\psi\psi,\mathcal E(\ketbra\psi\psi))$ over random pure states $\ket\psi$,
\begin{align}
  \label{eq:fidelityAvg}
  F_{\mathrm{avg}}(\mathcal E)
  = \int \mathrm{d}\!\ket\psi \braket{\psi|\mathcal E(\psi)|\psi}.
\end{align}
The integral is taken over pure states $\ket\psi$ whose $2d$ real coefficients are distributed
uniformly on the surface of a $2d$-dimensional unit
sphere. For simplicity, we use the notation that $\psi=\ketbra\psi\psi$ if
the meaning is clear from the context.
%for draft: say it can be evaluated by ... bengtsson2006geometryofquantumstates

The \emph{entanglement fidelity} of $\mathcal E$ is defined as the fidelity
$F(\Phi,(\mathcal E\otimes\mathds1)(\Phi))$, where $\ket\Phi=\sum_j\ket{jj}/\sqrt d$
is the maximally entangled state on an extended Hilbert space that is also of
dimension $d$. We have
\begin{align}
  \label{eq:fidelityEnt}
  F_{\mathrm{ent}}(\mathcal E)
  = \braket{\Phi|(\mathcal E\otimes\mathds1)(\Phi)|\Phi}
  = \sum_\alpha \frac{\abs{\mathrm{Tr}\,E_\alpha}^2}{d^2}.
\end{align}
%for draft: a relation between equref and equref is convenient since it prevents one from having to evaluate the integral...

\subsection{Quantum information theoretic proof}
\label{sec:prooffidelityalgebraic}

The following proof is based
on the algebraic proof given in \cite{nielsen2002gatefidelity}, generalized to
non-trace-preserving quantum operations and thus extending the work in
\cite{Gilchrist2005fidelities}.

We consider the so-called \emph{twirled} quantum operation
\begin{align}
  \label{eq:fidelityTwirl}
  \mathcal E_T(\rho) = \int \mathrm{d}U\,U^\dagger\mathcal E(U\rho U^\dagger)U,
\end{align}
where the integral is over the Haar measure on the group of
unitary matrices \cite{Spengler2012ParametrizationUnitaryHaar}. This operation
leaves both the average fidelity given by \equref{eq:fidelityAvg}
and the entanglement fidelity given by \equref{eq:fidelityEnt} invariant,
which can be shown by substitution \cite{nielsen2002gatefidelity}.
Therefore, we have
\begin{subequations}
  \begin{align}
    \label{eq:fidelityFavgTwirled}
    F_{\mathrm{avg}}(\mathcal E) &= F_{\mathrm{avg}}(\mathcal E_T), \\
    \label{eq:fidelityFentTwirled}
    F_{\mathrm{ent}}(\mathcal E) &= F_{\mathrm{ent}}(\mathcal E_T).
  \end{align}
\end{subequations}
%The proof in \cite{nielsen2002gatefidelity} continues by showing that
Furthermore,
for any $\rho$ and any unitary operator $V$, we have
\begin{align}
  \label{eq:fidelityTwirlCommute}
  V\mathcal E_T(\rho)V^\dagger = \int \mathrm{d}U\,VU^\dagger\mathcal E(U\rho U^\dagger)UV^\dagger = \mathcal E_T(V\rho V^\dagger),
\end{align}
which can also be shown by substitution, i.e., $W=UV^\dagger$. Let $\rho=P$ be a
rank-1 projector and $Q=\mathds1-P$ its orthogonal complement. We define the
space $S_P$ ($S_Q$) as the space onto which $P$ ($Q$) projects. For any
block-diagonal unitary matrix $V=V_P + V_Q$, where $V_P$ ($V_Q$) is only
non-zero on $S_P$ ($S_Q$), we have
$VPV^\dagger=V_PV_P^\dagger={\mathds1}_{S_P} = P$ and thus $V\mathcal
E_T(P)V^\dagger = \mathcal E_T(P)$. Since this holds for any block-diagonal
unitary matrix of the form $V=V_P+ V_Q$, $\mathcal E_T(P)$ must also be
block-diagonal and each block must be proportional to the identity. The identity
on $S_P$ is ${\mathds1}_{S_P}=P$, and the identity on $S_Q$ is
${\mathds1}_{S_Q}=Q=\mathds1-P$, so we have
\begin{align}
  \label{eq:fidelityTwirlAsSumOfProjectors}
  \mathcal E_T(P) &= \alpha P + \beta (\mathds1-P)
\end{align}
for some $\alpha$ and $\beta$.
Using \equref{eq:fidelityTwirlCommute} again for an arbitrary unitary operator $V$
transforms $P$ in this equation into any other rank-1 projector $P'=VPV^\dagger$,
so we see that $\alpha$ and $\beta$ are the same for each $P$.

Writing an arbitrary $\rho=\sum_i\rho_iP_i$ as a sum of rank-1 projectors $P_i$ with unit trace,
linearity of $\mathcal E_T$ yields that for any $\rho$, $\mathcal E_T(\rho) = \alpha \rho + \beta(\mathds1-\rho)$.
By replacing $\beta = p/d$ and $\alpha = t-p+p/d$, we obtain
\begin{align}
  \label{eq:fidelityDepolarizingAlmost}
  \mathcal E_T(\rho) = (t - p)\rho + p\mathds1/d,
\end{align}
where also the parameters $p$ and $t$ are independent of $\rho$.

The difference to the trace-preserving case considered in \cite{nielsen2002gatefidelity}
is now that in general, $\mathrm{Tr}\,\mathcal E_T(\rho)=t\neq1$. Consequently, \equref{eq:fidelityDepolarizingAlmost}
does not represent a depolarizing quantum operation anymore. Note that care must be taken
with the identity symbol $\mathds1$ in \equref{eq:fidelityDepolarizingAlmost}
since it only applies when the map $\mathcal E_T$ is restricted to density
matrices with $\mathrm{Tr}\,\rho=1$. If trace-decreasing quantum operations are applied
to operators with trace less than 1 (e.g.~if they are chained, $\mathcal E_T(\mathcal E_T(\rho))$),
the correct expression is $\mathcal E_T(\rho) = (t - p)\rho + p\mathds1\,\mathrm{Tr}\,\rho/d$.
To determine the trace parameter $t$, we evaluate \equaref{eq:fidelityTwirl}{eq:fidelityDepolarizingAlmost}
for $\rho=\mathds1/d$:
\begin{align}
  \label{eq:fidelityDetermineTraceParameter}
  t = \mathrm{Tr}\,\mathcal E_T(\mathds1/d)
  = \mathrm{Tr}\,\int \mathrm{d}U\,U^\dagger\mathcal E(UU^\dagger)U/d
  = \mathrm{Tr}\,\mathcal E(\mathds1/d).
\end{align}

Using the simple form of \equref{eq:fidelityDepolarizingAlmost}, we can
evaluate the fidelities in \equaref{eq:fidelityFavgTwirled}{eq:fidelityFentTwirled}
directly. For the average fidelity, we find
\begin{align}
  \label{eq:fidelityFavgEvaluated}
  F_{\mathrm{avg}}(\mathcal E) =  \int \mathrm{d}\!\ket\psi \braket{\psi|\mathcal E(\psi)|\psi} = \mathrm{Tr}\,\mathcal E(\mathds1/d) - p + \frac p d,
\end{align}
and the entanglement fidelity becomes
\begin{align}
  \label{eq:fidelityFentEvaluated}
  F_{\mathrm{ent}}(\mathcal E) =  \braket{\Phi|(\mathcal E_T\otimes\mathds1)(\Phi)|\Phi} = \mathrm{Tr}\,\mathcal E(\mathds1/d) - p + \frac p {d^2}.
\end{align}
Solving \equref{eq:fidelityFentEvaluated} for $p$ and inserting the result into
\equref{eq:fidelityFavgEvaluated} yields the desired relation given by
\equref{eq:fidelityAvgEntRelation}.

\subsection{Analytic proof}
\label{sec:prooffidelityanalytic}

In this version of the proof, we directly evaluate the integral for the average
gate fidelity given by \equref{eq:fidelityAvg}. First, by expanding the pure
state $\ket\psi=\sum_i c_i\ket{i}$ with $c_i\in\mathbb C$, we obtain
\begin{align}
  \label{eq:fidelityAvgIntegral1}
  F_{\mathrm{avg}}(\mathcal E)
  = \sum_\alpha \sum_{ijkl} \braket{i|E_\alpha|j}\braket{k|E_\alpha^\dagger|l} \int \mathrm{d}\!\ket\psi c_i^*c_jc_k^*c_l.
\end{align}
The integral at the end of this expression can be computed in the following way:
\begin{align}
  \label{eq:fidelityIntegralGeneral}
  \int \mathrm{d}\!\ket\psi c_i^*c_jc_k^*c_l
  = \frac{\int \mathrm{d}a_1\mathrm{d}b_1\cdots\mathrm{d}a_d\mathrm{d}b_d\,\delta(\sum_j (a_j^2+b_j^2)-1)c_i^*c_jc_k^*c_l}
  {\int \mathrm{d}a_1\mathrm{d}b_1\cdots\mathrm{d}a_d\mathrm{d}b_d\,\delta(\sum_j (a_j^2+b_j^2)-1)},
\end{align}
where we used the fact that the space of pure states of dimension $d$ is characterized by
$c_j=a_j+ib_j$ for $a_j,b_j\in\mathbb R$ with $\sum_j (a_j^2+b_j^2)=1$. The
first thing to note is that the integral is non-zero only if $i=j$ and $k=l$ or
$i=l$ and $j=k$, since otherwise the integrand in the numerator is an odd
function integrated over a symmetric interval. Hence, the required
integrals are $\int \mathrm{d}\!\ket\psi |c_i|^2|c_j|^2$ and $\int
\mathrm{d}\!\ket\psi |c_i|^4$.
These integrals have been evaluated in \cite{HamsDeRaedt2000RandomStateTechnology} and \cite{Jin2020RandomStateTechnology}.
We do not repeat the full calculation here but, for the sake of reference,
we outline three common strategies used to compute such integrals.

One way to evaluate \equref{eq:fidelityIntegralGeneral} is to use spherical
coordinates, i.e., polar coordinates for $(a_i,b_i)$ and $(a_j,b_j)$, and
hyperspherical coordinates for the remaining coefficients. Since the integrands
only depend on the radii in these coordinates, they reduce to single integrals
over these radii multiplied by the surface of the respective spheres. See
\cite{HamsDeRaedt2000RandomStateTechnology, Jin2020RandomStateTechnology} for
more information.

Another way to compute the integrals is to make use of the representation of
the $\delta$-function $\delta(x)=\int\mathrm{d}t\, e^{itx}/2\pi$
and closing the contour of integration in the complex plane with $\mathrm{Im}\,t=\varepsilon>0$.
See the supplementary material of \cite{Boixo2018quantumsupremacy} for an example using this approach.

A third option to obtain the result is particularly convenient for numerical work:
The coefficients of a random pure state
$\ket\psi$ can be generated from the normal distribution \cite{bengtsson2006geometryofquantumstates}
\begin{align}
  \label{eq:fidelityNormalDistribution}
  p(a_1,b_1,\ldots,a_d,b_d) = \frac 1 {(2\pi)^d} e^{-(a_1^2+b_1^2+\cdots+a_d^2+b_d^2)/2}.
\end{align}
This is a construction that has a long history (see Muller's method in \cite{muller1959anoteongeneratepointsuniformlyonsphere})
and is commonly used in random matrix theory \cite{edelman2005randomMatrixTheory}.
With this strategy, we set $c_j=(a_j+ib_j)/\sum_j (a_j^2+b_j^2)$ for $a_j,b_j\in\mathbb R$,
and the evaluation of the integral amounts to
\begin{align}
  \label{eq:fidelityIntegralStrategy2}
  \int \mathrm{d}\!\ket\psi f(c_1,\ldots,c_d) = \int \mathrm{d}a_1\mathrm{d}b_1\cdots\mathrm{d}a_d\mathrm{d}b_d\,p(a_1,b_1,\ldots,a_d,b_d)f(c_1,\ldots,c_d).
\end{align}
As shown in \cite{Jin2020RandomStateTechnology}, this expression basically
reduces to a set of Gaussian integrals.

Independent of the strategies used, one obtains
\begin{align}
  \int \mathrm{d}\!\ket\psi c_i^*c_jc_k^*c_l = \frac{1}{d(d+1)}(\delta_{ij}\delta_{kl} + \delta_{il}\delta_{jk}).
\end{align}
Inserting this expression into \equref{eq:fidelityAvgIntegral1}, we immediately
find
\begin{align}
 \label{eq:fidelityFavgEvaluatedClosedForm}
 F_{\mathrm{avg}}(\mathcal E) = \sum_\alpha \frac{\abs{\mathrm{Tr}\,E_\alpha}^2 +  \mathrm{Tr}\,E_\alpha^\dagger E_\alpha}{d(d+1)},
\end{align}
which is equivalent to the desired relation given by \equref{eq:fidelityAvgEntRelation}
after substituting $F_{\mathrm{ent}}(\mathcal E)=\sum_\alpha \abs{\mathrm{Tr}\,E_\alpha}^2/d^2$ (see \equref{eq:fidelityEnt})
and $\mathrm{Tr}\,\mathcal E(\mathds1) = \sum_\alpha \mathrm{Tr}\,E_\alpha^\dagger E_\alpha$. We remark that the result given by \equref{eq:fidelityFavgEvaluatedClosedForm} can also be found in
\cite{Pedersen2007FidelityGeneralQuantumOperations}.

\clearpage
\section{Diamond distance between unitary quantum operations}
\label{app:diamonddistanceunitary}

In \secref{sec:diamonddistance}, we obtained an expression for the diamond distance
$\eta_\Diamond$ between two quantum operations $\mathcal G_{id}(\rho)=U\rho U^\dagger$
and $\mathcal G_{ac}(\rho)=M\rho M^\dagger$, where $U$ is a unitary matrix
representing an ideal quantum gate, and $M$ denotes a matrix describing the actual
implementation. In this appendix, we evaluate this expression for the
case that both $M$ and $U$ are unitary. The result provides an explicit proof
for the statements given in \cite{Aharonov1998DiamondNorm} and \cite{Johnston2009ComputingStabilizedNormsQC}
and illustrates the construction.

The expression in \equref{eq:gatemetricsdiamondnormPureStateEvaluated} reads
\begin{align}
  \label{eq:appdiamondunitaryStart}
  \eta_\Diamond
  = \frac 1 2 \|M\cdot M^\dagger - U\cdot U^\dagger\|_\Diamond\,
  = \frac 1 2 \max_{\ket{x}} \sqrt{(\braket{x|W^\dagger W\otimes\mathds1|x}+1)^2-4\abs{\braket{x|W\otimes\mathds1|x}}^2},
\end{align}
where $W=MU^\dagger$.
If both $M$ and $U$ are unitary, we have $W^\dagger W=\mathds1$ and therefore
$\braket{x|W^\dagger W\otimes\mathds1|x}=1$.

\begin{figure}
  \centering
  \def\svgwidth{\textwidth}
  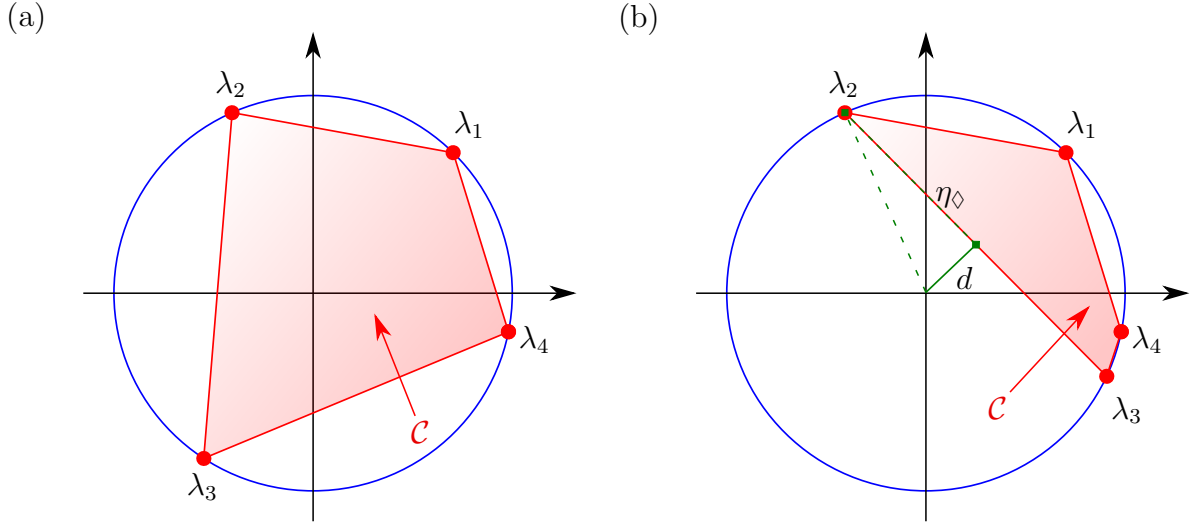
  \caption{Illustration of the diamond distance between unitary quantum
  operations. The red points represent the eigenvalues $\lambda_i$. They lie on
  the unit circle in the complex plane. The polygon is the convex hull $\mathcal
  C$ of these points (see \equref{eq:appdiamondunitaryConvexHull}). The two
  possible cases are: (a) $0\in\mathcal C$, i.e., the origin is inside the
  polygon such that $\eta_\Diamond=1$; (b) the origin
  is outside the polygon such that $\eta_\Diamond=\sqrt{1-d^2}$, where $d$ is
  the distance between the origin and the polygon.}
  \label{fig:diamondpolygon}
\end{figure}

The first step is to diagonalize $W$ so that $V^{-1}WV=\Lambda$, where $\Lambda=\mathrm{diag}(\lambda_i)$
denotes the eigenvalues of $W$. Since $W$ is unitary, it is a normal matrix, so also $V$
is unitary and preserves the norm.
Thus, we can substitute $\ket{x}\mapsto V\ket{x}$ in the maximization to obtain
\begin{align}
  \label{eq:appdiamondunitaryEvaluated}
  \eta_\Diamond = \max_{\ket{x}} \sqrt{1-\abs{\braket{x|\Lambda\otimes\mathds1|x}}^2}.
\end{align}
Obviously, the maximum is attained when $\abs{\braket{x|\Lambda\otimes\mathds1|x}}^2$
is minimal. We now expand $\ket x=\sum_{ij} x_{ij}\ket{i}\otimes\ket{j}$ with $i,j=0,\ldots,N-1$ and
$\sum_{ij}\abs{x_{ij}}^2=1$ (we can also think of $\ket i$ as the eigenbasis
of $W$; but the important thing is that $V$ is unitary since otherwise, in general, $\sum_{ij}\abs{x_{ij}}^2\neq1$).
Thus we have
\begin{align}
  \label{eq:appdiamondunitaryExpanded}
  \eta_\Diamond
  %= \sqrt{1 - \min_{\substack{x_{ij}\in\mathbb R\\\sum \abs{x_{ij}}^2=1}} \sum_{ijkl} \abs{x_{ij}}^2\lambda_i\lambda_k^*\abs{x_{kl}}^2}
  %=  \sqrt{1 - \min_{\substack{p_i\ge0\\\sum p_i=1}} \sum_{ik} p_i \lambda_i\lambda_k^* p_k}
  = \sqrt{1 - \min_{\substack{x_{ij}\in\mathbb R\\\sum \abs{x_{ij}}^2=1}} \left\vert\sum_{ij} \abs{x_{ij}}^2\lambda_i\right\vert^2}
  = \sqrt{1 - \min_{\substack{p_i\ge0\\\sum p_i=1}} \left\vert\sum_{i} p_i \lambda_i\right\vert^2},
  %= \sqrt{1 - \min_{\substack{p_i\ge0\\\sum p_i=1}} \left\Vert\sum_{i} p_i \begin{pmatrix} \cos\varphi_i \\ \sin\varphi_i \end{pmatrix}\right\Vert^2},
\end{align}
where $p_i=\sum_j\abs{x_{ij}}^2\ge0$ with $\sum_i p_i=1$.
The minimization is now over all convex combinations of the eigenvalues $\lambda_i$,
i.e., all points in the set
\begin{align}
  \label{eq:appdiamondunitaryConvexHull}
  \mathcal C = \left\{\sum_i p_i \lambda_i\:\::\:\:p_i\ge0\:\mathrm{and}\:\sum_i p_i=1\right\}.
\end{align}
By definition, this set is the convex hull of all $\lambda_i$, i.e., a polygon
whose vertices are given by $\lambda_i$. Since $W$ is unitary, all eigenvalues
satisfy $\abs{\lambda_i}=1$, so the  vertices $\lambda_i$ of the polygon lie on
the complex unit circle  (see \figref{fig:diamondpolygon}). The quantity
$\abs{\sum_i p_i\lambda_i}$ to be minimized represents the distance from the
point $\sum_i p_i\lambda_i\in\mathcal C$ to the origin.

There are now two possible cases: Either the origin is inside the polygon
($0\in\mathcal C$), or it is outside. If $0\in\mathcal C$ as shown in
\figref{fig:diamondpolygon}(a), there exists a set $\{p_i\}$ such that $\sum_i
p_i\lambda_i=0$, and we have $\eta_\Diamond=1$.

In the other case shown in \figref{fig:diamondpolygon}(b), the point in $\mathcal C$ with the closest
distance $d=\min\abs{\sum_i p_i \lambda_i}$ to the origin lies in the middle of
the line between $\lambda_j$ and $\lambda_k$, for which $\abs{\lambda_j-\lambda_k}$
is maximal. Inserting $d$ into \equref{eq:appdiamondunitaryExpanded}, we find that
the diamond distance is $\eta_\Diamond=\sqrt{1-d^2}$, which
is the result stated without proof in \cite{Aharonov1998DiamondNorm}.

Furthermore,
this result can be related to the result given in \cite{Johnston2009ComputingStabilizedNormsQC}
by noting that the hypotenuse of the right triangle indicated in \figref{fig:diamondpolygon}(b)
is $1$. Hence we have that $\eta_\Diamond = \sqrt{1-d^2}$ is half the line between $\lambda_j$ and $\lambda_k$,
i.e., $\eta_\Diamond = \max_{jk} \abs{\lambda_j - \lambda_k}/2$.
This means that $2\eta_\Diamond$ is the diameter of the smallest closed disc enclosing
all eigenvalues of $W$, as stated in \cite{Johnston2009ComputingStabilizedNormsQC}.

In summary, we have
\begin{align}
  \label{eq:appdiamondunitaryResult}
    \eta_\Diamond = \begin{cases}
      1 & (0\in \mathcal C) \\
      \frac{1}{2} \max\limits_{jk} \abs{\lambda_j - \lambda_k} & (\mathrm{otherwise})
    \end{cases}.
\end{align}
A practical way to check for $0\in \mathcal C$ is to compute all
arguments $\varphi_i=\arg(\lambda_i)\in[0,2\pi)$ and list them in increasing order
$0\le\varphi_0\le\cdots\le\varphi_{N-1}<2\pi$.
If all $N$ successive differences $\Delta\varphi_i=(\varphi_{i+1}-\varphi_{i})\,\mathrm{mod}\,[0,2\pi)$
(including the case $\varphi_N\equiv\varphi_0$) are in $[0,\pi]$,
the convex polygon $\mathcal C$ includes the origin.

Note that in the single-qubit case, $W$ has only two eigenvalues such that the
convex hull $\mathcal C$ is a straight line. In this case, the formula for the second case
in \equref{eq:appdiamondunitaryResult} is always valid such that
$\eta_\Diamond = |\lambda_1-\lambda_2|/2.$

\clearpage
\section{Proof of a diamond-distance bound for trace-decreasing operations}
\label{app:proofdiamondnormbound}

For trace-preserving quantum operations $\mathcal E$, the best known lower bound
for the diamond distance in terms of the average fidelity was proven in
\cite{Wallman2014RBwithConfidence}. In this appendix, we derive a new bound that
also applies to non-trace-preserving quantum operations
(cf.~\secref{sec:quantumoperations}). Furthermore, it reduces to the original
bound in the trace-preserving case, thereby generalizing the previous result.

The known lower bound for a completely positive trace-preserving map $\mathcal E$
reads \cite{Wallman2014RBwithConfidence, Kueng2016ComparingExperimentsToThreshold}
\begin{align}
  \label{eq:diamondnormBoundCPTP}
  \frac{d+1} d (1 - F_{\mathrm{avg}}(\mathcal E)) \le \eta_\Diamond.
\end{align}
Here, $d$ is the dimension of the Hilbert space, $F_{\mathrm{avg}}(\mathcal E)$ is the
average fidelity defined in \equref{eq:fidelityAvg}, and
$\eta_\Diamond=\|\mathcal E-\mathds1\|_\Diamond/2$ is the diamond distance
defined in \secref{sec:diamonddistance}. \sequref{eq:diamondnormBoundCPTP}
is only valid under the assumption that $\mathcal E$ is
trace-preserving.

In what follows, let $\mathcal E$ be a completely positive quantum operation that is not
necessarily trace-preserving.
We start from the definition of the diamond norm given by \equref{eq:gatemetricsdiamondnorm}
for $\mathcal T=\mathcal E-\mathds1$,
\begin{align}
  \label{eq:diamondnormBoundDefinition}
  \eta_\Diamond
  = \frac 1 2 \sup_{X\neq0} \frac{\|((\mathcal E-\mathds1)\otimes\mathds1)(X)\|_{\mathrm{Tr}}}{\|X\|_{\mathrm{Tr}}},
  %= \sup_{\|X\|_{\mathrm{Tr}}=1} \|(\mathcal E\otimes\mathds1)(X)\|_\mathrm{Tr},
\end{align}
where $\|X\|_{\mathrm{Tr}}\ =\mathrm{Tr}\,\abs{X}$ with $\abs{X}=\sqrt{X^\dagger
X}$ denotes the trace norm (sum of the singular values). Because of the supremum
in the definition of $\eta_\Diamond$, we have
\begin{align}
  \label{eq:diamondnormBoundStep1}
  \eta_\Diamond
  \ge \frac 1 2 \|((\mathcal E-\mathds1)\otimes\mathds1)(\Phi)\|_{\mathrm{Tr}}\,
  = \frac 1 2 \|J(\mathcal E) - \Phi\|_{\mathrm{Tr}},
\end{align}
where $\Phi=\ketbra\Phi\Phi$ with $\ket\Phi=\sum_j\ket{jj}/\sqrt{d}$ is the
maximally entangled state, and we used the definition of the Choi matrix
$J(\mathcal E)=(\mathcal E\otimes\mathds1)(\Phi)$. Note that the order of
tensor-product factors in this definition of the Choi matrix is reversed (as
compared to \equref{eq:choimatrix}) such that it complies with the definition of
the diamond norm. The result is independent of the order.

Using the definition of the trace norm given above and inserting $\mathds1 = \Phi + \mathds1 - \Phi$,
we obtain
\begin{align}
  \label{eq:diamondnormBoundStep2}
  \frac 1 2 \|J(\mathcal E) - \Phi\|_{\mathrm{Tr}}\,
  = \frac 1 2 \Big(\mathrm{Tr}\,\Phi\abs{J(\mathcal E) - \Phi}
  + \mathrm{Tr}\,(\mathds1-\Phi)\abs{J(\mathcal E) - \Phi}\Big).
\end{align}
Since the matrix $J(\mathcal E) - \Phi$ is Hermitian, its singular values
are the absolute values of its eigenvalues. For any positive semidefinite matrix $A$
and any Hermitian matrix $B$ with singular values
$\abs b$, eigenvalues $b$, and eigenvectors $\ket b$, we have
\begin{align}
  \label{eq:diamondnormBoundStepTraceNorm}
  \mathrm{Tr}\,A\abs B = \sum_b \abs b\, \mathrm{Tr}\,A\ketbra b b
  \ge \sum_b b\, \mathrm{Tr}\,A\ketbra b b
  = \mathrm{Tr}\,AB,
\end{align}
and similarly, $\mathrm{Tr}\,A\abs B\ge \mathrm{Tr}\,A(-B)$. %, so $\mathrm{Tr}\,A\abs B\ge\abs{\mathrm{Tr}\,AB}$.
Since both $\Phi$ and $\mathds1-\Phi$ are projectors, their eigenvalues are $1$ and $0$, so they
are positive semidefinite. Thus, \equref{eq:diamondnormBoundStep2} is
bounded by%
\begin{subequations}
\begin{align}
  \label{eq:diamondnormBoundStep3}
  \frac 1 2 \|J(\mathcal E) - \Phi\|_{\mathrm{Tr}}\,
  &\ge \frac 1 2 \Big( \mathrm{Tr}\,\Phi(\Phi - J(\mathcal E))
  +  \mathrm{Tr}\,(\mathds1-\Phi)(J(\mathcal E) - \Phi)\Big)\\
  &= \mathrm{Tr}\,\Phi(\Phi - J(\mathcal E))
  + \frac 1 2 \mathrm{Tr}\,(J(\mathcal E) - \Phi)\\
  \label{eq:diamondnormBoundStep4}
  % &= \frac 1 2 (1 - \braket{\Phi|J(\mathcal E)|\Phi})
  % + \frac 1 2 \abs{\mathrm{Tr}\,J(\mathcal E) - \braket{\Phi|J(\mathcal E)|\Phi}}.
  &= 1 - \braket{\Phi|J(\mathcal E)|\Phi}
  + \frac 1 2 (\mathrm{Tr}\,J(\mathcal E) - 1).
\end{align}
\end{subequations}
Note that for this step, the proof in \cite{Wallman2014RBwithConfidence}
made use of the Fuchs-van de Graaf inequality
$\|J(\mathcal E) - \Phi\|_{\mathrm{Tr}}/2 \ge 1 - \braket{\Phi|J(\mathcal E)|\Phi}$
\cite{FuchsVandDeGraaf1999Inequality}, which only works if $\mathrm{Tr}\,J(\mathcal E)=1$,
i.e., if $\mathcal E$ is trace-preserving.
The step from \equref{eq:diamondnormBoundStep2} to
\equref{eq:diamondnormBoundStep3}, however, is elementary and
also works in the non-trace-preserving case.

By identifying the term $\braket{\Phi|J(\mathcal E)|\Phi}$ in \equref{eq:diamondnormBoundStep4}
as the entanglement fidelity $F_{\mathrm{ent}}(\mathcal E)$ defined in \equref{eq:fidelityEnt},
we obtain
\begin{align}
\label{eq:diamondnormBoundStep5}
  \eta_\Diamond
  \ge 1 - F_{\mathrm{ent}}(\mathcal E)
  + \frac 1 2 (\mathrm{Tr}\,J(\mathcal E) - 1).
\end{align}
After using $\mathrm{Tr}\,J(\mathcal E)=\mathrm{Tr}\,\mathcal E(\mathds1/d)$
and inserting the relation between entanglement fidelity and average fidelity
derived in \appref{app:prooffidelity}, i.e.,
$F_{\mathrm{ent}}(\mathcal E) = (d+1)F_{\mathrm{avg}}(\mathcal E)/d - \mathrm{Tr}\,\mathcal E(\mathds1/d)/d$,
we finally obtain
\begin{align}
  \label{eq:diamondnormBoundResult}
  \eta_\Diamond
  \ge \frac{d+1}d(1-F_{\mathrm{avg}}(\mathcal E)) - \frac{d+2}{2d}(1 - \mathrm{Tr}\,\mathcal E(\mathds1/d)).
\end{align}
The first term in this expression is the result given in
\equref{eq:diamondnormBoundCPTP}, referred to as
$\eta_\Diamond^{\mathrm{Pauli}}$ in
\equref{eq:gatemetricsdiamonddistanceBoundPauliTP} and
\cite{Sanders2016ThresholdTheorem}. The second term is a new contribution that
represents the correction required for trace-decreasing quantum operations.

\clearpage
\section{Gate decompositions and effective Hamiltonians}
\label{app:gatedecomposition}

In this appendix, we outline a method to relate an arbitrary quantum operation
$\mathcal G$ (see \secref{sec:quantumoperations}) to an effective Hamiltonian
$H$. The method was used to obtain the axis-angle decompositions of the
two-qubit GST results studied in \secref{sec:gatesettomography} (see
\tabref{tab:gstgatedecompositions}), which are much more straightforward to interpret than
the corresponding Pauli transfer matrices shown in \figref{fig:gstptms}. A
similar method to obtain such decompositions is implemented by the \texttt{pyGSTi} package
\cite{Nielsen2018pyGSTi0944}.

A quantum operation $\mathcal G$ can be written in Kraus form (see
\equref{eq:krausrepresentation}),
\begin{align}
  \label{eq:GateDecompositionKrausForm}
  \mathcal G(\rho) = \sum_{\alpha=1}^R E_\alpha \rho E_\alpha^\dagger,
\end{align}
where $E_\alpha$ are the Kraus operators and $R$ is the Kraus rank of $\mathcal G$.
The aim of the method is to find a ``Hamiltonian'' $H$ that approximately generates the evolution
described by $\mathcal G$ according to
\begin{align}
  \label{eq:GateDecompositionApproximateGenerator}
  \mathcal G(\rho) \approx e^{-iH} \rho\, e^{iH}.
\end{align}
We explicitly left out a symbol for the time in the expression for the
generator.  Typically, one would rather write $H=\widehat H t$, where $\widehat
H$ is the  Hamiltonian and $t$ is the time, to express the characteristic structure of
time evolutions in quantum theory. However, since $\mathcal G$ describes only
one discrete evolution $\rho\mapsto\mathcal G(\rho)$, there is no notion of time
such that a separation into individual components $\widehat H$ and $t$ is
arbitrary. For convenience, we still refer to the symbol $H$ as the Hamiltonian.

The relation given in \equref{eq:GateDecompositionApproximateGenerator} can only be
exact if the Kraus rank $R$ in \equref{eq:GateDecompositionKrausForm} is $1$
and the operator $E_1$ is unitary. This means that the action of
$\mathcal G$ can be written as $\mathcal G(\rho) = U\rho U^\dagger$, where
$U$ is a unitary matrix. But even if this is not the case, the following
method can be used to produce an approximation to $\mathcal G$ in terms of
a Hamiltonian, which may be simpler to understand or provide insights into
potential errors. In the context of quantum gate optimization, this may provide
information on how to improve the gate's implementation.

If $\mathcal G$ is not unitary but has a Kraus rank $R$ of almost one (meaning
that all but one singular value of the Choi matrix $J(\mathcal G)$ given by
\equref{eq:choimatrix} are close to zero), the method yields an effective
Hamiltonian that approximates the quantum operation. This Hamiltonian may have
non-Hermitian components to model non-unitary or non-trace-preserving quantum
operations.

Note that a Hamiltonian generating $\mathcal G$ according to
\equref{eq:GateDecompositionApproximateGenerator} is not unique, since the
complex matrix exponential is a many-to-one function. One way to see this is
that by adding a multiple of $2\pi$ to any eigenvalue of $H$, the matrix
exponential $e^{-iH}$ does not change. In what follows, we aim for an expression
for $H$ in terms of Pauli matrices that can be readily interpreted as a rotation
by a certain angle around an axis specified by $H$, so this ambiguity can be
understood in the context of rotations.

\subsection{The matrix logarithm}

We start with the matrix representation $G$ of the superoperator $\mathcal G$
defined in \equref{eq:paulitransfermatrix} (the Pauli transfer matrix).
We denote the correspondence between the matrix representation $G$ and the
map $\mathcal G$ by
\begin{align}
  \label{eq:GateDecompositionMatrixCorrespondence}
  G\sket\rho \quad\leftrightarrow\quad \mathcal G(\rho),
\end{align}
where $\sket\rho$ denotes a vector representation of the density matrix $\rho$,
obtained by expanding $\rho=\sum_i\rho_i\widehat P_i$
in the normalized Pauli basis $\widehat P_i$ (see \equref{eq:paulibasis}).

If $\mathcal G$ preserves Hermiticity, the matrix $G$ is a real $d\times d$
matrix, where $d=N^2=4^{n}$ for an $n$-qubit system. We define the matrix
logarithm of $G=e^L$ as
\begin{align}
  \label{eq:GateDecompositionMatrixLog}
  L = \log G = V \log(\Lambda) V^{-1},
\end{align}
where $G=V\Lambda V^{-1}$ denotes the eigendecomposition of $G$
(see \cite{Meyer2009TopicsInLinearTheory} for a common alternative definition
of the matrix logarithm). Thus, $\Lambda=\mathrm{diag}(\lambda_0,\ldots,\lambda_{d-1})$ contains the
eigenvalues of $G$, the columns of $V=(v_0,\ldots,v_{d-1})$ contain the right eigenvectors,
and the columns of $(V^{-1})^\dagger=(w_0,\ldots,w_{d-1})$
contain the left eigenvectors. In this definition, $\log\lambda_i =
\log\abs{\lambda_i} + i \arg\lambda_i$ denotes the principal logarithm defined
by $\arg\lambda_i\in(-\pi,\pi)$. If all $\lambda_i\in\mathbb
C\setminus(-\infty,0]$, the principal matrix logarithm defined by
\equref{eq:GateDecompositionMatrixLog} is unique. If some $\lambda_i=0$, the
matrix $G$ is singular and the matrix logarithm does not exist.

If the matrix logarithm $L$ is real, there is a Hermiticity-preserving map
$\mathcal L$ such that
\begin{align}
  \label{eq:GateDecompositionLindbladCorrespondence}
  G\sket\rho = e^L\sket\rho \quad\leftrightarrow\quad \mathcal G(\rho) = e^{\mathcal L}\rho.
\end{align}
Unfortunately, some ideal quantum gates are special in the sense that their
Pauli transfer matrices have eigenvalues $-1$. This is the case for the
\textsc{CNOT} gate defined in \equref{eq:cnotgate}, for which the Pauli transfer
matrix corresponding to the map $\rho\mapsto
\textsc{CNOT}\,\rho\,\textsc{CNOT}^\dagger$ has eigenvalues $\lambda_i=\pm 1$.
In this case, however, one can still find a real matrix logarithm since the
negative eigenvalues occur in pairs \cite{Culver1966RealMatrixLogarithm}. It is
constructed by choosing, for each pair $\lambda_i=\lambda_j=-1$, two conjugate
branches of the logarithm $\log\lambda_{i/j}\leftarrow\pm i\pi$. Similarly, the
corresponding real eigenvectors $v_i$ and $v_j$ in the degenerate subspace have
to be replaced by conjugate pairs, i.e., $v_i\leftarrow(v_i+iv_j)/\sqrt 2$ and
$v_{j}\leftarrow(v_i-iv_j)/\sqrt 2$.

\subsection{Extracting the Hamiltonian}

The map $\mathcal L$ in \equref{eq:GateDecompositionLindbladCorrespondence} is
typically called the Lindblad operator or \emph{Lindbladian}. The goal is to
approximate $\mathcal L$ by a map of the form $\mathcal L^H(\rho) =-i[H,\rho]$. If
this is possible, the correspondence given by
\equref{eq:GateDecompositionLindbladCorrespondence}  becomes
\begin{align}
  \label{eq:GateDecompositionLindbladCorrespondenceHamiltonian}
  e^L\sket\rho
   \quad\leftrightarrow\quad
  e^{\mathcal L}\rho \approx e^{-i[H,\,\cdot\,]}\rho = e^{-iH} \rho\, e^{iH}.
\end{align}
We consider a Hamiltonian $H$ expressed in the Pauli basis (see \equref{eq:paulibasis}),
\begin{align}
  \label{eq:GateDecompositionHamiltonian}
  H = \sum_{k=0}^{d-1} h_k P_k/2.
\end{align}
The real coefficients $h_k$ are explicitly defined with respect to $P_k/2$ instead of
the normalized basis elements $\widehat P_k$. The reason is that in this way,
the action of $e^{-iH}$ can be interpreted as a rotation about an axis
specified by $\widehat h = h/\abs{h}$, where the angle of rotation is given by
$\varphi=\abs{h}=\sqrt{\sum_k h_k^2}$ (cf.~\equaref{eq:GSTHamiltonianAngle}{eq:GSTHamiltonianAxis}).

Using the form of the Hamiltonian given by \equref{eq:GateDecompositionHamiltonian}, we evaluate
the Pauli transfer matrix $L^H$ corresponding to the map $\mathcal L^H=-i[H,\,\cdot\,]$.
Its matrix elements are (cf.~\equref{eq:paulitransfermatrix})
\begin{align}
  \label{eq:GateDecompositionHamiltonianPTM}
  L^H_{ij}
  = \frac{1}{N}\mathrm{Tr}\,P_i(-i[H,P_j])
  = \sum_k h_k \frac{\mathrm{Tr}\,P_i[P_k,P_j]}{2Ni} = \sum_k h_k s_{kji},
\end{align}
where $s_{kji}=\mathrm{Tr}([P_k,P_j]P_i)/2Ni$. Note that, as each $P_i$ is an $n$-fold
tensor product of Pauli matrices, evaluating $s_{kji}$ analytically may be cumbersome.
However, $s_{kji}$ can be easily
evaluated with computer algebra systems such as Mathematica \cite{mathematica12}.
One finds $s_{kji}\in\{1,0,-1\}$ (see also \cite{rigetti2009quantumgates}, in which $s_{kji}$ is
called the $n$-qubit super-commutator). The typical structure of the matrix
$L^H$ for $n\ge1$ qubits is
\begin{align}
  \label{eq:GateDecompositionHamiltonianPTMExplicit}
  L^H = \begin{pmatrix}
      0 & 0 & 0 & 0 & \cdots \\
      0 & 0 & -h_3 & h_2 & \cdots \\
      0 & h_3 & 0 & -h_1 & \cdots \\
      0 & -h_2 & h_1 & 0 & \cdots \\
      \vdots & \vdots & \vdots & \vdots & \ddots
  \end{pmatrix}.
\end{align}
The upper left $4\times4$ block corresponds to $n=1$ qubit, the upper left
$16\times16$ block corresponds to $n=2$ qubits, and so on.

The map $\mathcal G$ can be approximately generated by the Hamiltonian
given in \equref{eq:GateDecompositionHamiltonian} if
the matrix logarithm $L=\log G$ given by \equref{eq:GateDecompositionMatrixLog}
has the form of the matrix $L^H$ in \equref{eq:GateDecompositionHamiltonianPTMExplicit}.
We construct a candidate Hamiltonian by projecting $L$ onto this form,
such that
\begin{align}
  \label{eq:GateDecompositionHamiltonianProjection}
  h_k = \begin{cases}
    \sum_k L_{ij} s_{kji} / \sum_k \abs{s_{kji}} & (\sum_k \abs{s_{kji}} \neq 0) \\
    0 & \mathrm{(otherwise)}
\end{cases}
\end{align}
for all $k=0,\ldots,d-1$. If the decomposition error defined by
\begin{align}
  \label{eq:GateDecompositionHamiltonianProjectionError}
  \gamma = \|G-e^{L^{H}}\|_F
\end{align}
is much smaller than $1$, we accept the Hamiltonian generator. This was the case
for almost all experiments studied in this thesis.

However, since the matrix logarithm given by
\equref{eq:GateDecompositionMatrixLog} is not unique, and because the
correspondence in \equref{eq:GateDecompositionLindbladCorrespondenceHamiltonian}
is only approximate, there may of course be other Hamiltonians generating the
evolution. One way to proceed is to optimize the entries of $L$ towards the
intended target gate's matrix logarithm $L^U=\log G^U$, where $G^U$ is the Pauli
transfer matrix of the map $\mathcal G^U=U\,\cdot\,U^\dagger$ and $U$ is the
target gate. This can be done by optimizing a joint objective function including
both $\|G-e^{L^{H}}\|$ and $\|L^U-L^{H}\|$. This option is implemented by the
\texttt{pyGSTi} package \cite{Nielsen2018pyGSTi0944} and was also used to obtain the
decomposition of the \textsc{CNOT} gate reported in
\tabref{tab:gstgatedecompositions}. The objective function was
$\|G-e^{L^{H}}\|_1+10\|L^U-L^{H}\|_F^2$, where
$\|A\|_1~=\sum_{ij}\abs{A_{ij}}$, using the L-BFGS-B algorithm
\cite{Zhu1997LBFGSBalgorithm, Morales2011LBFGSBalgorithmImprovement}. However,
for two qubits, $L$ already contains $d\times d=256$ real numbers, so this
approach becomes impractical for more qubits.

Another option is to optimize the $d$ coefficients $h_k$ defined by
\equref{eq:GateDecompositionHamiltonian}, starting from initial values given by
\equref{eq:GateDecompositionHamiltonianProjection} and using the objective
function defined in \equref{eq:GateDecompositionHamiltonianProjectionError}.
This approach directly relates the Hamiltonian $H$ to the map $\mathcal G$
without using an explicit matrix logarithm, so the approach is closer in spirit
to the relation anticipated by
\equref{eq:GateDecompositionApproximateGenerator}. The advantage is that this
approach does not rely on the matrix logarithm as an intermediate step.
Furthermore, it only requires an optimization of $d$ real numbers instead of
$d^2$. A similar approach was implemented in \cite{Willsch2016Master}.

A third option would be to resolve the ambiguity of the matrix logarithm by
adding integer multiples of $2\pi iv_j w_j^\dagger$ to $L$. This corresponds to
changing the eigenvalues $\lambda_j\mapsto\lambda_j+2\pi i$, which leaves $e^L$
invariant but may lead to imaginary parts in the Hamiltonian. If the resulting
Hamiltonian indeed describes the original map $\mathcal G$, as reflected by a low
decomposition error $\gamma$ (see
\equref{eq:GateDecompositionHamiltonianProjectionError}), this procedure may
provide an effective qubit model for decay in the system under investigation.

\end{appendices}

%% file: figs/diamond-polygon.pdf_tex
%% Accompanies image file 'diamond-polygon.pdf' (pdf, eps, ps)
\begingroup%
  \makeatletter%
  \providecommand\color[2][]{%
    \errmessage{(Inkscape) Color is used for the text in Inkscape, but the package 'color.sty' is not loaded}%
    \renewcommand\color[2][]{}%
  }%
  \providecommand\transparent[1]{%
    \errmessage{(Inkscape) Transparency is used (non-zero) for the text in Inkscape, but the package 'transparent.sty' is not loaded}%
    \renewcommand\transparent[1]{}%
  }%
  \providecommand\rotatebox[2]{#2}%
  \newcommand*\fsize{\dimexpr\f@size pt\relax}%
  \newcommand*\lineheight[1]{\fontsize{\fsize}{#1\fsize}\selectfont}%
  \ifx\svgwidth\undefined%
    \setlength{\unitlength}{595.27559055bp}%
    \ifx\svgscale\undefined%
      \relax%
    \else%
      \setlength{\unitlength}{\unitlength * \real{\svgscale}}%
    \fi%
  \else%
    \setlength{\unitlength}{\svgwidth}%
  \fi%
  \global\let\svgwidth\undefined%
  \global\let\svgscale\undefined%
  \makeatother%
  \begin{picture}(1,0.42784322)%
    \lineheight{1}%
    \setlength\tabcolsep{0pt}%
    \put(-0.00228525,0.41020433){\color[rgb]{0,0,0}\makebox(0,0)[lt]{\lineheight{1.25}\smash{\begin{tabular}[t]{l}(a)\end{tabular}}}}%
    \put(0.500824,0.40927609){\color[rgb]{0,0,0}\makebox(0,0)[lt]{\lineheight{1.25}\smash{\begin{tabular}[t]{l}(b)\end{tabular}}}}%
    \put(0,0){\includegraphics[width=\unitlength,page=1]{diamond-polygon.pdf}}%
    \put(0.36569634,0.32144727){\color[rgb]{0,0,0}\makebox(0,0)[lt]{\lineheight{1.25000012}\smash{\begin{tabular}[t]{l}$\lambda_1$\end{tabular}}}}%
    \put(0.16956911,0.35520086){\color[rgb]{0,0,0}\makebox(0,0)[lt]{\lineheight{1.25000012}\smash{\begin{tabular}[t]{l}$\lambda_2$\end{tabular}}}}%
    \put(0.14636297,0.02196063){\color[rgb]{0,0,0}\makebox(0,0)[lt]{\lineheight{1.25}\smash{\begin{tabular}[t]{l}$\lambda_3$\end{tabular}}}}%
    \put(0.41926723,0.14356082){\color[rgb]{0,0,0}\makebox(0,0)[lt]{\lineheight{1.25}\smash{\begin{tabular}[t]{l}$\lambda_4$\end{tabular}}}}%
    \put(0,0){\includegraphics[width=\unitlength,page=2]{diamond-polygon.pdf}}%
    \put(0.86966461,0.32144728){\color[rgb]{0,0,0}\makebox(0,0)[lt]{\lineheight{1.25}\smash{\begin{tabular}[t]{l}$\lambda_1$\end{tabular}}}}%
    \put(0.67353736,0.35520088){\color[rgb]{0,0,0}\makebox(0,0)[lt]{\lineheight{1.25}\smash{\begin{tabular}[t]{l}$\lambda_2$\end{tabular}}}}%
    \put(0.90652708,0.08508137){\color[rgb]{0,0,0}\makebox(0,0)[lt]{\lineheight{1.25}\smash{\begin{tabular}[t]{l}$\lambda_3$\end{tabular}}}}%
    \put(0.92323553,0.14356084){\color[rgb]{0,0,0}\makebox(0,0)[lt]{\lineheight{1.25}\smash{\begin{tabular}[t]{l}$\lambda_4$\end{tabular}}}}%
    \put(0.33064218,0.06494605){\color[rgb]{0.90196078,0,0}\makebox(0,0)[lt]{\lineheight{1.25000012}\smash{\begin{tabular}[t]{l}$\mathcal C$\end{tabular}}}}%
    \put(0,0){\includegraphics[width=\unitlength,page=3]{diamond-polygon.pdf}}%
    \put(0.7788023,0.19206114){\color[rgb]{0,0,0}\makebox(0,0)[lt]{\lineheight{1.25000012}\smash{\begin{tabular}[t]{l}$d$\end{tabular}}}}%
    \put(0.76188074,0.26546392){\color[rgb]{0,0,0}\makebox(0,0)[lt]{\lineheight{1.25000012}\smash{\begin{tabular}[t]{l}$\eta_\Diamond$\end{tabular}}}}%
    \put(0,0){\includegraphics[width=\unitlength,page=4]{diamond-polygon.pdf}}%
    \put(0.80544066,0.08470655){\color[rgb]{0.90196078,0,0}\makebox(0,0)[lt]{\lineheight{1.25000012}\smash{\begin{tabular}[t]{l}$\mathcal C$\end{tabular}}}}%
    \put(0,0){\includegraphics[width=\unitlength,page=5]{diamond-polygon.pdf}}%
  \end{picture}%
\endgroup%

%% file: bib.tex
\phantomsection
\addcontentsline{toc}{chapter}{Bibliography}
\printbibliography

%% file: publications.tex
\thispagestyle{empty}

\chapter{List of publications}

\begin{enumerate}[topsep=0pt, partopsep=0pt, itemsep=0pt, leftmargin=.8cm, label=(\arabic*)]
  \item
  \textsc{K.~Michielsen, M.~Nocon, D.~Willsch, F.~Jin, Th.~Lippert, H.~De Raedt},\\
  ``Benchmarking gate-based quantum computers'',\\
  \href{https://doi.org/10.1016/j.cpc.2017.06.011}{\emph{Comput.~Phys.~Commun.} \textbf{220}, 44 (2017)}
  \item
  \textsc{D.~Willsch, M.~Nocon, F.~Jin, H.~De Raedt, K.~Michielsen},\\
  ``Gate-error analysis in simulations of quantum computers with transmon qubits'',\\
  \href{https://doi.org/10.1103/PhysRevA.96.062302}{\emph{Phys.~Rev.~A} \textbf{96}, 062302 (2017)}
  \item
  \textsc{D.~Willsch, M.~Willsch, F.~Jin, H.~De Raedt, K.~Michielsen},\\
  ``Testing quantum fault tolerance on small systems'',\\
  \href{https://doi.org/10.1103/PhysRevA.98.052348}{\emph{Phys.~Rev.~A} \textbf{98}, 052348 (2018)}
  \item
  \textsc{H.~De Raedt, F.~Jin, D.~Willsch, M.~Willsch, N.~Yoshioka, N.~Ito, S.~Yuan, K.~Michielsen},\\
  ``Massively parallel quantum computer simulator, eleven years later'',\\
  \href{https://doi.org/10.1016/j.cpc.2018.11.005}{\emph{Comput.~Phys.~Commun.} \textbf{237}, 47 (2019)}
  \item
  \textsc{H.~De Raedt, M.~Katsnelson, D.~Willsch, K.~Michielsen},\\
  ``Separation of conditions as a prerequisite for quantum theory'',\\
  \href{https://doi.org/10.1016/j.aop.2019.01.012}{\emph{Ann.~Phys.~(N.~Y.)} \textbf{403}, 112 (2019)}
  \item
  \textsc{M.~Willsch, D.~Willsch, F.~Jin, H.~De Raedt, K.~Michielsen},\\
  ``Real-time simulation of flux qubits used for quantum annealing'',\\
  \href{https://doi.org/10.1103/PhysRevA.101.012327}{\emph{Phys.~Rev.~A} \textbf{101}, 012327 (2020)}
  \item
  \textsc{D.~Willsch, H.~Lagemann, M.~Willsch, F.~Jin, H.~De Raedt, K.~Michielsen},\\
  \mbox{``Benchmarking Supercomputers with the Jülich Universal Quantum Computer Simulator''},\\
  \href{https://hdl.handle.net/2128/24529}{\emph{NIC Symposium 2020,
  Publication Series of the John von Neumann Institute for Computing (NIC) NIC Series} \textbf{50}, 255 (2020)}
  \item
  \textsc{D.~Willsch, M.~Willsch, H.~De Raedt, K.~Michielsen},\\
  ``Support vector machines on the D-Wave quantum annealer'',\\
  \href{https://doi.org/10.1016/j.cpc.2019.107006}{\emph{Comput.~Phys.~Commun.} \textbf{248}, 107006 (2020)}
  \item
  \textsc{M.~Willsch, D.~Willsch, K.~Michielsen, H.~De Raedt},\\
  ``Discrete-Event Simulation of Quantum Walks'',\\
  \href{https://doi.org/10.3389/fphy.2020.00145}{\emph{Front.~Phys.} \textbf{8}, 145 (2020)}
  \item
  \textsc{H.~De Raedt, M.~Jattana, D.~Willsch, M.~Willsch, F.~Jin, K.~Michielsen},\\
  ``Discrete-Event Simulation of an Extended Einstein-Podolsky-Rosen-Bohm Experiment'',\\
  \href{https://doi.org/10.3389/fphy.2020.00160}{\emph{Front.~Phys.} \textbf{8}, 160 (2020)}
  \item
  \textsc{M.~Willsch, D.~Willsch, K.~Michielsen, F.~Jin, T.~Denkmayr, S.~Sponar, Y.~Hasegawa, H.~De Raedt},\\
  ``Long-time correlations in single-neutron interferometry data'',\\
  \href{https://doi.org/10.7566/JPSJ.89.064005}{\emph{J.~Phys.~Soc.~Jpn.} \textbf{89}, 064005 (2020)}
  \item
  \textsc{M.~Willsch, D.~Willsch, F.~Jin, H.~De Raedt, K.~Michielsen},\\
  ``Benchmarking the Quantum Approximate Optimization Algorithm'',\\
  \href{https://doi.org/10.1007/s11128-020-02692-8}{\emph{Quantum Inf.~Process.} \textbf{19}, 197 (2020)}
\end{enumerate}

%% file: declaration.tex
\thispagestyle{empty}

\chapter{Eidesstattliche Erklärung}

Ich, Dennis Willsch,
erkläre hiermit, dass diese Dissertation und die darin dargelegten Inhalte die
eigenen sind und selbstständig, als Ergebnis der eigenen originären Forschung,
generiert wurden.
\\\\\noindent
Hiermit erkläre ich an Eides statt
\begin{enumerate}[label=\arabic*.]
  \item Diese Arbeit wurde vollständig oder größtenteils in der Phase als
  Doktorand dieser Fakultät und Universität angefertigt;
  \item Sofern irgendein Bestandteil dieser Dissertation zuvor für einen
  akademischen Abschluss oder eine andere Qualifikation an dieser oder einer
  anderen Institution verwendet wurde, wurde dies klar angezeigt;
  \item Wenn immer andere eigene- oder Veröffentlichungen Dritter herangezogen
  wurden, wurden diese klar benannt;
  \item Wenn aus anderen eigenen- oder Veröffentlichungen Dritter zitiert wurde,
  wurde stets die Quelle hierfür angegeben. Diese Dissertation ist vollständig
  meine eigene Arbeit, mit der Ausnahme solcher Zitate;
  \item Alle wesentlichen Quellen von Unterstützung wurden benannt;
  \item Wenn immer ein Teil dieser Dissertation auf der Zusammenarbeit mit
  anderen basiert, wurde von mir klar gekennzeichnet, was von anderen und was
  von mir selbst erarbeitet wurde;
  \item % Kein Teil dieser Arbeit wurde vor deren Einreichung veröffentlicht. oder Ein Teil oder
  Teile dieser Arbeit wurden zuvor veröffentlicht und zwar in:
  \begin{enumerate}[align=parleft,labelwidth=*,leftmargin=2.2cm]
    \item[\cite{Willsch2017GateErrorAnalysis}]
    \textsc{D.~Willsch, M.~Nocon, F.~Jin, H.~De Raedt, K.~Michielsen},\\
    ``Gate-error analysis in simulations of quantum computers with transmon qubits'',\\
    \emph{Phys.~Rev.~A} \textbf{96}, 062302 (2017)
    \item[\cite{Willsch2018TestingFaultTolerance}]
    \textsc{D.~Willsch, M.~Willsch, F.~Jin, H.~De Raedt, K.~Michielsen},\\
    ``Testing quantum fault tolerance on small systems'',\\
    \emph{Phys.~Rev.~A} \textbf{98}, 052348 (2018)
  \end{enumerate}
\end{enumerate}

\vspace*{1.5cm}
\par\noindent\makebox[8cm]{\hrulefill}
\par\noindent\makebox[8cm][l]{Datum und Unterschrift}

%% file: ack.tex
\chapter{Acknowledgments}

There are many people to whom I am deeply grateful for supporting me during the
past three years (and also before that, of course).
\vfill\noindent
Kristel, first and foremost, I am immensely happy that you have given me the
opportunity to do my PhD in your group. I really appreciate the time you have
taken to guide me through this project, despite your full schedule of over 80
hours every week. I am truly grateful to you and Hans, you are great scientists
and I am genuinely looking forward to working together with you in the future
SDL at JSC so that we can make something out of all the interesting, unfinished
projects we have started recently. And Hans, besides simulating time-dependent
tantrum systems in D-dimensional Dilbert spaces (where clarity is an issue,
after all), I have not forgotten that we still need to meet at the
\emph{voormalig station Raeren}; we will definitely make it this year (you only
need to let me know a day in advance so that me and my grandfather's old bike
have a chance to make it in time). Toedeloe.
\vfill\noindent
David, thank you very much for proofreading my thesis and for the time you have
taken for discussions with me, especially during the final stage, and  for
always getting back to me with additional thoughts by email. I am excited to put
the results of our latest discussion on modeling electromagnetic environments to
work.
\vfill\noindent
Thomas, thanks a lot for finding the time to join my presentation on the
problems of quantum theory and for providing valuable ideas for discussion. I
very much appreciate you and Kristel offering me a position in the young quantum
computing SDL that is about to start at JSC.
\vfill\noindent
Seiji, thank you very much for the ideas you have given me during our regular
discussions that have taken place almost every, like, year. I recall that your
suggestions have helped me find an efficient scheme to manage all four-level
transformations on the supercomputer. Your comments were also very helpful in
finding the QUBO formulation of the SVM experiments that we did. The last time
we met was just a few weeks ago --- I hope your train was still coming, after 400
years.
\vfill\noindent
Special thanks go to my wife Madita. I truly appreciate that you spent so much
time proofreading my thesis, even though we both had less than a week left until
the final deadline,
and for spotting all those silly little errors that everyone else overlooked. I
thank you for the countless great moments that we have shared. Living, working,
and laughing with you makes me really happy.
\clearpage\noindent
Fengping, for you it's very simple --- it's also very hard. As our
kind-of-post-doc (I'm not saying \emph{senior researcher}), you have helped me
get into the business (also ITB) from the very start almost five years ago.  I
appreciate that, wherever we are, you always find the time to engage in funny
and stimulating discussions. I'm sure many friends and relatives of mine know
your stories by now even if they've never met you in person. ;)
\vfill\noindent
Hannes, thank you very much for bringing in new ideas and suggestions; I am
especially excited to see you continue to develop our simulation approach and
address many diverse situations that I have only barely been able to touch (if
at all).
\vfill\noindent
Manpreet, thanks a lot for your never-ending supply of lunch questions. I hope we
manage to keep up the good work, especially the BITB. By the way, I have hidden
some \emph{dips} in here.
\vfill\noindent
Carlos, you are a great addition to the group, especially when you \emph{have a
question}. I am really happy to be one of your \emph{quantum mechanics lunatics}
and I hope I can draw on your $\langle\!\langle$insert the right programming
language here$\rangle\!\rangle$ knowledge in the future again. Thank you!
\vfill\noindent
Vrinda, thanks for being a part of our group and bringing those delicious little
thingies (I forgot their name if I ever knew). You may not have noticed but I
was very happy when you asked me for help on \texttt{C++} stuff because this was
much more fun than what I had to do at the time (which I don't even recall).
\vfill\noindent
Miriam, thank you for
helping me with a lot of good advice on managing all that bureaucracy, and also
that you always almost \emph{instantly} managed to reply to my emails whenever
I had a question.
\vfill\noindent
I would also like to thank all colleagues from JSC for their friendliness and
technical know-how, and for providing a wonderful place to work. Special thanks
go to Thommy for rescuing my data when the file system played a trick on me, to
Sandipan, Benedikt, and Morris for their great courses on high-performance
programming and machine learning, to Gabriele for his recent work on qSVM, and
to Jenia for sharing his knowledge and experience in machine learning.
\vfill\noindent
Many thanks to Ioan and Dennis from KIT for many helpful discussions and the
first-hand knowledge about transmon experiments that you provided. I am glad you
could bring the qubit through its recent crisis, and I am very much looking
forward to continuing our cooperation.
\vfill\noindent
Thanks to Ren\'e for helpful discussions and for reading parts of my thesis. I
am sure we will set up some great workshops this year. Thanks also to your three
\emph{quantum schoolboys} Jakov, Jonathan, and Paul for having me keep myself up
to date on quantum annealing. You did a great job!
\clearpage\noindent
Marco and Berni, our \emph{Sauerbratengruppe} and \emph{Trauzeugen}, thanks a
lot for numerous bright and challenging discussions (in both space and time) and
countless great moments spent sometimes in far off places and sometimes in local
pubs around the corner. Thanks also to Nicolas for that gorgeous piece of
software (I don't think you will ever read this --- but Marco and Berni will
know what it's about).
\\\\
My sincere thanks go to my family, especially my parents, grandparents, and the
parents of my wife, for supporting us in every possible way, always trying to
get us back to life  when we were deeply lost in physics and work. Kelli, thank
you very much for reading my thesis and for always showing a keen interest in my
work. Chris, thanks a lot for reading and helping me with the most crucial
parts, as usual \textsf{:D}
\\\\
I would like to dedicate this work to the memory of my grandfather Willi, who
always, after some persuasion by grandma and me, enjoyed playing yet another
round of Skat with us. I thank you for being the first to supply me with
challenging games and complicated books on maths and physics (\emph{un ahle
Verzällcher us uns kölsche Sproch}) when I wasn't even able to read (let alone
understand) half of the words. I have always admired you.
\\\\
Finally, I would like to thank all my friends, especially  Andi,
Felix, Felix, Janni, Jojo, Kevin, Mel, Philipp, Philipp, Philipp, Philipp, and
all the others who have more diverse names ;) Thanks for the time
we spent together!
\\\\
Most of the simulations reported in this work were done on the supercomputers
JURECA and JUWELS. I gratefully acknowledge the computing time granted through
JARA on the supercomputer JURECA at Forschungszentrum Jülich, as well as the
Gauss Centre for Supercomputing e.V.~for funding this project by providing
computing time on the GCS Supercomputer JUWELS at Jülich Supercomputing Centre.
Furthermore, I acknowledge use of the IBM Q for this work; as usual, the views
expressed are my own and do not reflect the official policy or position of IBM
or the IBM Q team. This thesis was made in the context of the project ``Scalable
Solid State Quantum Computing'' of the Initiative and Networking Fund of the
Helmholtz Association.